%% file: thesis.tex
\pdfoutput=1
\documentclass[12pt, a4paper, headsepline, footsepline, oneside]{scrbook} % page headings in the middle

\input{Style/Style}
\makeglossary

\begin{document}

%\setpagewiselinenumbers
%\modulolinenumbers[5]
%\linenumbers

\pagenumbering{roman}
\input{Frontmatter/Title}
\input{Frontmatter/Abstract}

\input{Frontmatter/Acknowledgements}

\tableofcontents
\begingroup
%\chapter*{}\let\chapter=\section
%%\chapter*{Lists of Figures and Tables}\let\chapter=\section
\listoffigures
\listoftables
\endgroup

\pagenumbering{arabic}

\chapter{Introduction and Motivation} \label{chp:introduction}
 \input{Frontmatter/Introduction}

\part{Theoretical Aspects and Experimental Setup} \label{prt:setup}
\chapter{Theoretical Aspects} \label{chp:theory}
 \input{Theory/TheoryIntro}
 \section{The Standard Model} \label{sec:sm}
 \input{Theory/SM}
 \section{The Minimal Supersymmetric Extension of the Standard Model} \label{sec:susy}
 \input{Theory/SUSY}
\chapter{Experimental Setup} \label{chp:setup}
 \input{Setup/SetupIntro}
 \section{The Large Hadron Collider Accelerator Complex} \label{sec:lhc}
 \input{Setup/LHC}

 \section{The ATLAS Detector} \label{sec:atlasdetector}
 \input{Setup/ATLASDetector}

  %\subsection{Physics Goals of General Purpose Detectors at the LHC}
  %\input{Setup/PhysicsGoals}
  \subsection{Inner Detector} \label{ssec:innerdetector}
  \input{Setup/InnerDetector}

  \subsection{Calorimetry} \label{ssec:calorimetry}
  \input{Setup/Calorimetry}
  \subsection{Muon Spectrometer} \label{ssec:muonspectrometer}
  \input{Setup/MuonSpectrometer}

  \subsection{Trigger System} \label{ssec:trigger}
  \input{Setup/Trigger}
  \subsection{ATLAS Coordinate Frames} \label{ssec:frames}
  \input{Setup/Frames}
  \subsection{{\sc Athena}: the ATLAS Software Framework} \label{ssec:athena}
  \input{Setup/Athena}

\part{Search for Supersymmetry in Trilepton Final States} \label{prt:susy}
\chapter{Search for Supersymmetry in Trilepton Final States: Introduction}\label{chp:introSUSY}
 \input{IntroSUSY/IntroSUSY}
\chapter{Signal Signature and Backgrounds}\label{chp:signature}
 \input{Signature/Signature}

\chapter{Monte Carlo Samples Used}\label{chp:samples}
 \input{Samples/Samples}
\chapter{Preselection and Overlap Removal}\label{chp:preselection}
 \input{Preselection/Preselection}

\chapter{Signal Selection}\label{chp:selection}
 \input{Selection/Selection}
\chapter{Results}\label{chp:resultsSUSY}
 \input{ResultsSUSY/ResultsSUSY}

\chapter{Lepton Trigger Study}\label{chp:trigger}
 \input{Trigger/Trigger}
\chapter{Background Estimation Techniques}\label{chp:background}
 \input{Backgrounds/Backgrounds}
 \section{Classification of Systematic Uncertainty Sources} \label{sec:bgrClassification}
 \input{Backgrounds/Classification}
 \section{Assessing Instrumental Uncertainties} \label{sec:bgrInstrumental}
 \input{Backgrounds/UncertInstrumental}
 \section{Assessing Physics Uncertainties} \label{sec:bgrPhysics}
 \input{Backgrounds/UncertPhysics}
 \section{Secondary Leptons from $b$-Decays\newline Passing Isolation Criteria} \label{sec:secLept}
 \input{Backgrounds/SecondaryLeptons}
 \section{Estimation of Systematic Uncertainties}
 \input{Backgrounds/Systematics}
\chapter{Trilepton SUSY Search: Conclusion}\label{chp:conclusionSUSY}
 \input{Conclusion/Conclusion}
\part{The Alignment of the ATLAS Silicon Tracker} \label{prt:alignment}
\chapter{Introduction: Alignment at ATLAS} \label{chp:alignment}
 \input{Alignment/Alignment}

 \section{Hardware-Based Alignment: an Overview} \label{sec:alignmentHardware}
 \input{Alignment/AlignmentHardware}
 \section{Track-Based Alignment: an Overview} \label{sec:alignmentTracks}
  \input{Alignment/AlignmentTracks}
  \subsection{Important Track-Based Alignment Structures} \label{ssec:alignmentStructures}
  \input{Alignment/AlignmentStructures}
  \subsection{ATLAS Tracking Event Data Model} \label{ssec:trackingEDM}
  \input{Alignment/TrackingEDM}
\chapter{The \RA{} Algorithm} \label{chp:ra}
 \input{RA/RA}
 \section{Input to \RA{}: (Overlap) Residuals} \label{sec:residuals}
 \input{RA/Residuals}

 \section{Definition of the \RA{} Procedure} \label{sec:procedure}
  \input{RA/ProcedureRA}
  \subsection{Level 3} \label{ssec:l3}
  \input{RA/L3}

  \subsection{Pixel Stave Bow Alignment} \label{ssec:pixelStaveBow}
  \input{RA/PixelStaveBow}
  \subsection{Level 2} \label{ssec:l2}
  \input{RA/L2}
  \subsection{Level 1} \label{ssec:l1}
  \input{RA/L1}
  \subsection{\RA\ as an Iterative Procedure} \label{ssec:iterativeRA}
  \input{RA/IterativeRA}
  \subsection{Running of the \RA{} Algorithm} \label{ssec:running}
  \input{RA/Running}
  \subsection{Steering Options to \RA{}} \label{ssec:steering}
  \input{RA/Steering}

\chapter{Pixel End-Cap A Alignment with SR1~Cosmic Ray Data} \label{chp:pixelSR1}
 \input{PixelSR1/PixelSR1}
\chapter{Silicon Tracker Alignment with M8+~Cosmic Ray Data} \label{chp:m8plus}
 \input{M8plus/M8plus}
 \section{The Analysed Dataset} \label{sec:datasetM8plus}
 \input{M8plus/Dataset}

 \section{Selection of Residuals for Alignment} \label{sec:selectionRes}
 \input{M8plus/Selection}
  \subsection{Track Selection} \label{ssec:selectionTrack}
  \input{M8plus/SelTrack}
  \subsection{Hit Selection} \label{ssec:selectionHit}
  \input{M8plus/SelHit}

  \subsection{Residual Selection} \label{ssec:selectionRes}
  \input{M8plus/SelRes}
  \clearpage
 \section{The \RA\ Procedure for M8+} \label{sec:alignM8}
 \input{M8plus/Procedure}
  \subsection{Level 1 Alignment} \label{ssec:l1M8}
  \input{M8plus/L1}

  \clearpage
  \subsection{Level 2 Alignment} \label{ssec:l2M8}
  \input{M8plus/L2}

  \clearpage
  \subsection{Pixel Stave Bow Alignment} \label{ssec:pixelStaveBowM8}
  \input{M8plus/L4}

  \clearpage
  \subsection{Level 3 Alignment} \label{ssec:l3M8}
  \input{M8plus/L3}

 \section{Alignment Results with \RA\ in M8+} \label{sec:resultsM8}
 \input{M8plus/Results}

 \section{Validation of \RA\ Alignment Results in M8+} \label{sec:validationM8}
 \input{M8plus/Validation}
 \section{Differences in Residual and Overlap Residual Means\newline for $B$-field Off and On in M8+} \label{sec:B0minusB1M8}
 \input{M8plus/B0vsB1}
\chapter{Alignment: Conclusion and Outlook} \label{chp:conclusionAlignment}
\input{ConclusionAlign/ConclusionAlign}
%\chapter{Outlook:} \label{chp:outlook}
%\part{Conclusion} \label{prt:conclusion}

%\addcontentsline{toc}{chapter}{Glossary of Acronyms}
 %\setstretch{14cm}
 %\renewcommand\glossaryalignment{@{\hspace{\tabcolsep}\bfseries}lp{\descriptionwidth}p{12cm}}
 \setlength{\descriptionwidth}{0.85\linewidth}
 \printglossary
\addcontentsline{toc}{chapter}{Bibliography}
 \bibliography{bibliography}
\bibliographystyle{utphys} 

\appendix
 %\chapter{Implementation of \RA{} in Athena} \label{chp:implementation}
  %\input{Appendix/Implementation/Implementation}
  %\section{The Class Structure of the \RA\ Algorithm} \label{sec:classStructure}
  %\input{Appendix/Implementation/ClassStructure}
  %\section{\texttt{initialize()} Method} \label{sec:initialize}
  %\input{Appendix/Implementation/Initialize}
  %\section{\texttt{execute()} Method} \label{sec:execute}
  %\input{Appendix/Implementation/Execute}
  %\section{\texttt{finalize()} Method} \label{sec:finalize}
  %\input{Appendix/Implementation/Finalize}
 \chapter{Minimal Supergravity Sparticle Mass Spectra} \label{chp:spectra}
  \input{Appendix/Spectra/Spectra}
 \chapter{Complete Set of Pixel Stave Bow Fits} \label{chp:l4}
  \input{Appendix/L4/L4}

\end{document}

%% file: Style/Style.tex
\usepackage[latin1]{inputenc}
\usepackage{longtable}
\usepackage[sf,bf,small]{caption}
\usepackage{graphicx}
\usepackage{amsfonts}
\usepackage{amssymb}
\usepackage{amsmath}
\usepackage{epsfig}
\usepackage{color}
\usepackage{verbatim}
\usepackage[clearempty]{titlesec}
\usepackage{booktabs}
\usepackage{hhline}
\usepackage{array}
\usepackage{subfigure}
\usepackage{mathrsfs}
\usepackage{rotating}
\usepackage{enumerate}
\usepackage[urlcolor=black,   % farbige Hyperlinks
            colorlinks=false, % farbige Querverweise
            linkcolor=black,  % Blau als Farbe fuer Links
            bookmarks,       % Navigationsleiste im Acrobat
            bookmarksnumbered=true]{hyperref}
\usepackage{lineno}
\usepackage{multirow}
\usepackage[style=super,number=page,header=plain,border=none,cols=2,hypertoc=true,hyper=true]{glossary}
\usepackage{cancel}

\renewcommand{\floatpagefraction}{0.7}
\renewcommand{\textfraction}{0.1}
\newcommand{\cDist}{-0.8cm}
\newcommand{\cDistHalf}{-0.4cm}

\usepackage{multirow}

\newcommand{\bra}{\left\langle}
\newcommand{\ket}{\right\rangle}

\newcommand{\del}{\partial}
\newcommand{\eps}{\varepsilon}

\newcommand{\mum}{\ensuremath{\mu\textnormal{m}}}
\newcommand{\mm}{\ensuremath{\textnormal{mm}}}
\newcommand{\GeV}{\ensuremath{\textnormal{GeV}}}
\newcommand{\rad}{\ensuremath{\textnormal{rad}}}
\newcommand{\mrad}{\ensuremath{\textnormal{mrad}}}
\newcommand{\SU}{\ensuremath{\mathbb{SU}}}
\newcommand{\dif}{\ensuremath{{\rm d}}}

\newcommand{\Dzero}{D{\O}}

\newcommand{\ttbar}{\ensuremath{t\bar{t}}}

\newcommand{\chisq}{\ensuremath{\chi^{2}}}
\newcommand{\chisqNDF}{\ensuremath{\chi^{2}/N_{\rm DoF}}}
\newcommand{\met}{\ensuremath{/\!\!\!\!{E}_{\rm T}}}

\newcommand{\fb}{\ensuremath{{\rm fb}^{-1}}}

\newcommand{\lumi}{\ensuremath{\mathscr{L}}}
\newcommand{\intlumi}{\ensuremath{\int\!\!{\rm d}t\,\mathscr{L}}}
\newcommand{\instlumi}[1]{\ensuremath{\lumi=10^{#1}$\,cm$^{-2}$s$^{-1}}}
\newcommand{\Athena}{\textsc{Athena}}
\newcommand{\cha}{\ensuremath{\chi^\pm}}
\newcommand{\neu}{\ensuremath{\chi^0}}

\newcommand{\dR}{\ensuremath{\Delta R}}
\newcommand{\Lagr}{\ensuremath{\mathcal L}}
\newcommand{\mll}{\ensuremath{m_{\ell\ell}}}
\newcommand{\mossf}{\ensuremath{m_{\ell\ell}^{\rm OSSF}}}
\newcommand{\order}[1]{\ensuremath{\mathscr O(#1)}}
\newcommand{\Rsec}{\ensuremath{R_{\rm sec}}}
\newcommand{\sgnf}{\ensuremath{{\mathcal S}}}

\newcommand{\fdampover}{\ensuremath{f_{d}^{\rm ov}}}
\newcommand{\oneOverSqrt}{\ensuremath{{}^1\!\!\,{}/\!\!{}_{\sqrt2}}}
\newcommand{\oneOverTwo}{\ensuremath{{}^1\!\!\,{}/{}_{2}}}
\newcommand{\oneOverX}[1]{\ensuremath{{}^1\!\!\,{}/{}_{#1}}}
\newcommand{\threeOverTwo}{\ensuremath{{}^3\!\!\,{}/{}_{\!2}}}
\newcommand{\xOverY}[2]{\ensuremath{{}^{#1}\!\!\,{}/{}_{#2}}}
\newcommand{\qtrk}{\ensuremath{q^{\rm trk}}}
\newcommand{\qhit}{\ensuremath{q^{\rm hit}}}

\newcommand{\dqhit}{\ensuremath{\Delta q^{\rm hit}}}
\newcommand{\nclust}{\ensuremath{n_{\rm clust}}}
\newcommand{\incAngle}{\ensuremath{\theta_{\rm inc}}}
\newcommand{\incAngleT}{\ensuremath{\theta_{\rm T}}}
\newcommand{\lorentz}{\ensuremath{\alpha_{\rm Lorentz}}}
\newcommand{\pt}{\ensuremath{p_{\rm T}}}
\newcommand{\et}{\ensuremath{E_{\rm T}}}
\newcommand{\effPath}{\ensuremath{\ell_{\rm eff}}}

\newcommand{\RA}{{\sc Robust Alignment}}
\newcommand{\GX}{{\sc Global}~$\chi^2$}
\newcommand{\LX}{{\sc Local}~$\chi^2$}
\newcommand{\nill}{\ensuremath{0\!\!\!/}}
\newcommand{\rms}{\ensuremath{{\rm RMS}}}
\newcommand{\cov}{\operatorname{cov}}
\newcommand{\var}{\operatorname{var}}
\newcommand{\sigmaCut}{\ensuremath{A_{\sigma\mbox{-}\rm cut}}}
\newcommand{\rmean}[1]{\ensuremath{\langle r_{#1}\rangle}}
\newcommand{\rpull}[1]{\ensuremath{\frac{\langle r_{#1}\rangle}}{\delta r_{#1}}}
\newcommand{\ormean}[1]{\ensuremath{\langle o_{#1}\rangle}}
\newcommand{\rmeanstave}[1]{\ensuremath{\langle r_{#1}\rangle_{\rm stave}}}
\newcommand{\rsig}[1]{\ensuremath{\sigma(r_{#1})}}
\newcommand{\orsig}[1]{\ensuremath{\sigma(o_{#1})}}

\newcommand{\Mhalf}{\ensuremath{M_{\xOverY12}}}
\newcommand{\eOverP}{\ensuremath{\xOverY E p}}

%% file: Frontmatter/Title.tex
%%%%%%%%%%%%%%%%%%%%%%%%%%%%%%%%%%%%%%%%%%%%%%%%%%%%%%%%%%%%%%%%%%%%%%%%%%%%%
%                                                                           %
% Dieser File kann mit                                                      %
%                                                                           %
% \input instituts_deckblatt                                                %
%                                                                           %
% im Haupt-TeX-File aufgerufen werden.                                      %
%                                                                           %
% Er produziert vier Seiten mit den Seitennummern [0][0][0][0].             %
% Es mssen v i e r Seiten sein !                                           %
% Falls aufgrund eines gr�eren Titels (4 oder mehr Zeilen) mehr Seiten     %
% entstehen, muss man halt an den Zwischenr�men etwas drehen.              %
%                                                                           %
% Die zu ver�dernden Zeilen sind mit  %<--- gekennzeichnet.                %
%                                                                           %
%%%%%%%%%%%%%%%%%%%%%%%%%%%%%%%%%%%%%%%%%%%%%%%%%%%%%%%%%%%%%%%%%%%%%%%%%%%%%

%----------
% 1. Seite
%----------
\begin{titlepage}
    \begin{center}
        \font\GIANT=cmr17 scaled\magstep2
        {\GIANT U\kern0.8mm N\kern0.8mm I\kern0.8mm V\kern0.8mm %
         E\kern0.8mm R\kern0.8mm S\kern0.8mm I\kern0.8mm %
         T\kern0.8mm Y\kern5mm O\kern0.8mm F\kern5mm O\kern0.8mm 
	 X\kern0.8mm F\kern0.8mm O\kern0.8mm R\kern0.8mm D
          } \\[8mm]
        {\GIANT
         S\kern0.8mm u\kern0.8mm b\kern0.8mm -\kern0.8mm D\kern0.8mm
         e\kern0.8mm p\kern0.8mm a\kern0.8mm r\kern0.8mm t\kern0.8mm
	 m\kern0.8mm e\kern0.8mm n\kern0.8mm t\kern5mm
	 o\kern0.8mm f\kern5mm
	 P\kern0.8mm a\kern0.8mm r\kern0.8mm t\kern0.8mm i\kern0.8mm
	 c\kern0.8mm l\kern0.8mm e\kern5mm
	 P\kern0.8mm h\kern0.8mm y\kern0.8mm s\kern0.8mm i\kern0.8mm
	 c\kern0.8mm s
        } \\[20mm]
        {\Large \bf
          Search for Supersymmetry in Trilepton Final States\\
          with the ATLAS Detector and\\
          the Alignment of the ATLAS Silicon Tracker\\[10mm]
        }
     {\Large
     Oleg Brandt\\
     Corpus Christi College\\[3mm]
     \includegraphics[width=2cm]{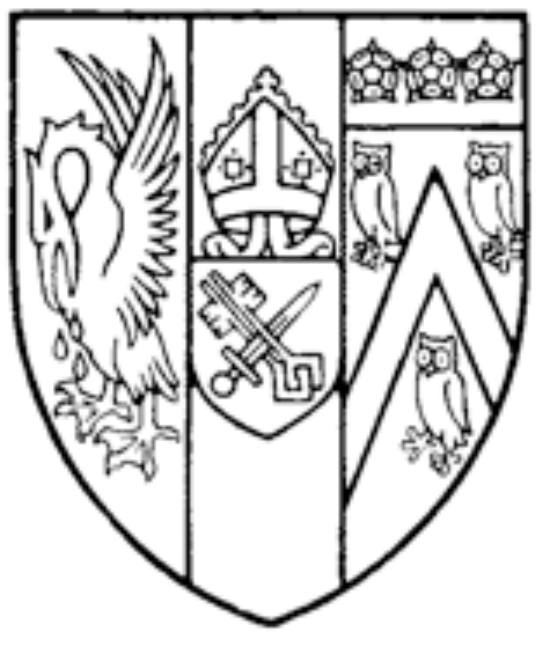}\\[5mm]
     \includegraphics[width=2cm]{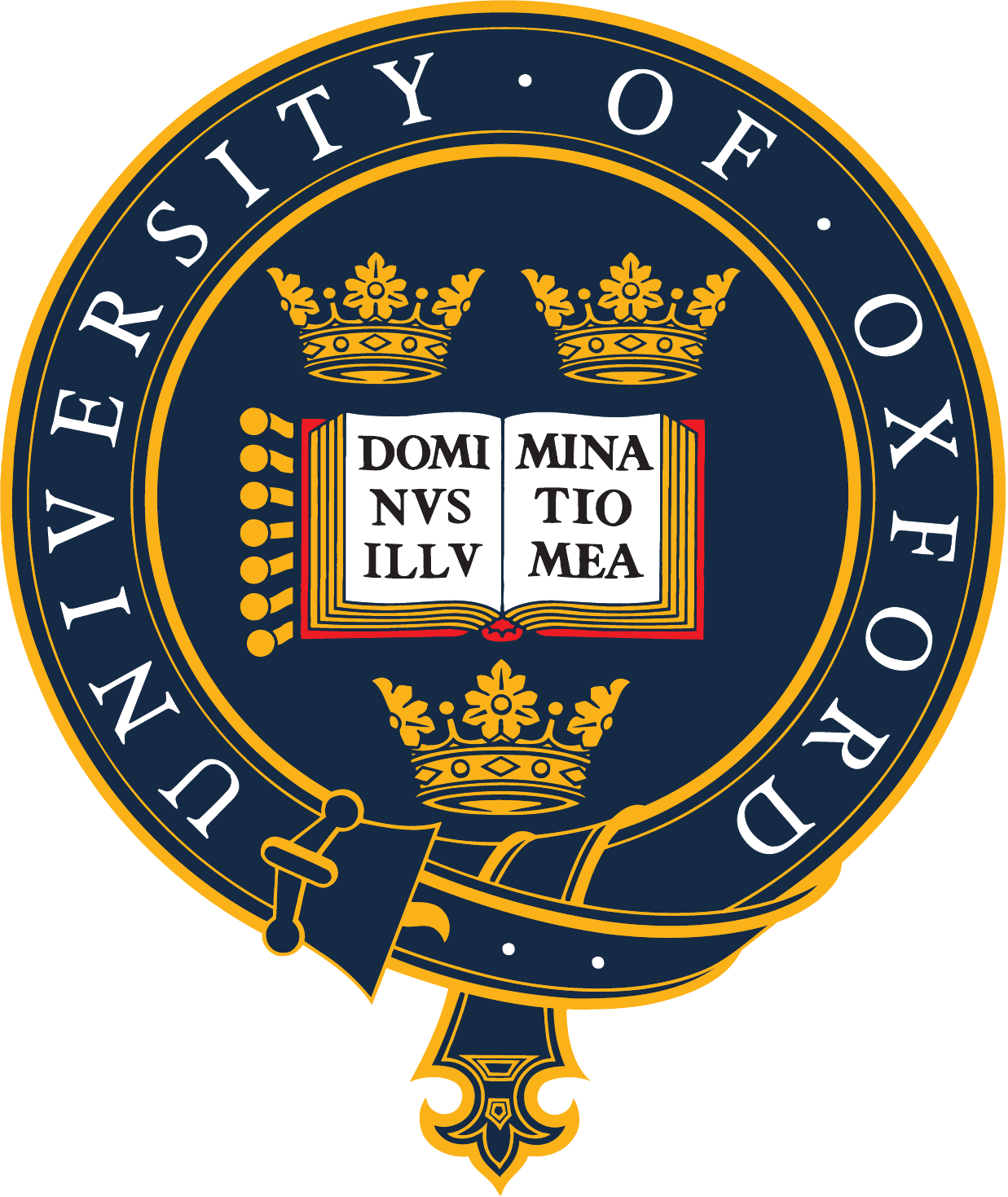}\\[10mm]
     Thesis submitted in fulfilment of the requirements\\
     for the degree of Doctor of Philosophy\\
     at the University of Oxford \\[7mm] Trinity Term, 2009 \\[10mm]
     %\\v0.9.0
     }  %<---
\end{center}

\begin{figure}[b]
    \begin{minipage}{\textwidth}
         \begin{raggedright}
         \begin{tabular}{@{}l@{}}
                 Postal address:  \\
                 Denys Wilkinson Building \\
                 Keble Road, Oxford \\
                 OX1 3RH, UK \\
         \end{tabular}
         \end{raggedright}
         \hfill
         \begin{raggedleft}
         \begin{tabular}{@{}l@{}}
             University of Oxford                 \\
             Sub-Dept. of Particle Physics        \\
             \today                        \\          %<---
         \end{tabular}
         \end{raggedleft}
      \end{minipage}
   \end{figure}
\end{titlepage}

\cleardoublepage

%% file: Frontmatter/Abstract.tex
%%%%%%%%%%%%%%%%%%%%%%%%%%%%%%%%%%%%%%%%%%%%%%%%%%%%%%%%%%%%%%%%%%%%%%%%%%%%%
%                                                                           %
% Dieser File kann mit                                                      %
%                                                                           %
% \input instituts_deckblatt                                                %
%                                                                           %
% im Haupt-TeX-File aufgerufen werden.                                      %
%                                                                           %
% Er produziert vier Seiten mit den Seitennummern [0][0][0][0].             %
% Es mssen v i e r Seiten sein !                                           %
% Falls aufgrund eines gr�eren Titels (4 oder mehr Zeilen) mehr Seiten     %
% entstehen, muss man halt an den Zwischenr�men etwas drehen.              %
%                                                                           %
% Die zu ver�dernden Zeilen sind mit  %<--- gekennzeichnet.                %
%                                                                           %
%%%%%%%%%%%%%%%%%%%%%%%%%%%%%%%%%%%%%%%%%%%%%%%%%%%%%%%%%%%%%%%%%%%%%%%%%%%%%

\thispagestyle{empty}

%----------
% 1. Seite
%----------
\begin{center}
   \font\GIANT=cmr17 scaled\magstep2
   {\GIANT U\kern0.8mm N\kern0.8mm I\kern0.8mm V\kern0.8mm %
    E\kern0.8mm R\kern0.8mm S\kern0.8mm I\kern0.8mm %
    T\kern0.8mm Y\kern5mm O\kern0.8mm F\kern5mm O\kern0.8mm 
    X\kern0.8mm F\kern0.8mm O\kern0.8mm R\kern0.8mm D
     } \\[8mm]
   {\GIANT
    S\kern0.8mm u\kern0.8mm b\kern0.8mm -\kern0.8mm D\kern0.8mm
    e\kern0.8mm p\kern0.8mm a\kern0.8mm r\kern0.8mm t\kern0.8mm
    m\kern0.8mm e\kern0.8mm n\kern0.8mm t\kern5mm
    o\kern0.8mm f\kern5mm
    P\kern0.8mm a\kern0.8mm r\kern0.8mm t\kern0.8mm i\kern0.8mm
    c\kern0.8mm l\kern0.8mm e\kern5mm
    P\kern0.8mm h\kern0.8mm y\kern0.8mm s\kern0.8mm i\kern0.8mm
    c\kern0.8mm s
   } \\[8mm]
   {\Large \bf
          Search for Supersymmetry in Trilepton Final States\\
          with the ATLAS Detector and\\
          the Alignment of the ATLAS Silicon Tracker\\[3mm]
   }
{\large
Oleg Brandt\\
Corpus Christi College\\[6mm]
{\large \sffamily{\bfseries{Abstract}}}\\\vspace{-1mm}%[-7mm]
}
\end{center}

\begin{small}
\noindent % 300 words maximum!
One of the main goals of the ATLAS detector at the Large Hadron Collider of CERN, a proton-proton collider with a nominal centre-of-mass energy of $\sqrt s = 14$ TeV, is to search for New Physics beyond the Standard Model. A widely favoured Beyond the Standard Model candidate is Supersymmetry (SUSY), which postulates a superpartner with the same quantum numbers, but a spin changed by $\xOverY12$ for each Standard Model particle. The {\em first part} of this thesis describes a strategy for an early discovery of SUSY using the trilepton signature, with a focus on gravity-mediated SUSY breaking, mSUGRA. The discovery potential for SUSY at the LHC for the case where strongly interacting supersymmetric particles are very massive is critically investigated. A possible choice of triggers for $\instlumi{31}$ is suggested by optimising the event yield at intermediate and final selection stages. A novel method to measure the rate of leptons from heavy flavour decays passing isolation requirements by isolating $\ttbar$ events in data is outlined.
\vspace{3mm}\newline
The task of the ATLAS silicon tracker is to track particles produced in proton-proton collisions in its centre, measuring their momenta and production vertices. The precise knowledge of the silicon tracker module positions and their orientation in space (alignment) down to some microns and fractions of a miliradian in the critical coordinates is of vital importance for large parts of the ambitious ATLAS physics program. In the {\em second part} of the thesis, the alignment of the ATLAS silicon tracker using the \RA\ algorithm and particle tracks is described. The algorithm is applied to align end-cap A of the pixel detector using cosmic ray particle tracks recorded during its on-surface commissioning in 2006. Finally, about 2M cosmic ray tracks collected by ATLAS in situ in autumn 2008 are utilised to provide a coherent alignment of the entire silicon tracker with the \RA\ algorithm. %The performance of the ATLAS silicon tracker with this alignment is critically investigated using the official ATLAS monitoring procedure.
\noindent
\end{small}

\begin{center}
{\large
Thesis submitted in fulfilment of the requirements\\
for the degree of Doctor of Philosophy\\
at the University of Oxford \\[4mm] Trinity Term, 2009 \\%[10mm]
}
\end{center}
\clearpage

%% file: Frontmatter/Acknowledgements.tex
%Acknowlegments
\thispagestyle{plain}
\begin{center}
%{\Large \emph{.}}\\
%{\Large \emph{.}}\\
\end{center}
\vspace{20mm}
\begin{flushright}
{
{\emph{Der Wissenschaftler muss durch sein Handeln immer wieder kund tun,\\ dass er zum humanen Teil der Menschheit geh\"ort.\\}}
\vspace{5mm}
{The scientist is to prove by his deeds\\ that he belongs to the human fraction of Mankind indeed.}\\
\vspace{5mm}
{\hspace{30mm}J. W. von G\"othe, ``Zur Farbenlehre''.}
}
\end{flushright}
%\end{center}
\vfill

\begin{flushleft}
This thesis was typeset with \LaTeXe.\\
\end{flushleft}
\vspace{8mm}
\copyright~Oleg~Brandt, 2008.\\
All rights reserved. No part of this publication may be reproduced, stored in a retrieval system, or transmitted, in any form or by any means, %%@
electronic, mechanical, photocopying, recording or otherwise, without express permission of the author.\\
Published at the University of Oxford, Oxford, United Kingdom.

\newpage
\thispagestyle{plain}
\noindent{\Large {\bfseries Acknowledgements}}\\

The last three years of research towards a D.Phil. degree certainly are among the most intense and interesting periods of my life. They shaped me in both scientific as well as in personal aspects. I would like to thank everyone who made these last three years so precious and worthwhile!

Firstly, I would like thank the Sub-Department of Particle Physics for accepting me into their excellent D.~Phil.\! in Particle Physics program, and their financial support throughout this period for attending innumerous meetings, workshops, conferences, and schools which is a highly important ingredient to scientific development. %In particular, I would like to thank for their support during the first months of my D.~Phil. 
I would also like to express my gratitude to the Deutscher Akademischer AustauschDienst (DAAD) for their generous financial and ideological support, and the Studienstiftung des Deutschen Volkes for their ideological support over the last 2.5 years. Last but not least I would like mention the Joint graduate Fellowship and the Senior Scholarship from Corpus Christi College, for which I am very grateful.

There are too many names to mention on the scientific side, and writing this just a couple of hours before submission I am sure I will forget to mention some -- please excuse my oversight.\\
In the first place, I would like to express my deep appreciation for the help and advice, innumerous highly interesting discussions, but also directness and personal support from my supervisors: Dr. Pawel Br\"uckman de Rentstrom and Dr. Alan James Barr. The same appreciation goes to Dr. Anthony Weidberg. They always were open for discussions and had an open ear for my questions. I would also like to thank Prof. Arnulf Quadt for his support during the initial orientation phase at Oxford, and Prof. Hans Kraus, my college advisor, for his steady encouragement. I highly value the engagement of Dr. Todd Huffman in student matters.\\
I owe a special word of gratitute to the 2006 ATLAS group students: Maria Fiascaris, Guillaume Kirsch, and Kristin Lohwasser. The tight collaboration with them, especially in our first year, was hightly fruitful and enjoyable at the same time. The same is true for Florian Heinemann, who helped me enormously to find a good starting point with the \RA\ algorithm.\\
Mentioning the Oxford ATLAS group, I cannot close this paragraph before mentioning very interesting discussions with Pierre-Hughes Beauchemin, Sinead Farrington, James Ferrando, Stephen Gibson, Chris Hays, Cigdem Issever, Richard Nickerson, and M\"uge Karag\"oz Unel.\\
I would like to thank my collaborators in the ATLAS Supersymmetry group lead by Giacomo Polesello, Davide Constanzo, and Paul de Jong. In particular, I would like to mention the names of people mainly involved in the trilepton analysis: Christina Potter, Katarina Peichel, Antonella de Santo. I appreciate the discussions with Tobias Golling, Chris Lester, Dan Tovey, and Giacomo Polesello.\\
I would also like to express gratitute to my collaborators in the Inner Detector alignment group lead by Salvador Marti i Garcia and Jochen Schieck. Particular thanks go to the Valencia group: Carlos Escobar, Sergio Gonzalez-Sevilla, Vicente Lacuesta, Regina Moles Valls; the Munich group: Giorgio Cortiana, Tobias G\"ottfert, and Roland H\"artel; John Alison and Andrea Bocci from the TRT group; Ben Cooper and Tobias Golling from the alignment monitoring group; Stephen Haywood from the Tracking Performance group; Anthony Morley for the tight collaboration on the TRT curvature constraint; and Daniel Froidevaux for keeping a watchful eye on the alignment activities.\\
My thanks go also to Sarah Allwood-Spiers, Ellie Dobson, Mark Owen for  the highly interesting collaboration in organising the Young Experimentalist and Theorist Institute 2009 (YETI), and to Nigel Glover and the IPPP for giving us this unique opportunity.\\
Last but not least, I would also like to thank our secretariat: Sue Geddes, Kim Proudfood, Laura Nevay and Faheem Khan, who were always extremely helpful.

My deepest gratitude goes to my family, who always supported me in my scientific and non-scientific endeavours. Same goes to my friends whom I don't want to mention here explicitly -- you know who you are!

\begin{comment}
\newpage
\vspace{3cm}
\subsubsection{Contact details:}
\begin{Large}
{\sf preferred:}\vspace{5mm}\\
{\tt o.brandt@physics.ox.ac.uk}\vspace{2mm}\\
{\tt obrandt@fnal.gov}\vspace{15mm}\\
{\sf postal address:}\vspace{5mm}\\
Oleg Brandt\vspace{2mm}\\
II. Physikalisches Institut\vspace{2mm}\\
Friedrich-Hund Platz 1\vspace{2mm}\\
37077 G\"ottingen\vspace{2mm}\\
GERMANY\\
\end{Large}
\end{comment}

%% file: Frontmatter/Introduction.tex
%\pagenumbering{arabic}
% \setcounter{page}{1}
\begin{quote}
$\nu\acute o\mu\omega\iota ~ \gamma\lambda\upsilon\kappa\acute\upsilon, ~ 
\nu\acute o\mu\omega\iota ~ \pi\kappa\rho\acute o\nu, ~ 
\nu\acute o\mu\omega\iota ~ \theta\varepsilon\rho\mu\acute o\nu, ~ 
\nu\acute o\mu\omega\iota ~ \psi\upsilon\chi\rho\acute o\nu, ~ 
\nu\acute o\mu\omega\iota ~ \chi\rho oi\acute\eta, ~ 
\newline
\textnormal{\`{}}\!\!\!\varepsilon\tau\varepsilon\tilde\eta\iota ~ \delta\grave\varepsilon ~ \textnormal{\`{}\!\!\'{}}\!\!\!\alpha\tau o\mu\alpha ~ \kappa\alpha\grave\iota ~ \kappa\varepsilon\nu\acute o\nu.$
\begin{flushleft}
\textit{By convention sweet, by convention bitter, by convention hot, by convention cold, by convention colour: but in
reality atoms and void.}
\end{flushleft}
\begin{flushright}
\footnotesize{Democritus, Fragments {\bf B125} (V-IV century b.C.)}
\end{flushright}
\end{quote}

For generations, Mankind is looking for answers on how our world is organised and what governs it in order to understand who we are by analysing \textit{our} reflection of the world. Evidence of ancestral cults indicating this continuous strive for explanations can be traced back to times as early as several tens of thousands of years ago. 
% It is remarkable that these dark ages of Mankind's birth were combined with the beginnings of the empirical observation of the surrounding environment.

A milestone to our modern view of the world was placed by Greek philosophers more than 3000 years ago. Besides bringing the idea of empiricism to a higher level, they contributed another essential element to Science as we know it today -- strict logic. An excellent example is the citation of Democritus above, who anticipated the main idea of Elementary Particle Physics by introducing the concept of ``the indivisible'' -- ``$\textnormal{\`{}\!\!\'{}}\!\!\!\alpha\tau o\mu o\varsigma$'' from the observation that stepstones would be abraded in not visible, infinitely small pieces.

This approach was carried to a scientific level by (post-) renaissance philosophers.
% or, as they were called, \textit{thinkers}, for example by Goethe, who besides his poetic oeuvre had also some articles published on the origin of color ("Farbenlehre"). 
For the first time experiments were {\em in\-ten\-tionally} and {\em systematically} designed to probe Nature. A milestone for the change of this paradigm is the works of Galileo Galilei. For instance, he derived the acceleration law $s=a/2\cdot t^2$ by measuring the acceleration due to Earth's gravitation using inclined surfaces and pendulums.

This naturally grown scientific approach has drastically changed our view of the world and our view of ourselves over the last millennia. The advancements of Science culminated in the great discoveries of the XX$^{\rm th}$ century, like the Theory of Relativity, Quantum Mechanics and Quantum Field Theory, the discovery of the role of the DNA, the ongoing investigation of the genome, and our furthered understanding of Universe's history to name a few.

% However, besides the crucial breakthroughs listed above the most intriguing question still remains: what are the most elementary building blocks our world is made of? Elementary Particle Physics attempts to answer this question. Of course, there are no final answers and, luckily, never will be. But now we have all tools in place to make another step in the search for them: over the XX-th century the so-called Standard Model of Elementary Particle Physics has emerged \cite{bib:weinberg, bib:glashow, bib:salam,bib:gross, bib:politzer, bib:gell-mann,bib:higgs}, which serves us tremendously well in interpreting experimental findings; on the other hand in the beginning of the 90-ies the Tevatron, the world's most powerful proton anti-proton collider with a center-of-mass energy of $\sqrt s=1.96\,$TeV, was launched.
% 
However, besides the crucial breakthroughs listed above the most intriguing question still remains: what are the most elementary building blocks our world is made of? Elementary Particle Physics attempts to answer this question. Of course, there is no final answer and, fortunately, never will be. 

Over the XX$^{\rm th}$ century the so-called Standard Model of Elementary Particle Physics has emerged \cite{bib:weinberg, bib:glashow, bib:salam,bib:gross, bib:politzer, bib:gell-mann,bib:higgs}, which served us tremendously well in interpreting experimental findings over the last decades. However, there are towering experimental indications that it is merely a low-energy approximation to yet another Beyond the Standard Model theory. A widely favoured Beyond the Standard Model candidate is Supersymmetry (SUSY), which postulates a superpartner with the same quantum numbers, but a spin different by $\xOverY12$ for each Standard Model particle. The Standard Model and its minimal supersymmetric extension are briefly reviewed in Chapter~\ref{chp:theory} of {\bfseries Part~\ref{prt:setup}} {\sc``\nameref{prt:setup}''} of this thesis. One of the main goals of the ATLAS detector at the Large Hadron Collider of CERN, a proton-proton collider with a nominal centre-of-mass energy of $\sqrt s = 14$ TeV, is to search for new physics Beyond the Standard Model. The Large Hadron Collider and the ATLAS detector are introduced in Chapter~\ref{chp:setup}.

{\bfseries Part~\ref{prt:susy}} {\sc``\nameref{prt:susy}''} of this thesis describes a strategy for an early discovery of SUSY using the {\em trilepton signature}, with a focus on gravity-mediated SUSY breaking, mSUGRA. The supersymmetric production of trilepton final states and Standard Model backgrounds are discussed in Chapter~\ref{chp:signature}. The discovery potential for SUSY at the LHC for the case where strong interacting supersymmetric particles are very massive -- the {\em massive sparton scenario} -- and at other benchmark points in mSUGRA phase space are critically investigated in:
\vspace{-3mm}
\begin{description}
\item[Chapter~\ref{chp:samples}:]
the description of simulated Monte Carlo event samples used in this analysis is given here;
\vspace{-1mm}
\item[Chapter~\ref{chp:preselection}:]
the preselection of physics objects: muons, electrons, and jets is discussed in this Chapter;
\vspace{-1mm}
\item[Chapter~\ref{chp:selection}:]
the selection of the supersymmetric signal using the trilepton signature is described. A special focus is placed on the massive sparton scenario;
\vspace{-1mm}
\item[Chapter~\ref{chp:resultsSUSY}:]
the discovery prospects of ATLAS for a discovery of supersymmetry with this trilepton search analysis are presented in this Chapter;
\vspace{-3mm}
\end{description}
A possible choice of triggers for $\instlumi{31}$ is suggested by optimising the event yield at intermediate and final selection stages, which is documented in Chapter~\ref{chp:trigger}. A~novel method to measure the rate of leptons from heavy flavour decays passing isolation requirements by isolating $\ttbar$ events in data is suggested in Chapter~\ref{chp:background}, alongside with an outline to measure other backgrounds to the trilepton search analysis from data. The SUSY part of the thesis is concluded in Chapter~\ref{chp:conclusionSUSY}.

The task of the ATLAS silicon tracker is to track particles produced in proton-proton collisions in its centre, measuring their momenta and production vertices. The precise knowledge of the silicon tracker module positions and their orientation in space{\em~(alignment)} down to some microns and fractions of a miliradian in the critical coordinates is of vital importance for large parts of the ambitious ATLAS physics program. 

In {\bfseries Part~\ref{prt:alignment}} {\sc``\nameref{prt:alignment}''} of the thesis, the alignment of the ATLAS silicon tracker using the \RA\ algorithm and particle tracks is described. After a general overview of alignment techniques at ATLAS in Chapter~\ref{chp:alignment}, the \RA\ algorithm is described in Chapter~\ref{chp:ra}. In particular, novel alignment techniques for a coherent alignment of parts of the silicon tracker by using topological hit--track residual distributions are introduced. The \RA\ algorithm is applied to align end-cap A of the pixel detector using cosmic ray particle tracks recorded during its on-surface commissioning in 2006, which is described in Chapter~\ref{chp:pixelSR1}. In Chapter~\ref{chp:m8plus}, about 2M cosmic ray tracks collected by ATLAS in situ in autumn 2008 are utilised to provide a coherent alignment of the entire silicon tracker with the \RA\ algorithm. %The performance of the 
The alignment part of the thesis is concluded in Chapter~\ref{chp:conclusionAlignment}.

Given the space limitations for this thesis and in order to preserve coherence, some parts of research done by the author in the last three years are not documented here. The most prominent deals with the so-called ``weak mode'' deformations of the silicon tracker, which can systematically bias the track parameter reconstruction. Two methods to detect and eliminate any bias on the track curvature measurement $\frac q{\pt}$ were elaborated, implemented, and validated by the author in collaboration with P.~Br\"uckman de~Rentstrom, B.~Cooper and A.~Morley. This is documented in~\cite{bib:weakModes}.\vspace{\cDistHalf}

\subsubsection{Author's Contribution}\vspace{\cDistHalf}
As is the requirement, the author's contribution is explicitly listed below. Unfortunately, the usage of some technical terms not introduced at this stage is unavoidable. For explanation refer to the glossary and the main text body of the thesis.\vspace{2mm}\\
{\sffamily{\bfseries Supersymmetry Part:}}
\vspace{-2mm}
\begin{itemize}
\item
The set-up of the preselection for the trilepton search analysis and its further development in \Athena\ release 12. Additional studies on overlap removal between physics objects like e.g. electrons and jets, cf.~Chapter~\ref{chp:preselection};
\item
The set-up of the selection for the tri-lepton search analysis and its co-development with other memebers of the ATLAS SUSY group. Suggestion of the baseline method to select the opposite sign same flavour lepton pair. Most relevantly, the introduction of a tight track isolation requirement, which improved the statistical significance by \order{10}\ in the part of mSUGRA phase space where the trilepton search analysis is particularly sensitive (id est SU2 and direct gaugino pair-production), cf.~Chapter~\ref{chp:selection};
\item
All the results shown in Chapter~\ref{chp:resultsSUSY};
\item
The study aimed at selecting a possible trigger strategy for the \instlumi{31}\ trigger menu, cf.~Chapter~\ref{chp:trigger};
\item
The background estimation techiques outlined in Sections~\ref{sec:bgrInstrumental} and~\ref{sec:bgrInstrumental} %were elaborated by the author 
using valuable input from discussions with P.~Br\"uckman de~Rentstrom and A.~J.~Barr;
\item
The method to measure the rate of secondary leptons from heavy flavour decays passing isolation criteria by isolating \ttbar\ events in data in Section~\ref{sec:secLept} after an initial discussion with A.~J.~Barr;
\end{itemize}
\vspace{-2mm}
{\sffamily{\bfseries Alignment Part:}}
\vspace{-2mm}
\begin{itemize}
\item
The \RA\ procedures to coherently align parts of the silicon tracker (``superstructure alignment''), their implementation and validation, as described in Chapter~\ref{chp:ra}. The overlap residual treatment was completely re-designed, as the existing procedure was prone to systematic biases, cf.~Chapter~\ref{chp:ra}. Some technical help in producing track-hit residual distributions for initial studies and alignment results monitoring was received from M.~Ahsan and S.-M.~Wang.;
\item
The alignment of the end-cap~A of the pixel detector using on-surface cosmic ray data, cf.~Chapter~\ref{chp:pixelSR1};
\item
The alignment of the ATLAS silicon tracker in the cavern using the \RA\ algorithm on cosmic ray data collected by ATLAS in autumn 2008:
	\begin{itemize}
	\item
Selection of tracks, hits, and hit--track residuals for the alignment procedure (in particular the implementation of the {\tt InDetAlignHitQualSelTool}), cf.~Section~\ref{sec:selectionRes};
	\item
	The actual alignment procedure: running the \RA\ algorithm, production and invesitagation of monitoring plots. Valuable input from P.~Br\"uckman de~Rentstrom and the ATLAS Inner Detector alignment community was received. Cf.~Sections~\ref{sec:alignM8},~\ref{sec:resultsM8}, and~\ref{sec:B0minusB1M8}.
	\end{itemize}
\vspace{-2mm}
\end{itemize}

%% file: Theory/TheoryIntro.tex
To our best experimental knowledge, %\footnote{``Knowledge'' in this context refers to experimentally proven results.}, 
the world is built of fundamental particles which are governed by four basic types of interactions. They are organised\footnote{other than the gravitational interaction.} in a scheme described by the so-called Standard Model of Elementary Particle Physics (SM\glossary{name=SM,description=Standard Model of Elementary Particle Physics}). A brief review of the SM shall be given in the following. There are indications that the SM is merely a low-energy approximation of yet another theory to be discovered. The Minimal Supersymmetric extension to the Standard Model (MSSM\glossary{name=MSSM,description=The Minimal Supersymmetric extension to the SM}), which postulates a so-called superpartner for each of the SM particles, is a promising candidate. It is briefly introduced in the second part of this section. Finally, the mechanisms to produce trilepton final states within the framework of the MSSM are reviewed.
%It will be subject of discussion in the following section, combined with a broad overview of the experimental techniques. 

%% file: Theory/SM.tex
Over the last decades, the SM has served us remarkably well as a description for the world's most
fundamental known processes. It emerged in the course of the last century, culminating in two hot
phases: in the 60's and 70's, as well as recently from about 1995 on. A wealth of canonical literature is available, for example \cite{bib:peskin,
bib:halzen, bib:griffith, bib:herrero}. After a brief review of the SM in the following, its known shortcomings are summarised.
%Therefore, only a brief overview of the
%Standard Model with references shall be given in the following paragraph. 

\subsection{Brief Overview of the Standard Model}
The SM describes the elementary matter particles observable in our world alongside three basic interactions governing them: {\em strong}, {\em weak}, and {\em electromagnetic} interactions. As of now, there is no canonical way to include gravitational interaction in the SM.

\begin{table}[tb]
  \centering
  \begin{tabular}{llccc}
    \hline
    \vspace{3pt}
      &  & \multicolumn{3}{c}{Generation (fermions only)}\\
      &  & {\bf I} & {\bf II} & {\bf III} \\
    \hline
    \multirow{4}{*}{{\bf Fermionic Sector:}}
                        & \multirow{2}{*}{leptons:}
                                    & $\nu_e$ (1953)  & $\nu_\mu$ (1962)  & $\nu_\tau$ (2000) \\
                        &           & $e$ (1897)      & $\mu$ (1936)      & $\tau$ (1975) \\
    \cline{2-5}
                        & \multirow{2}{*}{quarks:}
                                    & $u$ (1968)      & $c$ (1974)        & $t$ (1995) \\
                        &           & $d$ (1968)      & $s$ (1964)        & $b$ (1977) \\
    \hline
    \multirow{3}{*}{{\bf Gauge Sector:}}
                        & gluons:   &  & $g_1,...,g_8$ (1979) &  \\
                        & photon:   &  & $\gamma$ (1900)      &  \\
                        & EW massive bosons: &  & $W^{\pm}, Z^0$ (1983)&  \vspace{3pt} \\
    \hline
  \end{tabular}
\caption[The scheme of elementary particles described by the SM]{\label{tbl:sm}
The scheme of elementary particles described by the Standard Model. In parentheses, the year of discovery is given. % \cite{bib:halzen, bib:griffith, bib:top_discovery_d0,bib:top_discovery_cdf}. 
The postulated Higgs particle is not shown, since it has not been discovered yet.
\vspace{\cDistHalf}
}
\end{table}

\subsubsection{The Gauge Sector of the Standard Model}
The SM is a quantum field theory based on the principle of local gauge invariance, which, starting from the $\SU_C(3)\times\SU_L(2)\times\mathbb{U}_Y(1)$ symmetry, yields a formalism for the description of {\em strong}, {\em weak}, and {\em ElectroMagnetic} (EM\glossary{name=EM,description=ElectroMagnetic}) interactions in a natural way \cite{bib:weinberg, bib:glashow, bib:salam}. These interactions are mediated by force carriers called {\em gauge} bosons, which are %eigenstates of the field constructed to preserve gauge invariance. 
introduced to restore local gauge invariance. The gauge bosons are: eight gluons %\footnote{To be precise, the theory features nine gluons, but one of them must remain colourless and is thus irrelevant.}
for $\SU_C(3)$ and the colour charge gauge field associated with it; plus the $W^\pm$,\,$Z$,\,$\gamma$ bosons for $\SU_L(2)\times\mathbb{U}_Y(1)$ ElectroWeak (EW\glossary{name=EW,description=ElectroWeak}) interactions. All force carriers have an integer non-zero spin\footnote{Throughout this document, natural units are used: $\hbar\equiv c\equiv1$.}. Gauge bosons of the strong and the EW interaction have spin 1. This is why they are sometimes referred to as {\em vector} bosons. It is expected that the yet undiscovered graviton -- the gauge boson of the gravitational force -- is a tensor particle with a spin of 2.

\subsubsection{The Fermionic Sector of the Standard Model}
%The particles of the Standard Model can be divided up into two distinct groups with respect to their role in the theory. 
Besides the gauge sector, all remaining particles of the SM with the exception of the Higgs boson belong to the so-called fermionic sector. They all have spin $\xOverY12$.

%Fermions participate in at least one of the interactions provided by the Bosonic Sector. 
In the SM framework, fermions are organised in a scheme with respect to their masses and the interactions in which they can participate. Firstly, there are the {\em quark} and the {\em lepton} sectors. While quarks carry colour charge, leptons do not. Thus, the former participate in strong and EW interactions, and the latter can undergo only EW processes. Both quark and lepton sector particles can be organised in two categories regarding their electric charge: quarks can carry the charge $+\xOverY23$ or $-\xOverY13$, leptons $-1$ or $0$. Electrically neutral leptons are called neutrinos. In both fermionic sectors, there are three pairs of particles, called {\em generations}, which are ordered by increasing mass. In turn, every generation has two chiral manifestations: the {\em left-handed} and {\em right-handed} one. Only left-handed particles can participate in weak interactions via $W^{\pm}$ bosons. Both left-handed particles in a given quark or lepton generation are assigned a so-called {\em weak-isospin} quantum number, identifying them as partners of each other with respect to the weak interaction. Each particle of the fermionic sector has an {\em antiparticle}, featuring the same mass, but opposite internal quantum numbers like charge.

The particles of the Standard Model are summarised in Table~\ref{tbl:sm}. Fundamental publications on the unification of the weak and the EM interaction to the EW interaction placed the cornerstone of the Standard Model in the 60's~\cite{bib:weinberg, bib:glashow, bib:salam}. The theory of the strong interacion, Quantum ChromoDynamics (QCD\glossary{name=QCD,description=Quantum ChromoDynamics}), was formulated in the 70's \cite{bib:gross, bib:politzer, bib:gell-mann}. The theoretical framework of the SM is reviewed in~\cite{bib:peskin, bib:halzen, bib:griffith, bib:herrero} and many more.

\subsubsection{Electroweak Symmetry Breaking in the Standard Model}
The $\SU_C(3)\times\SU_L(2)\times\mathbb{U}_Y(1)$ symmetry is not a symmetry of the vacuum: for example $W^\pm$ and $Z$ bosons are massive in contrast to the photon. Similarly, particles of the Fermionic Sector have differing masses.
% into $\SU_C(3)\times\mathbb{U}_{\rm em}(1)$. 
One of the theoretical approaches to generate the mass spectrum of SM particles via Spontaneous Symmetry Breaking is the introduction of the so-called Higgs field. This field couples to the SM particles via its excitation quanta, the Goldstone bosons, as suggested by P.~Higgs~\cite{bib:higgs} et al. One of the Goldstone bosons -- the Higgs boson -- remains physical and can in principle be detected.  In the framework of the SM, it must be a non-charged scalar boson. Its existence remains to be experimentally proven, however its mass can be inferred from precision measurements of other parameters of the SM: for example the mass of $W$ boson. It receives logarithmic contributions from the mass of the hypothesised Higgs boson, and quadratic ones from the mass of the top quark, which is now known to a precision of better than 1\%~\cite{bib:pdg2008}. The concept of Spontaneous Symmetry Breaking was introduced by Ginzburg and Landau in the context of superconductivity \cite{bib:landau_ginzburg}.

\subsection{The Shortcomings of the Standard Model}
The SM has served us tremendously well over the last decades in explaining the observed phenomena. However, evidence is mounting that it is merely a low-energy approximation of yet another Beyond the Standard Model~(BSM\glossary{name=BSM,description=Beyond the Standard Model}) theory. Most prominent of these observations are the measurement of the cosmic microwave background with WMAP and the resulting estimation of the proportion of dark matter in Universe~\cite{bib:wmap,bib:wmap2}, the determination of the muon anomalous magnetic moment ($g_\mu-2$)~\cite{bib:pdg2008}, and the branching ratio $BR(b\rightarrow s\gamma)$~\cite{bib:ellisOlive}. %the tension in the $m_{\rm top}$-$m_W$~\cite{bib:lepEWWG} fit for the SM Higgs mass. 

Among {\em theoretical} shortcomings of the SM are: the absence of an explanation for gravitation, the unknown source for neutrino masses, and arguably the lack of unification of gauge couplings at any common scale, an appealing concept for a Grand Unified Theory. 

%The major problem of the Standard Model is arguably the Hierarchy problem: one-loop quantum mechanical corrections to the Higgs boson mass are quadratically divergent. If we assume that no BSM physics occurs up to the Planck scale, the corrections are some 30 orders of magnitude larger than the actual Higgs boson mass. 

The major problem of the Standard Model is arguably the Hierarchy problem: one-loop quantum mechanical corrections to the squared Higgs boson mass are quadratically proportional to the ultraviolet cut-off scale $\Lambda_{\rm{UV}}$ at which BSM physics enters. If we assume that no BSM physics occurs up to the Planck scale $M_P=(8\pi G_{\rm Newton})^{-1/2}\simeq2.4\times10^{18}$\,GeV, the corrections are some 60 orders of magnitude larger than the actual squared Higgs boson mass.%This immediately raises questions of naturalness and fine tuning.

% \subsection{Standard Model Physics at Hadron Colliders}
% The basis of experimental Particle Physics is the equivalence of energy and mass. Basically speaking, this concept is
% used in transforming the kinetic energy of colliding particles into energy in form of mass of newly produced
% particles,
% whose decay products can be consequently be studied with detectors. Besides of experiments employing particles
% originating from space (astroparticles), modern High Energy Particle Physics knows two types of experimental setup:
% electron and hadron colliders. In the former, electrons are accelerated and brought to collision, the latter use
% hadrons. Electron colliders have several conseptual advantages of  \cite{bib:fernow}

%% file: Theory/SUSY.tex
A promising BSM candidate which offers an elegant and \ae{}sthetic solution to many problems of the Standard Model is the Minimal Supersymmetric extension\footnote{In literature, the MSSM is often referred to as ``supersymmetry'' or SUSY\glossary{name=SUSY,description=SUperSYmmetry. In the context of this thesis it is synonym with the MSSM unless stated otherwise}, although strictly speaking supersymmetry does not impose the restriction of {\em one} superpartner per SM particle. In this thesis, the terms MSSM and supersymmetry will be used as synonyms unless stated otherwise.} to the Standard Model (MSSM)~\cite{bib:susy,bib:susy2}. It postulates a supersymmetric partner for each SM particle called ``sparticle'' or ``gaugino''\footnote{In this thesis, a superpartner naming scheme is adopted where: an ``s'' is prefixed to names of particles belonging to the fermionic sector of the SM; and ``ino'' is appended to names of particles belonging to the bosonic sector of the SM. For example, the superpartner of the electron is called {\bf s}electron, and the superpartner of the $W$ boson is the W{\bf ino}.} and denoted as $\tilde p$ for particle $p$. 
Matters are more complicated in the Higgs sector, which has to be extended to host five scalar SM Higgs particles (plus three Nambu-Goldstone bosons providing for EW symmetry breaking) and two supersymmetric Higgs doublets.
%The exception is the Higgs sector, where five Higgs particles with four supersymmetric counterparts are predicted. 
The superpartners of leptons, quarks, and gauge bosons are postulated to have respectively the {\em same} quantum numbers, and only the spin differs by $\xOverY12$. This assumption naturally solves the Hierarchy problem, since quadratic contributions to the Higgs boson mass are cancelled by the superpartners. For example, for a fermion $f$ and a sfermion $\tilde f$ one obtains:
\begin{equation}\label{eqn:hierarchy1}
\Delta m_h^2=-\frac{1}{8\pi^2}\left(\tilde\lambda_f-|\lambda_f|^2\right)\cdot\Lambda_{\rm{UV}}^2+...\,,
\end{equation}
where $\tilde\lambda_f~(\lambda_f)$ are the Yukawa coupling constants of sfermion (fermion) to the Higgs field. Note, that the relative sign between the $\tilde\lambda_f$ and $\lambda_f$ terms is due to the different spin nature of fermions and sfermions. According to the assumption of same quantum numbers
\begin{equation}\label{eqn:hierarchy2}
\tilde\lambda_f \equiv |\lambda_f|^2\,,
\end{equation}
and thus trivially:
\begin{eqnarray*}
\Delta m_h^2 &=& -\frac{1}{8\pi^2}\,\cancelto{_0}{ \left(\tilde\lambda_f-|\lambda_f|^2\right) }\cdot\Lambda_{\rm{UV}}^2+...\\
             &=& 0+...\,.
\end{eqnarray*}
The supersymmetric partners of SM particles are summarised in Tables~\ref{tab:susyFermion} and~\ref{tab:susyGauge}.

\begin{table}
\begin{center}
\begin{tabular}{l|c|c}
\hline
Name & Spin $0$& Spin $\xOverY12$\\
\hline\hline
squarks, quarks ($\times 3$ families) & $(\tilde{u}_L\;\tilde{d}_L),\,\tilde{u}_R,\,\tilde{d}_R$ & $(u_L\;d_L),\,u_R,\,d_R$\\
\hline
sleptons, leptons ($\times 3$ families) & $(\tilde{\nu}\;\tilde{e}_L),\,\tilde{e}_R$ & $(\nu\;e_L),\,e_R$\\
\hline
Higgs, higgsinos & $(H_u^+\;H_u^0),\,(H_d^0\;H_d^-)$ & $(\tilde{H}_u^+\;\tilde{H}_u^0),\,(\tilde{H}_d^0\;\tilde{H}_d^-)$\\
\hline
\end{tabular}
\end{center}
\vspace{\cDistHalf}
\caption{\label{tab:susyFermion}
Supersymmetric partners of the fermionic sector of the SM together with the MSSM Higgs sector.}
\end{table}

\begin{table}
\begin{center}
\begin{tabular}{l|c|c}
\hline
Name & Spin $\xOverY12$& Spin $1$\\
\hline\hline
gluino, gluon & $\tilde{g}$ & $g$\\
\hline
winos, $W$ bosons & $\tilde{W}^{\pm},\,\tilde{W}^0$ & $W^{\pm},\,W^0$\\
\hline
bino, $B$ boson& $\tilde{B}^0$ & $B^0$\\ 
\hline
\end{tabular}
\end{center}
\vspace{\cDistHalf}
\caption{\label{tab:susyGauge}
Supersymmetric partners of the gauge sector of the SM.}
%\vspace{\cDistHalf}
\end{table}

The assumption of same quantum numbers in particular implies the equality of masses, for example $m_{\tilde e_L}=m_{\tilde e_R}=m_e=511$\,keV. Since no supersymmetric particles have been observed so far, supersymmetry must be broken. In order to preserve Equation~\ref{eqn:hierarchy2} important for resolving the Hierarchy problem, we are led to consider an \textit{effective} supersymmetric Lagrangian of the form
\begin{equation*}
 \Lagr=\Lagr_{\rm SUSY}+\Lagr_{\rm soft}\,.
\end{equation*}
Here, $\Lagr_{\rm SUSY}$ represents all gauge and Yukawa interactions with exact supersymmetry invariance, and $\Lagr_{\rm soft}$ contains supersymmetry-violating additional terms with a {\em positive} mass dimension. It can be shown~\cite{bib:susy}, that this gives an at most logarithmic divergence
\begin{equation}\label{eqn:msoft}
 \Delta m_h^2 = m_{\rm soft}^2\cdot\left(\frac{\lambda}{16\pi^2}\ln(\Lambda_{\rm UV}/m_{\rm soft})+...\right),
\end{equation}
where $m_{\rm soft}$ is the mass scale associated with the $\Lagr_{\rm soft}$ part of the Lagrangian. %In case of the MSSM, this results in 105 free parameters~\cite{bib:susy} for $\Lagr_{\rm soft}$.

An elegant mechanism for spontaneous soft supersymmetry breaking is proposed by supergravity (SUGRA\glossary{name=SUGRA,description=Supergravity}). It postulates SUSY breaking in a ``hidden'' sector at a scale of around 10$^{10}$\,GeV, which is mediated to the ``visible'' sector by flavour-blind gravitational interactions.  The sparticle content of MSSM is shown in Table~\ref{tab:sugraParticles}.

\begin{table*}
\begin{center}
\begin{tabular}{l|c|c|c}
\hline
Names & Spin & Gauge eigenstates & Mass eigenstates\\
\hline\hline
Higgs bosons&$0$&$H_u^0\;H_d^0\;H_u^+\;H_d^-$&$h^0\;H^0\;A^0\;H^\pm$\\
\hline
&&$\tilde{u}_L\;\tilde{u}_R\;\tilde{d}_L\;\tilde{d}_R$&$\sim$ same\\
squarks&$0$&$\tilde{s}_L\;\tilde{s}_R\;\tilde{c}_L\;\tilde{c}_R$&$\sim$ same\\
&&$\tilde{t}_L\;\tilde{t}_R\;\tilde{b}_L\;\tilde{b}_R$&$\tilde{t}_1\;\tilde{t}_2\;\tilde{b}_1\;\tilde{b}_2$\\
\hline
&&$\tilde{e}_L\;\tilde{e}_R\;\tilde{\nu}_e$&$\sim$ same\\
sleptons&$0$&$\tilde{\mu}_L\;\tilde{\mu}_R\;\tilde{\nu}_{\mu}$&$\sim$ same\\
&&$\tilde{\tau}_L\;\tilde{\tau}_R\;\tilde{\nu}_{\tau}$&$\tilde{\tau}_1\;\tilde{\tau}_2\;\tilde{\nu}_{\tau}$\\
\hline
neutralinos&$\xOverY12$&$\tilde{B}^0\;\tilde{W}^0\;\tilde{H}_u^0\;\tilde{H}_d^0$&$\tilde\chi_1^0\;\tilde\chi_2^0\;\tilde\chi_3^0\;\tilde\chi_4^0$\\
\hline
charginos&$\xOverY12$&$\tilde{W}^{\pm}\;\tilde{H}_u^+\;\tilde{H}_d^-$&$\tilde\chi_1^{\pm}\;\tilde\chi_2^\pm$\\
\hline
gluino&$\xOverY12$&$\tilde{g}$&$\sim$ same\\
\hline
gravitino&$\xOverY32$&$\tilde{G}$&$\sim$ same\\
\hline
\end{tabular}
\end{center}
\vspace{\cDistHalf}
\caption[The sparticle content of the MSSM]{\label{tab:sugraParticles}
The sparticle content of the MSSM and the extended SM Higgs sector assuming no sfermion mixing for the first two families~\cite{bib:susy}. Charginos and neutralinos are collectively referred to as {\em gauginos}.}
\vspace{\cDistHalf}
\end{table*}

The supergravity parameter space is far too large to be experimentally investigated in its entirety. Therefore, a phenomenologically simplified model has been introduced, the minimal version of supergravity (mSUGRA\glossary{name=mSUGRA,description=Minimal supergravity}). It is characterised by mass unification at the so-called GUT Scale, defined as the common crossing point of the running coupling constants, $M_{\rm GUT}\simeq2\times10^{16}$\,GeV. This reduces the number of parameters governing mSUGRA to five values: 
%\vspace{-2mm}
\begin{itemize}
\vspace{-2mm}
\setlength{\itemsep}{0mm}
 \item $M_0$: the universal scalar soft mass term;
 \item \Mhalf: the uniform fermion soft mass term;
 \item $A_0$: the uniform tri-linear coupling;
 \item $\tan\beta$: the ratio of vacuum expectation values of the two Higgs doublets;
 %for the $\mathcal{CP}$-odd and $\mathcal{CP}$-even Higgs doublets, 
 \item $\arg\mu$: the sign of the Higgs mixing parameter $\mu$. 
\vspace{-2mm}
\end{itemize}
There have been many phenomenological efforts to investigate the implications of recent precision measurements~\cite{bib:ellisOlive,bib:msugra1,bib:msugra2,bib:msugra3,bib:msugra4} on the mSUGRA parameter space.

%Being favoured by theorists, mSUGRA is the focus of this analysis. However, one may not forget that mSUGRA is merely an additional constraint on SUGRA, reducing the number of free parameters from 105 to 5. Surely, this assumption of unified parameters at the GUT scale is an \ae{}sthetically appealing one, but nature might have chosen a different, more complex path. From this point of view, mSUGRA is nothing more but a simplification of the SUGRA model to enhance the ease of classification of possible scenarios. Characterisic points in the mSUGRA parameter space will reproduce basic features of that particular class of SUGRA models, but one should restrain from believing them in detail.

Lepton $L$ and baryon $B$ quantum number conservation is a strong phenomenological constraint on any extension to the SM, since we know experimentally that the lifetime of the proton $\tau_{\rm proton}>10^{31}$\,s~\cite{bib:pdg2008}. For the MSSM, rather than imposing $L$ and $B$ conservation by hand, an additional symmetry, $R$-parity, is usually introduced:
\begin{equation}
 P_R\equiv(-1)^{2S+3(B-L)}\,.
\end{equation}
Here, $S$ is the spin of the particle, so $P_R=+1$ for SM particles and $P_R=-1$ for squarks, sleptons, gauginos and higgsinos. 
%This casts the terms ``particle'' and ``sparticle'' into a precise definition. 
If $R$-parity violating (perturbative) processes are strictly forbidden and not just suppressed, a nice phenomenological property arises: not only are supersymmetric particles produced in pairs at colliders, but the Lightest Super\-symmetric Particle (LSP\glossary{name=LSP,description=Lightest Super\-symmetric Particle}) is absolutely stable and thus a perfect candidate for dark matter~\cite{bib:msugra1,bib:darkmatter}. In this thesis, an mSUGRA model with exact $R$-parity conservation is assumed.

%% file: Setup/SetupIntro.tex
The subjects of this Chapter is firstly CERN's Large Hadron Collider accelerator complex, and secondly the ATLAS detector, one of the two general purpose detectors at the LHC. %ATLAS' discovery potential for Supersymmetry in trilepton final states has been investigated in Part~\ref{prt:susy}.

%% file: Setup/LHC.tex
The Large Hadron Collider~\cite{bib:lhc} (LHC\glossary{name=LHC,description=Large Hadron Collider}) at CERN\glossary{name=CERN, description=Conseil Europ\'een pour la Recherche Nucl\'eaire (European nuclear research council)}, Geneva is designed to deliver proton-proton ($p$-$p$) collisions at an unprecedented centre-of-mass energy of $\sqrt{s}=14\,$TeV with an instantaneous luminosity of \instlumi{34}. The physics potential of this unchartered kinematic regime is discussed in Chapter~\ref{chp:theory}. The LHC is a superconducting two-ring accelerator and collider, comprising eight sections and four interaction points. Predominantly, both rings are accommodated in the same magnet line, adopting the so-called ``twin-bore'' design. The NbTi superconducting magnets are expected to provide a field of more than 8\,T. The LHC is installed in the former LEP ring tunnel of 26.7\,km circumference, roughly 90\,m under the Earth's surface, and with an inclination of about 1.4$^\circ$ towards Lac L\'eman.

\begin{figure}
\begin{center}
\vspace{\cDistHalf}
\includegraphics[width=9.5cm,height=6.5cm,clip=true]{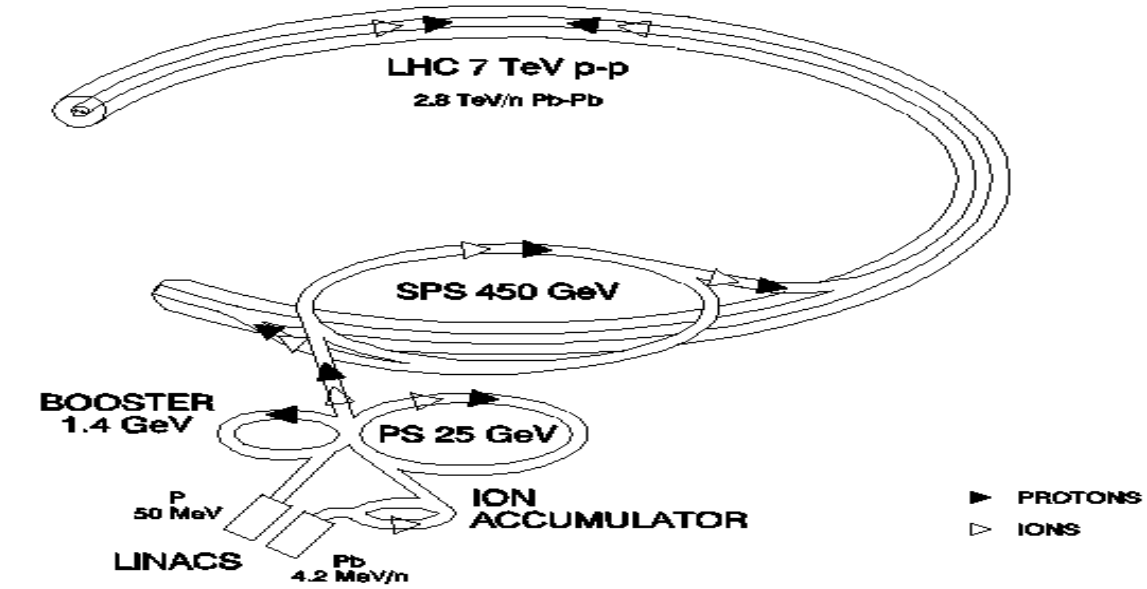}\qquad
\vspace{\cDist}
\end{center}
\caption[The LHC Accelerator Complex at CERN and its major components]{\label{fig:lhc}
The LHC Accelerator Complex at CERN and its major components.
\vspace{\cDistHalf}
}
\end{figure}%\nopagebreak[5]

The full acclelerator chain culminating in the LHC is schematically depicted in Figure~\ref{fig:lhc}. Protons are supplied by a linear acelerator with $E=50\,$MeV, accelerated to $E=1.4\,\GeV$ by the booster synchrotron, and injected into the Proton Synchrotron (PS), where they are bunched to the LHC time spacing of 25\,ns and accelerated to $E=25\,\GeV$. After that, the protons are given their LHC injection energy, $E=450\,\GeV$, by the Super Proton Synchrotron (SPS). The whole injector chain had to be significantly upgraded to meet the highly stringent beam quality requirements of the LHC and to accommodate the 2,088 high intensity $p$ bunches for each of its rings.

%% file: Setup/ATLASDetector.tex
\begin{figure}
\begin{center}
\includegraphics[width=23.5cm,clip=true,angle=90]{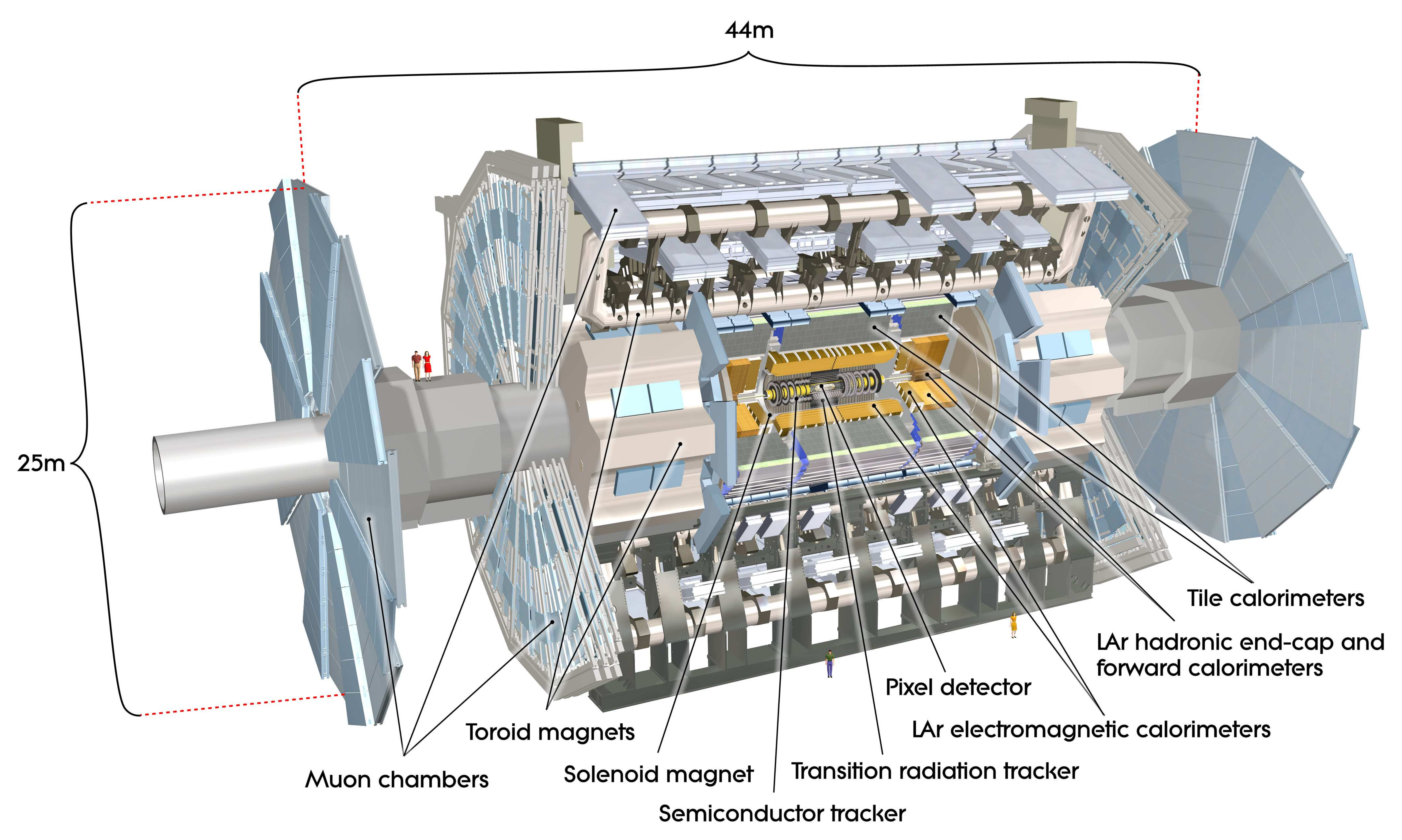}\qquad
\end{center}
\caption[Cut-away view of the ATLAS detector]{\label{fig:atlas}
Cut-away view of the ATLAS detector. Note the size of symbolic human figures depicted.}
\end{figure}

The ATLAS (A Toroidal {\sc Lhc} ApparatuS\glossary{name=ATLAS, description=A Toroidal {\sc Lhc} ApparatuS}) detector~\cite{bib:atlasTDR1,bib:atlasTDR2,bib:atlasJINST} is one of the two general purpose detectors at the LHC. It is a classical layered collider detector with almost full solid angle coverage, which extends up to $|\eta|\lesssim5$\footnote{For the definition of $\eta$ and other geometric variables, please refer to Subsection~\ref{ssec:frames}}. A schematic cut-away view of it can be found in Figure~\ref{fig:atlas}. In this section, I will %briefly review the main physics goals of ATLAS, and then 
give an overview about the main subdetectors of the ATLAS experiment, going from the Designed Interaction Point (DIP\glossary{name=DIP,description=Designed Interaction Point: the spatial point where $p$-$p$ are supposed to collide by design}) to the outer layers.
%as implied by the physics goals. 
Naturally, a particular focus will be placed on the silicon tracker, since its alignment is a substantial part of this dissertation (cf. Part~\ref{prt:alignment}).

%\nopagebreak[5]

%% file: Setup/InnerDetector.tex
\begin{figure}
\begin{center}
\includegraphics[width=12.5cm,clip=true]{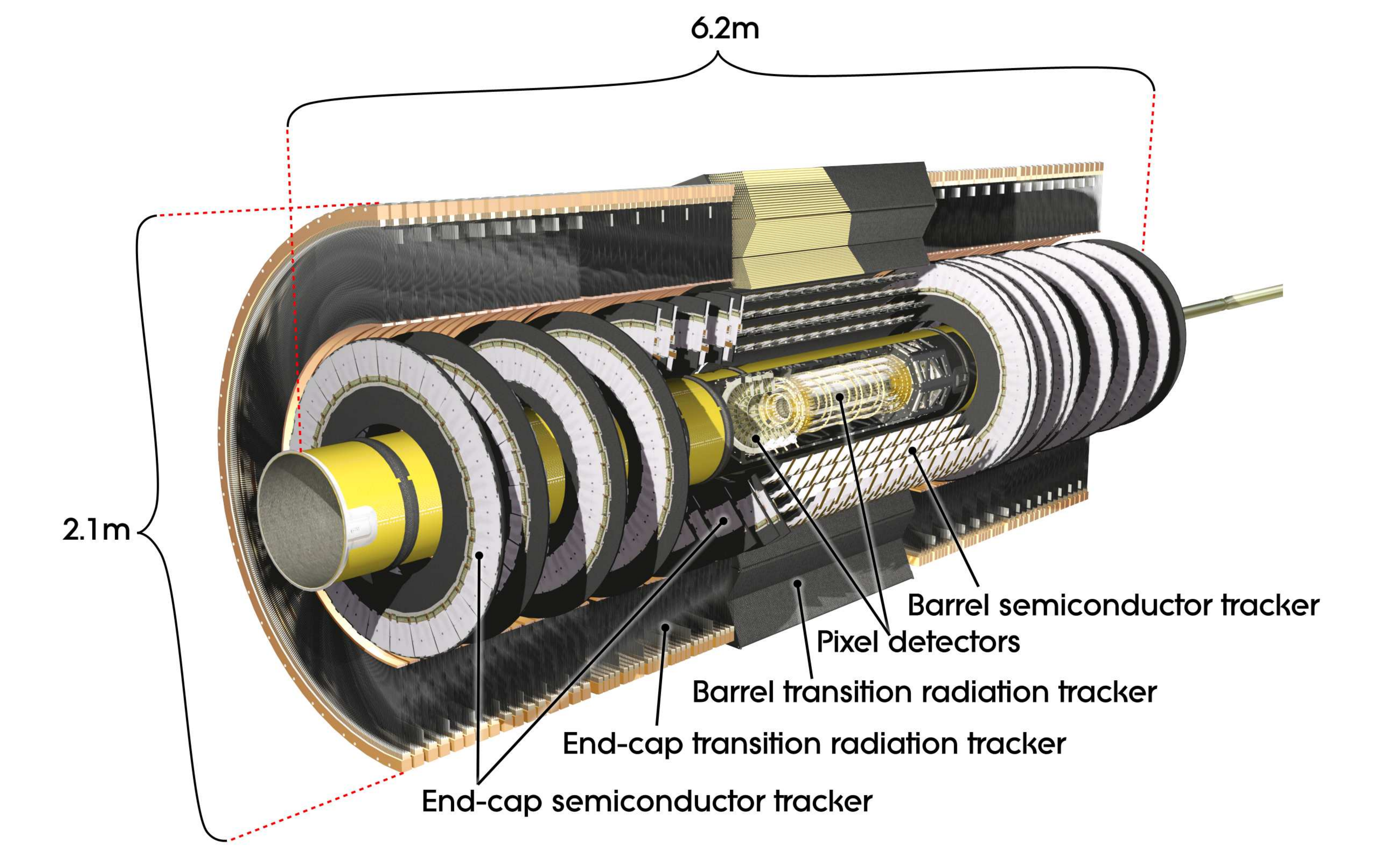}\qquad
\end{center}
\caption{\label{fig:inDet}
The ATLAS Inner Detector and its major components.
}
\end{figure}
\begin{figure}
\begin{center}
\includegraphics[width=15.2cm,clip=true]{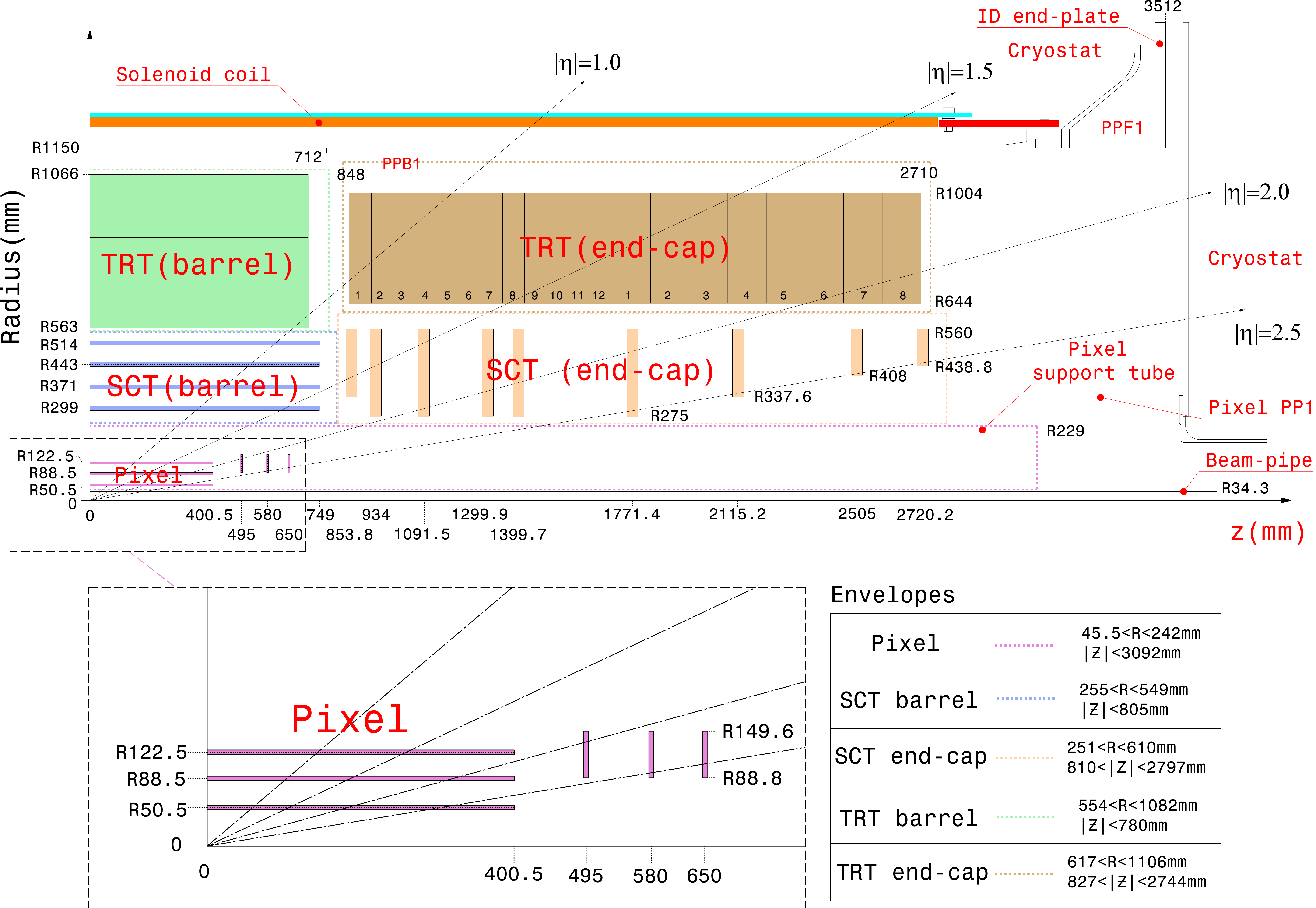}\qquad
\end{center}
\caption[A technical drawing of a quadrant of the ATLAS ID in the $R$-$Z$ plane]{\label{fig:inDetTechnical}
A technical drawing of a quadrant of the ATLAS ID in the $R$-$Z$ plane. All dimensions are in mm.
}
\end{figure}

The Inner Detector~\cite{bib:atlasJINST,bib:inDetTDR1,bib:inDetTDR2}~(ID\glossary{name=ID, description=Inner Detector}) is highly important for achieving the physics goals of the LHC: on average, \order{10^3} charged particles will be produced per bunch crossing in $|\eta|<2.5$ at the design luminosity of $\instlumi{34}$. They need to be tracked as they emerge from the interaction point until they enter the calorimetry, and their transverse momenta need to be measured from their bending radii in a magnetic field. Further, the primary interaction vertex and secondary vertices from particles with macroscopic lifetimes need to be reconstructed to allow for a life-time measurement of such particles and $b$- and \linebreak[5]$c$-tagging of jets. To meet these requirements in the enormous particle densities at the LHC, high-precision detectors with fine granularity are needed. On top of that, the sensors, especially the innermost ones, need to be radiation hard, but on the other hand should add as little as possible to the material budget in front of the calorimeters. This is achieved at ATLAS by the design as shown in Figure~\ref{fig:inDet}. The innermost detector is a silicon pixel detector with very high granularity for excellent pattern recognition and vertexing. It is followed by the SemiConductor Tracker (SCT\glossary{name=SCT, description=SemiConductor Tracker}) based on silicon strips for a precise transverse momentum measurent. The outermost is the Transition Radiation Tracker (TRT\glossary{name=TRT, description=Transition Radiation Tracker}) based on gaseous straw tube technology, enhancing the transverse momentum precision due to its large lever arm. The whole system is immersed in a 2\,T magnetic field. The subdetector dimensions are indicated in Figure~\ref{fig:inDetTechnical}. In order to exploit the full potential of the ID, a good knowledge of the individual module positions, i.e.\! their alignment, is crucial. This is discussed in Part~\ref{prt:alignment}. The ID is expected to reconstruct the transverse momenta of particles with a precision of $\frac{\sigma(\pt)}\pt=0.05\%\cdot\pt\,[\GeV]\oplus1\%$.

%\nopagebreak[5]

%\nopagebreak[5]

\begin{table}[ht]
\small
\begin{center}
%\begin{tabular*}{\textwidth}{@{\extracolsep{\fill}}ccccc} \hline
\begin{tabular*}{0.9\textwidth}{@{\extracolsep{\fill}}lcccc} \hline
%\multicolumn{5}{c}{{ATLAS Inner Detector Modules}}\\\hline\hline
{\bf~Detector type} & {\bf Layer} & {\bf Ring} & {\bf Sector} &
{\bf \# of modules}\\\hline\hline
\multirow{3}{*}{{\bf~Pixel barrel}} & 0 & -6,$\ldots$,0,$\ldots$,6 & 0,$\ldots$,21 & 286\\ 
                    & 1 & -6,$\ldots$,0,$\ldots$,6 & 0,$\ldots$,37 & 494\\ 
                    & 2 & -6,$\ldots$,0,$\ldots$,6 & 0,$\ldots$,51 & 676\\\hline
{\bf~Pixel end-cap} & 2 $\times$ 0,1,2 & 0 & 0,$\ldots$,47 & 288\\\hline 
\multirow{4}{*}{{\bf~SCT barrel}} & 0 & -6,$\ldots$,-1,1,$\ldots$,6 & 0,$\ldots$,31 & 384\\ 
                  & 1 & -6,$\ldots$,-1,1,$\ldots$,6 & 0,$\ldots$,39 & 480\\ 
                  & 2 & -6,$\ldots$,-1,1,$\ldots$,6 & 0,$\ldots$,47 & 576\\
	 	  & 3 & -6,$\ldots$,-1,1,$\ldots$,6 & 0,$\ldots$,55 & 672\\\hline
\multirow{3}{*}{{\bf~SCT end-cap}} & 2 $\times$ 0,$\ldots$,8 & 0 & 0,$\ldots$,51 & 936\\
		  & 2 $\times$ 0,$\ldots$,7 & 1 & 0,$\ldots$,39 & 640\\
 		  & 2 $\times$ 1,$\ldots$,5 & 2 & 0,$\ldots$,39 & 400\\\hline
%{Overall Total}      & 31 layers & 137 rings & \multicolumn{2}{c|}{5832 modules}\\\hline
\end{tabular*}
\caption[The arrangement of the ID modules]{\label{tab:inDet}The arrangement of the 5832 ID modules and their numbering scheme in the offline reconstruction software. Each SCT module consists of two back-to-back sides and is considered as one module. The definition of further substructures can be useful: staves, defined by modules with the same sector number (barrel only) and rings, defined by modules with the same ring number (barrel and end-caps).}
\end{center}
\end{table}

\subsubsection{Pixel Subdetector}

\begin{figure}
\begin{center}
\includegraphics[width=15.2cm,clip=true]{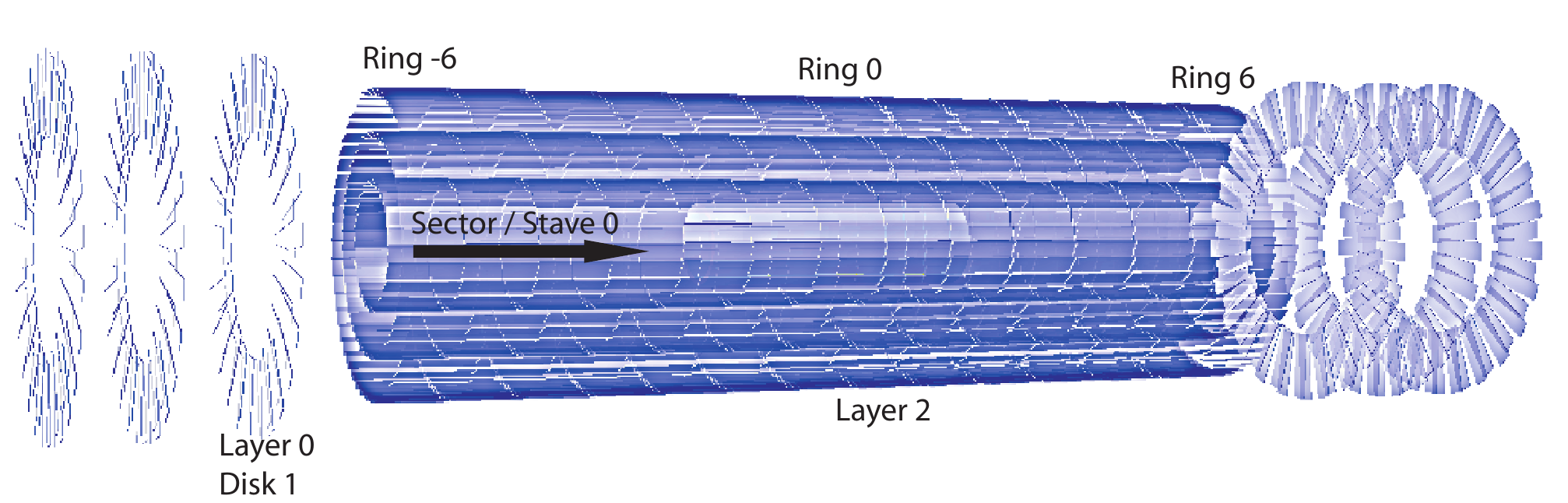}\qquad
\end{center}
\caption[Schematic view of the pixel subdetector]{\label{fig:pixel}
Schematic view of the pixel subdetector and the numbering scheme for its modules.
}
\end{figure}%\nopagebreak[5]

The pixel detector~\cite{bib:inDetTDR1,bib:inDetTDR2,bib:pixel,bib:pixelJINST} consists of 1744 identical modules, which are arranged in three barrel layers and two end-caps of three disk layers each, as detailed in Table~\ref{tab:inDet} and demonstrated in Figure~\ref{fig:pixel}. With this design, a typical track with $|\eta|<1.9$ produces three hits in the barrel region and tracking coverage is provided up to $|\eta|<2.5$. The innermost layer, also referred to as $b$-layer, is only 50.5\,mm away from the design interaction point to allow for a good (secondary) vertex reconstruction precision. There are plans to insert an additional layer of pixel sensors between the $b$-layer and the beryllium beampipe, which extends up to 39\,mm. In order to achieve an optimal coverage given the geometrical constraints, the barrel modules of the pixel detector are tilted by 20$^\circ$ with respect to the tangent to the support cylinder surface. Each 13 barrel modules are mounted on a stave which is made of a highly heat-conductive carbon-carbon laminate support material. It provides stiff mechanical support combined with a very low thermal expansion coefficient to guarantee a good reproducibility of alignment after thermal cycling between -20$^\circ$ and +20$^\circ$\,C. The staves are mounted on carbon fibre support structures. The end-cap disks are set together of eight sectors hosting six modules each. The sectors are carbon-carbon laminate structures with provision for cooling of the modules. Three modules are mounted on each side of a sector.

\begin{figure}
\begin{center}
\includegraphics[width=7.2cm,clip=true]{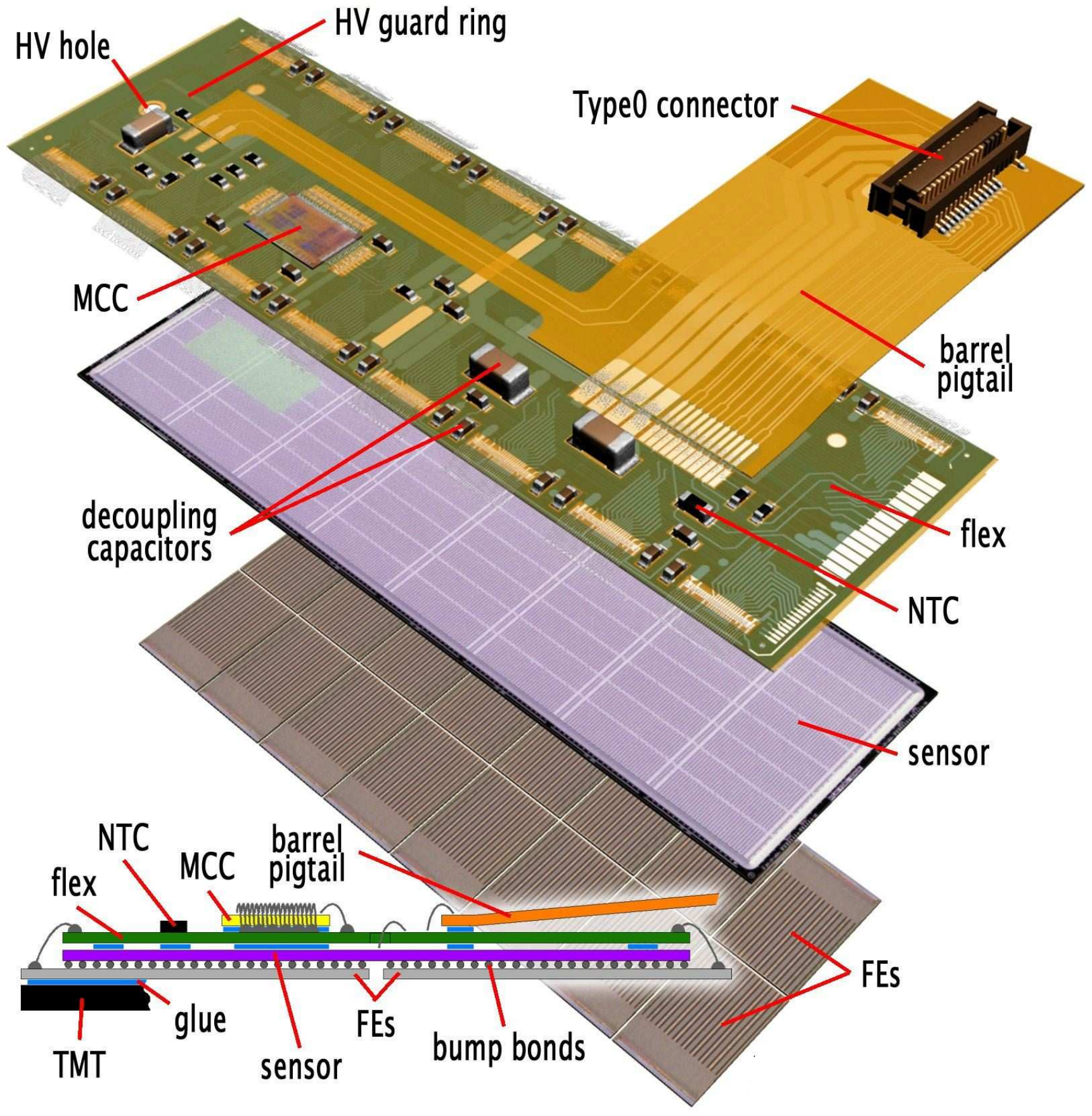}\qquad
\includegraphics[width=7.2cm,clip=true]{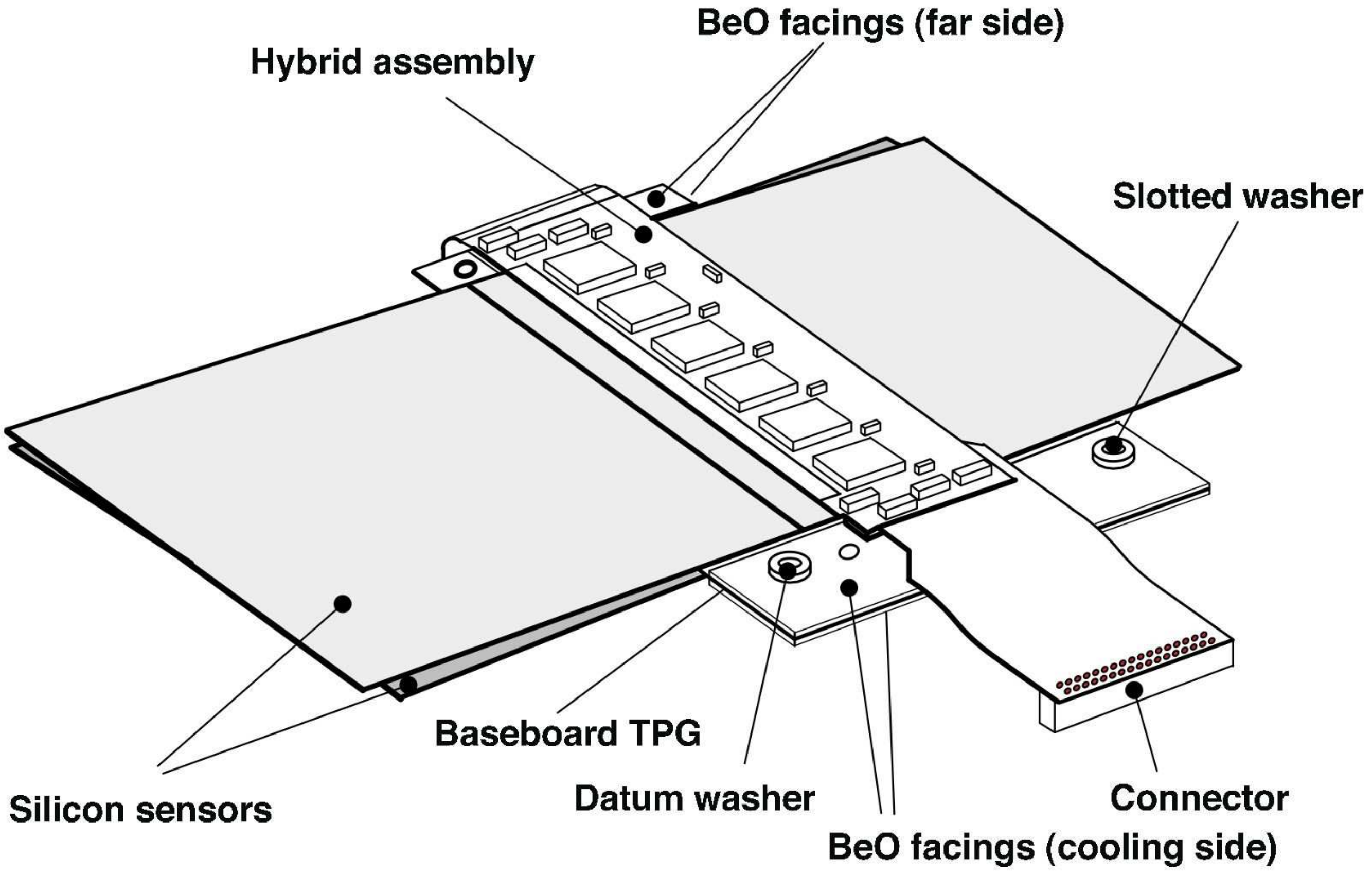}
\end{center}
\caption[Schematic view of a pixel and SCT barrel module]{\label{fig:modulePixelSCT}
Schematic view of a module of the pixel subdetector {\bf (left)} illustrating its major components: front-end chips (FE), sensors, polyamide circuit-board, Module Control Chip (MCC). % As an inset, the bump-bonded structure is shown. 
A barrel module of the SCT subdetector {\bf (right)} shows schematically its major components.
}
\end{figure}%\nopagebreak[5]

A module of the pixel detector, as depicted in Figure~\ref{fig:modulePixelSCT} (left), has the dimensions $60.8\times16.4$\,mm$^2$ in $x\times y$, and hosts 47268 pixels which are read out by 16 front-end chips. For space reasons, four pixels per front-end chip column are ``ganged'', id est connected to the same channel. This totals the number of channels to 46080 per module. The pixel detector accounts for 80.4\,M channels. 

The nominal size of a pixel is $80\times400\,\mu$m$^2$, however, 5284 pixels situated at the boundaries between front end chips are slightly larger, $80\times600\,\mu$m$^2$, which is necessary to cover the gap between the chips. This results in a nominal intrinsic resolution of about 10\,$\mu$m in local~$x$ and 115\,$\mu$m in local~$y$. The optimal resolution of 4.7\,$\mu$m and 6.0\,$\mu$m before and after irradiation, respectively is achieved for incidence angles of 10-15$^\circ$~\cite{bib:atlasJINST}, as then an optimal compromise between the cluster size and the amount of charge collected per pixel is reached. 
%When the charge is collected in the depletion zone, it exhibits a deflection in the magnetic field, i.e. 
The Lorentz angle varies between 12$^\circ$ and 6$^\circ$ before and after irradiation, respectively~\cite{bib:pixelJINST,bib:gorelov}. During the M8+ cosmic data taking in autumn 2008 described in Chapter~\ref{chp:m8plus}, a Lorentz angle of 12.25$^\circ\pm0.03^\circ$ was measured~\cite{bib:twikiResultsPixel}.

The technology used in pixel module sensors is cutting-edge: they are 250\,$\mu$m thick, and use oxygenated n-type sides with the readout pixels situated on the n$^+$~side of the detector. The oxygenation provides extraordinary radiation hardness in a hadronic environment and guarantees an improved charge collection efficiency
% even at lower depletion voltages 
after type inversion of the sensors, which occurs after about a tenth of their life time. The sensors will initially operate at a bias voltage of about 150\,V, increasing up to 600\,V. For the outer layers, this will happen approximately after ten years of operation at nominal luminosity. Despite the extraordinary specifications of the pixel modules, the $b$-layer will have to be replaced after three years at nominal luminosity. The bias voltage, together with the dark current, the hit efficiency and the noise occupancy, can be used for monitoring the radtion damage of the detector. 

\subsubsection{SCT Subdetector}

\begin{figure}
\begin{center}
\includegraphics[width=15.2cm,clip=true]{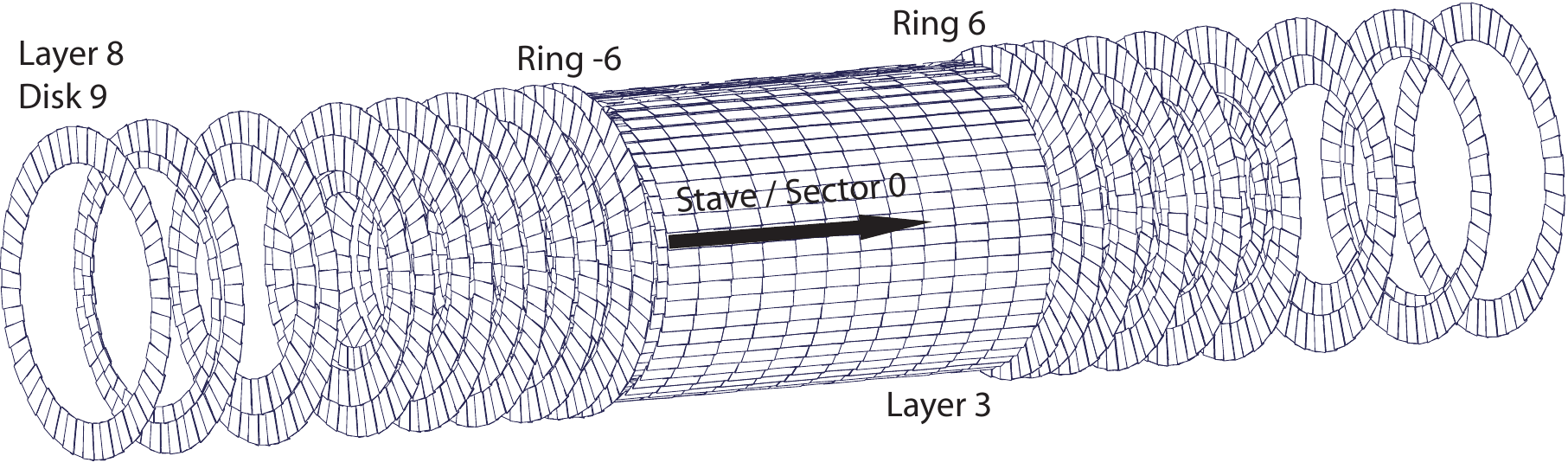}\qquad
\end{center}
\caption[Schematic view of the SCT subdetector]{\label{fig:sct}
Schematic view of the SCT subdetector and the numbering scheme for its modules.
}
\end{figure}%\nopagebreak[5]

The main task of the SCT subdetector~\cite{bib:inDetTDR1,bib:inDetTDR2,bib:sct,bib:sctBJINST,bib:sctEJINST} is the measurement of the transverse momentum of tracks from their bending radii in the magnetic field, and its geometrical design was optimised for this. The SCT extends radially from 250\,mm to 610\,mm, and its 4088 modules are arranged in four coaxial layers in the barrel part of the SCT and in two end-caps of nine disk layers each, as shown in Figures~\ref{fig:inDetTechnical} and \ref{fig:sct}. This provides for a hermetic tracking coverage up to $|\eta|<2.5$. Unlike in the pixel detector, the SCT modules in the barrel region were not mounted on staves, but rather were attached to a support frame, which was optimised for stiffness, reproducibility of alignment after thermal cycling and lightness. The SCT modules in the barrel region are tilted with respect to the tangent of the layer envelope by 11.0$^\circ$ in the two inner layers, and by 11.25$^\circ$ in the two outer layers. However, the tilt direction is opposite with respect to the one in the pixel subdetector due to the different charge carriers. The end-cap disk layers typically consist of three concentric rings with increasing radii which are mounted on alternating sides of a disk-like support structure. The exact number of rings per disk layer is summarised in Table~\ref{tab:inDet}, and is dictated by geometrical acceptance.

Unlike in the pixel detector, one type of module~\cite{bib:sctModuleBarrel} is used in the barrel (2112 modules), and three different types~\cite{bib:sctModuleEC} in the end-caps (1976 modules) of the SCT. A common feature of all modules is that they consist of two silicon strip sides glued back-to-back, each side comprised of an upper and a lower silicon wafer. The sides are rotated by 40\,mrad with respect to each other in order to provide a measurement in the local $y$ direction ``along'' the strips. The small magnitude of the stereo angle was chosen to minimise occupancy in the high-multiplicity experimental environment of the LHC. There are two types of SCT module side orientation: ``phi'' and ``stereo'', where the former has strips parallel to the global $Z$ axis, and the latter is rotated by 40\,mrad. The sign of the stereo strip rotation alternates between layers both in the barrel and end-caps in order to reduce systematic biases in tracking. Whereas the strips on the barrel modules are parallel, they exhibit a fan-out structure in the end-cap modules.

In the following, a description of a barrel module will be given, for end-cap module specific details see~\cite{bib:atlasJINST} and references therein. A schematic view of an SCT barrel module is shown in Figure~\ref{fig:modulePixelSCT} (right). It has two sides referred to as `side~0' and `side~1', and consists of four sensors in total: one on the top and bottom part of each side separated by a 2\,mm gap. Each sensor has 768 active strips, which are connected to binary readout chips, thus totalling the number of readout channels to 6.3\,M.

In the geometrical centre of all module types, the pitch between strips is 80\,$\mu$m, which results in a nominal intrinsic resolution of 17\,$\mum\times580\,\mu$m in $x\times y$, where the value in $y$ was obtained by combining the measurements of both sides of a module. The resolution has been measured in a testbeam~\cite{bib:sctTestbeam}, and was found to be somewhat better: 16\,$\mu$m in $x$, due to a small fraction of multi-strip hits. This value did not significantly degrade after full irradiation.

The technology used to produce SCT sensors is a classic single-sided p-in-n technology with AC-coupled readout chips for reasons of cost effectiveness and reliability. The initial operating voltage is about 150\,V, going up to 350\,V after 10 years, depending on the module position and the associated irradiation dose.

\subsubsection{TRT Subdetector}

The Transition Radiation Tracker~\cite{bib:inDetTDR1,bib:inDetTDR2} extends radially from 554\,mm to 1106\,mm with a tracking coverage in $|\eta|<2.0$. It provides only two-dimensional measurements in the $R$-$\Phi$ plane. Typically, 36 hits are produced along a high-$p_T$ track, which adds to the robustness of pattern recognition. Despite its limited intrinsic resolution compared to silicon detectors, the TRT significantly enhances the momentum measurement due to a twofold extension of the lever arm for tracking, and adds little to the material budget. 

The detection technology in the TRT is based on polyamide drift (straw) tubes of 4\,mm diameter~\cite{bib:strawTRT}. The straw tube walls are designed to enhance transition radiation production, such that with an appropriate gas mixture, i.e. 70\%~Xe, 27\%~CO$_2$, and 3\%~O, electron identification is facilitated. The intrisic per-straw accuracy is about 130\,$\mum$. Like the pixels and the SCT, the TRT detector is divided into a barrel region, where 144\,cm long straws run run parallel to $Z$, and two end-caps, where 37\,cm long straws are aligned parallel to $R$. The total number of TRT channels is about 351\,k.

%% file: Setup/Calorimetry.tex
The next downstream subdetector layers of ATLAS as seen from the DIP are the ElectroMagnetic (EM\glossary{name=EM, description=ElectroMagnetic}) and the hadronic calorimeters (see \cite{bib:atlasTDR1,bib:atlasJINST} {\em and references therein}). Their main role is to measure the energy and direction of incident particles. Further, they are crucial for the identification of EM objects -- electrons and photons, as well as hadronic ones -- jets and pions. From the imbalance of the transverse energy $\et$ the presence of neutrinos and other non-interacting particles can be inferred. A projective view of the EM and hadronic calorimeters is depicted in Figure~\ref{fig:calo}. They are briefly described in the following:

\begin{figure}
\begin{center}
\vspace{\cDistHalf}
\includegraphics[width=12.2cm,clip=true]{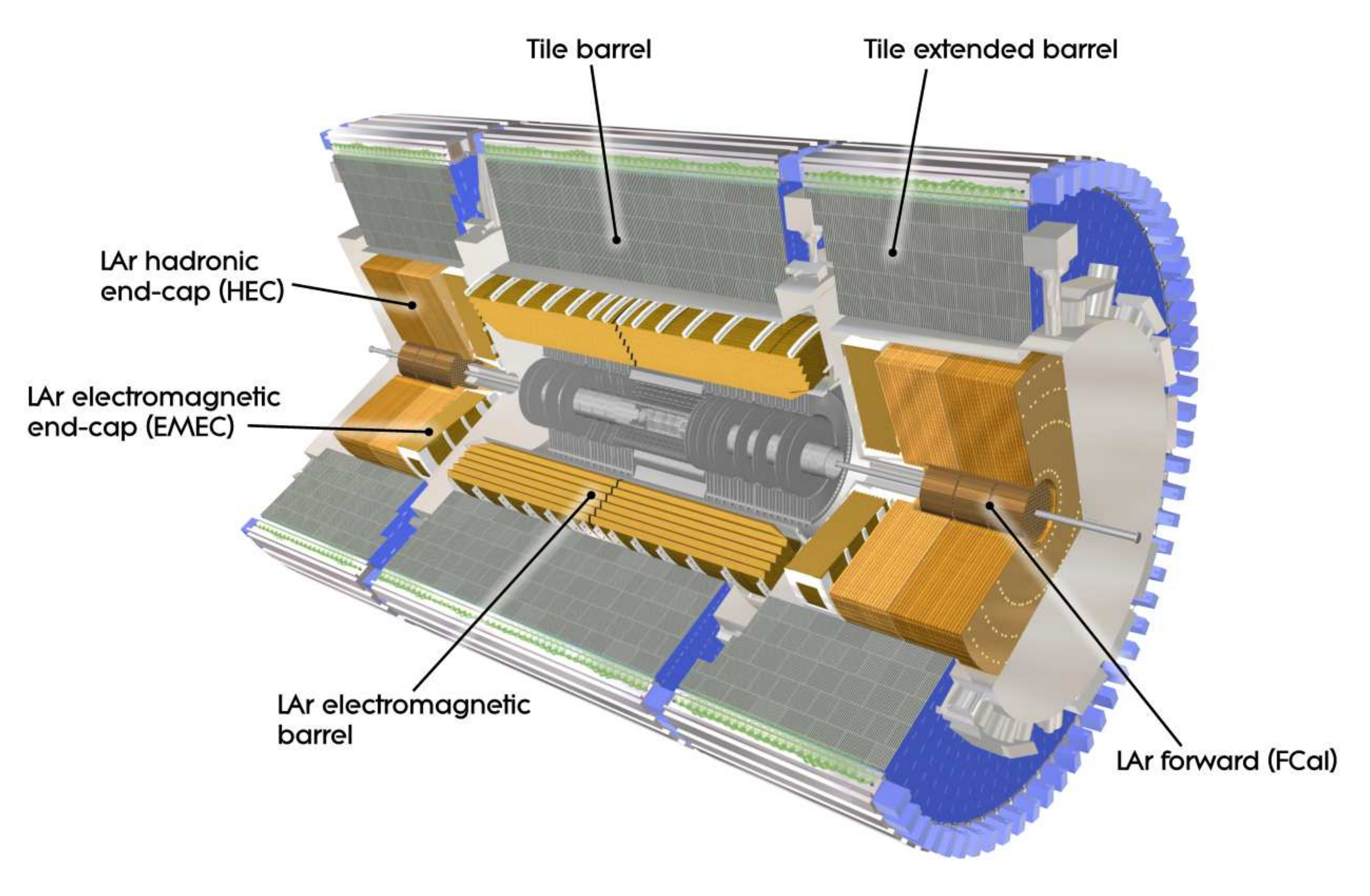}\qquad
\vspace{\cDist}
\end{center}
\caption[Schematic view of the ATLAS calorimeters]{\label{fig:calo}
Schematic view of the ATLAS calorimetery with its main components.
\vspace{\cDistHalf}
}
\end{figure}%\nopagebreak[5]

\begin{description}
\item[EM Calorimeter:]
this subdetector is split up into a barrel part ($|\eta|<1.475$) and two ECs ($1.375<|\eta|<3.2$), each housed in its own cryostat. The barrel part shares its vacuum vessel with the solenoid magnet in order to reduce dead material. All three parts feature the same detection technology: they are sampling calorimeters with a lead absorber and Liquid Argon (LAr\glossary{description=Liquid Argon, name=LAr}) as active medium. The signal is registered by accordion-shaped kapton electrodes. This smart design provides for a complete symmetry in $\Phi$ without any azimuthal cracks. In order to measure the longitudinal development of EM and hadronic showers, the barrel part is divided into three ($|\eta|<1.35$) or two ($1.35<|\eta|<1.475$) coaxial floors. Analogously, the ECs are realised with three ($1.5<|\eta|<2.5$) and two ($1.375<|\eta|<1.5$, $2.5<|\eta|<3.2$) floors. In order to account for the energy lost upstream the calorimeter, an  active-medium-only presampler is introduced in $|\eta|<1.8$. To provide for a measurement of the lateral shower development, the EM calorimeter is realised in projective towers with varying granularity of typically $0.0025\times0.0025$ in $\eta\times\Phi$. This results in an angular resolution of about 20\,mrad in $\theta$ for mildly non-projective photons, as found with 2004 test beam data by the author~\cite{bib:resolutionLAr,bib:resolutionLArTalk}.\\
The total thickness of the EM calorimeter is more than 22/24 radiation lenths ($X_0$) in the barrel/end-caps, respectively. It is expected to achieve a resolution of $\frac{\sigma(E)}E=\frac{10\%}{\sqrt E\,[\GeV]}\oplus0.7\%$.
\item[Hadronic Calorimeter:]
this detector is split into three parts: barrel ($|\eta|<1.7$), ECs ($1.5<|\eta|<3.2$), and forward ($3.1<|\eta|<4.9$). Similarly to the EM calorimeter, it features a multi-floor structure with an arrangement of laterally segmented blocks of typically $0.1\times0.1$ in projective towers. The barrel part is realised with steel as absorber and scintillator tiles as active material. Due to higher radiation levels owing to a higher particle flux, both end-cap and forward parts feature LAr as active medium. Relatively dense but comparably cheap copper was chosen as sampling material in the ECs. 
The forward calorimeter is set back by 1.2\,m along $Z$ towards the DIP in order to reduce radiation background levels for the muon spectrometer. Therefore, tungsten is utilised as sampling material in order to provide a uniform depth for strong interacting particles in all hadronic calorimeters.\\
%The utilisation of tungsten for the hadronic part of the forward calorimeter was inevitable to provide the required nuclear interaction depth, as well as to allow for its 1.2\,m set-back position along $Z$ with respect to the ECs in order to reduce radiation background levels for the muon spectrometer.\\
The total thickness of both the EM and hadronic calorimeters is more than 9.7/10 nuclear interaction lengths ($\lambda$) in the barrel/end-caps, respectively, with additionally more than 1.3$\lambda$ from the (inactive) support structure. This renders the punch-through probability for hadronic jets almost negligible, providing for a good \met\ reconstruction, which is highly important for BSM searches. The hadronic calorimeter is expected to reconstruct the energy of jets with a precision of $\frac{\sigma(E)}E=\frac{50\%}{\sqrt{E\,[\GeV]}}\oplus3\%$ in the barrel and  $\frac{\sigma(E)}E=\frac{100\%}{\sqrt{E\,[\GeV]}}\oplus10\%$ in the end-caps.
\end{description}

%% file: Setup/MuonSpectrometer.tex
\begin{figure}
\begin{center}
\vspace{\cDistHalf}
\includegraphics[width=12.2cm,clip=true]{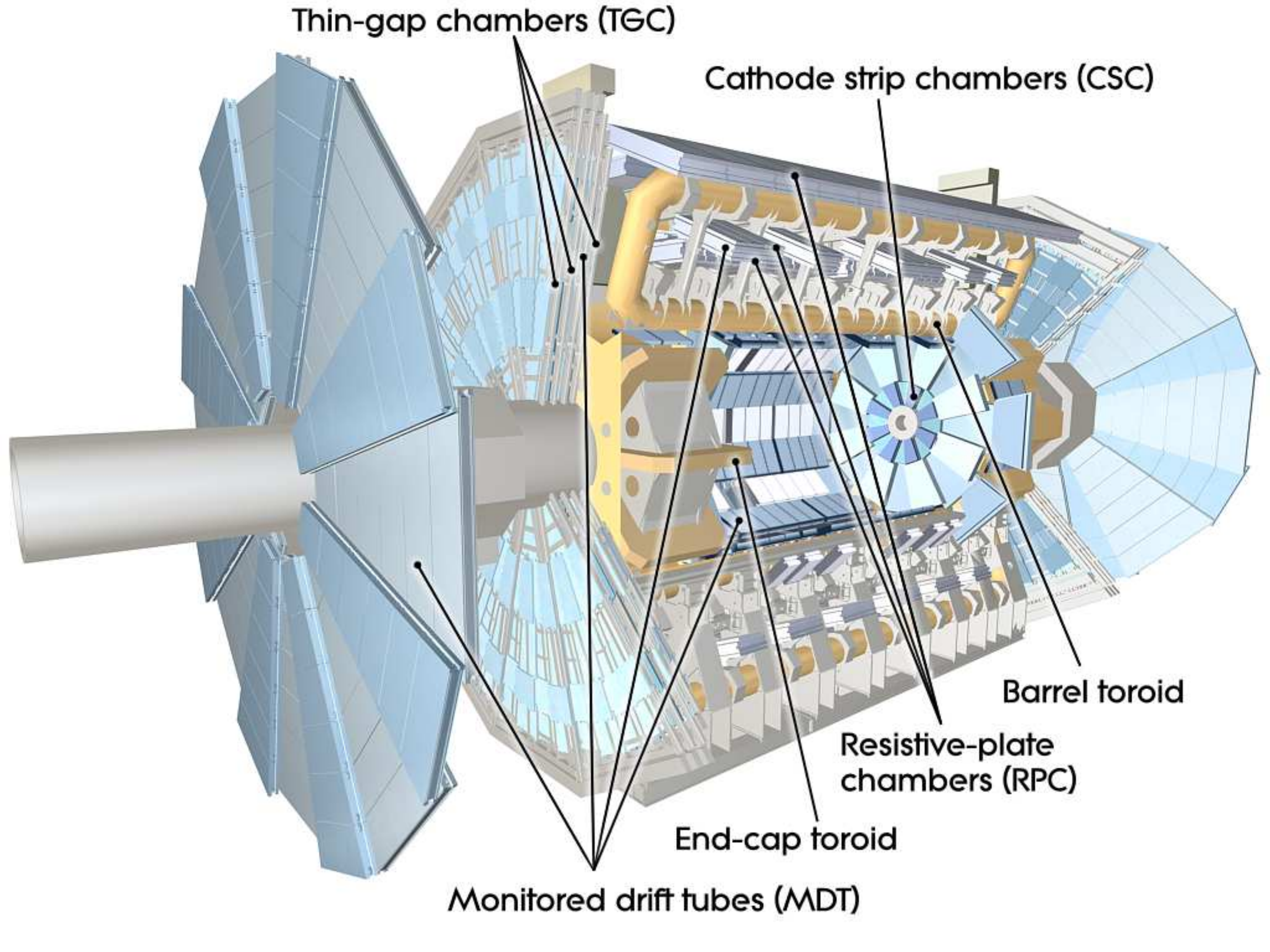}\qquad
\vspace{\cDist}
\end{center}
\caption[Schematic view of the Muon Spectrometer]{\label{fig:ms}
Schematic view of the Muon Spectrometer and its main components.
\vspace{\cDistHalf}
}
\end{figure}

The Muon Spectrometer (MS\glossary{name=MS, description=Muon Spectrometer}) is the outermost layer of the ATLAS detector (see \cite{bib:atlasTDR1,bib:atlasJINST} {\em and references therein}). Its main components are displayed in Figure~\ref{fig:ms}. The geometrical design of the MS features three instrumentation layers both in the barrel and in the end-caps. It was optimised such that a typical muon originating from the DIP with $|\eta|\lesssim2.5$ will be registered by each of the three layers. The key aspect of the MS are its $1+2$ toroidal magnetic fields in the barrel~($|\eta|<1.4$) and the two ECs~($1.6<|\eta|<2.7$), which provide for a measurement of charged particle momenta. With this configuration, a bending power $\int\!\dif\vec\ell\cdot\{\vec e_B\times(\vec e_\ell\times\vec B)\}$ of 1.5 to 5.5\,Tm is achieved in the barrel region, 1 to 7.5\,Tm in the EC region, and somewhat lower values in the transition region~($1.4<|\eta|<1.6$). Here, $\vec\ell$ is the path vector, and $\vec e_x\equiv\xOverY{\vec x}{|\vec x|}$.

%\nopagebreak[5]

The tasks of the MS are manyfold: high precision measurement of muon track parameters, triggering, bunch-crossing identification, and the identification of cosmic ray particles. This is achieved by a dual design: while the Monitored Drift Tubes (MDTs\glossary{name=MDT, description=Monitored Drift Tube}) and the Cathode Strip Chambers (CSC\glossary{description=Cathode Strip Chamber, name=CSC}) provide a high precision measurement of the track parameters, the Resistive Plate Chambers (RPCs\glossary{description=Resistive Plate Chamber, name=RPC}) and Thin Gap Chambers (TGCs\glossary{description=Thin Gap Chamber, name=TGC}) are used for triggering, bunch crossing identification, and the cosmic ray veto utilising their excellent time resolution of between 1.5 and 4\,ns. A yet another important task of the RPCs and TGCs is to measure the muon coordinate in the direction orthogonal to that determined by precision-tracking chambers. The main characteristics of the MS sensors as well as the regions of their coverage are summarised in Table~\ref{tab:ms}. 

Ultimately, the muon spectrometer is expected to provide a precision of $\frac{\sigma(\pt)}{\pt}\simeq10\%$ for muons with $\pt=1$\,TeV~\cite{bib:atlasJINST}.

\begin{table}
\small
\begin{center}
%\begin{tabular*}{\textwidth}{@{\extracolsep{\fill}}ccccc} \hline
%\begin{tabular*}{0.9\textwidth}{@{\extracolsep{\fill}}lcccc} \hline
%\multicolumn{5}{c}{{ATLAS Inner Detector Modules}}\\\hline\hline
\begin{tabular}{l|cc|cc} \hline
 & \multicolumn{2}{c|}{{\bf Barrel}} & \multicolumn{2}{c}{{\bf End-cap}}\\
{\bf Characteristic} & {\bf MDT} & {\bf RPC} & {\bf CSC} & {\bf TGC}\\\hline\hline
Coverage & $|\eta|<2.7$ & $|\eta|<1.05$ & $2.0<|\eta|<2.7$ & $1.05<|\eta|<2.7$\\
\# of chambers & 1,150 & 606 & 32 & 3,588\\
\# of channels & 354,000 & 373,000 & 31,000 & 318,000\\
Function & Precision tracking & Triggering & Precision tracking & Triggering \\
\hline
\end{tabular}
\caption[Main characteristics of Muon Spectrometer sensors and their coverage region]{\label{tab:ms}
Main characteristics of Muon Spectrometer sensors and their coverage region.}
\vspace{\cDistHalf}
\end{center}
\end{table}

%% file: Setup/Trigger.tex
At the design instanteneous luminosity of \instlumi{34}\ a total inelastic $p$-$p$ interaction rate of about 1\,GHz is expected at the LHC, whereas the ATLAS rate-to-tape is limited to some 200\,Hz. To cope with this disparity, a rejection of \order{10^{7}} against minimum bias events is needed, while preserving a high efficiency for relevant SM physics measurements and BSM searches. At ATLAS, this is realised in form of a three-level trigger system. It consists of
\begin{itemize}
\item 
Hardware trigger at level~1~(L1\glossary{name=L$x$, description=Trigger or alignment level~$x$});
\item 
Software trigger at level~2~(L2), which uses the full detector information in $\eta\times\phi$ Regions-of-Interest~(RoIs\glossary{name=RoI, description=Region-of-Interest}) defined in the preceding decision step;
\item
Offline-analysis-like EventFilter (EF\glossary{name=EF, description=EventFilter}), which can access the full detector information. 
\end{itemize}
The trigger system reduces the event rate from about 1\,GHz to some 75\,KHz\footnote{75\,KHz is the current hardware limit of the ATLAS read-out system.} after L1, to about 3.5\,KHz after L2, and finally to circa 200\,Hz after EF. Mind that the event rates given are average values, and each of the trigger levels has its own buffer in order to reduce dead time.

The L1 trigger uses a limited amount of detector information with coarse granularity to make a decision in less than 2.5\,$\mu$s. It can trigger on high-\pt\  muons using the RPCs and TGCs, EM objects, jets, hadronically decaying $\tau$-leptons, and large (missing) transverse energy, where analogue sums are used. The trigger flags from the calorimeters and the MS are combined and decided upon by a central trigger processor according to the so-called {\em trigger menu}. In case of a positive decision, the L1~trigger flags define the RoIs in $\eta\times\phi$. The full detector information in the ROIs (about 2\% of the total event size) is passed to the L2~trigger, which reaches a decision in about 40\,ms. Finally, the ultimate decision about the event is made by the EF in several seconds.

%% file: Setup/Frames.tex
At ATLAS, various dedicated coordinate frames are used. Here, the ones relevant for the alignment of the silicon tracker will be reviewed briefly~\cite{bib:atlasTDR1, bib:frames}. It makes sense to define a global ATLAS frame, for whose quantities capital letter variables\footnote{The variables $\theta$ and $\eta$ are the only exceptions from this rule.} will be used, and a local module frame with small letter variables:
\begin{description}
 \item[Global frame:] 
 its cartesian parameterisation is a right-handed coordinate system with its origin at the centre of the detector, the $X$-axis pointing towards the accelerator ring centre, the $Y$-axis pointing upwards perpendicular to the accelator plane\footnote{The LHC is tilted by 1.23$^\circ$ with respect to the horizont line due to sediment topology between Geneva and the Jura.}, and the $Z$-axis pointing along the beamline such that right-handedness is observed. Moreover, a cylindrical parameterisation is used, where coordinates in the $X$-$Y$ plane are given by their radial distance $R$ from the $Z$-axis, and the angle $\Phi$, which is counted positive from the $X$-axis towards the $Y$-axis. Finally, a spherical parameterisation is used, where the angle with respect to the $Z$ axis is denoted as $\theta$. Conveniently, $\theta$ is often parametrised as pseudorapidity $\eta\equiv-\ln(\tan\,\theta/2)$ to reflect the fact that the hadronic flux is to first order constant in intervals of $\eta$ at hadron colliders.
% When referring to axes, angles, quantities etc.\! in the \textit{global} frame, capital variables are used.
 \item[Local frame:] 
 %for each individual silicon module, a local coordinate frame can be defined in a natural way. In the barrel region, the $x$-axis is along the precise measurement direction of the module, $y$ is along the imprecise measurement direction and its positive direction coincides\footnote{For a misaligned module the axes, e.g. $y$ and $Z$ in the barrel, will not coincide \textit{exactly}.} with $Z$, whereas $z$ is orthogonal to the $x$-$y$ plane and points away from the interaction point, which at the same time defines the positive direction of the $x$-axis. In the end-caps,\\
 for each individual silicon module, a cartesian parameterisation of the local coordinate frame can be defined in a natural way. In the barrel region, the $x$-axis is along the precise measurement direction of the module, $y$ is along the imprecise measurement direction, whereas $z$ is normal to the module. In the end-cap modules of the SCT the axes are defined at the centre of module using the same convention. The positive directions of the axes are defined as the directions of increasing $\Phi$ for the $x$-axis, increasing $Z$ or $R$ for the $y$-axis in the barrel or end-caps, respectively, and as the direction of $\vec e_x\times \vec e_y$ for the $z$-axis.
% When referring to axes, angles, quantities etc.\! in the \textit{local} frame, small variables like are used.
\end{description}

In the context of the global ATLAS frame it is worthwhile noting that the convention used is an idealised one. %In reality, the centre of the detector and its orientation in space needs to be defined in the first place. At the time of writing, no procedure has been clearly established. The most widely discussed procedures suggest a common centre-of-gravity of either the entire ID, the silicon tracker, or the outer layer of the SCT. 
The actual global reference frame of ATLAS must be related to arbitrarily chosen sensitive devices within the spectrometer. The convention to be adopted is currently under discussion, but most likely choice is the centre-of-gravity of the entire ID. A more general discussion of ATLAS frames can be found in~\cite{bib:framesHawkings}.

In Euclidean space, each body has six degrees of freedom: three translations and three rotations. For alignment purposes, translations are conveniently defined along the cartesian coordinate frame axes, and rotations are defined about these axes following the right-hand convention. The rotation angles about $x$, $y$, $z$ are denoted as $\alpha$, $\beta$, and $\gamma$ (the same naming conventions hold for the global frame). In general, translations and rotations, as well as rotations among each other do not commute. However, angular alignment corrections almost never exceed 1\,mrad, which justifies the small angle approximation and restores commutativity.

By convention, the end-cap A of any subdetector of ATLAS is at $Z>0$, whereas end-cap C is at $Z<0$.

%\cite{bib:beamspotTwiki,bib:beamspotIndico}

%% file: Setup/Athena.tex
The ATLAS collaboration uses an extensive and versatile software framework called \Athena~\cite{bib:athena,bib:athenaUsr,bib:athenaDev}, which provides for:
\begin{itemize}
\item Operation and monitoring of the detector;
\item Event trigger;
\item Event reconstruction;
\item Event simulation;
\item Physics analysis tools.
\end{itemize}
Technically speaking, \Athena\ is a concrete implementation of an underlying architecture called Gaudi, originally developed by LHCb. Besides laying down how data is stored with the so-called Event Data Model (EDM\glossary{name=EDM, description=Event Data Model}), it defines:
\begin{itemize}
\item
{\em Algorithms} -- user application building blocks with three basic methods: {\tt initialize()}, {\tt execute()}, {\tt finalize()} called at defined times;
\item
Algorithm {\em tools}, which implement methods to be used by algorithms;
\item
Globally available {\em services} providing widely used framework capabilities, like the {\tt THistSvc};
\end{itemize}
and many more. They can be steered via configuration files called JobOptions (JO\glossary{name=JO, description=JobOptions: \Athena\ configuration file with Python syntax}).
%JOs can be used not only to initialise algorithms, tools and services, but also to pass variable values to them at initialisation time. 
The \Athena\ framework implements algorithms, tools, services in C++, and JOs in Python.
Highly relevant for data reconstruction and in particular alignment, \Athena\ also provides access to detector geometry and offline condition databases for algorithms and tools. 

Guidelines and further information to the \Athena\ framework are available in~\cite{bib:workBook} for end-users, and in~\cite{bib:athenaDev} for developers.

%% file: IntroSUSY/IntroSUSY.tex
One of the main goals of the ATLAS detector described in Chapter~\ref{chp:setup} is to search for signatures of BSM physics in the $pp$ collision data of up to $\sqrt s = 14$\,TeV produced at the LHC. A widely favoured BSM candidate is supersymmetry, which was briefly introduced in Section~\ref{sec:susy}.

Search strategies for {\em direct} experimental evidence of SUSY at a collider experiment include signatures from {\em multi-jet} and {\em multi-lepton} final states. Whereas the former typically offer a higher cross section, the latter yield a cleaner signal and can be less prone to systematic uncertainties. An brief overview about supersymmetry search strategies at ATLAS and CMS can be found e.g. in~\cite{bib:procHCP}.

The research documented in this part of the thesis was mainly aimed at elaborating a generic search strategy for supersymmetry using the striking signature of {\em trilepton} final states. It is documented in~\cite{bib:intNote}. Further, it posed a substantial contribution to~\cite{bib:csc7} and~\cite{bib:csc5}, two chapters of the supersymmetry part of the ATLAS publication on ``Expected Performance of the ATLAS Experiment - Detector, Trigger and Physics''~\cite{bib:cscbook}.

The presented analysis is not optimised for any one particular SUSY model point, but instead represents a rather generic strategy which should be sensitive to the trilepton signature over much of the SUSY parameter space. The resulting statistical significance tables and integrated luminosities for a 5$\sigma$ discovery at $\sqrt s = 14$\,TeV are presented. Clearly, the trilepton signature is also sensitive to many other BSM theories other than supersymmetry. However, their discovery potential at ATLAS is not evaluated explicitly in this document.

A special focus is placed on investigating the SUSY discovery potential in the case where all {\em strong} interacting supersymmetric particles {\em including} gluinos are very massive, and one cannot rely on multijet final states for a discovery. In this document, such a scenario will be referred to as the {\em massive sparton scenario}. It appears plausible, that direct gaugino pair-production and its trilepton decay modes will play a crucial role in discovering SUSY if this scenario is indeed realised in nature.

Multilepton final states offer various handles for measuring supersymmetric mass spectra, couplings, and other parameters of the theory~\cite{bib:massSpectrumMeas1,bib:massSpectrumMeas2,bib:massSpectrumMeas3}. A veto on hadronic activity in the event can isolate direct gaugino pair-production from other supersymmetric production mechanisms. The potential of this strategy was evaluated using the simulated MC event record.

%A special focus is placed on the difficult {\em massive sparton scenario}, where one can rely only on associated gaugino pair-production for a discovery. Arguably, the trilepton final state is optimally suited for such a scenario. 

%The main goal of this analysis is to investigate the discovery potential for SUSY in the case where all {\em strong} interacting supersymmetric particles \textit{including} gluinos are very heavy and one cannot rely on multijet final states for a discovery. In this document, such a scenario will be referred to as the {\em massive sparton scenario}. Moreover, the trilepton final state plays an important role for most of the ``typical'' supersymmetric scenarios, both for discovery and for measuring SUSY parameters. 

Further, one possible choice of triggers for initial running at \instlumi{31} is suggested, based on optimising their efficiency at two intermediate and the final stage of the selection. 

Moreover, several universal approaches to control instrumental backgrounds are critically investigated. 

Finally, a strategy is presented to measure the rate of leptons from semileptonic $b$-decays to pass isolation criteria, as they account for a major fraction of the background. It is suggested that this can be done by isolating a pure sample of semileptonically decaying $b$-jets from \ttbar{} events in data.

%% file: Signature/Signature.tex
In this Chapter, supersymmetric processes to produce trileptonic signatures at the LHC are reviewed, starting with a general discussion in Section~\ref{sec:signatureGeneral}, and highlighting the massive sparton scenario, id est direct gaugino pair-production in Section~\ref{sec:signatureDirectGaugino}. Finally, important background processes will be summarised in Section~\ref{sec:signatureBgr}. For naming conventions of supersymmetric particles, please refer to Tables~\ref{tab:susyFermion},~\ref{tab:susyGauge}, and~\ref{tab:sugraParticles} on pages~\pageref{tab:susyFermion},~\pageref{tab:susyGauge}, and~\pageref{tab:sugraParticles}, respectively.

\section{Supersymmetric Production of Trilepton Final States}\label{sec:signatureGeneral}
%In the following, the mechanisms by which multi-leptonic final states may arise in supersymmetric interactions will be discussed. 
Based on the currently available precision measurement data to constrain the mSUGRA parameter space~\cite{bib:ellisOlive,bib:msugra1,bib:msugra2,bib:msugra3,bib:msugra4}, the ATLAS collaboration has chosen a number of experimentally-different scenarios called SU$x$, $x=1,\,...,\,8$. The supersymmetric scenarios investigated in this analysis are summarised in Table~\ref{tab:susy_points} on page~\pageref{tab:susy_points}. For a more intuitive picture of the supersymmetric particle mass spectra see Appendix~\ref{chp:spectra}.

An mSUGRA point in accordance with WMAP data and other precision measurements is SU3, which is situated in the so-called Bulk Region. It features moderate $M_0$, \Mhalf{}, $\tan\beta$ parameters combined with a fairly high Leading Order (LO\glossary{name=LO,description=Leading Order}) total production cross-section of 18.59\,pb~\cite{bib:msugraPoints}. No particular mass degeneracies in the sparticle spectrum are observed, as displayed in Figure~\ref{fig:SU13} in Appendix~\ref{chp:spectra}. Below, the supersymmetric production of the trilepton final states is discussed using this ``typical'' SU3 example.

At the LHC, for an mSUGRA scenario similar to SU3, supersymmetric particles will be predominantly produced in strong processes via gluon-gluon fusion and quark-gluon initial states. This is due to the dominance of gluon and sea quark Parton Distribution Functions (PDFs{\glossary{name=PDF,description=Parton Distribution Function}}) in the relevant kinematic region and the relative strength of the strong coupling over the electroweak one. Direct slepton production is small at the LHC. The production of multileptonic final states is dominated by processes in which there are cascade decays involving charginos \cha{} and neutralinos \neu.

\newpage
Tree-level diagrams for the dominant production of supersymmetric particles via gluon-gluon fusion are reviewed below\footnote{In a similar fashion, tree-level diagrams for the quark-gluon initial states can be deduced.}.
%In the following, tree level diagrams for the dominant production of supersymmetric particles via gluon-gluon fusion will be reviewed. 
In the $s$-channel, gluon-gluon fusion can pair-produce supersymmetric particles via a gluon propagator:
\begin{center}
\vspace{\cDistHalf}
\includegraphics[height=30mm,clip]{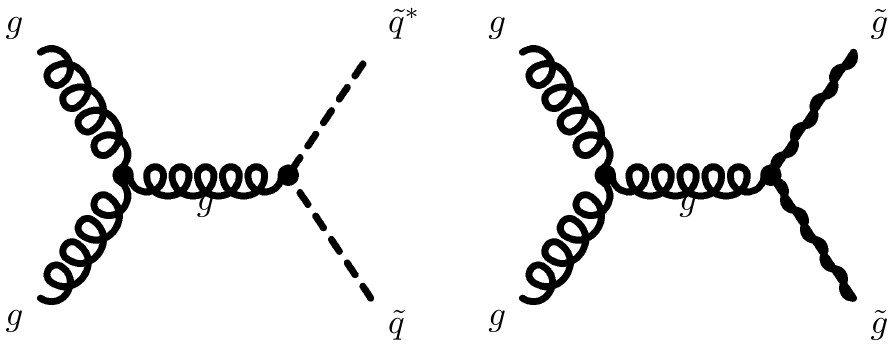}
\vspace{\cDistHalf}
\end{center}
which couples both to squarks and gluinos. Their relative production cross-sections will depend on the ratio of their masses, id est in case of SU3 squarks will dominate over gluinos: $m_{\tilde t_1}=424\,\GeV<m_{\tilde g}=717\,\GeV$~\cite{bib:msugraPoints}. Also important is the $t$-channel production:
\begin{center}
\vspace{\cDistHalf}
\includegraphics[height=30mm,clip]{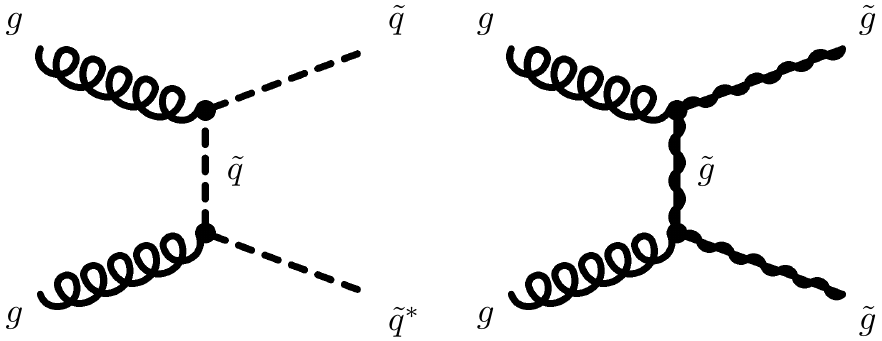}
\vspace{\cDistHalf}
\end{center}
which is mediated by a squark for squark pair-production and a gluino for gluino pair-production. Additionally, there are interference terms from the $u$-channel, which are obtained by applying crossing symmetry to the final state sparticle pair. Similarly to the $s$-channel, squark pair-production will dominate in the $t,\,u$ channels in case of SU3. Additionally, the pair-production of squarks has an extra tree level diagram with a quartic vertex with a gluon and a squark pair.

Charginos and neutralinos cannot couple to gluinos {\em directly} at the same vertex. Therefore, gluinos must decay to a quark and an eventually virtual squark in order to reach $\neu_1$ -- the LSP, which {\em must} stand at the end of any decay chain in $R$-parity conserving SUSY. Therefore, from a diagrammatic point of view, it is sufficient to consider solely squark decays involving chargino or neutralino production in order to understand how multileptonic final states are created via gluon-gluon fusion at the LHC. 

Charginos can be predominantly produced from decays of supersymmetric partners of left-handed SM quarks mediated by the Wino component of the chargino, with an admixture of squarks of any handedness mediated by the Yukawa coupling of the Higgsino component:
\begin{center}
\vspace{\cDistHalf}
\includegraphics[height=3.0cm,clip]{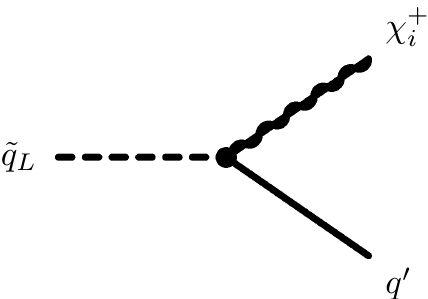}
\vspace{\cDistHalf}
\end{center}
\pagebreak
Neutralinos are produced in decays of squarks via the diagram:
\begin{center}
\vspace{\cDistHalf}
\includegraphics[height=3.0cm,clip]{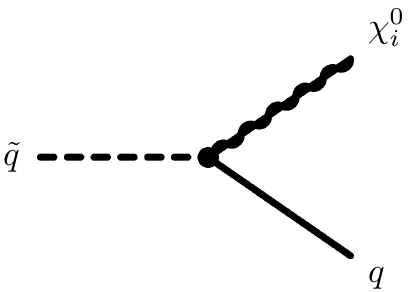}
\vspace{\cDistHalf}
\end{center}

The trilepton final state is produced, if:
\begin{itemize}
\item
in one of the sparticle decay chains a next-to-lightest or heavier neutralino is produced; and
\item
in the other chain a chargino is among the decay products; and 
\item
both gauginos decay leptonically.
\end{itemize}
The leptonic decays of the charginos can be mediated by a $W^\pm$ boson, a slepton, or a scalar charged Higgs boson $H^\pm$:
\begin{center}
\vspace{\cDistHalf}
\includegraphics[height=3.5cm,clip]{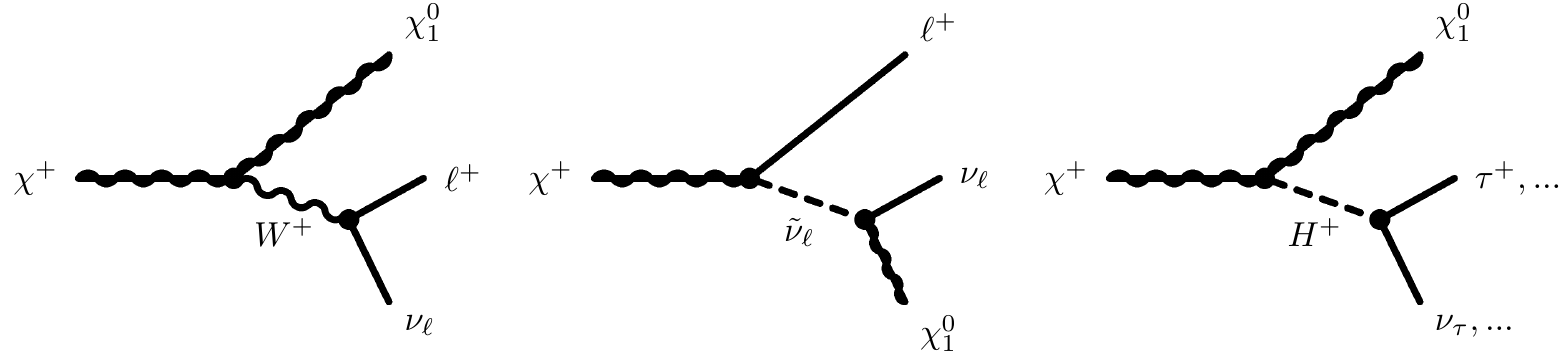}
\vspace{\cDistHalf}
\end{center}
The dominant diagram and the total fraction of leptonic decays will mostly depend on the mass of the mediating slepton\footnote{and of course on the $W^\pm$ boson mass.} as well as on the higgsino/wino composition of the lightest $\cha_1$ mass eigenstate. Similarly, leptonic decays of the next-to-lightest neutralino will be mediated either by the $Z$-boson, a slepton, or any of the scalar neutral Higgs bosons $h^0,\,H^0,\,A^0$:
\begin{center}
\vspace{\cDistHalf}
\includegraphics[height=3.5cm,clip]{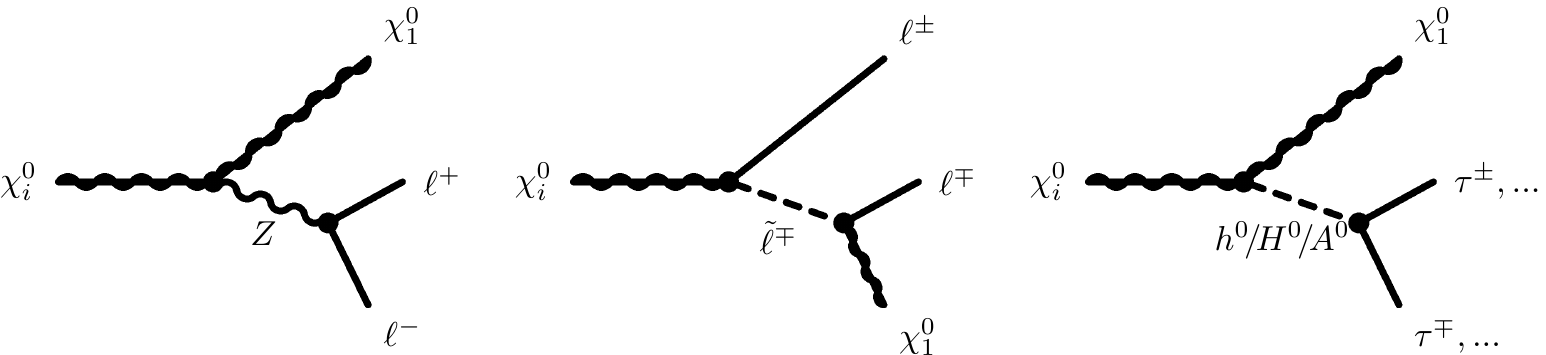}
\vspace{\cDistHalf}
\end{center}
Here again the decay rates will be determined by higgsino/bino/wino composition of the neutralino and the slepton mass.

The above diagrams\footnote{A full example diagram is presented in Figure~\ref{fig:SU2signal} on page~\pageref{fig:SU2signal}.} result in a highly prominent signature: three leptons in the final state with moderately high $\pt$, among them one Opposite Sign Same Flavour (OSSF\glossary{name=OSSF,description=Opposite Sign Same Flavour}) lepton pair from neutralino decay. Further, a significant amount of $\met$ is expected due to the two LSP's and the neutrino escaping detection. The importance of this ``gold-plated'' signature for the Tevatron has been pointed out as early as in 1985~\cite{bib:golden_channel} (LHC: 1994~\cite{bib:trileptonLHC}), and there are analyses based on it both at \Dzero~\cite{bib:trileptonD0, bib:trileptonD0_conf} and CDF~\cite{bib:trileptonCDF, bib:trileptonCDF_conf}.

For most of the mSUGRA parameter space the $\pt$ scale of the leptons, as well as the magnitude of $\met$ are determined by the mass relations $m_{\cha_1}\simeq m_{\neu_2}\simeq 2m_{\neu_1}\simeq 0.8\Mhalf$\,~\cite{bib:herbi}. 

At least two high-$\pt$ jets from gluino or squark decays are expected in the final state in case of strong sparton pair-production. However, the main focus of this analysis is to maintain a flat search performance in the trilepton final state not only for various mSUGRA configurations but also for other BSM scenarios. Therefore jets are not used in the signal selection.

The SUSY benchmark point SU4 is similar to SU3 (Figure~\ref{fig:SU48}), but with smaller scalar and fermion masses, and a high cross-section of 262\,pb~\cite{bib:msugraPoints}. It is just above the Tevatron discovery limit for strong sparticle masses of around 350\,GeV~\cite{bib:searcHeavySUSY_D0,bib:searcHeavySUSY_CDF} and the LEP limit for an LSP mass of approximately 55\,GeV~\cite{bib:searchLEP}. 
%Due to the lower scale and the resulting softness of lepton and jet spectra, the lepton and jet properties of this point will be similar to \ttbar{}. 
Thus, a counting experiment will be the preferred method for a discovery in view of the high total SUSY cross-section.

The point SU1 (Figure~\ref{fig:SU13}) is situated in the so-called Coannihilation region. It is characterised by a high degree of degeneracy between neutralino and slepton masses: 
\[M_{\neu_2}-M_{\tilde \ell_L},\, M_{\tilde \ell_R}-M_{\neu_1},\, M_{\neu_2}-M_{\tilde\tau_2},\, M_{\tilde\tau_1}-M_{\neu_1}
=\mathscr O(10\,{\rm GeV})\,,{\rm~where~}\ell=\mu,\,e\,.\]
As a consequence of this degeneracy, a large fraction of the leptonic signal will be comprised of taus from stau decays. Therefore a high efficiency for reconstructing soft leptons will be required.

The point SU8 (Figure~\ref{fig:SU48}) is similar to SU1: it also belongs to the Coannihilation region and features similar mass degeneracies. The main difference is the higher mass scale of squarks and sleptons, which results in a lower cross-section and larger mass differences. Further, SU8 features a much higher $\tan\beta$ parameter and a low stop mass, which enhances top quark production.

The mSUGRA parameters and LO cross-sections~\cite{bib:msugraPoints} of all SU$x$ points are summarised in Table~\ref{tab:susy_points}.

\begin{table}
\begin{center}
\begin{small}
\begin{tabular}{lccccccr}
\hline
Process & $M_0$ [GeV]	& \Mhalf [GeV]	& $A_0$ [GeV]	& $\tan\beta$	& $\arg\mu$	& $\sigma$ [pb] & Region \\
\hline\hline	
SU1 	& 70		& 350		& 0 		& 10 	 	& +		& 7.43	& Coannihilation\\
SU2 	& 3550  	& 300		& 0 		& 10 	 	& +		& 4.86	& Focus\\
SU3 	& 100  		& 300		& $-300$	& 6 	 	& +		& 18.59 & Bulk\\
SU4 	& 200  		& 160		& $-400$	& 10 	 	& +		& 262 	& Low Mass\\
SU8 	& 210  		& 360		& 0		& 40 	 	& +		& 6.44 	& Coannihilation\\
\hline
\end{tabular}
\end{small}
\end{center}
\vspace{\cDistHalf}
\caption[The SU$x$ ATLAS benchmark points, together with mSUGRA parameters and total production cross-sections at LO.]{\small\label{tab:susy_points}
The SU$x$ ATLAS benchmark points, together with mSUGRA parameters and total production cross-sections~\cite{bib:msugraPoints} at LO.
\vspace{\cDistHalf}
}
\end{table}

\section{Trilepton Final States\newline in the Massive Sparton Scenario}\label{sec:signatureDirectGaugino}
As already touched upon in the introduction, this analysis is particularly geared towards investigating the discovery potential for SUSY in the case where \textit{all} strongly interacting supersymmetric particles are very massive. Arguably, multilepton final states will be favoured to multijet final states for a discovery in such a scenario. Among mSUGRA ATLAS benchmark points, the one which most closely matches this scenario is SU2 (Figure~\ref{fig:SU2}). This point is characterised by a very high sfermion mass scale $M_0$ and moderate $\Mhalf$ (Table~\ref{tab:susy_points}). The configuration of $M_0$ and $\Mhalf$ in SU2 dramatically changes the balance of the production mechanisms compared to SU3: whereas squarks will be strongly suppressed due to the high $M_0$ scale, gluino pair-production will retain approximately the same fraction of the total SUSY cross-section\footnote{Gluino masses are fairly similar in SU2 and SU3: $m_{\tilde g}^{\rm SU2}=856\,\GeV \simeq m_{\tilde g}^{\rm SU3}=717\,\GeV$}, and electroweak direct gaugino pair-production from quark-antiquark annihilation will become important ($\gtrsim$50\% of the total SUSY cross-section). As a result, strong SUSY production will be suppressed, and associated chargino-neutralino production, mainly $\cha_1\neu_2$ and $\cha_1\neu_3$, become significant:
\begin{center}
\vspace{\cDistHalf}
\includegraphics[height=3.5cm,clip]{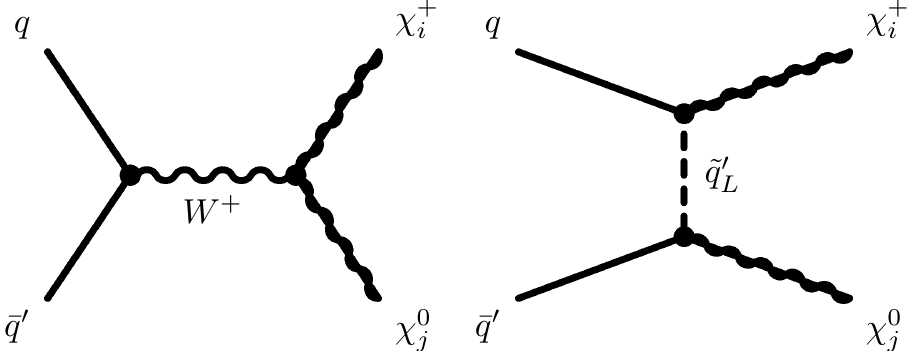}
\vspace{\cDistHalf}
\end{center}
The $t$-channel production is suppressed due to the high squark and slepton masses. 

\begin{figure}
\begin{center}
\vspace{\cDistHalf}
\includegraphics[width=7.5cm,clip=true]{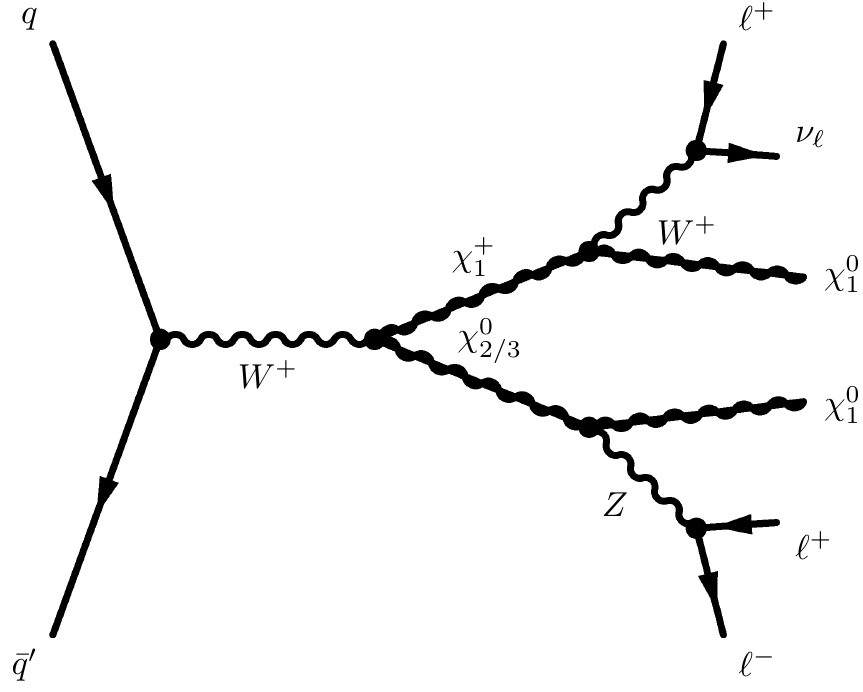}\qquad
\vspace{\cDist}
\end{center}
\caption[A possible production diagram of the trilepton final state via direct gaugino pair-production]{\label{fig:SU2signal}
A possible production diagram of the trilepton final state via direct gaugino\linebreak pair-production.
\vspace{\cDistHalf}
}
\end{figure}

% Since the effective mass scale defined as $M_{\rm eff}\equiv\sum_{i=1}^4 \pt^{j_i}+\sum_{i} \pt^{\ell_i}+\met$ of such events is rather low, the best strategy to discover them is to look for fully leptonic decays mediated by off-shell electroweak bosons or sleptons as already discussed in the subsection above. 
% Slepton mediated processes will be highly suppressed due to the high slepton mass. 
The resulting signature for chargino-neutralino pair-production is very similar to the one discussed in Section~\ref{sec:signatureGeneral}: an opposite sign same flavour (OSSF) lepton pair, an additional lepton, and significant \met\ is expected. A possible production diagram of the trilepton final state via direct gaugino pair-production is illustrated in Figure~\ref{fig:SU2signal}.

%\nopagebreak[5]

However, despite high squark masses, SU2 is not exactly the massive sparton scenario, as gluinos are still moderately light ($\sim$800\,GeV). 
%This 
% and the dominance of gluon and sea quark probability density functions in the Bj$\ddot{\rm o}$rken's~$x$ region of $x\simeq0.1$ are 
%is the reason for a significant fraction of the trilepton SU2 signal to be comprised of decay chains starting with a gluino pair.
%
In order to quantify the performance of the outlined search strategy for the case where gluinos are very massive ($\gtrsim3$\,TeV), the discovery potential for {\em only} direct gaugino pair-production is examined, which is extracted from the inclusive SU2 SUSY sample by filtering at MC truth level. In the following, this will be synonymously referred to as {\em massive sparton scenario}. 

Associated chargino-neutralino production is also important {\em outside} of the massive sparton scenario, as it has the potential to isolate a particular decay chain. This offers a possibility to determine many SUSY parameters, like the relative gaugino mass spectrum, the composition of their mass eigenstates and their couplings, and most importantly, the mass of the LSP which defines an absolute mass scale~\cite{bib:massSpectrumMeas1,bib:massSpectrumMeas2,bib:massSpectrumMeas3}. The leptonic decay modes of associated chargino-neutralino production can be isolated by adding a jet veto to the selection criteria.

% In order to quantify the performance of our search strategy for the case where gluinos are very heavy ($\gtrsim3$\,TeV) we examine the discovery potential for direct gaugino production, which is isolated from our inclusive SUSY sample by adding a jet veto to the selection criteria. In the following, this will be referred to as ``massive sparton scenario''. Associated chargino-neutralino production is also important for another reason: it has the potential to isolate a particular decay chain, from which subequently many SUSY parameters, like the gaugino mass spectrum, the composition of their mass eigenstates and their couplings, can be determined.

\section{Backgrounds to the Trilepton Final State}\label{sec:signatureBgr}

Based on the discussion above several backgrounds can be identified. They can be classified in two categories -- {\em physics} and {\em instrumental} backgrounds. Physics backgrounds mimic the signal signature due to the generically similar final state, whereas instrumental ones do so due to mis-measurements. The backgrounds investigated in this analysis are briefly described below:
\begin{description}
\vspace{-2mm}
\setlength{\itemsep}{0mm}
 \item[$\mathbf{WZ:}$] This is the only irreducible physics background. It can be controlled by vetoing events with $\mossf$ close to $m_{Z}$ or $m_{\gamma}\equiv0$, where $\mossf$ is the four-vector mass of the OSSF lepton pair from the $Z/\gamma^*$ decay. However, this strategy has limitations due to the Drell-Yan interference structure;

 \item[$\mathbf{WW:}$] The dileptonic decay mode contains two leptons and significant missing transverse energy due to escaping neutrinos. A third lepton may be faked from initial state radiation;
 %While providing for significant $\met$ due to escaping neutrinos, even for fully leptonic decays this background lacks a third lepton. It can be faked by a jet from initial state radiation.

 \item[$\mathbf{ZZ:}$] The situation with $ZZ$ is opposite to $WW$: in fully leptonic decay modes of the $Z$'s four leptons are produced, while no $\met$ is expected. Both leptons and $\met$ can be provided by decays to $\tau$ leptons, however, the lepton $\pt$ will typically be rather low;

 \item[$\mathbf{Z\gamma:}$] In the high luminosity environment of ATLAS electrons can be faked by photons and random tracks, or by narrow photon conversions;

 \item[$\mathbf{Zb:}$] The production of $Z$-bosons in combination with jets from the hard matrix element or initial state radiation is one of the major instrumental backgrounds. A lepton from semileptonic heavy quark ($c,b$) decay may enter the selection if it passes the track and calorimeter isolation criteria. Leptons from such decays will be referred to as {\em secondary} leptons in the following, as opposed to {\em prompt} leptons, which come from the hard matrix element. Additionally, light quark ($u,d,s$) jets with a significant $\pi^0$ content may mimic electromagnetic showers. Simulations suggest that the probability of this is $\mathscr O(10)$ smaller than for a heavy quark jet to produce an isolated secondary lepton, so only the latter is simulated in this analysis\footnote{An alternative sample with associated production of both light quark and heavy quark jets is not included, as it cannot be combined with the $Zb$ background because of double-counting. At the time as this analysis was performed (and more critically, in the \Athena\ release used), there was no $Zc$ MC sample available at ATLAS.};
 %the probability for this is by a factor of 10-100 smaller than for a heavy jet to produce an isolated secondary lepton.
 
 %For a lepton coming from a $b$-jet decay,  the probability to fake its isolation is by a factor of 10-100 higher than for a light jet to fake a lepton. Therefore, only the $Zb$ background is considered in this analysis. An alternative sample with associated production of both light and heavy jets is not included, as it cannot be combined with the $Zb$ background because of double-counting.

 \item[$\mathbf{t\bar{t}:}$] Two leptons come from semi-leptonic decays of the top quarks, and an additional lepton can enter the selection if a secondary lepton from a $b$-decay passes the isolation criteria or if it is faked by a jet.
\vspace{-2mm}
\end{description}

%% file: Samples/Samples.tex
Since the LHC did not deliver any $pp$ collisions so far, this analysis is based on simulated Monte Carlo (MC\glossary{name=MC,description=Monte Carlo (sometimes used synomymously with {\em simulated Monte Carlo events})}) events with full detector simulation. The cross sections with and without generation filter efficiencies, the $k$-factors for taking into account of most recent available calculations, the average weights, the available number of MC events, and the corresponding integrated luminosities for the investigated SUSY benchmark points together with backgrounds are listed in Table~\ref{tab:samples}. The references for this information are given in the last column of the table. It has been checked~\cite{bib:kfactorSUSY}, that for SU2 and SU3 the $k$-factors for direct gaugino production are around $k=1.25-1.3$. The $k$-factors for other significant SUSY processes typically lie between $k=1.3$ and $k=1.7$. Therefore, conservatively, for all SU$x$ points the value of $k\equiv1.25$ has been used.

\begin{table}
\begin{footnotesize}
\hspace{-0.5cm}
\begin{tabular}{lcccccccc}
\hline
Process 	& sample \# 
			& $\sigma$ [pb]
				& $\sigma\!\times\!\eps_{\rm gen}$ [pb] 
		        		& $k$-factor
						& $\langle$w$\rangle$
							& MC events     
								& $\intlumi$ [\fb]
										& Reference \\
\hline\hline
SU1 		& 5401	& 7.43	& 7.43  & 1.25	& 1  & 196,350       & 21.1  & \cite{bib:msugraPoints, bib:kfactorSUSY} \\
SU2 		& 5402	& 4.86	& 4.86  & 1.25	& 1  & 49,700        & 8.18  & \cite{bib:msugraPoints, bib:kfactorSUSY} \\
SU3 		& 5403	& 18.59	& 18.59 & 1.25	& 1  & 483,250       & 20.8  & \cite{bib:msugraPoints, bib:kfactorSUSY} \\
SU4 		& 5404	& 262	& 262   & 1.25	& 1  & 194,850       & 0.59  & \cite{bib:msugraPoints, bib:kfactorSUSY} \\
SU8 		& 5408	& 6.44	& 6.44  & 1.25	& 1  & 47,750        & 5.93  & \cite{bib:msugraPoints, bib:kfactorSUSY} \\\hline
%SU3$\chi$	& 6402	& 18.59	& 1.8	& 1.25	& 1  & 27,600        & 12.3  & \cite{bib:msugraPoints, bib:kfactorSUSY} \\
%SU2 (Pile Up)	& 5402	& 4.86  & 1	&     ,000  4   & 9.9 \\
$WW$ 		& 5985	& 70	& 24.5  & 1.67	& 1  & 50,000        & 1.22  & \cite{bib:gaugeBosonPairs, bib:dibosonHN, bib:zhou, bib:xsecRecomm} \\
$WZ$ 		& 5986	& 27	& 7.8   & 2.05	& 1  & 47,900        & 2.98  & \cite{bib:gaugeBosonPairs, bib:dibosonHN, bib:zhou, bib:xsecRecomm} \\
$ZZ$ 		& 5987	& 11	& 2.1   & 1.88	& 1  & 49,800        & 12.7  & \cite{bib:gaugeBosonPairs, bib:dibosonHN, bib:zhou, bib:xsecRecomm} \\
$Z\gamma$	& 5900	& 3.8	& 2.58  & 1.30	& 1  & 10,000        & 2.98  & \cite{bib:zgamma, bib:dibosonHN, bib:zhou} \\
$Zb$		& 5178	& 205	& 154	& 1	& 0.66	& 115,000       & 0.49 & \cite{bib:zb} \\
$\ttbar$	& 5200	& 461	& 461	& 1	& 0.73	& 564,350       & 0.89 & \cite{bib:bonciani, bib:kidonakis, bib:talkTPkarl, bib:topDataset} \\
\hline
\end{tabular}
\end{footnotesize}
\caption[Main parameters of the simulated MC events used in this analysis]{\label{tab:samples}
Cross sections, cross sections including generation efficiencies, $k$-factors, average weights, number of MC events available for analysis, corresponding integrated luminosities (including generation efficiencies, $k$-factors, negative weights), and references for signal and background processes. The cross section $\sigma$ given for \ttbar{} is at NLO including leading logarithmic corrections, other cross sections are LO.
%% ``SU3$\chi$'' stands for SU3 direct gaugino and gluino pair production.
\vspace{\cDistHalf}
}
\end{table}
 
\section{Software Used}
As already mentioned above, full ATLAS detector simulation with GEANT~4~\cite{bib:geant} was used, together with the ATLAS analysis framework, \textsc{Athena} in release~12.0.6~\cite{bib:athena}. In the case of the signal, the initial event was generated using the Herwig~6.5 MC event generator~\cite{bib:herwig}.
%, followed by Pythia~6.4~\cite{bib:pythia} for showering. 
The additional proton-proton interactions which do not constitute the hard scattering process of interest -- the {\em underlying event} -- were simulated with the Jimmy plug-in to Herwig~\cite{bib:jimmy} using the `ATLAS tune'~\cite{bib:Moraes:2007rq}. Separate samples with additional pile-up interactions ($\bra n\ket\cong2$) were simulated for $\instlumi{33}$. Photon production and $\tau$-lepton decays were implemented with PHOTOS~\cite{bib:photos} and TAUOLA~\cite{bib:tauola}. The sparticle mass spectrum together with the coupling constants and the decay branching ratios was generated using Isajet 7.71~\cite{bib:isajet}, and then interfaced to Herwig~\cite{bib:herwigInterface}.
% via the Les Houches Accords format~\cite{bib:lesHouches}. 
The $WW$, $WZ$ and $ZZ$ backgrounds were generated in the same way as the signal (except for Isajet). The latter two feature the full Drell-Yan interference structure, however with a lower bound on the mass of the  OSSF lepton pair of $\mossf=20$\,GeV. In case of $ZZ$, the pair can come from the decay of any $Z$ boson in the event. The $Z\gamma$ background was entirely generated by Pythia~6.4~\cite{bib:pythia}. The main background, \ttbar, is simulated using MC@NLO~\cite{bib:mc@nlo} combined with Herwig, and includes Next-to-Leading Order (NLO\glossary{name=NLO,description=Next-to-Leading Order}) corrections to the matrix element. The simulation of the $Zb$ process was done with the AcerMC~3.3~\cite{bib:acerMC} generator, followed by Pythia~6.4 for showering. Unlike many others, the AcerMC generator simulates $b$-quarks coming either from the hard matrix element or from initial state radiation. Like $WZ$, $Zb$ contains the $Z/\gamma^*$ interference, however with an $\mossf$ lower bound of 30\,GeV.

The high level part of the analysis was carried out outside of \Athena{}, employing solely the ROOT framework~\cite{bib:root}. This analysis pioneered at ATLAS to become completely Structured Athena-Aware Ntuple (SAN\glossary{name=SAN,description=Structured Athena-Aware Ntuple}) based. The SAN format~\cite{bib:san} is a physics-object oriented Derived Physics Data format to decouple the final analysis stage from \Athena{}. The SAN production was done on the Grid, using the backend applications Ganga~4.3~\cite{bib:ganga} and pAthena~\cite{bib:pathena}.

%The previous version version of this analysis~\cite{bib:seshadri} used the High Level Analysis package~\cite{bib:OxfordHLA}, a physics-object oriented Derived Physics Data format devised at Oxford to decouple the final analysis stage from \Athena{}. With the transition to release 12 of \Athena, and the emergence of the centrally maintained Structured Athena-Aware Ntuple (SAN) format~\cite{bib:san} following the same strategy as High Level Analysis package, this analysis pioneered at ATLAS to become completely SAN based. The SAN production was done on the grid, using the backend applications Ganga~4.3 and pAthena.
%Ganga~4.3~\cite{bib:ganga} and pAthena~\cite{bib:pathena}.

%% file: Preselection/Preselection.tex
Before any analysis of the trilepton final state can be done, physics objects like electrons, muons et cetera need to be carefully defined. This step is detailed below.

\section{Muon Preselection}\label{sec:presMu}
To reconstruct muons, the \texttt{Staco}~\cite{bib:atlasTDR2,bib:muonsCSC,bib:muonsCSC2} muon reconstruction algorithm was used. This algorithm first builds two-dimensional track segments in each of the three Muon Spectrometer layers, and then proceeds to form MS tracks from them. These tracks are extrapolated towards the beamline by the {\tt Muonboy}~\cite{bib:muonsCSC2} algorithm, and can be expressed using the canonical five track parameters $q$ at the perigee~(cf.~Subsection~\ref{ssec:trackingEDM}). In the extrapolation, both MCS and the expected enegry loss in the calorimeter are taken into account. The {\tt Staco} algorithm matches MS tracks with tracks independently reconstructed in the ID using
\[\chi^2_{\rm match}\equiv\left(q_{\rm ID}-q_{\rm MS}\right)^T\left(C_{\rm ID}-C_{\rm MS}\right)^{-1}\left(q_{\rm ID}-q_{\rm MS}\right)\Big|_{\rm perigee}\]
as a figure of merit. Here, $C$ is the covariance matrix of $q$. Once the best match is found, the track parameters of the full muon track are calculated by statistically combining $q_{\rm ID,\,MS}$:
\[q_{\rm Staco}\Big|_{\rm perigee}\equiv\left(C_{\rm ID}^{-1}-C_{\rm MS}^{-1}\right)^{-1}\left(C_{\rm ID}^{-1}q_{\rm ID}-C_{\rm MS}^{-1}q_{\rm MS}\right)\Big|_{\rm perigee}\,.\]

To improve the momentum measurement and obtain a clean muon sample, only muons inside the coverage region of the SCT and the pixel detector, $|\eta_\mu|<2.5$, were considered. To find a good compromise between a clean selection and a high efficiency, $\pt^{\mu}>10$\,GeV is required for all muons. %Only muons which have two or more matching track segments reconstructed in the inner detector and in the muon spectrometer are considered. 
For the matching of the ID and MS tracks, $\chisq_{\rm match}>0$ is required\footnote{Cases with $\chisq_{\rm match}=0$ are the ones where the \texttt{Staco} algorithm did not converge.}. It has been shown that the fake rate is sufficiently small even without cutting on the maximum value of $\chisq_{\rm match}$~\cite{bib:stacoMuonsFake}. For this analysis, demanding $\chisq_{\rm match}<100$ would reduce the statistical significance by approximately~5\%, and this cut is therefore not applied. For a small fraction of muons ($\sim$1\%) there are several inner detector track segments which do match with the muon spectrometer track within the allowed $\chisq_{\rm match}$ region. In such a case, only muons with the best match of the segments are considered. To suppress secondary muons from heavy quark decays, leptons are required to be calorimeter-isolated, with nearby energy $I_{0.2}^{\rm cal}<10$\,GeV inside a $\Delta R=0.2$ cone\footnote{\dR{} is defined as distance in $\eta\times\phi$, id est $\dR\equiv\sqrt{\Delta\eta^2+\Delta\phi^2}.$}.

\begin{figure}
\begin{center}
\includegraphics[width=7.9cm,clip]{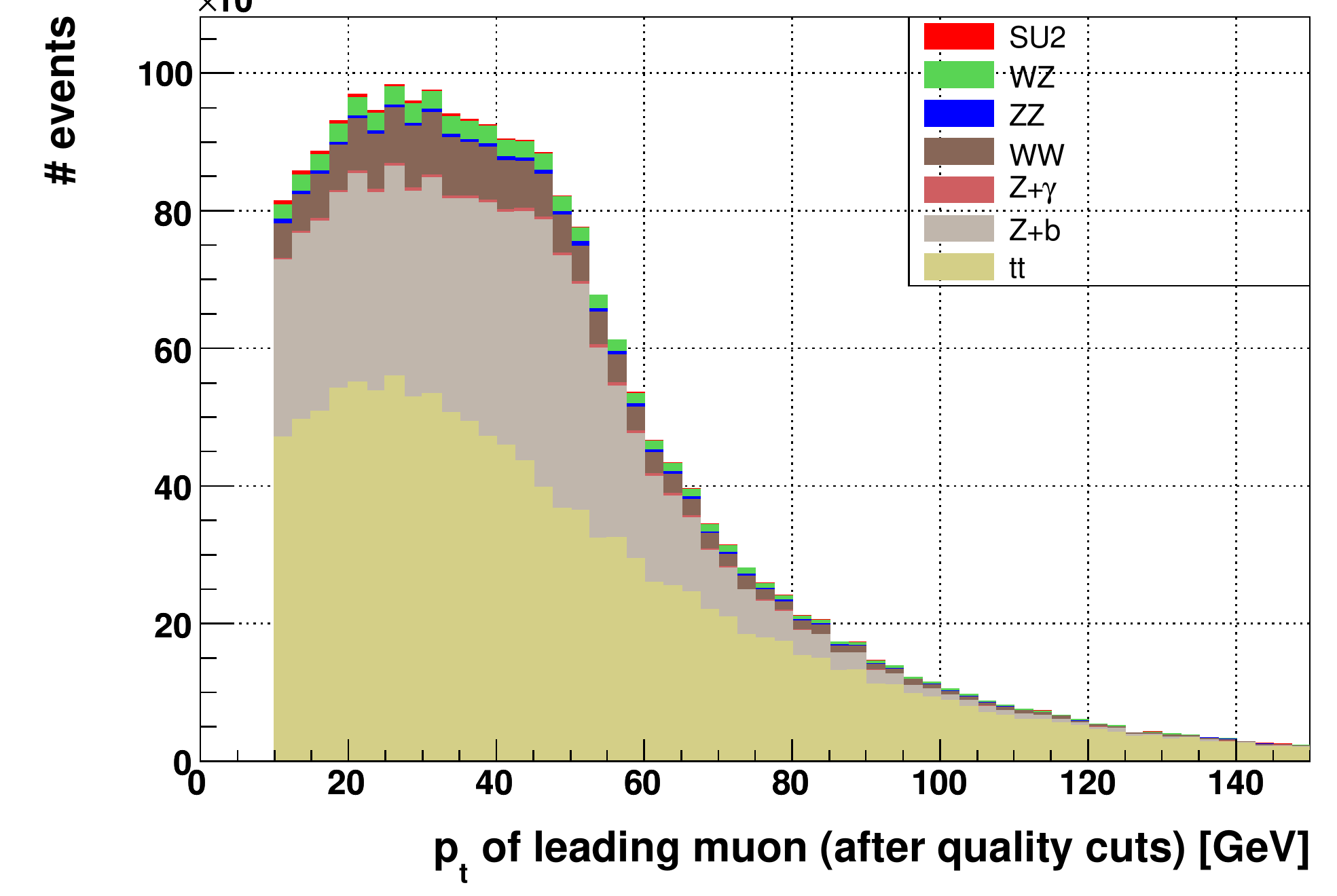}
\includegraphics[width=7.9cm,clip]{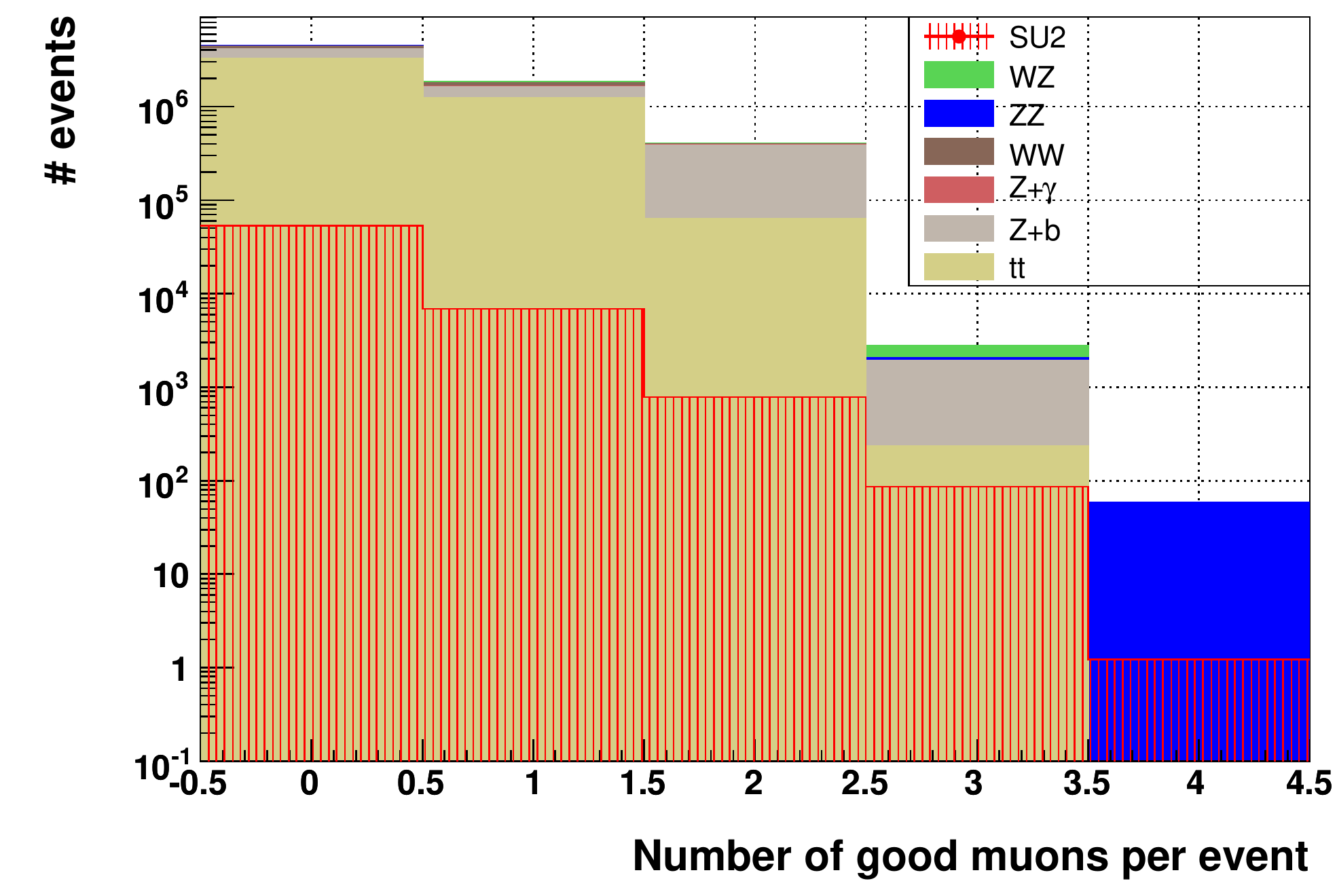}
\vspace{\cDist}
\vspace{\cDistHalf}
\end{center}
\caption[Leading muon $\pt^{\mu_1}$ and multiplicity of muons per event for SU2]{\label{fig:mPreselection}
Transverse momentum distribution of the leading muon $\pt^{\mu_1}$ {\bf(left)} and multiplicity of muons per event {\bf(right)} for SU2 after quality cuts summarised in Section~\ref{sec:presMu}. The individual backgrounds are colour-coded and ``stacked'' on top of each other. The signal is shown ``stacked'' on top of the background in the \pt-distribution, and separately in the foreground for the muon multiplicity distributions.
\vspace{\cDistHalf}
}
\end{figure}

% Up to this point, the standard lepton definition as agreed upon by the ATLAS SUSY working group has been used~\cite{bib:preselectionSUSYWG}. 
On top of that, this analysis uses further cuts to account for the special requirements on leptons in the trilepton channel. In order to reject muons from {\em heavy flavour} jets\footnote{In this document, {\em heavy flavour} jets refer to $c$- and $b$-jets, unless stated otherwise.}, muons which are found within a cone of $\dR=0.4$ of any reconstructed jet are not used in this analysis, however the event as a whole is kept. Further, all selected muons in an event are to be separated from each other by at least $\dR=0.2$. Muon pairs where this requirement is not fulfilled are not considered further, as they are likely to come from decays of heavy mesons like $J/\psi$ and $\Upsilon$, or the $\gamma$ resonance.

The transverse momentum distribution of the leading muon $\pt^{\mu_1}$ and the muon multiplicity per event after all preselection cuts detailed above are shown in Figure~\ref{fig:mPreselection} for the signal and background processes.

\section{Electron Preselection}\label{sec:presEl}
In this analysis, electrons reconstructed by the \texttt{eGamma}~\cite{bib:eGamma,bib:electronsCSC} algorithm are used. This algorithm starts from EM towers with $\et\gtrsim3\,\GeV$ as seeds to reconstruct EM clusters. These clusters are then matched within $\eta\times\phi=0.05\times0.1$ to ID tracks which were not explicitly reconstructed as gamma conversion tracks. Electrons reconstructed {\em only} by the \texttt{SoftE}~\cite{bib:eGamma,bib:electronsCSC} algorithm  are not considered in this analysis. The reason is that {\tt SoftE} is optimised for reconstructing electrons with transverse momenta of a few GeV by using ID tracks as seeds, which results in a higher fake rate from {\em light flavour} jets\footnote{In this document, {\em light flavour} jets refer to $u$-, $d$-, $s$-, and gluon-jets, unless stated otherwise.}. Similarly to the muons, electrons are required to be inside the coverage region of the electromagnetic calo\-rimeter (and thus the tracker), $|\eta_e|<2.5$. The transverse momentum threshold is $\pt^{e}>15$\,GeV for all electrons. To ensure a high rejection power for light flavour quark jets, sophisticated electron definition criteria developed by the eGamma group need to be passed. These criteria are summarised in form of a bit-word referred to as \texttt{isEM()}~\cite{bib:eGamma}. In this analysis, ``medium'' electrons are used, which are to satisfy all of the \texttt{isEM()} criteria but two: the requirement of a minimum number of TRT hits and pattern; and the \eOverP{} cut\footnote{In technical terms, this means they must satisfy the (hexadecimal) bit pattern 3FF.}. The unused TRT criterion provides little rejection power for fakes, but has a serious impact on efficiency of genuine electrons ($\sim$10\%). The dropped \eOverP{} requirement also reduces efficiency without decreasing the electron fake rate at sufficient level to be useful for this analysis. The remaining criteria used in the selection of medium electrons are summarised in Table~\ref{tab:eGamma}.
%connected to systematic effects for a misaligned tracking system in the first years of running of the LHC. This is due to a possible $\eta$-, $\phi$-assymmetry in the measurement of the momentum with the tracking system due to weak deformation modes of the inner detector.
Since some form of calorimeter isolation is provided by the medium electron definition, no {\em additional} calorimeter isolation was required for electrons. For a detailed discussion see Chapter~\ref{chp:selection}. It should be mentioned, that the calculation of $I_{0.2}^{\rm cal}$ for electrons is affected by a bug in \Athena{} software: the $\eta\times\phi$ window summed over to obtain the energy around the electron is enlarged from $3\times5$ to $3\times10$ cells. Nevertheless, the $I_{0.2}^{\rm cal}$ variable still provides a useful handle to reject electrons from semileptonic heavy flavour jet decays.

\begin{table}
\begin{footnotesize}
\begin{center}
\hspace{-0.5cm}
\begin{tabular}{lcl}
\hline
Type & Name & Description \\
\hline\hline
Acceptance & n/a & $|\eta_{\rm det}|<2.47$ \\
\hline
hadronic leakage & n/a & Ratio of $\et^{\rm 1st\,sampling\,hadronic}$ over total $\et^{\rm EM\,cluster}$ \\
\hline
\multirow{6}*{1$^{\rm st}$ EM sampling} & \multirow{2}*{$\Delta E_s$} & Difference of $\et^{\rm second\mbox-highest\,strip}$ and \\
&& \hspace{2.7cm} $\min_{\rm all\,strip\,btw.\,highest\,and\,second\mbox-highest\,strip}\left\{\et^{\rm cell}\right\}$ \\
& $R_{\rm max2}$ & Ratio of $\et^{\rm second\mbox-highest\,strip}$ over $\et^{\rm EM\,cluster}$ \\
& $w_{\rm tot}$ & Total shower width\\
& $w_{s3}$ & Shower width for three strips around strip with $\et^{\rm highest\,strip}$ \\
& $F_{\rm side}$ & Ratio of $\sum_{\rm 3\,central\,strips}\et^{\rm strip}$ over $\sum_{\rm7\,central\,strips}\et^{\rm strip}$ \\
\hline
 & n/a & lateral shower width \\
2$^{\rm nd}$ EM sampling & $R_\eta$ & Ratio of $\sum_{3\times7}\et^{\rm cell}$ over $\sum_{7\times7}\et^{\rm cell}$ (sum range defined in $\eta\times\phi$) \\
& $R_\phi$ & Ratio of $\sum_{3\times3}\et^{\rm cell}$ over $\sum_{3\times7}\et^{\rm cell}$ (sum range defined in $\eta\times\phi$) \\
\hline
Isolation & n/a & Ratio of $I_{0.2}^{\rm cal}\equiv\sum_{\Delta R<0.2}\et$ over total $\et^{\rm EM\,cluster}$ \\
\hline
\end{tabular}
\end{center}
\end{footnotesize}
\vspace{\cDistHalf}
\caption[The medium electron selection criteria used in this analysis.]{\label{tab:eGamma}
The medium electron selection criteria used in this analysis. If no explicit cut value is given, the cut value is parametrised in terms is $\eta$ and \pt. A more detailed description can be found in~\cite{bib:electronsCSC}.
\vspace{\cDistHalf}
}
\end{table}

%Finally, the same cut on calorimeter isolation as for the muons, $I_{0.2}^{\rm cal}<10$\,GeV, is applied. It should be mentioned, that in the calculation of $I_{0.2}^{\rm cal}$ a bug was introduced in the \Athena{} software, enlarging the $\eta\times\phi$ window summed over to obtain the energy around the electron from $3\times5$ to $3\times10$ cells. Despite this bug, calorimeter isolation $I_{0.2}^{\rm cal}$ still provides good discrimination power to reject secondary leptons from $b$ decays. %, and its further usage is encouraged by the ATLAS SUSY working group.

\begin{figure}
\begin{center}
\includegraphics[width=7.9cm,clip]{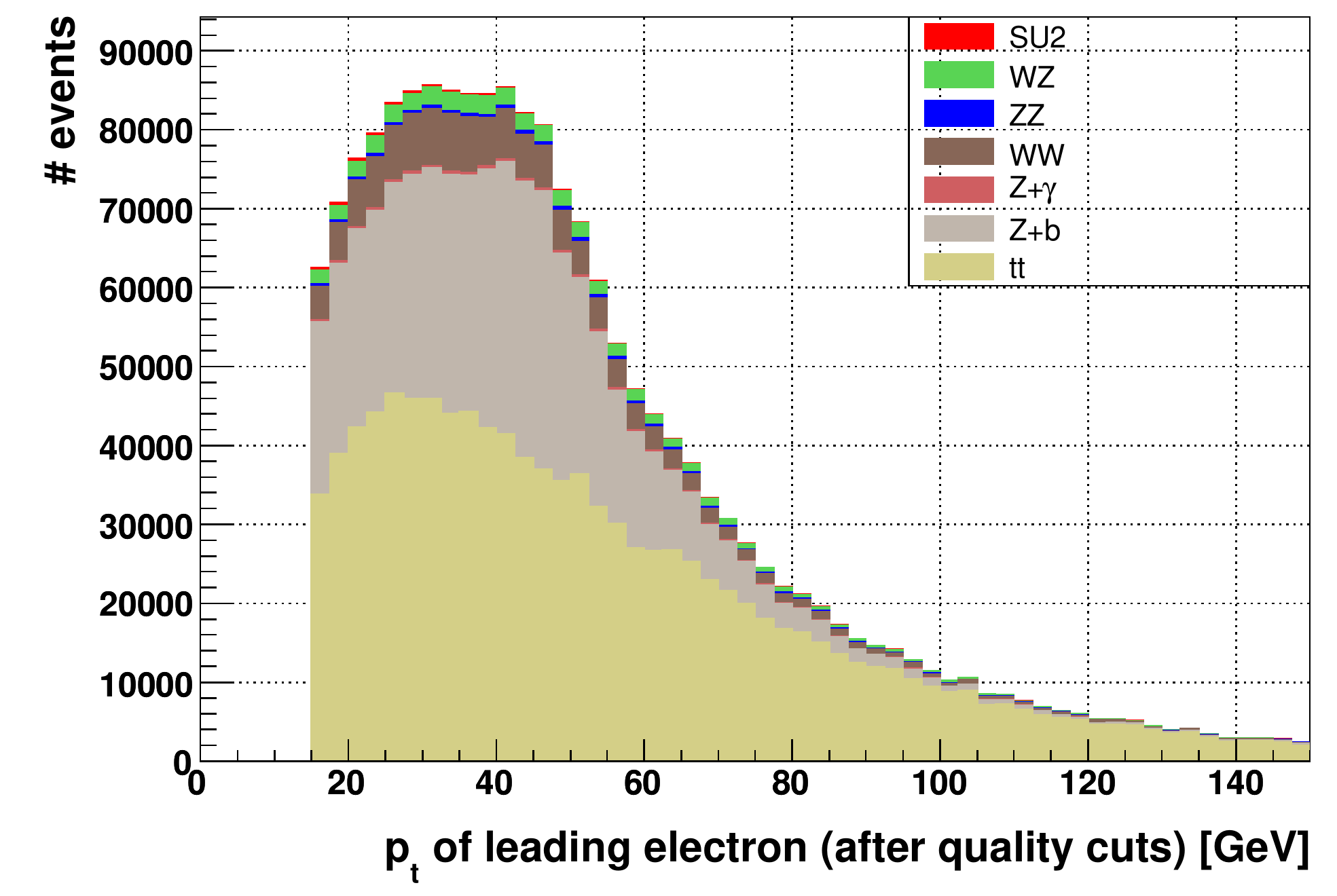}
\includegraphics[width=7.9cm,clip]{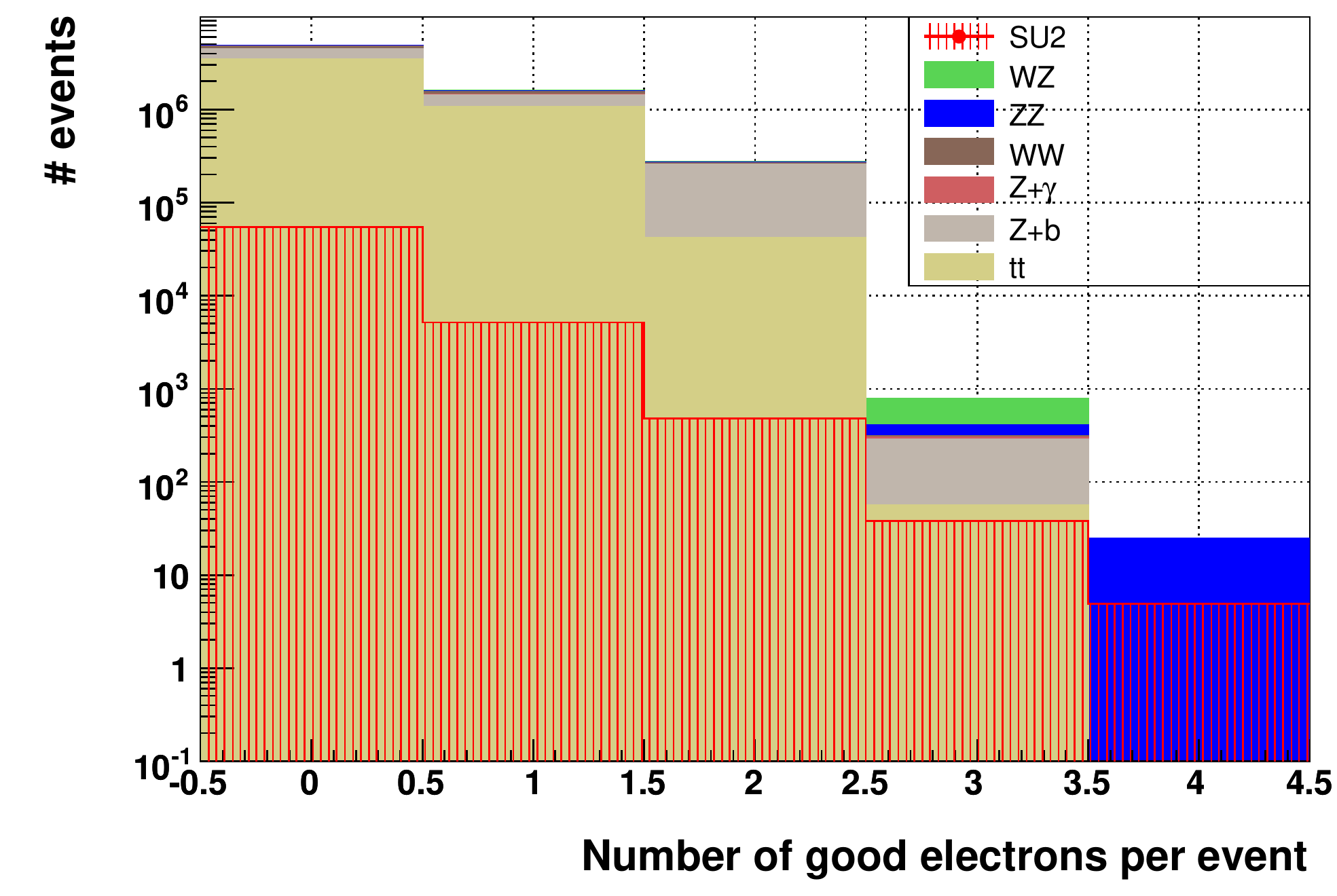}
\vspace{\cDist}
\vspace{\cDistHalf}
\end{center}
\caption[Leading electron $\pt^{e_1}$ and multiplicity of electrons per event for SU2]{\label{fig:ePreselection}
Transverse momentum distribution of the leading electron $\pt^{e_1}$ {\bf(left)} and multiplicity of electrons per event{\bf~(right)}  for SU2. Both are after quality cuts summarised in Section~\ref{sec:presEl}.
\vspace{\cDistHalf}
}
\end{figure}

Electrons are required to be separated from each other by at least $\dR=0.2$\,. Cases where 2 electrons have no separation, id est $\dR\equiv0$, are treated separately: due to a subtle bug in the electron reconstruction~\cite{bib:doubledElectrons}, it can happen that two slightly different clusters (typically they share more than 90\% of their cells) are associated with the same track. Thus, they are reconstructed as {\em two} electrons with slightly different energies, but identical $\eta,\phi$-coordinates which are given by the (same) associated track. Therefore, if $\Delta R\equiv0$ and $(\pt^{e_1}-\pt^{e_2})/(\pt^{e_1}+\pt^{e_2})<0.1$, only the electron with the higher transverse momentum is kept. In case the latter requirement is not met, both electrons are rejected, as such cases are not understood yet. As was done for the muons, electrons closer than $\Delta R=0.4$ to a jet are vetoed in order to discriminate against secondary leptons from heavy flavour decays. 

Figure~\ref{fig:ePreselection} depicts the resulting transverse momentum distribution of the leading electron $\pt^{e_1}$ and the electron multiplicity per event after the above cuts.

\section{Jet Preselection}\label{sec:presJt}
In this analysis, \texttt{towerCone4} jets~\cite{bib:jetsCSC} are used. They are defined as the calibrated sum of energy in cells belonging to projective calorimeter towers defined by $\dR=0.4$ around the seed. %FIXME: towerCone4 reference!
%Our jet veto introduced above is intended for jets from the hard scattering process rather than Initial State Radiation (ISR), and so is restricted to $|\eta_j|<2.5$. Therefore we do not select jets in the forward regions where the infrared divergence of QCD leads to a high probability of ISR.
%For the reasons described in Section~\ref{sec:signatureDirectGaugino}, we are interested in the case where there is no hard jet in the final state coming from the decay of a supersymmetric particle. Since jets from Initial State Radiation (ISR) begin to dominate at high rapidity, any jets are required to satisfy $|\eta_j|<2.5$.
As outlined in Section~\ref{sec:signatureDirectGaugino}, one version of this analysis aims to select for events where there is no hard jet in the final state coming from the decay of any {\em supersymmetric} particle. Since jets from Initial State Radiation (ISR\glossary{name=ISR,description=Initial State Radiation}) tend to dominate at high rapidity, only jets in $|\eta_j|<2.5$ are considered. Finally, jets are required to satisfy  $\et>10$\,GeV. 

%Since the jet veto mentioned above does not aim at initial state radiation but rather jets from hard SUSY processes, they are required to come from $|\eta_{j}|<2.5$, as initial state radiation will be found rather in the high pseudorapidity region due to infrared divergence of QCD. 

\begin{figure}
\begin{center}
\includegraphics[width=7.9cm,clip]{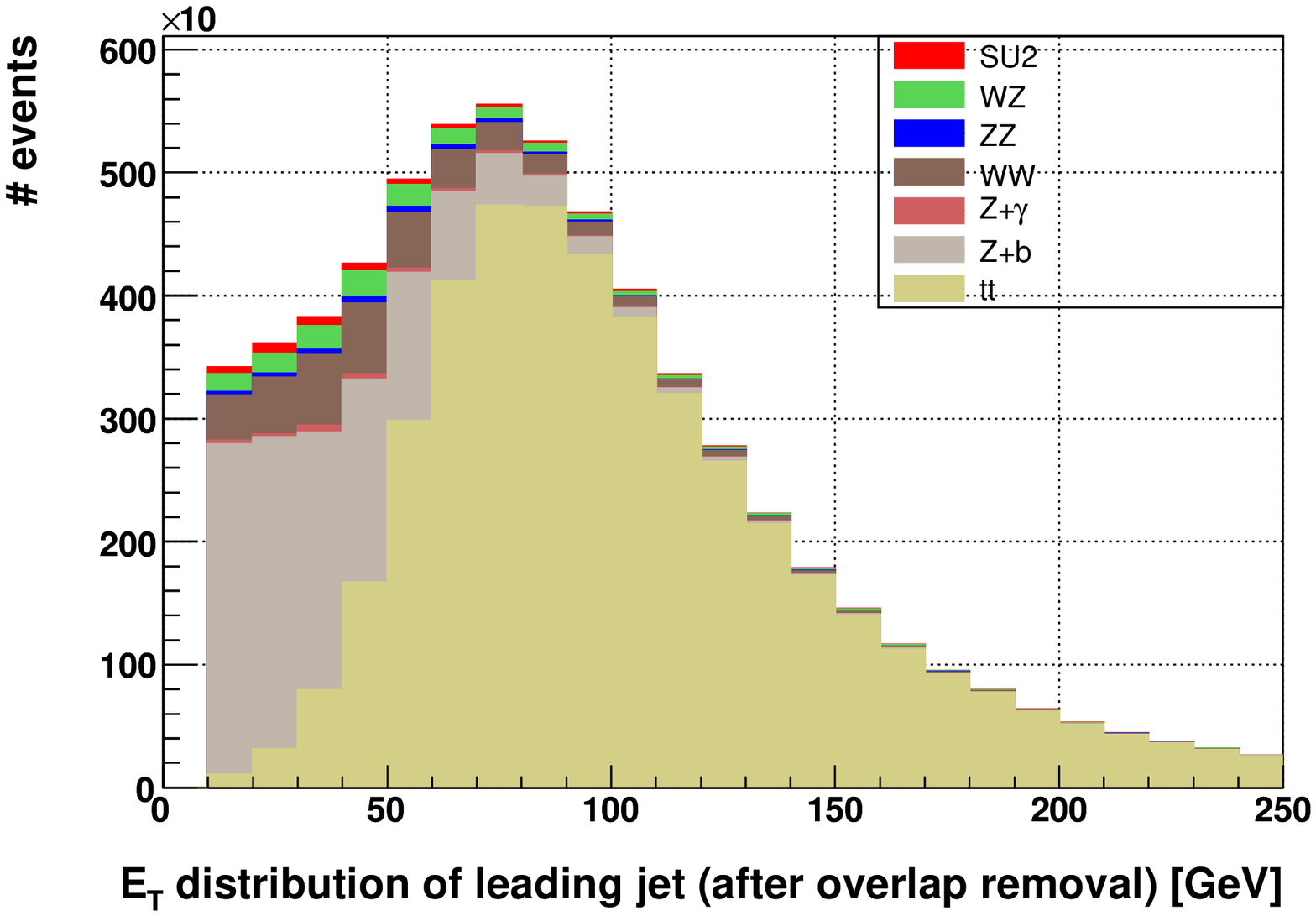}
\includegraphics[width=7.9cm,clip]{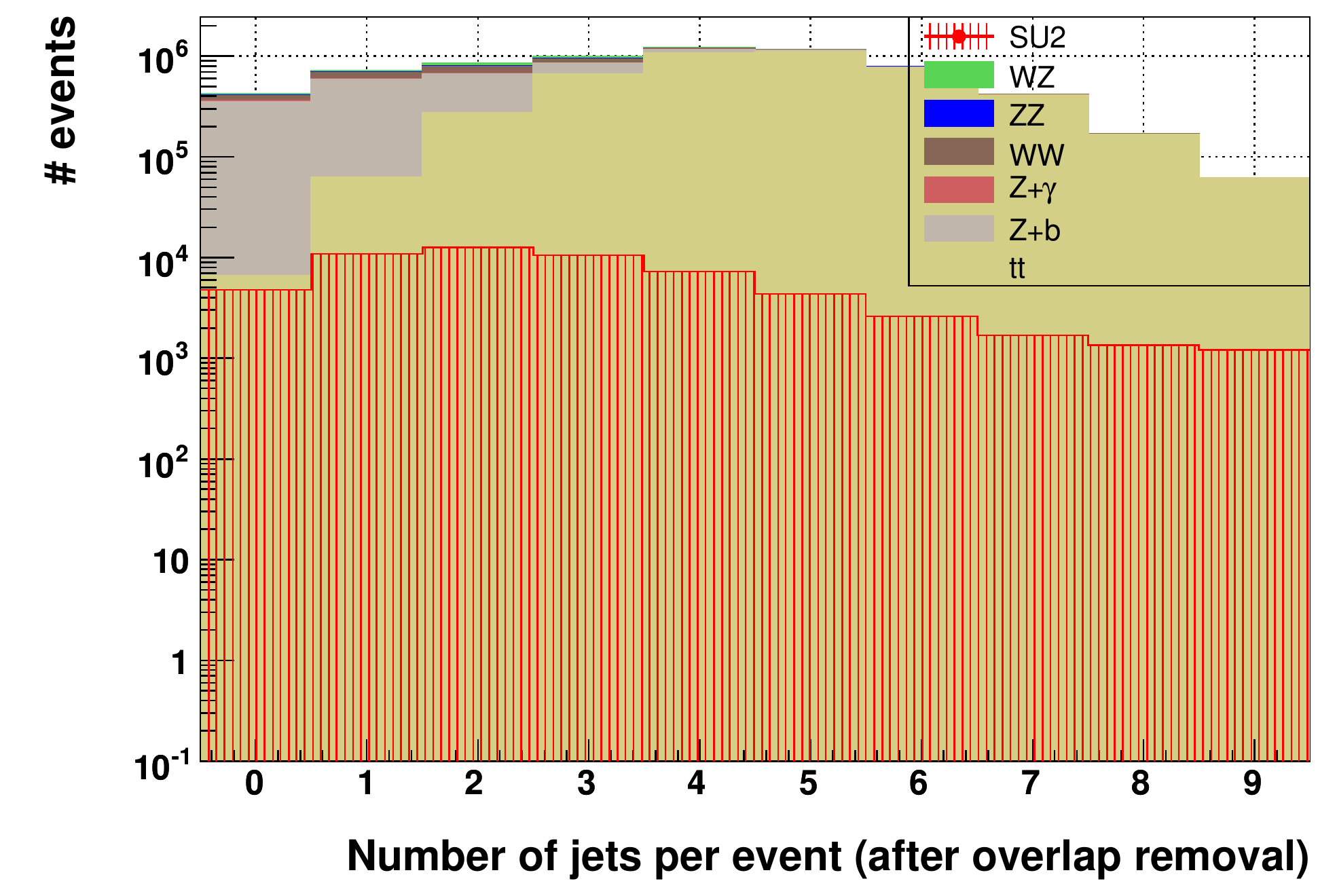}
\vspace{\cDist}
\vspace{\cDistHalf}
\end{center}
\caption[Transverse energy of the leading jet $\et^{j_1}$ and multiplicity of jets per event for SU2]{\label{fig:jPreselection}
Transverse energy of the leading jet $\et^{j_1}$ {\bf(left)} and multiplicity of jets per event{\bf~(right)} for SU2. Both are after basic jet quality cuts summarised in Section~\ref{sec:presJt} and overlap removal detailed in Section~\ref{sec:presOR} for SU2.
%%\vspace{\cDistHalf}
}
\end{figure}

The transverse energy distribution of the leading jet $\et^{j_1}$ and the number of jets per event {\em after} overlap removal with electrons -- which is explained in the following Section -- are shown in Figure~\ref{fig:jPreselection}.

\section{Overlap Removal Between\newline  Electrons or Photons and Jets}\label{sec:presOR}
A feature of \Athena{} is that it does not remove overlap between reconstructed objects, id est medium quality electons as described in Section~\ref{sec:presEl} are more than 95\% likely to be {\em independently} reconstructed as jets. The main philosophy behind this approach is that each analysis should remove the overlap between contents of physics object containers itself, depending on its goals and the actual physics object definitions. The trilepton analysis presented here is either ``jet inclusive'' (no requirement on the number or properties of any jets is made), or ``jet exclusive'' (events with significant hadronic activity are vetoed against at the final selection step, cf. Chapter~\ref{chp:selection}). In the latter case, it is crucial to apply sufficiently tight criteria when removing the overlap between jets and electromagnetic objects\footnote{It should be mentioned that in principle there might be some overlap between muons and electromagnetic objects, but it was shown using Monte Carlo simulations to be less than 0.1\% in busy events (at \instlumi{34})~\cite{bib:nabil} and will not be considered further.}. 

\begin{figure}
\begin{center}
\includegraphics[width=7.9cm,clip]{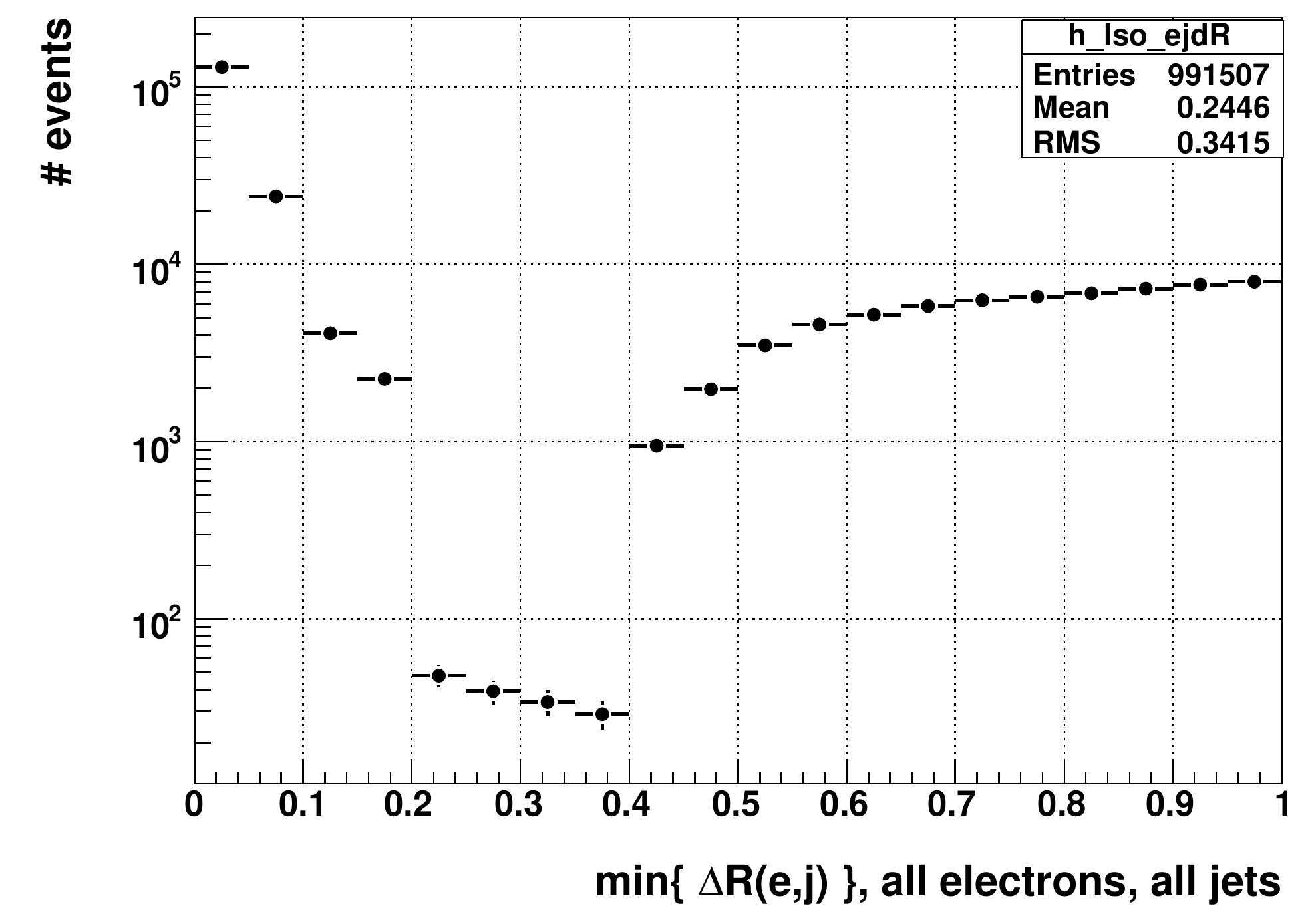}
\includegraphics[width=7.9cm,clip]{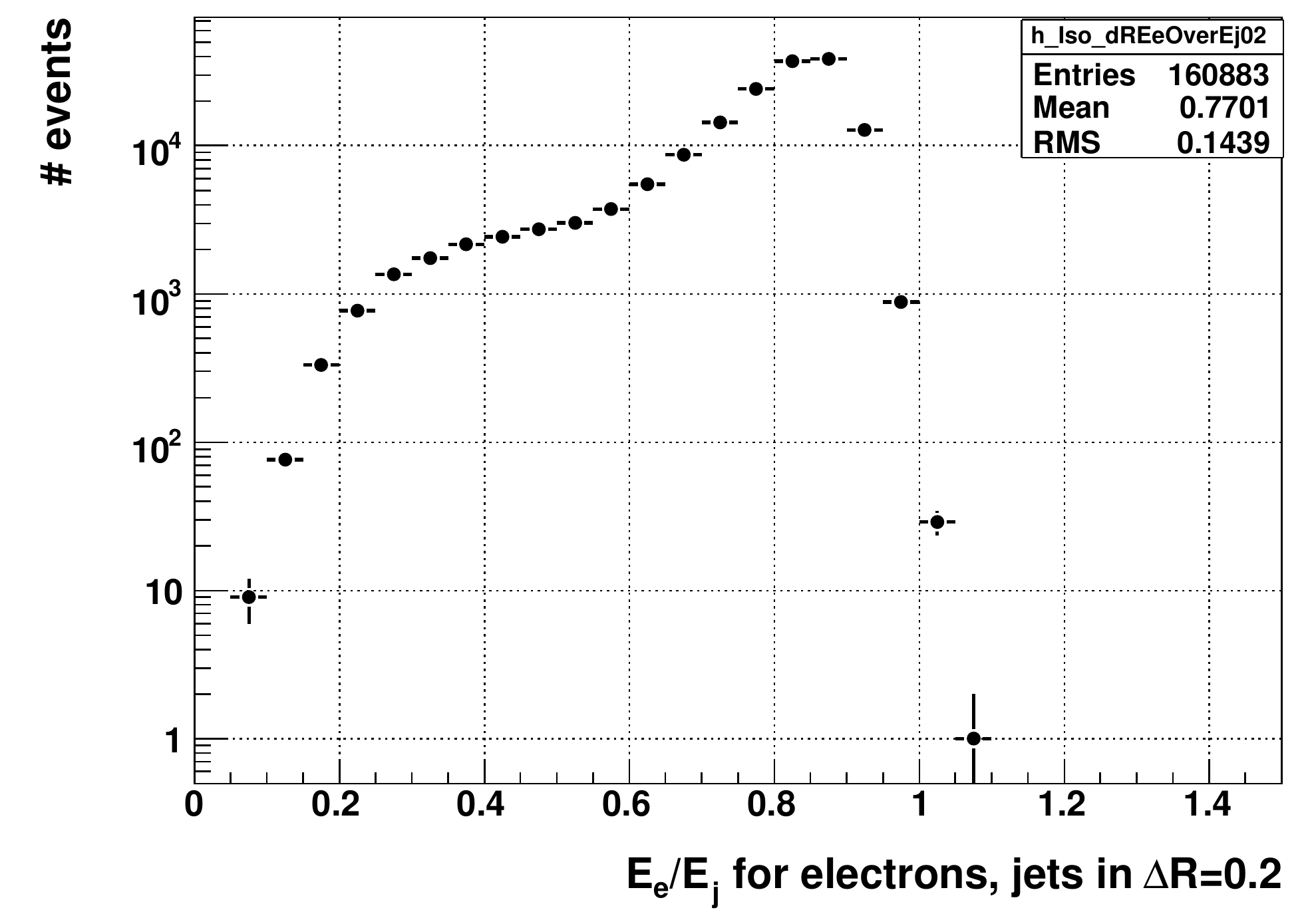}
\vspace{\cDist}
\vspace{\cDistHalf}
\end{center}
\caption[Distance in $\Delta R$ between jets and electrons before overlap removal as found in \ttbar~events and $\et^e/\et^j$ for $\Delta R(e,j)<0.2$]{\label{fig:ejdR}
Distance in $\Delta R$ between jets and electrons before overlap removal as found in \ttbar~events {\bf(left)} and $\et^e/\et^j$ for $\Delta R(e,j)<0.2$ {\bf(right)}.
\vspace{\cDistHalf}
}
\end{figure}

If an electron is independently reconstructed as a jet, both objects are likely to be closer than $\Delta R=0.2$ in $\eta\times\phi$. Indeed, the $\Delta R(e,j)$ distribution in Figure~\ref{fig:ejdR} shows a prominent enhancement for $\Delta R<0.2$. Furthermore, a smooth rise beyond $\Delta R>0.4$ due to the increasing phase space for {\em prompt} electrons and {\em genuine}\footnote{In this context, {\em genuine} means that reconstructed jets indeed correspond to actual quarks and/or gluons, rather than electrons.} hadronic jets is observed, whereas the region in-between shows the slightly falling bulk of secondary electrons from heavy flavour decays. One would of course expect that the bulk of secondary electrons continues below $\Delta R<0.2$. The right hand side of Figure~\ref{fig:ejdR} shows the ratio ${\et^e}/{\et^j}$ for $\Delta R(e,j)<0.2$. Despite the smearing due to different electromagnetic and hadronic energy scales, a clear kink at $\et^e/\et^j\simeq0.6$ is visible. %We interpret it as separating electrons which are likely to be  {\em secondary}, id est coming from heavy flavour decays with $\et^e/\et^j\lesssim0.6$; and electrons which are likely to be {\em prompt}, id est originating from the hard scattering process and independently reconstructed as jets with $\et^e/\et^j\gtrsim0.6$.\\
It can be interpreted as separating:
\begin{itemize}
\item$\et^e/\et^j\lesssim0.6$:~
Electrons which are likely to be  {\em secondary}, id est coming from heavy flavour decays;
\item$\et^e/\et^j\gtrsim0.6$:~
Electrons which are likely to be {\em prompt}, id est originating from the hard scattering process and independently reconstructed as jets.
\end{itemize}

Based on the above observation, an object found by the jet algorithm is {\em not} considered to be a {\em genuine} hadronic jet if:
\begin{itemize}
\vspace{\cDistHalf}
\item there is a photon or an electron satisfying all the criteria used in the preselection of this analysis closer than $\dR=0.2$ to it; \textit{and}
\item $\et^{e,\gamma}/\et^j>0.6$ is fulfilled. 
\vspace{\cDistHalf}
\end{itemize}
This way secondary electrons with $\et^{e,\gamma}/\et^j<0.6$, even if passing the preselection, will be rejected by the $\Delta R(\ell,j)>0.4$ veto. Adopting the simplistic {\sc EventView}-type of overlap removal~\cite{bib:eventView} relying on \dR{} only leads to a decrease of the statistical significance by 11\% for SU2, and a couple of per cent for other SU$x$ points except SU4, where the opposite is the case due to the abundance of jets and its enormous cross-section.

%% file: Selection/Selection.tex
After the preselection, the presence of at least 2 leptons is required. In the following, several further selection requirements will be applied. The main philosophy of this analysis is to suppress known Standard Model backgrounds, while avoiding optimisation for any particular SU$x$ point in the SUSY parameter space. The selection is necessarily not completely general, and where it has been unavoidable to favour particular parts of parameter space, the preference is given to the focus point region (SU2) and the difficult massive sparton scenario introduced in Chapter~\ref{chp:signature}.

\section{Opposite Sign Same Flavour Lepton Pair Selection}
After the dilepton requirement, an OSSF lepton pair is selected. This is done because the $WZ,\,ZZ,\,Z\gamma$ and $Zb$ backgrounds have a peak in their OSSF dilepton mass $m_{\ell\ell}^{\rm OSSF}$ distributions at the $Z$ boson mass, which can later be used as a handle for their suppression. On the SUSY side, an OSSF lepton pair from neutralino decay is expected. Moreover, once data comes, the $m_{\ell\ell}^{\rm OSSF}$ distribution can be used for validating the preselection, as it should describe the Drell-Yan interference together with the $Z$-mass peak, and the \ttbar{} continuum. 

In case there are 3 leptons of the \textit{same} flavour, the OSSF pair with its dilepton mass \mossf\ closest to $m_Z$ is chosen. The main idea behind this approach is to minimise com\-bi\-na\-to\-rial background in case of $WZ$ or $ZZ$ production, as one might combine leptons from different bosons to pairs. It is sufficient to reduce the case of 4 leptons to the case of 3 leptons, since for the only significant SM background to produce it, $ZZ$, 2 out of 3 leptons considered \textit{must} come from the decay of the same boson. More than 4 leptons would not constitute a significant SM background in any case. If there is both an $ee$ and a $\mu\mu$ pair, the one closest to $m_Z$ is chosen. This is the baseline OSSF pair selection method.

\begin{table}
\begin{center}
\begin{small}
\begin{tabular}{lccccccccc}
\hline                                                                                              
OSSF selection algorithm               &SU2    &SU3    &SU4    &$WZ$   &$ZZ$   &$WW$   &$Z\gamma$      &$Zb$   &$\ttbar$ \\
\hline\hline                                                                                        
$\min_{i,j}|m_{\ell_i\ell_j}-m_Z|$     & 1.00   & 1.00   & 1.00  & 1.00  & 1.00  & 1     & $\times$       & 1.0    & 1.00   \\
$\min_{i,j}\Delta R(\ell_i\ell_j)$     & 1.10   & 1.15   & 1.07  & 2.41  & 2.86  & 1     & $\times$       & 2.0    & 1.19   \\
$\max_{i,j}\{\pt^i,\pt^j\}$            & 0.92   & 1.02   & 1.03  & 1.23  & 1.38  & 1     & $\times$       & 1.0    & 1.00   \\
\hline       
\end{tabular}
\end{small}
\vspace{\cDistHalf}
\end{center}
\caption[The performance of three algorithms to reject diboson backgrounds involving $Z$]{\label{tab:selectionOSSF}
The performance of the three algorithms described in the text to reject diboson backgrounds involving the $Z$ boson normalised to the $\min_{i,j}|m_{\ell_i\ell_j}-m_Z|$ method. ``$\times$'' means that in none of the simulated MC events an ambigous OSSF lepton pair combination was found.}
\end{table}

Two alternative prescriptions to select the OSSF lepton pair from three leptons of the same flavour were investigated:
\begin{itemize}
\vspace{-2mm}
\setlength{\itemsep}{0mm}
 \item the lepton pair with minimal distance in $\Delta R$ is chosen, id est $\ell_i\ell_j$ with $\min_{i,j}\Delta R(\ell_i\ell_j)$, where $i\neq j$ run over all possible OSSF lepton combinations in a given event; 
 \item the lepton pair with the highest $\pt$ (per lepton) is taken.
\vspace{-2mm}
\end{itemize}
The performance of the three methods listed above to select the OSSF lepton pair was evaluated with the full selection chain. As a figure of merit the number of events passing the selection, as a proportion of the baseline method, was used. The results are presented in Table~\ref{tab:selectionOSSF}. As demonstrated in the Table, the baseline method used in this analysis yields a high signal efficiency for SU2, SU3, and SU4, whilst being most successful at suppressing $WZ$ and $ZZ$ backgrounds. 

% It is true that the favoured algorithm based on the minimisation of the difference of the OSSF dilepton mass to the mass of the $Z$ leads to a bias in the \mossf{} distribution, as it ``shifts'' the masses towards the $Z$ peak. For phenomenological studies like the dilepton mass edge, the second algorithm may be a better choice. It selects the OSSF leptons \textit{topologically} closest together, as expected in a decay of a mother particle, irrespective of its mass. The advantage of investigating the distribution with ``shifted'' OSSF masses is the consistency of normalisation of the $WZ$ background to data.

One disadvantage of the baseline method in which the OSSF combination with the smallest difference between \mossf\ and the $Z$ boson mass is selected is that the \mossf{} mass distribution will be biased towards the $Z$ mass. This method may therefore be less useful if one were trying to extract mass information. Nevertheless, it is used here because it allows to easily compare the normalisation of the \mossf{} distribution with respect to backgrounds involving $Z$-boson production, in particular $WZ$.

When selecting the OSSF lepton pair, a minimal dilepton mass, $\mossf>15$\,GeV, is required for all possible OSSF pairings in a given event. This is done in order to suppress backgrounds from dilepton production via a photon propagator, as well as $J/\Psi$ and $\Upsilon$ backgrounds. Because the $\Delta R>0.2$ separation requirement is demanded for electrons and muons earlier in the preselection, the fraction of signal events passing this cut is more than 90\%.

\section{Selection of Three Isolated Leptons}

After the OSSF lepton pair has been selected, the presence of at least 3 leptons is required. As a third lepton, the muon or electron not being part of the OSSF pair and with the highest $\pt$ is chosen.

\begin{figure}
\begin{center}
\includegraphics[width=7.9cm,clip]{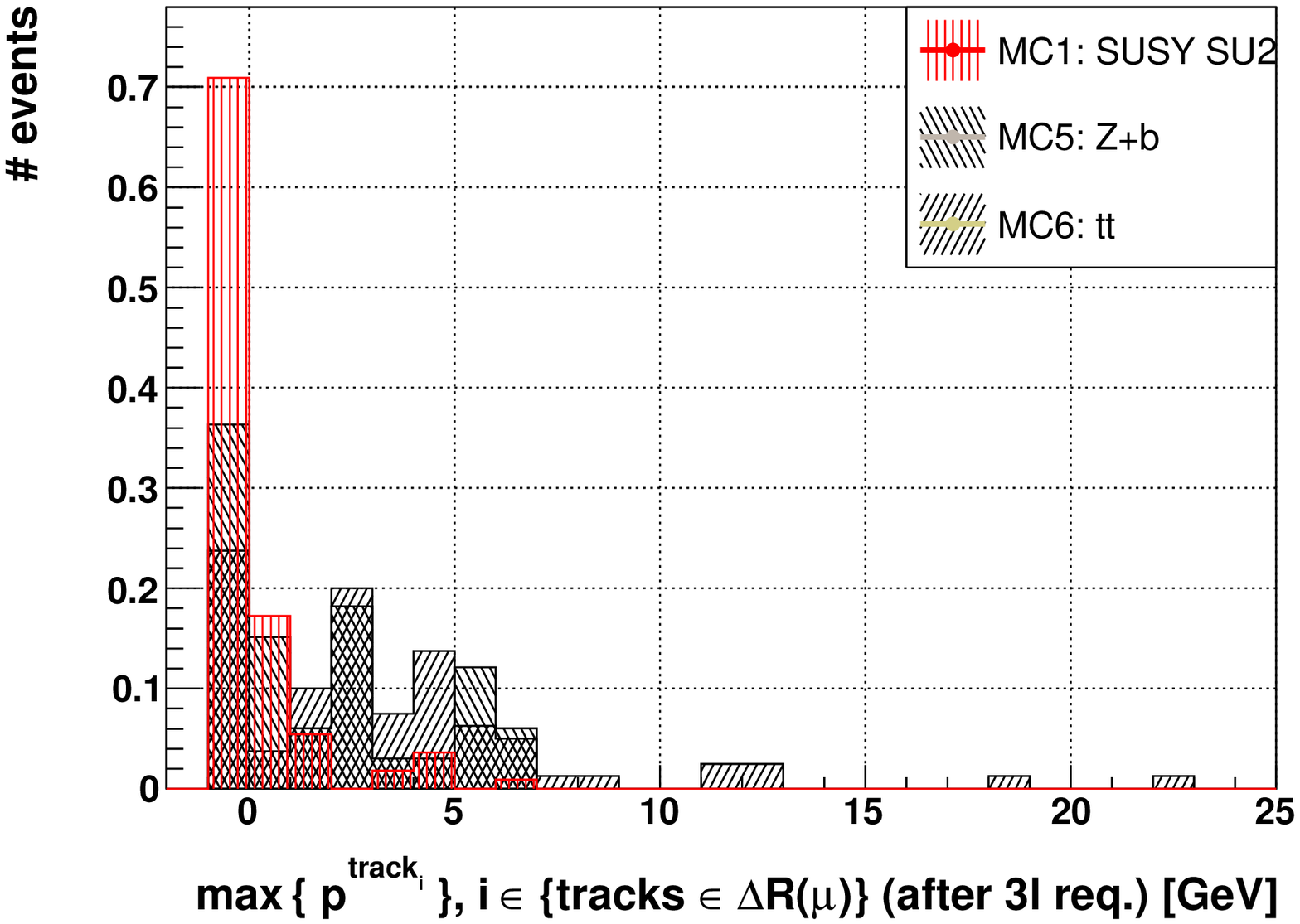}
\includegraphics[width=7.9cm,clip]{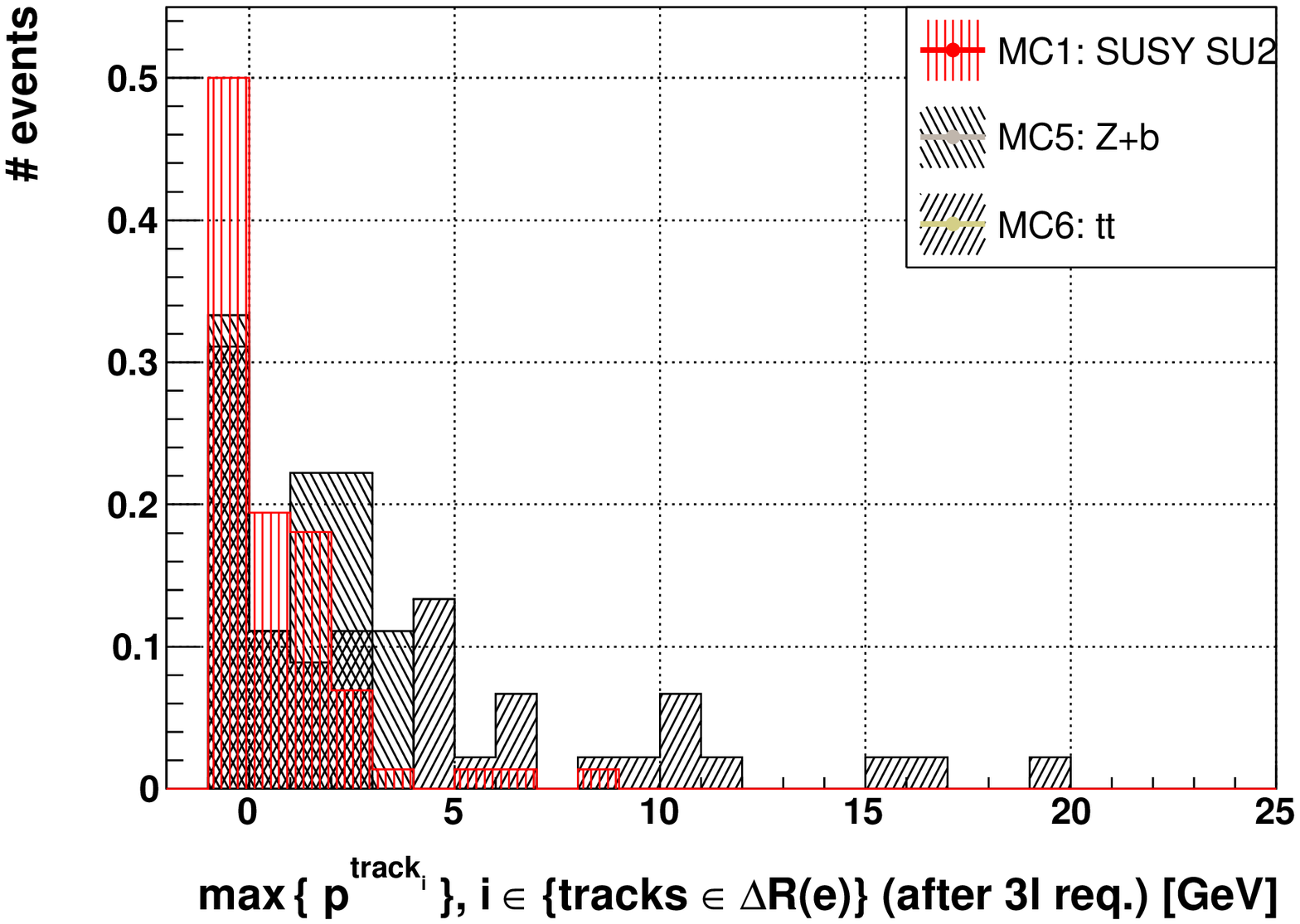}
\includegraphics[width=7.9cm,clip]{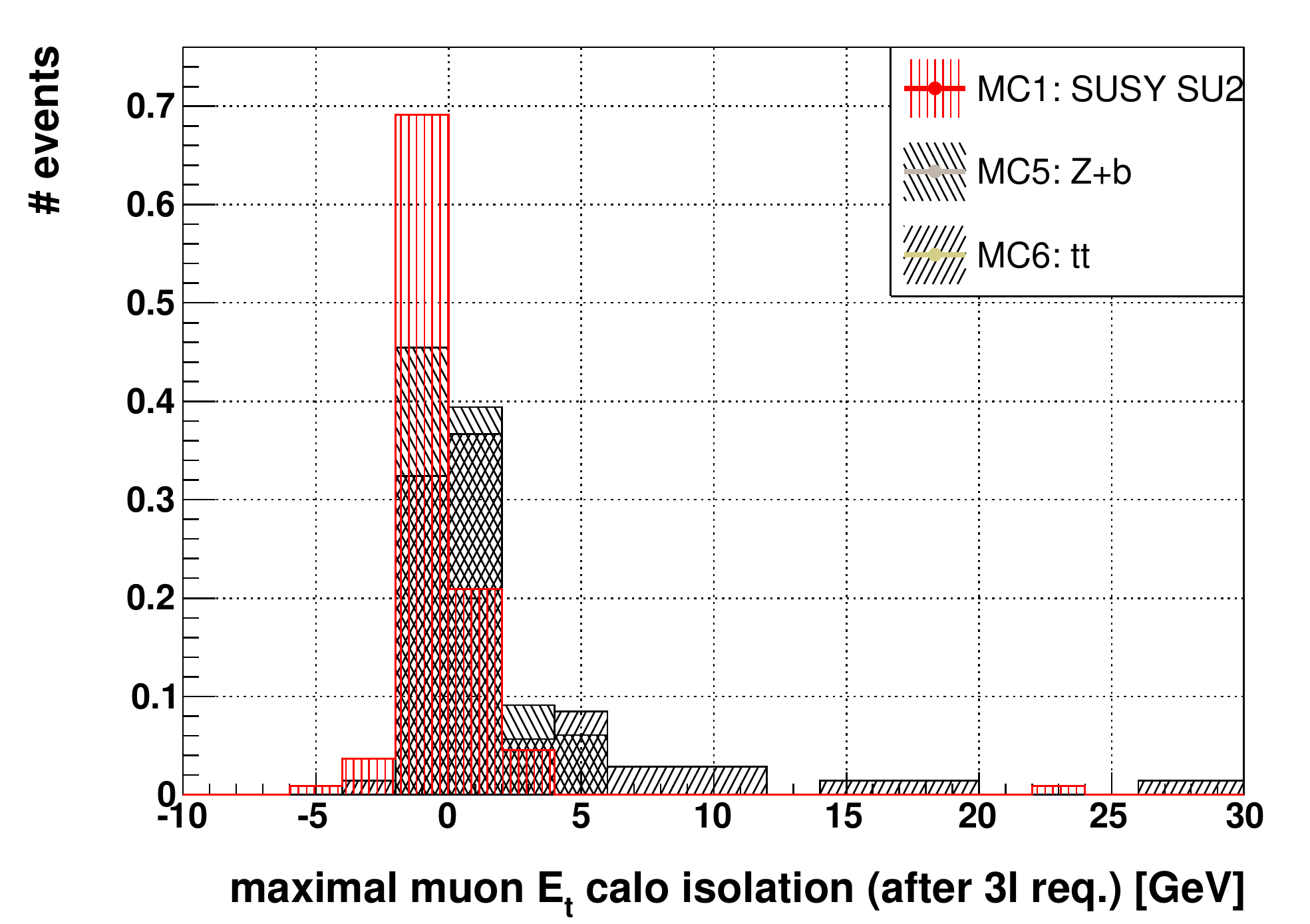}
\includegraphics[width=7.9cm,clip]{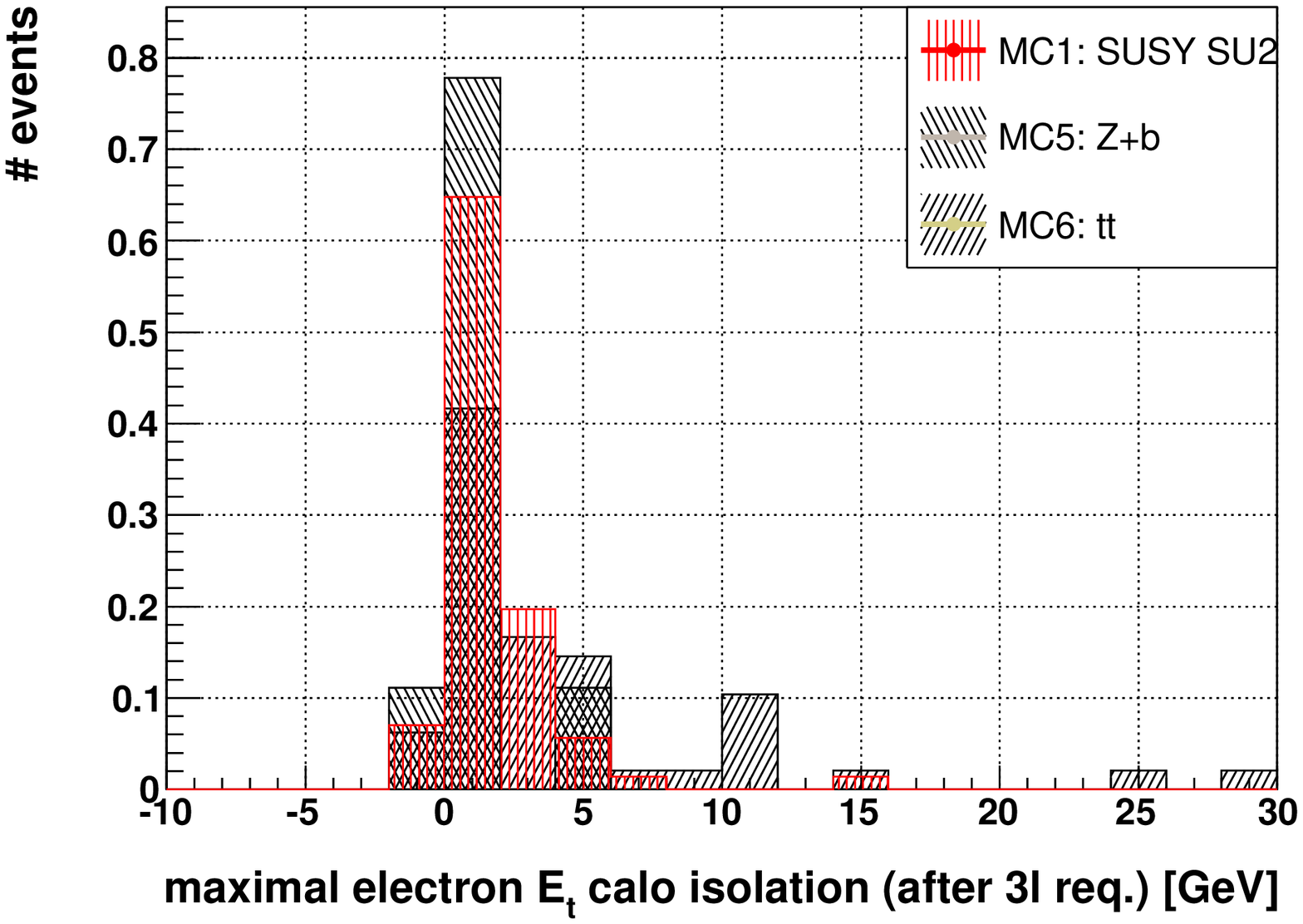}
\vspace{\cDist}
\vspace{\cDistHalf}
\end{center}
\caption[Track $I_{0.2}^{\rm trk}$ and calorimeter isolation $I_{0.2}^{\rm cal}$ distribution for muons and electrons]{\label{fig:trackIsolation}
Track isolation $I_{0.2}^{\rm trk}$ distribution for muons {\bf(top left)} and electrons {\bf(top right)}. Calorimeter isolation $I_{0.2}^{\rm cal}$ for muons {\bf(bottom left)} and electrons {\bf(bottom right)}. Only SU2, \ttbar{}, and $Zb$ are shown. Each process is normalised to 1.
\vspace{\cDistHalf}
}
\end{figure}

There are very few significant backgrounds from the SM with three prompt leptons originating from the hard matrix element, they all come from diboson production. However, as previously mentioned, there is a possibility that backgrounds with less than three prompt leptons enter the selection, where the third lepton comes from a misidentification of a light flavour jet, or if a secondary lepton from a semileptonic $c$- or $b$-quark decay passes the isolation requirements. The strategy outlined below aims mainly at controlling the latter category, since its rate is $\mathscr O(10)$ higher than the former, as found by the author and~\cite{bib:scndLeptonRate} in \ttbar{} events. Considering the above, \ttbar{} and $Zb$ are the most important backgrounds in the trilepton channel due to their high cross sections and the presence of 2 leptons plus one ore more $b$-jets. 

%To ensure a high purity of the leptons
%We have examined using both calorimeter and track isolation to try to select against leptons from heavy quark decays. 
In order to select against secondary leptons from heavy quark decays, both the {\em calorimeter} and {\em track} isolation have been studied.
Here, track isolation is defined as:
\begin{equation}
 I_{0.2}^{\rm trk}(\ell) \equiv \max{}_{i,j}\{\pt^{{\rm track}_i}|{\rm track}_i\in\dR(\ell_j)\}~{\rm where}~\ell=\mu,e\,,
\end{equation}
that is $I_{0.2}^{\rm trk}(\ell)$ for flavour $\ell$ is defined as the $\pt$ of the track with the highest transverse momentum inside a $\dR=0.2$ cone around {\em any} of the preselected leptons of flavour $\ell$ in the event. It is understood that the tracks of the actual leptons are not considered. 
%Defined this way, $I_{0.2}^{\rm trk}(\mu)$, $I_{0.2}^{\rm trk}(e)$ become global discrimination variables for an event. 
The top row of Figure~\ref{fig:trackIsolation} shows the track isolation for signal (SU2) and backgrounds ($Zb$, \ttbar) after the 3$^{\rm rd}$ lepton requirement. Since electrons are more likely to produce bremsstrahlung and undergo early conversions, $I_{0.2}^{\rm trk}$ is shown separately for muons (left) and electrons (right). Based on Figure~\ref{fig:trackIsolation}, a cut on $I_{0.2}^{\rm trk}>1\,{\rm GeV}~(2\,{\rm GeV})$ is chosen for muons and electrons, respectively. 

\begin{figure}
\begin{center}
\includegraphics[width=7.9cm,clip]{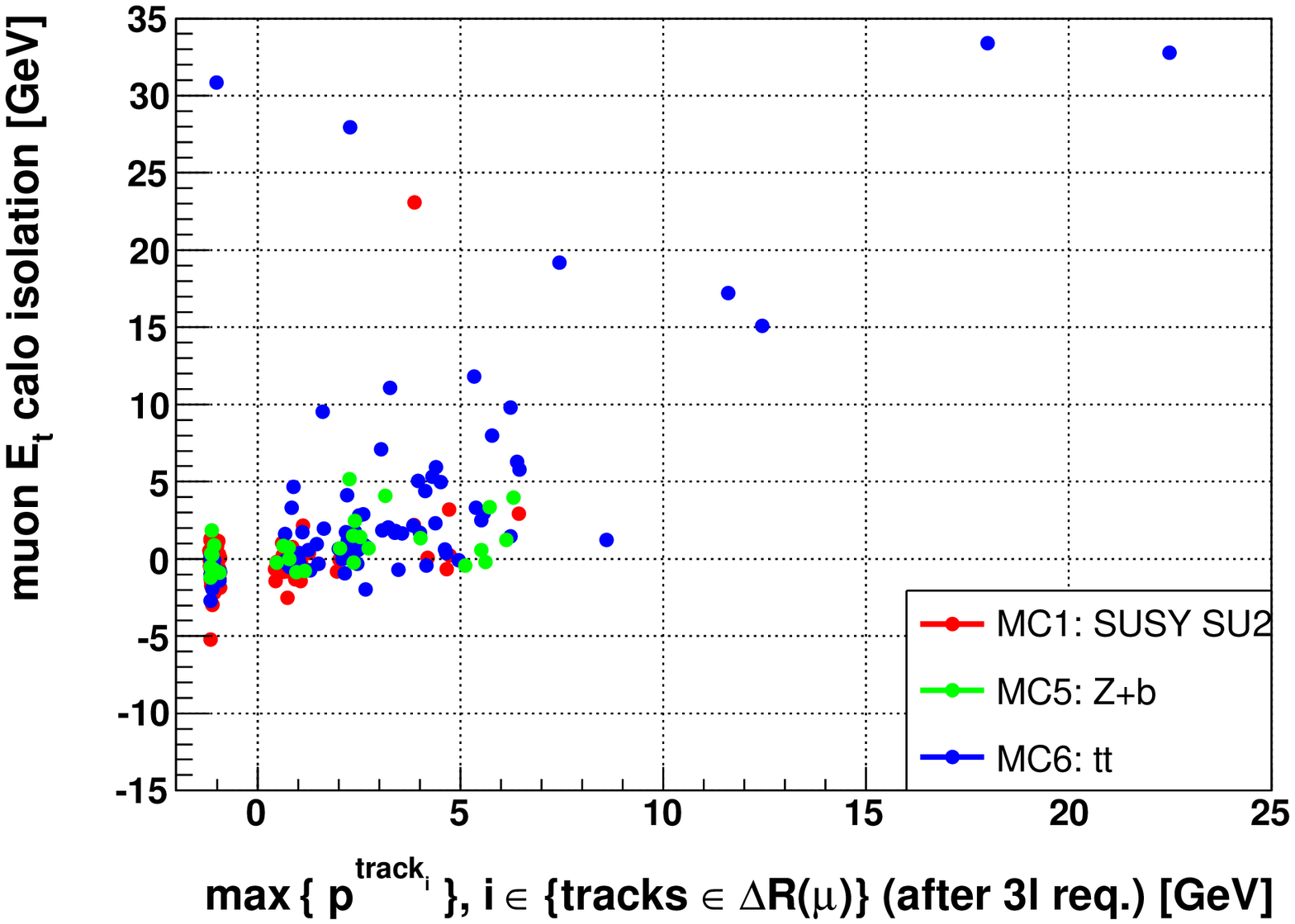}
\includegraphics[width=7.9cm,clip]{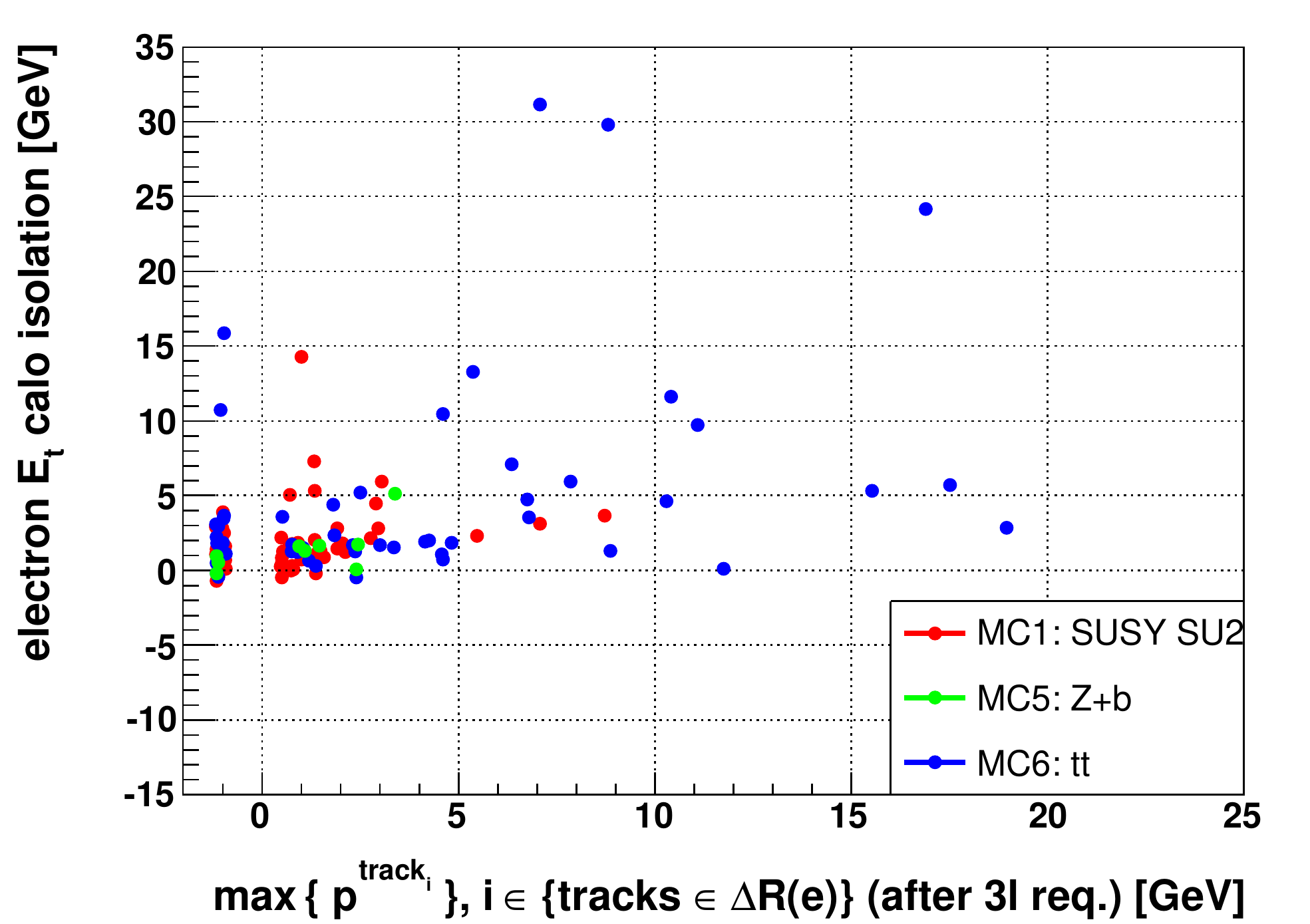}
\vspace{\cDist}
\vspace{\cDistHalf}
\end{center}
\caption[Track versus calorimeter isolation for muons and electrons]{\label{fig:TrackCaloIsolation}
Track versus calorimeter isolation for muons {\bf(left)} and electrons {\bf(right)}. For SU2, \ttbar{}, and $Zb$ all available simulated MC events are shown.
\vspace{\cDistHalf}
}
\end{figure}

In a similar fashion, calorimeter isolation $I_{0.2}^{\rm cal}$ has been studied, but found to yield a smaller rejection power against secondary leptons from heavy quark decays, as demonstrated in the bottom row of Figure~\ref{fig:trackIsolation}. The correlation between track and calorimeter isolation is shown in Figure~\ref{fig:TrackCaloIsolation}. As can be seen from the Figure, with the tight cuts on track isolation used in this analysis, the importance of calorimeter isolation is reduced. For muons, loose calorimeter isolation must be kept. Dropping it reduces the statistical significance after all cuts by 20\% for SU2 and similar values for other SU$x$ benchmark points. Since a degree of calorimeter isolation is implied by the medium electron quality requirement, $I_{0.2}^{\rm cal}$ is dropped for electrons, which results in an increase of the statistical significance by $\sim$7\% for SU2 and a couple of per cent for the other SU$x$ benchmark points. 

The conclusion is, that both calorimeter and track  isolation can be used for this type of analysis, with a preference towards track  isolation. For the early days of ATLAS, the discrimination variable which is understood and validated first can be used.

Besides $I_{0.2}^{\rm cal}$ and $I_{0.2}^{\rm trk}$, the normalised transverse track impact parameter, $d_0/{\sigma(d_0)}$, has been studied~\cite{bib:d0discrimination}. It was found to yield a much smaller rejection power against secondary leptons compared to the discrimination variables presented above and is therefore not included in the analysis.

Ideally, once all aspects of the ATLAS detector are understood with collision data, a combined likelihood isolation variable constructed from track isolation, calorimeter isolation, the normalised transverse impact parameter, and other quantities appears as the logical next step.

\section{Final Cuts Against Standard Model Backgrounds}
To further suppress backgrounds without or with small genuine \met{} like $Zb$, $Z\gamma$, and $ZZ$, a requirement of $\met>20$\,GeV has been imposed. Additionally, in order to suppress backgrounds involving $Z$-boson production, all events where the mass of the OSSF lepton pair fulfils $|\mll^{\rm OSSF}-m_Z|<10$\,GeV, $m_Z\equiv91.2\,\GeV$ are removed. Event numbers for various stages of the main selection are presented in Table~\ref{tab:selection}.

% FIXME describe jet veto and why it is used: 
% 1) jet inclusive search analysis
% 2) jet exclusive analysis to measure SUSY params

\begin{sidewaystable}[h]
% \setlength{\parskip}{0.2em} % distance btw 2 paragraphs
% \setlength{\topsep}{0em} % distance btw top of page and text
%\begin{scriptsize}
%\input{NumCutFlow}
\begin{tabular}{l|rr|rr|rr|rr|rr|rr|rr}
\hline                                                                                              
& \multicolumn{14}{c}{{\bf Signal}} \\
Cut applied               & \multicolumn{2}{c|}{SU1} 
						&\multicolumn{2}{c|}{SU2}         
                                                			&\multicolumn{2}{c|}{SU3}        
                                                                        			&\multicolumn{2}{c|}{SU4}          
                                                                                                			&\multicolumn{2}{c|}{SU8}       
                                                                                                                        		   &\multicolumn{2}{c|}{SU2$\chi$} 
                                                                                                                                           		      &\multicolumn{2}{c}{SU3$\chi$}  \\
\hline\hline                                                                                                                   
$\geq 2\ell$              & 4761.8     & 47.5	& 1935        & 48.6	& 11124.2    & 73.1	& 106545      & 1338.2	& 1471.8    & 49.8 & 1338.5    & 40.4 & 838.1     & 20.1\\
OSSF pair                 & 2381.6     & 33.6	& 1169.8      & 37.8	& 7785.6     & 61.2	& 52087.4     & 935.7	& 625.5     & 32.5 & 905.7     & 33.3 & 629       & 17.4\\
OSSF+$3^{\rm rd}\ell$     & 321.6      & 12.3	& 233.5       & 16.9	& 1025.7     & 22.2	& 6504.6      & 330.6	& 87.7      & 12.2 & 144.2     & 13.3 & 78.9      &  6.2\\
Track isol.               & 265.8      & 11.2	& 204.1       & 15.8	& 837.7      & 20.1	& 5277.6      & 297.8	& 74.2      & 11.2 & 129.6     & 12.6 & 70.2      &  5.8\\
\met{}                    & 263.9      & 11.2	& 191.9       & 15.3	& 828.5      & 20.0	& 5092.8      & 292.6	& 74.2      & 11.2 & 118.6     & 12.0 & 66.8      &  5.7\\
\mll                      & 192.5      & 9.5	& 154         & 13.7	& 529.9      & 16.0	& 4185.1      & 265.2	& 43.8      &  8.6 & 89.2      & 10.4 & 39.9      &  4.4\\
\mll,\,3\,bin             & {\bf139.1} & 8.1	& {\bf129.6}  & 12.6	& {\bf441.4} & 14.6	& {\bf3109.4} & 228.6	& {\bf25.3} &  6.5 & {\bf81.9} & 10.0 & {\bf36.5} &  4.2\\
Jet veto                  & 5.2        & 1.6	& 35.         & 6.6	& 21.6       & 3.2	& 100.8       & 41.2	& 3.4       &  2.4 & 35.4      &  6.6 & 19.2      &  3.0\\
Jet veto,\,3\,bin         & {\bf4.7}   & 1.5	& {\bf31.8}   & 6.2	& {\bf19.7}  & 3.1	& {\bf50.4}   & 29.1	& {\bf0.0}  &  1.7 & {\bf31.8} &  6.2 & {\bf17.8} &  2.9\\
\hline       
& \multicolumn{14}{c}{{\bf Background}}\\
                          &\multicolumn{2}{c|}{$WZ$}
                          			&\multicolumn{2}{c|}{$ZZ$}
                                                			&\multicolumn{2}{c|}{$WW$}
                                                                        			&\multicolumn{2}{c|}{$Z\gamma$}
                                                                                                			&\multicolumn{2}{c|}{$Zb$}
                                                                                                                        			&\multicolumn{2}{c}{$t\bar t$} && \\
\hline\hline                            
$\geq 2\ell$              & 17542.7    & 242.5	& 10741.4   & 92.1	& 22789.7  & 431.8	& 7650.5   & 160.2	& 541839    & 2693.7	& 289464 	& 1797.5 && \\
OSSF pair                 & 15009.3    & 224.3	& 10357.9   & 90.4	& 10973.4  & 299.7	& 7365.4   & 157.2	& 533001    & 2671.6	& 100551 	& 1059.4 && \\
OSSF+$3^{\rm rd}\ell$     & 2037.4     & 82.6	& 595.7     & 21.7	& 16.4     & 11.6	& 90.6     & 17.4	& 3441.6    & 214.7	& 926.5  	& 101.7 && \\
Track isol.               & 1802.9     & 77.7	& 513.6     & 20.1	& 8.2      & 8.2	& 26.8     & 9.5	& 1499.8    & 141.7	& 167.4  	& 43.2 && \\
\met{}                    & 1598.4     & 73.2	& 179.1     & 11.9	& 8.2      & 8.2	& 3.4      & 3.4	& 160.7     & 46.4	& 189.8  	& 46.0 && \\
\mll                      & 315.0      & 32.5	& 22.9      & 4.3	& 8.2      & 8.2	& 0.0      & 3.4	& 26.8      & 18.9	& 156.3  	& 41.8 && \\
\mll,\,3\,bin             & {\bf184.3} & 24.9	& {\bf11.8} & 3.1	& {\bf8.2} & 8.2	& {\bf0.0} & 3.4	& {\bf13.4} & 13.4	& {\bf111.6}  	& 35.3 && \\
Jet veto                  & 234.6      & 28.0	& 11.0      & 2.9	& 8.2      & 8.2	& 0.0      & 3.4	& 13.4      & 13.4	& 11.2   	& 11.2 && \\
Jet veto,\,3\,bin         & {\bf127.3} & 20.7	& {\bf4.7}  & 1.9	& {\bf8.2} & 8.2	& {\bf0.0} & 3.4	& {\bf13.4} & 13.4	& {\bf11.2}   	& 11.2 && \\
\hline       
\end{tabular}
%\end{scriptsize}
\vspace{1.5mm}
\caption[Expected event numbers after successive cuts at various selection stages for 10\,\fb{} for signal and background processes.]{\label{tab:selection}
Expected event numbers after successive cuts at various selection stages for 10\,\fb{} for signal and background processes. For each process, the left column gives the expected event numbers, while the right column gives the statistical uncertainty on it based on the number of simulated MC events available for this study. ``SU$x\chi$'' stands for SU$x$ direct gaugino pair production, id est the massive sparton scenario. In bold face rows marked with ``3 bin'', events with $\mll\in[21.2,\,81.2]$\,GeV have been considered only. See text for details.}
\end{sidewaystable}

%% file: ResultsSUSY/ResultsSUSY.tex
In this Section, the OSSF dilepton mass distributions and the statistical significances, defined as $\mathcal{S}\equiv S/\sqrt{S+B}$, where $S$ is the expected number of signal and $B$ of background events, will be shown for various mSUGRA scenarios\footnote{A small additional contribution to $B$ is expected from $Zc$ production (cf.~Section~\ref{sec:bgrPhysics}).}. All figures and plots have been normalised to~10\,\fb{}.

\begin{figure}
\begin{center}
\includegraphics[width=7.9cm,clip=true]{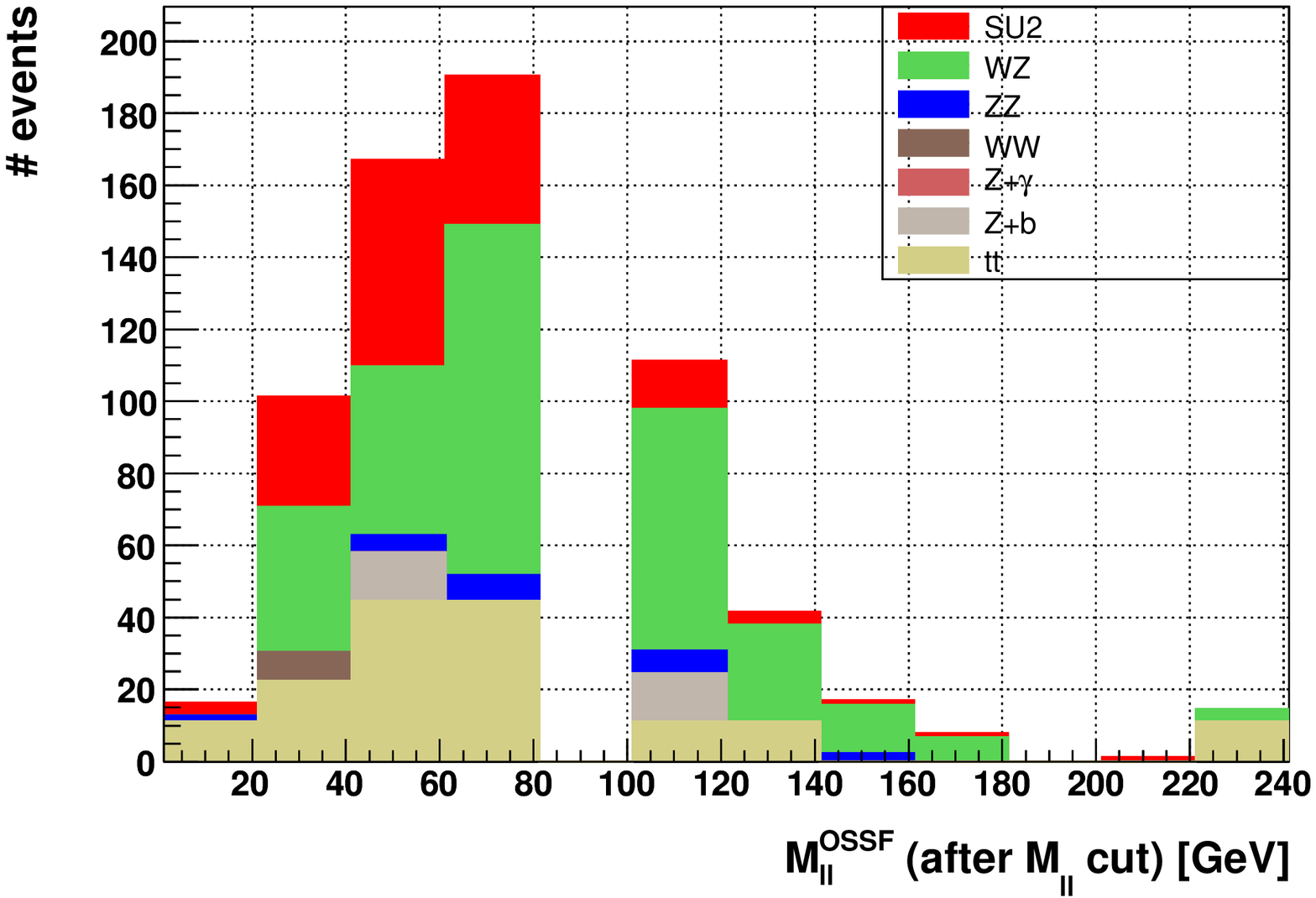}
\includegraphics[width=7.9cm,clip=true]{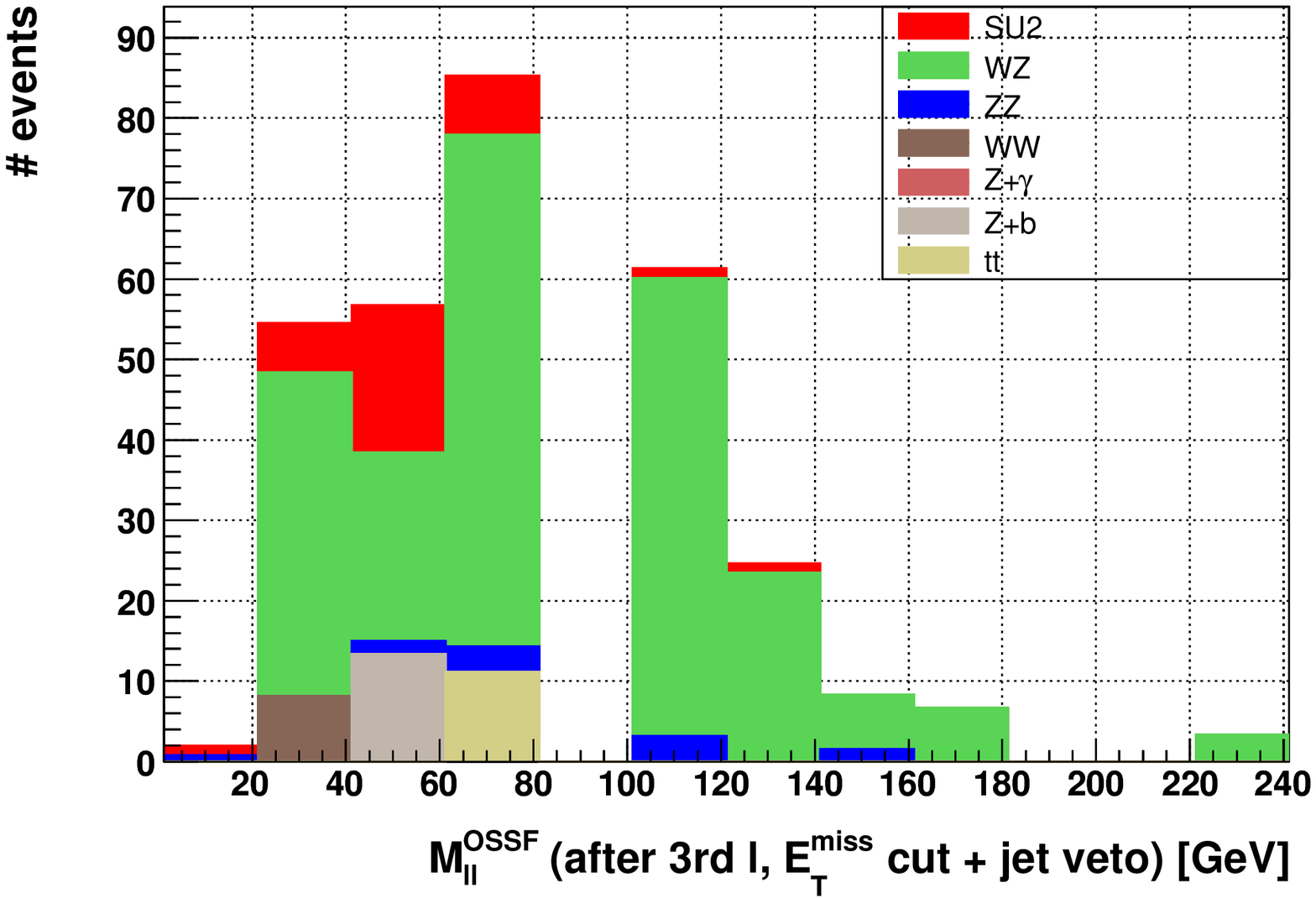}
\includegraphics[width=7.9cm,clip=true]{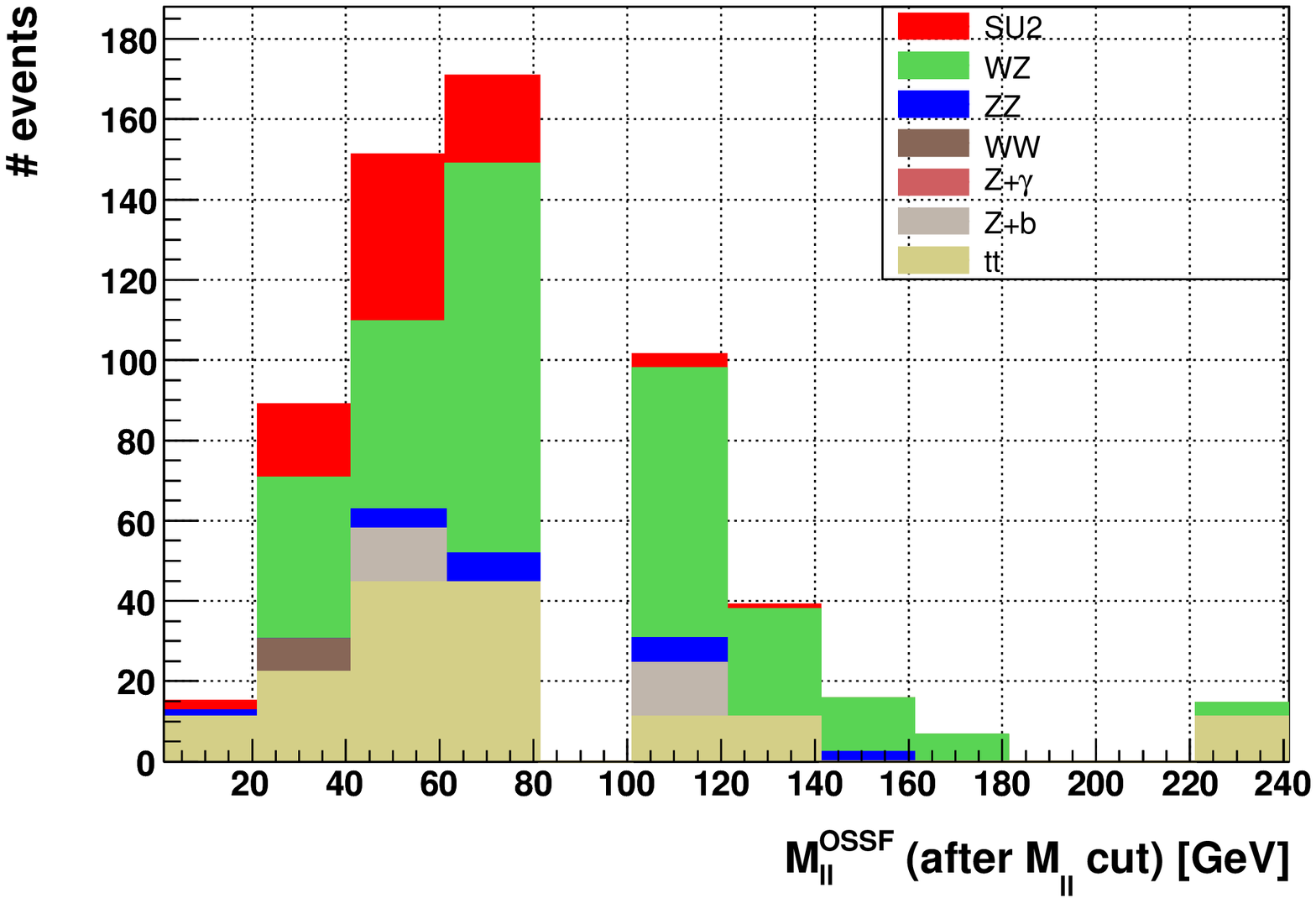}
\includegraphics[width=7.9cm,clip=true]{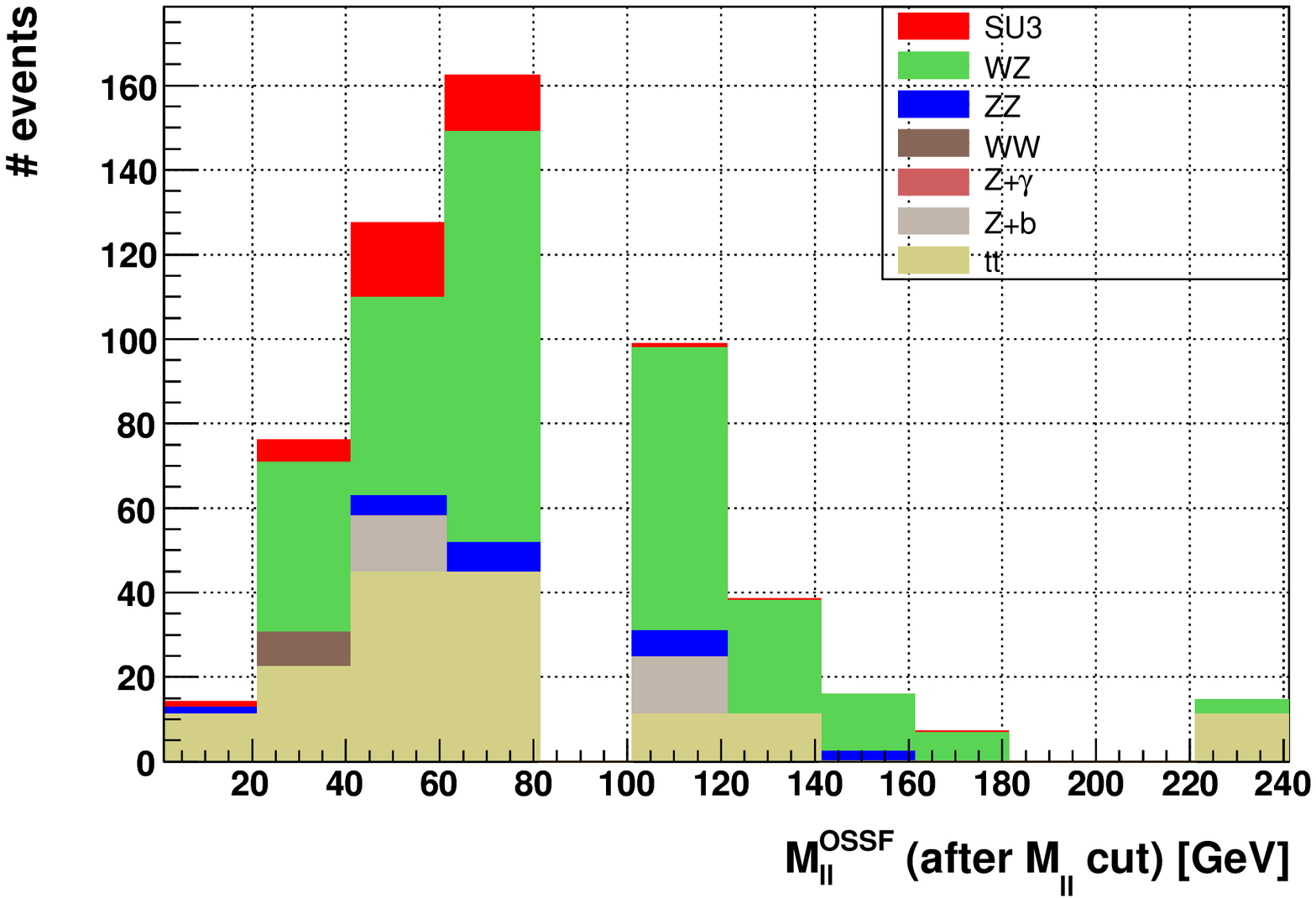}
\vspace{\cDist}
\vspace{\cDistHalf}
\end{center}
\caption[OSSF dilepton mass distribution for 10\,\fb{} for inclusive SU2 signal with and without jet veto, as well as direct gaugino production in SU2 and SU3 without any jet veto]{\label{fig:mllDirect}
OSSF dilepton mass \mossf\ distribution for 10\,\fb{} for inclusive SU2: after all cuts{\bf~(top~left)} and with the jet veto added {\bf(top~right)}. The same distribution after all cuts but without the jet veto for the massive sparton scenario: direct gaugino pair-production in SU2 {\bf(bottom left)} and in SU3 {\bf(bottom right)}. The individual backgrounds are colour-coded and ``stacked'' on top of each other. The signal is shown ``stacked'' in red on top of the background. The individual signal and background contributions are summarised in Table~\ref{tab:selection} on page~\pageref{tab:selection}.
\vspace{\cDistHalf}
}
\end{figure}

The results after all cuts for the so-called Focus Point region benchmarked by SU2 are shown in Figure~\ref{fig:mllDirect}. The statistical significance for 10\,\fb{} is $\sgnf=6.0$. This implies an expected discovery at a 5$\sigma$ level taking into account statistical uncertainties only with $\intlumi\simeq 6.8\,\fb$. %For comparison, using {\em standard} leptons reconstruction and overlap removal procedures, as well as a somewhat altered selection, $\sgnf=$ and $\intlumi\simeq$ for discovery are obtained~\cite{bib:csc7}.

\begin{figure}
\begin{center}
\includegraphics[width=7.9cm,clip=true]{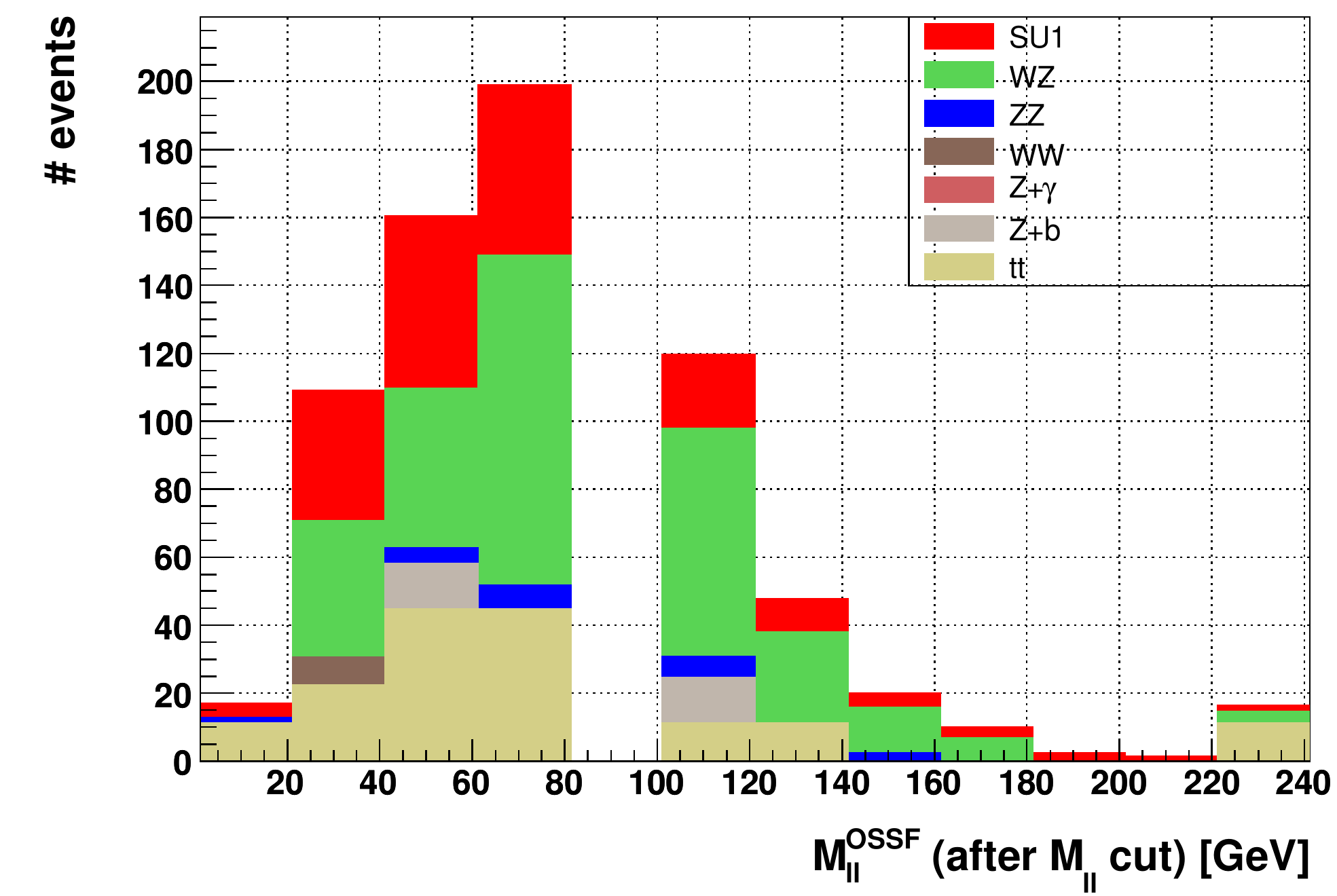}
\includegraphics[width=7.9cm,clip=true]{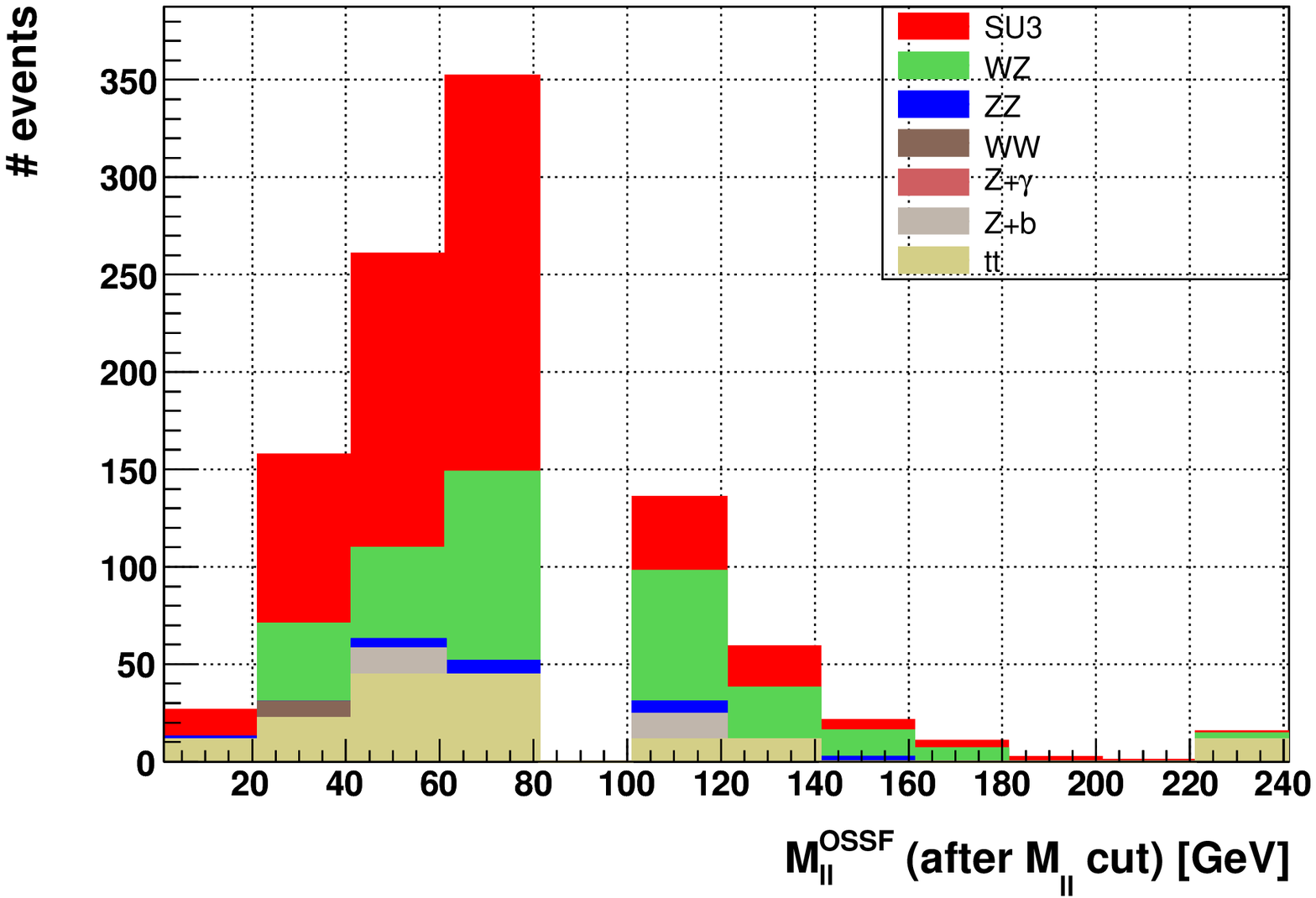}
\includegraphics[width=7.9cm,clip=true]{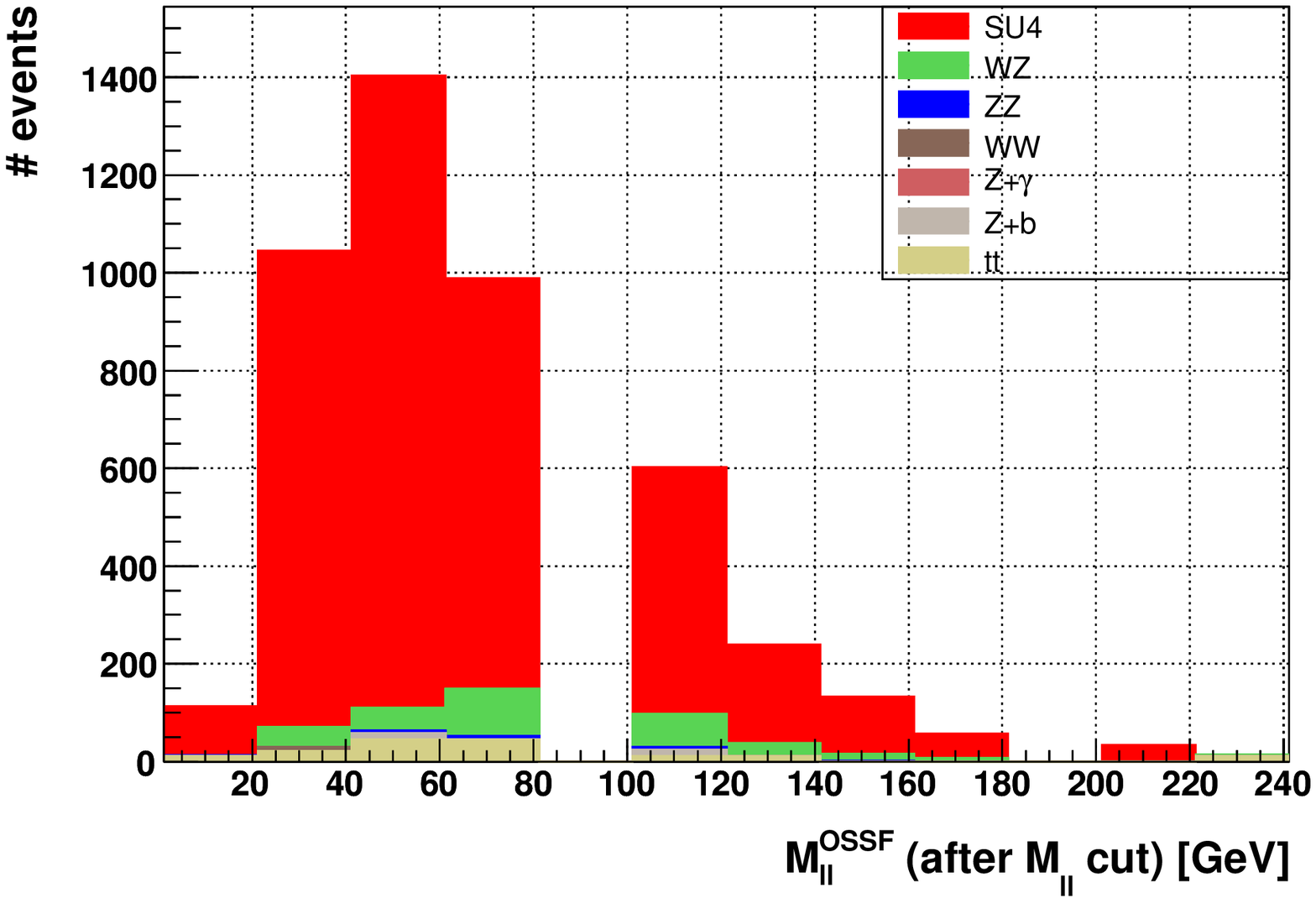}
\includegraphics[width=7.9cm,clip=true]{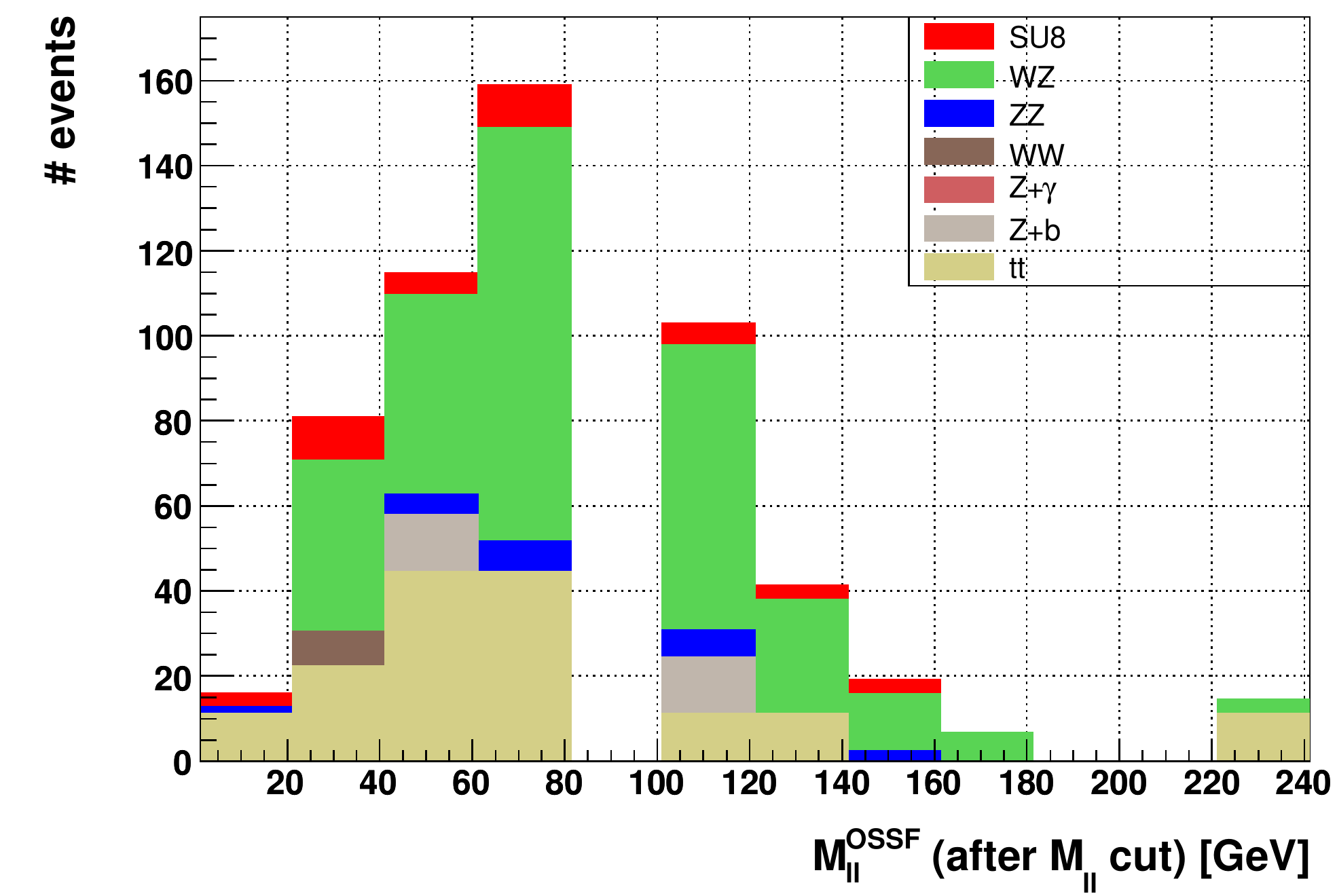}
\vspace{\cDist}
\vspace{\cDistHalf}
\end{center}
\caption[OSSF dilepton mass distribution for 10\,\fb{} after all cuts and without any jet veto for SU1, SU3, SU4, and SU8]{\label{fig:mllInclusive}
OSSF dilepton mass \mossf\ distribution for 10\,\fb{} and after all cuts and without the jet veto for SU1 {\bf(top left)}, SU3 {\bf(top right)}, SU4 {\bf(bottom left)} and SU8 {\bf(bottom right)}. The individual signal and background contributions are summarised in Table~\ref{tab:selection} on page~\pageref{tab:selection}.
\vspace{\cDistHalf}
}
\end{figure}

To quantify how well this general trilepton search strategy performs for other mSUGRA benchmark points SU1, SU3, SU4, and SU8, the same selection was applied. Deliberately, no optimisation of the selection algorithm based on special features of any of those four points\footnote{For example extremely high $\met$ for SU3.} was made. This was done in order not to be biased by specific features of any given benchmark point of the limited mSUGRA model in the broad search for Supersymmetry, according to the general philosophy of this analysis.
The results for SU1, SU3, SU4, and SU8 with 10\,\fb{} are shown in Figure~\ref{fig:mllInclusive}.% (top left/right, bottom left/right, respectively).

The statistical significances $\sgnf$ and the corresponding integrated luminosities $\intlumi$ for a discovery of supergravity at the SU$x$ benchmark points are summarised in Table~\ref{tab:significancies} on page~\pageref{tab:significancies}. It is striking that the point SU8 yields a rather small statistical significance compared to the other benchmark points, which cannot be explained with its slightly higher $M_0$, \Mhalf{} scales. The reason is its high $\tan\beta$ parameter which leads to a high signal branching ratio into taus rather than muons and electrons, and a mass degeneracy between the top and the stop resulting in a great enhancement of the Higgs-mediated gaugino decays into top quarks.

The discovery potential for the difficult {\em massive sparton scenario} defined by {\em direct gaugino pair-production} at the benchmark point SU2 (cf.~Chapter~\ref{chp:introSUSY}) is similar to the inclusive SU2 results, since $\sim$90\% of its total cross section are comprised of direct gauigino pair-production\footnote{However, it $S/\!B$ is not expected to decrease by 10\%, because the lepton $\pt$ spectrum for direct gaugino production is typically softer than for other processes.}. The statistical significance with 10\,\fb{} is $\mathcal S=4.0$, resulting in $\intlumi\simeq 15.3\,\fb$ for a discovery. The corresponding numbers for direct gaugino pair-production in SU3 are $\mathcal S=1.9$ and $\intlumi\simeq 68.5\,\fb$. Figure~\ref{fig:mllDirect} shows the \mossf{} distribution for direct gaugino pair-production in SU2 and SU3 after all cuts but the jet veto.

The trileptonic decay mode of direct chargino-neutralino pair-production can be used to measure various SUSY parameters. One possible approach to isolate it is by applying a veto %at the final stage of the selection 
on events with at least one jet within $|\eta|<2.5$ which fulfils  $\et^{\rm jet}>20$\,GeV and/or $b$-${\rm jet~likelihood}>0.3$. This strategy is pursued since few high-$\et$ and heavy quark jets are expected from initial state radiation or pile-up. Moreover, it removes the $\ttbar$ background almost completely, leading to different systematic uncertainties. The resulting dilepton mass distribution for SU2 with 10\,\fb{} is shown in Figure~\ref{fig:mllDirect}. The statistical significance after all cuts and the jet veto is 2.3~(1.5) for SU2~(SU3). A check at MC truth level shows that the actual fraction of direct chargino-neutralino production over all SUSY processes is $\sim$100\% for SU2 and $\sim$90\% for SU3 after all cuts and the jet veto. 

For reference, the $\mossf$ distribution for mSUGRA benchmark points SU1, SU3, SU4, and SU8 without any jet veto are shown in Figure~\ref{fig:mllInclusive}.

%\nopagebreak[5]

%% file: Trigger/Trigger.tex
The massive sparton scenario has strong implications on the ATLAS SUSY search strategy. If it is realised in nature, one cannot rely on hadronic processes and a~fortiori jet triggers for its discovery. It also appears unlikely that any $\met$ triggers could be reliably used right from the start of ATLAS. Therefore, it was investigated which lepton triggers might be expected to satisfy the requirements of this trilepton search analysis during the first years of LHC running at \instlumi{31-32}.

\begin{sidewaysfigure}
\vspace{-0.9cm}
\includegraphics[height=11.8cm,angle=90,clip]{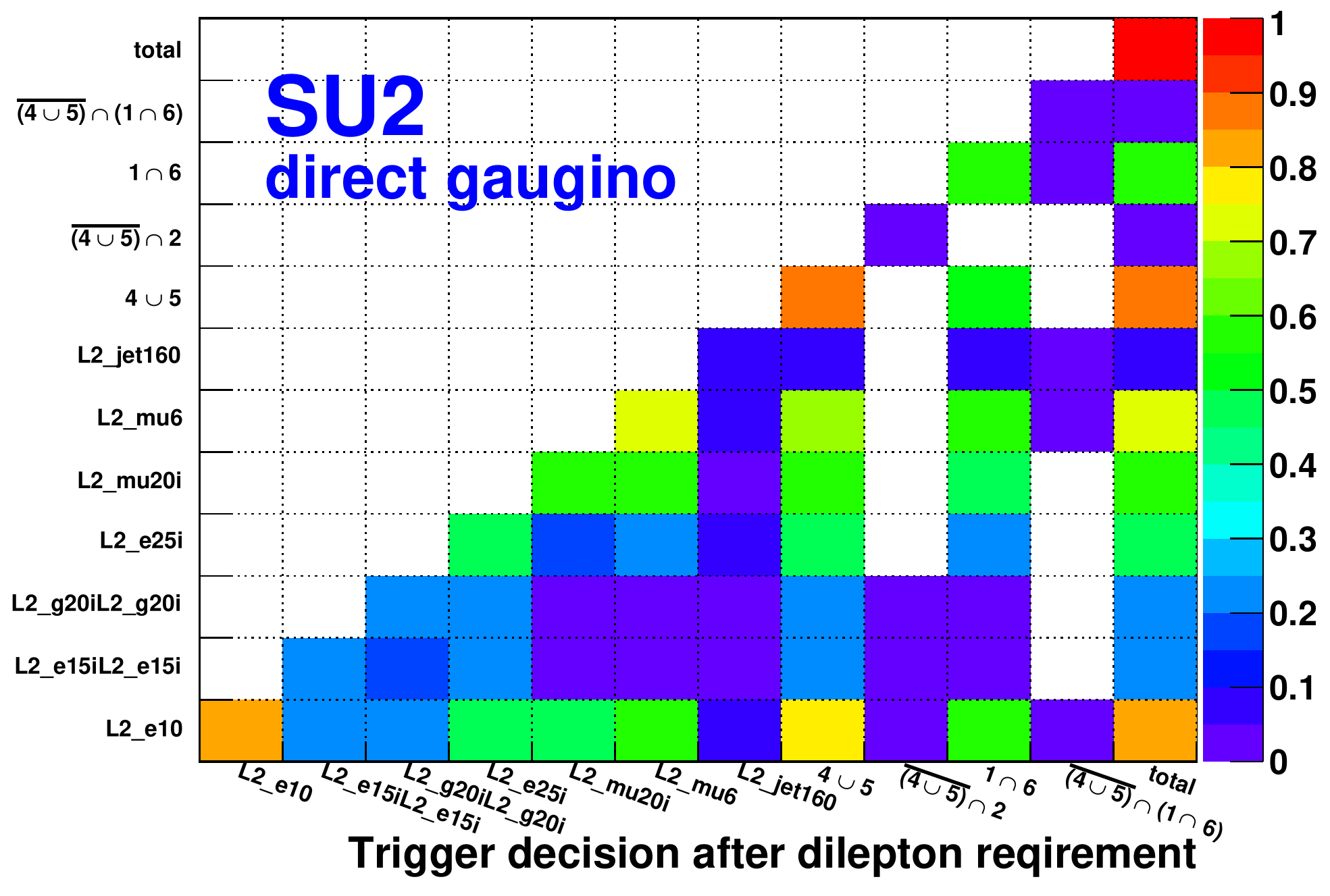}
\includegraphics[height=11.8cm,angle=90,clip]{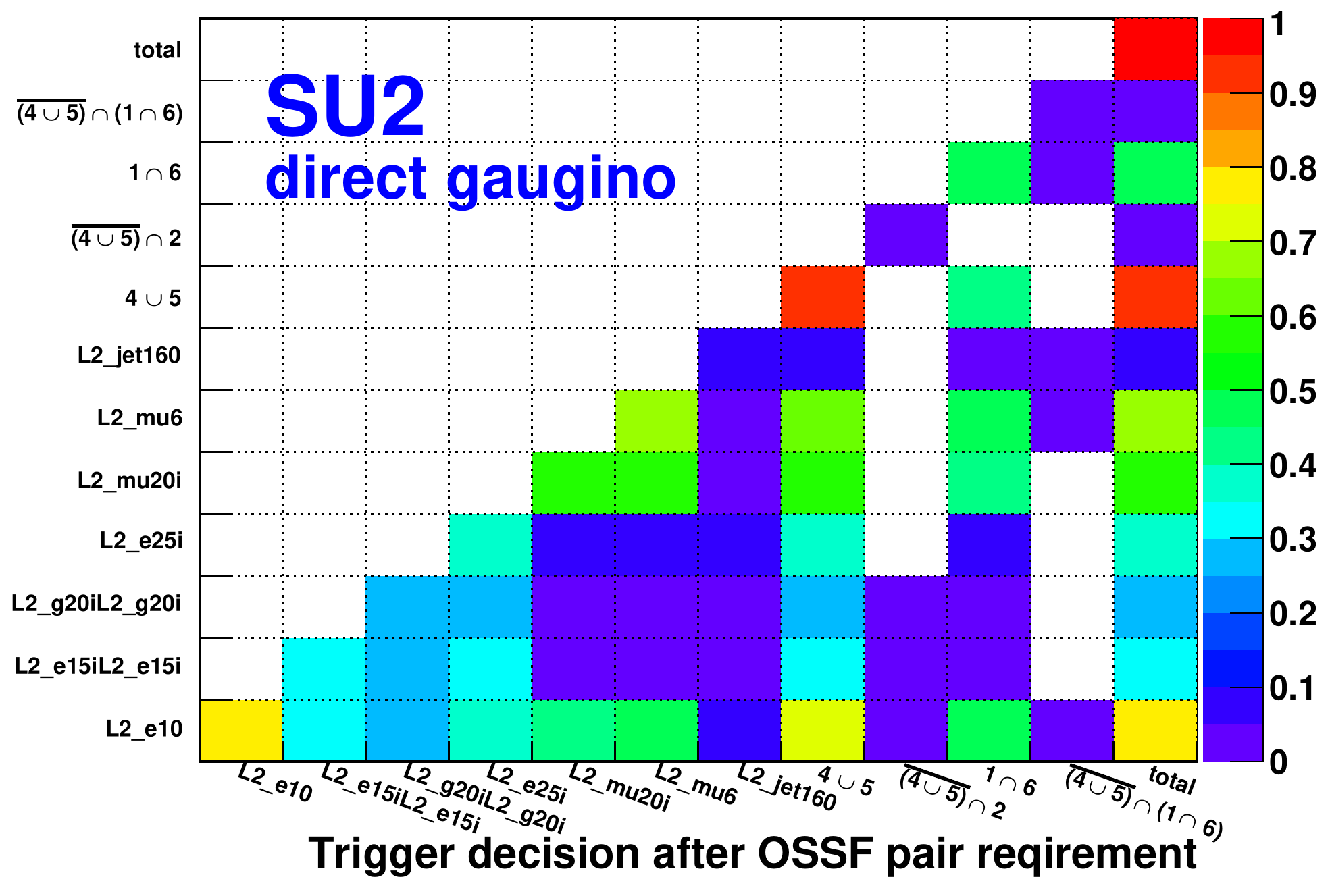}\\
\includegraphics[height=11.8cm,angle=90,clip]{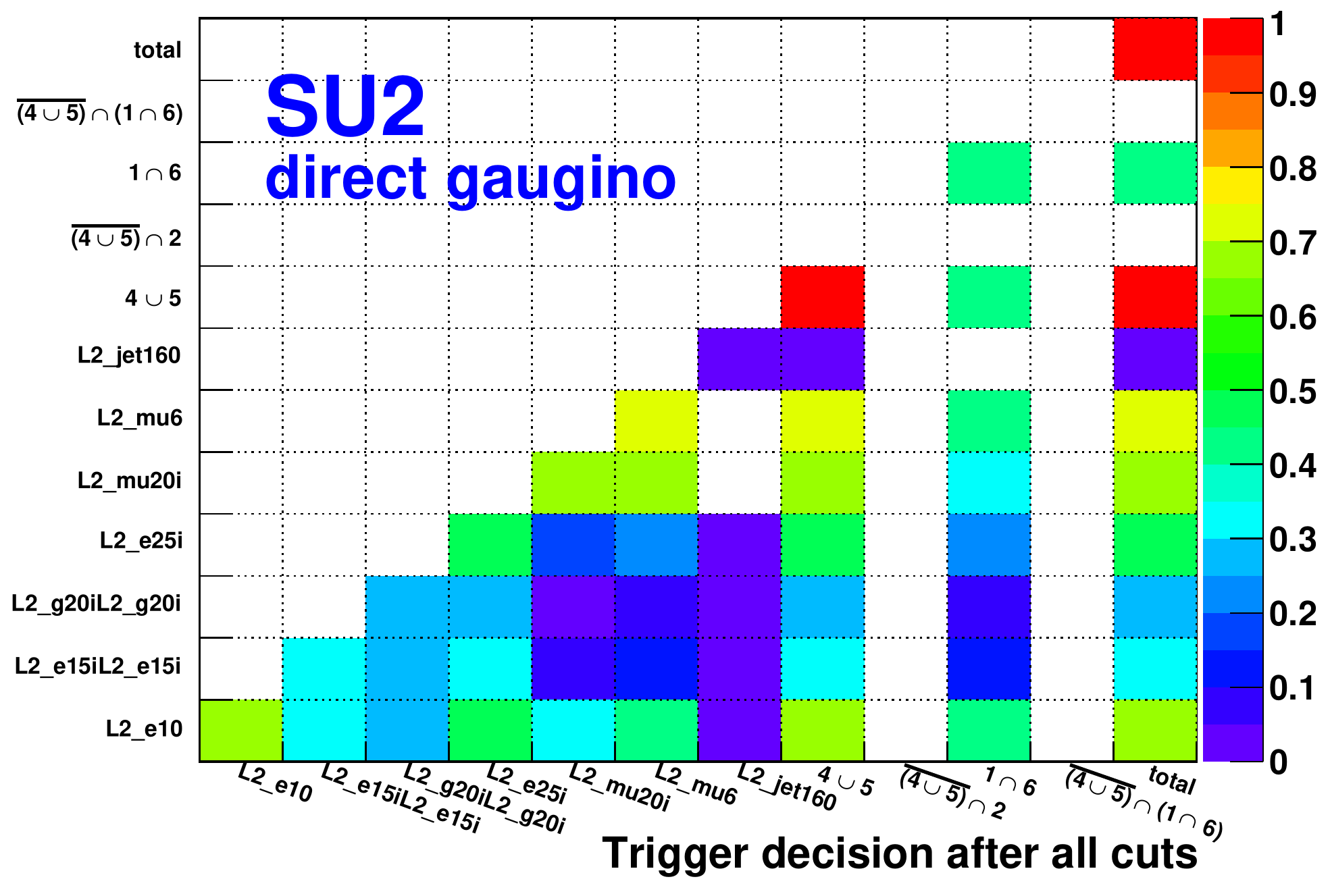}
\includegraphics[height=11.8cm,angle=90,clip]{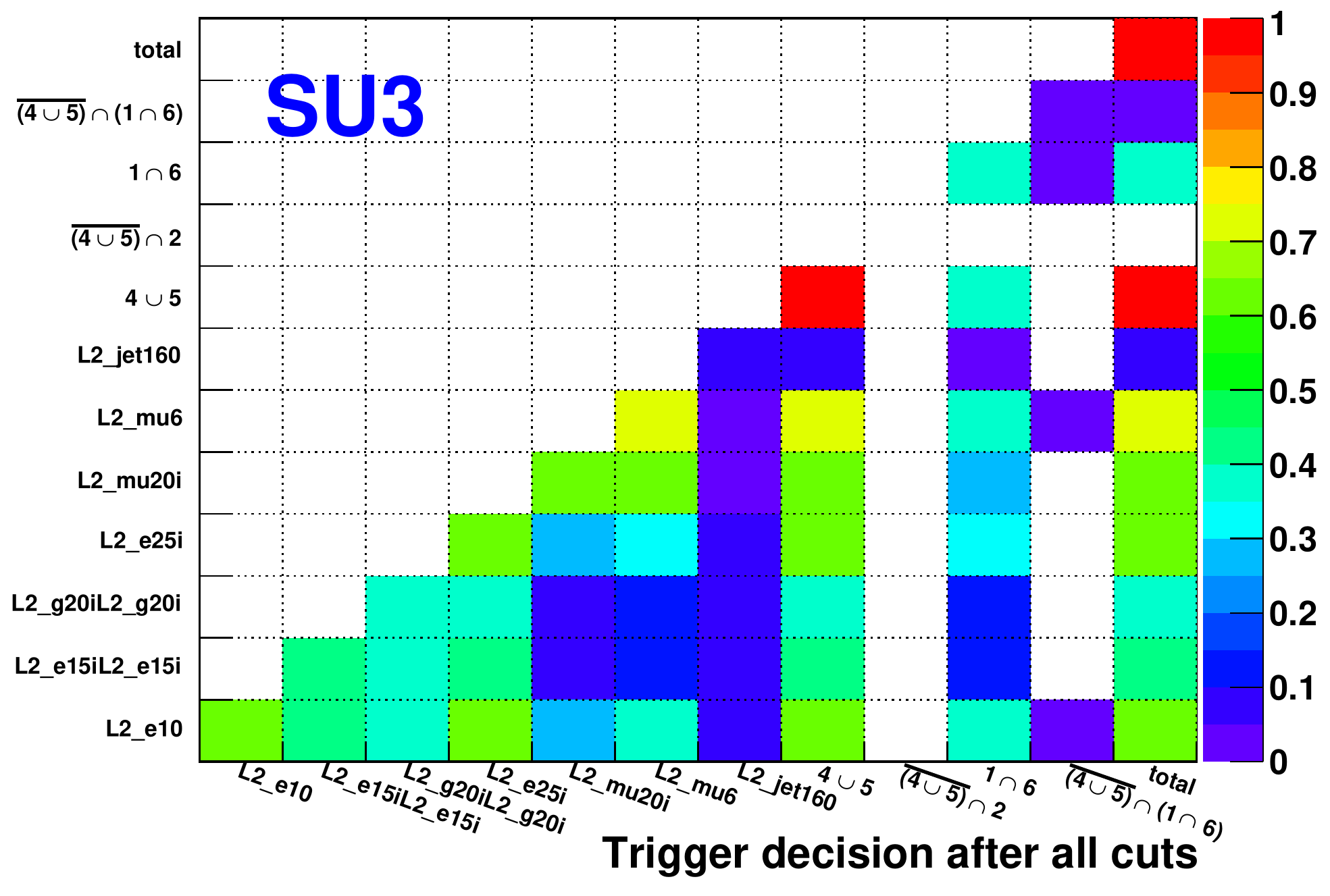}
\caption[Trigger efficiencies for AND-combinations of various triggers for the massive sparton scenario and SU3]{\label{fig:triggerJetVeto}
Trigger efficiencies for AND-combinations of various triggers for the massive sparton scenario (SU2, gaugino pair-production only) is shown at three selection stages: after the dilepton requirement {\bf(top left)}, after the OSSF pair + 3$^{\rm rd}$ lepton requirement  {\bf(top right)} and after all cuts {\bf(bottom left)}. The same table after all cuts is presented for the inclusive SU3 SUSY production {\bf(bottom right)}. The efficiency is shown colour-coded, with the scale indicated next to each of the figures. All figures are without any jet veto. For a detailed description see text.
\vspace{\cDistHalf}
}
\end{sidewaysfigure}

As a figure of merit, the fraction of triggered events at three steps of the analysis was used: at the dilepton stage, after the OSSF pair + 3$^{\rm rd}$ lepton requirement, and after all cuts (except for the jet veto). An inalienable condition was to have a {\em non-prescaled} trigger. 

%The jet veto was applied right after the preselection stage.

Figure~\ref{fig:triggerJetVeto} shows the performance of various triggers and their AND-combinations at three selection stages for the massive sparton scenario, id est direct gaugino pair-production in the SU2 sample. Each box $b_{ij}$ shows the efficiency of the AND-combination of the two triggers $i,j$ corresponding to its row ($i$) and column ($j$) coordinates, id est the diagonal boxes $b_{ii}$ give {\em single} trigger efficiencies. 
%The last row is a dummy trigger with 100\% efficiency. 
Rows marked as $i\cup j$, $i\cap j$ show OR, AND combinations of triggers $i,j$, respectively. $\bar A$ stands for the negation of $A$. All triggers have been studied at L2, in order to be independent of the Event Filter, which may not be well-understood in the early days of ATLAS.

From the table presented in Figure~\ref{fig:triggerJetVeto}, two single lepton triggers, \texttt{L2\_e25i} and \texttt{L2\_mu20i}, have been {\em identified} as well-suited for the trilepton analysis at an instanteneous luminosity of~\instlumi{31-32}:
\begin{itemize}
\vspace{\cDistHalf}
\item
The {\tt  L2\_e25i} trigger is optimised in $|\eta|$-bins of $\{0,\,0.75,\,1.5,\,1.8,\,2.0,\,2.5\}$ reflecting the calorimeter geometry. It is seeded by the {\tt  L1\_e25i} trigger at L1, which is formed from $2\times2$-tower EM-clusters and $4\times4$-tower EM and hadronic calorimeter clusters. Each tower has a granularity of $0.1\times0.1$ in $\eta\times\phi$, and is formed by analog summation of calorimeter cells with an effective threshold of about 1\,GeV on the total sum. The requirements of {\tt  L2\_e25i} trigger and its L1 seed are briefly summarised in Table~\ref{tab:triggerMuEm}.
\item
The {\tt  L2\_mu20i} trigger combines the $\pt$-based selection of the {\tt MuFast} and {\tt MuComb} algorithms, and the calorimeter isolation-based selection of the {\tt MuIso} algorithm. The three algorithms are executed in sequence: first, {\tt MuFast} is run on all L1 seeds in the MS; subsequently, tracks are reconstructed around the muon in the ID with the {\tt SiTrack} and {\tt IDSCAN} algorithms; in the next step, the ID and MS tracks are combined using the  {\tt MuComb} algorithm; and finally, muon isolation in the calorimeter is checked by the {\tt MuIso} algorithm, which uses the $\eta\times\phi=0.2\times0.2$ calorimeter towers around the muon. The \pt-thresholds of the {\tt MuFast}~({\tt MuComb}) algorithms are $\pt>18.95\,\GeV\,(19.33\,\GeV)$.\\
The {\tt  L2\_mu20i} trigger is seeded by the {\tt L1\_mu20} trigger at L1. This L1 muon trigger uses so-called coincidence windows in the innermost~(outermost) RPC layer for low~(high)-\pt\ tracks. A coincidence window is defined in the $Z-\eta$ plane of the detector: it is the distance between the hit measured in the respective RPC layer with respect to the intersect of an infinite momentum track, and is thus a direct measure of the muon momentum. The infinite momentum track is given by the straight line connecting the DIP and the pivotal RPC layer (i.e. the middle layer in the barrel).
\vspace{\cDistHalf}
\end{itemize}

\begin{table}
\vspace{\cDistHalf}
\begin{footnotesize}
\begin{center}
\begin{tabular}{l|ll}
\hline
Trigger & Variable & Cut \\
\hline
\hline
\multirow{4}*{\tt L1\_e25i} & Maximum \et\ in a $\eta\times\phi=2\times1$ or $1\times2$ cluster in the $2\times2$ region of interest  & $>18\,\GeV$\\
& $\sum\et^{\rm cell}$ in the 12 cell ring around the $2\times2$ region of interest & $\le3\,\GeV$\\
& $\sum\et^{\rm tower}$ in the 4 hadronic towers behind  the $2\times2$ region of interest & $\le2\,\GeV$\\
& $\sum\et^{\rm tower}$ in the 12 hadronic tower ring around the $2\times2$ region of interest & $\le2\,\GeV$\\
\hline
\multirow{5}*{\tt L2\_e25i} & \et\ of the cluster & $>24\,\GeV$ \\
 & $\et$ of leakage into hadronic calorimeter (only if $\et^{\rm cluster}<90\,\GeV$ & $>3\,\GeV$ \\
 & Ratio of $\sum_{3\times3}\et$ over $\sum_{7\times7}\et$ in the $\rm2^{nd}$ sampling of EM calorimeter & $>0.89$\\
 & $(E_1-E_2)/(E_1+E_2)$, where $E_i=\et^{i\mbox{-}\rm highest\,cell}$ in $\rm1^{st}$ sampling of EM calor. & $<0.6$ \\
 & \pt\ of matched track in $\eta\times\phi(\mbox{w/r/t\,cluster})=0.03\times0.15$ & $> 3\,\GeV$ \\
\hline
\end{tabular}
\end{center}
\end{footnotesize}
\vspace{\cDistHalf}
\caption[Trigger requirements for the \texttt{L2\_e25i} trigger]{\label{tab:triggerMuEm}
Trigger requirements for the \texttt{L2\_e25i} trigger and its L1 seed {\tt L1\_e25i}.
\vspace{\cDistHalf}
}
\end{table}

OR-combined, the \texttt{L2\_e25i} and \texttt{L2\_mu20i} trigger provide $\sim$89,\,90,\,97\% efficiency for the massive sparton scenario at the three selection stages, repectively (Table~\ref{tab:trigger}, first block). With L2 rates of $\sim$5 and $\lesssim$1\,Hz they are unprescaled~\cite{bib:triggerReferenceEl, bib:triggerReferenceMu}, and have sufficiently high $\pt$ thresholds to be easily studied with leptonic $Z$ decays. If operated in pass-through mode at L1, \texttt{L2\_mu20i} is expected to fire at a rate of 14\,Hz~\cite{bib:triggerReferenceMu}.

Additionally, the same trigger set was studied for the direct gaugino pair-production in~SU3. Figure~\ref{fig:triggerJetVeto} (bottom right) shows the efficiencies for the final selection stage without the jet veto. Also here the OR-combination of \texttt{L2\_e25i} and \texttt{L2\_mu20i} demonstrates a convincing performance of $\sim$91,\,93,\,98\%. Their individual efficiencies are summarised in the middle block of Table~\ref{tab:trigger}.

\begin{table}[b]
\vspace{\cDistHalf}
\begin{footnotesize}
\hspace{-3mm}
\begin{tabular}{l|ccc|ccc|ccc}
\hline
Selection & \multicolumn{3}{c|}{\bf SU2$\chi$} & \multicolumn{3}{c|}{\bf SU3$\chi$} & \multicolumn{3}{c}{\bf SU3 incl.} \\
stage & \texttt{L2\_e25i} & \texttt{L2\_mu20i} & $\bigcup$ & \texttt{L2\_e25i} & \texttt{L2\_mu20i} & $\bigcup$ & \texttt{L2\_e25i} & \texttt{L2\_mu20i} & $\bigcup$ \\
\hline\hline
$\geq2\ell$		& 45\% & 60\% & {\bf89\%} & 46\% & 55\% & {\bf91\%} & 53\% & 56\% & {\bf93\%} \\
OSSF+3$^{\rm rd}\ell$	& 38\% & 59\% & {\bf90\%} & 41\% & 56\% & {\bf93\%} & 48\% & 55\% & {\bf94\%} \\
after all cuts		& 48\% & 66\% & {\bf97\%} & 60\% & 63\% & {\bf98\%} & 60\% & 64\% & {\bf98\%} \\
\hline
\end{tabular}
\end{footnotesize}
\vspace{\cDistHalf}
\caption[Trigger efficiencies at three selection stages for direct gaugino pair-production in SU2, SU3, and the inclusive SU3 production]{\label{tab:trigger}
Trigger efficiencies at three selection stages for the direct gaugino pair-production in SU2 (first block, ``SU2$\chi$''), the direct gaugino production in SU3 (second block, ``SU3$\chi$''), and the inclusive SU3 signal (third block, ``SU3 incl.''). ``$\bigcup$'' stands for the OR-combination of \texttt{L2\_e25i} and \texttt{L2\_mu20i}.
\vspace{\cDistHalf}
}
\end{table}

One important question is how well the \texttt{L2\_e25i} and \texttt{L2\_mu20i} triggers perform for the trilepton analysis outside of the massive sparton scenario. They have been studied for the inclusive SU3 production, and were found to provide a high efficiency of around $\sim$93,\,94,\,98\%, as shown in the last block of Table~\ref{tab:trigger}.

One might be puzzled why the efficiency is so high for lepton triggers with relatively high thresholds compared to the offline cuts on lepton transverse momenta. The answer is given by simple combinatorics: since the three leptons are ordered by their transverse momenta, it is likely that the leading lepton has a high $\pt$ if the third one passes the offline threshold. On the other hand, cases where \textit{all} the three leptons are below 20-30\,GeV but above 10/15\,GeV are unlikely.

It can be concluded that the \texttt{L2\_e25i} and \texttt{L2\_mu20i} single lepton triggers provide a good performance for the trilepton analysis in the early days of ATLAS running at\linebreak[4]\instlumi{31}. 

For higher luminosities, the event filter stage can be included, and the trigger object definition can be tightened. Furthermore, dileptonic triggers with lower thresholds like \texttt{e15ie15i}, \texttt{mu10imu10i}, and \texttt{e15imu10i} can additionally be used to recover events where all three leptons have low transverse momenta around 20\,GeV. In the same spirit, {\em trilepton} triggers of all possible combinations of \texttt{g10} and \texttt{mu6} were implemented in the \instlumi{31} trigger menu~\cite{bib:trileptonTriggers} for further studies, aiming at recovering leptons with low transverse momenta, leptons failing the trigger isolation criteria, and electrons whose track has been missed by the trigger.

%% file: Backgrounds/Backgrounds.tex
So far, general analysis strategies have been presented. However, one must keep in mind that they were elaborated using simulated Monte Carlo events, and that many assumptions, for example about the cross sections, have been made. This Chapter aims at outlining experimental techniques to determine these assumed quantities from data, as well as to measure the background contributions in a way least relying on Monte Carlo simulations.

%% file: Backgrounds/Classification.tex
In general, systematic uncertainties can be divided in two categories:% {\em physics} and {\em instrumental}.
\begin{description}
\item[Instrumental uncertainties:]
they account for our limited understanding of the {\em experimental} environment. The most relevant factors are:
\begin{itemize}
 \vspace{-2mm}
 \setlength{\itemsep}{0mm}
 \item Modelling of the detector and its readout electronics response;
 \item Pile up;
 \item Secondary effects like beam-gas and beam halo interactions, cosmics, cavern background;
 \item Uncertainty on the integrated luminosity \intlumi.
 \vspace{-2mm}
\end{itemize}
\item[Physics uncertainties:]
they typically mirror our limited theoretical understanding of the {\em underlying physics}. The most important ones for this analysis are:
\begin{itemize}
 \vspace{-2mm}
 \setlength{\itemsep}{0mm}
 \item Uncertainties on the total cross sections of background\footnote{Since this search analysis is a counting experiment in its nature, the total cross sections of the signal processes play a minor role, as they are used to estimate discovery prospects only.} processes;
 \item Parton Density Function (PDF) uncertainties;
 \item Uncertainties on differential distributions like \met, \pt{}-spectra of leptons, etc.;
 \item Underlying event uncertainties.
 \vspace{-2mm}
\end{itemize}
\end{description}

%% file: Backgrounds/UncertInstrumental.tex
The first three sources for instrumental uncertainties listed above mainly affect the Jet Energy Scale (JES\glossary{name=JES,description=Jet Energy Scale}), the missing transverse energy, as well as the efficiencies and fake rates of electrons and muons. 
All these three effects can be measured from data.\\
For this analysis, the JES is only relevant for the jet veto. It can be studied in $Zj\rightarrow\ell\ell j$ events as described in~\cite{bib:jetsCSC}.\\
Typically, the uncertainties on \met{} are high, as it includes measurements of all major components of the detector. However, this analysis is in the advantageous situation to require a relatively relaxed cut on \met. Therefore a detailed understanding of the tails of the \met{} distribution is of minor importance. Thus, this uncertainty can be safely estimated from the $Zj\rightarrow\ell\ell j$ processes, for which no missing transverse energy is expected, using experimental techniques outlined in~\cite{bib:jetsCSC}.\\
The efficiencies and fake rates of leptons at trigger and offline reconstruction levels can be determined from data using the so-called ``tag-and-probe'' method in $Z\rightarrow\ell\ell$ events. Their careful understanding is particularly important, because this analysis requires three leptons\footnote{Typically, the uncertainties on each of the three leptons are fully and positively correlated.}. Deliberately, the actual event selection begins from an OSSF lepton pair requirement. This strategy will provide a possibility to validate the preselection by verifying the matching of the \mossf{} distribution in simulated MC events against data. Preliminary results on lepton efficiencies and fake rates based on MC simulations can be found in~\cite{bib:leptonEff}.

The trilepton analysis relies on the identification of isolated prompt leptons. Preliminary studies based on MC simulations showed, that secondary leptons from $b$-jets are $\mathscr O(10)$ more likely to mimic an isolated lepton than light quark jets. 
%As detailed in Chapter~\ref{chp:selection}, track and calorimeter isolation were found to be suitable tools to reduce this background. 
Therefore it is essential to understand the expected rate of such secondary leptons passing isolation criteria. Experimental techniques for this are outlined in Section~\ref{sec:secLept}.

The effects of pile-up interactions on lepton identification, track and calorimeter isolation can be studied using luminosity blocks of data with similar instantaneous luminosity $\mathscr L$. This can be done using $Z\rightarrow\ell\ell$ events for prompt and \ttbar{} for secondary leptons.

So far, the fake rate of electrons from photon conversions has not been thoroughly investigated in the context of multi-lepton searches. The preliminary results of this analysis and~\cite{bib:hays} suggest that the $Z\gamma$ process gives a negligible contribution. 
%On the other hand it should be mentioned, that a generator level cut of $\pt^\gamma>25$\,GeV was applied to photons in this sample, whereas for reconstructed electrons at least $\pt^e>15$\,GeV is required.
The sample used for those studies had a generator-level cut of $\pt^\gamma > 25$\,GeV. Since this analysis requires electrons with $p^e_T > 15$\,GeV, it will be necessary to determine (in a future study) the sensitivity of this analysis to $Z\gamma$ with lower $\pt$ photons. It should be also mentioned that for a very small fraction of signal events the leading lepton has a $\pt<30\,\GeV$, which requires a detailed knowledge of the trigger turn-on for the {\tt L2\_e25i} and {\tt L2\_mu20i} triggers.

An important source of systematic uncertainty for this search -- which is a counting experiment -- can be attributed to the measurement of \intlumi. It is expected to reach a precision of around 10-15\% in early days of the LHC, and about 5\% by the time when $\intlumi\simeq10\,\fb$ are collected. To avoid any bias from the normalisation of background MC simulations due to \intlumi{} uncertainties, the same technique as discussed in the context of cross section uncertainties in the next Section can be applied.

%% file: Backgrounds/UncertPhysics.tex
Strategies to control some of the physics uncertainties relevant for this analysis are outlined in this Section. The modelling of some of the phenomena giving rise to physics uncertainties -- such as the underlying event -- will be addressed by an LHC-wide effort, and so are not discusssed here in any detail.

At the moment, the uncertainties on differential distributions like the $b$-jet $\pt$ in $Zb$ events would have to come from theoretical calculations. However, they could be measured in data once the LHC starts up. To some extent, the effects of some of these uncertainties could be reduced using {\em control regions}. Besides the lepton $\pt$, the lepton pseudorapidity distribution is of major importance, as it changes the yield of both signal and background processes inside the detector acceptance. In case of signal, $\eta$ has a strong dependance on the mass scale of supersymmetry.

One can attempt to estimate the uncertainty on the cross section due to imperfectly known parton distributions\footnote{Sometimes the uncertainty on PDF's is included in the cross section uncertainty.} by using a PDF set with ``error eigenvectors'' which allow one to calculate an uncertainty band. However it must be borne in mind that this band only accounts for uncertainties originating from the experimental data which were used in the PDF determination, and thus does not take into account theoretical assumptions\footnote{Strictly speaking, this is true for the PDF sets available at the time as this analysis was performed. There has been some recent progress to account for uncertainties from theoretical assumptions in a quantitative way.}.

Another important source of systematics comes from theoretical uncertainties on the total cross sections of background processes. Most of the MC simulations used in this analysis were generated using a LO matrix element, cf.~Chapter~\ref{chp:samples}. If available, $k$-factors are used to take into account the latest NLO calculations\footnote{In fact, for \ttbar{} an NLO cross section was corrected for the leading logarithmic resummation effects.}. Their magnitude can be significant, like for example $k=2.05$ for $WZ$ production~\cite{bib:gaugeBosonPairs, bib:dibosonHN, bib:zhou, bib:xsecRecomm}. One must also bear in mind that higher order calculations do not only change the total cross section, but can also affect the shape of distributions, so genuine NLO calculations are preferable to LO calculations multiplied by correction factors.

To reduce errors due to the uncertainties on total cross sections and PDF's, the contribution of the most important backgrounds can be determined using control regions. For this analysis, $\met<20$\,GeV and $\mossf\in[81.2,\,101.2]$\,GeV appear to be a logical choice. In general, the former criterion will isolate backgrounds without (or with small) genuine \met{} like $ZZ$, $Z\gamma$, and $Zb$, whereas the latter will predominantly isolate backgrounds involving $Z$ boson production: $WZ$, $ZZ$, $Z\gamma$, $Zb$. Therefore, it is advantageous to define the $\met$ control region first to estimate the $ZZ$ and $Zb$ contribution before focusing on the dominating $WZ$ background in the $\mossf\in[81.2,\,101.2]$\,GeV control region. With the less important $Z\gamma$ background it is planned to fully rely on MC for estimating its contribution.\\
The normalisation of the $ZZ$ background can be estimated by counting events with two OSSF pairs giving $\mossf\simeq m_Z$ and taking into account the reconstruction efficiency and acceptance for the 4$^{\rm th}$ lepton.\\
After the trilepton requirement, the contribution of the $Z+{}$heavy flavour background processes can be estimated by comparing the area under the fit to the $m_Z$ peak in the control region $\mossf\in[81.2,\,101.2]$\,GeV for $\met<20$\,GeV and $\met>20$\,GeV, taking from MC simulations only the {\em fraction} of $Z+{}$heavy flavour in the respective \met{} region. It should be mentioned that besides $Zb$, a contribution from the $Zc$ process is expected, which is much smaller in magnitude because of the softer $\pt$-spectrum of the leptons from $c$-quark decays. This process is not included in this analysis due to the lack of simulated Monte Carlo events at ATLAS at the time as it was performed. However the effect is expected to be small due to aforementioned reasons.\\
After the trilepton and the \met{} requirements have been applied, the dominating backgrounds are $WZ$ and \ttbar{} (see Table~\ref{tab:selection}). Using the \mossf{} distribution, their contribution can be estimated by fitting the $m_Z$ peak with a Voigt function and the continuum with a function appropriately describing the \mossf{} profile of \ttbar. Eventually, a third fit function describing the \mossf{} distribution of the signal can be included, and the event yield of the signal can be evaluated using a maximal likelihood method or similar techniques.

%% file: Backgrounds/SecondaryLeptons.tex
Even after the trilepton requirement, the supersymmetric signal is dominated by the main background:~\ttbar. Since top-antitop production can provide only two \textit{prompt} leptons from dileptonic \ttbar{} events, this process can enter the selection if \textit{secondary} leptons from semileptonic $b$ decays pass the lepton definition and isolation criteria. 
%In fact, the introduction of the track isolation, aimed at identifying such secondary leptons, reduces the \ttbar{} background by an order of magnitude.

In order to reduce the background from secondary leptons from $b$-decays in \ttbar\ and $Zb$ events, a track isolation requirement was introduced for  leptons in the ATLAS baseline trilepton search analysis by the author~\cite{bib:csc7}. It  reduced the \ttbar{} background by almost an order of magnitude. Given this, it is essential to determine the rate \Rsec{} of secondary leptons from $b$'s passing track and calorimeter isolation criteria, defined as:
\begin{equation}\label{eqn:rsec}
 \Rsec \equiv \frac{\textnormal{\# of leptons from $b$-jets passing the isolated lepton definition}}{\textnormal{\# of $b$-jets}}\,.
\end{equation}
In order to measure this rate from ATLAS data using a tag-and-probe method, a clean sample of $b$-jets is needed for the definition of the \Rsec\ denominator -- the {\em tag}, and a suitable definition of the {\em probe}. This can be done using $b\bar b$ and \ttbar{} events:
\begin{description}
%\vspace{-2mm}
\setlength{\itemsep}{0mm}
\item[$\mathbf{b\bar b}$ events:]
since many years, a topological tag-and-probe method is successfully used at the Tevatron. The approach is simple: one of the $b$'s is used as the {\em tag}; the object opposite in $\phi$ is the {\em probe}. Straightforward, the former defines the denominator of \Rsec\ in Equation~\ref{eqn:rsec}, and the frequency of isolated leptons in the probe $\phi$-region defines the numerator.\\
Despite its long success story at the Tevatron, this method has several disadvantages at the LHC: the rate of backgrounds from e.g. associated $Wb$ or single top production is much higher than at the Tevatron. Further, despite their high production cross section at the LHC, only a limited fraction of $b\bar b$ events will be recorded to tape. Moreover, most $b$-triggers are dedicated to heavy flavour physics, and a possible bias, which might come from e.g. the presence of a prompt lepton in the event, will be difficult to disentangle. Finally, there are high uncerainties on the production rates of $c$- and $b$-quark jets events from inelastic parton--$c$-quark or parton--$b$-quark scattering. This is because of the poor knowledge of $c$- and $b$-quark PDFs in the kinematic regime of the LHC;
 %\item One uses the impact parameter of the track to tag a lepton which is likely to come from a $b$ decay, and investigates the track isolation of such leptons. The disadvantage of this method is that there is a strong correlation between the track impact parameter and the track isolation: if a strong cut on the impact parameter is required in order to calibrate the track isolation, it is not straight forward to extend the calibration to $b$ decays with smaller impact parameters. Experimentally, decay kinematics of $b\bar q$ systems are still poorly understood, and theoretical calculations currently available have a big uncertainty of several tens of percent.
\item[$\mathbf{\ttbar}$ events:]
given the caveats regarding the implementation of the $b\bar b$ method at the LHC, the author and A.~J.~Barr have elaborated a suggestion to employ $\ttbar$ events. They offer various advantages: the \ttbar\ process has a high cross section at the LHC, and will be triggered upon extensively. Further, $b$-jets from \ttbar{} production will inherently feature the correct $\et$ spectrum, whereas $b\bar b$ will have to be re-weighted, which is a source for systematic uncertainties. The main advantage is however that -- as will be detailed in Subsection~\ref{ssec:rateEstimation} -- the \ttbar\ process offers various handles to define clean and unbiased tag and probe samples.\\
In the context of the trilepton search analysis, {\em semileptonic} \ttbar{} decays appear as a locial choice, not only because of their $\order{10}$ higher branching ratio, but also because {\em dileptonic} \ttbar{} event sample will make a non-negligible contribution as a background to the trilepon search analysis.
\end{description}
From the arguments above, it appears that it is favourable to use semileptonic \ttbar{} events to obtain a clean sample of $b$-jets in the experimental environment of the LHC. However, both methods should be used as they provide fully independent cross-checks to each other. 

\subsection{Estimation of the Rate of Secondary Leptons\newline Passing Isolation Criteria Using \ttbar\ Events} \label{ssec:rateEstimation}
The \ttbar\ method to estimate the rate of secondary leptons passing isolation criteria is outlined in the following. Before the tag and the probe are defined, consider that in the tag selection only {\em one} of the $b$-jets -- referred to as the ``tag-$b$-jet'' in the following -- can be used. This is to avoid any possible bias to the probe selection. For the same reason, both leptons in the event are to be sufficiently far separated from the tag-$b$-jet in $\eta\times\phi$.

Considering the above, the following criteria could be used to define the {\em tag} without using one of the $b$-jets:
\begin{enumerate}[1)]
\item a tight $b$-tag requirement on the tag-$b$-jet;
\item two jets with with an invariant mass consistent with that of a $W$ boson: $m_{jj}\simeq m_W$;
\item missing transverse energy;
\item kinematic properties of one full leptonic or hadronic top decay branch, depending on which of the $b$-jets serves as the tag-$b$-jet in 1).
\end{enumerate}
This tag definition can be used directly as the \Rsec\ denominator in Equation~\ref{eqn:rsec} with a reasonably small background contamination, which will be addressed briefly in Subsection~\ref{ssec:rateEstBgr}. Logically, the {\em probe} is the $b$-jet\footnote{This is irrespective of whether it has been reconstructed or not.} which was {\em not} used in the tag definition.

\begin{figure}
\begin{center}
\includegraphics[width=7.9cm,clip]{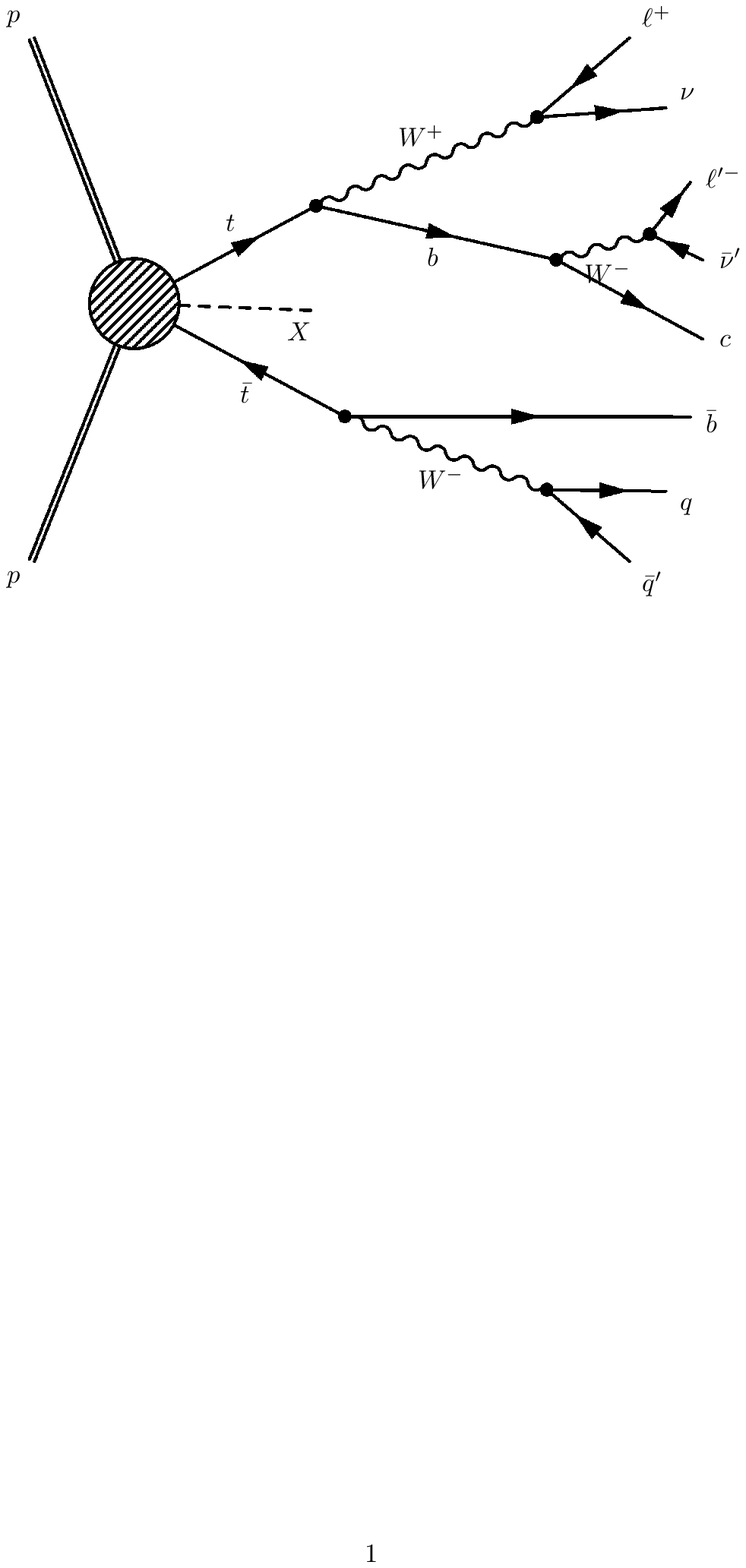}
\includegraphics[width=7.9cm,clip]{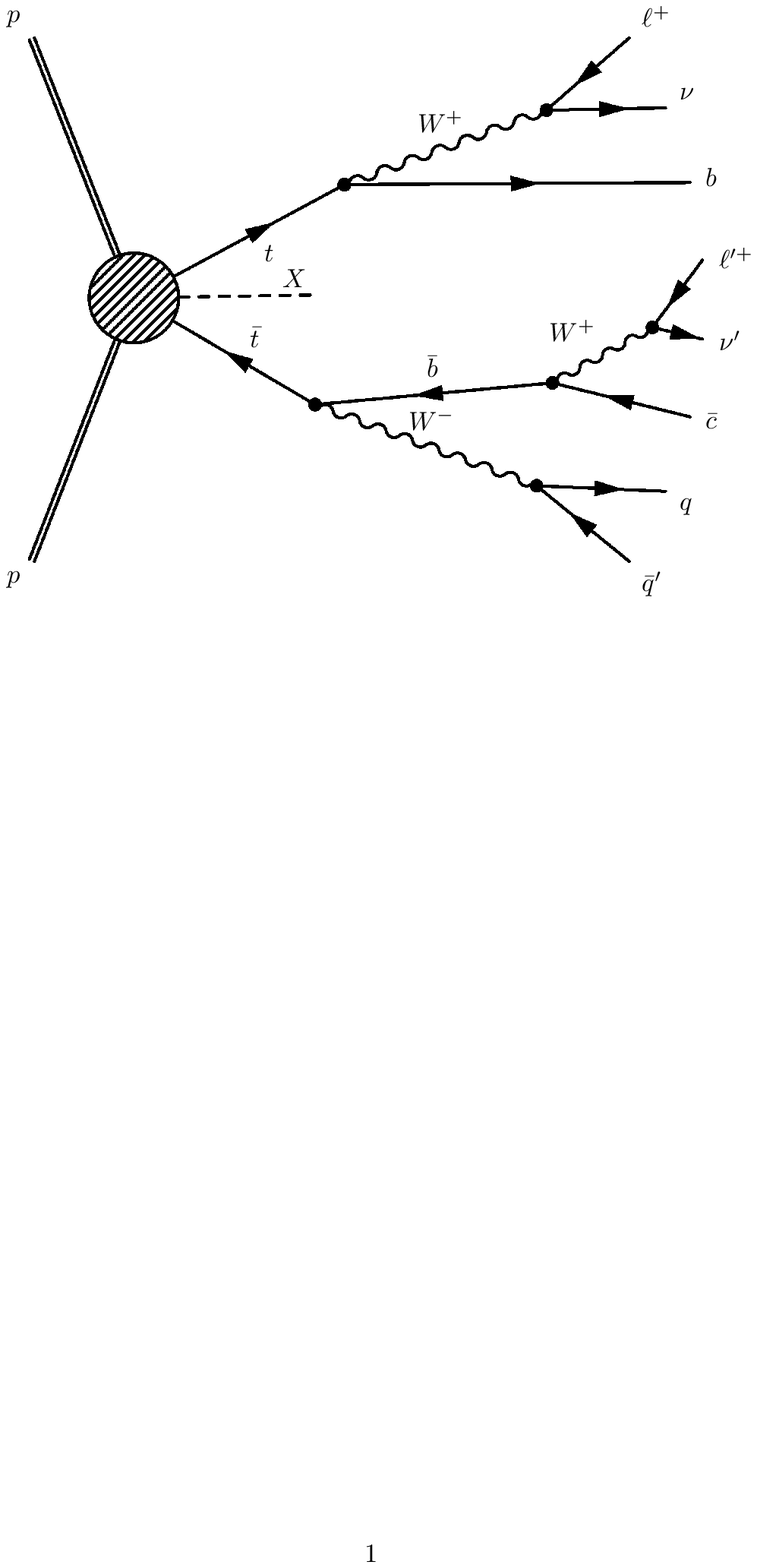}
\end{center}\vspace{-4mm}
\caption[Two representative \ttbar{} decay scenarios, where the secondary lepton is of opposite sign or of like sign to the prompt lepton]{\label{fig:ttbarSS}
Two representative \ttbar{} decay scenarios, where the secondary lepton is of opposite sign {\bf(left)} or of same sign {\bf(right)} to the prompt lepton.
}\vspace{-2mm}
\end{figure}

The more challenging part is to define the numerator of \Rsec\ in Equation~\ref{eqn:rsec}, as one needs to reliably identify events with {\em exactly one} prompt and {\em exactly one} secondary lepton. Otherwise one will not measure the rate of \textit{secondary} leptons to pass isolation criteria, but also add an admixture of \textit{prompt} leptons. This inalienable requirement is why it is not possible to use semileptonic \ttbar{} events where the secondary lepton from a $b$-decay has an \textit{opposite} sign to the prompt one, like for example on the diagram in Figure~\ref{fig:ttbarSS} (left). A high bias on the \Rsec\ numerator measurement would result from dileptonic \ttbar{} events with {\em two prompt} leptons. On the contrary, \textit{like-sign} events, like on Figure~\ref{fig:ttbarSS} (right), provide for a good identification of an event with {\em one prompt} and {\em one secondary} lepton. 

Strictly speaking, it is not possible to relate the secondary lepton either to a $b$ or a $\bar b$ decay using its charge, since a bottom quark can form a bound state and thus oscillate, or dominantly decay to a charm quark, which can produce a lepton of opposite charge in its decay. However, such a situation will occur {\em symmetrically} for both the $b$ and the $\bar b$ in a \ttbar{} event, so an overall combined estimate can be provided. %Further, one does not aim at disentangling the individual contributions from bottom and anti-bottom to the rate of secondary leptons passing isolation criteria.%, but rather considers the event with both of them as a whole. 

Thus, the numerator for \Rsec{} in Equation~\ref{eqn:rsec} is given by the rate of {\em like-sign} lepton pairs passing lepton isolation criteria in {\em tagged} events multiplied by 2:
\[
 \Rsec \equiv \frac{2\times\{\textnormal{\# of events fulfilling criteria 1)...4) with a {\em like-sign} isolated lepton pair}\}}{\textnormal{\# of events fulfilling criteria 1)...4)}}\,.
\]
The factor of~2 accounts for the equal contribution to the total \Rsec\ rate from opposite-sign lepton pair events. To draw this conclusion, the assumption was made that the probability to become the tag-$b$-jet (or the probe) is 50\% both for the $b$ and $\bar b$.

In the entire argumentation above, it is implicitly assumed that the kinematic properties of $b$-jets in dileptonic and semileptonic \ttbar{} events are the same.

\subsection{Backgrounds to the Estimation of the Rate of Secondary Leptons Passing Isolation Criteria Using \ttbar\ Events} \label{ssec:rateEstBgr}
As argued in the previous Subsection, one is interested in events with \textit{exactly one} prompt and \textit{exactly one} secondary lepton for the \Rsec\ measurement. This category of events shall be defined as the ``signal''\footnote{The terms ``signal'' and ``background'' strictly refer to this Chapter, and are not to be confused with signal and background of the SUSY search analysis.} here. In this sense, {\em tag} events with \textit{two} prompt leptons are to be considered as ``background''. The possible sources are:
\begin{description}
 \item[{\bf Dileptonic~}$\mathbf{\ttbar+j:}$] due to the high cross section and the similarity to semileptonic \ttbar{} events this is probably the main background for this study. It can enter the numerator of \Rsec{} by lepton charge misidentification. The only tag criterion not fulfilled generically is 2), id est $m_{jj}\simeq m_W$, so this jet pair has to come from initial or final state radiation. The contribution of this process to the ``signal'' can be neglected if events with {\em exactly} two leptons  are considered;
 \item[$\mathbf{Single~top+j:}$] For this process, in the three tree-level production diagrams one has to differentiate whether they contribute to the ``signal'', id est have a secondary lepton from a $b$ to enter the numerator of \Rsec{}, or the ``background'', id est have two prompt leptons in the final state. For example, associated $Wt$ production in the case where one of the $W$'s decays hadronically is to be considered ``signal'', whereas leptonic decays of both the $W$ and $t$ will provide two prompt leptons and is ``background''. In both cases additional jets are needed to pass the tag selection;
 \item[$\mathbf{Z+j:}$] This process can provide two prompt like-sign leptons from the $Z$ decay and charge misidentification. Additional jets, of which at least one with a tight $b$-tag are needed to pass the tag selection;
 \item[$\mathbf{WZ,ZZ+j:}$] These processes can provide up to four prompt leptons. $WZ$, where the $Z$ decays leptonically and the $W$ hadronically, provides for the more important contribution. For this, at least one jet with a tight $b$-tag is needed to pass 1).
%  \item[$\mathbf{b\bar b+j:}$] This process is here for completeness, but does not pose a background contribution, as both leptons from the $b$ and $\bar b$ decays are secondaries. However, it will surely pose a contribution to both the numerator and denominator of \Rsec, which cannot be neglected due to the high cross section. The same sign of the lepton pair can come about in the same processes as discussed above for the signal. Additional jets of which one needs to have a $b$-tag are needed.
\end{description}

There are other processes, which cannot be strictly classified as ``signal'' or ``background'', but should be mentioned here for their importance:
\begin{description}
 \item[$\mathbf{W+j:}$] this is the main background to any semileptonic \ttbar{} selection. It can pass the tag selection by a prompt lepton from the $W$ decay, and additional jets can mimic the full top event kinematics. Moreover, if an associated $b\bar b$ pair is produced, one of those can pass the $b$-tag and the other provide a secondary lepton, making it a contribution to the ``signal'';
 \item[$\mathbf{b\bar b+j:}$] it does not pose a background contribution in the strict sense as defined above, since both leptons from the $b$ and $\bar b$ decays are secondaries. However, it will surely enter both the numerator and denominator of \Rsec, which cannot be neglected due to the high production cross section. A like-sign lepton pair can come about via the same mechanism as discussed for semileptonic \ttbar. Additional jets, one $b$-tagged, are needed to pass the tag selection.
\end{description}

% \item[$\mathbf{W+j:}$] This background is similar to $Z+j$, but provides only 1 prompt lepton. The second lepton can be a secondary one.

\subsection{Conclusion}
A method to measure the rate of secondary leptons passing track- and calorimeter-isolation criteria from data using $\ttbar$ events has been suggested by the author in collaboration with A.~J.~Barr. Although it offers various advantages over the traditional $b\bar b$ method successfully employed at the Tevatron, the final proof of principle is still pending. Initially, it was planned to be provided by the author. However, these plans were put on ice given the tight ATLAS start-up schedule and the author's intense involvement in the activities around the alignment of the ATLAS silicon tracker, which is the subject of Part~\ref{prt:alignment} of this thesis.

%% file: Backgrounds/Systematics.tex
The systematic uncertainties most relevant for this analysis have been estimated by the author in tight collaboration with A.~J.~Barr and P.~Br\"uckman de~Rentstrom in the context of~\cite{bib:cscbook} and can be found in Section~6 ``Systematic uncertainties'' of~\cite{bib:csc7}.

%% file: Conclusion/Conclusion.tex
In the analysis presented, ATLAS' discovery potential for different mSUGRA scenarios using trilepton final states was investigated, with a particular focus on the massive sparton scenario. The statistical significances for $\mossf\in[21.2,\,81.2]$\,GeV are summarised in Table~\ref{tab:significancies}:

\begin{table}[h]
%\vspace{1mm}
% \setlength{\parskip}{0.2em} % distance btw 2 paragraphs
% \setlength{\topsep}{0em} % distance btw top of page and text
\begin{center}
\begin{tabular}{l|rrrrr|rr|rr}
\hline                                        
& \multicolumn{5}{c|}{\bf Jet inclusive} & \multicolumn{2}{c|}{\bf Direct $\chi\chi$} & \multicolumn{2}{c}{\bf Jet Veto}\\
                          &SU1      &SU2      &SU3      &SU4    &SU8       & ~SU2 &~SU3 &~SU2  &~SU3  \\
\hline\hline                                                                        
$\sgnf$                   & 6.4     & 6.0     & 15.9    & 53.0    & 1.3    & 4.0     & 1.9      & 2.3     & 1.5 \\
$\intlumi$ for 5$\sigma$ (in $\fb$)  & 6.1     & 6.8     & 1.0     & 0.1     & 138.6  & 15.3    & 68.5     & 48.6    & 118.9 \\
\hline       
\end{tabular}
\end{center}
\vspace{\cDistHalf}
\caption[Expected statistical significances for various SUSY benchmark points with 10\,\fb{} and projected integrated luminosities $\intlumi$ for a 5$\sigma$ discovery] {\label{tab:significancies}
Expected statistical significances $\sgnf$ for various SUSY benchmark points with 10\,\fb{}. The line below shows projected integrated luminosities $\intlumi$ for a 5$\sigma$ discovery taking into account statistical uncertainties only. ``Direct $\chi\chi$'' stands for direct gaugino pair-production, id est the massive sparton scenario in case of SU2. In the ``Jet Veto'' column a jet veto was applied to the inclusive SU$x$ signal, as explained in Chapter~\ref{chp:selection}.
\vspace{-1mm}
}
\end{table}

As expected, the trilepton jet inclusive search performs uniformly well for most of the SU$x$ points, with the exception of those points with particular mass degeneracies which reduce the branching ratio for the process of interest (SU8). While the discovery potential for the jet inclusive trilepton signal appears feasible with less than $\intlumi=10\,\fb$ for all points but SU8, the discovery of SUSY in the massive sparton scenario is expected to require twice as much integrated luminosity.

The possibility to isolate direct gaugino pair-production with a jet veto was investigated. The veto was found capable of selecting about $\sim$100\%~(90\%) of direct gaugino pair-production over {\em all} supersymmetric processes in SU2~(SU3). With the jet veto, the signal was found  measurable with $\sim$50\,\fb\ in the SU2 scenario.

\begin{figure}
\vspace{\cDistHalf}
\begin{center}
\includegraphics[width=7.9cm,clip=true]{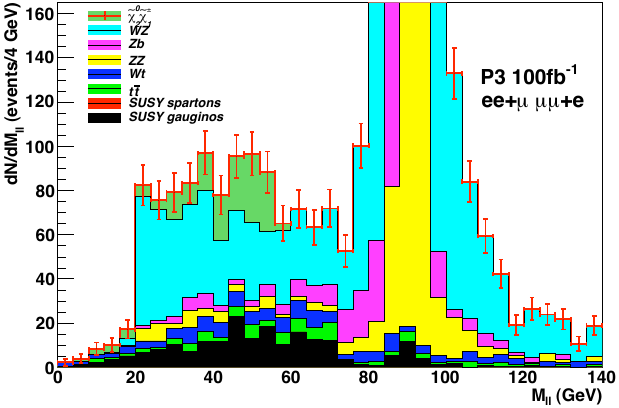}
\end{center}
\vspace{\cDist}
\caption[OSSF dilepton mass distribution for SU2 for 100\,\fb{} from an \textsc{Atlfast} study]{\label{fig:mllcompare}
% {\bf(left):~~} 
OSSF dilepton mass distribution for SU2 in the $ee+\mu$ and $\mu\mu+e$ channels with 100\,\fb{} from an \textsc{Atlfast} study~\cite{bib:wainerTesi}.
% {\bf(right):} same distribution for a full-simulation study in \Athena{} 11.0.42~\cite{bib:seshadri}
\vspace{\cDistHalf}
}
\end{figure}

This analysis more precisely renders previous results~\cite{bib:wainerTesi}, obtained with parameterised (``fast'') simulation of the ATLAS detector with \textsc{Atlfast} and parameterised mis-tagging rates of jets as leptons, shown in Figure~\ref{fig:mllcompare} for the $ee+\mu$ and $\mu\mu+e$ channels in SU2 and $\intlumi=100\,\fb$. The statistical significance is $S/\sqrt B\simeq4$ for the SU2 benchmark point and $\intlumi=30\,\fb$ (not shown in the Figure). A similar study based on full simulation with \Athena{} release 11.0.42 was performed~\cite{bib:seshadri}. Unfortunately, quantitative comparison is not possible due to extremely limited statistics for backgrounds involving heavy flavour jets decaying to leptons: $\ttbar$ and $Zb$\footnote{In fact, only $Zj$ production where $j$ is a light flavour jet was simulated, and heavy flavour jets can come only from initial of final state radiation.}.

The work presented in this part of the thesis posed a substantial contribution to~\cite{bib:csc7} and the trilepton chapter of~\cite{bib:csc5}. The main difference is that here a more sophisiticated and tailored preselection of electrons and muons as well as $e$-$j$ overlap removal procedures are used. As a result, in the massive sparton scenario only $\sgnf\simeq3.3$ and $\intlumi\simeq22.4\,\fb$ for 5$\sigma$ are obtained in~\cite{bib:csc7}. The picture is similar for the jet veto analysis and inclusive SUSY production at the SU2 benchmark point: only $\sgnf\simeq1.9$ and $\intlumi\simeq66.9\,\fb$ for 5$\sigma$ are achieved in~\cite{bib:csc7}. Contrarily, jet-inclusive search performance is only {\em marginally} lower for SU2 in~\cite{bib:csc7}: $\sgnf\simeq5.9$, $\intlumi\simeq7.1\,\fb$ for 5$\sigma$. The fact that the performance is so similar in this case can be attributed to a higher cut on missing transverse momentum of\linebreak $\met>30\,\GeV$ applied in~\cite{bib:csc7}, which is rather advantageous for hadroproduction of trilepton final states via decay chains. This is further supported by the fact, that~\cite{bib:csc7} performs somewhat better than this analysis for other benchmark points than SU2, where hadroproduction is dominating the total SUSY production cross section. Given the general philosophy of this analysis -- to provide an orthogonal sensitivity to SUSY searches based on multi-jet final states, the $\met$ cut is left at 20\,GeV to maintain maximal sensitivity to the difficult massive sparton scenario.

%In the massive sparton scenario, the $\met$ cut plays a notably less important role, and the more advanced physics object definition used in this analysis results in a substantial increase in sensitivity. Employing the SU2 benchmark point for comparison, the jet inclusive search performance is marginally lower in~\cite{bib:csc7}: $\sgnf\simeq5.9$, $\intlumi\simeq7.1\,\fb$ for 5$\sigma$. The fact that the performance is so similar can be attributed to a higher cut on missing transverse momentum of $\met>30\,\GeV$ in~\cite{bib:csc7}.

For the first LHC running period at \instlumi{31}, the \texttt{L2\_e25i} and \texttt{L2\_mu20i} triggers have been identified as optimal for this analysis, providing a combined trigger efficiency of $\gtrsim$97\%.

A coherent strategy to measure backgrounds from data using control regions and other experimental techniques have been outlined in the context of the trilepton search analysis presented. A novel method to measure the rate of secondary leptons from heavy flavour decays passing isolation criteria by isolating $\ttbar$ events in data has been proposed.

Overall, the trilepton search analysis is in an excellent shape for first LHC collisions which are expected later this year.

%% file: Alignment/Alignment.tex
\begin{figure}
\begin{center}
\includegraphics[width=12cm,clip=true]{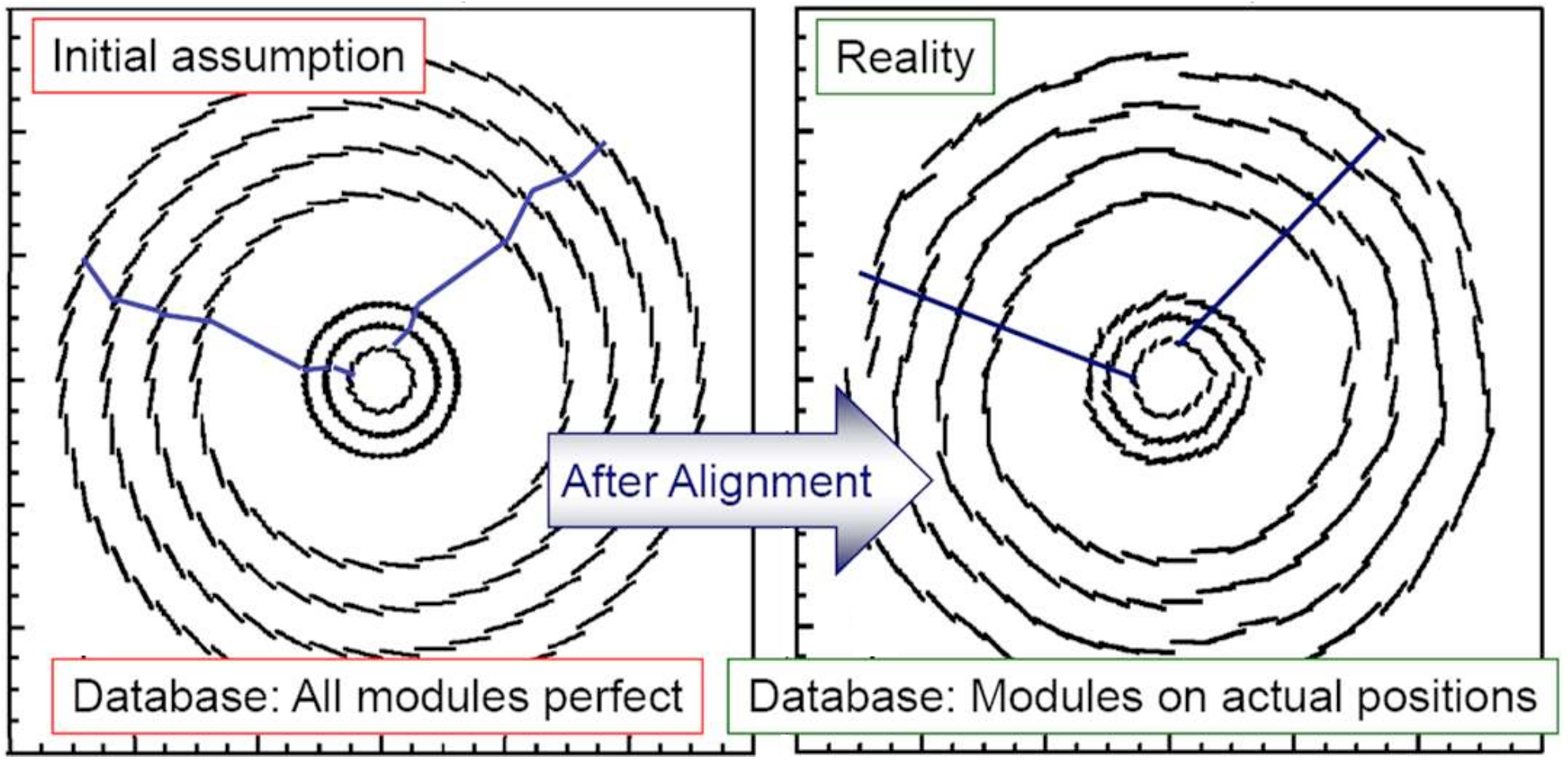}
\end{center}
\vspace{\cDistHalf}
\caption[Schematic illustration for the alignment of the Silicon Tracker.]{\label{fig:alignmentPrinciple}
The alignment of the ATLAS silicon tracker is schematically illustrated: {\em before} alignment {\bf (left)}, modules are at their nominal positions and seemingly ``kinky'' tracks are observed, while {\em after} alignment {\bf (right)} the modules are at their actual positions, and tracks are perfect helices. The misalignments are grossly exaggerated. Figure courtesy K.~St\"orig.
\vspace{\cDistHalf}
}
\end{figure}

As detailed in Subsection~\ref{ssec:innerdetector}, the Inner Detector is an essential ingredient to achieve the physics goals of the ATLAS collaboration. However, its full potential can only be exploited with the exact knowledge of the positions of tracker modules and their orientation in space, id est $3+3$ Degrees of Freedom (DoF\glossary{name=DoF,description=Degree of Freedom}) per module. This is schematically illustrated in Figure~\ref{fig:alignmentPrinciple}. With the modules at their nominal positions~(left), seemingly ``kinky'' tracks are be observed, and the measurement of their track parameters is imprecisise and is likely to be biased. Once the position and the orientation of the modules are determined in the process of {\em alignment}~(right), tracks become perfect helices\footnote{Up to Coulomb multple scattering, energy loss, and hit resolution effects.}.

%\nopagebreak[5]

Based on its physics goals, the ATLAS collaboration aims to achieve an alignment quality sufficient as not to let the resolution of the track parameters be degraded by more than 20\% due to misalignments~\cite{bib:atlasTDR1}. This is commonly referred to as ``TDR alignment goal''. Studies with simulated MC events show that this requirement translates into an alignment precision of about 7\,\mum\ for pixel detector modules, and about 12\,\mum\ for SCT modules~\cite{bib:misalTDRGoal}. A classical example is the measurement of secondary vertices from decays of long-lived particles such as $B$-mesons, which strongly depends on a precise measurement of the transverse impact parameter $d_0$. A random misalignment of about 10\,\mum\ in the precise measurement coordinate will result in a 10\% reduction of the $b$-tagging efficiency at the same fake rate~\cite{bib:bTagging}. More recent studies exist which explore the impact of misalignments on physics analyses~\cite{bib:impactMisalOnPhys,bib:bTaggingNote} using Monte Carlo simulations.

To achieve the level of alignment precision of the Inner Detector required by the TDR alignment goal, two basic strategies can be outlined: 
\begin{description}
\item[Hardware-based alignment:] all optical and mechanical survey-type measurements, both off-line and on-line, fall in this category. It is briefly discussed in Section~\ref{sec:alignmentHardware};
\item[Track-based alignment:] tracks fitted to particle trajectories, as measured by the Inner Detector in situ, are used to determine the position and orientation of its modules. This category is further discussed in Section~\ref{sec:alignmentTracks}. The \RA\ algorithm which was used to obtain the alignment results presented in Chapters~\ref{chp:pixelSR1} and \ref{chp:m8plus} falls into this category. It is described in Chapter~\ref{chp:ra}.
\end{description}

The alignment requirement defined above should not be regarded as an ultimate goal, but rather as a benchmark sufficient for most of analyses. A prominent exception where a better alignment is needed is the measurement of the $W$ boson mass using the tracker to calibrate the calorimeters along the lines of the analyses presented by the CDF collaboration~\cite{bib:wmass1, bib:wmass2}. To achieve  a 25\,MeV precision on the $W$ boson mass, this analysis requires not only an uncertainty on the track curvature of $\sigma(p^{-1})\!\cdot\!p<0.1\%$, but also the absence of any systematic effects to a level of about 1\,\mum. The detection and removal of systematic biases to the track curvature measurement was studied by the author, and two approaches have been validated using simulated Monte Carlo events~\cite{bib:weakModes}.

%% file: Alignment/AlignmentHardware.tex
The individual modules of both the pixel and the SCT detector were carefully surveyed upon production. Mask tolerances for module fabrication are in the region of ca. 1\,\mum~\cite{bib:privateHaywood}. SCT modules comprise two stereo sides. The in-plane tolerance for positioning a stereo sensor pair was measured, and its Root Mean Square (RMS\glossary{name=RMS,description=Root mean square}) in transverse direction is 2.1\,/\,1.6\,\mum\ for barrel/end-cap modules, while 2.7\,/1.3\,\mum\ was found in longitudinal direction~\cite{bib:atlasJINST}. Given these specifications, no independent alignment of the sides of SCT detector modules is foreseen in the next years to come. The out-of-plane positioning precision is much less relevant for the momentum measurement. It was measured to be 33\,/\,15\,\mum~(RMS) for barrel/end-cap modules~\cite{bib:atlasJINST}. The RMS of out-of-plance distortions changed only by some microns after thermal cycling.

The out-of-plane bending profile of individual SCT modules was surveyed and a common deformation profile was established at the level of a few microns. The modules of the pixel detector were glued to a stave support structure using robotic tools in the barrel. The precision of the module-to-stave alignment was estimated to be below 10\,\mum~(RMS) in $R\cdot\dif\Phi$ and 20\,\mum\ in $\dif Z$~\cite{bib:hawkingsAlignment}. In the end-caps, modules were glued to each of the eight sectors comprising a disk separately using a robot, after which the sectors were assembled to a disk. The positioning of the modules on the sectors was found to be better than 5\,\mum~(RMS) in-plane and 15\,\mum\ out-of-plane~\cite{bib:atlasJINST}. The mounting positions of the individual SCT barrel layers were only coarsely surveyed using photogrammetry~\cite{bib:photogrammetry1,bib:photogrammetry2,bib:photogrammetry3,bib:photogrammetry4}, since it was planned to measure the individual positions of SCT barrel modules in an absolute coordinate frame to a precision of some microns using an X-ray source~\cite{bib:thesisNick}. However, these plans were abondened in view of a tight ATLAS startup schedule. The mounting precision of SCT barrel modules by robotic tools~\cite{bib:mountingSCTmodules} was estimated not to exceed 100\,\mum~\cite{bib:hawkingsAlignment}, but in fact stayed well below as will be discussed in Chapter~\ref{chp:m8plus}. The mounting pins of SCT end-cap modules were surveyed with a precision of about 10\,\mum~\cite{bib:snowSurvey}.

The barrel staves to which the modules of the pixel detectors are glued were estimated to show lateral displacements of up to 100\.\mum\ after mounting on the support structure~\cite{bib:hawkingsAlignment}. However, track-based alignment has shown that these misalignments can be as large as 0.5\,mm, cf.~Subsection~\ref{ssec:pixelStaveBowM8}. The assembly precision pixel end-cap discs comprising eight sectors is about 25\,\mum~(RMS) in the plane of the disk. The mounting on larger support structures has been monitored to a precision of 20\,\mum~\cite{bib:hawkingsAlignment}. The deformation of barrel layers from a perfect cylindrical shape was found to be below~90\,\mum~(RMS), and the barrels are aligned concentrically with their axis within the measurement precision of 20\,\mum\ in $R$ and 40\,\mum\ in $Z$~\cite{bib:atlasJINST}. The disk mounting positions in the SCT~ECs are known to a precision of 100\,\mum\ in $R$ and about 1\,mm in $Z$~\cite{bib:atlasJINST}.

\begin{figure}
\begin{center}
\includegraphics[width=12cm,clip=true]{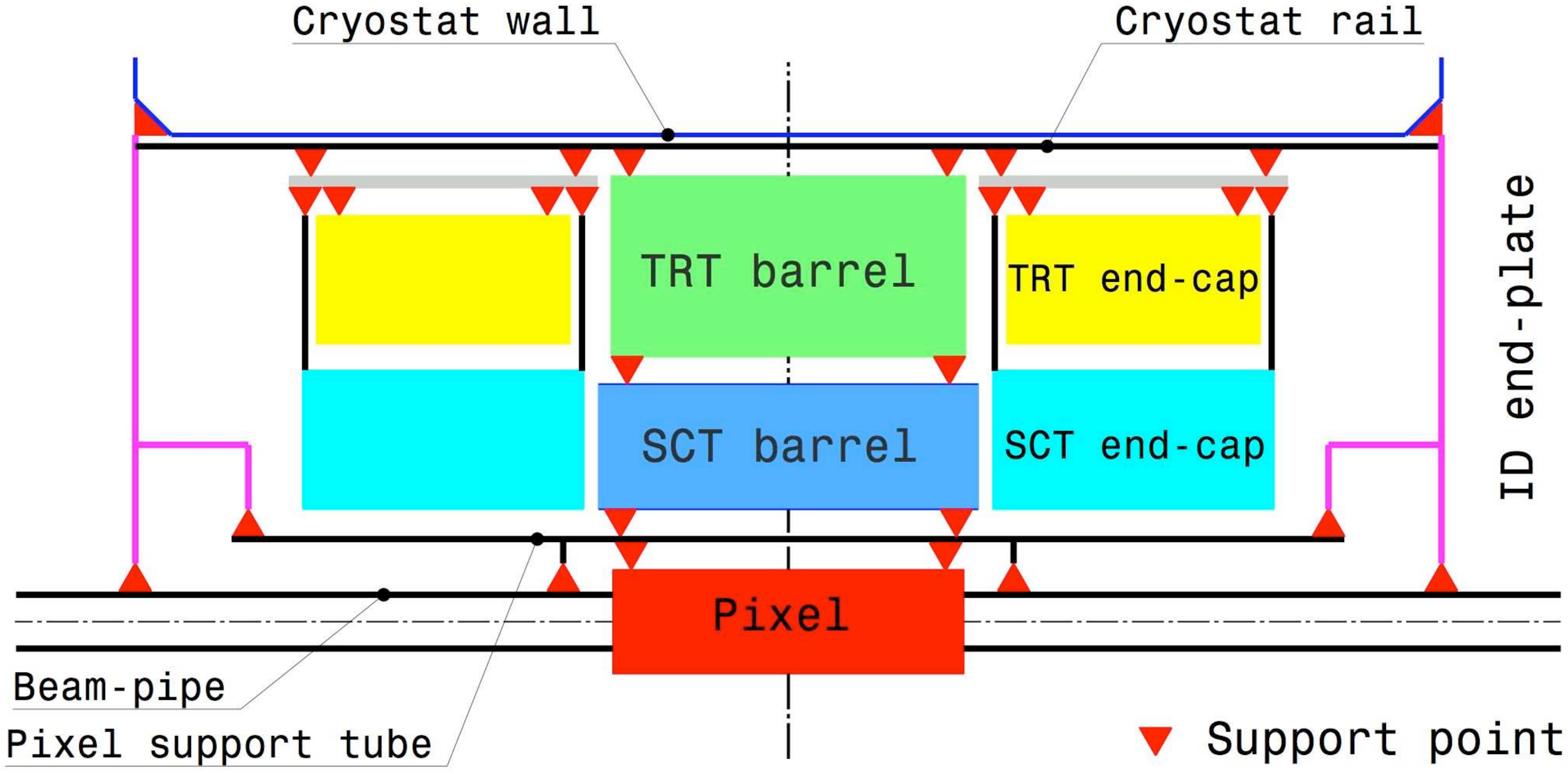}
\end{center}
\vspace{\cDistHalf}
\caption[The support structure of the Inner Detector]{\label{fig:supportID}
Schematic cross-section view of the Inner Detector sub-systems with their vertical support points. All supports are in the $R$-$Z$ plane and symmetric about the $Z$-axis.
\vspace{\cDistHalf}
}
\end{figure}%\nopagebreak[5]

The entire ID is supported by rails fastened to the inner wall of the barrel cryostat. The support scheme for the subdetectors is illustrated in Figure~\ref{fig:supportID}. The barrel of the SCT and TRT detectors are mechanically supported by a carbon fibre structure designed for high stiffness and stability, with $<$10\,\mum\ displacements under thermal cycling and humidity variations~\cite{bib:atlasJINST}. The positions of the sub-detectors relative to the inner wall of the barrel cryostat were mechanically measured with a precision of about $\xOverY14$\,mm~\cite{bib:atlasJINSTcoord}. However, the results of track-based alignment do not fully agree~\cite{bib:trackAlignCoord1,bib:trackAlignCoord2} which may be due to maintenance interference and the imprecision of the mechanical survey. 

For space reasons, end-cap~A of the SCT has been displaced by 4.88\,\mm\ away from the interaction point, while end-cap~C was moved by 5.35\,\mm. These changes are included in the nominal geometry with the (same) value of 5\,mm.

More details on the survey and assembly tolerances of the silicon tracker can be found in~\cite{bib:surveyRingberg}, Section~4.7 of~\cite{bib:atlasJINST} and references therein. The consideration of survey constants as a ``soft'' constraint in \GX\ is briefly described in~\cite{bib:surveyImplementation}.

\begin{figure}
\begin{center}
\includegraphics[width=15.2cm,clip=true]{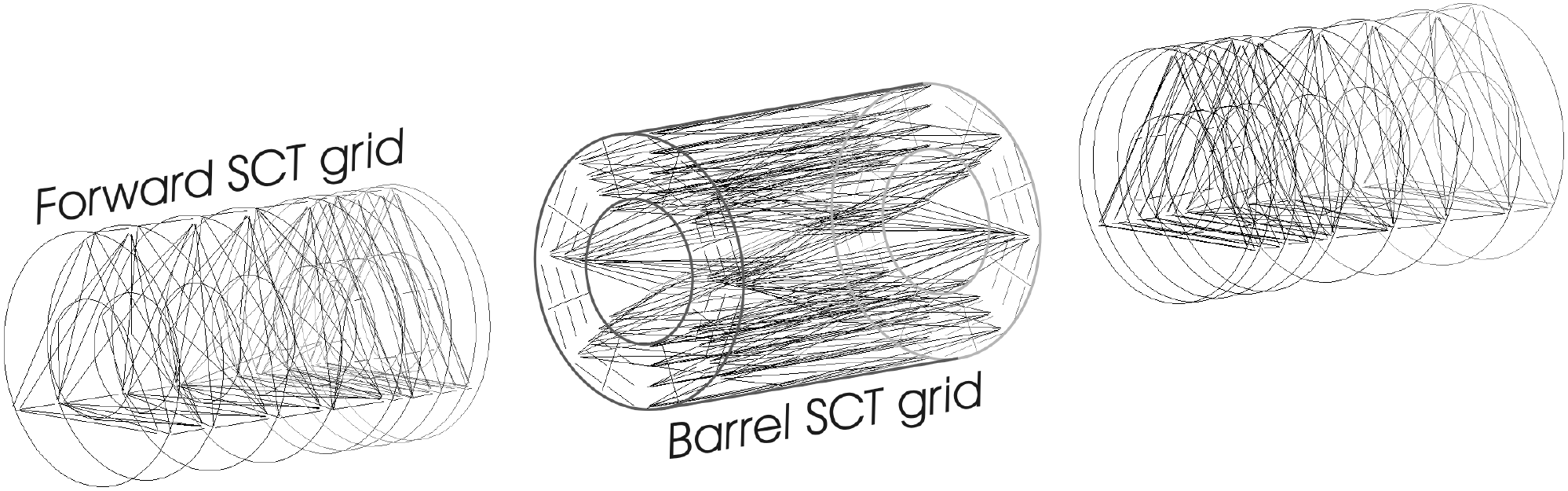}
\end{center}
\vspace{\cDistHalf}
\caption[The geodetic grid of the FSI system]{\label{fig:gridFSI}
The geodetic grid of the FSI system (straight lines). The circular lines are for eye-guidance only.
\vspace{\cDistHalf}
}
\end{figure}%\nopagebreak[5]

The geometry of the barrel and each of the end-caps of the SCT can be monitored {\em on-line} at the scale of a few microns using a geodetic grid of 842 nodes shown in Figure~\ref{fig:gridFSI} and Frequency Scanning Inerferometry~(FSI\glossary{name=FSI,description=Frequency Scanning Interferometry})~\cite{bib:fsi,bib:fsiLaser,bib:fsiThesis,bib:fsiDocumentation}. The main principle of the FSI system is to use tunable lasers to generate a frequency-dependent interferometry pattern, which can be translated into a length measurement using a thermally and mechanically stabilised reference interferometer. The FSI system is capable of simultaneously measuring the distances between nodes to a precision of $<$1\,\mum. This translates into a precision of better than 5\,\mum\ in the critical direction (typically $R\cdot\dif\Phi$) after a three-dimensional reconstruction of the grid has been performed. It is foreseen to repeat the measurement of node positions every 10~minutes~\cite{bib:privateGibson} in order to monitor any time-dependent deformations of the SCT geometry. Such deformations may be introduced by temperature and humidity changes induced by e.g. varying power consumption of the SCT module electronics.\\
The FSI system is not designed to determine the positions of individual modules, and the position of FSI grid nodes with respect to the modules and the global reference frame are not known with a precision required for an absolute alignment. Rather, its information is useful to detect and compensate for any time-dependent deformations of the SCT. This will complement the determination of its geometry by track-based alignment algorithms, which can only access time scales extending substantially beyond 10 minutes because of the need to integrate a sufficient number of residuals as to obtain the required precision.\\ %Moreover, the FSI may prove to be very helpful to constrain low spatial frequency deformation modes of the SCT detector geometry, which typically can be only weakly constrained by track data.\\
The FSI is now fully installed and was operated in 24/7 mode since 5 May~\cite{bib:fsiPresentation} in the ATLAS cavern. Unfortunately, it was not fully exercised in M8+ because few interferometry lines were causing high-voltage trips in some of the SCT EC modules~\cite{bib:privateGibson} due to extremely conservative trip limits having been set.%\footnote{The increased high-voltage current was shown not to compromise the detector noise performance.}. 

%% file: Alignment/AlignmentTracks.tex
The philosophy of track-based alignment is quite different from the hardware-based one: tracks reconstruced by the respective detectors are used as an input to the alignment procedure. %, rather than dedicated hardware used. 
This is the main advantage and disadvantage of the method at the same time. On the one hand, input collected {\em in-situ} by an {\em operating} detector is used. %, in fact the very same data as used for analysis. 
On the other hand, depending on the type of the input data, track-based alignment may have a reduced sensitivity to some deformations of the detector, the so-called {\em weak modes}, the removal of some of which was studied by the author~\cite{bib:weakModes}. Another limitation of track-based alignment is that its (statistical) precision is determined by the number of tracks collected. This deems the method blind to small changes of the ID geometry on a short time scale $\lesssim$1\,h. However, the very same fact  can be used to reduce the statistical error to a negligible degree by integrating over many tracks\footnote{Under the condition that the detector remains stable over the considered period.}. Depending on the track-based alignment algorithm used, several iterations may be needed to obtain a satisfactory convergence of the alignment procedure. This defines the minimum time needed before any sensible feedback about the quality of alignment can be obtained. 

\begin{figure}
\begin{center}
\includegraphics[width=12cm,clip=true]{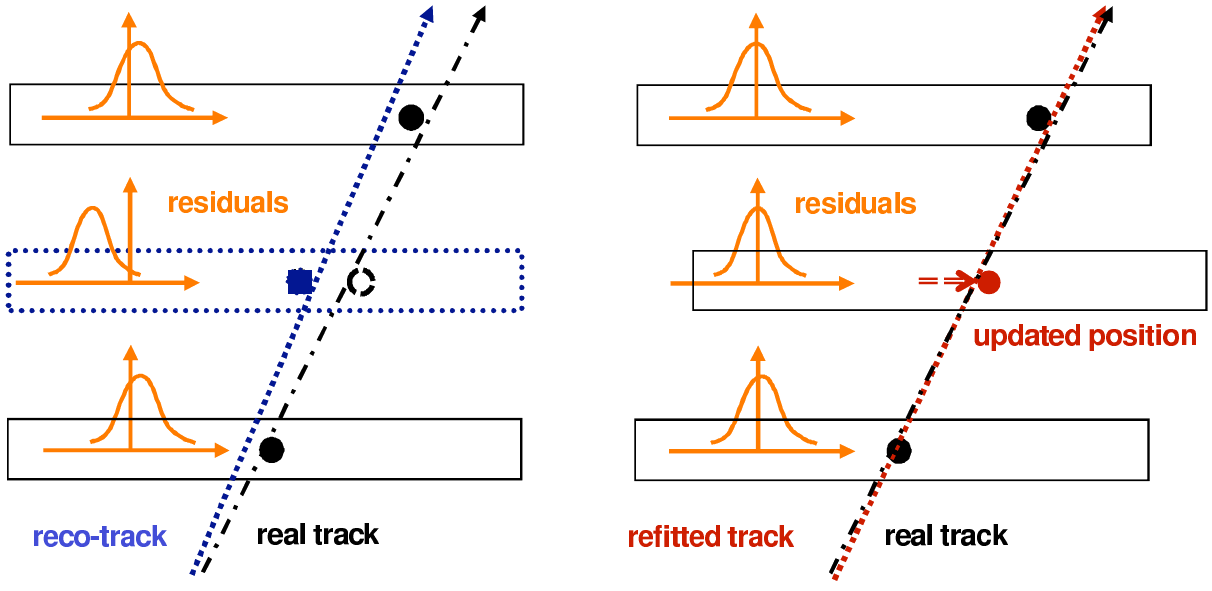}
\end{center}
\vspace{\cDist}
\caption[Schematic illustration for the principle of track-based alignment of the Silicon Tracker.]{\label{fig:alignmentResiduals}
The behaviour of track-hit residuals in the process of track-based alignment is schematically illustrated for the ATLAS silicon tracker: {\em before} alignment {\bf (left)}, the residual distributions are wide and not centred about 0, while {\em after} alignment {\bf (right)} the modules are at their actual positions, and narrower residual distributions centred about 0 are observed. %Clearly, the reconstruction of tracks and thus the residuals will be strongly affected by misalignments. 
Figure courtesy K.~St\"orig.
\vspace{\cDistHalf}
}
\end{figure}%\nopagebreak[5]

The basic principle of track-based alignment is illustrated in Figure~\ref{fig:alignmentResiduals}. It is based on hit--track residuals\footnote{{\em Hit--track residuals} will be referred to as ``residuals'' throughout this document.} $r$, which are canonically defined as the distance between the measured hit position and the intersection of the track with a given module. 
%so-called track-hit residuals\footnote{Track-hit residuals will be referred to as ``residuals'' throughout this document.} $r$ are formed from the positions of the hits associated with the given track $\qhit$ as measured by the respective modules and the tracks themselves:
%\[r\equiv\qhit-\qtrk\,,\]
%where $\qtrk$ stands for the position where the track intersects the module. 
In case of a perfectly aligned detector, id est a detector with perfectly known geometry, the residual distributions of all modules will be centred about zero and will have widths determined only by Coulomb multiple scattering, energy loss, and the hit resolution, as depicted on the right-hand side of Figure~\ref{fig:alignmentResiduals}. However, in case of a misaligned detector which is illustrated on the left-hand side of the Figure, the distribution of residuals for a given module will have a width determined by the degree of misalignments, and a mean typically deviating from zero. Thus, the ``quality'' of the residual distributions can be used as a figure of merit for the quality of alignment, e.g. by constructing a $\chisq$.

At ATLAS, there are three officially supported algorithms implemented in \Athena\ which use residuals to align the silicon tracker, and one to align the TRT:
\begin{description}
\item[Global \chisq\ Algorithm:]
this approach~\cite{bib:gx2twiki,bib:globalChi2,bib:globalChi2_2} is based on the minimisation of
\begin{equation} \label{eqn:chisqGX}
  \chi^2\equiv{\sum}_{\rm tracks}\,r^TV^{-1}r\,,
\end{equation}
with respect to alignment constants $c$. Here, $r$ is the residual vector for a given track, and $V$ its covariance matrix. Certainly, this quantity depends on the fitted track parameters $\pi$ via the residuals $r$. Linearising the expression in Equation~\ref{eqn:chisqGX} around the minimum under the assumption of small alignment corrections, the general solution for a variation\footnote{Beware, that it is only this Section where the $\delta$ symbol has this particular meaning -- normally it stands for an uncertainty of a quantity.} $\delta$ of $c$ is given by:
\begin{equation*}
  \delta c = -{\Big (}{\sum}_{\rm tracks}\frac{\dif r^T}{\dif c}V^{-1}\frac{\dif r}{\dif c}{\Big )}^{-1}{\sum}_{\rm tracks}\frac{\dif r^T}{\dif c}V^{-1}r
\end{equation*}
with
\begin{equation} \label{eqn:drdc}
  \frac{\dif r}{\dif c} = \frac{\del r}{\del c} + \frac{\del r}{\del\pi}\frac{\dif \pi}{\dif c}\,.
\end{equation}
In a similar fashion, the solution for a single track fit can be obtained:
\begin{equation*} \label{eqn:varPi}
  \delta\pi = -{\Big (}\frac{\del r^T}{\del \pi}V^{-1}\frac{\del r}{\del \pi}{\Big )}^{-1}\frac{\del r^T}{\del \pi}V^{-1}r\,.
\end{equation*}
The main advantage of the \GX{} algorithm is the proper treatment of correlations between residuals of the same track in different modules. The technical difficulty of finding the (mathematically) exact solution with the \GX{} algorithm lies with the necessity to solve a system of $n_{\rm DoF}=n_{\rm DoF\,per\,mod}\times n_{\rm mod}=34,992$ linear equations for the entire silicon tracker, which is inherently singular. Various preconditioning and fast solving techniques like MA27~\cite{bib:ma27} have been implemented, as documented in~\cite{bib:procSiena}. Despite the fact that \GX\ is conceptually a ``single-shot'' approach, \order{10} iterations are needed for the full alignment of the silicon tracker due to approximations used in the track reconstruction by \Athena;
\item[Local \chisq\ Algorithm:]
this method~\cite{bib:lx2twiki,bib:localChi2,bib:localChi2_2} takes a similar approach to the \GX\ algorithm with one modification: rather than using a full derivative for $\frac{\dif r}{\dif c}$ like in Equation~\ref{eqn:drdc}, a partial derivative is used:
\begin{equation*} \label{eqn:delrdelc}
  \frac{\dif r}{\dif c} \equiv \frac{\del r}{\del c} + \cancelto{_0}{ \frac{\del r}{\del\pi}\frac{\dif \pi}{\dif c} }\,.
\end{equation*}
This breaks up the matrix $\frac{\dif r^T}{\dif c}V^{-1}\frac{\dif r}{\dif c}$ into a block-diagonal form where all entries but the ones in $n_{\rm DoF\,per\,mod}\times n_{\rm DoF\,per\,mod}$ blocks along the diagonal are 0 per constructionem. This brings the advantage that the problem becomes mathematically easily solvable, however at the cost of not explicitly considering the correlations between modules in the alignment procedure. They have to be taken into account via iterations, and typically several tens of iterations are needed for a full silicon tracker alignment;
\item[Robust Alignment\ Algorithm:]
the main philosophy of this approach~\cite{bib:ratwiki,bib:dannyThesis,bib:florianThesis,bib:noteRA,bib:olegNote} is transparency and robustness. In this spirit, individual modules are aligned by re-centring residual distributions to obtain in-plane translational corrections. Moreover, for alignment of superstructures like for example barrel layers, topological distributions of residual means in $\eta$ and $\Phi$ are used. This new intuitive concept was introduced and implemented by the author~\cite{bib:olegNote}. The new \RA\ procedure is defined in Chapter~\ref{chp:ra}, and its applications cosmic ray data are presented in Chapters~\ref{chp:pixelSR1} ``\nameref{chp:pixelSR1}'' and \ref{chp:m8plus}~``\nameref{chp:m8plus}''. 
\item[TRT Alignment:]
this dedicated algorithm is used to align the TRT detector both internally -- at the level of TRT modules, and also externally -- with respect to the silicon tracker, which is normally aligned first. It is a $\chisq$-based algorithm following the same formalism as \GX. More details about TRT alignment can be found in~\cite{bib:trttwiki,bib:alignTRT} and references therein.
\end{description}
All the above algorithms have demonstrated a convincing performance when aligning the Combined Test Beam setup~\cite{bib:ctb}, parts of the inner detector operated in the so-called SR1 run to take cosmics data on surface~\cite{bib:sr1,bib:pixelSR1}, and Monte Carlo simulations of a misaligned detector in the so-called CSC-challenge~\cite{bib:alignmentCSC} as well as in the Full Dress Rehearsal exercise~\cite{bib:alignmentFDR}. Most importantly, all $3+1$ algorithms were utilised to align the inner detector in situ in the ATLAS cavern in the so-called M8+ run~\cite{bib:m8plus}, where about 2M tracks of cosmic ray particles have been recorded by the silicon tracker. This is about to be documented in~\cite{bib:m8plusNote}, and the results found with the \RA\ algorithm are the subject of Chapter~\ref{chp:m8plus} of this thesis.

Efforts are underway to create a common alignment Event Data Model (EDM\glossary{name=EDM, description=Event Data Model}) for track-based alignment of the inner detector and of the muon spectrometer, which will lead to a common framework for the $3+1$ approaches above.

In the following, a brief introduction to some terms relevant for alignment shall be given: the commonly used superstructures grouping modules of the silicon tracker in categories, and the tracking EDM implemented in \Athena.

%% file: Alignment/AlignmentStructures.tex
The modules of the silicon tracker can be grouped together into {\em superstructures} in order to optimally reflect %respect
the specifics of its geometry in the alignment procedure. These superstructures are initially treated and aligned as rigid bodies. A classical example for a superstructure class is an individual barrel layer of the SCT detector, which is treated as a cylinder. 

The concept of superstructures only makes sense as long as they reflect the mechanical rigidity and the assembly tolerances of the given detector parts they correspond to as outlined in Section~\ref{sec:alignmentHardware}. For example, it would not make sense to group EC~A of the pixel detector and EC~C of the SCT together. Thus, a loose hierarchy of alignment superstructure classes is established, starting from individual modules %over barrel layers and end-cap disks 
and culminating in sub-detectors like the entire pixel detector. Usually, for superstructure classes $S_i$ with $S_1\subset S_2\subset ...$ the relation
\begin{equation} \label{eqn:magMisalLevel}
  M_1\lesssim M_2\lesssim M_3\lesssim\,...
\end{equation}
can be established, where $M_i$ is the magnitude\footnote{more precisely, the average of the moduli of the magnitude of translations and rotations of {\em individual} modules comprising the given superstructure.} of misalignments averaged over all superstructures belonging to a superstructure class $S_i$, e.g. all SCT barrel layers. In other words, the bigger the superstructure, the ``larger'' the expected alignment correction.

The concept of superstructures brings several advantages with it. Firstly, the number of DoF of the system is greatly reduced from about 35k to
	\[n_{\rm DoF\,per\,super}\times n_{\rm super}\,,\]
where trivially $n_{\rm super}\le n_{\rm mod}$ and typically $n_{\rm DoF\,per\,super}\le6$. This can be very helpful in a statistically limited situation like the start-up of ATLAS in late summer 2008 and the imminent re-launch of ATLAS after maintenance. Therefore, the alignment will normally be performed for the largest superstructures first, and proceed via smaller ones to the individual module level. Another advantage of the superstructure concept is the numerical stability of the calculation of alignment constants resulting from Equation~\ref{eqn:magMisalLevel}: the absolute magnitude of alignment corrections is smaller for smaller-in-size structures with larger (statistical) uncertainties. Moreover, a coherent alignment of all modules comprising a superstructure ensures that modules which did not collect any residuals are not ``left behind'', so that they are not too far away from their neighbours and do not diverge once they are switched on and start to take data.

Historically, three canonical superstructure classes are defined at ATLAS in the context of the silicon tracker along the lines of Section~\ref{sec:alignmentHardware}:
\begin{description}
\item[Level 1:]
comprises 4 bodies: the barrel and the two end-caps of the SCT plus the entire pixel detector;
\item[Level 2:]
consists of 31 bodies: 3 layers for the pixel barrel, $2\times3$ end-cap disks for the pixel end-caps, 4 layers for the SCT barrel, and $2\times9$ disks for the SCT end-caps;
\item[Level 3:] comprises 5832 bodies: 1456 pixel barrel modules, $2\times144$ pixel end-cap modules, 2112 SCT barrel modules, and $2\times988$ SCT end-cap modules.
\end{description}
These {\em alignment levels} are frequently abbreviated as ``L$x$''. The ATLAS silicon tracker alignment database is structured in L1, L2, L3 -- exactly as explained above.

Besides the canonical alignment levels used in the ATLAS alignment database, additional superstructure classes can be defined along the lines of the discussion above. A famous example are the pixel barrel staves\footnote{Barrel staves are comprised of modules with the same $\Phi$-identifiers and are described in Subsection~\ref{ssec:innerdetector} for the pixel detector case.}, many of which were found to display parabola-shaped distortions of up to 0.5\,mm in the local~$x$ coordinate. Their distortions and the derived alignment corrections are discussed in Subsection~\ref{ssec:pixelStaveBowM8}. To our best knowledge, there is no need for a dedicated alignment of SCT barrel %modules with the same $\Phi$-identifier -- the equivalent for pixel barrel staves. 
staves. This is because SCT barrel modules were mounted individually on a common support structure of a given layer, as detailed in Section~\ref{sec:alignmentHardware}. However, because this mounting was done using robotic tools by row, modules of the same ring\footnote{Rings are comprised of modules with the same $\eta$-identifiers.} may display a common bias in $R\cdot\dif\Phi$, cf.~Section~\ref{sec:resultsM8}. SCT end-cap disks comprise up to three rings. The cosmic ray data collected by ATLAS so far supports the definition of these rings as a separate superstructure class. Finally, the \RA\ algorithm uses the barrel and the end-caps of the pixel detector as three separate bodies when performing the first stage of alignment. There are more substructure classes of somewhat smaller importance in use across the algorithms for silicon tracker alignment. They are described in the respective references and shall not be listed here.

%The superstructure classes reflected in the structure of the \RA\ algorithm are summarised in Table~\ref{tab:superstructures} together with the corresponding number of degrees of freedom. This table is tightly related to Table~\ref{tab:inDet}.

%% file: Alignment/TrackingEDM.tex
Since residuals and therefore the tracks themselves are of central imporance for track-based alignment, the tracking EDM -- the way tracks are reconstructed and parameterised in \Athena\ -- shall briefly be recapitulated here. 

The central object of the ATLAS tracking EDM is a flexible {\tt Track} class. It is basically a container of {\tt TrackStateOnSurface} (TSOS\glossary{name=TSOS, description=A {\tt TrackStateOnSurface} object}) objects. They hold polymorphic information which can serve as input to the track fit or other purposes. The polymorphic flexibility of the TSOS object is one of the main advantages of the new EDM: it can hold hit information as a common {\tt MeasurementBase} class, the traversed material and its effect on the track, a simple track expression with respect to a given surface as a {\tt ParametersBase} class, and many more.

\begin{figure}
\begin{center}
\includegraphics[width=15.2cm,clip=true]{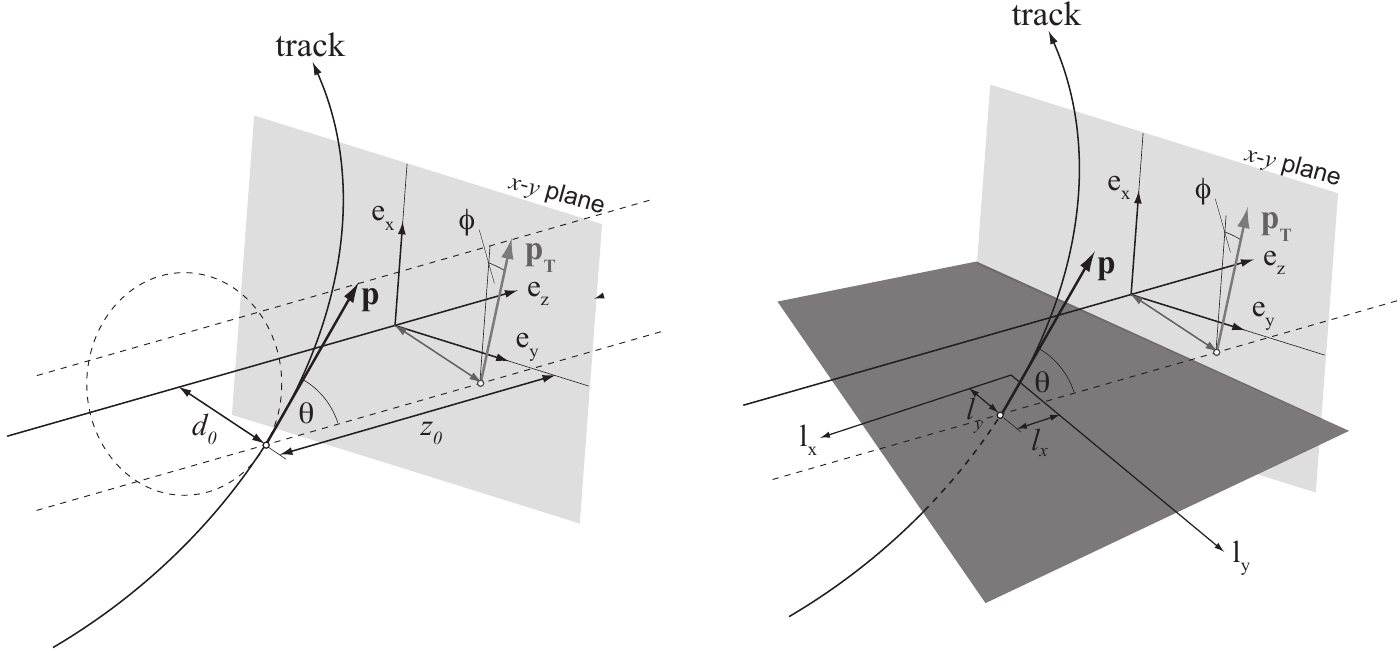}
\end{center}
\vspace{\cDist}
\caption[The canonical track representations with respect to the perigee and a planar surface.]{\label{fig:trkParam}
Two canonical track representations: with respect to the nominal $Z$-axis{\bf~(left)} -- at the perigee; and to a planar surface{\bf~(right)} -- for example a silicon tracker module. The representations differ only in the first two parameters: $d_0,\,z_0$ versus $\ell_x,\,\ell_y$. The latter correspond to $\qtrk_x,\,\qtrk_y$ defined in Section~\ref{sec:residuals}. The common track parameters are $\phi,\,\theta,\,\frac qp$.% The Figure is from~\cite{bib:trkEDM4}.
\vspace{\cDistHalf}
}
\end{figure}%\nopagebreak[5]

%As the name says, a TSOS is always expressed with respect to a surface or an axis. 
The two most commonly used  track parameterisations in a TSOS via {\tt ParametersBase} are sketched in Figure~\ref{fig:trkParam}:
\begin{description}
\item[Perigee representation:]
in this representation, the track is expressed with respect to the nominal $Z$-axis, which is characterised by the point of closest approach of the track to the axis -- the perigee. This is sometimes also referred to as {\em track expression at the perigee}. For the case with solenoidal magnetic field, the five canonical track parameters $(d_0,\,z_0,\,\phi,\,\theta,\,\frac qp)$ are used to describe the track;
\item[Planar surface represenation:]
the perigee point is not well-defined in this situation, and therefore the $d_0,\,z_0$ parameters are replaced by the coordinates of the intersect of the track with that surface: $\ell_x,\,\ell_y$. Note, that $\ell_x,\,\ell_y$ correspond to the $\qtrk_x,\,\qtrk_y$ introduced in Section~\ref{sec:residuals}. The other three parameters are kept, and are given at the intersection point.
\end{description}
In case without magnetic field, the charge-signed curvature $\frac qp$ cannot be measured and is dropped.

In some detectors, a track is not fully constrained due to one or several track parameters which cannot be measured in the given setup. A classical example are the $z_0$ and $\theta$ track parameters in the barrel of the TRT, which can determine the global $Z$ coordinate of a hit only with a very coarse precision of \order{1\,\rm m}. In order to allow for a full track fit, that extra information has to be provided externally. In the ATLAS tracking EDM, this is implemented as a generic {\tt PseudoMeasurementOnTrack} class. It derives from {\tt MeasurementBase} and can hold any measurements together with their covariance matrix to construct the $\chisq$ of the track fit. %This will be put to extensive use in the approaches discussed in Chapter~\ref{chp:weakMode}.

More information on the ATLAS tracking EDM can be found in~\cite{bib:trkEDM1,bib:trkEDM2,bib:trkEDM3,bib:trkEDM4}.

\begin{comment}
Tracking EDM:

First, basic track description at perigee + figure from rel. 12 note

more than just tracking, extra concepts

heavy construction zone for some years until stabilisation in rel. 11 and final changes in rel. 12

Track: collection of TSOSs

TSOS: various classes describing the geometrical correspondence at ATLAS

most important for us: PseudoMeasurement

citations: 
\end{comment}

%% file: RA/RA.tex
In this chapter, the \RA{} algorithm, one of the 3+1 track-based alignment algorithms at ATLAS introduced in Section~\ref{sec:alignmentTracks}, will be descibed in detail. Firstly, the basic input to Robust Alignment, residuals and overlap residuals, will be defined in Section~\ref{sec:residuals}, after that the actual alignment procedure will be described in Section~\ref{sec:procedure}. %, followed by a brief overview of the implementation in the ATLAS software framework \Athena{} in Appendix~\ref{chp:implementation}. 
The Robust Alignment algorithm is an Oxford-only product, which was started by D.~Hindson~\cite{bib:dannyThesis}, developed by F.~Heinemann~\cite{bib:florianThesis,bib:noteRA}, and augmented to its current state by the author, as documented in~\cite{bib:olegNote} and in the following.

%% file: RA/Residuals.tex
The basic inputs to any track-based alignment algorithm are track-hit residuals. In the \RA{} algorithm, overlap residuals also enter the alignment procedure explicitly. Residuals and overlap residuals, as well as associated quantities will be defined in this Section.

\subsection{Residuals}

\begin{figure}
\begin{center}
\includegraphics[width=10.5cm,clip=true]{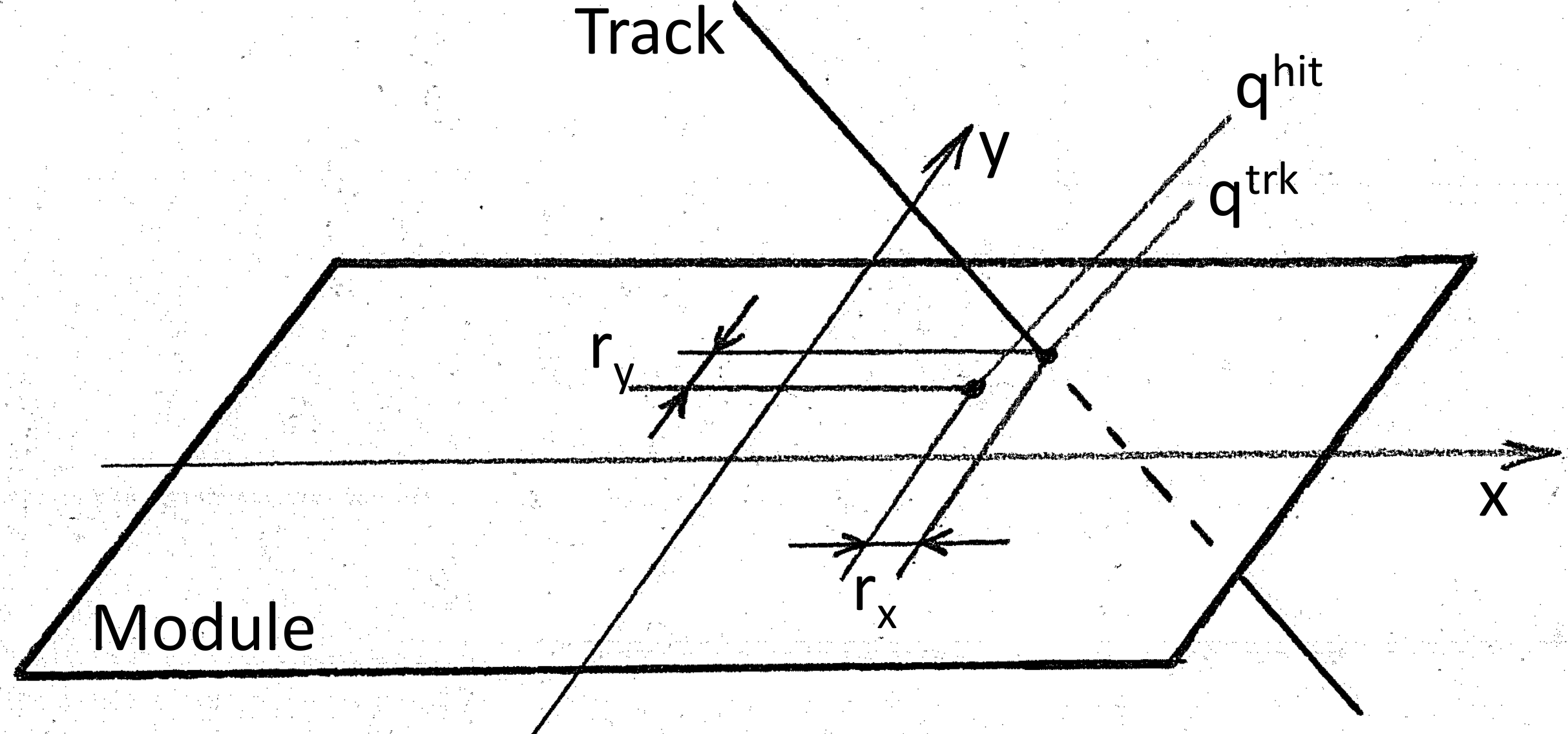}\qquad
\end{center}
\caption[The definition of a track-hit residual $r$ in $x$ and $y$ direction]{\label{fig:residual}
The definition of a track-hit residual $r$ in $x$ and $y$ direction. See text for details.
}
\end{figure}%\nopagebreak[5]

A track-hit residual $r$ is the distance in the module plane between the measured position $\qhit$ of the hit on the module and the intersection point $\qtrk$ of the track with the module plane, as sketched in Figure~\ref{fig:residual} and in Figure~\ref{fig:residualL3}~(b) on page~\pageref{fig:residualL3}. A residual for measurement direction $\zeta=x,y$ is thus defined as:
\begin{equation} \label{eqn:residual}
 r_{\zeta}\equiv\qhit_\zeta-\qtrk_\zeta\,.
\end{equation}
This definition is straight forward for the {\bf pixel} detector, as both $x$ and $y$ have a one-to-one correspondence in the measurement directions of the modules. For the {\bf SCT barrel} $r_x$ is naturally defined as being perpendicular to the strip in the context of alignment, id est the stereo angle of 40\,mrad between two wafers of the same SCT module is neglected\footnote{This can be done since the bias introduced equals $1-\cos(40\,\rm mrad)\simeq0.001$. Typical alignment corrections at module level stay well below 100\,\mum. Thus the bias is of $\order{0.1\mum}$, which is far beyond the alignment precision required (cf. Chapter~\ref{chp:alignment}). At any rate, no bias is introduced since the \RA\ is an iterative algorithm in the sense of Section~\ref{ssec:iterativeRA}.}. Due to the fan-out structure in the {\bf SCT end-caps}, the definition in Equation~\ref{eqn:residual} has to read:
\begin{equation*}
 r_{x}^{\rm SCT\,EC}=\cos\psi\cdot(\qhit_x-\qtrk_x) - \sin\psi\cdot\qtrk_y\,,
\end{equation*}
where $\psi$ is the fan-out angle of the strip (or cluster) that registered the hit with respect to the local $y$ coordinate of the given module. The reason for the ``missing'' $\qhit_y$ term in the second summand is the fact that $\qhit_x$ is given at the $y$-centre of the strip in \Athena, which is not aware of the $\qhit_y$ coordinate at the hit reconstruction stage, since it requires the track to be reconstructed.
 
The uncertainty on individual residuals is defined as a Gaussian sum of the individual errors on the relevant terms in Equation~\ref{eqn:residual}:
\begin{equation} \label{eqn:residualErr}
 \Delta r_{\zeta}\equiv\sqrt{ \left(\Delta\qhit_\zeta\right)^2 + \left(\Delta\qtrk_\zeta\right)^2 }\,,~~\zeta=x,y\,,
\end{equation}
where $\Delta\qhit_\zeta$ and $\Delta\qtrk_\zeta$ are the errors on the hit and track intersection positions. Again, for the SCT ECs Equation~\ref{eqn:residualErr} has to be altered to:
\begin{eqnarray} \label{eqn:residualErrSCTEC}
 \Delta r_{x} &=& \sqrt{ \left(\Delta\tilde q^{\rm hit}_x\right)^2 + \left(\Delta\tilde q^{\rm trk}_x\right)^2 }\,,{~\rm where}\nonumber\\
 \Delta\tilde q_x &=& \var_x(q)-\frac{\cov_{x,y}^2(q)}{\var_y(q)}\,,
 %\Delta r_{x} &=& \sqrt{ \left(\Delta\tilde q^{\rm hit}_x\right)^2 + \left(\Delta\qtrk_x\right)^2 }\,,{~\rm where}\nonumber\\
 %\Delta\tilde q^{\rm hit}_x &=& \var_x(\qhit)-\frac{\cov_{x,y}^2(\qhit)}{\var_y(\qhit)}\,,
\end{eqnarray}
and $\tilde~$ denotes the strip frame\footnote{Mind that the strip frame is rotated by the fan-out angle $\psi$ with respect to the local frame of the module.}. As can be seen from Equation~\ref{eqn:residualErrSCTEC}, the error $\Delta q^{\rm hit}_x$, which is defined in the {\it module} frame, is rotated to the {\it strip} frame to compensate for the fact that $\Delta q^{\rm hit}_x$ contains a significant contribution from the projection of $\Delta\tilde q^{\rm hit}_y = (\rm strip~length)/\sqrt{12}$ onto~$x$. Similarly, $\Delta q^{\rm trk}_x$ is rotated: it contains a substantial contribution from $\Delta\tilde q^{\rm trk}_y$, which is typically large.% The corresponding error

For SCT modules, $r_y$ cannot be formalised in a straight forward way since the modules do not provide a hit measurement in $y$. However, $r_y$ can be defined by constructing a two-dimensional hit point in the module plane using the information from both wafers of a given module, provided each of them registered a hit signal. This is done by finding the intersection point of the strips (or clusters of strips) from the two modules projected onto a plane, which is normal to the track momentum vector at the track intersection point with the module surface. This construction can be performed by using the {\tt makeSCT\_SpacePoint(.)} method of {\tt SiSpacePointMakerTool}~\cite{bib:siSPMakerTool} in \Athena. A more detailed account of the procedure is given in~\cite{bib:florianThesis}. It should be mentioned that although $y$-residuals can be calculated for SCT modules, their utilisability for alignment is rather limited, since any misalignments in local~$x$ will be reflected at a magnified scale in local~$y$ due to the strip stereo angle.
%This will be discussed somewhat more quantitatively in Subsection~\ref{ssec:l3}.

In the ATLAS alignment community, three types of residuals are used:
\begin{description}
 \item[Biased:] the track intersection with the module $\qtrk$ is defined for the full track, i.e. with all\footnote{In this context, ``all'' hits means all {\it good} hits, i.e. no outliers, edge channel hits, etc.} hits participating in the fit;
 \item[Unbiased:] $\qtrk$ is defined for a refitted track which has all the hits of the original full track except for the hit used to calculate the given residual;
 \item[Fully Unbiased:] $\qtrk$ is defined in a similar way to the unbiased case, with the difference that in the SCT the module for which the residual is to be calculated does not participate in the track refit at all, i.e. hits in both module sides are removed.
\end{description}
Clearly, the difference between the three definitions approach zero for the number of hits on a track going to infinity, and the most grave differences are found for a situation where the number of degrees of freedom for the track is close to the number of hits participating in the track fit, necessitating the introduction of damping factors when calculating alignment constants. This situation occured and was studied in detail~\cite{bib:pixelSR1_unbiased} by the author in the pixel end-cap A alignment excercise with SR1 cosmics and is briefly summarised in Chapter~\ref{chp:pixelSR1}. %On a parallel note, it is important to point out, that with perfectly known detector geometry the residual pull distribution  $\frac{r_\zeta}{\Delta r_\zeta}$ will {\it per definitionem} have a width different from unity for the unbiased or fully unbiased cases.

The \GX\ algorithm uses exclusively biased residuals, whereas the \LX\ uses unbiased residuals. In the context of \LX, fully unbiased residuals have been studied, but found to lead to numerically unstable alignment results due to convergence issues~\cite{bib:privateRoland}. To facilitate comparisons between the algorithms, the \RA{} algorithm is designed to work with any of the three residual types. However, most of the alignment results presented in this thesis were obtained with biased residuals since they provide the most reliable and stable convergence in the wide majority of scenarios.
%, where for the calculation of the latter two the {\tt SiRobustAlignTools}

The uncertainty on the mean of all residuals $\delta r_\zeta$ in direction $\zeta$ collected by a given module is canonically defined as the estimator of the uncertainty on the mean, that is:
\begin{equation}\label{eqn:residualErrAll}
\delta r_\zeta=\frac{ \sigma(r_\zeta) }{ \sqrt{n_\zeta} }\,,
\end{equation}
where $\sigma(r_\zeta)$ is the standard deviation of $r_\zeta$, and $n_\zeta$ is the number of residuals collected by the module.

\subsection{Overlap Residuals}

As will be detailed in Subsection~\ref{ssec:l3}, overlap residuals, denoted with an $o$, play a special role in the \RA\ algorithm: they are monitored and explicitly used for the alignment procedure. Their particular merit for aligning the silicon tracker has been convincingly studied with the \LX\ algorithm~\cite{bib:overlapStoerig}, too. 

An overlap residual can be defined in the region where two modules have a geometrical overlap when a track produces hits in two neighbouring modules 
%with identifiers $i$ and $i$+1 
in the same layer. Depending on whether the overlap occurs between two modules neighbouring each other in $\Phi$ or $\eta$, we speak about an overlap residual of $x$ or $y$ type, respectively. Like a conventional residual, an overlap residual can have two measurement directions: $x$ and $y$. Considering this, we denote overlap residuals as $o_{\zeta\xi}$, where $\zeta$ stands for the measurement direction, and $\xi$ defines the type of overlap.

Whenever there is an overlap residual between two modules with $\Phi$- or $\eta$-identifiers $i$ and $i+1$, the overlap is assigned to the one with the larger identifier $i+1$ to avoid double-counting. Detector rings are circular, and therefore the module with a $\Phi$-identifier $i=0$ is assigned its overlap residual with the last module in the ring, i.e. the one with the largest $\Phi$-identifier. 

In \RA, an overlap residual is simply defined as the difference between two residuals, so that for two modules $i-1$ and $i$ neighbouring each other in $\xi$ we have:
\begin{eqnarray} \label{eqn:overlap}
 o_{\zeta\xi}^i &\equiv& r_\zeta^i-r_\zeta^{i-1}\\
                &=& \left\{(\qhit_\zeta)^i-(\qtrk_\zeta)^i\right\} 
                 - \left\{(\qhit_\zeta)^{i-1}-(\qtrk_\zeta)^{i-1}\right\}\,,\nonumber
\end{eqnarray}
which is sketched in Figure~\ref{fig:ovresL3}~(b) on page~\pageref{fig:ovresL3}. Note, that the residuals $r^i$ and $r^{i-1}$ in Equation~\ref{eqn:overlap} are defined in the local frame of their {\it respective} module. This means the two neighbouring modules are simplistically regarded as having coordinate frames aligned to each other. The bias introduced per iteration is small and never exceeds 10\%\footnote{The bias is largest for modules in pixel ECs, which are rotated with respect to each other by 7.5$^\circ$.}, which can be safely neglected given that \RA\ is an iterative algorithm in the sense of Subsection~\ref{ssec:iterativeRA}.
%however, since \RA\ is an iterative algorithm with the alignment correction magnitudes per iteration asymptotically approaching zero\footnote{The ideal case with good convergence is meant here. However, this does not imply loss of generality: in the opposite case the alignment constant set should not be regarded as valid in the first place.}, this bias can be safely neglected.

Similarly to the residual case, the uncertainty on overlap residuals is defined in parallel to Equation~\ref{eqn:residualErrAll} as 
\begin{equation*}
\delta o_{\zeta\xi}=\frac{ \sigma(o_{\zeta\xi}) }{ \sqrt{n_{\zeta\xi}} }\,,
\end{equation*}
where $\sigma(o_{\zeta\xi})$ is the standard deviation of $o_{\zeta\xi}$, and $n_{\zeta\xi}$ is the number of overlap residuals of type $\xi$ collected in the overlap region of two respective modules.

%% file: RA/ProcedureRA.tex
In this Section, the core of the \RA\ algorithm will be described: the exact formulae used to calculate alignment corrections for misalignments at level 1 (L1), L2, and L3, as well as pixel stave bow misalignments.
% will be introduced and discussed. 
Typically, misalignments at L1 will be largest, decreasing in magnitude\footnote{In the sense that typical corrections resulting from superstructure alignment translated into L3 corrections are significantly larger than the misalignment of individual modules constituting the superstructre with respect to it.} over L2, pixel stave bow and L3. To reflect this, the \RA\ procedure will usually start at L1 and finish at L3. However, for didactical reasons the calculation of alignment constants will be described here in the following sequence: L3, pixel stave bow, L2, and L1.

The possibility to correct for misalignments at L1 and L2, as well as for pixel stave bow were introduced by the author in autumn and winter of 2008-2009 in the context of the M8+ excercise, and the reader is invited to consult Chapter~\ref{chp:m8plus}, where the application of the \RA\ algorithm to M8+ data is presented in detail.

%% file: RA/L3.tex
The most basic alignment one can imagine is an alignment {\it solely} with regular residuals at L3. In the \RA\ algorithm, both residuals and overlap residuals are used in the alignment procedure at L3. Firstly, I will discuss alignment corrections from regular residuals, then from overlap residuals, and finally present the complete formula to calculate L3 alignment corrections.

\subsubsection{L3 Alignment Corrections from Residuals}

\begin{figure}
\begin{center}
\includegraphics[width=14.5cm,clip=true]{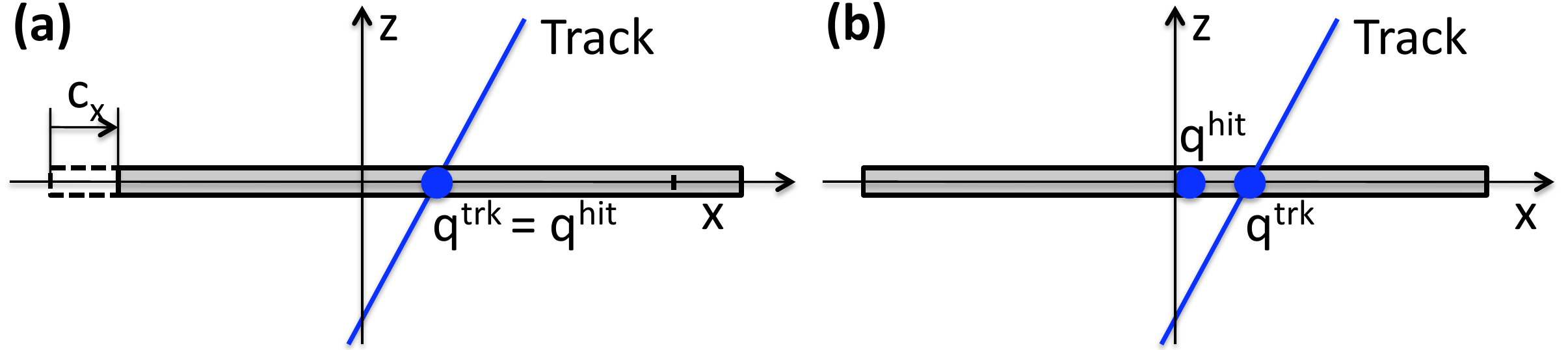}\qquad
\end{center}
\caption[Residuals: reality vs. situation as ``seen'' by reconstruction]{\label{fig:residualL3}
Residuals: situation as it is in reality {\bf (a)} and as ``seen'' by the reconstruction {\bf (b)} for one single track with infinitely many hits in absence of Coulomb multiple scattering. A misalignment of the given module by $c_x>0$ is assumed, whereas the the rest of the detector geometry is perfectly known.
}
\end{figure}%\nopagebreak[5]

As already touched upon in Section~\ref{sec:alignmentTracks}, the basic working principle of track-based alignment is to approach ideal residual distributions for all modules of the detector: their mean should be centered at zero, and their width should be mostly determined by the effect of Coulomb multiple scattering and the hit resolution\footnote{It should be mentioned that there may be some deformations of the ID which leave the residuals almost unchanged, but bias the track parameter reconstruction. This is addressed in~\cite{bib:weakModes}.}. To motivate the alignment correction formula from residuals used by the \RA\ algorithm, assume a detector with a perfectly known geometry, where only {\it one} module is misaligned by $\vec c$ in the local $x$-$y$ plane of the module. Without loss of generality, let us consider only the $x$ projection of the module and $c_x>0$. In absence of Coulomb multiple scattering and a track with infinitely many hits (in a detector with infintely many layers), the hit will be {\it physically} produced at the very point where the track intersects the module: $\qhit_x=\qtrk_x$, as indicated in Figure~\ref{fig:residualL3}~(a). However, as the reconstruction software is not aware of the {\it real} position of the module which is shifted by $c_x$, the hit will be ``seen'' as produced at $\qhit_x\neq\qtrk_x$, which is symbolically depicted in Figure~\ref{fig:residualL3}~(b). From the sketch it is clear, that in our case $r_x<0$. Moreover,  with our ideal assumptions above we know: $r_x=-c_x$. Of course, in presence of multiple scattering and hit resolution this relation will not be fulfilled for each individual residual. However, the mean of many residuals can be used:
\begin{equation} \label{eqn:corrL3res}
 c_\zeta = -\langle r_\zeta\rangle\,.
\end{equation}
Since the uncertainty on $r_x$ is given by Equation~\ref{eqn:residualErrAll}, we end up with an uncertainty for alignment corrections given by:
\begin{equation} \label{eqn:corrL3resErr}
 \delta c_\zeta=\frac{ \sigma(r_\zeta) }{ \sqrt{n_\zeta} }\,.
\end{equation}

\subsubsection{L3 Alignment Corrections from Overlap Residuals}

\begin{figure}
\begin{center}
\includegraphics[width=14.5cm,clip=true]{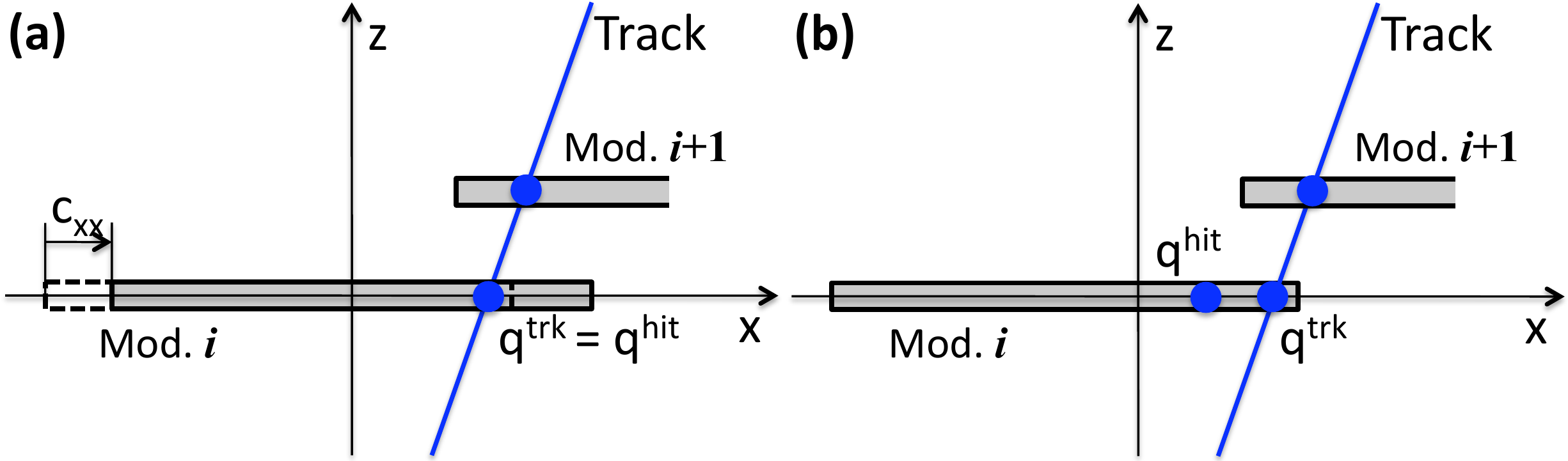}\qquad
\end{center}
\caption[Overlap residuals: reality vs. situation as ``seen'' by reconstruction]{\label{fig:ovresL3}
Overlap residuals: situation as it is in reality {\bf (a)} and as ``seen'' by the reconstruction {\bf (b)} for one single track with infinitely many hits in absence of Coulomb multiple scattering. A misalignment of module $i$ by $c_{xx}>0$ is assumed, whereas the the rest of the detector geometry is perfectly known.
}
\end{figure}%\nopagebreak[5]

Overlap residuals are a highly valuable input to the alignment procedure, since the effect of Coulomb multiple scattering is reduced by about $\order{10}$ compared to regular residuals due to the much smaller distance along the track between a pair of residuals forming an overlap residual. This provides a rather tight constraint on the distance in local $x$ (and $y$ in case of the pixel detector) between neighbouring modules in the same layer.

To motivate the alignment correction formula from overlap residuals, let us once again assume a detector with perfect geometry except for one misaligned module with\linebreak[4]$\Phi$-identifier $i$, and consider without loss of generality only the $x$ projection of the module and $c_{xx}^i>0$. Under the same idealised conditions assumed for the regular residual discussion, the hit will again be {\it physically} produced at the very point where the track intersects the module $i$: $\qhit_x=\qtrk_x$, as indicated in Figure~\ref{fig:ovresL3}~(a). It will be ``seen'' by the reconstruction as produced at $\qhit_x\neq\qtrk_x$, as shown in Figure~\ref{fig:ovresL3}~(b), and $o_{xx}^{i+1}=r_x^{i+1}-r_x^i=0-(-c_{xx}^i)=c_{xx}^i$. For a similar situation in the overlap region between modules $i-1$ and $i$, we will have $o_{xx}^{i}=-c_{xx}^i$. Regarding this, we can write for the alignment correction $c_{xx}^i$ of the module $i$ from $o_{xx}^{i}$, $o_{xx}^{i+1}$ overlap residuals by combining them weighted by their respective uncertainties:

%Thus, for a correction $c_{xx}^i$ for the module $i$ from $o_{xx}^{i}$, $o_{xx}^{i+1}$ overlap residuals we use the expression:
\begin{equation}
  \label{eqn:c_xx}
  c_{xx}^i=\frac12\cdot\fdampover\cdot 
           \frac{
   - \frac{\bra o_{xx}^i\ket}{(\delta o_{xx}^i)^2} 
   + \frac{\bra o_{xx}^{i+1}\ket}{(\delta o_{xx}^{i+1})^2}
                }{ 
     \frac{1}{(\delta o_{xx}^{i})^2} 
   + \frac{1}{(\delta o_{xx}^{i+1})^2}
                }
  \,.
\end{equation}
\fdampover{} is a tuning factor typically in the range of $[0.5,\,1]$, which is supposed to correct for slight overcompensations of misalignments when $\fdampover\equiv1$, resulting in a slightly sub-optimal convergence. A value of $\fdampover\simeq0.7$ was found to be a good choice. 

As is evident from the factor \oneOverTwo{} in Equation~\ref{eqn:c_xx}, corrections resulting from $o_{xx}^i$ are equally shared between modules $i-1$ and $i$. Similarly, $o_{xx}^{i+1}$ contributes to alignment corrections of modules $i$ and $i+1$. This is due to the fact, that measuring an overlap residual mean which deviates from zero indicates a relative misalignment of two neighbouring modules, but does not provide information which of the modules is more misaligned than the other in absolute terms: this can be determined with regular residuals only. ``Sharing'' alignment corrections from overlap residuals between modules puts all modules on equal footing. It also is a robust and convergent procedure in the sense that:
\begin{itemize}
 \item The alignment corrections asymptotically approach zero; 
 \item No over-all bias to the residual mean of a given ring or stave is introduced, i.e. there is no interference with absolute alignment, which can be performed using regular residuals only.
\end{itemize}
It was checked that performing an alignment at L3 using {\it only} overlap residuals leads to improved distributions of both overlap residuals {\it and} regular residuals. This points out the merit of using overlap residuals not only for relative module-to-module alignment, but also in absolute terms.
 %To verify that the sign of the  is correct, keep in mind that the $\Phi$ direction and thus the direction of increasing $\Phi$-identifiers of modules coincide with the positive $x$ direction of each module.

The uncertainty on the $c_{xx}^i$ correction is defined as:
\begin{equation}
  \label{eqn:dc_xx}
  \delta c_{xx}^i=\left\{
     \frac{1}{(\oneOverSqrt\cdot\delta o_{xx}^{i})^2} 
   + \frac{1}{(\oneOverSqrt\cdot\delta o_{xx}^{i+1})^2} 
                  \right\}^{-\oneOverTwo}\,,
\end{equation}
in accordance to Equation~\ref{eqn:c_xx}. The only difference to what Equation~\ref{eqn:c_xx} suggests, is an additional factor of \oneOverSqrt{}, which accounts for the fact, that corrections are shared between two neighbouring modules, otherwise their alignment correction pull distribution will not be a unit Gaussian. 
%On the contrary, using the full error $\delta o_{xx}$ would imply a larger uncertainty on overlap residuals when considering all the modules. 
$\fdampover$, being a tuning factor, is not considered in the error calculation\footnote{The appropriateness of this can be seen by considering (an unreasonably) small value of $\fdampover \ll 1$, which would result in unphysically small uncertainties on the correction $c_{xx}$.}.

Alignment corrections from overlap residuals of type $\xi=x,y$ in measurement direction of $\zeta=x,y$ are defined analogously to the above discussion.

It should be noted that before the above procedure to calculate alignment corrections from overlap residuals was introduced by the author, a somewhat sub-optimal and in some circumstances even biased approach was used. There, the alignment corrections from $o_{xx}$ overlap residuals for a module with $\Phi$-identifier $i$ would be calculated cumulatively by adding all overlap residuals starting from the one between the last and 0$^{\rm th}$ module: \begin{equation} \label{eqn:c_xxFlorian}
 c_{xx}^{i} \propto - \sum_{k=0}^io_{xx}^k\,,
\end{equation}
as detailed in~\cite{bib:florianThesis,bib:noteRA}. This procedure creates an artificial asymmetry in the calculation of alignment corrections for modules with different $\Phi$- or $\eta$-identifiers: since the error on the alignment correction grows with an increasing identifier $i$ as more and more terms have to be considered in the Gaussian error sum, the contribution of alignment corrections from overlap residuals gets less and less significant compared to the alignment correction from regular residuals. Clearly, this is sub-optimal, as the information from overlap residuals is not fully used. Moreover, consider a Gedankenexperiment where one module in a ring is grossly misaligned, but all the others are at their nominal positions. If the misaligned module happens to be the one with the largest $\Phi$-identifier $i_{\rm max}$, then the very first overlap residual in the sum in Equation~\ref{eqn:c_xxFlorian} will be different from zero: $o_{xx}^{i=0}\neq 0$, whereas all the other overlap residuals (up to $o_{xx}^{i_{\rm max}}$) are identically zero. After one iteration, this will introduce a rotational bias to the considered ring of the detector given by the magnitude of the misalignment of the last module $i_{\rm max}$. One can think of many other geometrical scenarios where such an alignment procedure will be suboptimal and biasing, which was the main motivation to implement the new procedure.

\subsubsection{Full Formula for L3 Alignment Corrections}
Taking into account Equation~\ref{eqn:corrL3res} for corrections $c_x^i$ from residuals and Equation~\ref{eqn:c_xx} for corrections $c_{xx}^i$ from $o_{xx}^{i}$, $o_{xx}^{i+1}$ overlap residuals, the corresponding expression for $c_{xy}^j$ from $o_{xy}^{j}$, $o_{xy}^{j+1}$, where $i,j$ are the $\Phi$- and $\eta$-identifiers of the module considered, the full expression for the module corrections can be constructed:
\begin{equation}
  \label{eqn:c_xtotal}
  c_{x,\rm total}^i= \frac{ 
   - \frac{\bra r_x^i\ket}{(\delta r_x^i)^2}
   + \frac12\cdot\fdampover\cdot\left\{
   - \frac{\bra o_{xx}^i\ket}{(\oneOverSqrt\cdot\delta o_{xx}^i)^2} 
   + \frac{\bra o_{xx}^{i+1}\ket}{(\oneOverSqrt\cdot\delta o_{xx}^{i+1})^2}
   - \frac{\bra o_{xy}^j\ket}{(\oneOverSqrt\cdot\delta o_{xy}^j)^2} 
   + \frac{\bra o_{xy}^{j+1}\ket}{(\oneOverSqrt\cdot\delta o_{xy}^{j+1})^2}
     \right\}
                          }{ 
     \frac{1}{(\delta r_x^i)^2}
   + \frac{1}{(\oneOverSqrt\cdot\delta o_{xx}^{i})^2} 
   + \frac{1}{(\oneOverSqrt\cdot\delta o_{xx}^{i+1})^2}
   + \frac{1}{(\oneOverSqrt\cdot\delta o_{xy}^{j})^2} 
   + \frac{1}{(\oneOverSqrt\cdot\delta o_{xy}^{j+1})^2}
                          }
  \,.
\end{equation}

Consequently, the error on $c_{x,\rm total}^i$ is given by:
\begin{equation}
  \label{eqn:dc_xtotal}
  \delta c_{x,\rm total}^i=\left\{ 
     \frac{1}{(\delta r_x^i)^2}
   + \frac{1}{(\oneOverSqrt\cdot\delta o_{xx}^{i})^2} 
   + \frac{1}{(\oneOverSqrt\cdot\delta o_{xx}^{i+1})^2}
   + \frac{1}{(\oneOverSqrt\cdot\delta o_{xy}^{j})^2} 
   + \frac{1}{(\oneOverSqrt\cdot\delta o_{xy}^{j+1})^2}
                           \right\}^{-\oneOverTwo}
  \,.
\end{equation}

%% file: RA/PixelStaveBow.tex
\begin{figure}
\begin{center}
\includegraphics[width=10.5cm,clip=true]{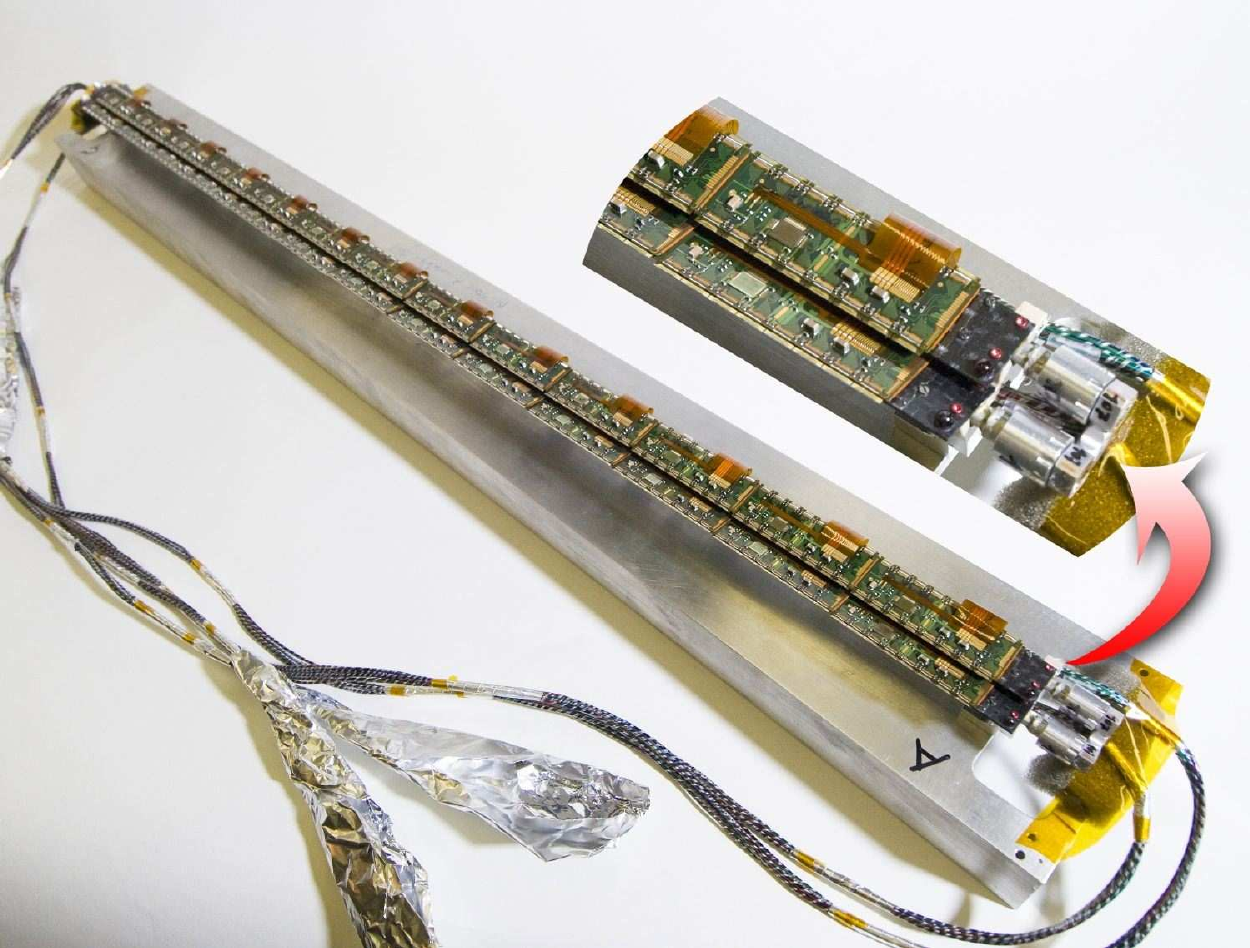}
\end{center}
\caption[Two fully equipped pixel staves with modules]{\label{fig:pixelStave}
Two fully equipped pixel staves, mounted nearly on top of each other. Each stave comprises 13 pixel modules. The insert shows the U-link cooling connection between staves. The local $x$ direction is aligned with the staves, whereas local $y$ is in the plane of the stave modules and perpendicular to $x$.
}
\end{figure}%\nopagebreak[5]

The pixel staves introduced in Subsection~\ref{ssec:innerdetector} are the smallest alignment superstructures necessitating a targeted alignment approach. 
%\footnote{In the sense that typical corrections resulting from superstructure alignment translated into L3 corrections are significantly larger than the misalignment of individual modules constituting the superstructre with respect to it.} is a pixel stave, which was already described briefly in Subsection~\ref{ssec:innerdetector}. 
A picture of two pixel staves with 13 modules each is shown in Figure~\ref{fig:pixelStave}. A determining factor for the alignment of pixel staves is that the mechanical stability of the pixel stave support rail is significantly better in the local~$z$ direction than in local~$x$. Due to the mechanical stresses in local~$y$ when mounting the pixel staves on the pixel subdetector frame, the staves tend to exhibit deformations in local~$x$ which can be well described by a polynomial of second order, i.e. of the form 
\begin{equation} \label{eqn:staveBow}
 f(\kappa)=a_0+a_1\!\cdot\!\kappa+a_2\!\cdot\!\kappa^2\,,
\end{equation}
where $\kappa\in\mathbb{Z}$ is the $\eta$-identifier of modules. This was pointed out by~\cite{bib:privatePixelStaveBow}. As of now, despite investigations~\cite{bib:privatePixelStaveBowInPlane} it is not completely clear whether these deformations tend to stay within the local~$x$-$y$ plane of the stave, or rather in the $(R_{\rm stave}\Phi)$-$Z$ plane, which is rotated by 20$^\circ$ with respect to the former. The reason for this is the statistically limited number of tracks with pixel hits collected by ATLAS to date.
%The attention of the ATLAS alignment community was drawn to this factor by~\cite{bib:privatePixelStaveBow}.

\begin{figure}
\begin{center}
\includegraphics[width=7.5cm,clip=true]{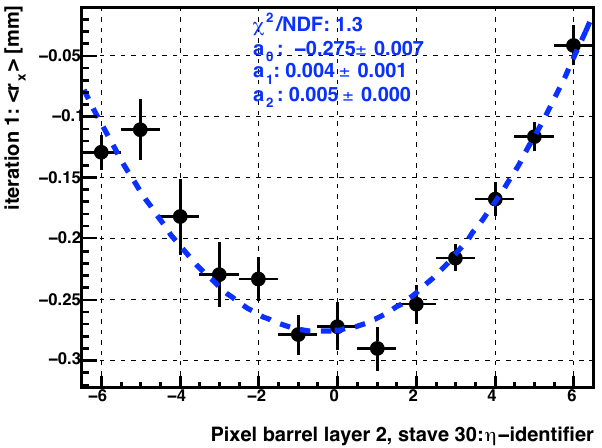}\qquad
\includegraphics[width=7.5cm,clip=true]{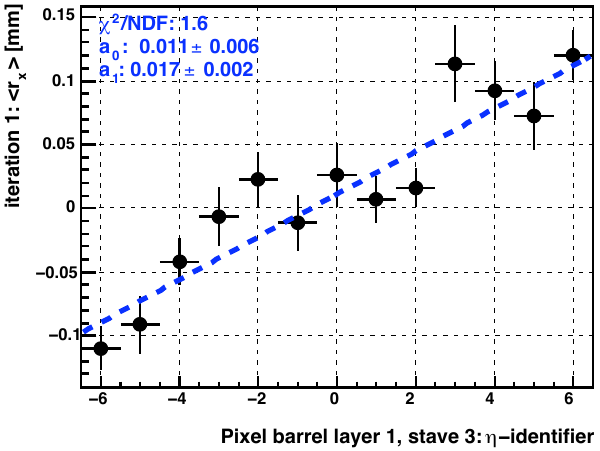}\qquad
\end{center}
\caption[Residual mean versus sector number distribution for two pixel staves]{\label{fig:pixelStaveBow}
Residual mean $\langle r_x\rangle$ versus sector number distribution for two typical pixel staves. While the left picture shows a clear parabolic dependance, the right can be better described by a linear function. The fit results are shown in blue.
}
\end{figure}%\nopagebreak[5]

Clearly, a pixel stave misalignment of the form indicated by Equation~\ref{eqn:staveBow} will be reflected in a similar dependence of the local $x$ residual mean $\langle r_x\rangle$ on the $\eta$-identifier of modules $\kappa$ constituting a stave for straight tracks in absence of any magnetic field. This is examplified in Figure~\ref{fig:pixelStaveBow} for two typical pixel staves for M8+ cosmic data after performing L1 and L2 alignment, as described in Chapter~\ref{chp:m8plus}. Misalignments due to the pixel stave bow can be as large as $\mathscr O(500\,\mu{\rm m})$ for some of the modules.

To align for any pixel stave bow misalignments, the approach as described below was adopted in the \RA\ algorithm:
\begin{enumerate}
 \item Three fits to the $\langle r_x\rangle$ versus $\kappa$ distrubution are performed using MINUIT~\cite{bib:minuit}:
  \begin{itemize}
   \item parabolic function: as in Equation~\ref{eqn:staveBow};
   \item linear function: as in Equation~\ref{eqn:staveBow} but with $a_2\equiv0$;
   \item offset function: as in Equation~\ref{eqn:staveBow} but with $a_1,a_2\equiv0$;
  \end{itemize}
 \item For each of the above fits, any modules with $\langle r_x\rangle$ further away from the fit than $N\!\cdot\!\sigma$ with $N=|\kappa|+4$ are excluded from the fit. This is done in order not to bias the fit results by modules with huge random misalignments, and to improve the numerical stability and reproducibility of the alignment results. $N$ is linearly dependent on $|\kappa|$ since linear and parabolic fits are more strongly determined by outer points in the $\langle r_x\rangle(\kappa)$ distribution, and their removal can have considerable consequences, as for example with the first point at $\kappa=-6$ on the left plot of Figure~\ref{fig:pixelStaveBow};
 \item The three functions in step 1. are refitted using only the non-excluded points, and the function with the best $\chi^2/N_{\rm DoF}$ is retained. The normalisation with $N_{\rm DoF}$ is to detect the preference of the fit for a function with less degrees of freedom, id est cases where the parabolic fit defaults to $a_2\equiv0$ or even $a_2,\,a_1\equiv0$;
 \item For the function selected in the preceding step, a check is performed whether the pull of the leading order parameter, for example $a_2$ for a parabola, is larger than the sigma cut value $\sigmaCut$: 
 \begin{equation} \label{eqn:l4sigmaCut}
 \frac{a_x}{\delta{a_x}}{>}\sigmaCut\,.
 \end{equation}
In case the condition is not met, the next lower fit order is tried. If all three fits fail, no alignmend is performed for the given stave.
\end{enumerate}

The alignment corrections in $x$ from pixel stave bow $c_{x,\,\rm stave\,bow}$ are determined and written to the alignment database at individual module level by calculating the value of the fit function for the given $\kappa$:
\begin{equation*}
 c_{x,\,\rm stave\,bow}=-f(\kappa)\,.
\end{equation*}
Since parabolic and linear displacements also change the orientation of stave modules in space, alignment corrections in local~$\gamma$, i.e. in-plane rotations of a module about the local~$z$ axis (cf.~Subsection~\ref{ssec:frames}), can optionally be calculated:
\begin{equation*}
 c_{\gamma,\,\rm stave\,bow}=-\frac{{\rm d}f}{{\rm d}\kappa}(\kappa)\,.
\end{equation*}

It is worthwhile mentioning that parabolic pixel stave bow misalignments were not anticipated by the alignment community: in particular, they have not not been simulated in the misaligned datasets of the CSC exercise~\cite{bib:alignmentCSC}. However, after initial studies by~\cite{bib:pixelStaveBowMax}, the pixel stave bow alignment was quickly implemented explicitly in the \RA\ algorithm~\cite{bib:pixelStaveBowRA}, and corrected for implicitely at L3 by the \GX\ and \LX\ algorithms.

It was checked that the SCT does not display deformations similar to the bowing of the pixel staves. This is because individual SCT modules were mounted {\it directly} onto the support frame of each barrel layer, rather than affixed to a stave which in turn was mounted on the barrel layer frame, as was done in case of the pixel detector. Nevertheless, the technical provision for stave bow alignment was implemented for both the pixel and the SCT detectors.

%% file: RA/L2.tex
After a pixel stave, the next-in-size superstructure is a {\it layer} in the barrel part of the detector, and a {\it disk} in the ECs, which is sometimes generically referred to as ``layer''. Barrel layers and EC disks have a direct correspondence in the database: they are L2 superstructures, as detailed in Subsection~\ref{ssec:alignmentStructures}. Their alignment procedure is defined below. 
%, first for the barrel, and then for the ECs.
It should be mentioned, that at the time of writing, the possibility to correct for misalignments in the $X$-$Y$ plane, rotations about $Z$, and translations along $Z$ are implemented in the \RA\ algorithm.% move it down?

\begin{figure}
\begin{center}
\includegraphics[width=7.5cm,clip=true]{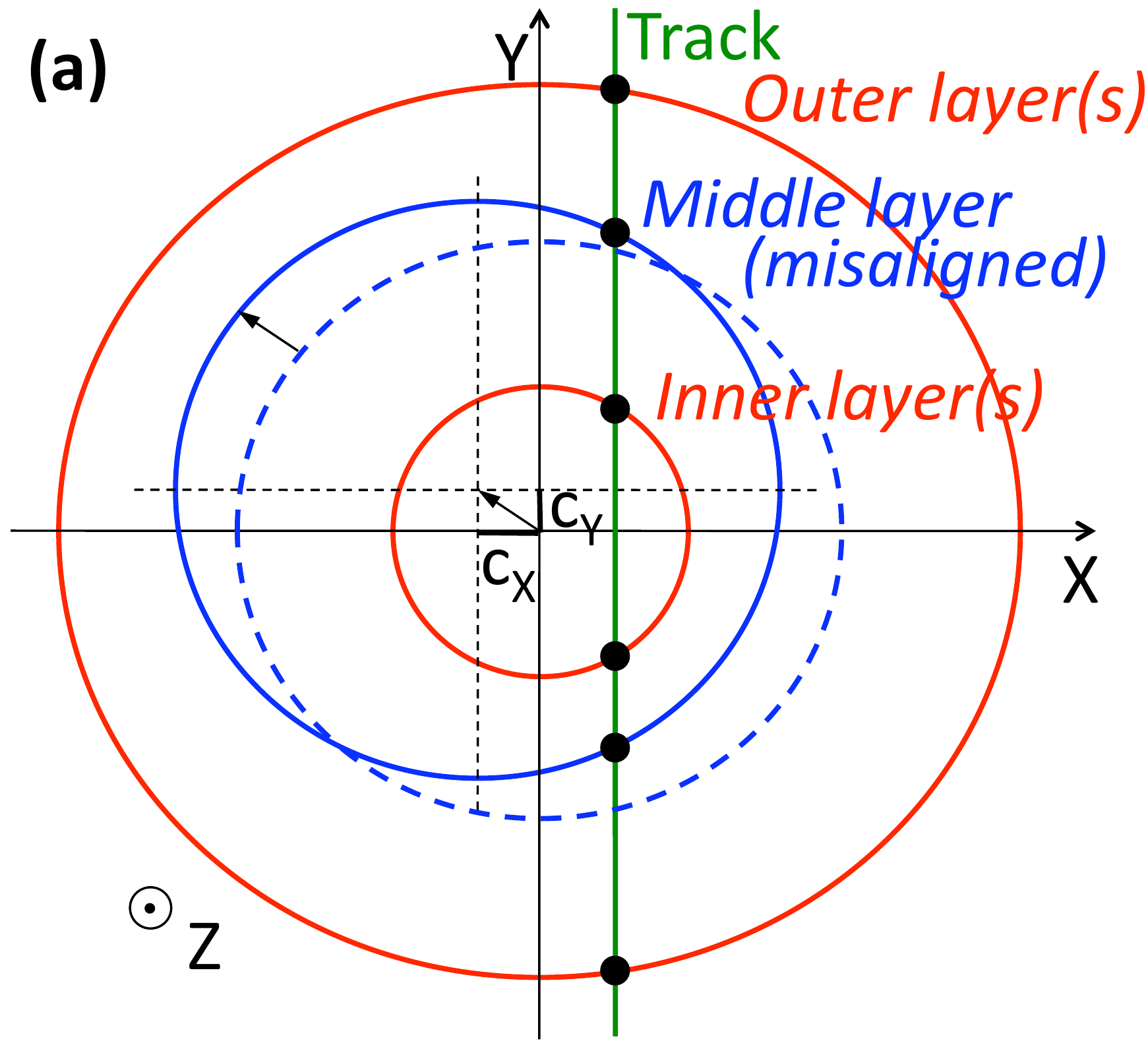}\qquad
\includegraphics[width=7.5cm,clip=true]{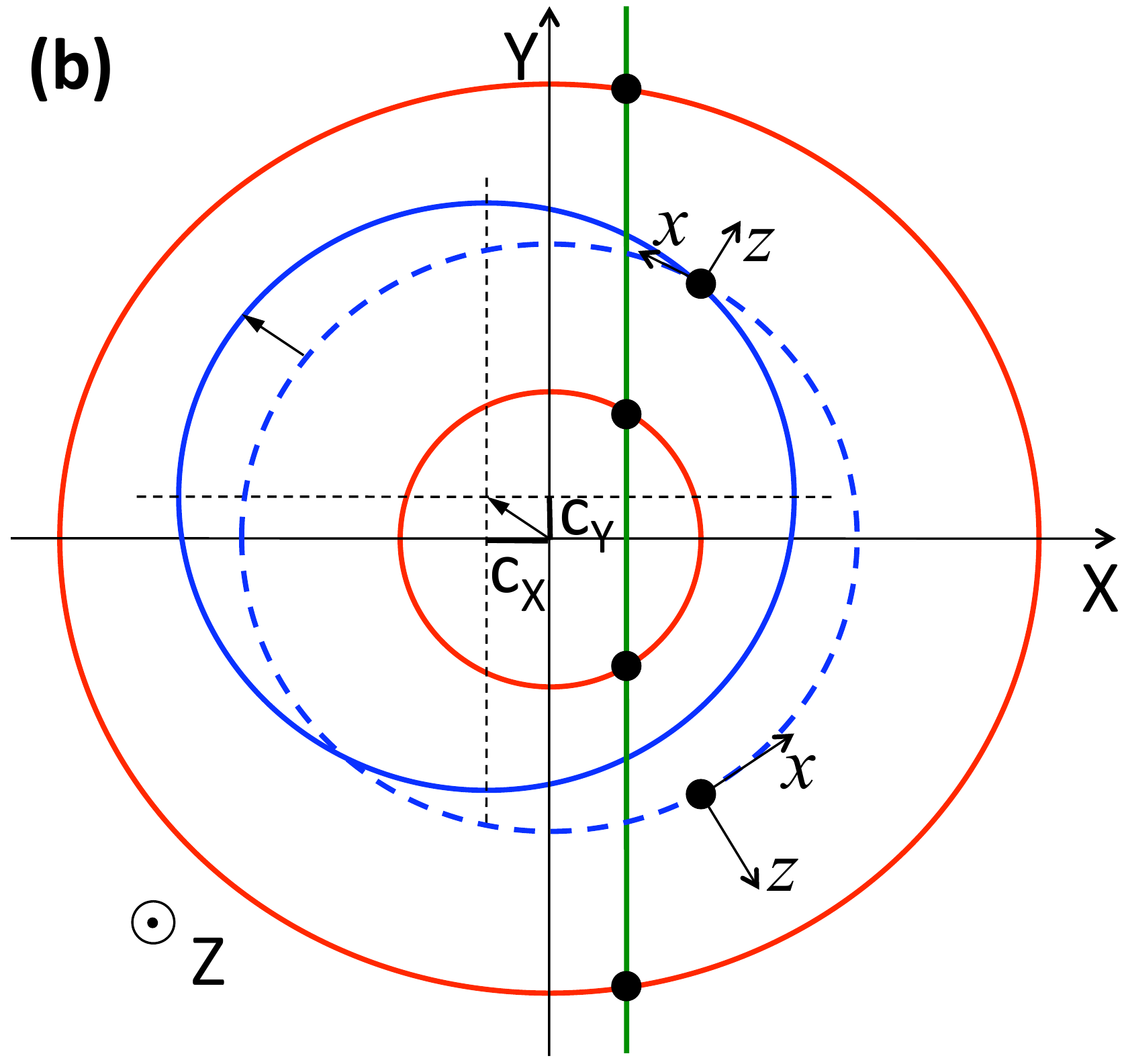}
\end{center}
\caption[L2 misalignments: reality vs. situation as ``seen'' by reconstruction]{\label{fig:residualsL2}
A symbolic cross-section through the barrel of an idealised detector with perfectly known geometry, infinitely many layers, and in absence of Coulomb multiple scattering. Only the middle layer (in blue) is assumed to be misaligned in the $X$-$Y$ plane by $\vec c$, $c_X<0$, $c_Y>0$. The situation as it is in reality {\bf (a)} and as ``seen'' by the reconstruction {\bf (b)} for one single track is depicted. In (b), the local frame for the hits is indicated.
}
\end{figure}%\nopagebreak[5]

The alignment of {\bf barrel layers} is geometrically somewhat more intuitive and will be addressed first. To understand the principle, let us assume an idealised situation where there is a detector with a barrel-like geometry which is perfectly known, and infinitely many layers. Only one layer in the middle\footnote{``Middle'' means in this context: with infinitely many layers at both smaller and larger radii than the layer considered.} shall be misaligned in the $X$-$Y$ plane by the vector $\vec c$, and without loss of generality let $c_X<0$, $c_Y>0$. Further, let there be no magnetic field, and also no Coulomb multiple scattering. This situation is sketched in part (a) of Figure~\ref{fig:residualsL2}. However, the reconstruction software, assuming that there are no misalignents, will rather ``see'' a picture as depicted in part (b) of the same Figure, such that the modules traversed by the track will measure $r_x$ residuals deviating from zero. With the frame conventions described in Subsection~\ref{ssec:frames}, local $x$ will be positive counterclockwise in the global frame for any given module. Thus, for the middle layer hit in the upper hemisphere of the detector we will have $r_x<0$, whereas $r_x>0$ will be true for the other middle layer hit. 

In order to obtain the exact magnitude of the corrections needed to compensate for the misalignments, let us consider the module at $\Phi=\threeOverTwo\pi$ and the case of $c_X<0$, $c_Y\equiv0$. For this module, the frame axes $x\Big|_{\Phi=\threeOverTwo\pi}$ and $X$ coincide, and therefore 
\begin{equation} \label{eqn:r_x_c_X_L2}
  r_x\Big|_{\Phi=\threeOverTwo\pi} = -\frac{c_X}{\cos\xi}\,,
\end{equation}
where $\xi$ is the tilt angle of the module plane with respect to the tangent of the layer envelope. Abandoning the idealised situation introduced in the preface, we can deduce for the alignment correction:
\begin{equation*} \label{eqn:c_X_L2_oneModule}
  c_X\equiv-\cos\xi\cdot\langle r_x\rangle_{\rm stave}{\Big|_{\Phi=\threeOverTwo\pi}}\,.
\end{equation*}
Here, the mean $\langle .\rangle_{\rm stave}$ is taken over all the modules in the stave, as they by definition have the same\footnote{Of course, the $\Phi$ positions of modules in a given stave are the {\it same} only up to misalignments, which can be safely neglected for our purposes here.} $\Phi$ coordinate. The stave considered is the one closest to $\Phi=\threeOverTwo\pi$. Similarly, we obtain for $c_Y>0$, $c_X\equiv0$ at $\Phi=0$:
\begin{eqnarray}
  r_x\Big|_{\Phi=0} &=& -\frac{c_Y}{\cos\xi} \label{eqn:r_x_c_Y_L2} \\
  \Longrightarrow c_Y &\equiv& -\cos\xi\cdot\langle r_x\rangle_{\rm stave}{\Big|_{\Phi=0}}\,.\nonumber%\label{eqn:c_Y_L2_oneModule}
\end{eqnarray}

Clearly, the above prescription to calculate alignment corrections would be suboptimal, as it uses only a small fraction of the residuals collected by modules at $\Phi=\threeOverTwo\pi,\,0$. In fact, as one can see from the geometry of the setup, any misalignments in the $X$-$Y$ plane will result in a sinusoidal dependence of $\langle r_x\rangle$ on $\Phi$ for a perfectly cylinder-shaped barrel layer. This way, residuals collected by all the modules can be used in a fully coherent and correlated way, and the alignment corrections $c_X,~c_Y$ can be extracted from the amplitude and the phase of the fitted sine. Let the sine function $f(\Phi)$ to fit the $\langle r_x\rangle(\Phi)$ distribution be of the form:
\begin{equation} \label{eqn:sineL2}
 f(\Phi) \equiv S\cdot\sin(\Phi+\Phi_0)+O\,
\end{equation}
with $O=0$ for the time being. Then the alignment corrections in the $X$-$Y$ plane are given by
\begin{eqnarray}
  c_X &=& ~~\,S\cdot\cos\Phi_0\cdot\cos\xi\,, \label{eqn:c_X_L2}\\
  c_Y &=& -    S\cdot\sin\Phi_0\cdot\cos\xi\,. \label{eqn:c_Y_L2}
\end{eqnarray}
This can be verified by evaluating Equation~\ref{eqn:sineL2} for $f(\Phi)\Big|_{\Phi=\threeOverTwo\pi,\,0}$ and requiring that it give the same result as Equations~\ref{eqn:r_x_c_X_L2},~\ref{eqn:r_x_c_Y_L2} for $\Phi_0=\pi,\,\threeOverTwo\pi$, which corresponds to the cases where $c_x<0,~c_y=0$ and $c_y>0,~c_x=0$, respectively.

With a similar argumentation and assumptions as above, one can deduce that the $r_x$ residuals will exhibit the {\it same} shift for any $\Phi$ in case a layer is rotatated by the angle $c_\Gamma$ around the global $Z$ axis. With this in mind, we can write for the alignment correction:
\begin{equation} \label{eqn:c_Gamma_L2_mean}
 c_\Gamma=-\frac{\langle r_x\rangle_{\rm layer}}{R_{\rm layer}}\cdot\cos\xi\,,
\end{equation}
where $\langle .\rangle_{\rm layer}$ is taken over all the modules of the given layer, and $R_{\rm layer}$ is the radius of that layer. In order to remain consistent when correcting for misalignments in the $X$-$Y$ plane and in~$\Gamma$ simultaneously, one can use a common fit with the function defined in Equation~\ref{eqn:sineL2} {\it without} the constraint $O=0$. Then, the alignment correction becomes:
\begin{equation} \label{eqn:c_Gamma_L2_sine}
 c_\Gamma=\frac O{R_{\rm layer}}\cdot\cos\xi\,.
\end{equation}

\begin{figure}
\begin{center}
\includegraphics[width=7.5cm,height=7.5cm,clip=true]{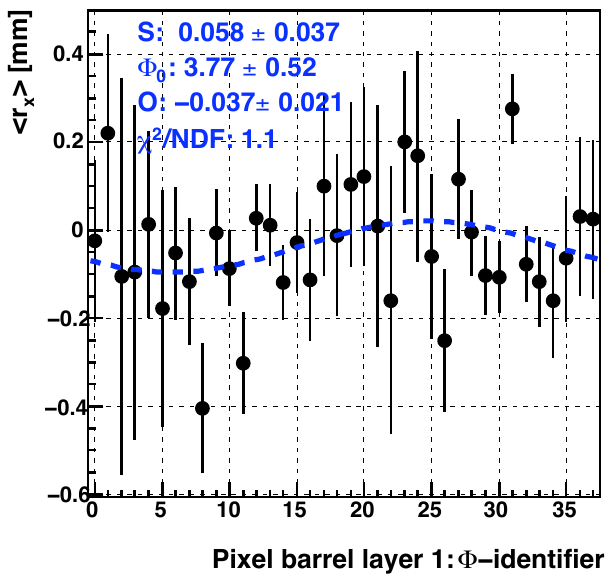}\qquad
\includegraphics[width=7.5cm,height=7.5cm,clip=true]{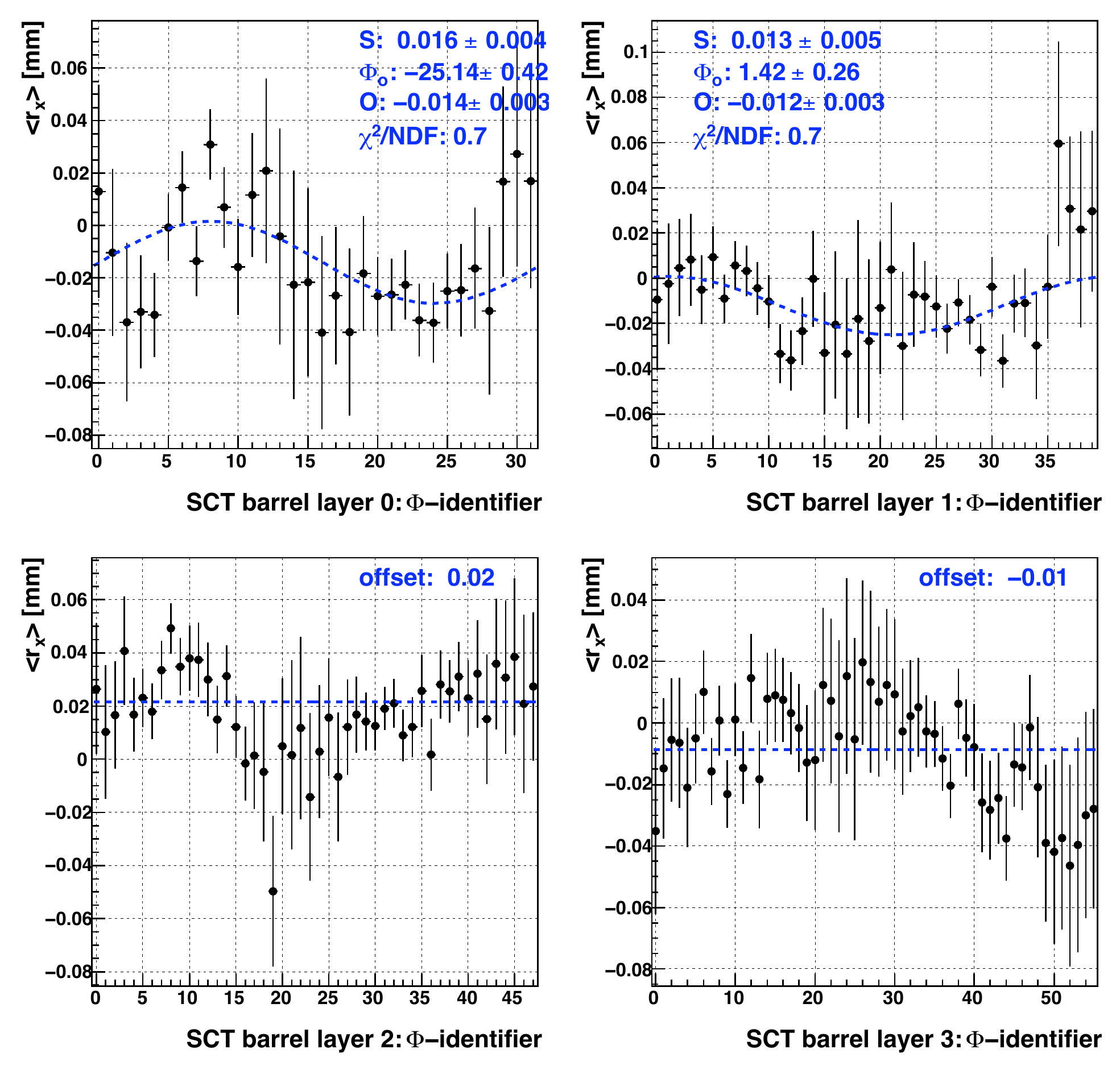}
\end{center}
\caption[The distribution $\langle r_x\rangle_{\rm stave}(\Phi)$ for a typical barrel layer of the pixels and SCT each]{\label{fig:r_x_vs_Phi_L2_BRL}
The distribution $\langle r_x\rangle_{\rm stave}(\Phi)$ for a typical barrel layer of the pixel detector {\bf (left)} and of the SCT {\bf (right)} using the full $B$-field off M8+ dataset after L1 alignment. Distributions for the entire ID can be found in Chaper~\ref{chp:m8plus}. The fit results with a sine of the form specified in Equation~\ref{eqn:sineL2} are shown in blue.
}
\end{figure}%\nopagebreak[5]

Figure~\ref{fig:r_x_vs_Phi_L2_BRL} shows the $\langle r_x\rangle_{\rm stave}(\Phi)$ distribution for a typical barrel layer of the pixel detector (left) and of the SCT (right), which were obtained using the full \linebreak[4]$B$-field off M8+ dataset after L1 alignment, as described in Chaper~\ref{chp:m8plus} (ibid., the corresponding distributions for the entire ID can be found). A clear sinusoidal dependance is evident.\\
Nominal errors on the residual means of each stave, $\delta r_x={\sigma_{\rm stave}(r_y)}/{\sqrt{n_y^{\rm stave}}}$, are statistical only. However, since the individial modules have misalignments and thus the shape of the barrel layer is not a {\it perfect} cylinder, they need to be scaled up in order to make the $\chisqNDF$ a reasonable estimate for the quality of the track fit. The scaling factors have to be tuned for each experimental setup individually. In case of ATLAS nominal geometry and cosmic ray data mentioned above, 15 and 10 have been found appropriate for the pixel and the SCT subdetector, respectively.

The fit procedure to obtain the $c_X$, $c_Y,$ and $c_\Gamma$ alignment constants is the following: 
\begin{enumerate}
 \item A fit with the function of the form defined by Equation~\ref{eqn:sineL2} is done using MINUIT;
 \item A fit with a simple offset function, i.e. Equation~\ref{eqn:sineL2} with $S\equiv0$, is performed. This is done to account for the fact that there may be a significant misalignment around the $Z$ axis, but not in the $X$-$Y$ plane;
 \item Only the function with the least \chisqNDF\ is considered further;
 \begin{itemize}
  \item In case the offset function is chosen, the alignment corrections are calculated using Equation~\ref{eqn:c_Gamma_L2_mean} with the corresponding uncertainty 
    \[\delta c_\Gamma=\frac{\delta r_x}{R_{\rm layer}}\,,
    \quad{\rm with}~\delta r_x = \frac{\sigma_{\rm layer}(r_x)}{\sqrt{n_x^{\rm layer}}}\,;\]
  \item In case the sine fit function is chosen, the errors as reported by MINUIT are used for all three parameters;
 \end{itemize}
 \item It is checked whether the condition $\chisqNDF<25$ is fulfilled. If not, the fit is considered as failed, and no alignment corrections are calculated;
 \item If the pulls of the fit parameters are larger than the \sigmaCut\ cut:
   \begin{eqnarray*}
    S/\delta S &>& \sigmaCut \\
    \Phi_0/\delta \Phi_0 &>& \sigmaCut\,,
   \end{eqnarray*}
 alignment corrections in the $X$-$Y$ plane are calculated using Equations~\ref{eqn:c_X_L2}, \ref{eqn:c_Y_L2};
 \item If the pull of the fit parameter $O$ {\bf or} the residual mean $\delta r_x$ are greater than the \sigmaCut\ cut, the alignment correction for rotations about $Z$ is calculated using Equation~\ref{eqn:c_Gamma_L2_mean} {\bf or} Equation~\ref{eqn:c_Gamma_L2_sine}, whichever is appropriate after step 3.
   %\begin{eqnarray*}
   % O/\delta O &\stackrel!<& \sigmaCut \\
   % \langle r_x\rangle_{\rm layer}/\delta r_x &\stackrel!<& \sigmaCut
   %\end{eqnarray*}
\end{enumerate}
There is also a possibility to force an alignment for the $c_\Gamma$ degree of freedom {\it only}, in which case Equation~\ref{eqn:c_Gamma_L2_mean} is used and only the check described in step 6 is made.

Finally, any misalignments from translations in global $Z$ can be corrected for by evaluating $r_y$ residuals directly:
\begin{eqnarray}
 c_Z &=& -\langle r_y\rangle_{\rm layer}\,, \label{eqn:c_Z_L2}\\
 \delta c_Z &=& \frac{\sigma_{\rm layer}(r_y)}{\sqrt{n_y^{\rm layer}}}\,,\nonumber
\end{eqnarray}
It was found~\cite{bib:divergenceZ} that although this is a valid approach in principle, it does not work well with the ATLAS geometry and track reconstruction: 
%not only the SCT with its space point residuals in $y$ did not provide sensible results, but also the layers of the pixel detector diverged when performing global $Z$ alignment. 
the layers of both the SCT and the pixel sub-detector diverged when performing global $Z$ alignment. 
It was verified that the problem is not due to a sign error in Equation~\ref{eqn:c_Z_L2} by inverting its sign and running several iterations of the \RA\ algorithm, which yielded an even worse divergence. The lack of convergence in $Z$ can probably be attributed to the instability of the track fit in $Z$ because of the small strip stereo angle of 40\,mrad in the SCT, which makes the fit very sensitive to any misalignments in~$x$ and thus the $X$-$Y$ plane. Since a {\it simultaneous} track fit through the pixel and the SCT detectors is performed, the resulting biases would affect both subdetectors.

As will be detailed in Chapter~\ref{chp:m8plus}, the alignment of the $c_X$, $c_Y$, and $c_\Gamma$ degrees of freedom in the barrel of the silicon tracker has proven to yield stable and reliable results with an exponential-like convergence.

Currently, the number of collected cosmics events is not sufficient for an alignment of the tilt degrees of freedom of a barrel layer: $c_A$, $c_B$, where $A$ and $B$\footnote{Note that $A$ and $B$ are the capital Greek variables $\alpha$ and $\beta$} are the rotation angles about the $X$ and $Y$ axes. Therefore, no alignment of these degrees of freedom is implemented at the time of writing. However, it would be straightforward to implement it in the \RA\ framework: instead of performing a single fit of $f(\Phi)$ to the $\langle r_x\rangle_{\rm stave}(\Phi)$ distribution in the entire barrel layer,
% where $\langle .\rangle_{\rm stave}$ is evaluated per-stave, 
one could perform a fit to each of the detector rings with $\langle .\rangle_{\rm mod}$ evaluated per-module, and deduce $c_A$, $c_B$ from the dependance $S(Z)$ and $\Phi_0(Z)$. Alternatively, a $g(\Phi,Z)$ fit to the two-dimensional $\langle r_x\rangle_{\rm mod}(\Phi,Z)$ distribution could be performed, and the $c_X,\,c_Y,\,c_A,\,c_B,\,c_\Gamma$ constants extracted simultaneously.

The alignment of {\bf end-cap disks} is conceptually very similar to the alignment of the barrel layers: one can perform an alignment in the $c_X$, $c_Y$, and $c_\Gamma$ degrees of freedom using the sinusoidal dependence $\langle r_x\rangle(\Phi)$ and a procedure as described above. However, there also are some fundamental differences due to the altered geometry, which have to be reflected in the alignment precedure:
\begin{itemize}
 \item SCT EC disks can comprise up to three concentric rings of modules, as detailed in Table~\ref{tab:inDet}, and each point of the $\langle r_x\rangle(\Phi)$ distribution corresponds to exactly one module side. On the contrary, a barrel layer has 13 or 12 (pixel/SCT) modules organised in a stave structure for each $\Phi$ coordinate, which provides $\order{10}$ times more statistics for each point of the $\langle r_x\rangle(\Phi)$ distribution;
 \item The mean $\langle .\rangle$ and any other statistical quantities are always taken over one single module side in the EC context;
 \item Because of the one-to-one correspondence between module sides and $\langle r_x\rangle(\Phi)$ distribution entries, only the statistical errors on the residual mean are considered, i.e. the error scaling factors introduced in the barrel context are unity;
 \item Under the assumption that the rings of each SCT EC disk are perfectly concentric, each of them should display the {\it same} sinusoidal dependance. Therefore, a sine fit of the form given by Equation~\ref{eqn:sineL2} is performed to the $\langle r_x\rangle(\Phi)$ distribution of {\it all} rings in a disk simultaneously. The $n_{\rm ring}$ histograms are conjoined to form a single histogram with range $[0,n_{\rm ring}\!\cdot\!2\pi)$, where $n_{\rm ring}$ is the number of rings of a given disk;
 \item \label{bul:rigid_body_assumption}The argument in the previous bullet is viable only as long as the ring-to-ring misalignments in the same disk are smaller than the disk-to-disk misalignments. Preliminary tests with the cosmic ray dataset collected by ATLAS to date indicate that this condition is fulfilled for the majority of the SCT EC disks. However, there are also some where this is not the case. More details can be found in Chapter~\ref{chp:m8plus};
 \item The EC disks in pixel detector comprise one ring each, i.e. there is a direct correspondence disk $\Leftrightarrow$ ring;
 \item All the modules of the ECs are oriented vertically, which gives rise to a rather different hit topology compared to the barrel region for cosmic ray data: {\it all} the modules of a disk will collect about the same number of residuals, comparable with the barrel layer modules at $\Phi=0,\,\pi$. This will affect the statistical uncertainty on the alignment constants;
 \item In order to translate any global shift in the $\langle r_x\rangle_{\rm disk}$ value into a $c_\Gamma$ alignment correction via Equation~\ref{eqn:c_Gamma_L2_mean}~or~\ref{eqn:c_Gamma_L2_sine}, the radius $R_{\rm disk}$ is needed. The value used is
  \[R_{\rm disk} = \frac{\Sigma_{i}n_{i}R_i}{\Sigma_{i}n_{i}}\,,\]
 where $i$ runs over all rings constituting a given disk, $n_i$ is the number of residuals collected by ring~$i$, and $R_i$ is the $R$-coordinate of the middle of the modules of ring~$i$ in nominal geometry. The above is a valid procedure since \RA\ is an iterative algorithm in the sense of Subsection~\ref{ssec:iterativeRA};
 %with corrections asymptotically approaching zero in case of exponential-like convergence;
 \item The cut on the maximum \chisqNDF\ is lowered to 20;
 \item The minimum number of residuals to be collected by a module side in order to participate in the sinusoidal fit and thus contribute to the determination of alignment constants is 4. While this may seem a rather low value for each individual module side, this is acceptable for alignment of an entire EC disk, as we are fitting $n_{\rm ring}\cdot\mathscr O(50)$ {\it correlated} data points.
 %\item Keep in mind, that in the EC context, $\langle .\rangle$ and any other statistical quantities are always taken over one single module;
 %\item While a barrel layer has several modules organised in a stave structure for each $\Phi$ coordinate, the EC disks can comprise up to three concentric rings of modules, as detailed in Table~\ref{tab:inDet}. Besides providing $\order{10}$ times more statistics for each point of the $\langle r_x\rangle$ distribution in case of the barrel layers, this has more profound consequences: under the assumption that the rings of each EC disk are perfectly concentric, they should display the same sinusoidal dependance. Therefore, a sine fit of the form given by Equation~\ref{eqn:sineL2} is performed to the $\langle r_x\rangle(\Phi)$ distribution of {\it all} rings in a disk simultaneously. Technically speaking, they are ``glued'' together to a single histogram with a range $[0,n_{\rm ring}\cdot2\pi)$ where $n_{\rm ring}$ is the number of rings of a given disk;
\end{itemize}

\begin{figure}
\begin{center}
\includegraphics[width=7.5cm,clip=true]{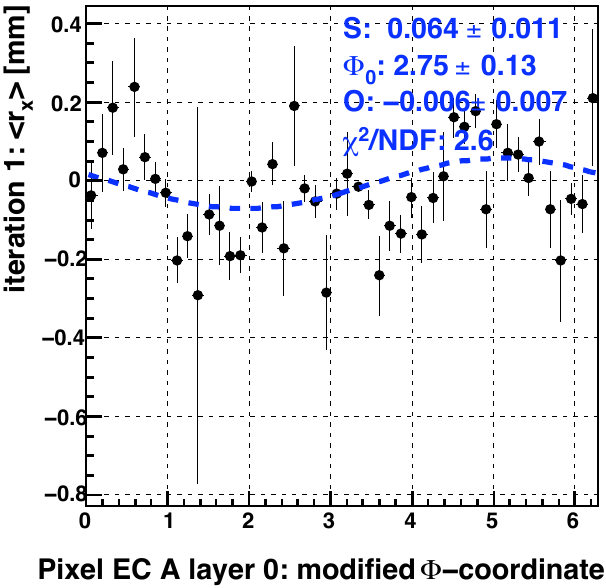}\qquad
\includegraphics[width=7.5cm,clip=true]{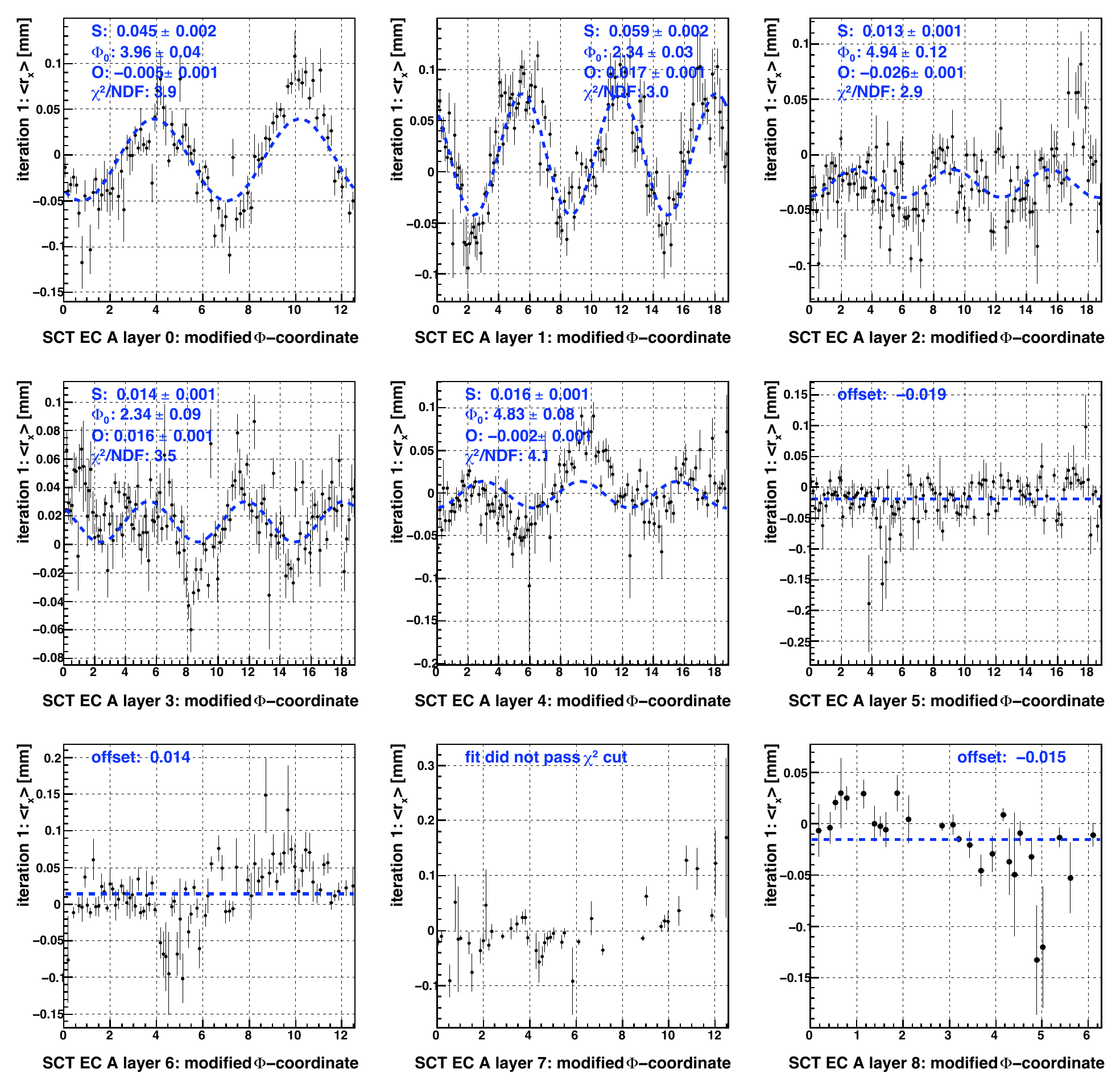}
\end{center}
\caption[The distribution $\langle r_x\rangle_{\rm mod}(\Phi)$ for an EC disk of the pixels and SCT each]{\label{fig:r_x_vs_Phi_L2_EC}
The $\langle r_x\rangle_{\rm mod}(\Phi)$ distribution for a typical end-cap disk of the pixel detector {\bf (left)} and of the SCT {\bf (right)} using the full $B$-field off M8+ dataset after L1 alignment. Distributions for the entire ID can be found in Chaper~\ref{chp:m8plus}. The fit results with a sine of the form specified in Equation~\ref{eqn:sineL2} are shown in blue.
}
\end{figure}%\nopagebreak[5]

The alignment procedure for the $c_Z$ degree of freedom is not directly transferrable from the barrel to the end-caps case. This is because the $r_y$ residual in the EC modules is contained in the $X$-$Y$ plane\footnote{$r_y\in (X,Y)$ up to misalignments.} and therefore cannot provide any insights about misalignments along the $Z$ axis in the sense of Equation~\ref{eqn:c_Z_L2} since $Z\perp X,\,Y$. Certainly, other approaches to align for $c_Z$, as well as $c_A$ and $c_B$ can be tried in the future once collision data become available. Currently, there is a consensus that cosmic ray tracks collected by ATLAS to date are limited in a statistical and topological sense, which would deem any alignment efforts in these three degrees of freedom unreliable.

In principle, the SCT\footnote{In the pixel detector there is a one-to-one correspondence between a disk and a ring.} EC rings can be aligned in a similar fashion as the EC disks: the only difference would be that the $\rmean{x}(\Phi)$ distribution of each ring constituting an SCT EC disk would have to be fitted individually. However, in absence of collision mode data and with the statistical limitations of the cosmic ray dataset collected by ATLAS to date it is felt that this challenge cannot be attacked yet. The provision for EC ring alignment is already implemented in the \RA\ algorithm: the \verb|Ring| class exists and the fitting routine parallel to the one in the \verb|Disk| class is fully implemented. The only remaining part to be introduced is the translation of sine fit parameters $S$, $\Phi_0$, and $O$ into alignment corrections.

%% file: RA/L1.tex
In terms of the procedure, % and the extraction of alignment corrections, 
the alignment at L1 is a special case of alignment at L2 introduced above, since the basic features of the geometry are the same. However, there are some minor differences which will be highlighted here.

\begin{figure}
\begin{center}
\includegraphics[width=15.5cm,clip=true]{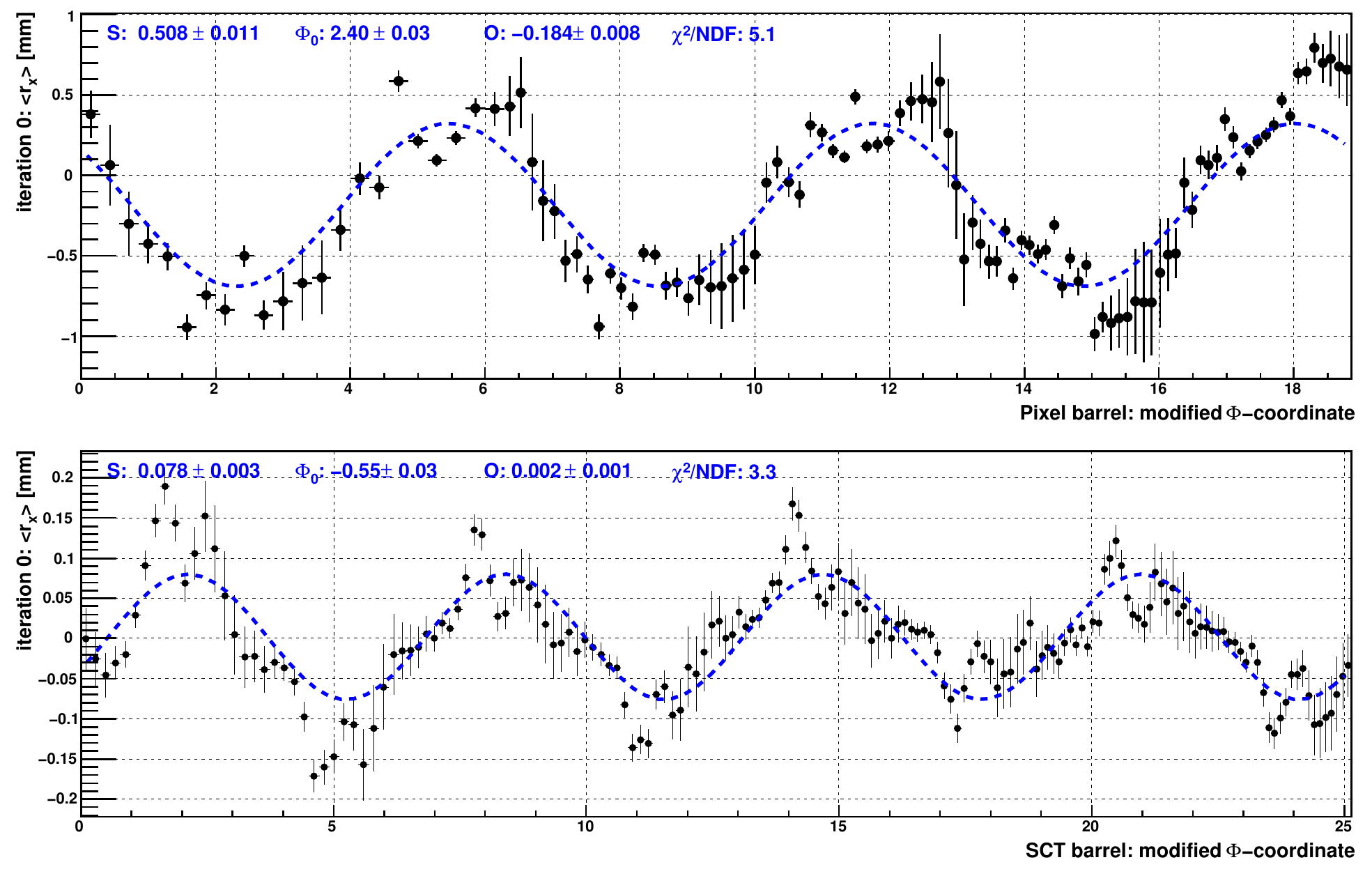}
\end{center}
\caption[The distribution $\langle r_x\rangle_{\rm stave}(\Phi)$ for the barrel of pixels and SCT]{\label{fig:r_x_vs_Phi_L1_brl}
The $\langle r_x\rangle_{\rm stave}(\Phi)$ distribution for the barrel part of the pixel detector {\bf (top)} and of the SCT {\bf (bottom)} using the full $B$-field off M8+ dataset before any alignment. The corresponding distributions {\it after} L1 aligment can be found in Chaper~\ref{chp:m8plus}. The fit results with a sine of the form specified in Equation~\ref{eqn:sineL2} are shown in blue.
}
\end{figure}

In the {\bf barrel} region, the main difference between the alignment of individual layers and the entire barrel is that the $\rmean{x}(\Phi)$ distributions of all layers of a given subdetector are fitted {\it simultaneously} by ``glueing'' them together to one single histogram, as illustrated in Figure~\ref{fig:r_x_vs_Phi_L1_brl}, as was done with the SCT EC disks comprised of rings. The histogram has the range $[0,\,n_{\rm layer}\!\cdot\!2\pi)$, where $n_{\rm layer}=3\,(4)$ in case of the pixel (SCT) subdetector. The procedure for defining the points of the $\rmean{x}(\Phi)$ distribution is very similar to the one used for individual layers at L2, the only difference being that the errors are scaled up by a factor of 15 and 25 for pixels and SCT, respectively. The fitting procedure is exactly the same up to the fact that the initialisation of the fit was augmented to take into account that the $\rmean{x}(\Phi)$ distribution is expected to have $n_{\rm layer}$ local maxima and as many local minima in case of a sine hypothesis, rather than just one. Finally, the prescription to extract alignment constants at L1 corresponds exactly to L2.

%\nopagebreak[5]

The alignment procedure of the {\bf end-caps} is tightly related to the alignment of the end-cap disks. The 
%basic principle is again that the 
$\rmean x(\Phi)$ distributions of the individual disks --- which in turn are comprised of $\rmean x(\Phi)$ distributions of individual rings in case of the SCT --- are fitted simultaneously in one single histogram. Its range is $[0,\,\Sigma_i n_i\!\cdot2\!\pi)$, where $i$ runs over all the disks participating in the common EC fit, and $n_i$ is the number of rings constituting disk~$i$. 

\begin{figure}
\begin{center}
\includegraphics[width=15.5cm,clip=true]{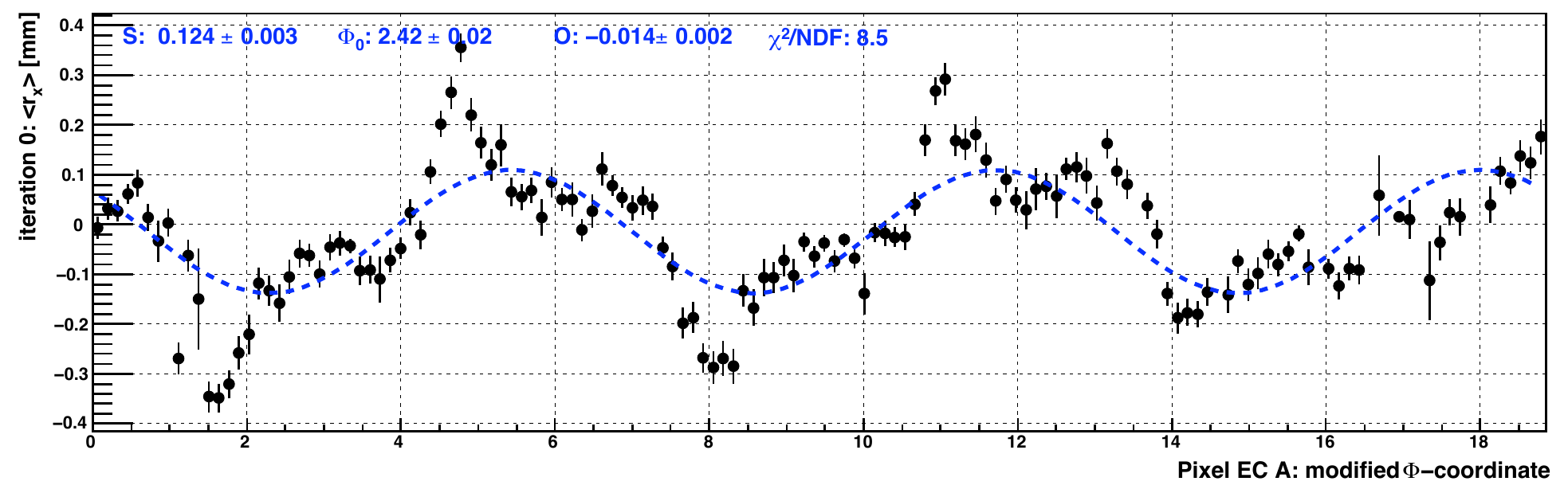}\\
\includegraphics[width=15.5cm,clip=true]{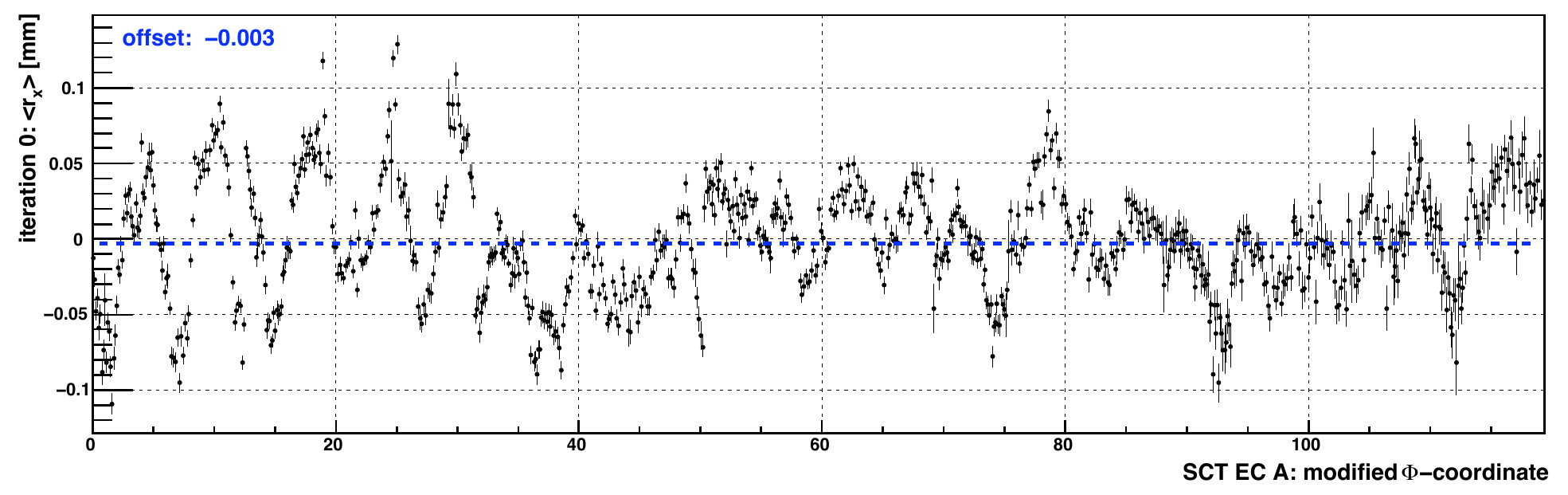}
\end{center}
\caption[The distribution $\langle r_x\rangle_{\rm mod}(\Phi)$ for the pixel end-caps]{\label{fig:r_x_vs_Phi_L1_EC}
The $\langle r_x\rangle_{\rm mod}(\Phi)$ distribution for the end-cap A of the pixel {\bf (top)} and the SCT {\bf (bottom)} detector using the full $B$-field off M8+ dataset before any alignment. The corresponding distributions {\it after} L1 aligment can be found in Chaper~\ref{chp:m8plus}. The fit results with a sine of the form specified in Equation~\ref{eqn:sineL2} (which defauts to $S\equiv0$ in case of the SCT) are shown in blue. Note that for the SCT EC fits only the 6 innermost disks are considered, i.e. $\Phi\in[0,\,17\cdot2\pi)$.
}
\end{figure}%\nopagebreak[5]

For the {\bf pixel} subdetector $n_i\equiv1$, and the resulting distribution for end-cap A is shown in Figure~\ref{fig:r_x_vs_Phi_L1_EC} (top). A clear sinusoidal dependence with a common amplitude and phase in each of the three disks constituting an EC is evident. This verifies that the pixel end-caps indeed can be considered rigid bodies along the lines of discussion in the SCT EC disk alignment part of the preceding Subsection on page~\pageref{bul:rigid_body_assumption}.\\
It should be mentioned that for historical reasons there is no one-to-one correspondence between a pixel end-cap and the L1 structures defined in the alignment constant database. Therefore, alignment corrections obtained for pixel end-caps at L1 are written to the L2 section of the database. Consequently, all three disks constituting an EC will have identical alignment corrections written at L2.\\ \label{par:pixelECL2}
An additional caveat is that the L1 pixel superstructure defined in the alignment constant database corresponds to the entire pixel subdetector. Therefore, alignment corrections for the pixel {\it barrel}, being written to the database at L1, will change the position of the {\it entire} subdetector. This is compensated for when performing pixel end-cap  alignment by applying the inverse of the transformation for the pixel barrel at L1.

The application of the {\bf end-cap} alignment procedure to the {\bf SCT} is somewhat less straight forward~\cite{bib:sctECL1}: Figures~\ref{fig:inDetTechnical} and~\ref{fig:sct} illustrate that the extension of an SCT EC in $Z$ is about twice as large as the diameter of its envelope in $X$-$Y$. The predominant incidence angle of cosmic rays with respect to the ATLAS cavern surface is roughly $\theta_{\rm incid}\simeq0$\,rad, and the flux falls steeply as $\theta_{\rm incid}$ increases. This means that few cosmic ray tracks produce hits in more than one SCT EC disk, and almost none traversing more than three. As a consequence, disks which are far apart are ``interconnected'' with only few tracks, and their relative alignment is rather weakly constrained, especially for the innermost and outermost disks. So, only a {\it local} alignment can be considered as credible due to this problem. The problem is evident from Figure~\ref{fig:r_x_vs_Phi_L1_EC} (bottom), where the $\rmean x(\Phi_0)$ distribution is shown for the SCT EC~A: no common sinusoidal dependence according to Equation~\ref{eqn:sineL2} can be established. Consequently, the fit defaults to an offset function, id est $S\equiv0$, as intended. Due to the locality of alignment with cosmic rays, it is not possible to render a stringent judgement on whether the rigid body assumption of the SCT ECs is valid using the dataset collected by ATLAS so far.\\
Certainly, the situation can be somewhat improved by recording more cosmic ray tracks, but even then the residual distributions for each module will be dominated by the vast majority of hits from tracks with small angles $\theta_{\rm incid}$ traversing only a few disks. Moreover, the trigger efficiency for cosmic ray data decreases dramatically for large $|Z|$, and the high-rate fast TRT-OR trigger provides coverage for the barrel region only. This results in a decreased data collection efficiency in the SCT end-caps, especially in the far end disks at large $|Z|$.\\
Irrespective of whether cosmic ray or collision mode data are used, a credible SCT EC alignment at L1 should display a strong dependence on the positions of the SCT barrel and of the pixel detector, since it is performed with respect to the rest of the silicon tracker per definitionem. In the context of track-based alignment, this dependance can only be provided by tracks producing hits in more than one of the respective sub-detectors. To enhance the contribution of this track category to the alignment procedure at L1, only the innermost six SCT EC disks are used.

As will be extended in Chapter~\ref{chp:m8plus}, no alignment of the SCT ECs at L1 was produced by the \RA\ algorithm. However, an alignment at L2 was performed, since local disk-to-disk alignment is considered to be feasible with the cosmics data collected by ATLAS to date. A reliable alignment of the SCT ECs at L1 is likely to be possible with only a few thousands of high quality collision-mode tracks, provided that a rigid-body assumption can be established from this data.

%% file: RA/IterativeRA.tex
The \RA\ is an {\it iterative} algorithm: in order to obtain alignment constants, it is run in several consecutive iterations until the alignment constants converge towards their final values. This has various implications, the most imporant being that convergence criteria need to be defined. This will be discussed below, preceded by the description of the technical reconstruction-related aspects. The technical aspects involved in the actual running of each single \RA\ iteration are described in Subsection~\ref{ssec:running}.

\subsubsection{Reconstruction-Related Aspects}

Technically speaking, the iterative nature of the \RA\ algorithm means that in each iteration the entire track reconstruction is rerun using the alignment constants from the preceding iteration (or null-constants whcih correspond to the nominal geometry of the ATLAS detector). These ``updated'' tracks serve as an input to the \RA\ algorithm, so that the residuals will reflect the alignment corrections calculated in previous iterations. Based on this, new alignment constants are calculated and written out to a file\footnote{To be precise, new alignment corrections are {\it added} to the ones from the previous iteration, and their sum is written out.}, which can be used to update the ATLAS ID geometry in the subsequent iterations. Once the alignment procedure has converged, one has the choice of committing the derived alginment constants to the global ATLAS conditions database. This entry can serve for bookkeeping purposes, but most importantly, can be retrieved by any analysis run within the \Athena\ framework.

In each alignment iteration the {\it full} track reconstruction including pattern recognition is rerun, rather than solely refitting the already reconstructed tracks with the updated ID geometry. Therefore, a successful alignment procedure will result in an increase of the number of reconstructed tracks, especially in the first iterations with nominal geometry. Also hits, which were previously labelled as ``outliers'' because they were too far away from the track and therefore did not participate in the alignmet procedure, may become reintegrated onto the track. This way not only the number of reconstructed tracks increases, but also the number of residuals on the existing tracks. Both criteria can be used to some extent to monitor the convergence of the alignment procedure, though not in a quantitative way: the increase of the number of reconstructed tracks and residuals per iteration will strongly depend on how loose the track reconstruction acceptance cuts are chosen.

\subsubsection{Convergence Criteria in the \RA\ Algorithm}

An indespensable requirement on the alignment procedure is exponential-like convergence, which manifests itself in the magnitude of alignment corrections per iteration asymptotically approaching zero. Thus, in the idealised case of infinite statistics, one could in principle run infinitely many iterations, gradually reducing the uncertainty on the alignment constants.\\
In a typical real-life situation, where the number of residuals available for alignment is statistically limited, a clear exponential-like convergence will be observed at the beginning of the alignment procedure, followed by quasi-oscillatory behaviour of alignment corrections around zero. Since the latter does not improve the alignment quality, a suitable criterion needs to be introduced to terminate the alignment procedure for a given module or substructure. Contrariwise, the better the statistical power of a dataset, the smaller will the magnitude of the quasi-oscillatory behaviour usually be, and the more iterations can be run before it occurs. Ideally, this should be taken into account by the termination criterion. In the \RA\ algorithm, this is done by imposing a cut on the pull of the calculated alignment correction $c$:
\begin{equation}\label{eqn:termination}
 \frac c{\delta c} >\sigmaCut\,,
\end{equation}
id est if this cut is not passed, the given alignment correction is not applied. Cut values of $\sigmaCut\in[0.5,\,1]$ were found to yield good results.\\
In case of superstructure alignment, similar requirements to Equation~\ref{eqn:termination} are imposed on the fit parameters, which is detailed explicitly in preceding Subsections.\\
When performing alignment of single modules at L3, the criterion defined in Equation~\ref{eqn:termination} will suffer from an uncertainty on the its denominator, $\delta c$, defined in Equation~\ref{eqn:corrL3resErr} for the case of regular residuals. Therefore, modules which did not collect a minimum number of residuals (typically 50)\label{pag:minNumHitsPerMod} can be excluded from the alignment procedure. In a similar fashion and using the full formula for $\delta c^{\rm L3}$ in Equation~\ref{eqn:dc_xtotal} one can argue that overlap regions which did not collect a minimum number of overlap residuals (typically 25) should be disallowed from the alignment constant calculation. The reason for a smaller minimum number of overlap residuals compared to the minimum number of regular residuals comes from a much reduced amount of Coulomb multiple scattering for the latter.

The iterative nature of the alignment and the exponential-like convergence requirement imply an important merit: the alignment corrections {\it per iteration} do not need to be perfectly exact. On the contrary, one can afford somewhat {\it suboptimal} alignment corrections per iteration: if they are reasonably close\footnote{(Empirically, a deviation by $\mathscr O(20\%)$ empirically) has been found as ``reasonably close''} to the correct values for any given iteration, the final result will still be approached asymptotically over the iterations, at the cost of a somewhat decreased convergence speed. The main philosophy of the \RA\ algorithm is to extensively use this fact in order to obtain the alignment constants in the most robust and transparent way. The major approximations made are mentioned explicitely in the definition of the \RA\ procedure above.\\
Admittedly, in a statistically limited situation 
%in the sense of the preceding paragraph 
the \sigmaCut\ criterion may terminate the suboptimal alignment procedure at a slightly prior stage compared to the optimal one. However, if the approximation made is not significantly larger than $\mathscr O(10\%)$, any biases from the interplay with the \sigmaCut\ criterion are much smaller than the statistical uncertainty on the alignment correction $\delta c$ because $\sigmaCut\in[0.5,\,1]$.

For obvious reasons, diverging alignment corrections should be discarded under any circumstances.

%% file: RA/Running.tex
Each iteration of the \RA\ algorithm requires several hundreds of CPU hours on a typical LXBATCH~\cite{bib:lxbatch} computer already for the relatlively small M8+ dataset described in Section~\ref{sec:datasetM8plus}. This is because the full track reconstruction has to be performed in the ID in order to account for the updated alignment constants. % from the preceding iteration.
The contribution of the \RA\ algorithm itself is with about 1\% of the total CPU time rather marginal. To cope with the high CPU time demand, {\em parallel processing} was introduced. In each iteration, multiple instances (typically of \order{100}) of the \RA\ algorithm referred to as {\em subjobs} are run. Each of the subjobs collects residuals and overlap residuals from its part of the dataset and writes out a pair of text files: one for the pixel, and one for the SCT detector. These files contain the residuals and overlap residuals ordered by modules. After all subjobs terminate, a {\em superjob} concludes the iteration of the \RA\ algorithm: it reads in the output text files of all subjobs, ``adds'' the residual and overlap residuals, and calculates the alignment constants based on them. That way, one iteration of the \RA\ algorithm can be run in some hours rather than weeks.

A common framework to steer the track-based inner detector alignment algorithms --  the {\tt InDetAlignExample} package -- was developed by the ATLAS alignment community since the beginning of 2008~\cite{bib:inDetAlignExampleTwiki,bib:inDetAlignExampleHowTo,bib:inDetAlignExample}. Its main purpose is to coordinate the sub- and superjobs, as parallel processing is a feature common to all algorithms. The interface of the \RA\ algorithm was adopted to this new framework and debugged by the author. The {\tt InDetAlignExample} package offers also the possibility to configure the \RA\ algorithm. The configuration parameters are the subject of the next Subsection.

%% file: RA/Steering.tex
One of the key features of the \Athena\ interface are the JobOptions, which offer the possibility to pass parameters to a given algorithm or tool at initialisation time. The \RA\ algorithm takes full advantage of it, and most of its steering parameters can be modified via the JobOptions. In the context of the {\tt InDetAlignExample} framework, this is done by modifying the file {\tt InDetAlignExample\_SiAlignAlgs.py}. Below, the full summary of the steering parameters to the \RA\ algorithm at the time of writing is given: 

\begin{description}
\vspace{-3mm}
\item[AlignLevel:] selects the level of alignment to {\bf1}~(L1,~Subsection~\ref{ssec:l1}), {\bf2}~(L2,~Subsection~\ref{ssec:l2}), {\bf3}~(L3,~Subsection~\ref{ssec:l3}, {\bf4} (pixel stave bow, ~Subsection~\ref{ssec:pixelStaveBow}); 
\vspace{-1mm}
\item[ConstantsTextFile$^\star$:] name of the text file with the {\em new} alignment constants;
\vspace{-1mm}
\item[DoReadTextFiles$^\star$:] read in subjob output text files (act as superjob, cf. Subsection~\ref{ssec:running};\vspace{-1mm}
\item[DoWriteConstantsTextFile:] write out a text file with the {\em new} alignment constants;
\vspace{-1mm}
\item[DoWriteTextFile$^\star$:] write out sub job text file output (act as subjob, cf. Subsection~\ref{ssec:running});\vspace{-1mm}
\item[GetResidualTool:] instance of the {\tt GetResidualTool} to be used;\vspace{-1mm}
\item[InDetAlignDBTool:] instance of the {\tt InDetAlignDBTool} to be used;\vspace{-1mm}
\item[InDetAlignHitQualSelTool:] instance of the {\tt InDetAlignHitQualSelTool} to be used;\vspace{-1mm}
\item[NSigmaCut:] the value of the $\sigmaCut$ to be used, cf. Equation~\ref{eqn:termination}.% $\sigmaCut=0.7$ is used in Chapter~\ref{chp:m8plus};
\vspace{-1mm}
\item[OutputLevel:] the verbosity level of the \RA\ algorithm: {\tt VERBOSE}, {\tt DEBUG}, {\tt INFO}, {\tt WARNING}, {\tt ERROR}, {\tt FATAL};\vspace{-1mm}
\item[RefitWithVertexTool:] instance of the {\tt RefitWithVertexTool} to be used;\vspace{-1mm}
\item[ShiftOnlyPIXEL$^\diamond$:] apply the {\tt ShiftVertex} option to the pixel detector only;\vspace{-1mm}
\item[ShiftVertex$^\diamond$:] shift the silicon tracker with respect to the vertex (no alignment involved);\vspace{-1mm}
\item[TextFileNameBasisRead$^\star$:] basis part of the name for sub job text output files to be read in, cf. {\tt DoReadTextFile};\vspace{-1mm}
\item[TextFileNameBasisWrite$^\star$:] basis part of the name for sub job text output file to be written out, cf. {\tt DoWriteTextFile};\vspace{-1mm}
\item[TextFileReadEndIndex$^\star$:] end index for sub job text output files to be read in;\vspace{-1mm}
\item[TextFileReadStartIndex$^\star$:] start index for sub job text output files to be read in;\vspace{-1mm}
\item[TextFileWriteIndex$^\star$:] index of the sub job text output file to be written out;\vspace{-1mm}
\item[TrackCol:] name of the track collection to be used;\vspace{-1mm}
\item[TrackSummaryTool:] instance of the {\tt TrackSummaryTool} to be used;\vspace{-1mm}
\item[VertexXShift$^\diamond$:] for the {\tt ShiftVertex} option: shift magnitude in $X$;\vspace{-1mm}
\item[VertexYShift$^\diamond$:] for the {\tt ShiftVertex} option: shift magnitude in $Y$;\vspace{-1mm}
\item[VertexZShift$^\diamond$:] for the {\tt ShiftVertex} option: shift magnitude in $Z$;\vspace{-1mm}
\item[locXPIXResidualMax:] maximum $|r_x|$ in the pixel detector to be used for alignment, cf.~Equation~\ref{eqn:resCut};\vspace{-1mm}
\item[locXSCTResidualMax:] maximum $|r_x|$ in the SCT detector to be used for alignment;\vspace{-1mm}
\item[locYPIXResidualMax:] maximum $|r_y|$ in the pixel detector to be used for alignment;\vspace{-1mm}
\item[locYSCTResidualMax:] maximum $|r_y|$ in the SCT detector to be used for alignment;\vspace{-1mm}
\item[minTrkPt:] minimum transverse momentum of tracks  to be used for alignment, cf.~Equation~\ref{eqn:p_TCut};\vspace{-1mm}
\item[minimumHitsPerModule:] minimum number of hits to be collected by a module to become alignable, cf.~Page~\pageref{pag:minNumHitsPerMod};\vspace{-1mm}
\item[minimumHitsPerTrackSi:] minimum number of silicon tracker hits on track, cf.~Equation~\ref{eqn:nHitCut};\vspace{-1mm}
\item[minimumOvHitsPerModule:] same as {\tt minimumHitsPerModule} for overlap hits of type $\zeta$;\vspace{-1mm}
\item[overlapResWeightCTB$^\diamond$:] ad-hoc overlap residual weight in Combined Test Beam~(CTB\glossary{name=CTB,description=Combined Test Beam cf.~\cite{bib:ctb,bib:florianThesis}}) alignment, cf.~\cite{bib:ctb};\vspace{-1mm}
\item[refitWithVertex:] refit tracks using a vertex constraint via {\tt RefitWithVertexTool};\vspace{-1mm}
\item[removeEdgeChannels$^\dagger$:] remove edge channels, cf. Subsection~\ref{ssec:selectionHit};\vspace{-1mm}
\item[residualWeightCTB$^\diamond$:] ad-hoc residual  weight in CTB alignment, cf.~\cite{bib:ctb,bib:florianThesis};\vspace{-1mm}
\item[useCTB$^\diamond$:] use CTB alignment procedure, cf.~\cite{bib:ctb,bib:florianThesis};\vspace{-1mm}
\item[useOutliers$^\dagger$:] remove hits marked as {\em outliers} by the track fit, cf. Subsection~\ref{ssec:selectionHit};\vspace{-1mm}
\item[useOverlapRes:] include overlap residuals in alignment;\vspace{-1mm}
\item[useOverlapResOnly:] use {\em only} overlap residuals for alignment;\vspace{-1mm}
\item[usePIX:]     set the pixel detector alignable;\vspace{-1mm}
\item[usePIX\_rZ:] align the $c_\Gamma$ DoF of the pixel detector;\vspace{-1mm}
\item[usePIX\_tX:] align the $c_x\,(c_X)$ DoF of the pixel detector;\vspace{-1mm}
\item[usePIX\_tY:] align the $c_y\,(c_Y)$ DoF of the pixel detector;\vspace{-1mm}
\item[usePIX\_tZ:] align for the radial expansion of the pixel detector using overlap residuals;\vspace{-1mm}
\item[useSCT:]     set the SCT detector alignable;\vspace{-1mm}
\item[useSCT\_rZ:] align the $c_\Gamma$ DoF of the SCT detector;\vspace{-1mm}
\item[useSCT\_tX:] align the $c_x\,(c_X)$ DoF of the SCT detector;\vspace{-1mm}
\item[useSCT\_tY:] align the $c_y\,(c_Y)$ DoF of the SCT detector;\vspace{-1mm}
\item[useSCT\_tZ:] align for the radial expansion of the SCT detector using overlap residuals;\vspace{-1mm}
\item[useSCTresX\_SP:] construct $r_y$ residuals from space point ``hits'' in the SCT, cf.~Section~\ref{sec:residuals};\vspace{-1mm}
\item[xShiftDamping:] damping factor for $c_x\,(c_X)$ alignment corrections from residuals;\vspace{-1mm}
\item[xShiftOvDamping:] same as {\tt xShiftDamping} for {\em overlap} residuals;\vspace{-1mm}
\item[yShiftDamping:] damping factor for $c_y\,(c_Y)$ alignment corrections from residuals;\vspace{-1mm}
\item[yShiftOvDamping:] same as {\tt yShiftDamping} for {\em overlap} residuals;\vspace{-1mm}
\item[zShiftDamping:] damping factor for radial expansion alignment corrections;\vspace{-1mm}
\end{description}
Note that all items marked with $^\star$ are set automatically by the {\tt InDetAlignExample} framework, the ones with $^\diamond$ are only relevant in the context of the older \RA\ version described in \cite{bib:noteRA,bib:florianThesis}, and the ones with $^\dagger$ are factually obsolete since the introduction of the {\tt InDetAlignHitQualSelTool}\footnote{They are kept for debugging purposes and will be phased out in future releases of the \RA\ algorithm}.

%% file: PixelSR1/PixelSR1.tex
In the process of commissioning the ATLAS pixel detector, its end-cap~A collected cosmic ray data~\cite{bib:pixelSR1} in December 2006 in the SR1 building which was widely used for commisioning of ATLAS on the surface. The \RA\ algorithm has been used to align the pixel end-cap~A modules using this data, which shall be described in the following. A  detailed description of the setup and other commisioning analyses performed on %the December 2006 pixel EC~A SR1 data 
this dataset can be found in~\cite{bib:pixelSR1}.

\section{SR1 Pixel End-Cap A Experimental Setup}

\begin{figure}
\begin{center}
\includegraphics[width=7.5cm,clip=true]{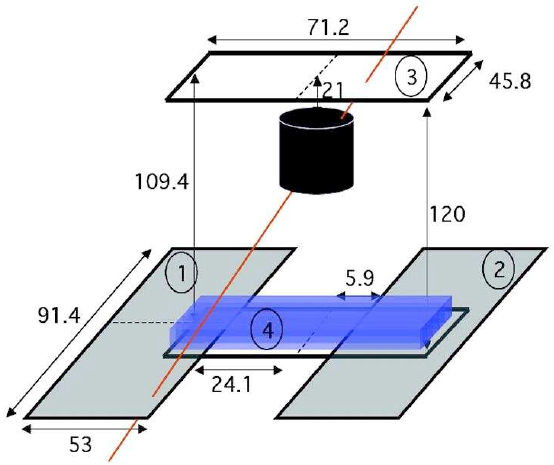}\qquad
\includegraphics[width=6.7cm,clip=true]{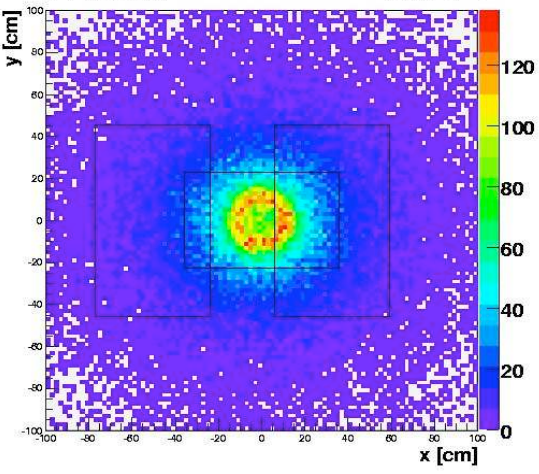}
\vspace{\cDist}
\end{center}
\caption[SR1 pixel end-cap~A experimental setup]{\label{fig:setupSR1}
The SR1 pixel end-cap~A experimental setup {\bf (left)} with the dry box hosting the pixel detector shown in black. All dimensions are in cm. The density distribution{\bf~(right)} of intersection points of tracks producing hits in all three EC disks with the plane of the three lower scintillators {\it1}, {\it2}, and {\it4} as obtained with a toy monte carlo study. Also shown are the positions of the three lower trigger scintillators.
}
\end{figure}

The setup used for cosmic ray data taking by pixel EC~A in the SR1 building is illustrated in Figure~\ref{fig:setupSR1}~(left). The EC together with dry air and electronics services is hooked up inside a dry box shown as a black cylinder. To maximise its acceptance for cosmic rays, the end-cap is oriented such that its symmetry axis $Z$ is perpendicular to the ground.

The trigger system consists of four scintillators labelled {\it1} through {\it4} on the same Figure. The two small scintillators {\it3} and {\it4} with identical dimensions of $71.2\,{\rm cm}\times45.8\,{\rm cm}$ provide a minimum triggering coverage, which is extended by the two big scintillators {\it1} and {\it2}. The density distribution of intersection points of tracks producing hits in all three EC disks with the plane of the three lower scintillators {\it1}, {\it2}, and {\it4} is presented in Figure~\ref{fig:setupSR1}~(right) as obtained in a MC study. The positions of the three lower scintillators are also indicated. The two small scintillators are placed exactly above each other. In order to reduce the trigger rates for low momentum cosmic ray particles and thus reduce the average amount of Coulomb multiple scattering, a 12\,cm thick layer of iron is added between the two small scintillators, directly below the end-cap. This provides a momentum cut-off at about 140\,MeV. A higher cut-off was not possible due to the load limitations of the support structure for the setup.

In the context of this Chapter, it is important to recall that in each pixel end-cap disk layer modules with odd and even $\Phi$-identifiers\footnote{Modules with odd and even $\Phi$-identifiers will be referred to as odd/even modules in the following.} are mounted on different sides of the carbon-carbon laminate support strucutre, resulting in a distance $\Delta z$ between them.

\section{\RA\ procedure in SR1 Pixel End-Cap A Setup}

The alignment of the pixel end-cap~A modules with the \RA\ algorithm has been performed in \textsc{Athena} release 13.0.20. A patched 13.0.30 version of {\tt SiRobustAlignAlgs} and the 13.0.30 default tag of {\tt SiRobustAlignTools} were used. These releases correspond to a bare L3-version of \RA~\cite{bib:florianThesis,bib:noteRA}, id est before any global superstructure alignment and the new overlap residual treatment were introduced. Unless indicated otherwise, the alignment was started from nominal geometry.

\subsection{Input to the \RA\ procedure}
Due to the missing magnetic field, the track fit was done with the {\tt GlobalChi2Fitter} algorithm forced to straight line mode. Accounting for multiple scattering effects and track energy loss have been switched off. Noisy pixels have been removed.

For the endcaps of the pixel detector, only the $x$-type overlaps (long edge) are present: $o_{xx},~o_{yx}$. When calculating $o_{\zeta x}$, the rotation of adjacent modules with respect to each other by $7.5^\circ$ can be neglected, as argued in Subsections~\ref{ssec:l3} and \ref{ssec:iterativeRA}. %since the \RA\ is an iterative algorithm, so the small bias introduced goes exponentially to 0, as do the shift magnitudes per iteration. Mixing $r_x$, $r_y$ residuals, one would combine a precise measurement in local $x$ with $y$ featuring an uncertainty $\mathscr O(10)$ larger, which may not be desirable. 

Two types of (overlap) residuals have been used: biased and unbiased. The alignment is performed for the two major degrees of freedom for each module -- translations in local $x$ and local $y$.

\subsection{Convergence of the \RA\ procedure}

\begin{figure}
\begin{center}
\includegraphics[width=7.5cm,clip=true]{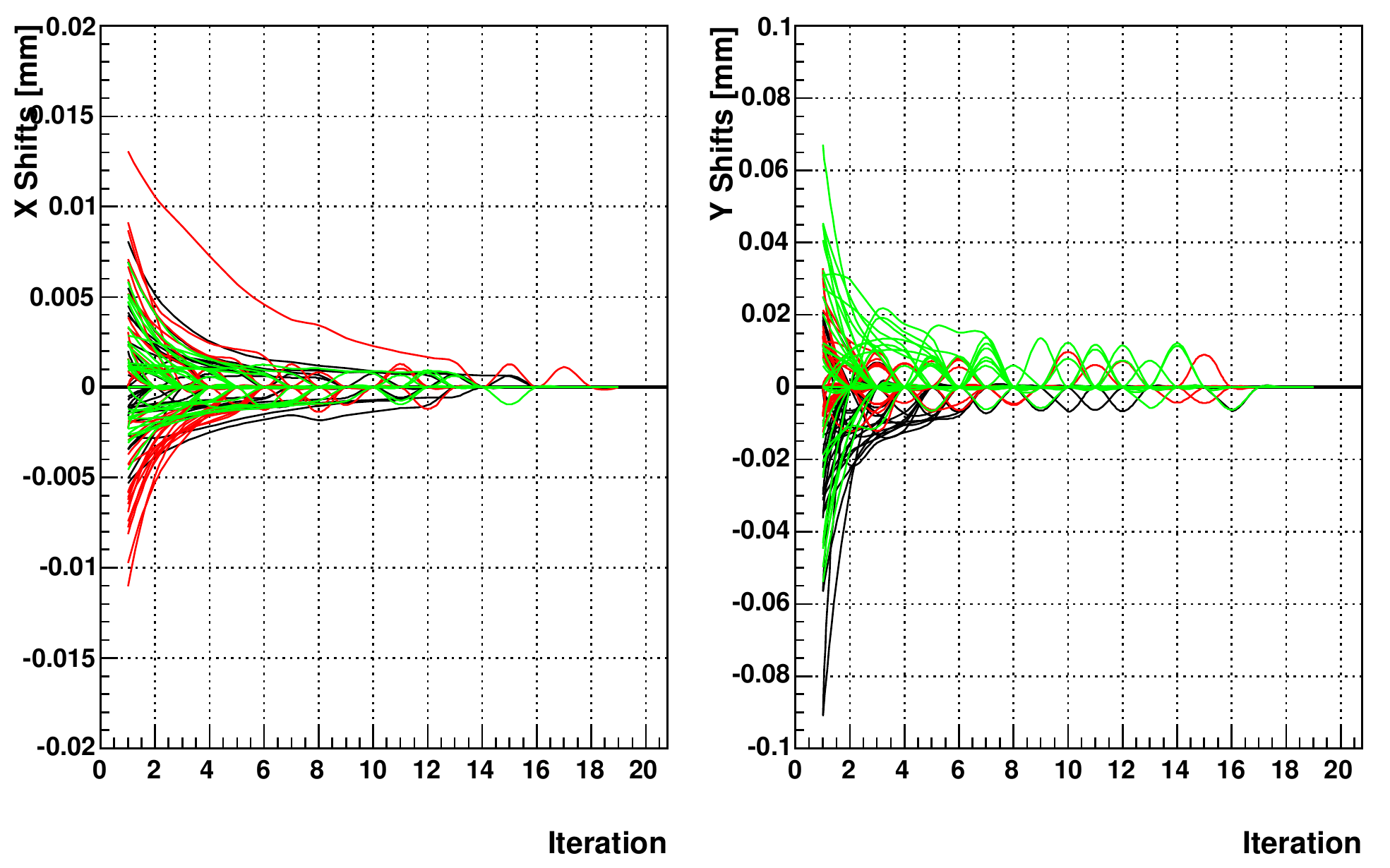}\qquad
\includegraphics[width=7.5cm,clip=true]{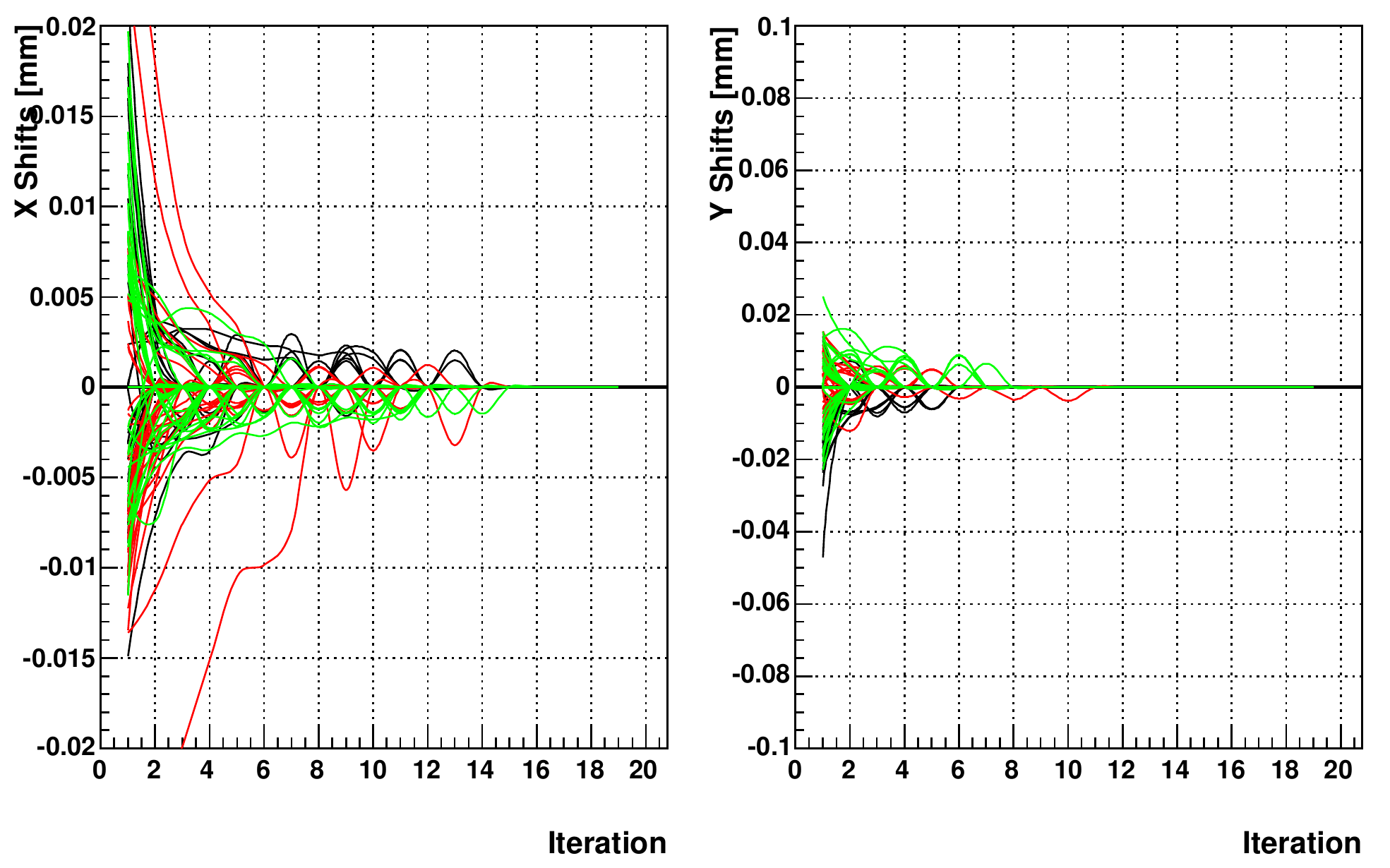}
\vspace{\cDist}
\end{center}
\caption[Differential shifts applied to modules per iteration]{\label{fig:shifts}
Differential shifts applied to modules per iteration for biased residuals {\bf (left)} and unbiased residuals  {\bf (right)}. Each line corresponds to a module. The colours red, green and blue stand for layer 0, 1, and 2, respectively.
}
\end{figure}%\nopagebreak[5]

In Figure~\ref{fig:shifts}, the differential shifts applied to modules per iteration are shown for 20 iterations for biased (left) and unbiased (right) residuals. The differences between biased and unbiased residuals will be highlighted later in the text. Each line corresponds to one module, and the colour-coding is red, green, and blue, which stands for disk layer 0, 1, and 2, respectively. The alignment results of the \RA\ algorithm converge: the shifts applied to modules per iteration decrease over the number of iterations. For most of the modules they go to 0, as the condition for a module to be shifted is not fulfilled any more:
\begin{equation}\label{eqn:shiftCondition}
 %\delta r > \frac{1}{N_a\cdot(\frac1{{\rm RMS}^2_{R_r}} + \frac1{G\cdot{\rm RMS}^2_{O_r}})}
 c_\zeta > \frac{1}{n_{\rm hit}}\cdot\frac1{\frac1{\langle r_\zeta^2\rangle-\langle r_\zeta\rangle^2} + \frac1{G\cdot(\langle o_{\zeta x}^2\rangle-\langle o_{\zeta x}\rangle^2)}}\,,
\end{equation}
id est the shift $c_\zeta$ calculated for the given module is smaller than the statistical error estimated from its residual $r_\zeta$ and overlap residual $o_{\zeta x}$ distributions. Here, $n_{\rm hit}$ stands for the number of hits collected, and $G$ is a geometrical weighting factor typically in the range of 10-300 defined in~\cite{bib:florianThesis,bib:noteRA}. 

\begin{figure}
\begin{center}
\includegraphics[width=7.5cm,clip=true]{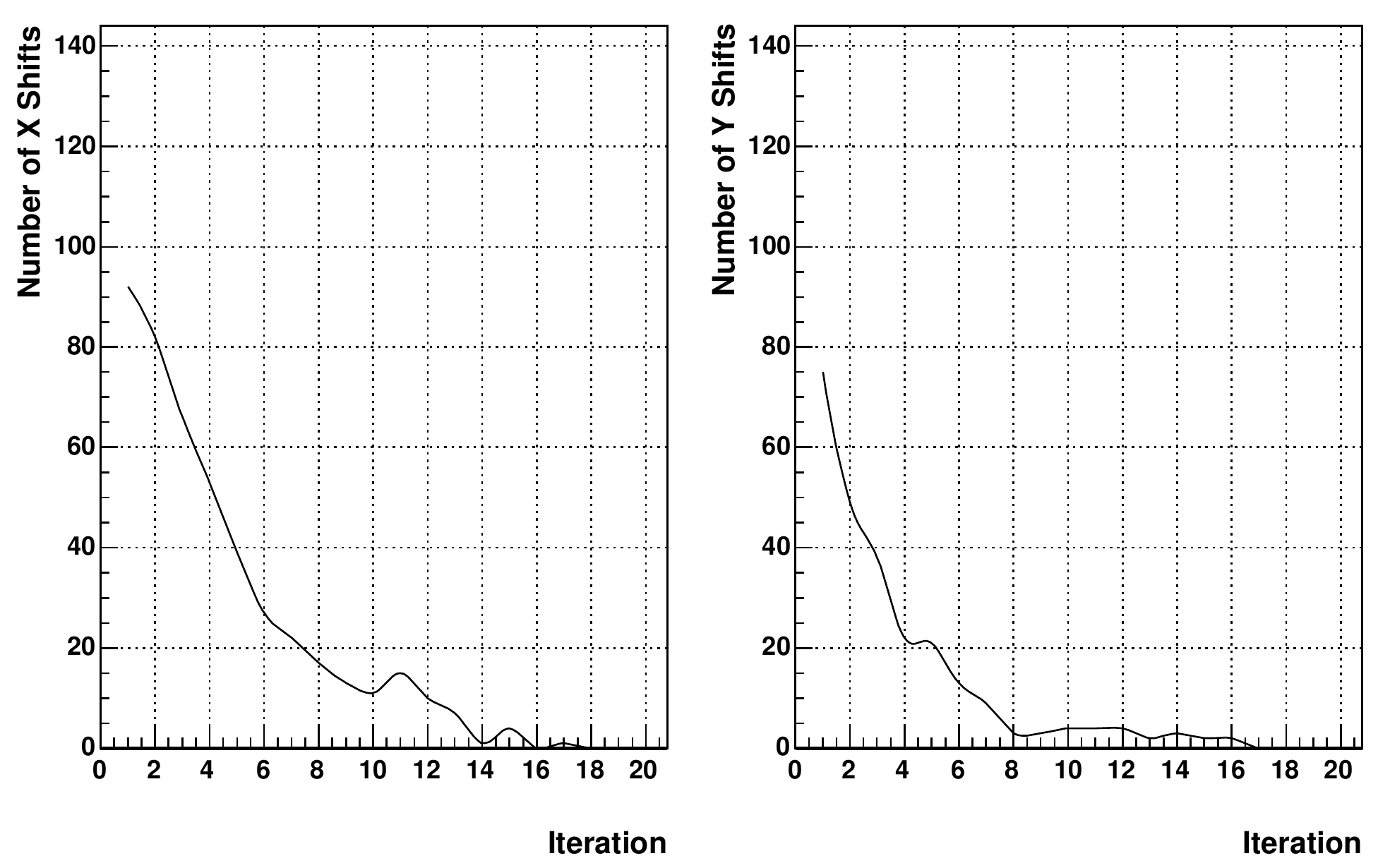}\qquad
\includegraphics[width=7.5cm,clip=true]{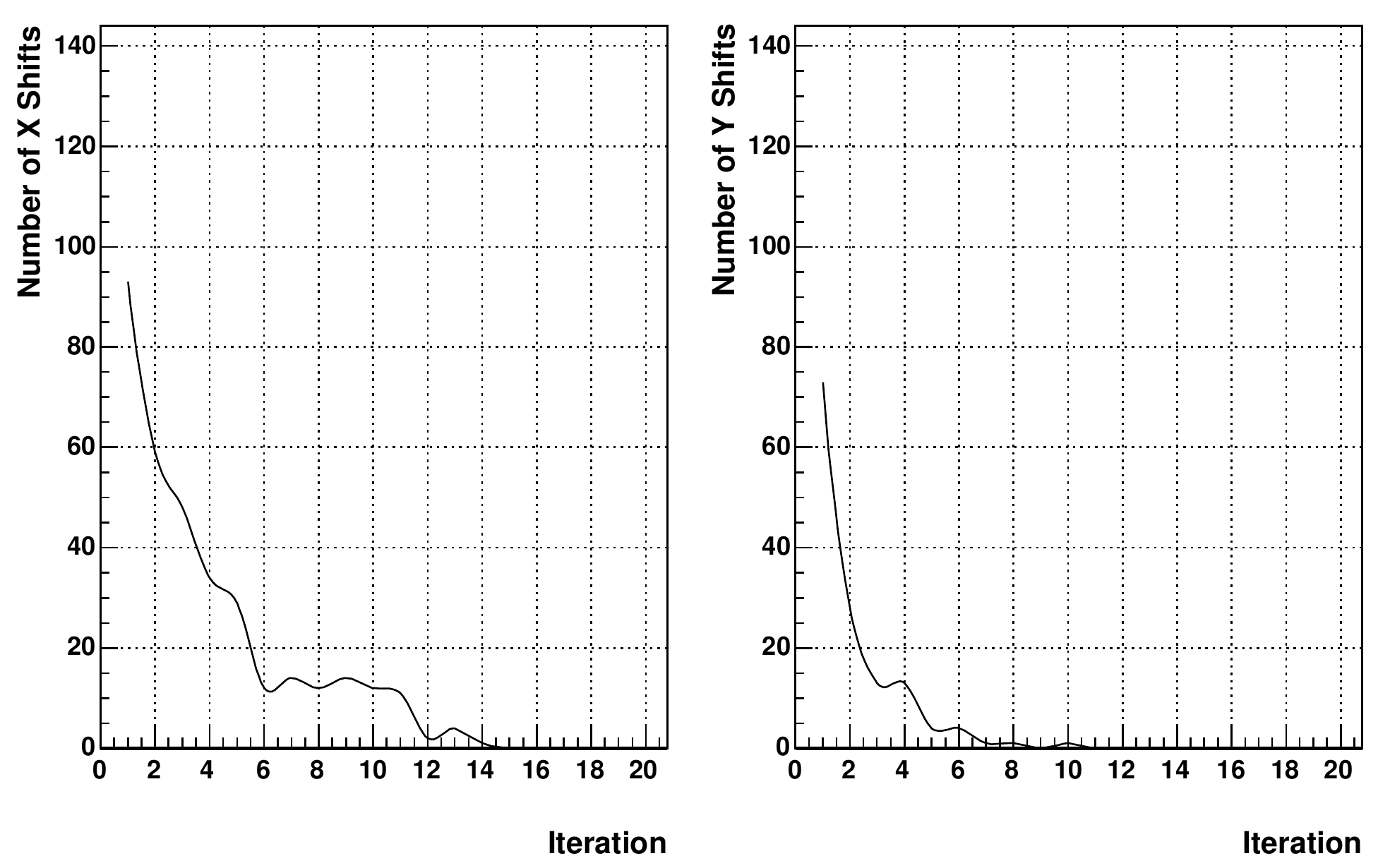}
\vspace{\cDist}
\end{center}
\caption[Number of modules shifted per iteration]{\label{fig:nShifts}
Number of modules shifted per iteration for biased residuals {\bf (left)} and unbiased residuals  {\bf (right)}. For shift conditions see text and Equation~\ref{eqn:shiftCondition}.
}
\end{figure}

Since the alignment of the detector improves with iterations, the number of modules to be shifted should decrease, in accordance with the shift condition defined in Equation~\ref{eqn:shiftCondition}. As demonstated in Figure~\ref{fig:nShifts}, this number asymptotically approaches~0. The initial number of modules to be shifted does not correspond to 144 -- the total number of pixel end-cap~A modules. This is due to the fact, that 23 modules were not read out, as indicated in figures~1-6 in~\cite{bib:pixelSR1}. On top of that, a condition on a minimum number of hits per module in order for it to be considered by the \RA\ algorithm was required. It was set to 100 hits for the biased and, as a check, to 25 hits for the unbiased case. 
%This condition proved very helpful to avoid nonsensical shifts from pathologic residual distributions with too few hits.

\begin{figure}
\begin{center}
\includegraphics[height=5cm,width=7.5cm,clip=true]{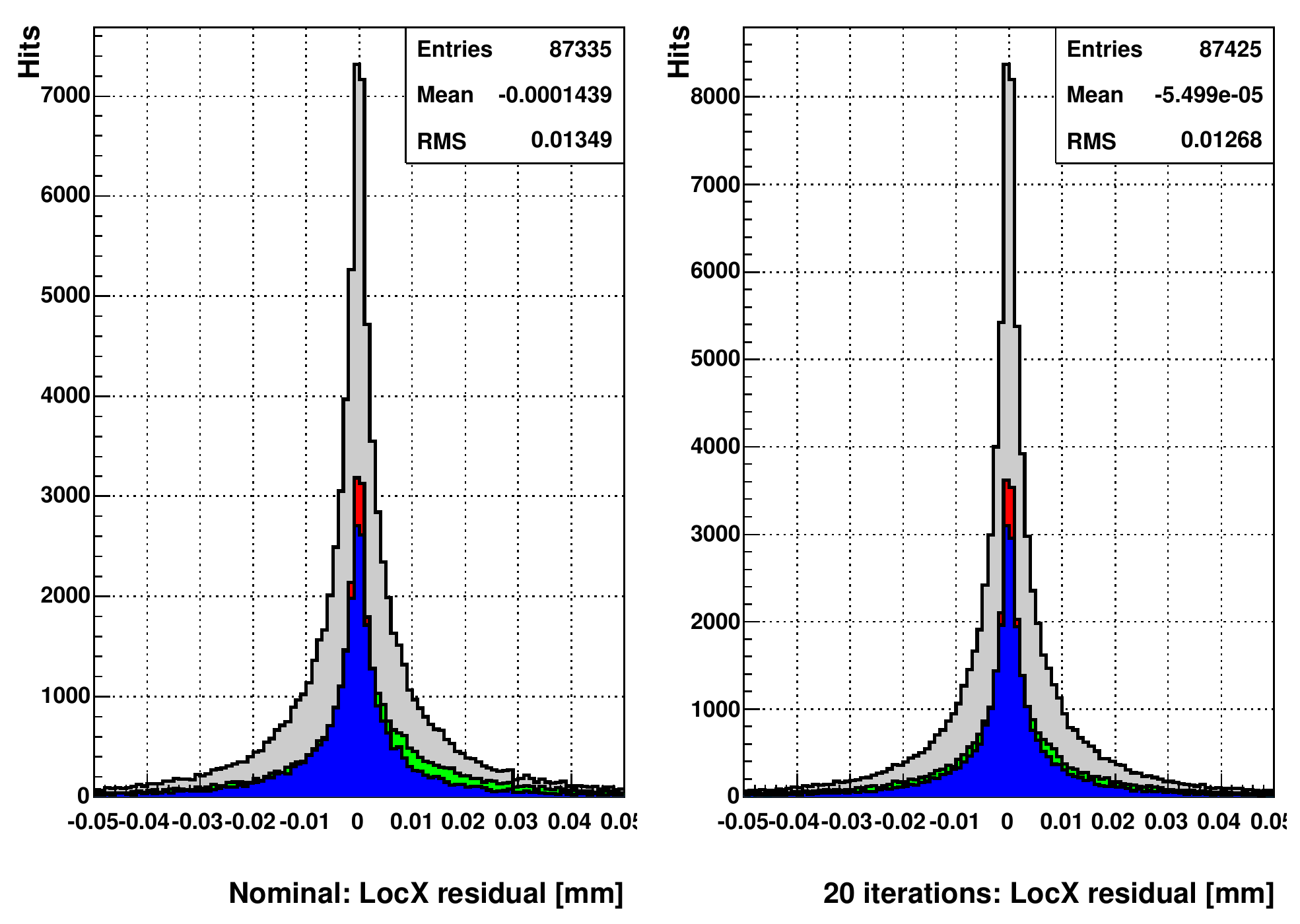}\qquad
\includegraphics[height=5cm,width=7.5cm,clip=true]{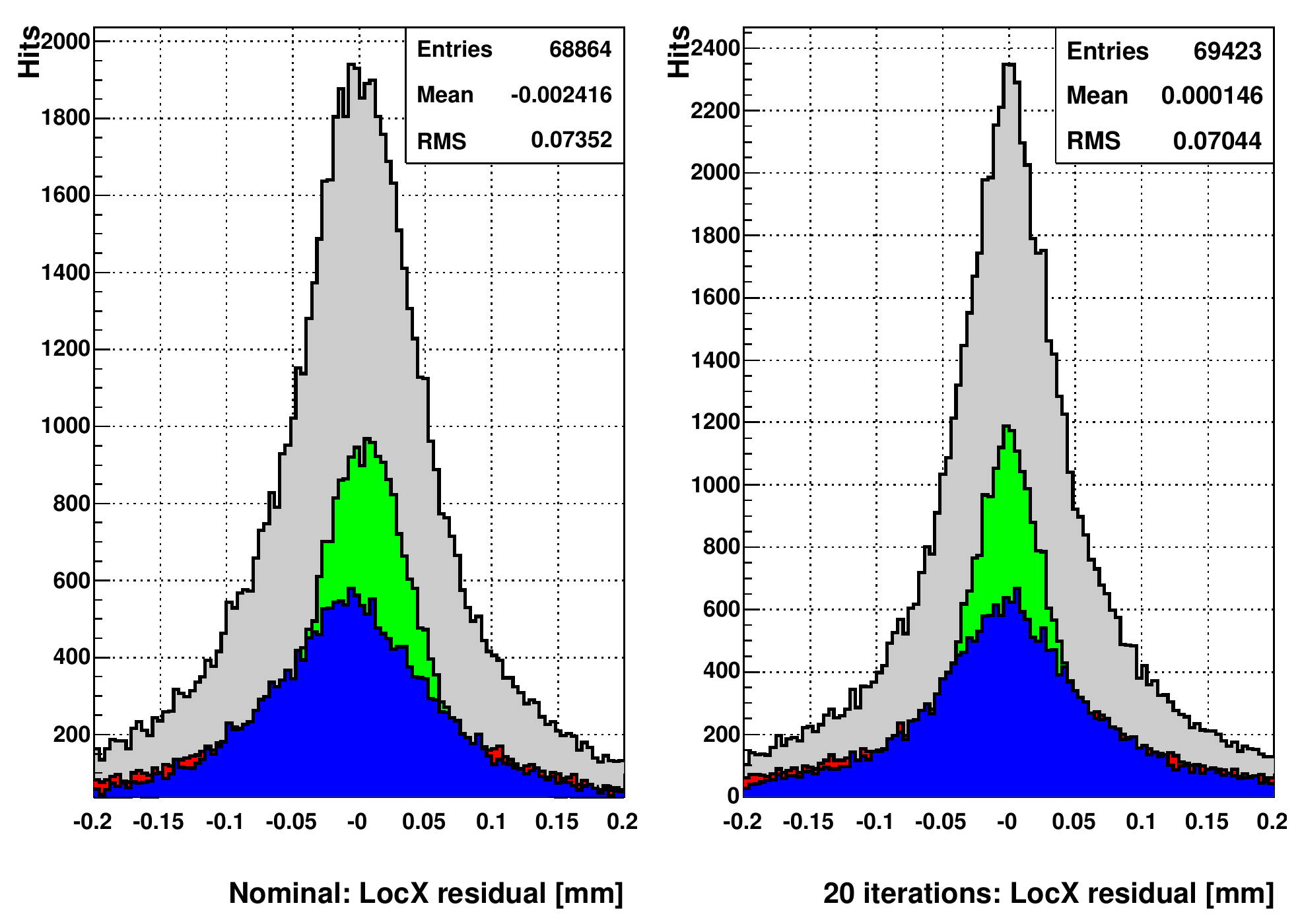}
\vspace{\cDist}
\end{center}
\caption[Residual distributions in local $x$ direction for nominal alignment and after 20 iterations]{\label{fig:resLocX}
Residual distributions in local $x$ direction for the biased {\bf(left)} and the unbiased {\bf(right)} case for nominal alignment and after 20 iterations of the \RA\ algorithm. The colour-coding is red, green, blue for layer 0, 1, 2, and grey for the sum of the residual distributions of the three layers.
}
\end{figure}

% \begin{figure}
% \begin{center}
% %\includegraphics[width=7cm,height=5cm,clip=true]{PixelSR1/fig/2007-10-05_Biased_x0.2_y1.0_minHit100/}\qquad\qquad
% \includegraphics[width=7.5cm,clip=true]{PixelSR1/fig/residuals/2007-10-05_Biased_x0.2_y1.0_minHit100/PixelEndcap_LocY_N}\qquad
% \includegraphics[width=7.5cm,clip=true]{PixelSR1/fig/residuals/2007-09-30_PixelUnbiased_x0.2_y1/PixelEndcap_LocY_N}
% \end{center}
% \caption{\label{fig:resLocY}
% Residual distributions in local $y$ direction for the biased {\bf(left)} and the unbiased {\bf(right)} case for nominal alignment and after 20 iterations of the \RA\ algorithm. The colour-coding is as in Figure~\ref{fig:resLocX}.
% }
% \end{figure}

With the improving alignment of the detector, the residual and overlap residual distributions should improve, id est their RMS should decrease and their mean should approach~0. For the local $x$ residual, this is demonstrated in Figure~\ref{fig:resLocX} for the biased (left) and the unbiased case (right). The three layers 0, 1, and 3 are colour-coded as red, green and blue. The grey histogram is the sum of the residual distributions from all three layers. 
% The residual in the local $y$ direction is depicted in Figure~\ref{fig:resLocY}. 
The general difference in the width of the residual distribution for the biased and the unbiased case results from the fact, that some 70\% of the tracks considered have 3 hits. Therefore, typically, for unbiased residuals a fit through only 2 points will be made, resulting in larger residuals for the module with the non-fitted hit. Additionally, for biased residuals, one might face a pathological situation, where the $\chi^2$ of the track fit is minimised for a track going {\em exactly} through one or even two hits. This will result in entries close to 0 in the residual distribution.

As the alignment of the detector improves, more tracks should be found and reconstructed by the track fitting algorithm, resulting in an increase of the number of residuals. This can be seen from the number of entries in the respective histograms in Figure~\ref{fig:resLocX}.

\begin{figure}
\begin{center}
\includegraphics[height=5cm,width=7.5cm,clip=true]{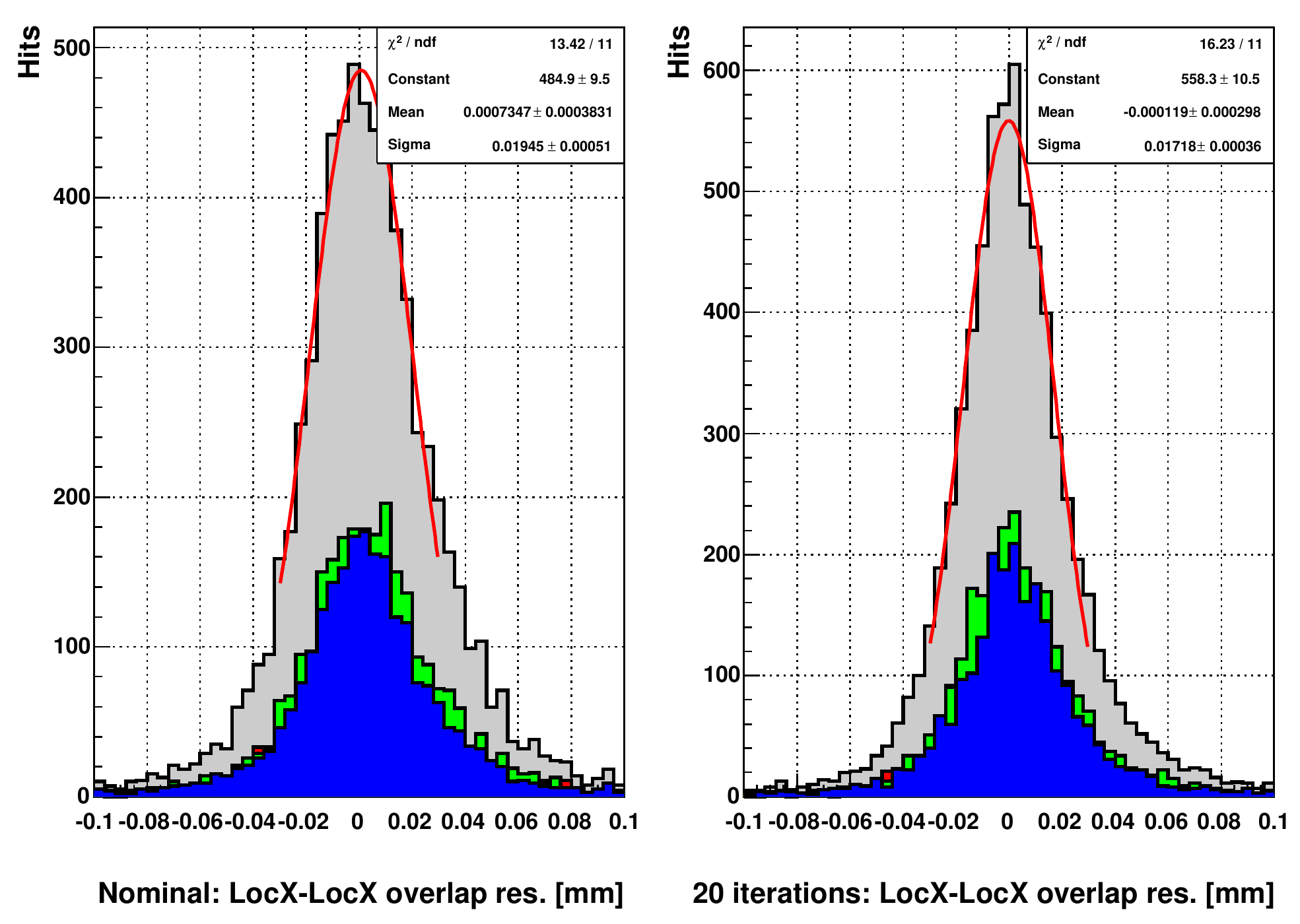}\qquad
\includegraphics[height=5cm,width=7.5cm,clip=true]{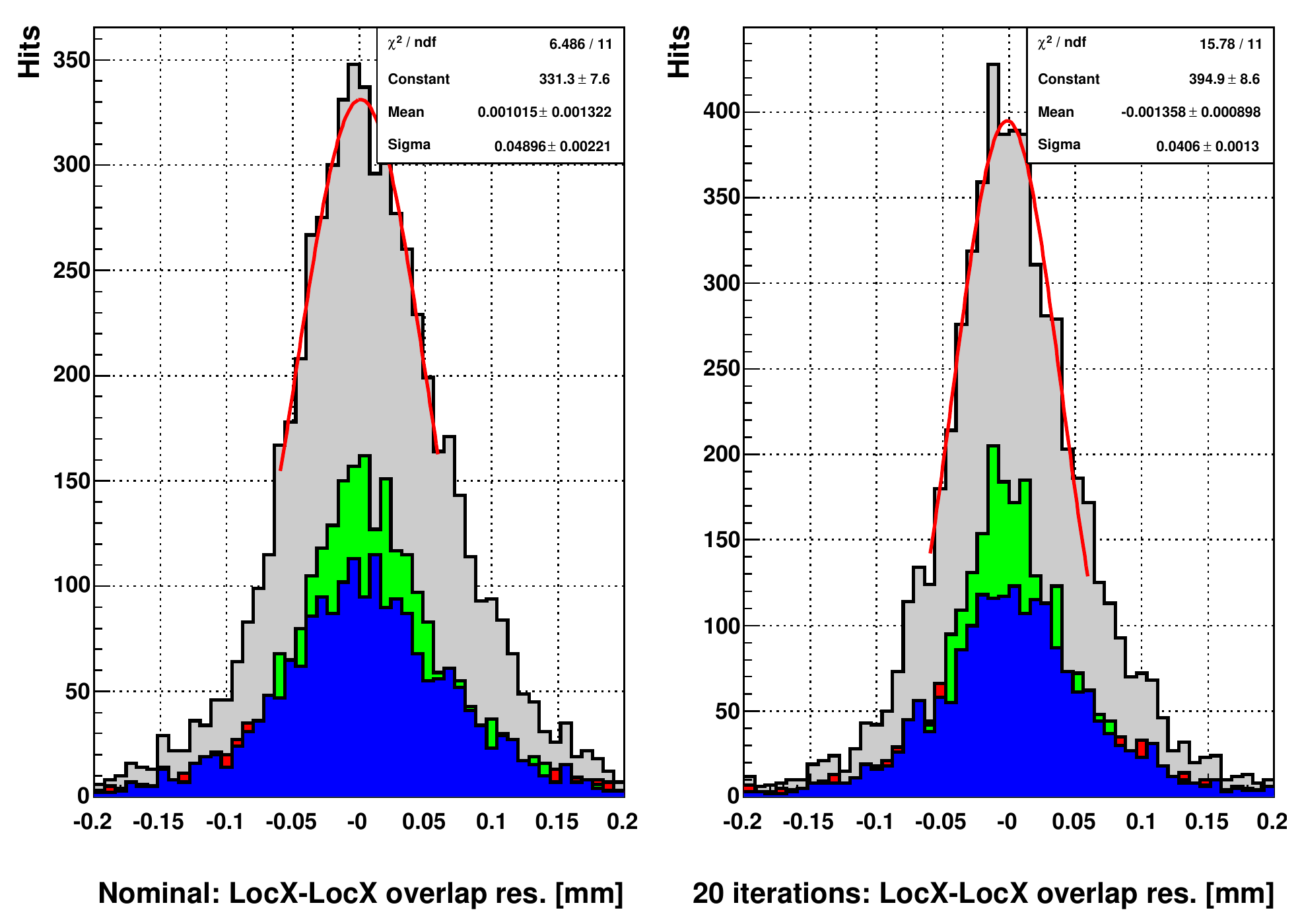}
\vspace{\cDist}
\end{center}
\caption[Overlap residual distributions $o_{xx}$ for nominal alignment and after 20 iterations]{\label{fig:ovResLocXLocX}
Overlap residual distributions $o_{xx}$ of type $x$ in local $x$ direction for the biased {\bf(left)} and the unbiased {\bf(right)} case for nominal alignment and after 20 iterations of the \RA\ algorithm. The colour-coding is as in Figure~\ref{fig:resLocX}.
}
\end{figure}

The improvement in the residual distributions discussed above is rather limited. This is mainly due to high Coulomb multiple scattering effects:
% which are immense due to the $1/p$ dependence, 
the momentum spectrum of cosmic muons starts as low as $p^{\rm min}\simeq140$\,MeV for particles triggered by scintillators  {\em3} and {\em4} because of the limited thickness of the iron block below the EC; and is practically not bound from below for particles triggered by scintillators {\em1} and {\em2} in coincidence with~{\em3}\footnote{with the exception of small regions of scintillators {\em1} and {\em2} covered by the iron block.}. For $p^{\rm fix}\equiv\langle p\rangle\simeq4\,\GeV$ at sea level~\cite{bib:pixelSR1},  back-of-the-envelope estimates yield $\rms(r_x^{\rm unbiased})\simeq45\,\mum$ assuming an interaction length per layer of $X_0\simeq3.5\%$ from~\cite{bib:atlasJINST}. Since the RMS is mostly determined by the flanks of the distribution, even higher $\rms(r_x^{\rm unbiased})$ are expected due to the low $p$ end of the spectrum, consistent with the observation in Figure~\ref{fig:resLocX}.

Given the sizable Coulomb multiple scattering effects, it is favourable to consider overlap residuals, since the distance between odd and even side modules of the same layer and thus the multiple scattering effects are smaller. %They are assigned a higher weight with the \RA\ algorithm compared to regular residuals, being 20 for the biased and 10 for the unbiased case. 
The $o_{xx}$ overlap residual distribution is shown in Figure~\ref{fig:ovResLocXLocX} for the biased and the unbiased case on the left or right hand side, respectively. For the same reasons as detailed in the previous paragraph, the unbiased overlap residuals are approximately two times wider than the biased ones. The improvement in the RMS of the overlap residuals is $\sim$12\% for the biased and $\sim$17\% for the unbiased case, which can be attributed to dominating multiple scattering effects, limited statistics and the high mounting precision of the pixel end-cap~A modules.

\section{Determination of the Layer Thickness}
\begin{figure}
\begin{center}
\includegraphics[width=7.5cm,clip=true]{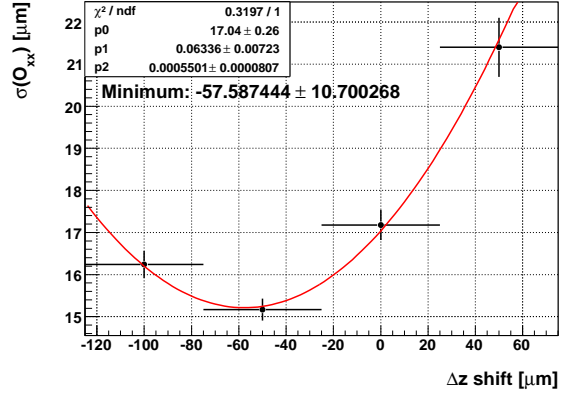}
\vspace{\cDist}
\end{center}
\caption[The $\sigma$-parameter of a Gaussian fit to the $o_{xx}$ residual distribution]{\label{fig:alignLocZ}
The $\sigma$-parameter of a Gaussian fit to the $o_{xx}$ residual distribution for different $\Delta z$ between odd and even side modules of a given layer.
}
\end{figure}

%In each pixel end-cap disk layer, the odd and even modules are mounted on different sides of the carbon-carbon laminate support strucutre, resulting in a distance $\Delta z$ between them. 
The distance $\Delta z$ between odd and even side modules has been measured with the \RA\ algorithm using the fact that (overlap) residuals are optimised for the correct $\Delta z$. As a figure of merit, the $\sigma$-parameter of a Gaussian fit to the $o_{xx}$ residual distribution and its error has been used. Residual and overlap residual distributions were obtained for the different $\Delta z$ values by changing the corresponding entries in the alignment data base and reconstructing tracks with this geometry. The resulting distribution of $\sigma(o_{xx})$ versus $\Delta z$  is shown in Figure~\ref{fig:alignLocZ} averaged over all three layers. From a parabolic fit, a mean correction to the nominal $\Delta z$ distance of 4200\,\mum\ has been derived:
 \[\delta(\Delta z)=-57.9\pm10.7\,\mu{\rm m}\]
This translates into $\Delta z=4257.9\pm10.7\,\mu{\rm m}$ between odd and even side modules of a given layer, averaged over all 3 layers.

\section{Derivation of $c_x,\,c_y$ Alignment Constants}
\begin{figure}
\begin{center}
\includegraphics[height=5cm,width=5.0cm,clip=true]{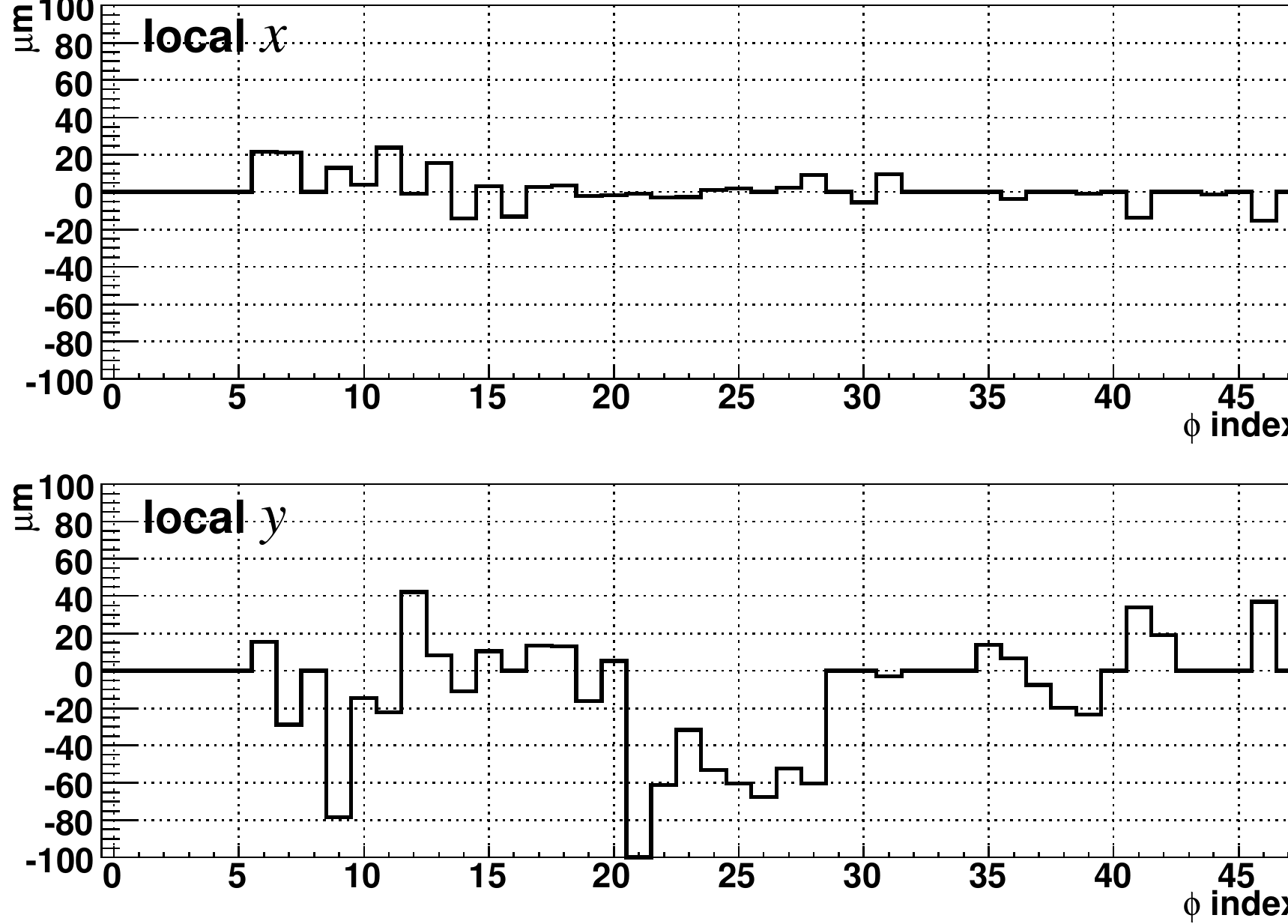}~\quad
\includegraphics[height=5cm,width=5.0cm,clip=true]{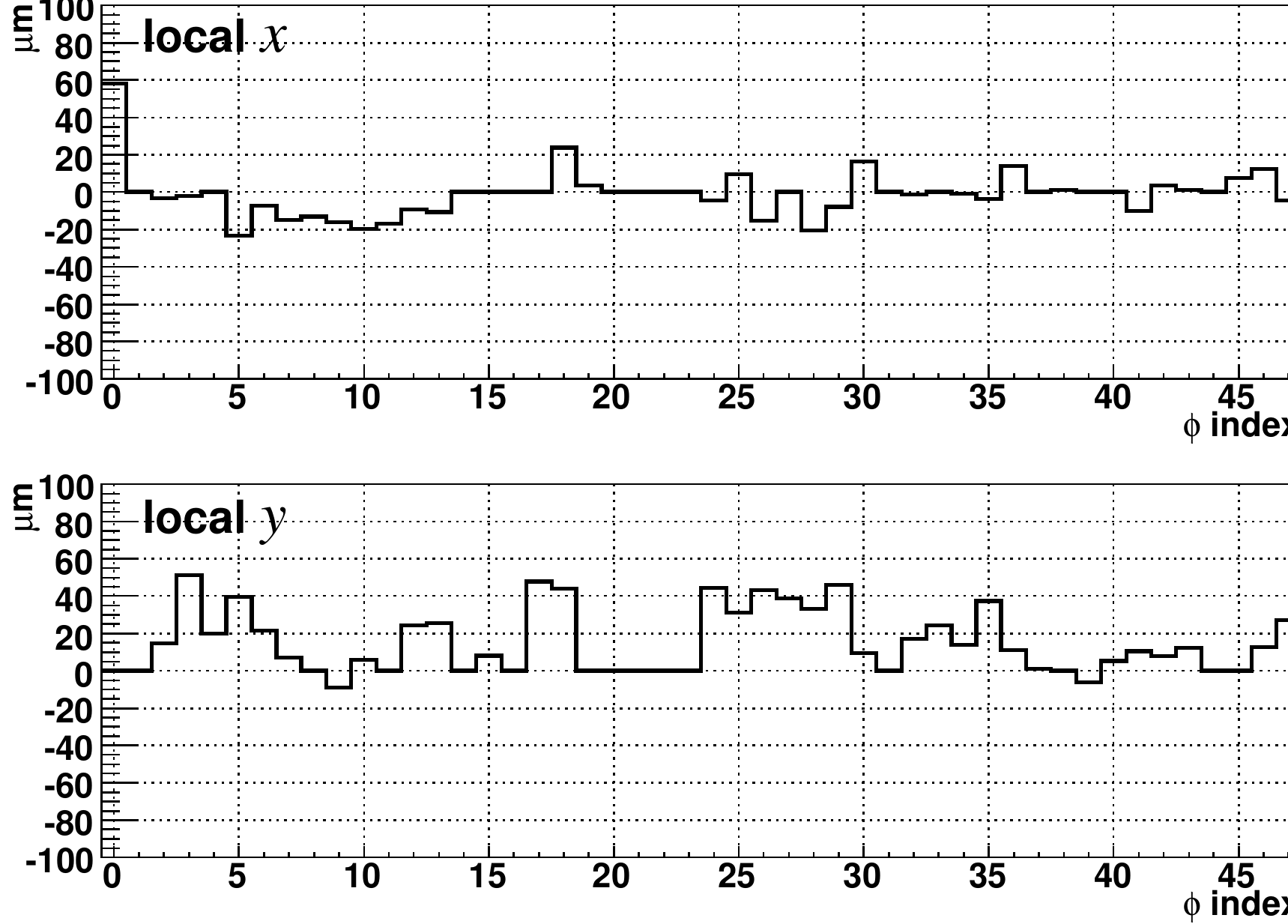}~\quad
\includegraphics[height=5cm,width=5.0cm,clip=true]{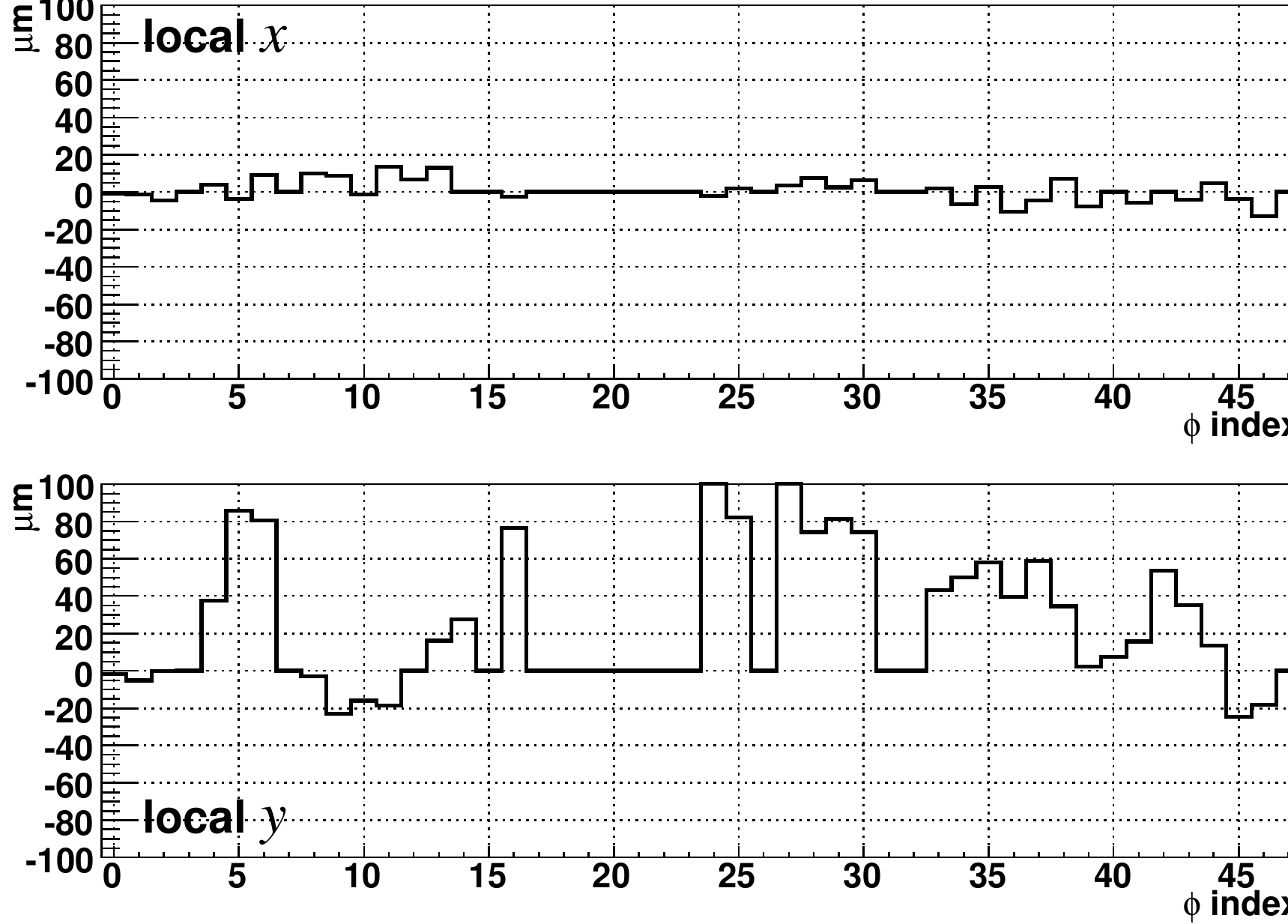}
\vspace{\cDist}
\end{center}
\caption[Alignment constants in local $x,y$ after 20 iterations with respect to nominal geometry]{\label{fig:alignConstL0}
Alignment constants in local $x,y$ after 20 iterations for layer 0 {\bf (left)}, layer 1 {\bf (middle)} and layer 2 {\bf (right)} of the pixel end-cap~A subdetector with respect to nominal geometry, derived with {\bf biased} residuals and $\Delta z=4250$\,\mum{} between odd and even modules for a given layer.
}
\end{figure}

Two sets of alignment constants for the pixel end-cap~A modules have been derived with respect to the nominal module positions in the local $x,y$ coordinates, using {\em biased} and {\em unbiased} residuals over 20 iterations, and a mean $\Delta z$ distance of 4250\,\mum{} between odd and even modules in each layer. %The details of the calculation can be found in~\cite{bib:noteRA}. 
The constants found with biased residuals are visualised in Figure~\ref{fig:alignConstL0} for layer 0, 1, and 2.

The average magnitude of corrections for the local $x$ direction is $\mathscr O(10\mu\rm m)$, with good agreement between the two sets of constants. For the local $y$ direction, the corrections are $\mathscr O(50\mu\rm m)$, and the agreement between the two sets of alignment constants is less pronounced than for local $x$. This can be attributed to the fact, that the  $\rmean y$ measurement is about one order of magnitude less precise than $\rmean x$, and so are the derived $c_y$ corrections. 

The mean correlation of the alignment constants in all 3 layers between the $\Delta z=4250$\,\mum{} and the nominal $\Delta z=4200$\,\mum{} geometry is $\langle\rho\rangle\simeq92$\% in local $x$, and $\langle\rho\rangle\simeq73$\% in local $y$ (first row of Table~\ref{tab:correlation}). Similary, the correlation between results obtained using biased residuals and $\Delta z=4250$\,\mum{} with ones using unbiased residuals and $\Delta z=4200$\,\mum{} is $\langle\rho\rangle\simeq77$\%/51\% in local $x/y$ (second row of Table~\ref{tab:correlation}).

The conclusion is, that with the limited track statistics available and the low momentum scale of the tracks, the local $y$ precision of track-based alignment is lower than the optical survey. For the local $x$ direction, the \RA\ algorithm yields numerically stable and reliable results.

\begin{figure}
\begin{center}
\includegraphics[height=5cm,width=5.4cm,clip=true]{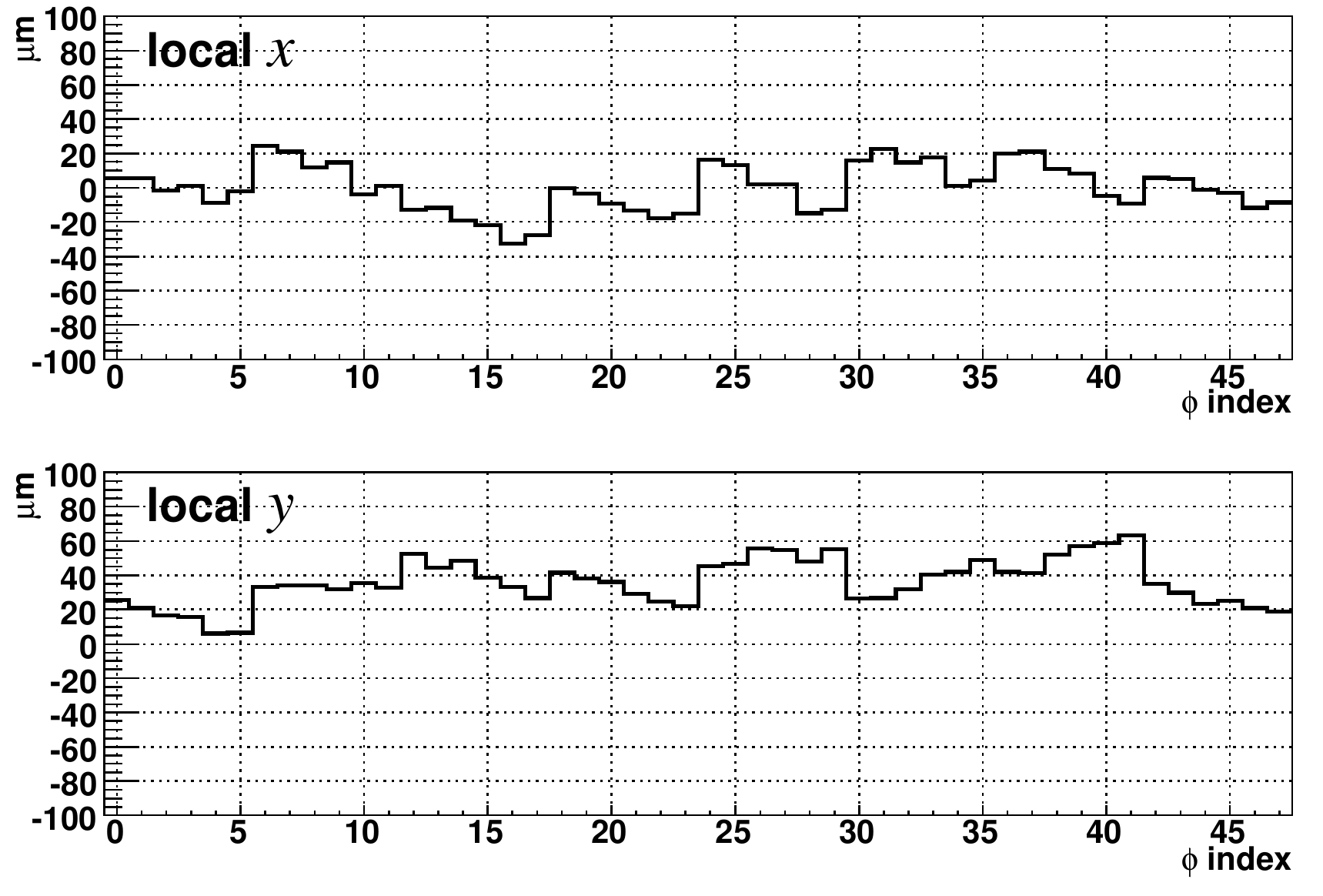}\qquad
\includegraphics[height=5cm,width=7cm,clip=true]{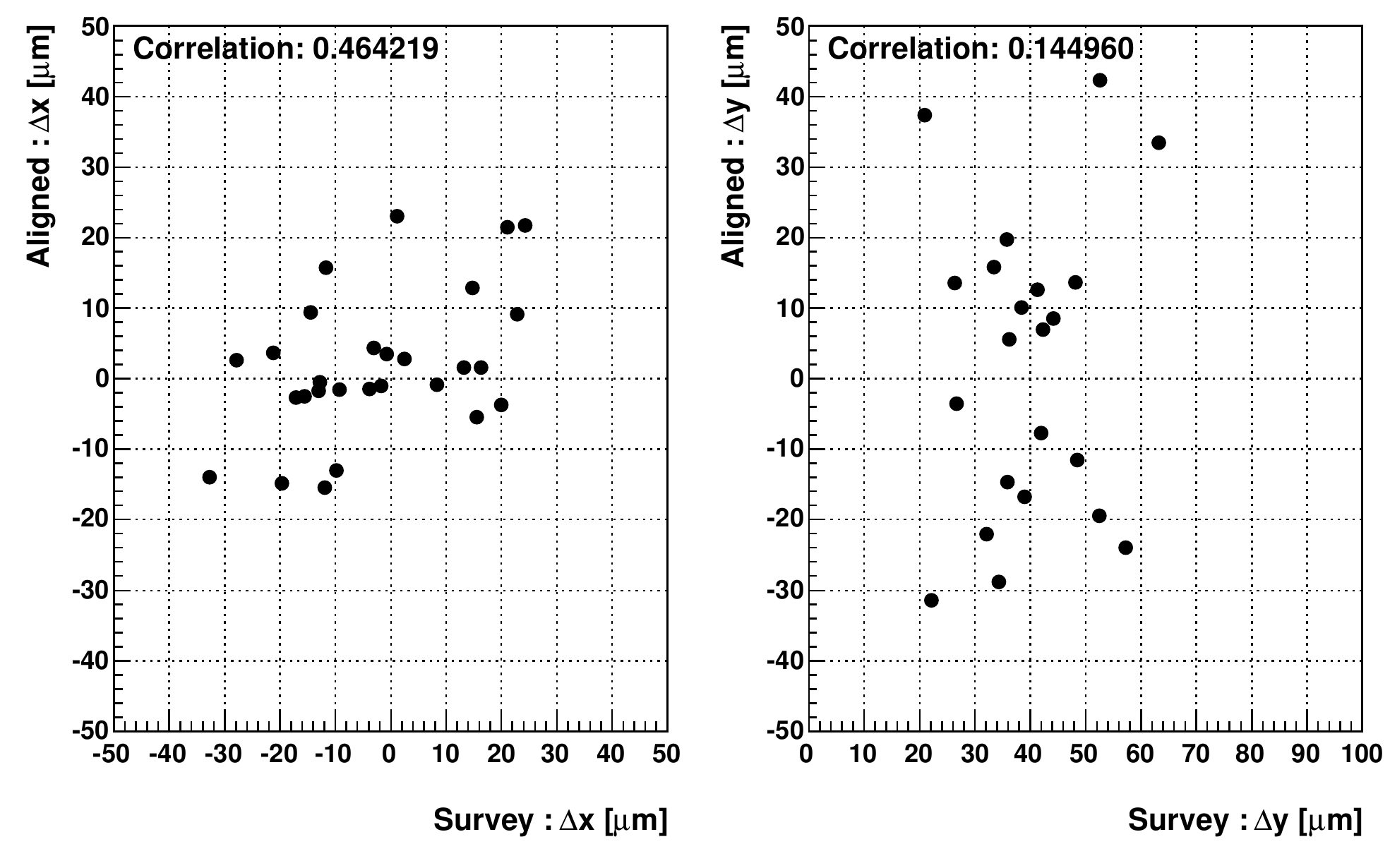}
\vspace{\cDist}
\end{center}
\caption[Alignment constants in local $x,y$ for layer 0 of the pixel end-cap~A subdetector as obtained with the survey]{\label{fig:surveyL0}
Alignment constants in local $x,y$ for {\bf layer 0} of the pixel end-cap~A subdetector as obtained with the {\bf survey} without any track-based alignment {\bf(left)}. The correlation between alignment constants derived with the \RA\ for $\Delta z=4250$\,\mum{} and the survey{\bf~(right)}.
}
\end{figure}

\begin{figure}
\begin{center}
\includegraphics[height=5cm,width=5.4cm,clip=true]{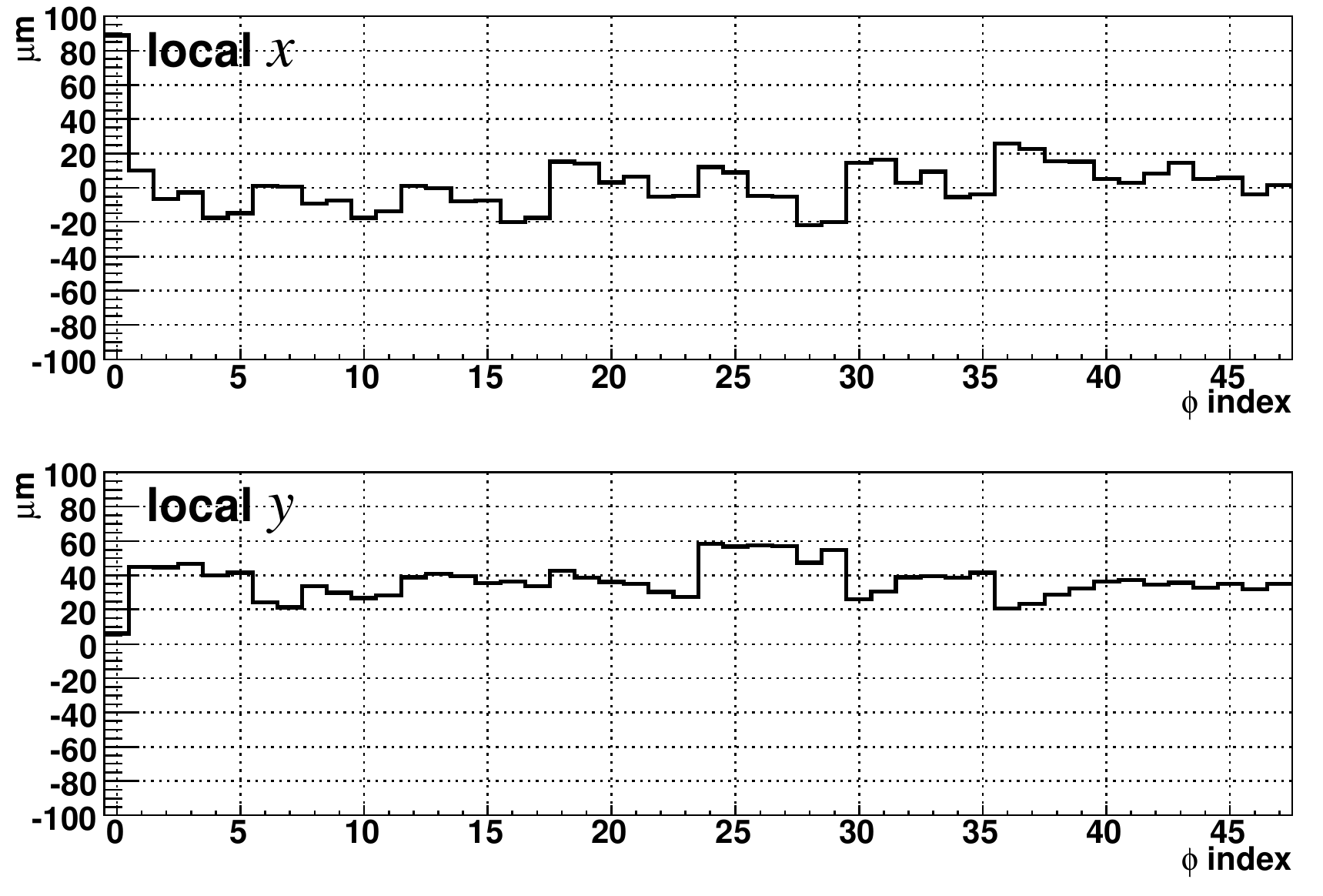}\qquad
\includegraphics[height=5cm,width=7cm,clip=true]{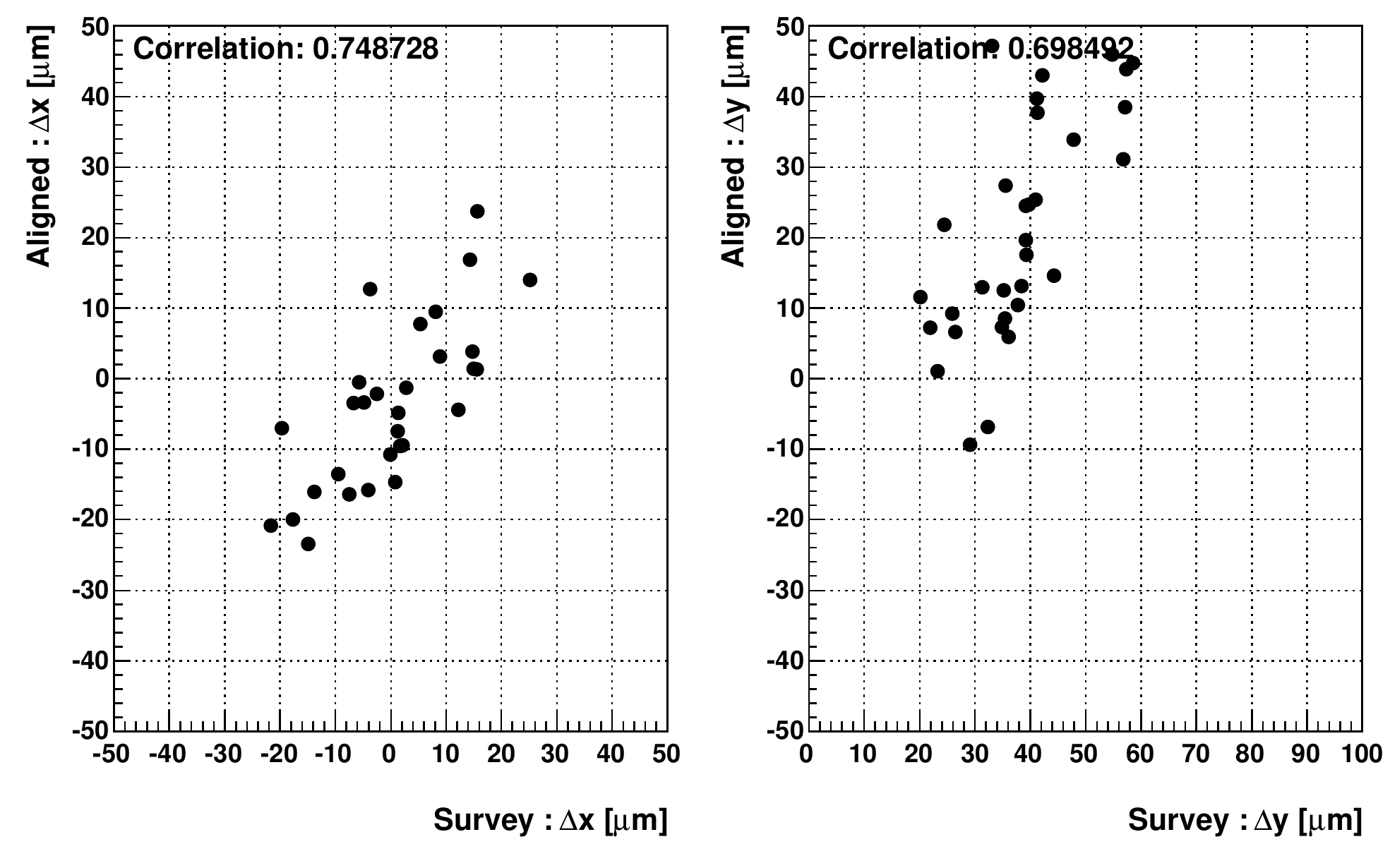}
\vspace{\cDist}
\end{center}
\caption[Alignment constants in local $x,y$ for layer 1 of the pixel end-cap~A subdetector as obtained with the survey]{\label{fig:surveyL1}
Alignment constants in local $x,y$ for {\bf layer 1} of the pixel end-cap~A subdetector as obtained with the {\bf survey} without any track-based alignment {\bf(left)}. The correlation between alignment constants derived with the \RA\ for $\Delta z=4250$\,\mum{} and the survey{\bf~(right)}.
}
\end{figure}

\begin{figure}
\begin{center}
\includegraphics[height=5cm,width=5.4cm,clip=true]{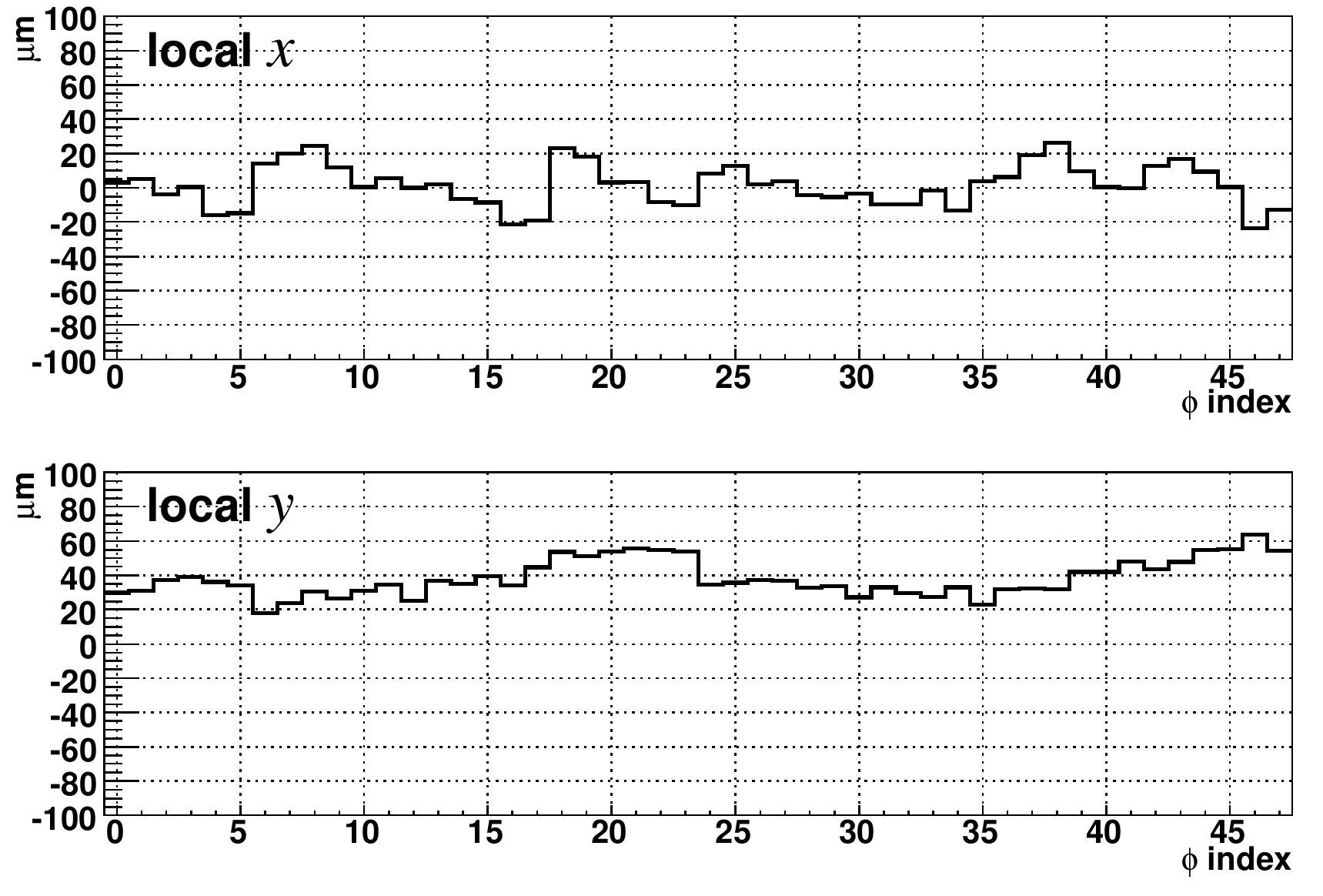}\qquad
\includegraphics[height=5cm,width=7cm,clip=true]{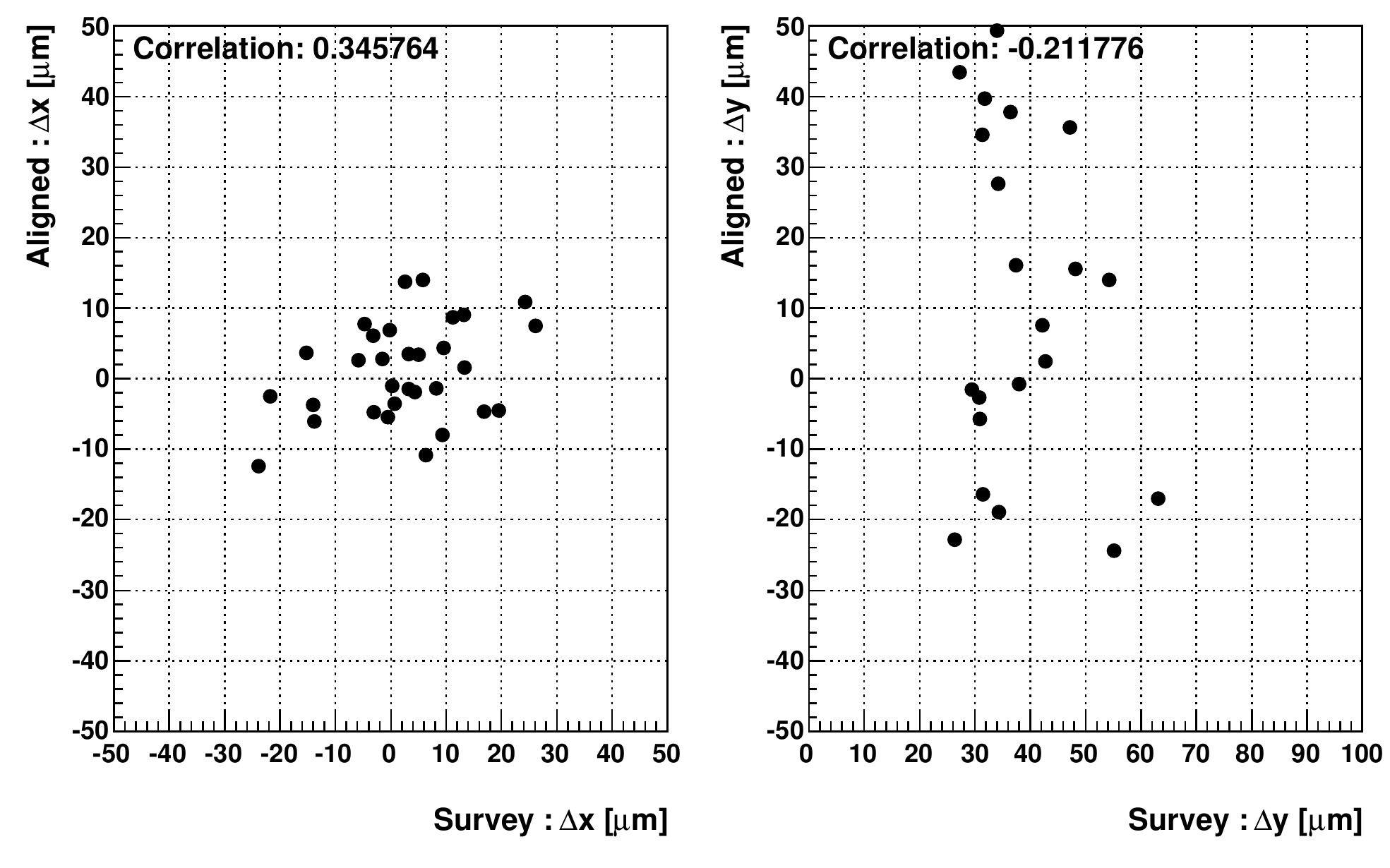}
\vspace{\cDist}
\end{center}
\caption[Alignment constants in local $x,y$ for layer 2 of the pixel end-cap~A subdetector as obtained with the survey]{\label{fig:surveyL2}
Alignment constants in local $x,y$ for {\bf layer 2} of the pixel end-cap~A subdetector as obtained with the {\bf survey} without any track-based alignment {\bf(left)}. The correlation between alignment constants derived with the \RA\ for $\Delta z=4250$\,\mum{} and the survey{\bf~(right)}.
}
\end{figure}

\begin{table}
\begin{center}
\begin{small}
\begin{tabular}{ll|rr|rr|rr}
\hline
\multicolumn{2}{c|}{\bf Alignment Set}& \multicolumn{2}{c|}{\bf Layer~0}& \multicolumn{2}{c|}{\bf Layer~1}& \multicolumn{2}{c}{\bf Layer~2}\\
{\bf set 1} & {\bf set 2} & $x$ & $y$ & $x$ & $y$ & $x$ & $y$ \\
\hline\hline
Biased $\delta z$& Biased
					& 0.92 & 0.67 & 0.90 & 0.93 & 0.93 & 0.59 \\
Biased $\delta z$& Unbiased
					& 0.84 & 0.46 & 0.72 & 0.63 & 0.76 & 0.45 \\
Biased $\delta z$& Survey
					& 0.46 & 0.14 & 0.75 & 0.70 & 0.35 & -0.21 \\
% Biased		& Unbiased		& 0.86 & 0.34 & 0.90 & 0.70 & 0.76 & 0.41 \\
Biased		& Survey		& 0.36 & 0.27 & 0.64 & 0.65 & 0.25 & -0.01 \\
% Unbiased	& Survey		& 0.48 & 0.38 & 0.49 & 0.32 & 0.22 & 0.24 \\
Biased from Survey & Survey		& 0.90 & 0.27 & 0.76 & 0.64 & 0.92 & -0.01 \\
% Biased from Survey & Biased		& 0.52 & 0.03 & 0.68 & 0.65 & 0.46 & -0.06 \\
Biased from Survey & Biased $\delta z$	& 0.66 & 0.28 & 0.84 & 0.65 & 0.58 & 0.00 \\
Biased w. $\gamma$ from Survey & Survey	& 0.40 & 0.22 & 0.63 & 0.65 & 0.21 & -0.24 \\
Biased w. $\gamma$ from Survey & Biased	& 0.97 & 0.94 & 0.99 & 0.96 & 0.97 & 0.94 \\
\hline
\end{tabular}
\end{small}
\end{center}
\caption[Correlation between various sets of alignment constants in local $x$ and local $y$]{\small\label{tab:correlation}
Correlation between various sets of alignment constants in local $x$ and local $y$. The addendum $\delta z$ indicates that $\Delta z=4250$\,\mum{} geometry was used. The last two rows are explained in the text.
}\end{table}

\subsection{Comparison of the Alignment Constants Derived using \RA\ with the Survey}

As already mentioned in the preceding Subsection, an optical survey of the pixel end-cap~A modules has been made. The alignment constants determined with the survey and without any track-based alignment are presented for layer 0, 1, and 2 on the left hand side of Figure~\ref{fig:surveyL0}, Figure~\ref{fig:surveyL1}, and Figure~\ref{fig:surveyL2}. The alignment constants obtained with \RA\ for $\Delta z=4250$\,\mum{} have been compared with the results of the survey, as shown on the right hand side of the respective figures. The agreement between the results obtained using \RA\ and the optical survey is limited for local $x$ ($\langle\rho\rangle\simeq52$\%, biased). The situation in local $y$ is exacerbated ($\langle\rho\rangle\simeq21$\%, biased). Similarly, for $\Delta z=4200$\,\mum{} $\langle\rho\rangle\simeq42$\%/30\% were obtained. The correlation of alignment constants and the survey is detailed in the third and fourth row of Table~\ref{tab:correlation} by layers. Notably, the by far best correlation is achieved for the middle disk layer~1. This indicates that the quality of alignment constants in the outer disks is mostly limited by the quality of the track fit, which is poorly constrained there.

There is a significant positive offset of alignment constants obtained with the survey with respect to \RA\ in local $y$. This is due to the fact that the pixel end-cap~A rings have $\mathscr O(30\mu\rm m)$ larger radii than nominal geometry~\cite{bib:pixelSR1}. With the circle of operating modules on a given layer not closed, such correlated movement of the modules is beyond detection with the \RA\ algorithm.

Because the agreement between track based alignment and the optical survey is limited, several cross-checks have been made. The \RA\ algorithm has been employed to align the subdetector over 20 iterations starting from the survey, rather than nominal geometry. The resulting alignment constants in local $x$ and the \RA\ results obtained starting from the $\Delta z=4250$\,\mum{} geometry are much alike, with a mean correlation of $\langle\rho\rangle\simeq69$\%. The situation in local $y$ is similar, but less pronounced with a mean correlation of $\langle\rho\rangle\simeq31$\%. This is shown in fifth row of Table~\ref{tab:correlation}. The correlation between the survey and \RA\ starting from the survey is 86\%/30\% (local $x/y$, sixth row of Table~\ref{tab:correlation}). To cross-check which role is played by the alignment for local $\gamma$ rotations in this comparison, a set of alignment constants has been derived over 20 iterations using the nominal geometry for all 6 degrees of freedom per module but the rotations around local $z$ axis. For these, survey results have been used. Again, the resulting alignment constants are very similar to the \RA\ results shown above, with a correlation coefficient of $\langle\rho\rangle\simeq98$\% in local $x$, and $\langle\rho\rangle\simeq95$\% in local $y$ (last row of Table~\ref{tab:correlation}). All cross-checks have been made with biased residuals.

\section{Conclusion}
End-cap~A of the pixel detector was successfully aligned with the \RA\ algorithm using tracks from cosmic ray particles, collected on the surface in a commissioning run in December 2006. Due to the limited track statistics availabe and the low $p>140$\,MeV momentum cut-off, the local~$y$ precision is lower than the precision of the optical survey. For the local~$x$ direction, numerically stable and reliable results were obtained, which show strong correlation with the optical survey. The disk layer thickness was found to be $\Delta z=4,257.9\pm10.7\,\mum$ averaged over all disks.

%% file: M8plus/M8plus.tex
The ATLAS Detector has been collecting cosmic ray data between September and the beginning of December 2008 in the so-called M8+\glossary{name=M8+,description=Global ATLAS run collecting cosmic rays in the cavern cf. Chapter~\ref{chp:m8plus}} run~\cite{bib:m8plus}. These data have proven highly valuable to gain a better understanding of the detector in view of collision data, which is expected to come later this year. The entire Inner Detector has participated in a substantial fraction of these runs, and its data have been used to align the ATLAS silicon tracker with the \RA\ algorithm. This will be the subject of this chapter. It is organised as follows: firstly, the dataset used for the alignment is described. Secondly, the basic hit and track selection is introduced. Thirdly, the alignment procedure is detailed, split by alignment levels. After that, the alignment results are presented, and compared to the findings of the \GX\ and \LX\ algorithms. Lastly, an anomaly unresolved at the time of writing -- the discrepancy of residuals for $B$-field off and on data -- will be sketched.\\
In this chapter, the term ``$B$-field'' refers to the solenoid magnetic field enclosing the Inner Detector rather than the toroidal field of the Muon Spectrometer, unless explicitly stated otherwise.
A document on inner detector alignment with track-based algorithms utilising M8+ cosmic ray data is in preparation~\cite{bib:m8plusNote}.

%% file: M8plus/Dataset.tex
The cosmic ray dataset collected by the ATLAS Inner Detector during M8+ is almost equally split between $B$-field on and off. Table~\ref{tab:statsM8plus}~\cite{bib:approvedPlotsID} summarises the number of tracks collected by the ID in M8+ and reprocessessed in December~2008. It is split in three categories: tracks with $\geq$1 pixel, tracks with $\geq$1 SCT hit, and all tracks. A large fraction of the former two can be used for the alignment of the pixel and SCT detector, respectively. In total, more than 400k tracks with pixel hits and 2M tracks with SCT hits were collected and used as input for alignment.

\begin{table}
\small
\begin{center}
\begin{tabular}{lrrr}
\hline
 & $B$-field {\bf on} & $B$-field {\bf off} & {\bf total}
\\\hline\hline
{ Tracks with $\geq$1 pixel hit}~~~ & 230,000 & 190,000 & 420,000 \\
{ Tracks with $\geq$1 SCT hit} & 1,150,000 & 880,000 & 2,030,000 \\
{ All tracks}                    & 4,940,000 & 2,670,000 & 7,610,000 \\
\hline
\end{tabular}
\caption[Number of cosmic ray tracks collected by the ATLAS ID in M8+]{\label{tab:statsM8plus}
The number of cosmic ray tracks collected by the ATLAS Inner Detector in M8+ and reprocessessed in December split by categories~\cite{bib:approvedPlotsID}.
}
\end{center}
\end{table}

\begin{figure}
\begin{center}
\includegraphics[height=10.5cm,clip=true,angle=90]{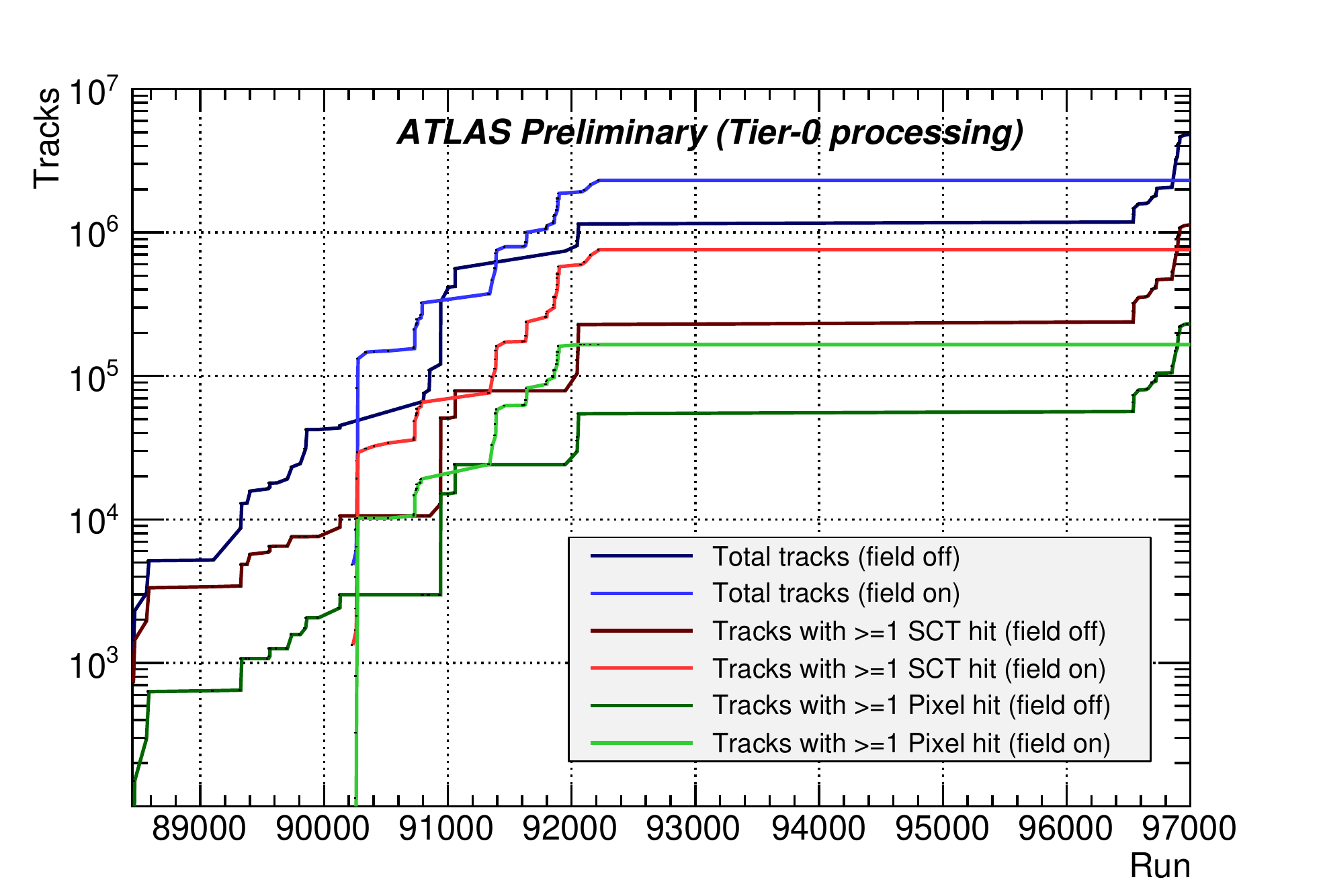}%\quad
\vspace{\cDist}
\end{center}
\caption[Number of cosmic ray tracks vs. run number collected by the ATLAS ID in M8+]{\label{fig:statsM8plus}
The number of cosmic ray tracks versus run number collected by the ATLAS ID in M8+ split by categories~\cite{bib:approvedPlotsID}. See text for details.%The steep increase in the number of collected tracks up to run \# $\sim$92k demonstrates a steep learning curve in the understanding of the detector. The flat section between run \#s $\sim$92k and $\sim$96.5 is due to the ID cooling plant failure. Runs beyond $\sim$96.5 show an almost doubled track rate per day due to the introduction of the fast TRT-OR L1 trigger.
}
\end{figure}

The number of tracks collected in M8+ versus run number is shown in Figure~\ref{fig:statsM8plus}. The dramatic increase up to run number 92082 demonstrates a steep learning curve in the understanding of the detector and of the ID geometry. The alignment constants were updated twice during data taking: at run number 89740, and 91396. Both increased the tracking efficiency of the track-based L2 triggers. The second set of alignment constants is considered to be good enough within the limitations of the L2 trigger, and was not updated any more. The flat section between run numbers 92082 and 96538 is due to an ID cooling plant failure. Runs beyond run number 96538 show an almost doubled track rate per day due to the introduction of the fast TRT-OR trigger at L1~\cite{bib:fastTRTOR}. It has both a high efficiency and purity~\cite{bib:triggerNote} combined with a rate of about 10\,Hz, which can be written to tape without any additional L2 trigger selection. The L1 trigger with the highest track-to-tape contribution besides the fast TRT-OR trigger is the RPC-based muon barrel trigger~\cite{bib:muonTriggerTwiki,bib:muonTrigger}, which was used in a large fraction of the M8+ run. However, its rate of $\sim$400\,Hz~\cite{bib:triggerNote} is too high to be written to tape directly. This necessitated the introduction of track-based trigger algorithms at L2 -- {\tt IDScan}, {\tt SiTrack} and {\tt TrigTRTSegFinder} -- which constitute the {\tt IDCosmic} stream. They are described in~\cite{bib:triggerNote}.

%\nopagebreak[5]

M8+ was not the first occasion for the ATLAS ID to take cosmic ray data in situ: the M6~\cite{bib:m6} cosmic ray runs were highly important for the alignment. In M6, the SCT barrel and the TRT were the only operated parts of the ID, and the number of events available is only a per-mille fraction of what was collected in M8+. Nevertheless, these data were sufficient for a coarse pre-alignment of the TRT barrel with respect to the SCT barrel, and of the SCT barrel at L2. This was highly important for a reasonable efficiency of the track-based L2 triggers at the {\it beginning} of M8+. It should be stressed that the alignment constants presented here were derived starting from nominal geometry rather than the one found in M6 in order to avoid any biases.
%$\textperthousand$

The full list of M8+ runs used for the alignment of the silicon tracker with the \RA\ algorithm is given in Table~\ref{tab:runList}. Additional information on runs taken until mid-November can be found in~\cite{bib:m8plus}, and the data quality information can be obtained in the ATLAS Data Quality database~\cite{bib:dqdb}. 
The streams {\tt TGCwBeam}, {\tt RPCwBeam}, {\tt L1Calo}, {\tt CosmicMuon}, and {\tt IDCosmic} were used, with a major contribution from the latter~\cite{bib:m8plus}.%, which contains all events with at least one track reconstructed in the ID at L2~\cite{bib:m8plus}.

\begin{table}
\small
\begin{center}
\begin{tabular}{l|rrrrrrrrrr}
\hline
$B$-field {\bf on}:
& 90260	& 90262	& 90264	& 90270	& 90272	& 90275	& 90345	& 90413	& 90525	& 90633\\
& 90731	& 90732	& 90733	& 91338	& 91359	& 91361	& 91387	& 91389	& 91390	& 91391\\
& 91398	& 91399	& 91400	& 91464	& 91561	& 91613	& 91627	& 91636	& 91639	& 91790\\
& 91799	& 91800	& 91801	& 91802	& 91803	& 91808	& 91828	& 91860	& 91861	& 91862\\
& 91884	& 91885	& 91888	& 91890	& 91891	& 91893	& 91897	& 91900	& 92065	& 92069\\
& 92072	& 92074	& 92077	& 92078	& 92079	& 92080	& 92081	& 92082	& 92092	& 92095\\
& 92098	& 92099	& 92100	& 92107	& 92112	& 92134	& 92157	& 92159	& 92160	& 92223\\
& 92226\\
\hline
$B$-field {\bf off}:
& 92057	& 92058	& 92059	& 92063	& 96516	& 96527	& 96535	& 96538	& 96541	& 96542\\
& 96543	& 96544	& 96582	& 96644	& 96659	& 96688	& 96696	& 96716	& 96717	& 96718\\
& 96721	& 96722	& 96732	& 96851	& 96858	& 96884	& 96895	& 96903	& 96906	& 96913\\
& 96916	& 96925	& 96929	& 96950	& 96973	& 96982	& 96988\\
\hline
\end{tabular}
\caption[List of runs used for the alignment of the ATLAS silicon tracker by the \RA\ algorithm in M8+]{\label{tab:runList}
List of runs used for the alignment of the ATLAS silicon tracker by the \RA\ algorithm in M8+ split in $B$-field on and off categories.
}
\end{center}
\end{table}

\begin{figure}
%\vspace{-0.7cm}
\begin{picture}(15.5,8.0)(0.1,0.1) \large
\put(-0.6,0.4)	{\includegraphics[height=7.0cm,clip=true]{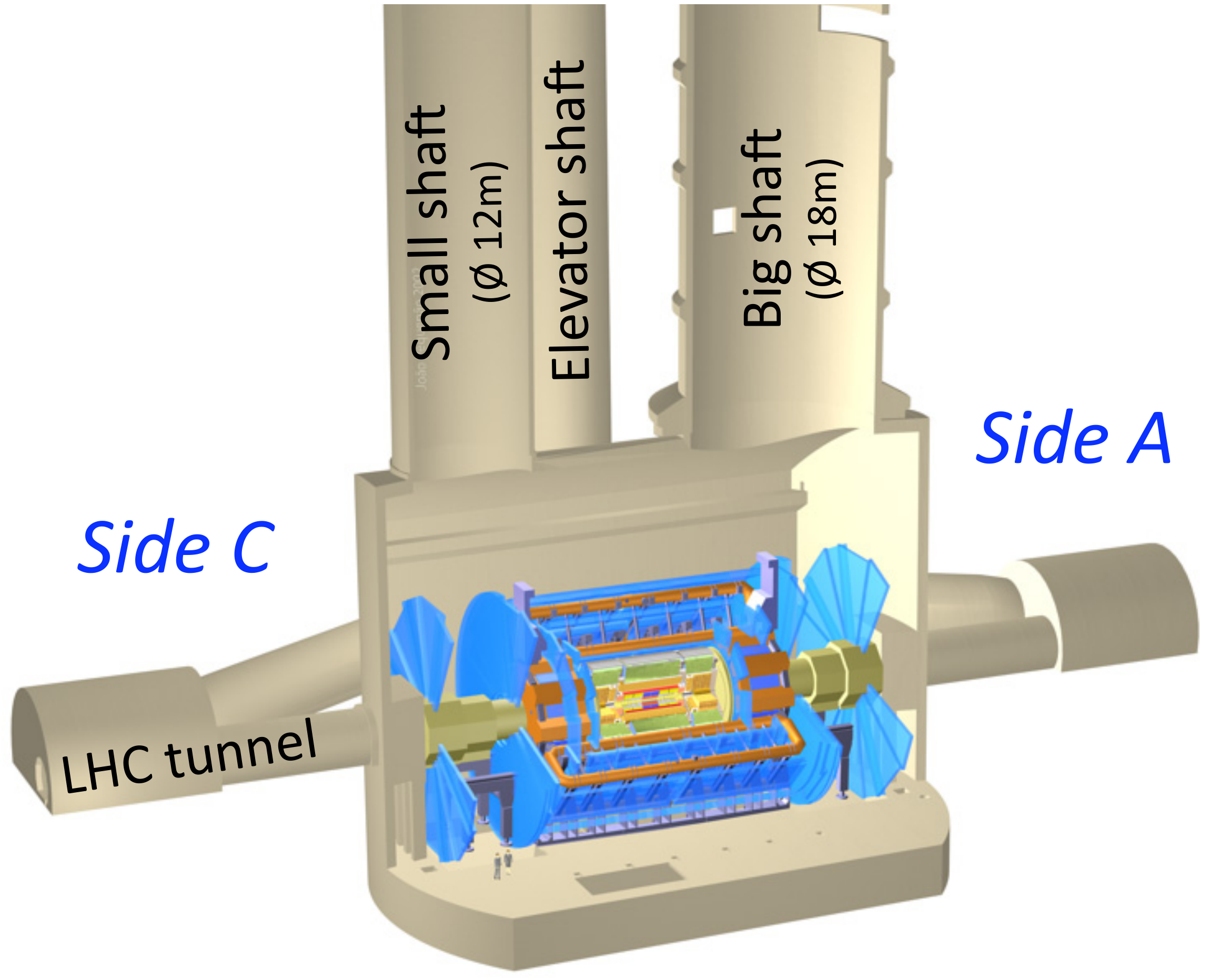}}
\put(8.0,0.0)	{\includegraphics[height=6.8cm,height=7.5cm,clip=true]{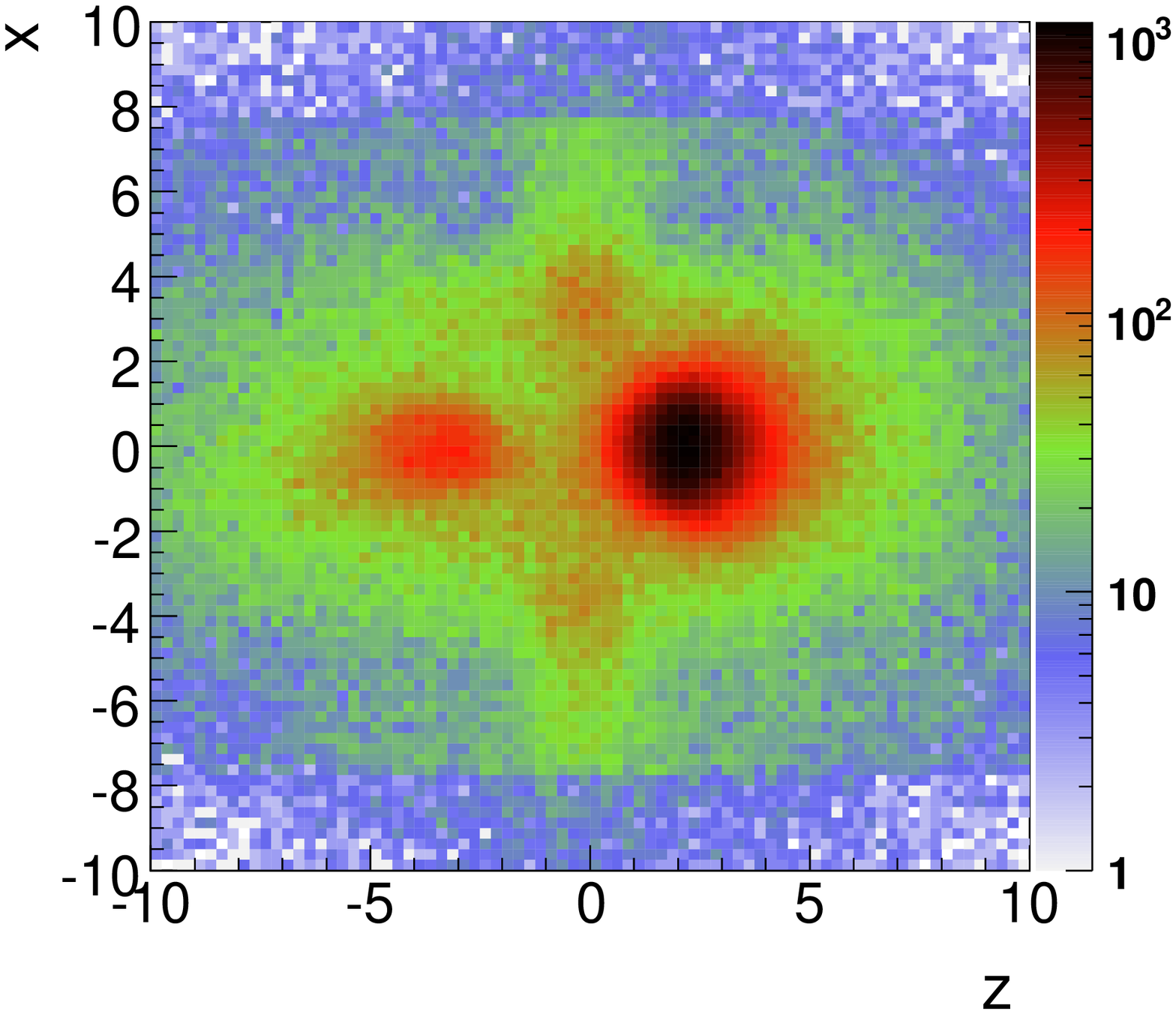}}
\put(0.,6.8){\sf\textcolor{red}{\textbf{(a)}}}
\put(9.3,6.8){\sf\textcolor{red}{\textbf{(b)}}}
\normalsize
\end{picture}
\vspace{-0.3cm}
\caption[ATLAS detector in the pit and topological incidence of ID tracks in M8+]{\label{fig:trackDirM8plus}
The ATLAS detector in the pit{\bf~(a)}: the excavated area of the pit, the two access shafts, and the LHC tunnel are indicated in light brown. This defines the predominant incidence of cosmic ray tracks: the intersection points of ID tracks extrapolated to a plane 10.5\,m above the ATLAS centre, i.e. just above the MS, is shown in{\bf~(b)}~\cite{bib:privateKorn}. $X$ and $Z$ are given in m. The two spots at $X=0$, $Z=-3\,\rm{m},\!+3\,\rm{m}$ correspond to the smaller and bigger access shaft in (a). The two spots at $X=-4\,\rm{m},\!+4\,\rm{m}$, $Z=0$ correspond to the elevator shafts PM15 and PX15 (not shown in (a)). See text for details. Figure (b) courtesy A.~Korn.
}
\end{figure}%\nopagebreak[5]

Figure~\ref{fig:trackDirM8plus}~(a) shows a sketch of the ATLAS detector in the cavern together with the access shafts and the LHC tunnel. This geometry defines the incidence of cosmic rays: instead of a dependence $\propto\cos^2\!\theta$ of the flux on the incidence angle $\theta$ at the surface, a predominantly vertical incidence is observed in the ATLAS cavern, 81\,m under the surface (measured from the centre of ATLAS). More quantitatively, Figure~\ref{fig:trackDirM8plus}~(b)~\cite{bib:privateKorn} shows the intersection points of ID tracks extrapolated to a plane 10.5\,m above the ATLAS centre, id est just above the muon spectrometer, for two representative M8+ runs: 96884 and 96903. The two spots with increased intersection point density at $X=0$, $Z=-3\,\rm{m},\!+3\,\rm{m}$ correspond to the smaller and bigger access shaft in (a), while the two spots at $X=-4\,\rm{m},\!+4\,\rm{m}$, $Z=0$ are a clear indication of the elevator shafts PM15 and PX15. The boundaries at $X=\pm7.75$\,m are believed to be due to the projection of the shaft positions onto the ATLAS acceptance, which is supported by MC simulations. The figure was produced using ID tracks with more than one hit in the silicon tracker in order to reject TRT-only tracks, whose $z_0$~track parameter is poorly determined. The {\tt IDCosmics} stream was used and the TRT fast-OR L1 trigger signal required to avoid any biases from the RPCs.

The geometry of the ATLAS cavern and the resulting predominant incidence direction of cosmic ray particles has a shaping impact on the illumination of the modules of the silicon tracker, and thus on the quality of their alignment: the modules at the top ($\Phi\simeq\xOverY\pi2$) and the bottom ($\Phi\simeq-\xOverY\pi2$) of the silicon tracker barrel collected the most hits\footnote{Here and in the remainder of the chapter, the term ``hit'' refers to hits associated with tracks, also called as hits-on-track, unless stated otherwise.} in M8+, whereas the modules at its sides ($\Phi\simeq0,\pi$) collected the fewest.

\begin{figure}
\begin{center}
\vspace{\cDistHalf}
\includegraphics[height=15.6cm,width=9.5cm,angle=-90,clip=true]{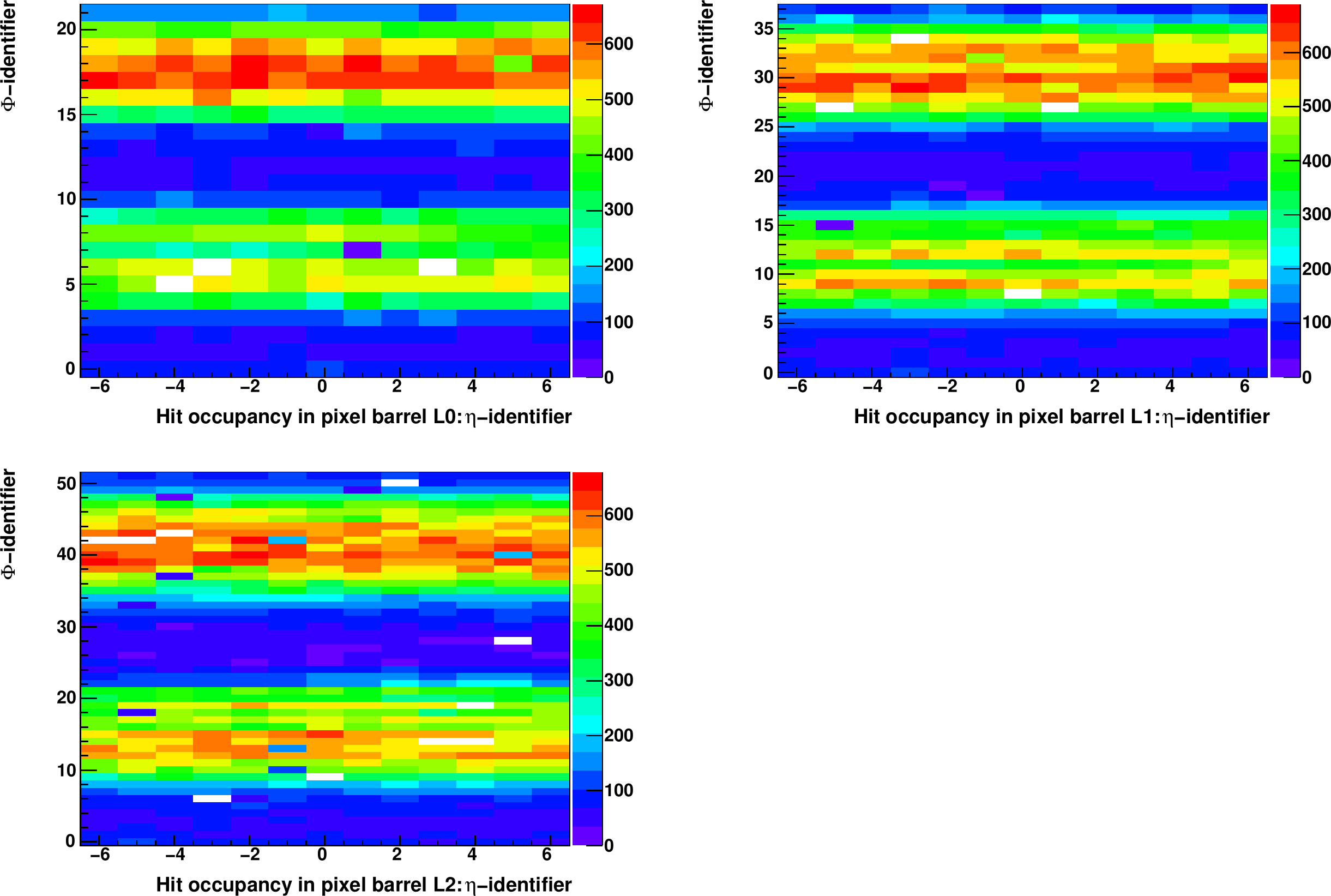}
\vspace{\cDistHalf}
\end{center}
\caption[Number of hits-on-track for $B$-field off in the barrel of the pixel detector in M8+ before any alignment]{\label{fig:hitOcc_PIXB}
Number of hits-on-track for $B$-field off in the barrel of the pixel detector in M8+ before any alignment. A clear periodic dependance in $\Phi$ with maxima near $\Phi$-identifiers $I_\Phi=\oneOverX4I_{\rm max},\,\xOverY34I_{\rm max}$, where $I_{\rm max}$ is the maximum $\Phi$-dentifier of a given layer, is visible. This corresponds to modules at $\Phi=+\xOverY\pi2,\,-\xOverY\pi2$ and reflects the dominant incidence of cosmic ray particles from above. Modules which did not collect any hits are shown in white.
}
\end{figure}

The number of hits per module collected in M8+ with solenoid off is shown for the {\bf pixel barrel} in Figure~\ref{fig:hitOcc_PIXB} by layers. All three layers display a strong periodic dependece as discussed above, with about 500 hits per module at $\Phi\simeq\pm\xOverY\pi2$, and circa 50 (id est $\order{10}$ fewer) hits per module at $\Phi\simeq0,\pi$. Modules at the top of the detector tend to collect about 20\% more hits than the ones at the bottom, since some tracks from the low-$p_T$ end of the spectrum do not reach the lower hemisphere of the detector. The alignment precision per module at L3 is roughly given by Equation~\ref{eqn:residualErrAll}. Therefore, it will show a $n^{-\xOverY12}$-like mapping of the number of hits per module distribution. Some ``dead'' modules which did not collect any hits (due to cooling, readout, etc. failures) are shown in white.
%{\it This demonstrates why L1 and L2 alignment is important: all modules of a subdetector/layer are moved coherently to correct for gross misalignments, whereas in case of pure L3 alignment dead modules whould stay at their nominal positions. Once the dead modules are turned on again (upon a repair of a cooling loop for example), they may be too far from the other modules and their alignment constants might diverge. Such pathological situations are avoided by superstructure alignment. {\bf (move this to the superstructure discussion with reference to the figure?)}}

%\nopagebreak[5]

\begin{figure}
\begin{center}
\vspace{\cDistHalf}
\includegraphics[height=15.6cm,width=9.5cm,angle=-90,clip=true]{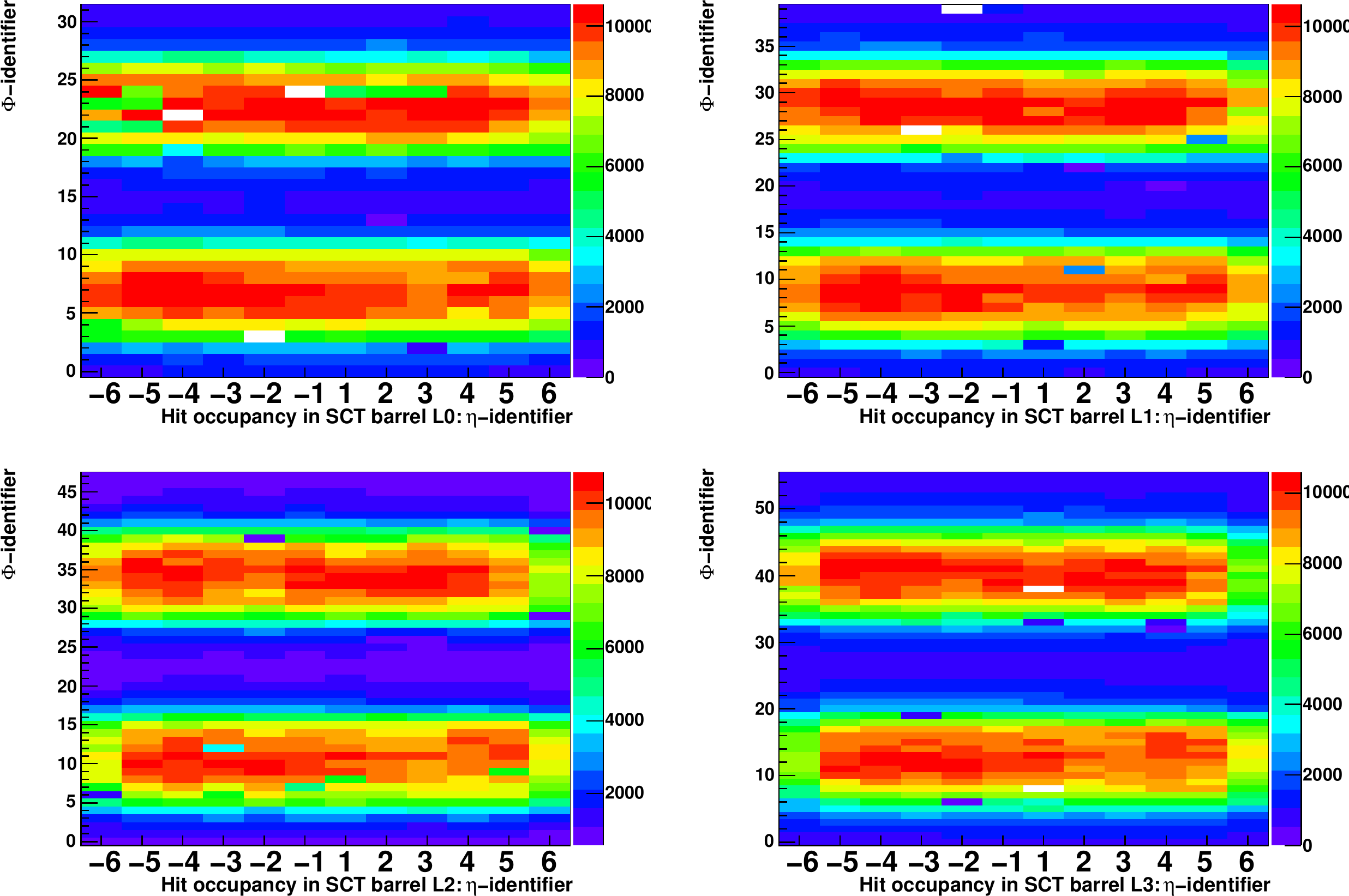}
\vspace{\cDistHalf}
\end{center}
\caption[Number of hits-on-track for $B$-field off in the barrel of the SCT detector in M8+ before any alignment]{\label{fig:hitOcc_SCTB}
Number of hits-on-track for $B$-field off in the barrel of the SCT detector in M8+ before any alignment. Each entry corresponds to one SCT module, id est hits collected by its two sides are added. Modules which did not collect any hits are shown in white.
\vspace{\cDist}
}
\end{figure}%\nopagebreak[5]

A qualitatively very similar picture to the pixel barrel is observed in the {\bf barrel} of the {\bf SCT} detector, which is shown in Figure~\ref{fig:hitOcc_SCTB}. Given the larger area of the SCT modules and a larger cross-section of the entire SCT compared to the pixel detector, the typical number of hits per module at the top and bottom of the detector is about 10k, whereas fewer than 1k is typical for the side modules. One qualitative difference compared to the pixel barrel is that the edge modules with $\eta$-identifiers -6 and 6 are notably less illuminated than the ones in the middle, which has to do with the acceptance of the ID-based trigger and the decreased tracking efficiency for cosmic ray tracks in the transition region between the barrel and ECs of the SCT.

\begin{figure}
\begin{center}
\vspace{\cDistHalf}
\includegraphics[height=11cm,width=6.8cm,angle=-90,clip=true]{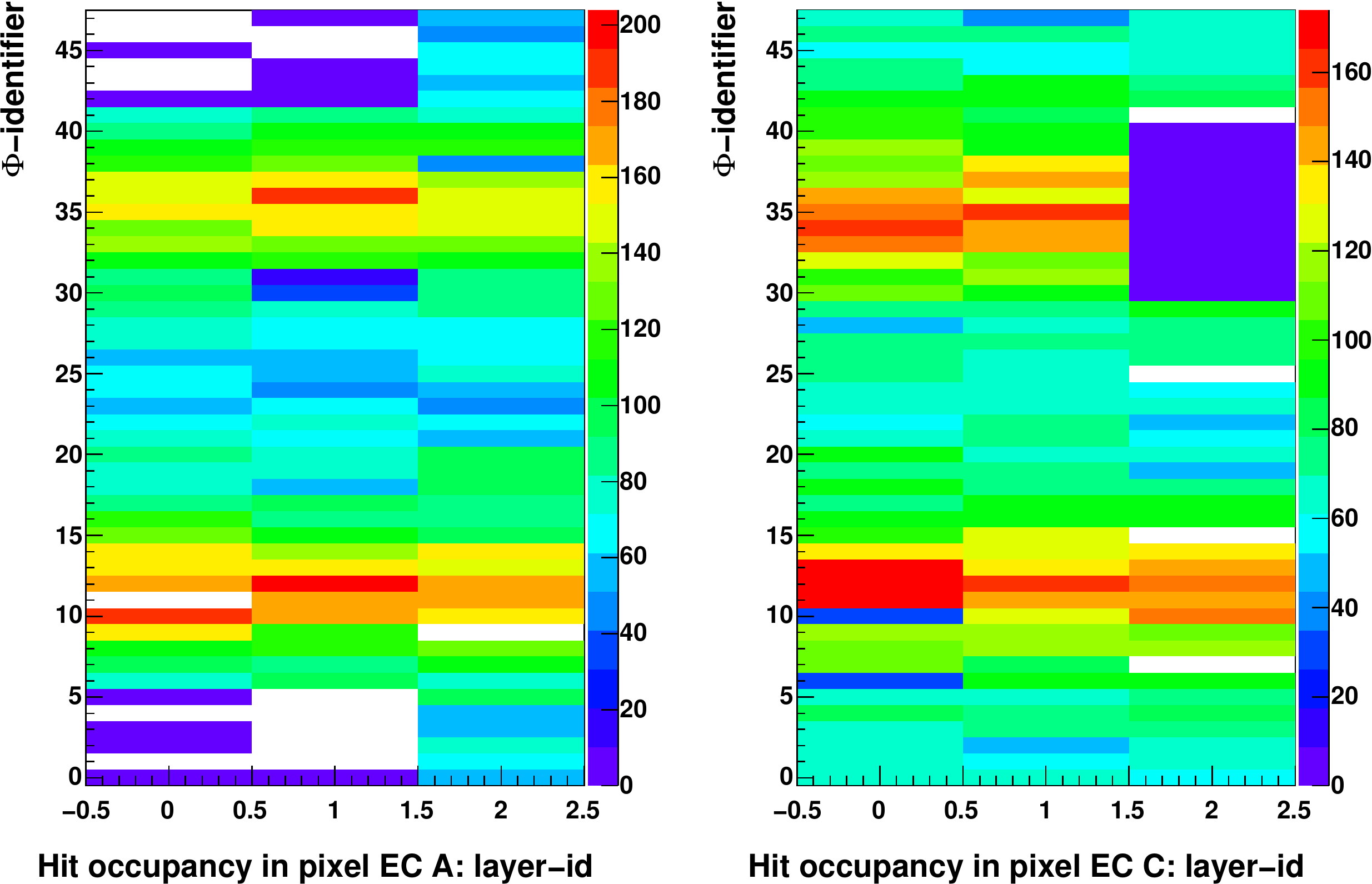}
\vspace{\cDistHalf}
\end{center}
\caption[Number of hits-on-track for $B$-field off in the ECs of the pixel detector in M8+ before any alignment]{\label{fig:hitOcc_PIXE}
Number of hits-on-track for $B$-field off in pixel end-caps in M8+ before alignment. A weakly pronounced periodic dependance in $\Phi$ with maxima at $\Phi=+\xOverY\pi2,\,-\xOverY\pi2$ is visible, which is explained in the text. Modules which did not collect any hits are shown in white.
\vspace{\cDistHalf}
}
\end{figure}

The modules in the {\bf end-caps} of the {\bf pixel} detector are vertically oriented, and collected about the same number of hits per module (about 100) as the sides of the pixel barrel during M8+, as shown in Figure~\ref{fig:hitOcc_PIXE}. This is due to the fact that their acceptance convoluted with the acceptance of the ID-based triggers is quite similar to the barrel side modules, as can be seen from Figure~\ref{fig:inDetTechnical} (this condition is a fortiori fulfilled for the muon and calorimeter stream triggers). A less pronounced periodial modulation is present in both ECs, which is a result of the hit quality cut on the transverse track incidence angle projected on the $x$-$z$ frame of the module: it will remove almost vertically incident tracks for the modules at $\Phi=0,\,\pi$, while keeping them for modules at $\Phi=+\xOverY\pi2,\,-\xOverY\pi2$ (cf. Subsection~\ref{ssec:selectionHit}).

The picture observed for the modules in the {\bf end-caps} of the {\bf SCT} detector is qualitatively similar to that in the pixel end-caps, and is therefore not shown here explicitly: on average, the modules collect between about 0.1k and 1k of hits for the outermost and innermost disks, respectively. Also here a slight modulation in $\Phi$ is observed. EC~A collected about 10\% more hits than EC~C due to its spatial proximity to the larger of the two shafts, cf. Figure~\ref{fig:trackDirM8plus}.

%\nopagebreak[5]

It should be mentioned that all the the distributions of the number of hits per module correspond to the number of residuals used in the \RA\ procedure: the same track, hit, and residual selection cuts are applied, as detailed in Subsections~\ref{ssec:selectionTrack}, \ref{ssec:selectionHit}, and \ref{ssec:selectionRes}. The figures were obtained with nominal ID geometry, id est before any alignment.

\begin{comment}
Please find attached the plot, that Pippa asked for during the discussion in todays  cosmics meeting.  You see the intersection of tracks extrapolated to a plane 10.5 m above ATLAS center  (e.g just above the muon system).  There is a clear indication of the two  elevator shafts PM15 and PX15 causing the satellite peaks in the phi distribution.  This is data from runs 96884 and 96903. I use the IDcosmics stream and require the TRT L1 trigger bit, so no RPC bias. I also require a silicon hit to remove TRT only tracks with no Z information..\\
Hi Oleg,The origin is understood.The tracks come from the 4 shafts above ATLAS (there are some nice pictures around).The 7.75 meters coem about, when one projects the shaft position onto the ATLAS acceptance.Nobody has verified that analytically, but the MC shows the same  thing ...
\end{comment}

%% file: M8plus/Selection.tex
%Like any other data analysis, track-based alignment stands and falls with the quality of the input to the alignment procedure: the reconstructed tracks, hits associated with tracks, and, at the bottom line, the track-hit residuals. 
% and overlap residuals in case of L3 alignment. 
%Therefore, due care needs to be excercised in the selection of tracks, hits, and residuals for the alignment procedure. This shall be the subject of this Section.
Due care needs to be excercised in the selection of {\em tracks}, {\em hits}, and {\em residuals} for the alignment procedure. This shall be the subject of this Section.

%% file: M8plus/SelTrack.tex
%As argued above, some minimum quality requirements on the quality of tracks need to be imposed in order to provide a sensible input for alignment. This is discussed below.

The alignment strategy at ATLAS is to perform an internal alignment of the silicon tracker first, and then to align the TRT detector with respect to it\footnote{After this, the relative alignment of the ID and the rest of ATLAS needs to be established. However, this is not relevant for the discussion here.}. Therefore, tracks which were reconstructed by the silicon tracker {\em only} are used for silicon tracker alignment with the \RA\ algorithm.

Currently, there are two tracking algorithms available at ATLAS for cosmic ray reconstruction: the so-called {\tt CTBTracking}~\cite{bib:CTBT} and {\tt NewTracking}~\cite{bib:newT}. While the latter is more advanced from the computing point of view and also has a smarter pattern recongnition and optimised track fitting, the former is more robust in the sense that it is better understood and debugged, since it was used from the start in M8+ and in earlier milestone runs. This robustness was the pivotal reason to use tracks reconstructed with the {\tt CTBTracking} algorithm for the \RA\ procedure\footnote{This decision was further supported by the fact that a reasonably debugged and validated {\tt NewTracking} version was available in \Athena\ release family $15.X.0$ only. However, $15.X.0$ incorporated many changes in ID reconstruction-related software, and was in turn a potential source of software bugs.}.

The most basic criterion for the quality of a track is the number of hits to which it is fitted: generally, the more associated hit measurements there are, the better constrained the track fit is. As detailed in Subsection~\ref{ssec:trackingEDM}, tracks can be fully described by five parameters in presence of a magnetic field, and four parameters without it. Certainly, the number of hits associated with a track should be well above these critical thresholds. Therefore, a cut on the minimum number of hits in the silicon tracker is placed in the \RA\ procedure:
\begin{equation} \label{eqn:nHitCut}
 n_{\rm hit}^{\rm Si} \geq 7\,.
\end{equation}

\begin{figure}
\begin{center}
\vspace{\cDistHalf}
\includegraphics[width=7.9cm,clip=true]{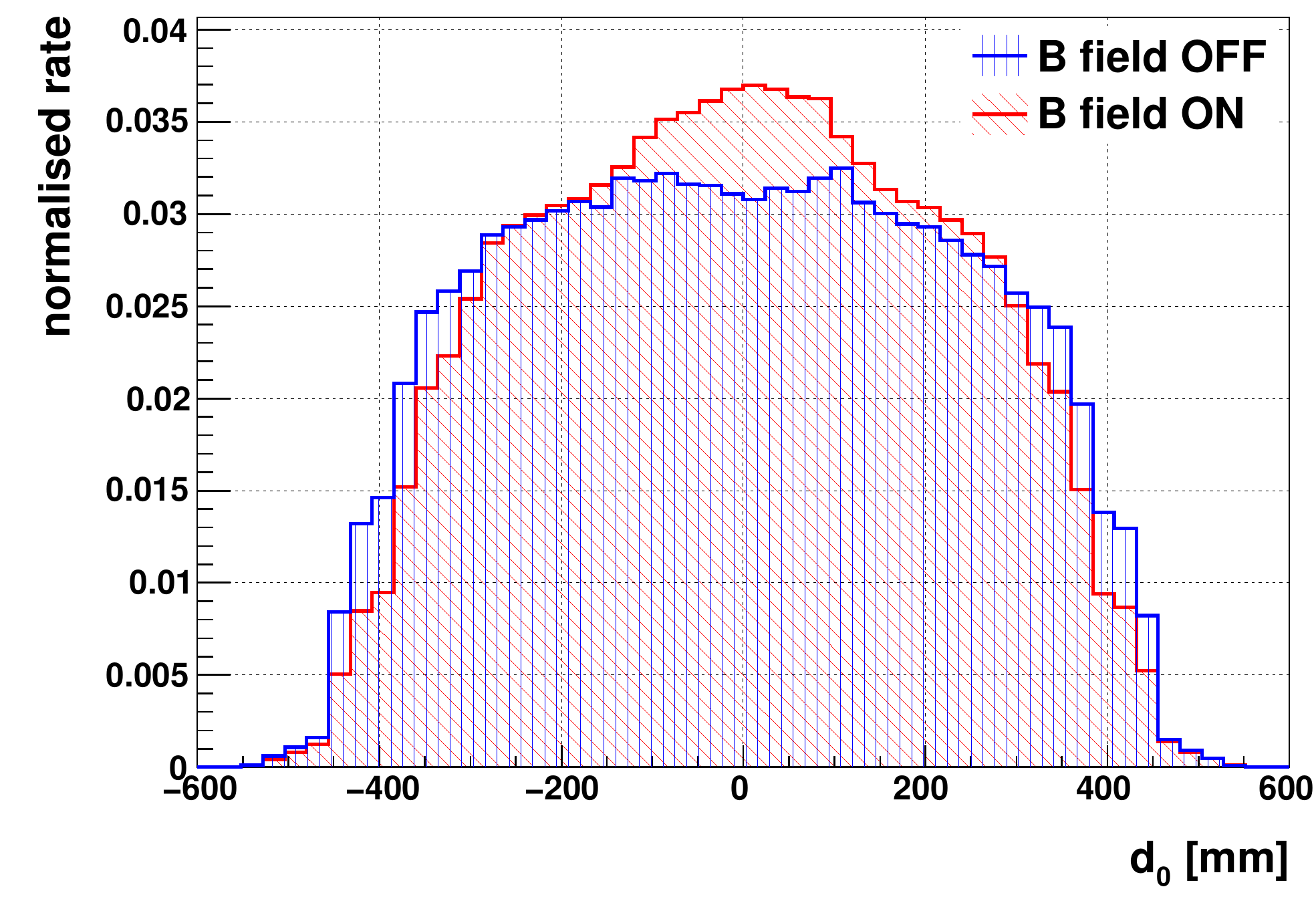}
\includegraphics[width=7.9cm,clip=true]{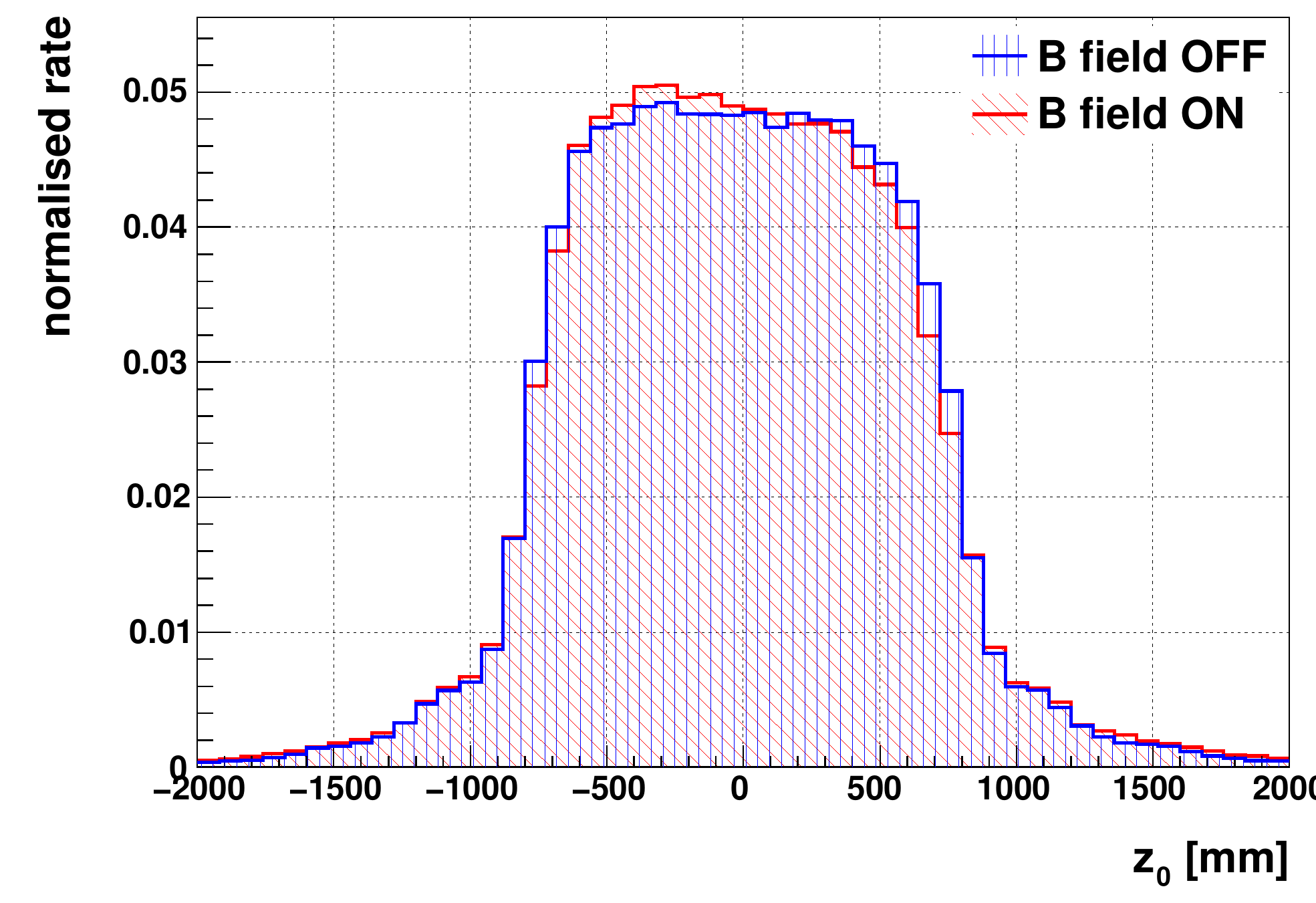}
\includegraphics[width=7.9cm,clip=true]{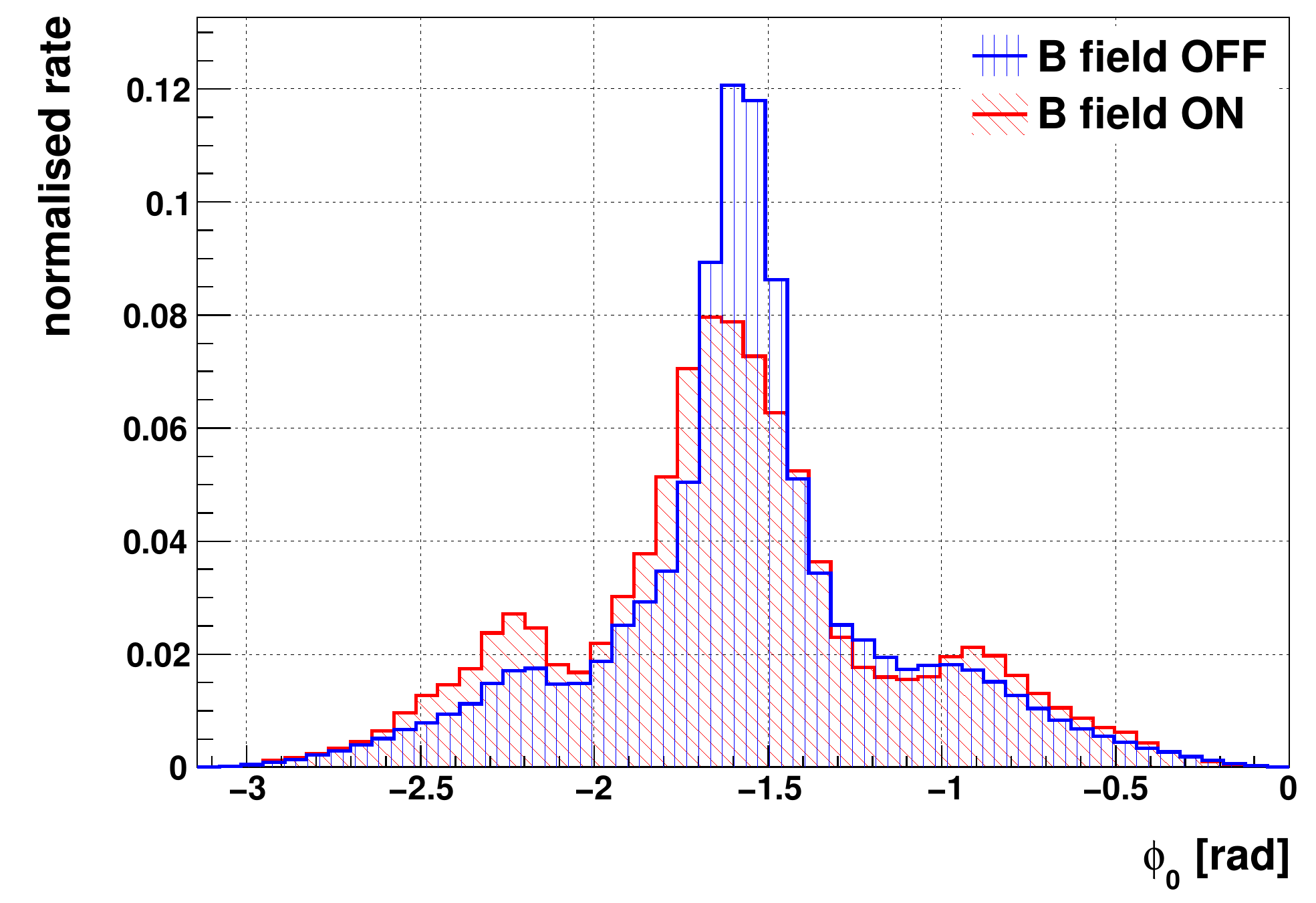}
\includegraphics[width=7.9cm,clip=true]{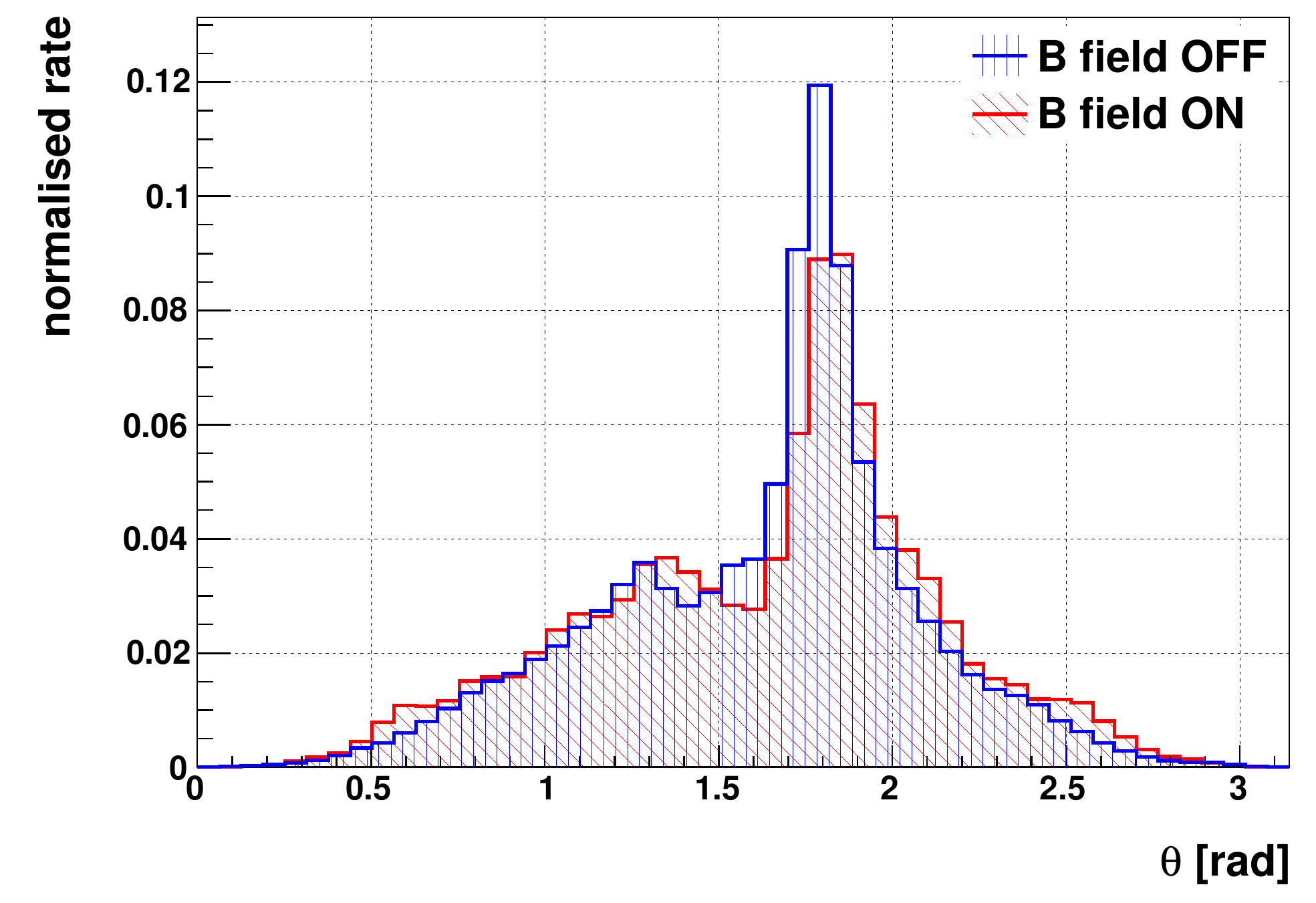}
\includegraphics[width=7.9cm,clip=true]{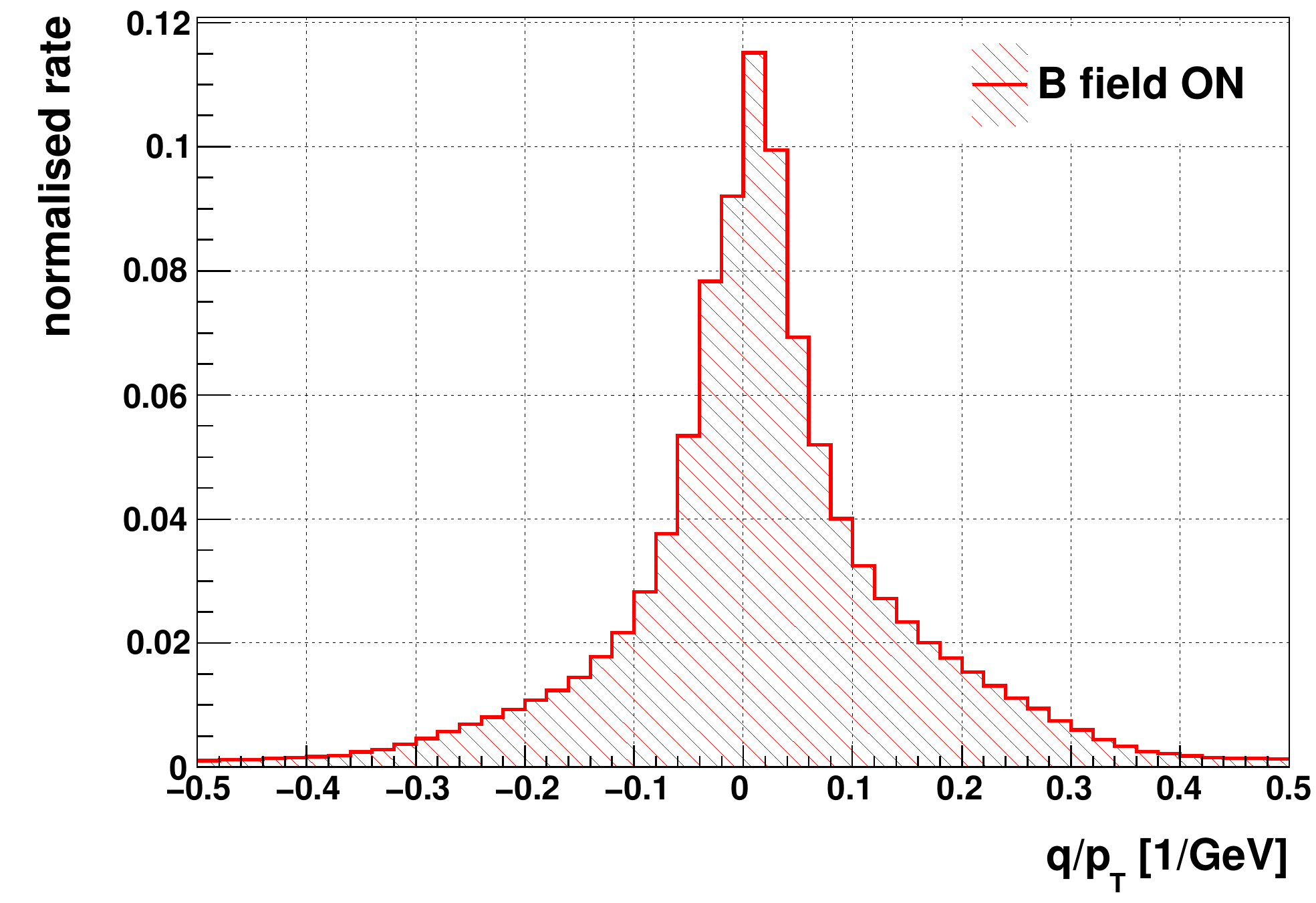}
\includegraphics[width=7.9cm,clip=true]{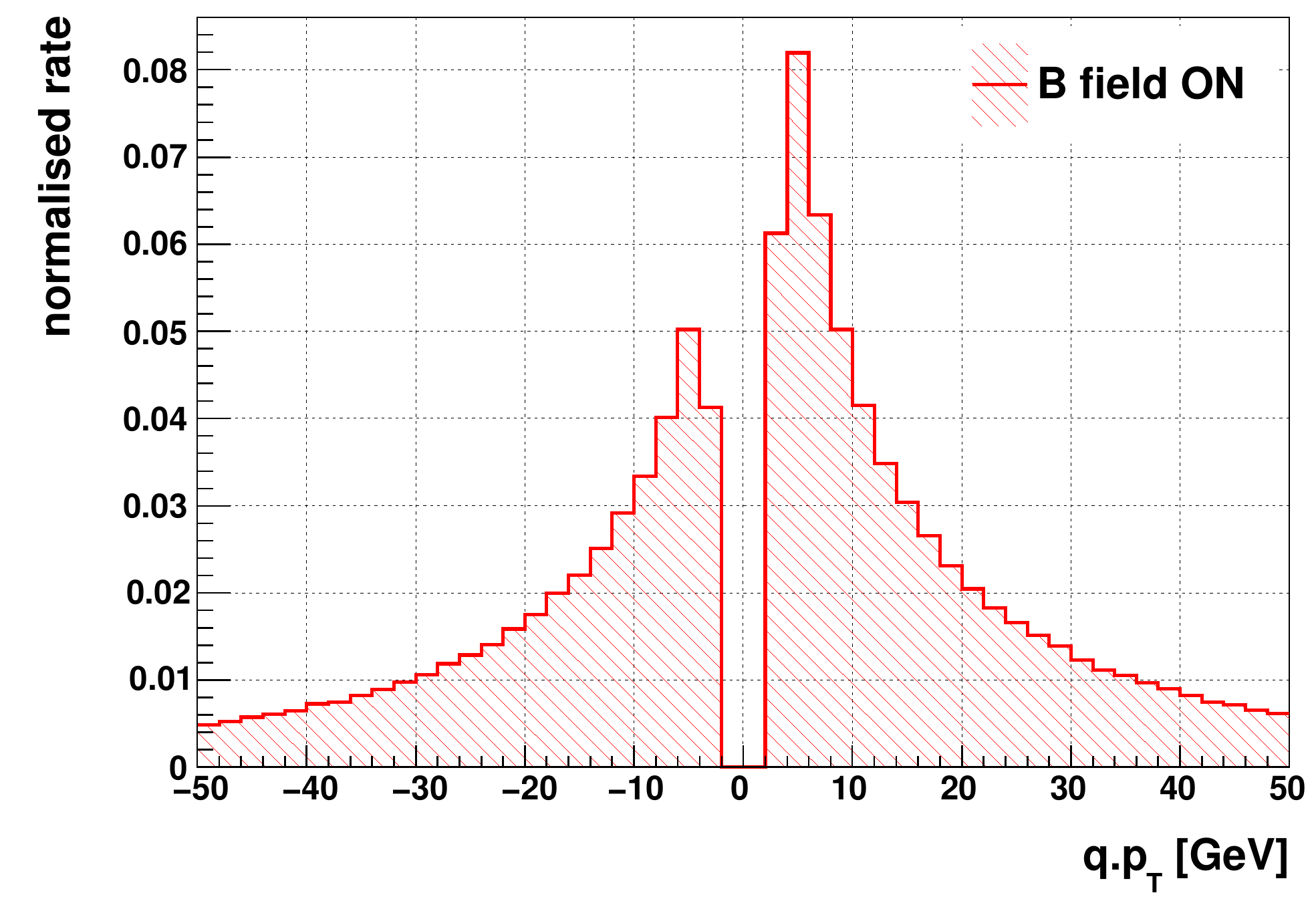}
\end{center}
\vspace{\cDist}
\caption[Track parameter distributions in M8+]{\label{fig:trkPar}
Track parameter distributions for the transverse impact parameter $d_0$ {(\bf top left)}, longitudinal track impact parameter $z_0$ {(\bf top right)}, $\phi_0$ {(\bf middle left)}, $\theta$ {(\bf middle right)}, $q/{p_T}$ {(\bf bottom left)}, and $q\cdot{p_T}$ {(\bf bottom right)} in M8+ after alignment with the \RA\ algorithm. For details see text.
}
\end{figure}%\nopagebreak[5]

\begin{figure}
\begin{center}
\vspace{\cDistHalf}
\includegraphics[width=5.2cm,clip=true]{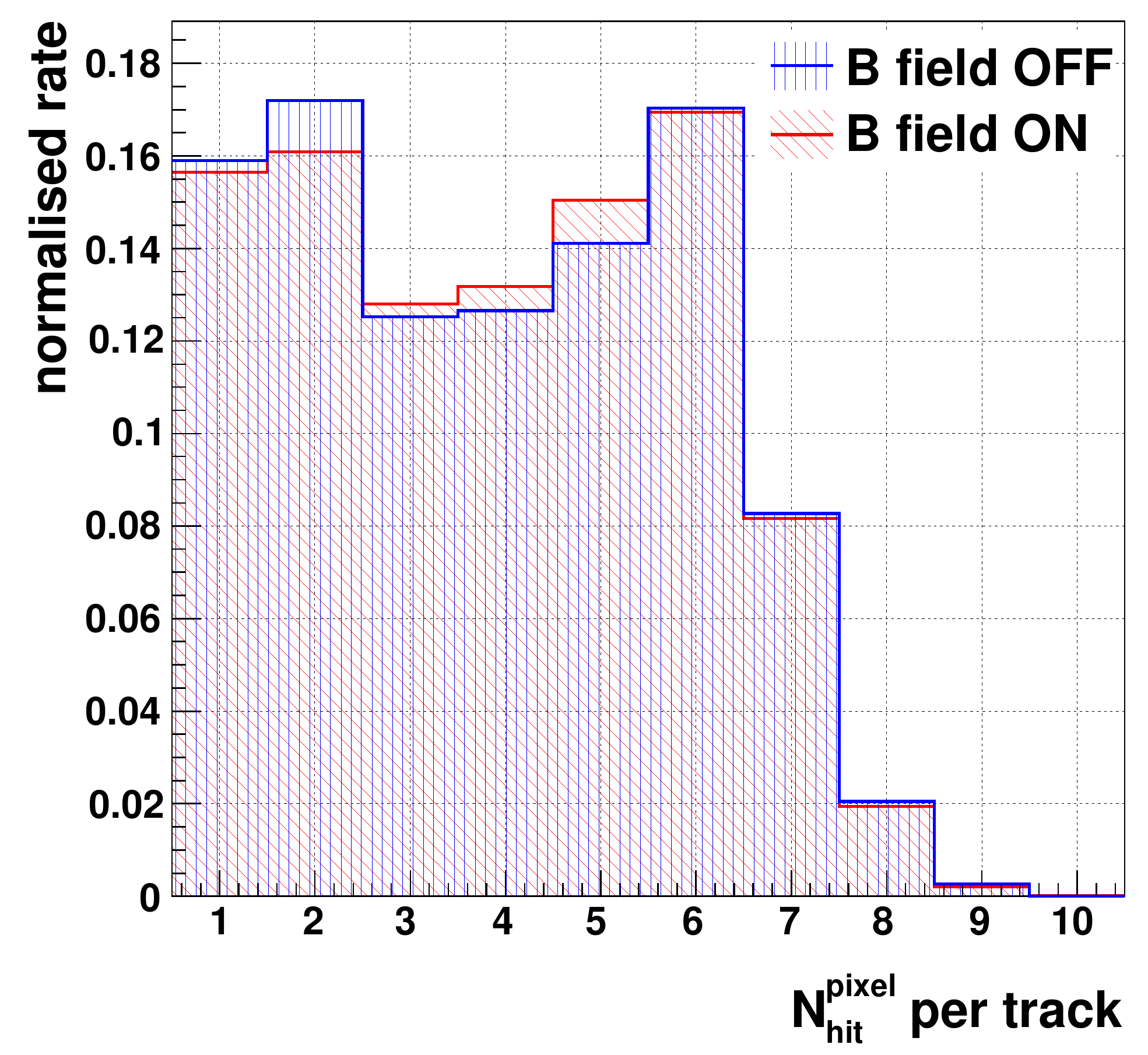}
\includegraphics[width=5.2cm,clip=true]{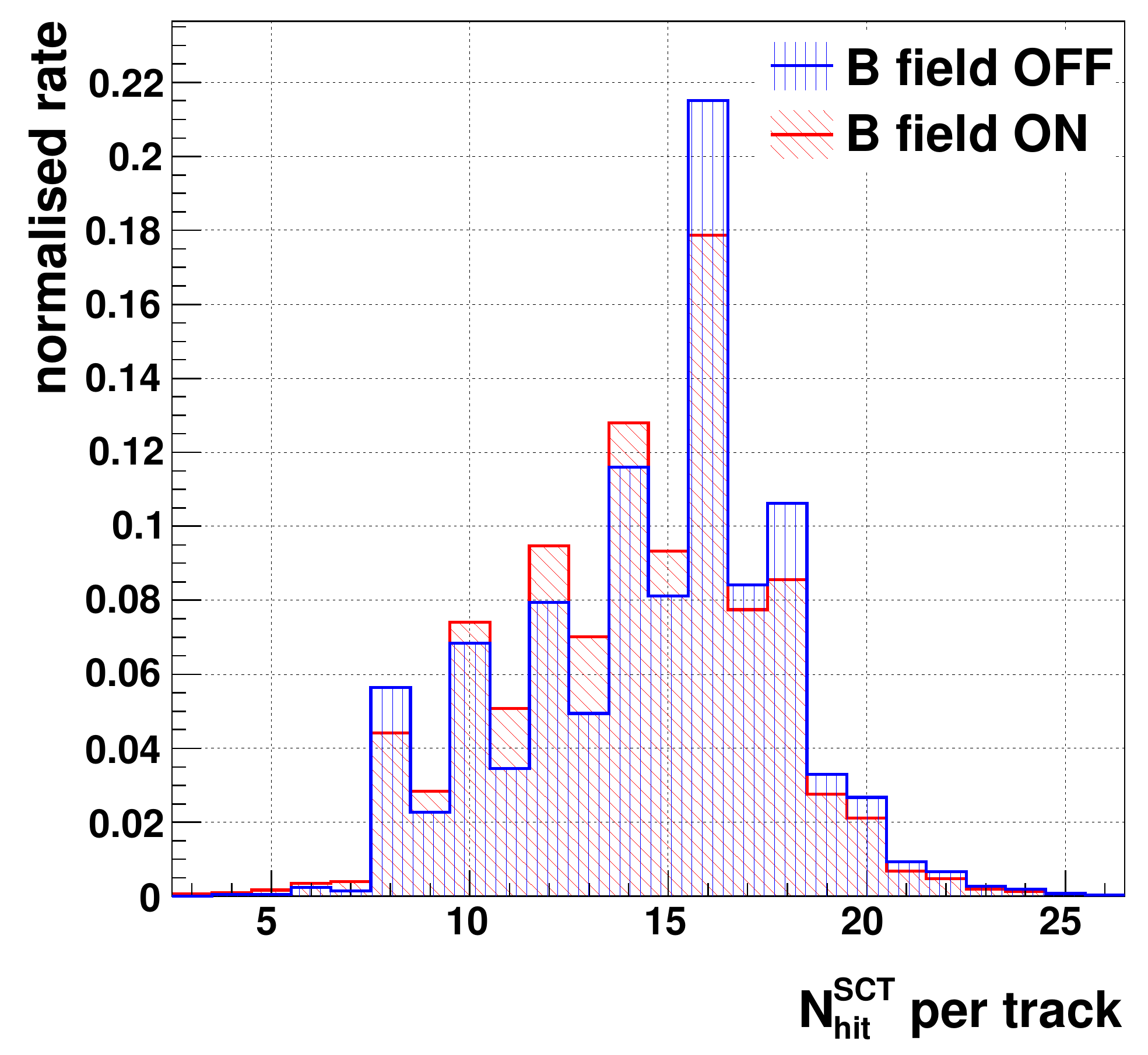}
\includegraphics[width=5.2cm,clip=true]{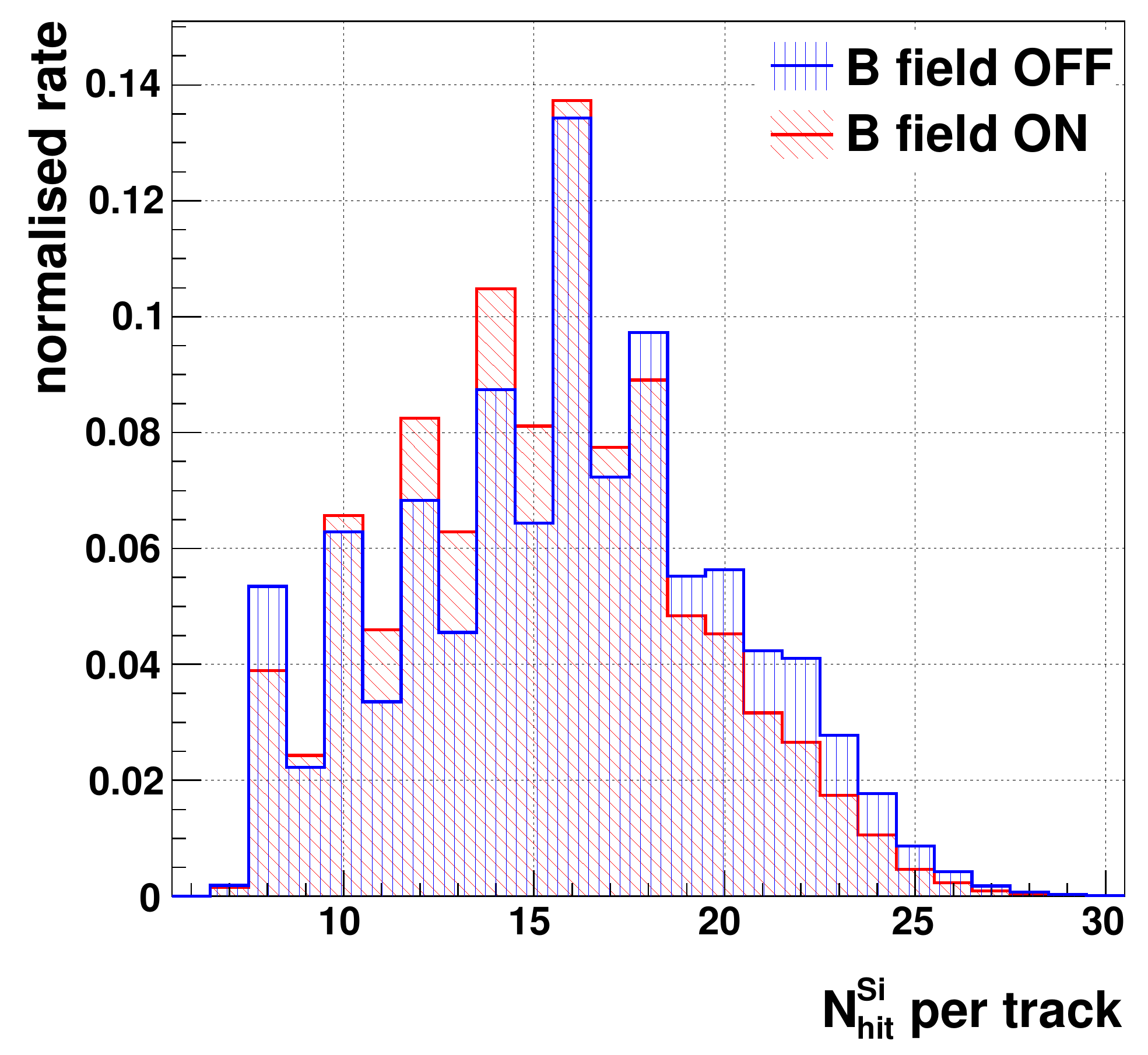}
\vspace{\cDist}
\end{center}
\caption[Number of hits per track in the pixel and SCT detectors, as well as in the entire Si tracker in M8+]{\label{fig:nHits}
Number of hits per track in the pixel and SCT detectors, as well as in the entire Si tracker in M8+ after alignment with the \RA\ algorithm. For details see text.
\vspace{\cDistHalf}
}
\end{figure}%\nopagebreak[5]

In a statistically limited sample such as that of M8+, some large residuals for low-$p_T$ tracks with large Coulomb multiple scattering may adversely bias the residual mean measurement of individual modules, which is the basic input to alignment. Therefore, for the $B$-field on case, a cut on the transverse momentum of the track is placed:
\begin{equation} \label{eqn:p_TCut}
 p_T^{\rm trk} > 2\,\GeV\,.
\end{equation}

Figure~\ref{fig:trkPar} shows the canonical track parameter ($d_0,z_0,\phi_0,\theta,q/p$) distributions at the perigee for all tracks collected in M8+ which were used in the \RA\ procedure. All distributions are after alignment, which not only refined the track parameter resolutions, but also increased the track reconstruction efficiency despite the very loose cuts used in M8+. All track parameters look as expected: the $d_0$ and $z_0$ distributions have clearly visible bulks at $[-50,\,50\,\rm cm]$ and $[-80,\,80\,\rm cm]$, respectively, which are defined by the SCT barrel acceptance. On top of that, $z_0$ has long tails extending up to about $\pm3$\,m due to tracks reconstructed in the SCT ECs. The $\phi_0$ parameter has a clear peak at about $-\xOverY\pi2$ corresponding to the two access shafts, and two less pronounced peaks due to the two elevator shafts, cf. Figure~\ref{fig:trackDirM8plus} (a). There are two peaks in the $\theta$ distribution at about $\xOverY\pi2\pm0.3$ reflecting the two access shafts, the more pronounced one at $\theta\simeq1.8$ corresponding to the larger shaft on side~A. The $q/p$ and $q\!\cdot\!p_T$ distributions are sharply peaked and notably skewed towards positive values, which is expected because of a  charge assymmetry of about 1.4 towards positive cosmics muons~\cite{bib:pdg2008}. Finally, the $q\cdot p_T$ histogram has no entries in $[-2\,\GeV,\,2\,\GeV]$ range due to the $p_T>2\,\GeV$ cut.

Generally, the distributions in Figure~\ref{fig:trkPar} look very similar for the $B$-field on and off cases. However, there are some subtle differences. These are due to a complex interplay of several factors. One imporant thing to keep in mind is that the (natural) charge assymmetry may result in a different trigger acceptance in case the toroidal magnetic field of the muon spectrometer is on. Moreover, the solenoidal $B$-field and the toroidal $B$-field of the muon spectrometer tended to be on at the same time in M8+ (albeit with some exceptions). This explains why the $\theta$-distribution is shifted towards positive values for $B$-field on tracks. Matters are more complicated by the fact that different triggers have been used for the datasets with and without magnetic field: for example, about $\xOverY34$ of the solenoid $B$-field off data were taken {\em after} the introduction of the fast TRT-OR L1 trigger, which contributed a significant fraction of the track-to-tape rate. This may well explain why the $d_0$ distribution is somewhat less sharply peaked for $B$-field off. %, and the $d_0$ distribution slightly skewed towards negative values for the $B$-field on case. 
As expected, the $\phi_0$ distribution is notably shifted towards negative values for the $B$-field on case. %due to bending of tracks in the solenoidal magnetic field of the ID. 
This is due to the sharply defined primary incidence angle of cosmic ray particles, and the fact that the track parameters are defined at the perigee {\em with respect to} the nominal detector centre: $\delta\phi_0(\pt)<0$ is added for tracks with solenoidal magnetic field on.

The number of hits in the pixel detector, the SCT, and the entire silicon tracker after the track selection discussed above, and after the hit and residual quality cuts discussed in Subsections~\ref{ssec:selectionHit}, \ref{ssec:selectionRes} is shown in Figure~\ref{fig:nHits}.\\
Due to the limited geometrical acceptance of the {\bf pixel} detector about 90\% of all tracks reconstructed in the silicon tracker have no pixel hits, and so this dominating bin is not shown. The histogram displays about the same number of tracks with 1,2,...6 pixel hits. Beyond, a rapid drop-off is observed, since overlap hits are needed in the barrel of the pixel detector to get 7 or more hits from cosmic ray particles.\\
In {\bf SCT} modules, each side can produce an independent hit measurement. Therefore, even numbers of hits per track are more likely than odd ones: if a particle is registered by one side of the module, it is very likely to be measured by the other side as well. This is also the reason why a distinct rise is observed at 8 rather than 7 hits per track, which corresponds to the cut in Equation~\ref{eqn:nHitCut}. The maximum is observed at 16~hits: 4~layers $\times$ 2~sides per module $\times$ 2~hemispheres. It is interesting to note that tracks with $B$-field off have on average more SCT hits, which could be due to a missing $p_T$ cut: muons below about 1\,GeV are not MIPs and produce more ionisation as they leave the Bethe-Bloch and approach the Anderson-Ziegler regime~\cite{bib:pdg2008}.\\
The distribution for the entire {\bf silicon tracker} is mostly determined by the SCT histogram and displays its characteristic odd-even structure, with some additional hits from the pixel detector for some tracks. The effect of the cut in Equation~\ref{eqn:nHitCut} is clearly visible.

%\clearpage

%% file: M8plus/SelHit.tex
In the context of \RA, residuals are formed directly from the hits measured by silicon tracker modules --- a problematic hit will directly result in a problematic residual. Therefore it is highly important to safeguard a high quality of hits selected for the alignment procedure. 

\subsubsection{Hit Quality Selection with the {\tt InDetAlignHitQualSelTool}}

%\vspace{\cDistHalf}
%\begin{scriptsize}
%\begin{verbatim}
%|=/***** Private AlgTool InDetAlignHitQualSelTool/InDetAlignHitQualSelTool ***************************
%| |-MaxClusterSize     = 5  (default: 5)
%| |-MaxIncidAngle      = 0.80000000000000004  (default: 0.80000000000000004)
%| |-RejectEdgeChannels = True
%| |-RejectGangedPixels = False
%| |-RejectOutliers     = True  (default: True)
%| \----- (End of Private AlgTool InDetAlignHitQualSelTool/InDetAlignHitQualSelTool) ------------------
%\end{verbatim}
%\end{scriptsize}
%\vspace{\cDistHalf}

\begin{table}
\small
\begin{center}
\begin{tabular}{lc}
\hline
Steering parameter name & Default value \\
\hline\hline
{\tt MaxClusterSize}     & 5 \\
{\tt MaxIncidAngle}      & 0.8\,rad \\
{\tt RejectEdgeChannels} & True \\
{\tt RejectGangedPixels} & False \\
{\tt RejectOutliers}     & True \\
\hline
\end{tabular}
\caption[Steering parameters of the {\tt InDetAlignHitQualSelTool}]{\label{tab:hitQualOpt}
Steering parameters of the {\tt InDetAlignHitQualSelTool} with their default values.
}
\vspace{\cDist}
\end{center}
\end{table}

A dedicated tool, the so-called {\tt InDetAlignHitQualSelTool}~\cite{bib:hitQualNote,bib:hitQualSelTool}, was written by the author in the context of M8+ alignment to select for high quality hits. This tool can can be used universally for all alignment algorithms by dynamically loading it via \Athena\ job options. The relevant steering options of the {\tt InDetAlignHitQualSelTool} are summarised in Table~\ref{tab:hitQualOpt} together with their default values. The default values were the ones used in M8+. Besides the non-trivial cuts on the maximum cluster size and the maximum transverse track incidence angle which will be discussed in detail below, some ``canonical'' cuts were implemented in this tool:
\begin{description}
 \item[Rejection of edge channels:] if a hit is registered by an edge channel\footnote{The term ``edge channel'' refers to a channel which is {\em geometrically} situated at the edge of the module, id est has no neighbouring channels to one of its sides.} of a module, one does not know the full extension of the region where ionisation charge was deposited. On average, this results in a significant bias on the measured hit position $\qhit$ towards the middle of the module. Therefore, hit clusters which contain an edge channel should be discarded from the alignment procedure. This is implemented for $\qhit_x$ only, since a potential bias for $\qhit_y$ is significantly smaller than the intrinsic resolution of the pixel detector;
 \item[Rejection of ganged pixels:] as detailed in Subsection~\ref{ssec:innerdetector}, four pixels sensors in each column of a module are ganged, id est connected to the same output channel. These account for $\sim$2.5\% of all pixels. This can lead to ambiguities in tracking, especially in a dense tracking environment at high luminosities. In such circumstances, hits registered {\em only} by ganged pixels may adversely affect the alignment procedure. However, given the very low noise levels and the small hit occupancy in M8+ cosmic ray data taking, this cut was not used and is off by default;
 \item[Rejection of outlier hits:] the track fitter has an internal logic to reject hits which are ``too far'' away. Such hits are labeled as {\em outliers} and do not participate in the track fit, but remain associated with the track. In more technical terms, ``too far'' means that the contribution of the given hit to the $\chisq$ of the track is larger than a chosen cut value, where the $\chisq$ contribution is calculated from the prediction of the track scattering angle based on the track momentum and material effects.
% in basic terms ``not consistent with the amount of Coulomb multiple scattering found in other hits on the given track''. 
Depending on whether the detector is considered to be reasonably pre-aligned or not, one may choose to enable this rejection. In {\tt CTBTracking}, which was used to produce the \RA\ results shown in this document, the internal outlier logic of the track fitter is off, and thus the cut has no effect. This is considered appropriate given the large initial misalignments of $\order{1\,\mm}$ at L1 between the pixel and the SCT detectors.
\end{description}

\subsubsection{Interplay of the Transverse Track Incidence Angle and the Cluster Size}

The ATLAS Inner Detector was optimised to provide excellent tracking of particles emerging from $p$-$p$ collisions at the DIP~(Designed Interaction Point). In particular, this means that the modules were oriented such as to {\em minimise} average hit cluster sizes given the manyfold external constraints. The hit cluster size \nclust\ depends on the transverse incidence angle $\incAngleT$ defined as:
\begin{equation}
 \incAngleT \equiv \arctan\left(\frac{\vec e_x\cdot\vec p_{\rm loc}}{\vec e_z\cdot\vec p_{\rm loc}}\right)\,,
\end{equation}
where $\vec p_{\rm loc}$ is the momentum vector of the track at its intersection point with the module surface, and $\vec e_{x,z}$ are the unit vectors corresponding to the $x,z$ coordinates in the local frame of the module. 

\begin{figure}[h]
\begin{center}
\includegraphics[width=7.5cm,clip=true]{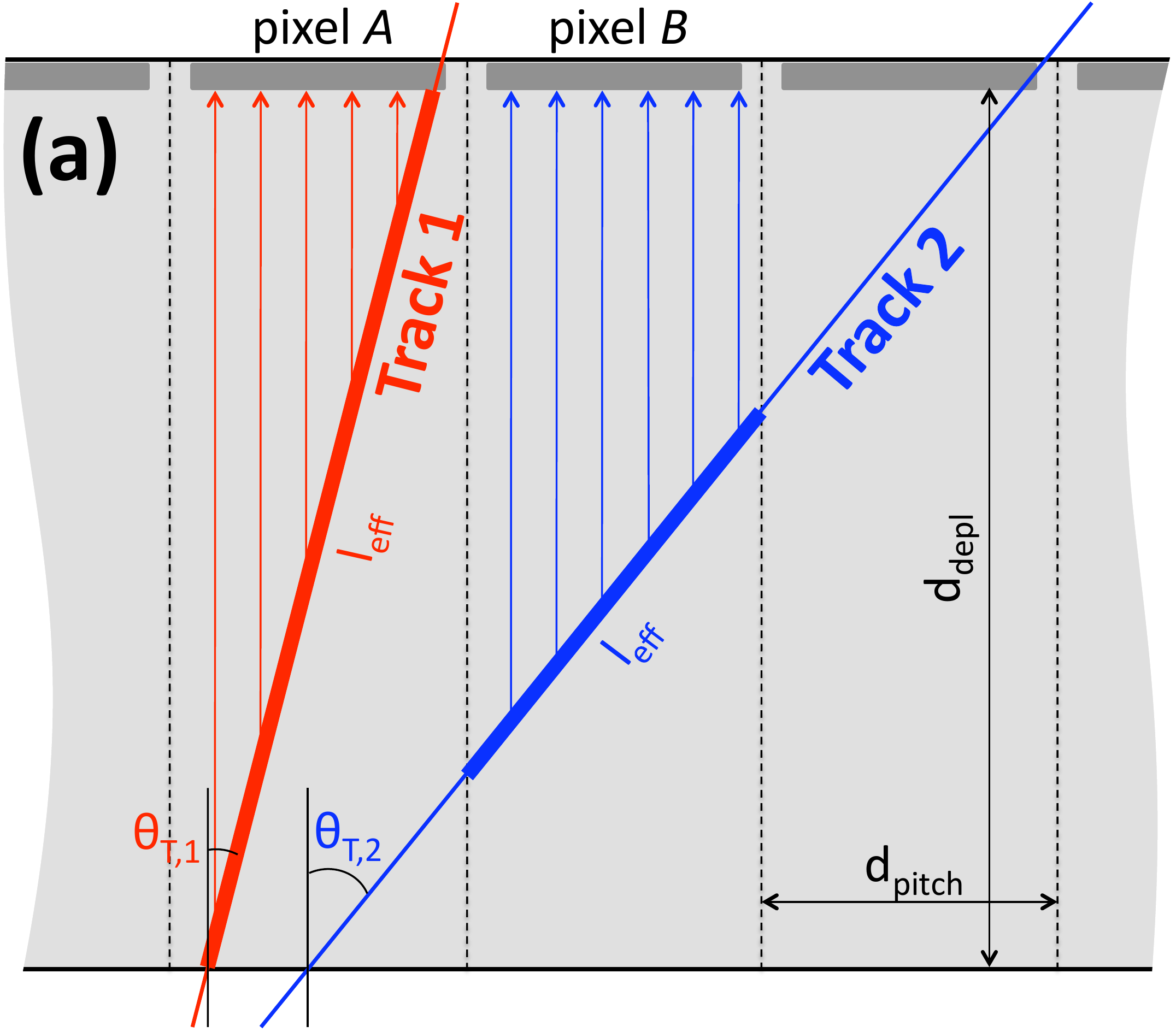}\qquad
\includegraphics[width=7.5cm,clip=true]{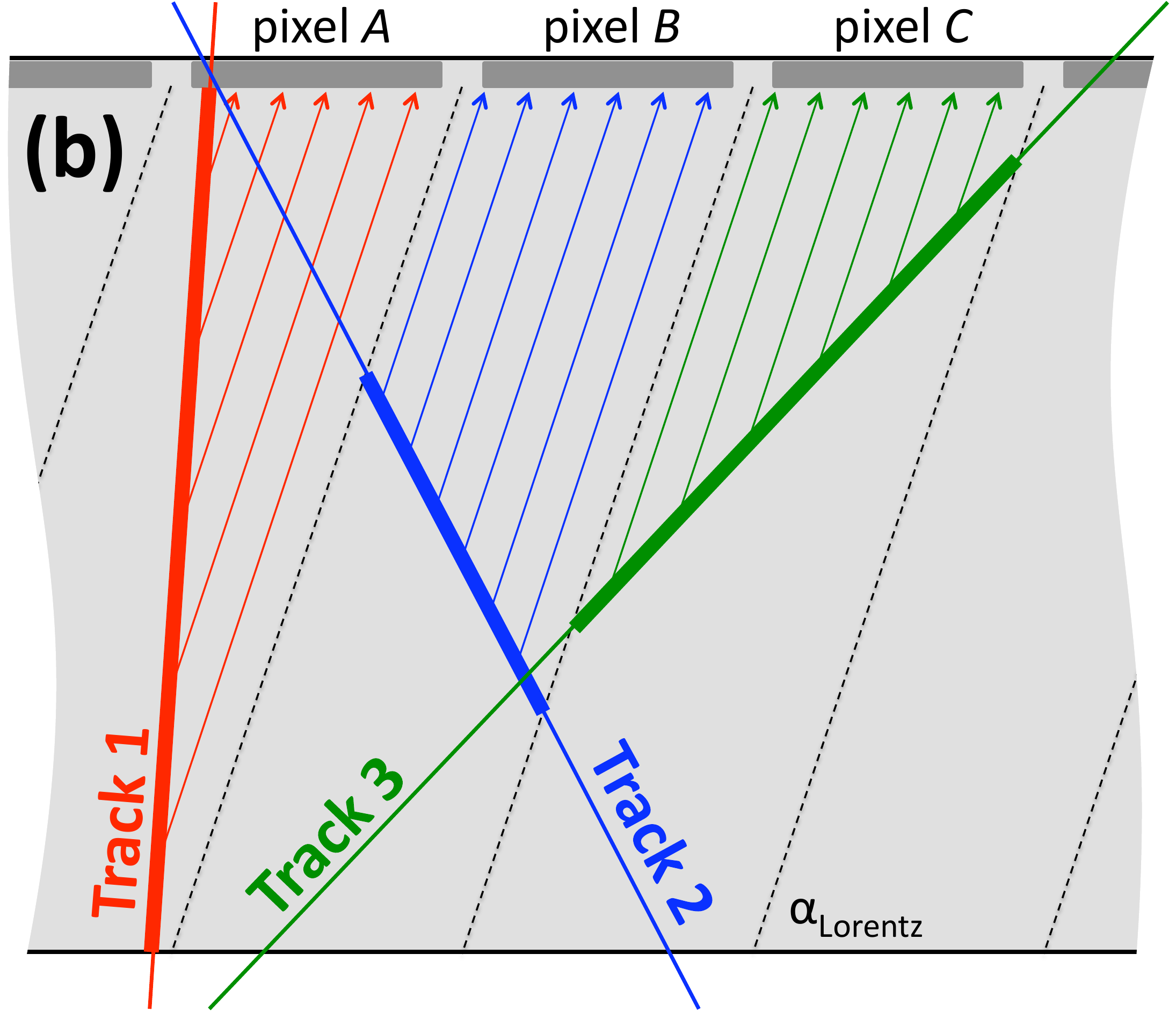}
\end{center}
\vspace{\cDist}
\caption[Sketch of the effective path length of an ionising particle in a pixel-type silicon sensor for $B$-field on and off]{\label{fig:pathLength}
The cross-section through a silicon sensor with two tracks from ionising particles is shown schematically for magnetic field off {\bf (a)} and on {\bf (b)}. The light arrows indicate the ionisation charge travel direction, while the light dashed lines define the boundaries between pixels. The effective path lengths for pixel~$A$/$B$ and the respective tracks is shown as a thick segment on the track. See text for the discussion of the figure.
}
\end{figure}%\nopagebreak[5]

The reason to optimise for a small transverse incidence angle is sketched in Figure~\ref{fig:pathLength}~(a) for a pixel-type silicon module {\bf in absence of any magnetic field}. The charge carrier path is perpendicular to the module surface, and thus each pixel will collect the ionisation charge deposited in the cuboid defined by the pixel surface area\footnote{Up to charge diffusion effects which shall be neglected for the sake of the argument.} and~$\vec e_z$. Thus, the {\em effective path length} $\effPath$ of an ionising particle for a given pixel is defined as the path inside of its charge collection cuboid. In this na\"ive model, the amount of charge collected per pixel $Q_{\rm pixel}$ will be {\em directly} proportional to $\effPath$. Clearly, the probability for a pixel to register a signal by exceeding the threshold for the binary readout will be strongly dependant on $Q_{\rm pixel}$ given adverse effects like electroncs noise, dark currents, etc. Two typical situations are illustrated in Figure~\ref{fig:pathLength}~(a):
\begin{description}
 \item[Case ({\em i})] Consider pixel~$A$ and track~1~(red): the effective path length, which is indicated as a thick segment on the track, is trivially given by the charge depletion area depth~$d_{\rm depl}$:
 \begin{equation} \label{eqn:effPathCase1}
 Q_{\rm pixel}\propto \effPath = \frac{d_{\rm depl}}{\cos\incAngleT}\,.
 \end{equation}
Thus {\em all} ionisation charge deposited in that module will be collected by pixel~$A$, which will result in a very high response probability for this pixel;
 \item[Case ({\em ii})] On the contrary, for pixel~$B$ and track~2~(blue) \effPath\ is given by
 \[Q_{\rm pixel}\propto \effPath = \frac{d_{\rm pitch}}{\sin\incAngleT}\,,\]
where $d_{\rm pitch}$ is the pixel pitch distance. As is evident from the figure, less charge will be collected than in Case~{\bf (i)}, such that pixel~$B$ is somewhat less likely to register a signal.
\end{description}

A qualitatively very similar picture to the one above is observed {\bf in presence of a magnetic field}, which is illustrated in Figure~\ref{fig:pathLength}~(b). The major difference is that now the charge carriers travel with the Lorentz angle $\lorentz$ with respect to the normal to module surface. Let us investigate three typical cases:
\begin{description}
 \item[Case (i)] Consider pixel~$A$ and track~1 (red): the effective path length is the same as in Equation~\ref{eqn:effPathCase1} without magnetic field;
 \item[Case (ii)] On the contrary, for pixel~$B$ and track~2~(blue) \effPath\ is given by
 \begin{eqnarray*}
  \frac{\sin(\xOverY\pi2-\lorentz)}{\effPath} &=& \frac{\sin(\incAngleT+\lorentz)}{d_{\rm pitch}}\\
  \Rightarrow\quad Q_{\rm pixel} \propto \effPath &=& \frac{\sin(\xOverY\pi2-\lorentz)}{\sin(\incAngleT+\lorentz)}\cdot d_{\rm pitch}\,.
 \end{eqnarray*}
 Again, less charge will be collected than in Case {\bf (i)}, such that pixel~$B$ is somewhat less likely to register a signal;
 \item[Case (iii)] Similarly, for pixel~$C$ and track~3~(green), the effective path length is
 \begin{eqnarray*}
  Q_{\rm pixel} \propto \effPath = \frac{\sin(\xOverY\pi2+\lorentz)}{\sin(\incAngleT-\lorentz)}\cdot d_{\rm pitch}\,.
 \end{eqnarray*}
\end{description}
It is important to note that
%because the quotients in Equations~(XX) and (XX) are different, 
Case~{\bf (ii)}-like and Case~{\bf (iii)}-like behaviour will result in a $Q_{\rm pixel}$ distribution which is asymmetric around \lorentz.

The ATLAS pixel detector charge collection thresholds are set such that the ionisation charge deposited by a muon with a $\pt$ of $\order{1\,\GeV}$ will suffice to produce a hit signal in about 65\% of the cases for
 \[\effPath \simeq \sqrt2\cdot d_{\rm pitch} 
            \simeq \frac{d_{\rm pitch}}{\sin\incAngleT}\Big|_{\incAngleT\equiv\xOverY\pi4,\,B\mbox{-}\rm field\,off}\,,\]
id est for a transverse incidence angle of about 45$^\circ$~\cite{bib:privateTobiG} in absence any magnetic field. 

The discussion above illustrates why large \incAngleT\ angles are less favourable: they imply a smaller effective path length $\effPath$ and thus $Q_{\rm pixel}$. In turn, the probability for a given pixel to register a signal decreases. Therefore, large transverse incidence angles are more prone to reconstruction pathologies like irregular cluster sizes, split clusters, etc.\\
%Moreover, one needs to keep in mind that the quantity primarily used for alignment at the bottom line are $r_x$ residuals. 
Moreover, for purely geometrical reasons, large \incAngle\ may result in undesirably large correlations between out-of-plance and in-plane misalignments, while we are primarily interested in constraining the latter for good measurements of  transverse momenta of particles produced in $p$-$p$ collisions.

Regarding the above arguments, the cut
\begin{equation} \label{eqn:incAngleT}
 |\incAngleT| < 0.8\,\rad \simeq 45^\circ
\end{equation}
was placed in the {\tt InDetAlignHitQualSelTool} for the pixel detector. The same cut value is used for the SCT modules, since the argumentation presented above works analogously.

\begin{figure}
\begin{center}
\includegraphics[width=7.9cm,clip=true]{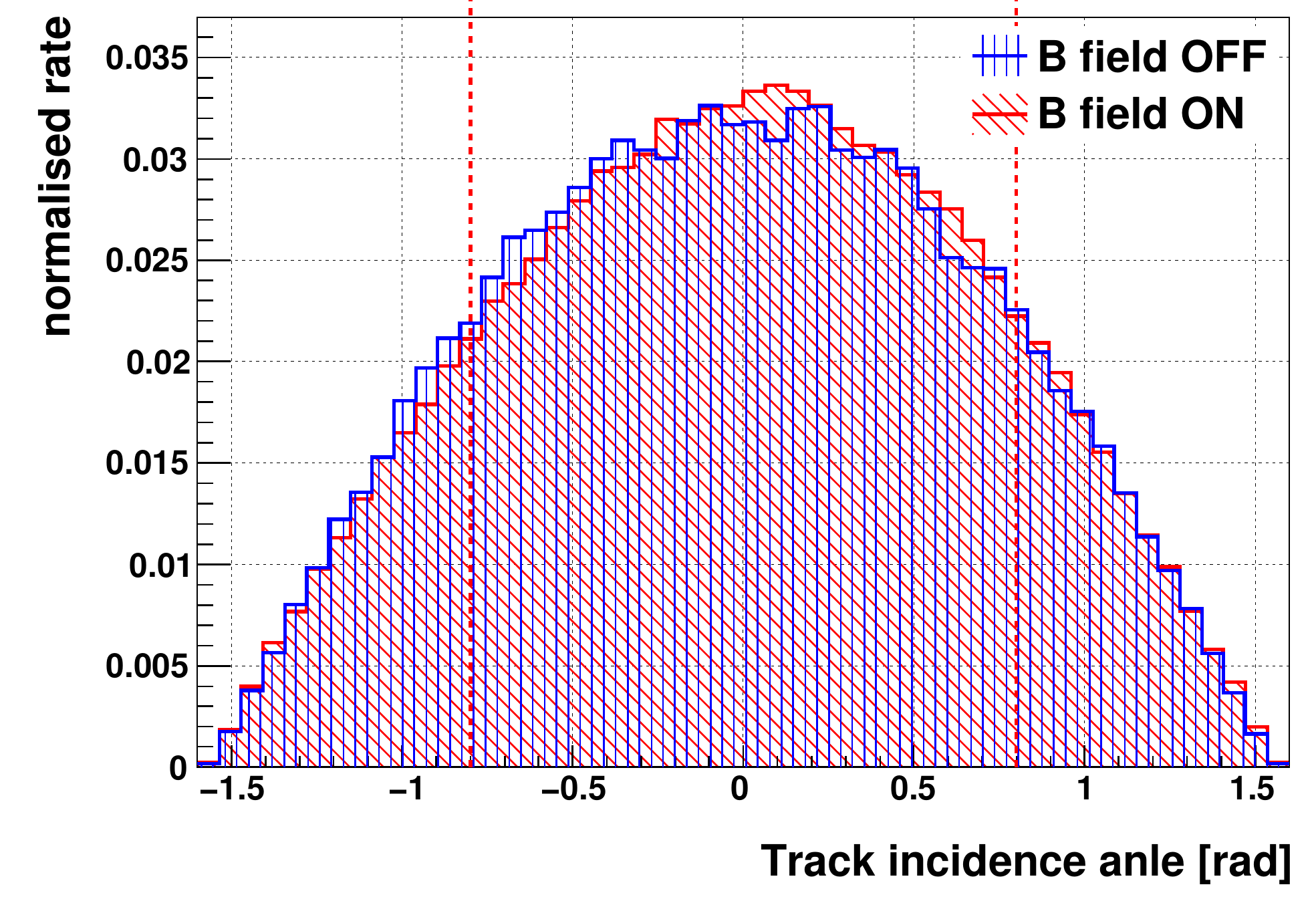}
\includegraphics[width=7.9cm,clip=true]{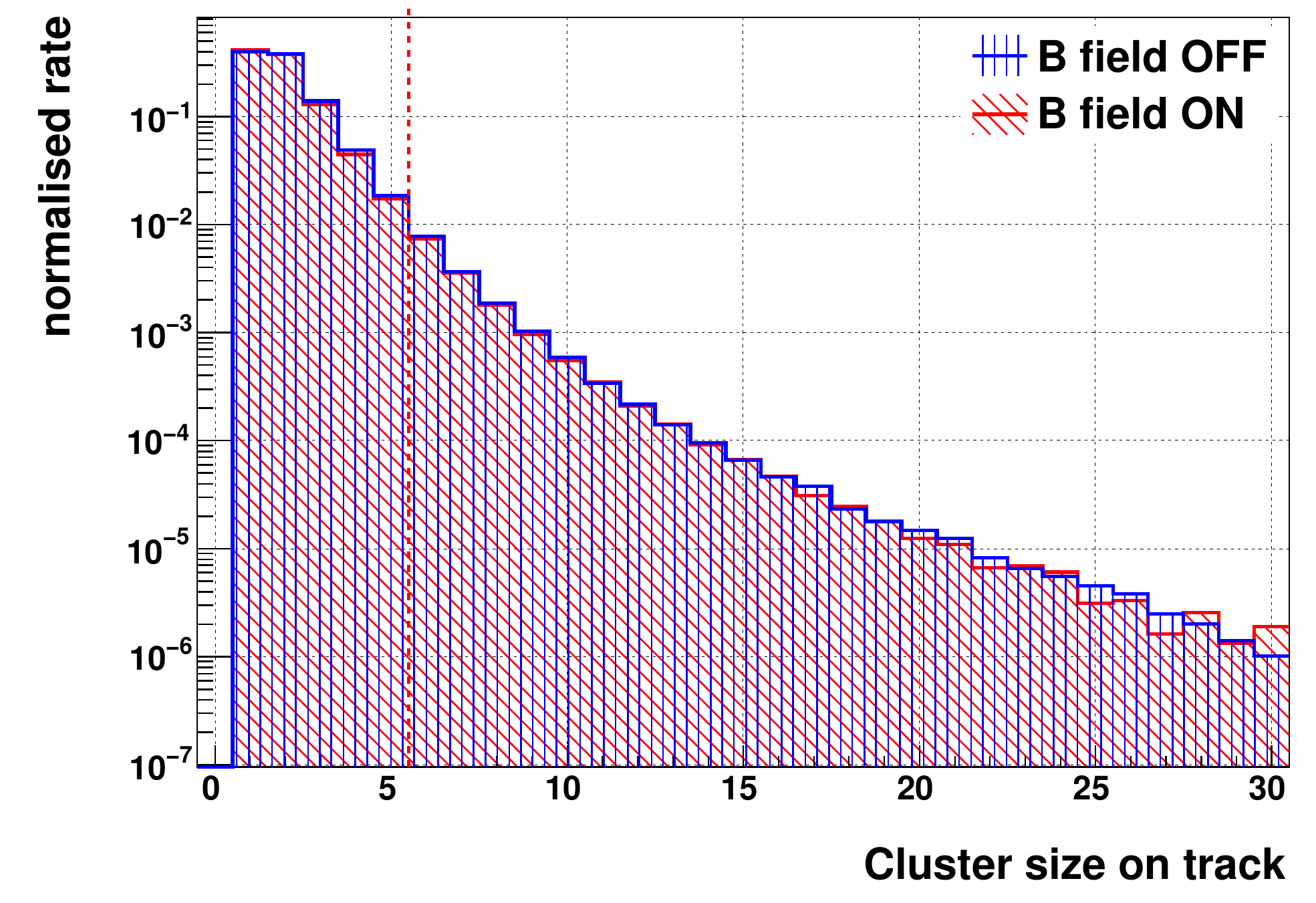}
\end{center}
\vspace{\cDist}
\caption[Transverse track incidence angle \incAngleT\ and hit-on-track cluster size \nclust\ in M8+]{\label{fig:clusterSz_incidAngle}
The transverse track incidence angle $\incAngleT$ (see text for definition) for the pixel subdetector {\bf (top left)} and the hit-on-track cluster size \nclust\ for the SCT {\bf (top right)} in M8+. The cuts $|\incAngleT|<0.8$\,rad and $\nclust<6$ are indicated as red dashed lines.
}
\end{figure}%\nopagebreak[5]

While high transverse incidence angles are not expected for tracks from LHC collisions\footnote{Particles from the extreme low-\pt\ end of the spectrum are an exception.}, the track topology of {\em cosmic ray} particles is fundamentally different from the collision scenario. In Figure~\ref{fig:clusterSz_incidAngle}~(left) the \incAngleT\ distribution is shown for M8+ together with the cut value in Equation~\ref{eqn:incAngleT}. It demonstrates, that the cut is indeed well-motivated.

For geometrical reasons, large cluster sizes $\nclust\gtrsim6$ are not expected for the selected range $\incAngleT\in[-0.8\,\rad,\,0.8\,\rad]$. If nonetheless large \nclust\ values are observed, they can be an indication of Bremsstrahlung or detector noise. In both cases, such hits do not provide a precise measurement of the $\qhit$ coordinate and should be discarded. Therefore, the cut
\begin{equation} \label{eqn:nclust}
 |\nclust| \leq 5
\end{equation}
was applied in the {\tt InDetAlignHitQualSelTool}. The cut value is a compromise between modules of the pixel and the SCT detector: if all pixels/strips traversed by an ionising particle with a maximum allowed value of $\incAngleT=0.8\,\rad$ register a signal in case of $B$-field off, we expect $\nclust\simeq7$\,pixels and $\nclust\simeq5$\,strips given the respective pitch and the depletion depth. Figure~\ref{fig:clusterSz_incidAngle}~(right) shows the cluster size found in M8+: large tails extending beyond $\nclust=30$ are observed, and the cut applied is indeed well-justified.

\begin{figure}
\begin{center}
\includegraphics[width=7.9cm,clip=true]{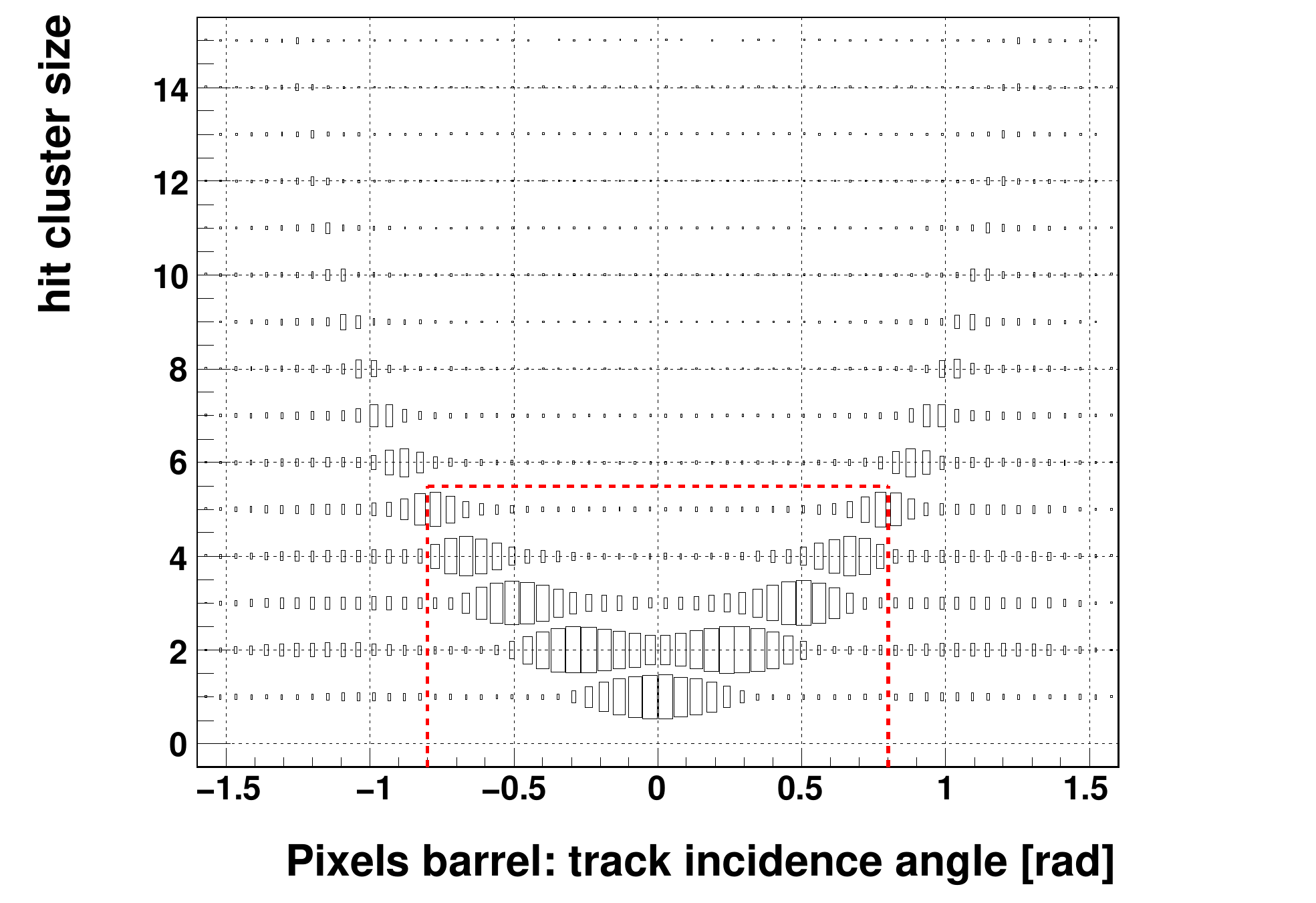}
\includegraphics[width=7.9cm,clip=true]{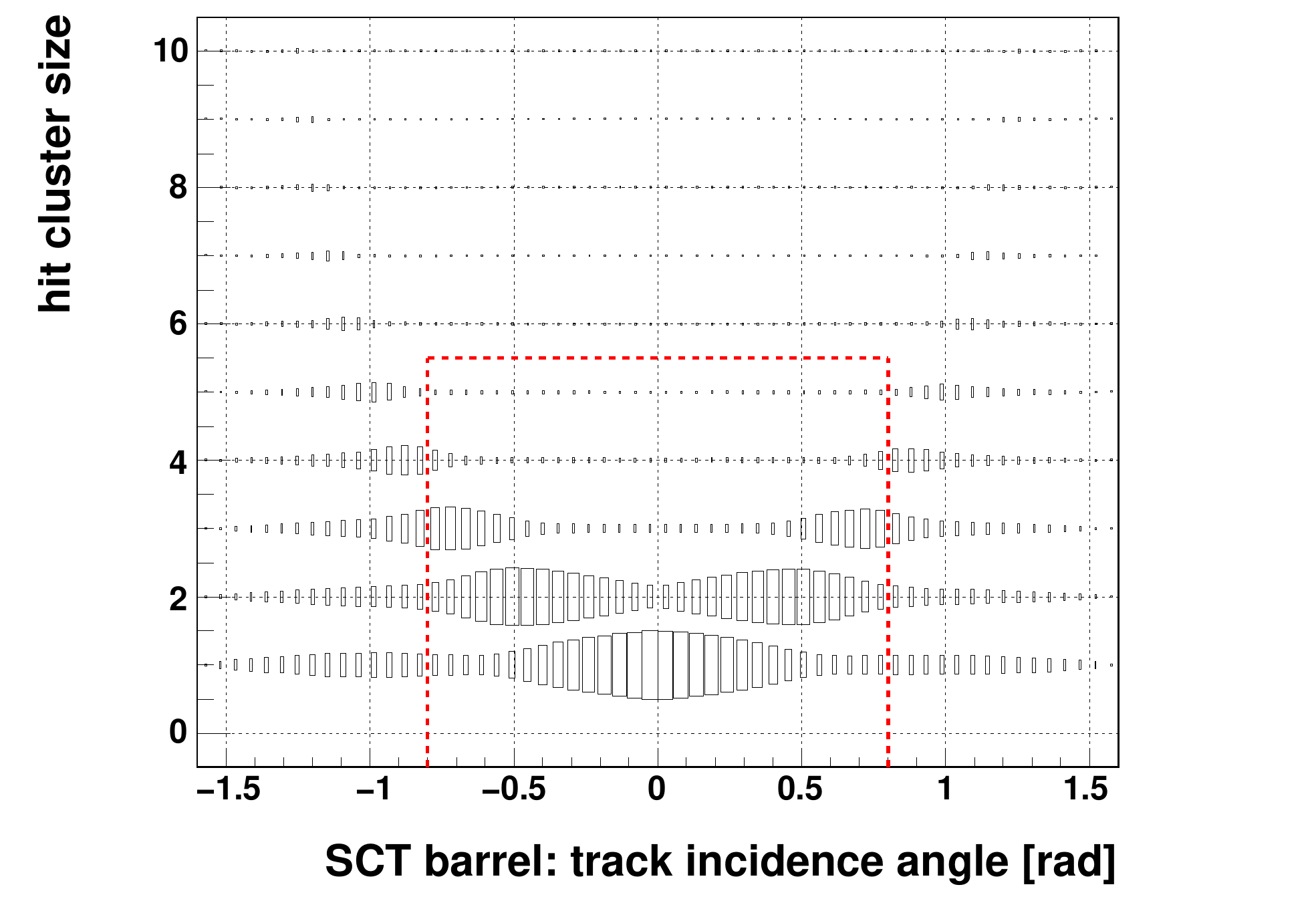}
\includegraphics[width=7.9cm,clip=true]{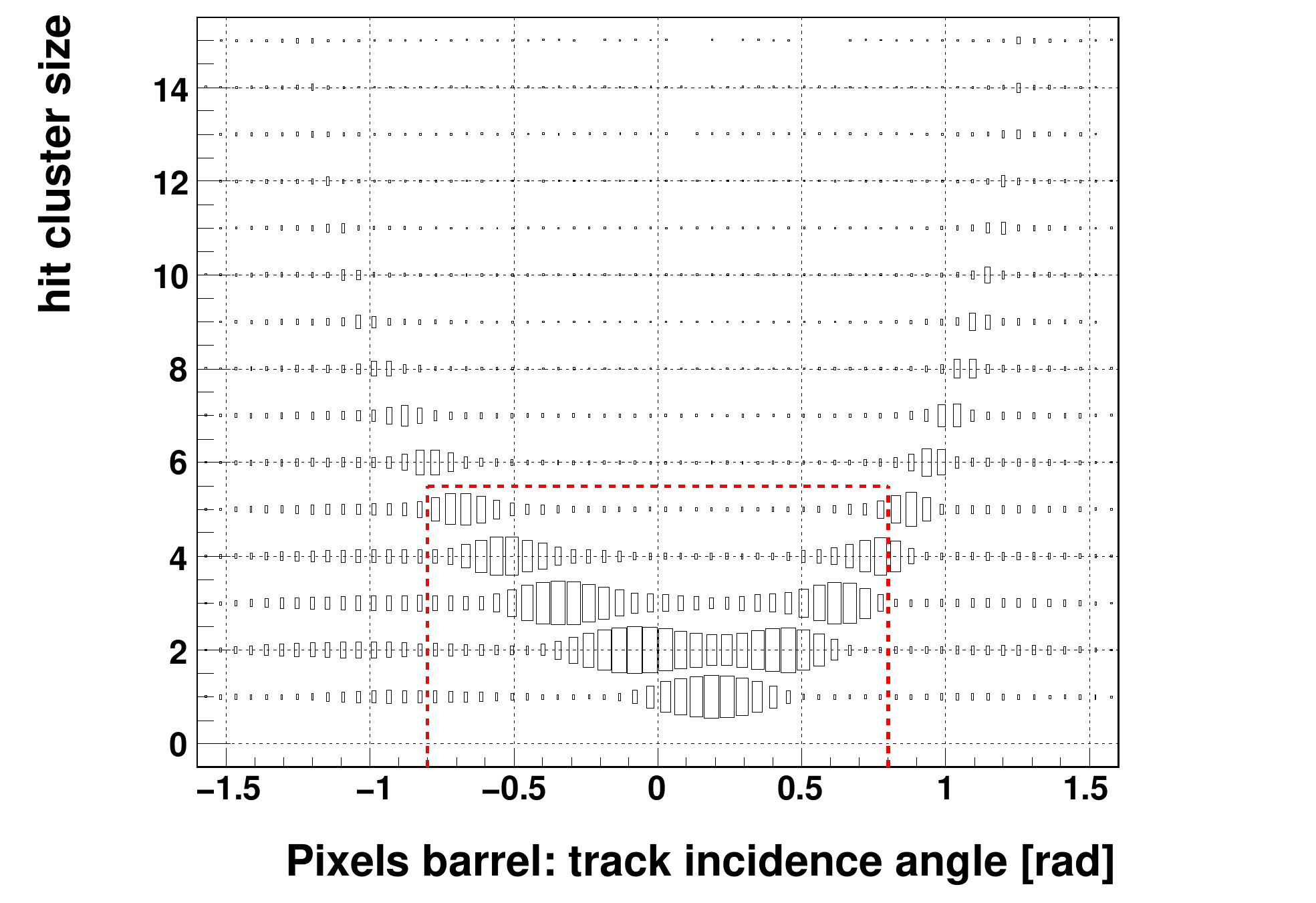}
\includegraphics[width=7.9cm,clip=true]{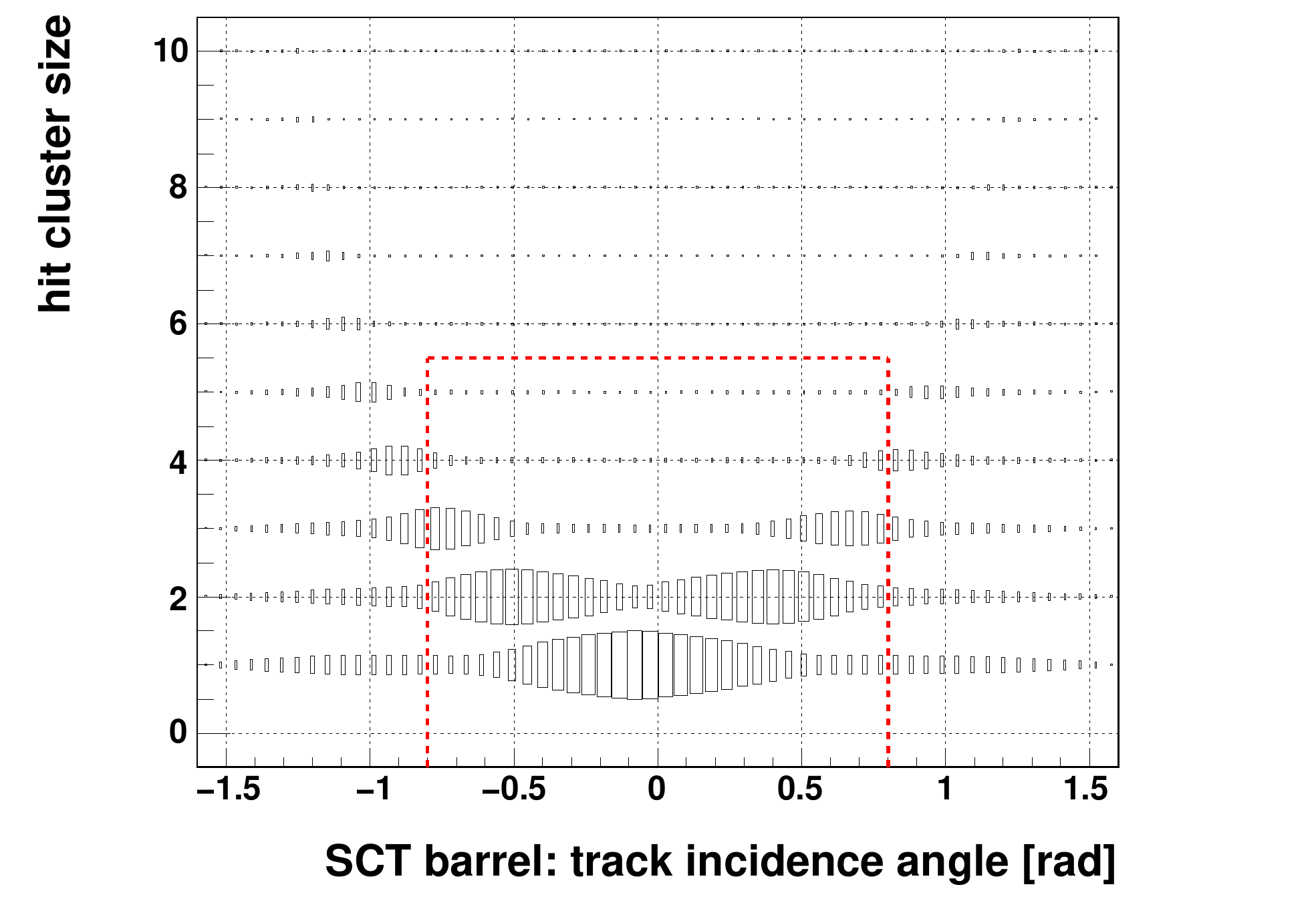}
\end{center}
\vspace{\cDist}
\caption[Hit-on-track cluster size \nclust\ versus transverse track incidence angle \incAngleT\ in M8+]{\label{fig:clusterSz_VS_incidAngle}
The hit-on-track cluster size \nclust\ versus transverse track incidence angle \incAngleT\ distribution as found in M8+ for the pixel  {\bf (top left)} and SCT {\bf (top right)} barrel detectors with $B$-field off. Analogous distributions for $B$-field on are shown below for pixel barrel {\bf (bottom left)} and SCT barrel {\bf (bottom right)}. The cuts $|\incAngleT|<0.8$\,rad and $\nclust<6$ are indicated as red dashed lines. For details see text.
}
\end{figure}%\nopagebreak[5]

A population plot in the \incAngleT\ versus \nclust\ plane is shown in Figure~\ref{fig:clusterSz_VS_incidAngle} for the pixel and SCT detector modules:
\begin{description}
 \item[$B$-field off:] 
 %as shown in the top row of the Figure, 
 a parabola-shaped area of increased population centred about $\incAngleT\equiv0$ is observed. This area corresponds to the cases where all pixels/strips fired which were actually traversed by the ionising particle. A relevant observation is that for high $|\incAngleT|$ values beyond $\sim$1\,rad an increase in the population of low $\nclust$ values under the parabola is found, which is due to split clusters. Moreover, for $\incAngleT\in[-0.8\,\rad,\,0.8\,\rad]$ the bins of large $\nclust$ are almost uniformly populated in $\incAngleT$, reflecting the nearly uniform probability to produce Bremsstrahlung;
 \item[$B$-field on:] 
 %as shown in the bottom row of the Figure, 
 this case displays a qualitatively similar picture to the $B$-field off scenario. The major difference is that now the parabola-shaped area of increased population is skewed, and its $\nclust=1$ bin is centred about $\incAngleT\simeq0.2\,\rad$/$-0.07\,\rad$ for the pixel/SCT detector modules, respectively. This is due to the Lorentz angle, which has an opposite sign for pixel and SCT sensors because of different charge carriers.
\end{description}
Ideally, rather than applying the somewhat simplistic cuts in Equations~\ref{eqn:incAngleT} and \ref{eqn:nclust}, a dynamic cut on \incAngleT\ depending on \nclust\ should be applied, which could be defined as containing 65\% of the entries around the two parabola branches in the given \nclust\ bin. This suggestion was discussed by the author with the ATLAS colleagues working on the relevant subdetectors, and positive feedback was received~\cite{bib:privateAttilio,bib:privateRichard}. Since a  preliminary alignment set was needed before Christmas 2008, these plans could not be implemented due to the extremely tight time scale. However, there are plans to work on it by a student from Krakow in the near future~\cite{bib:privatePawel}. 

\subsubsection{Hit Error $\dqhit_x$ Dependance on Transverse Incidence Angle and Cluster Size}

\begin{figure}
\begin{center}
\includegraphics[width=7.9cm,clip=true]{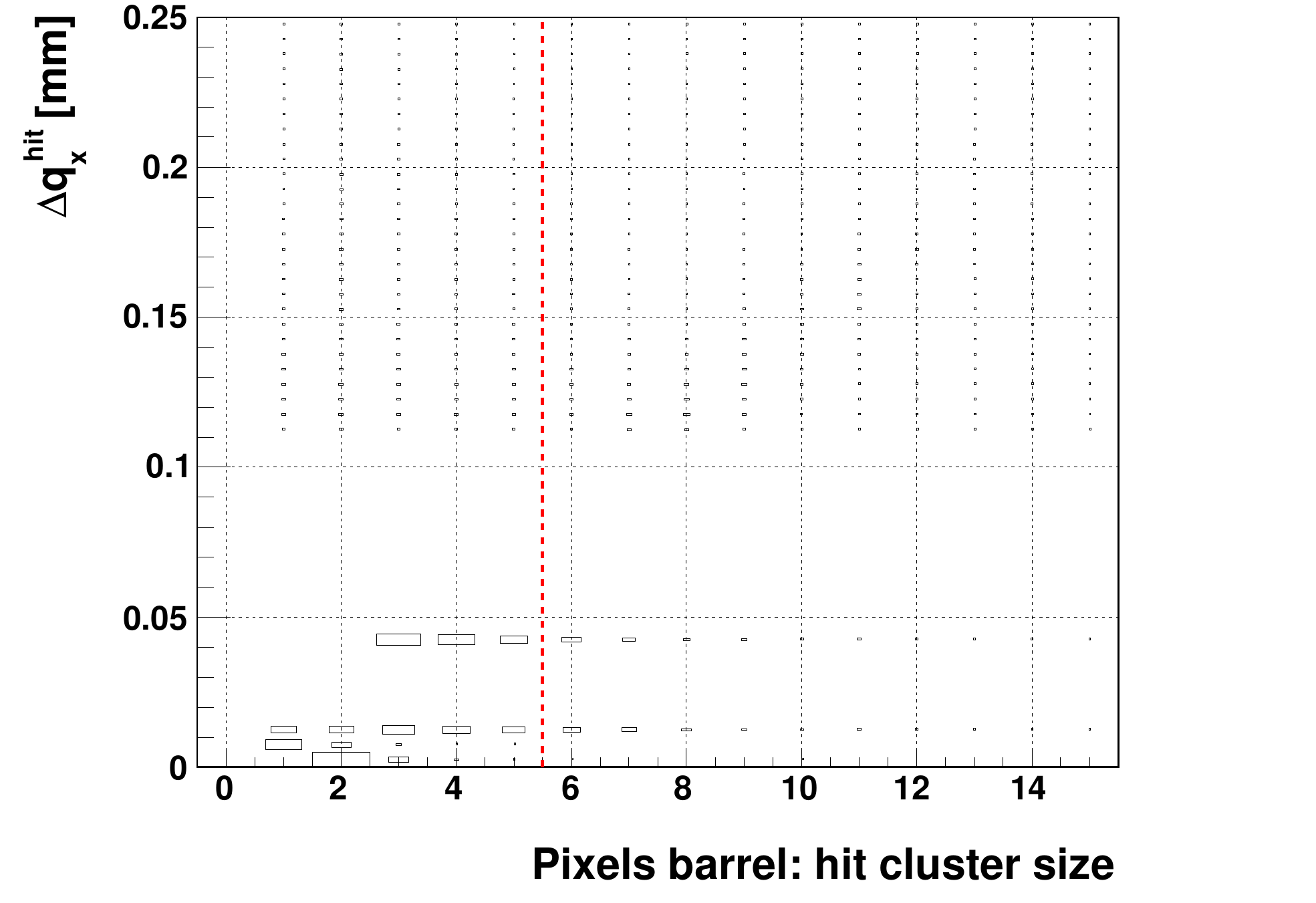}
\includegraphics[width=7.9cm,clip=true]{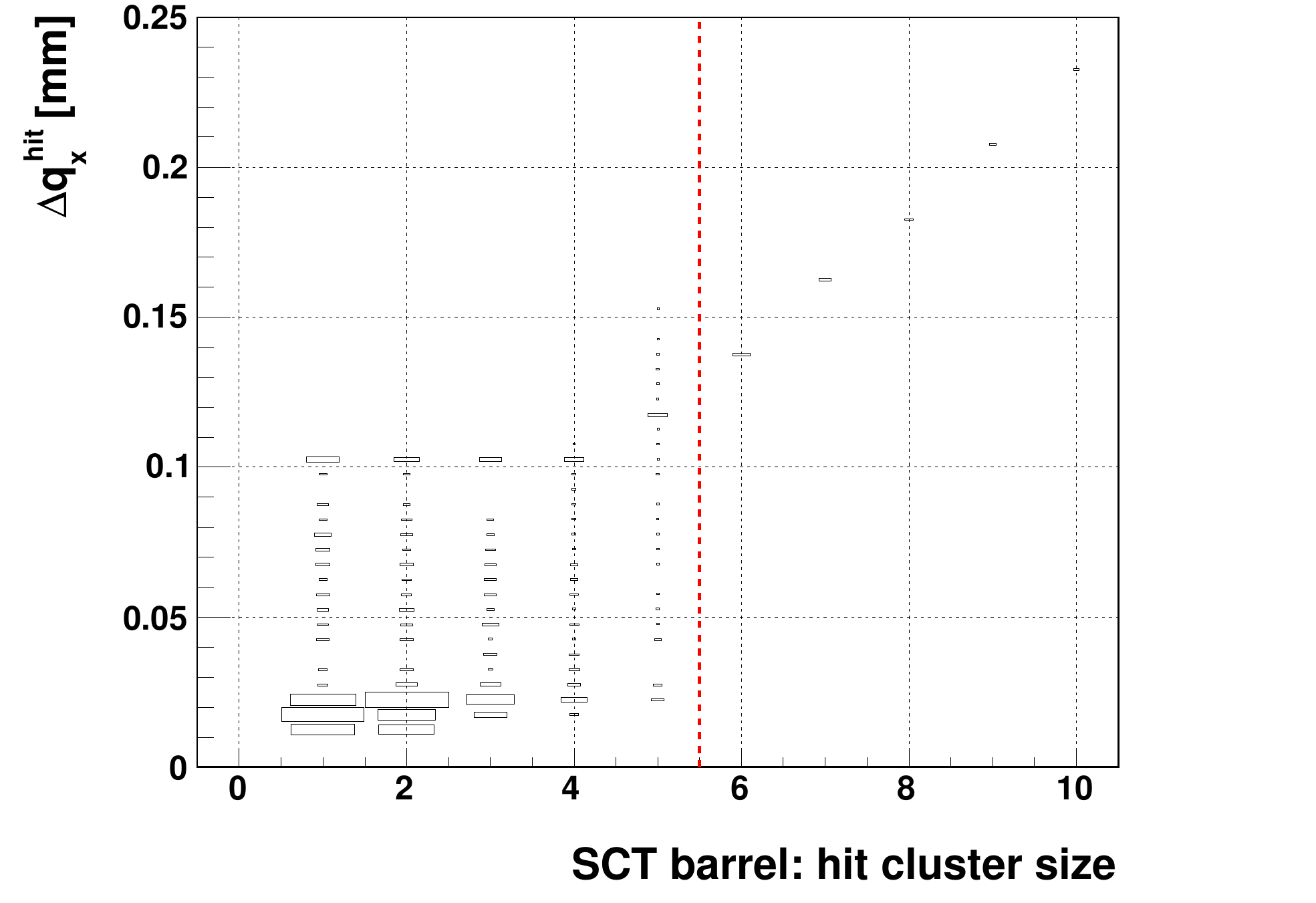}
\includegraphics[width=7.9cm,clip=true]{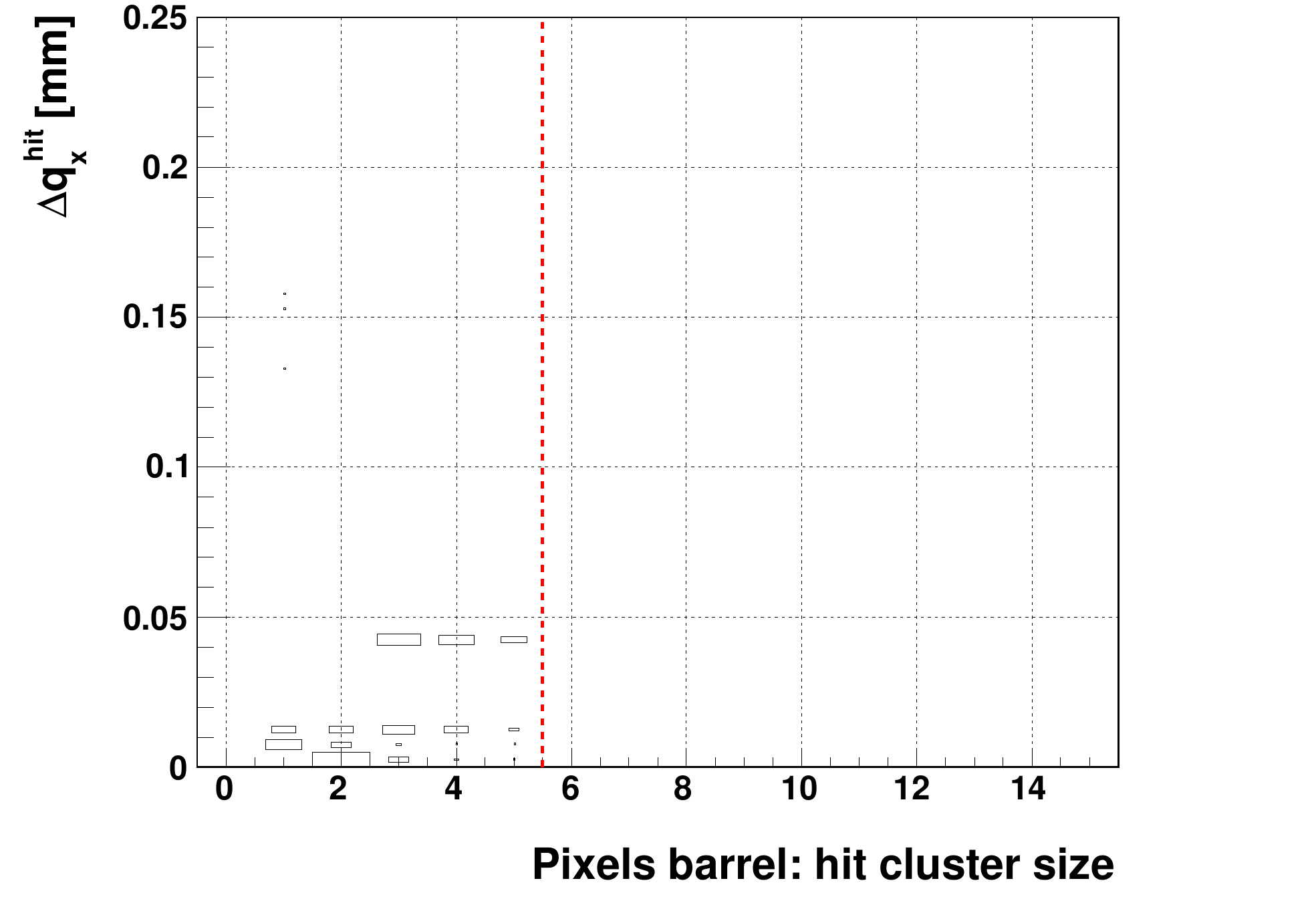}
\includegraphics[width=7.9cm,clip=true]{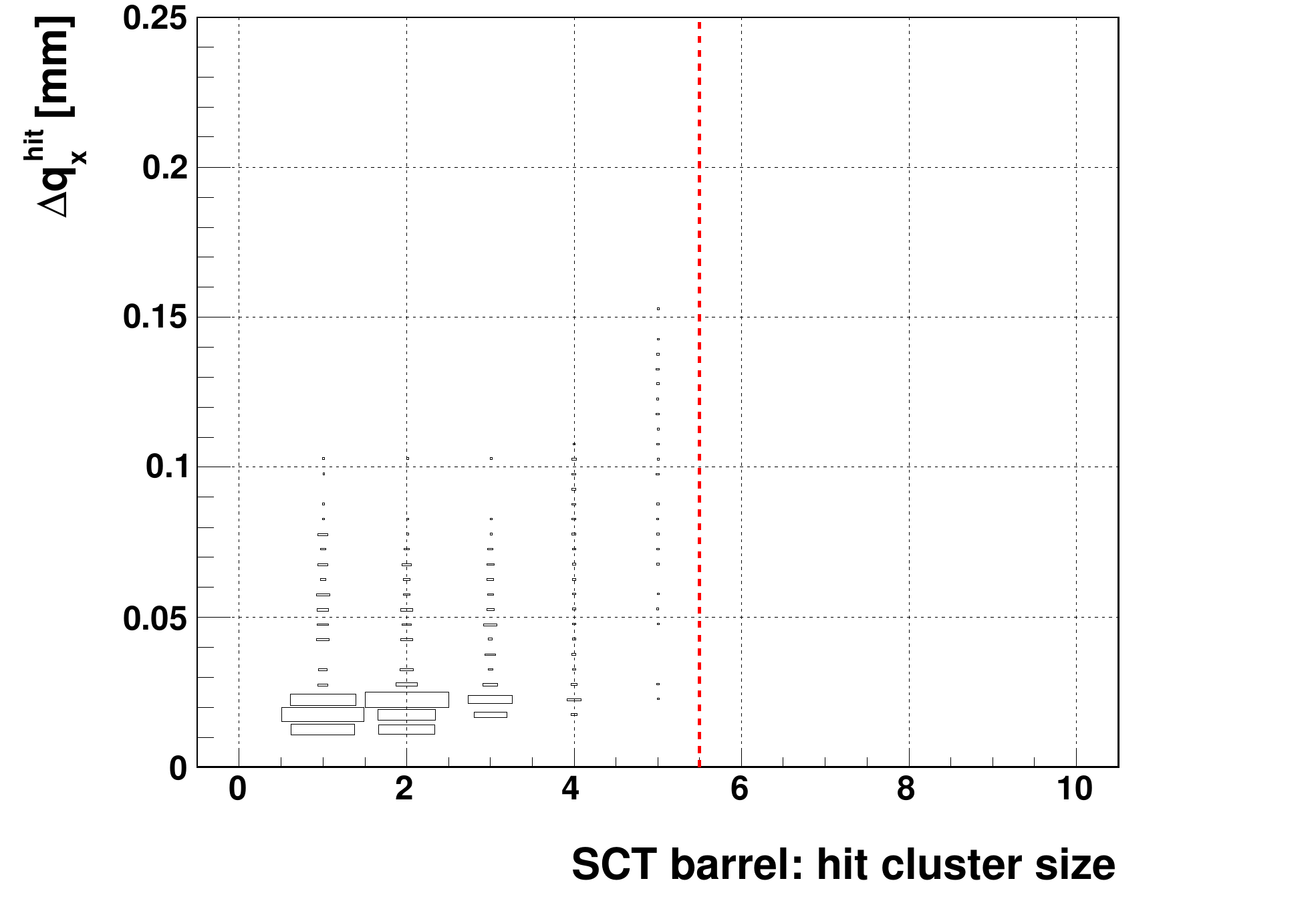}
\vspace{\cDist}
\end{center}
\caption[Hit-on-track uncertainty $\dqhit_x$ versus cluster size \nclust\ with $B$-field on in M8+]{\label{fig:hitQual_VS_clusterSz}
Hit-on-track uncertainty $\dqhit_x$ versus hit-on-track cluster size \nclust\  before any hit quality requirements as found in M8+ for the pixel {\bf (left)} and SCT {\bf (right)} barrel detectors with $B$-field on. Analogous distributions with hit quality requirements are shown below for pixel barrel {\bf (bottom left)} and SCT barrel {\bf (bottom right)}. The cut $\nclust<6$ is indicated as a red dashed line. For details see text.
}
\end{figure}%\nopagebreak[5]

The main goal of the {\tt InDetAlignHitQualSelTool} is to provide a selection of ``clean'' hits in order to reject pathological cases not suitable for track-based alignment. A nice side effect is that it {\em indirectly} reduces the hit uncertainty $\dqhit_x$ provided by the reconstruction. The dependance of $\dqhit_x$ on both the cluster size $\nclust$ and the transverse incidence angle $\incAngleT$ shall be discussed here. 

\begin{figure}
\begin{center}
\includegraphics[width=7.9cm,clip=true]{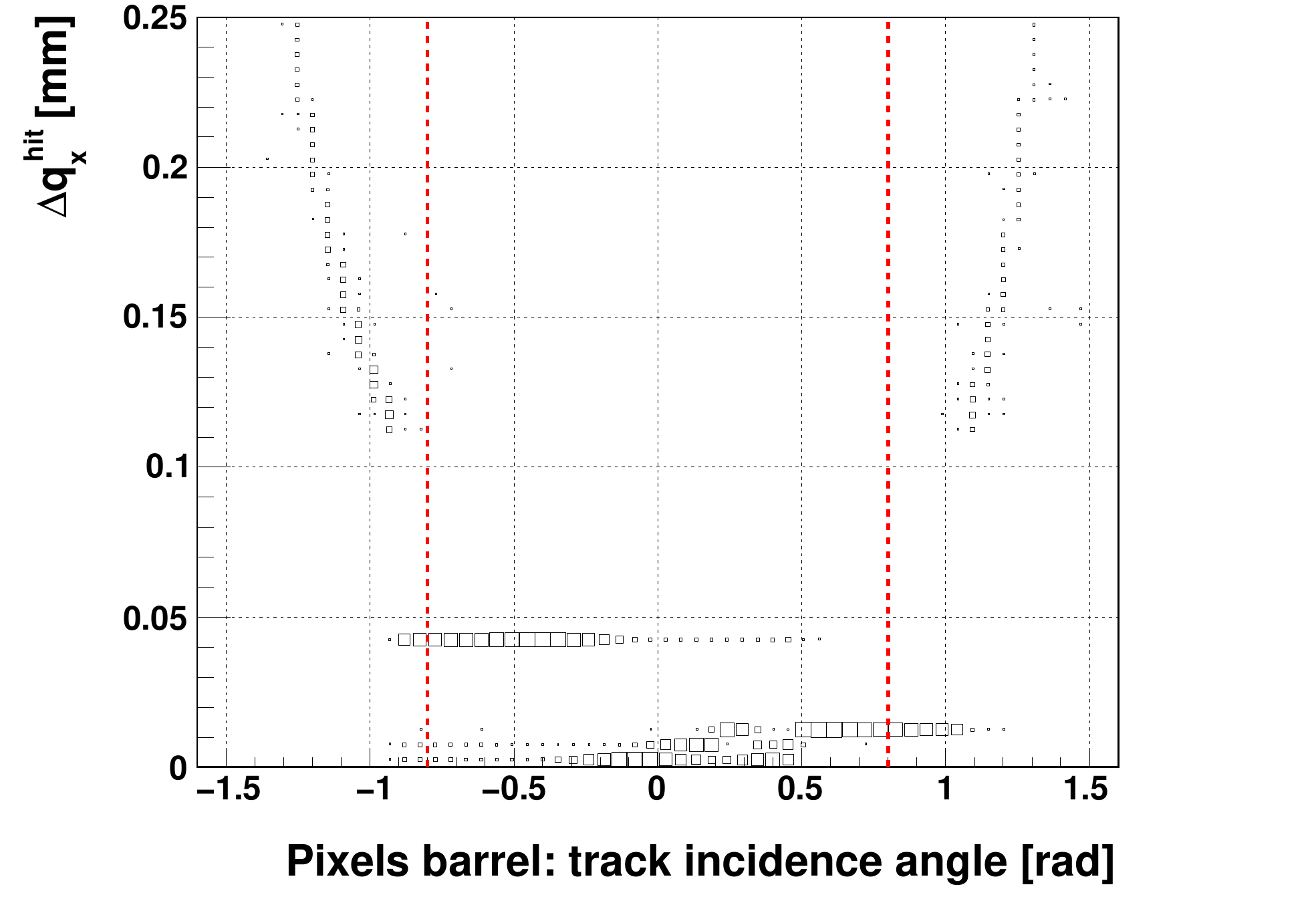}
\includegraphics[width=7.9cm,clip=true]{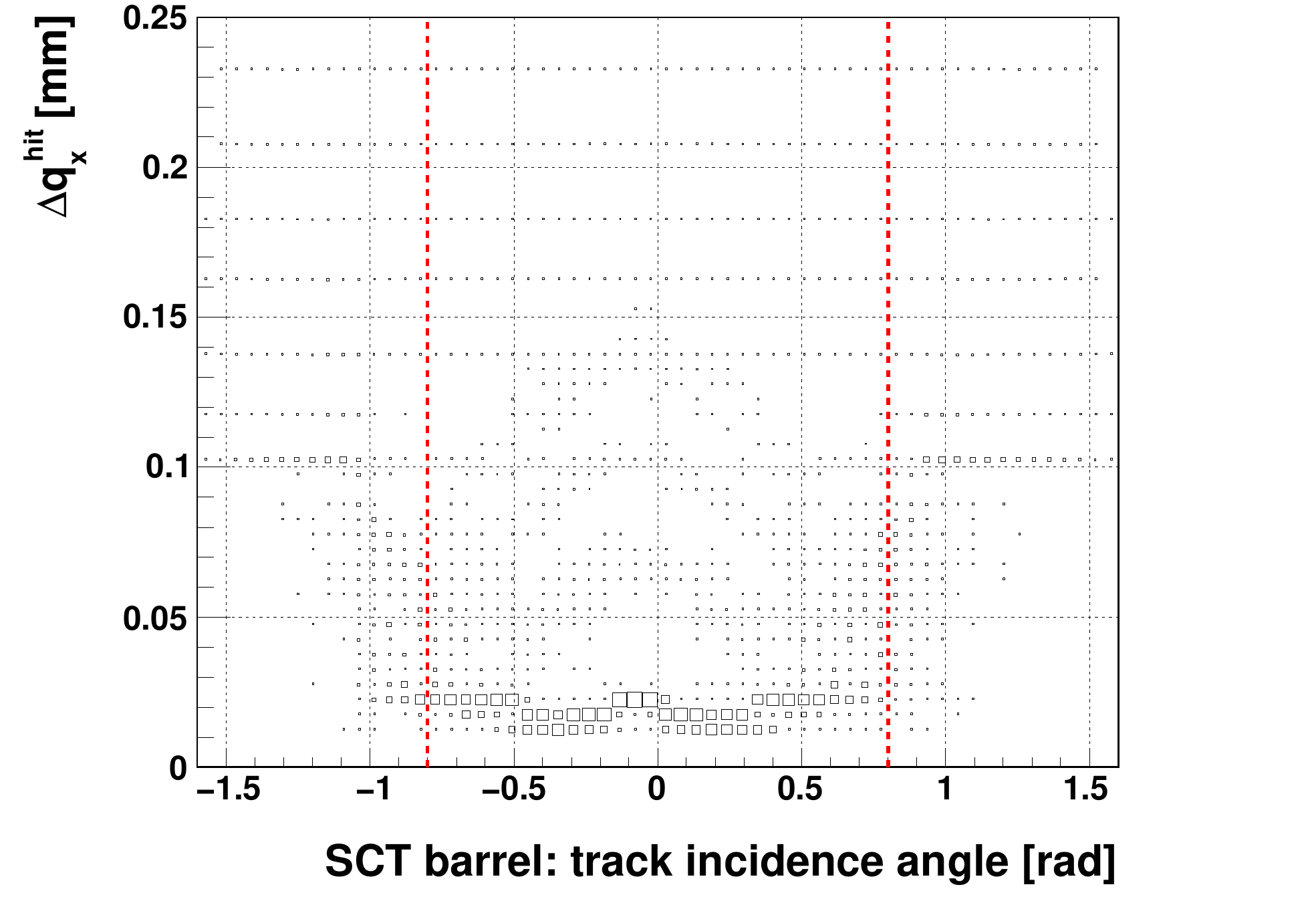}
\includegraphics[width=7.9cm,clip=true]{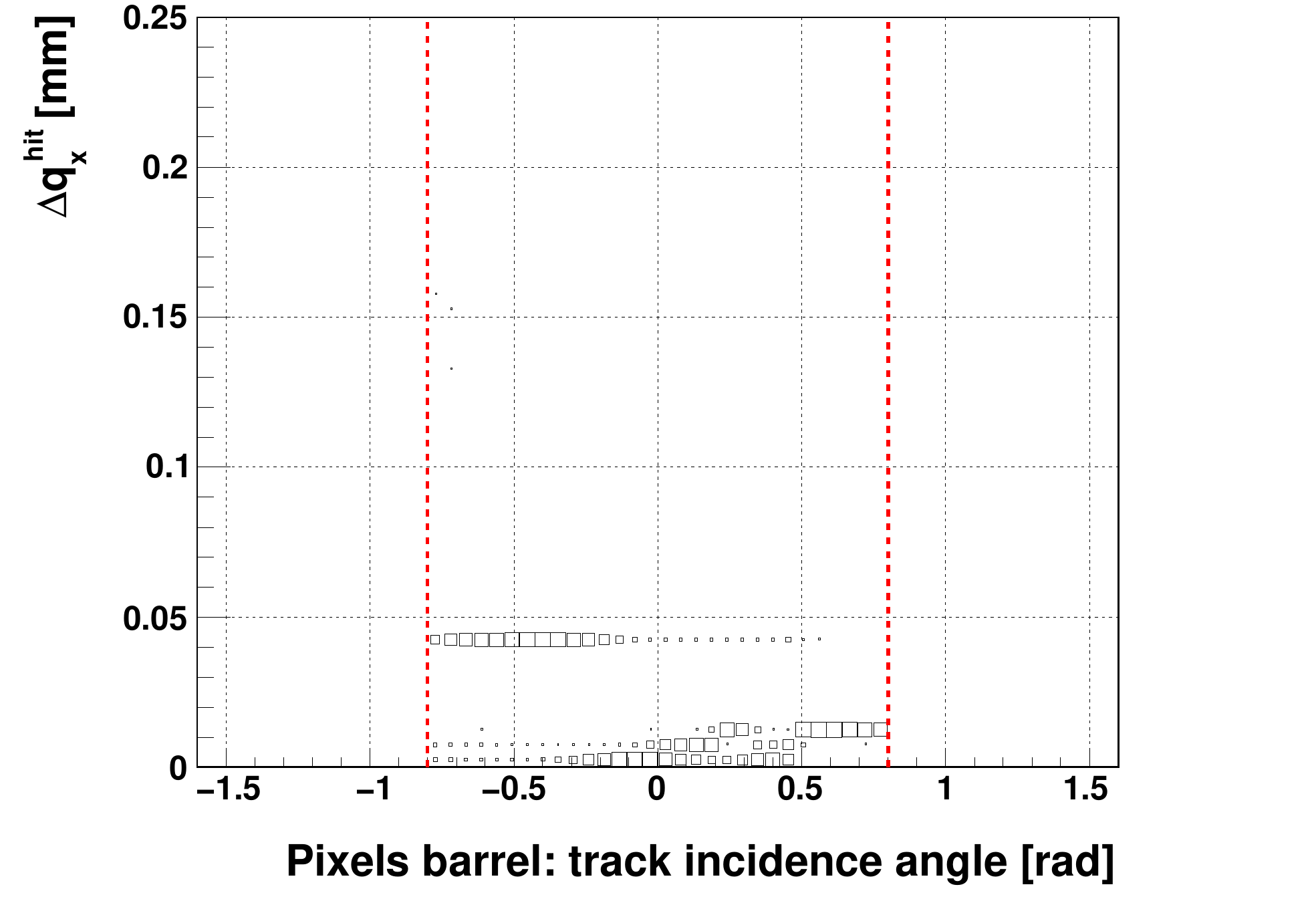}
\includegraphics[width=7.9cm,clip=true]{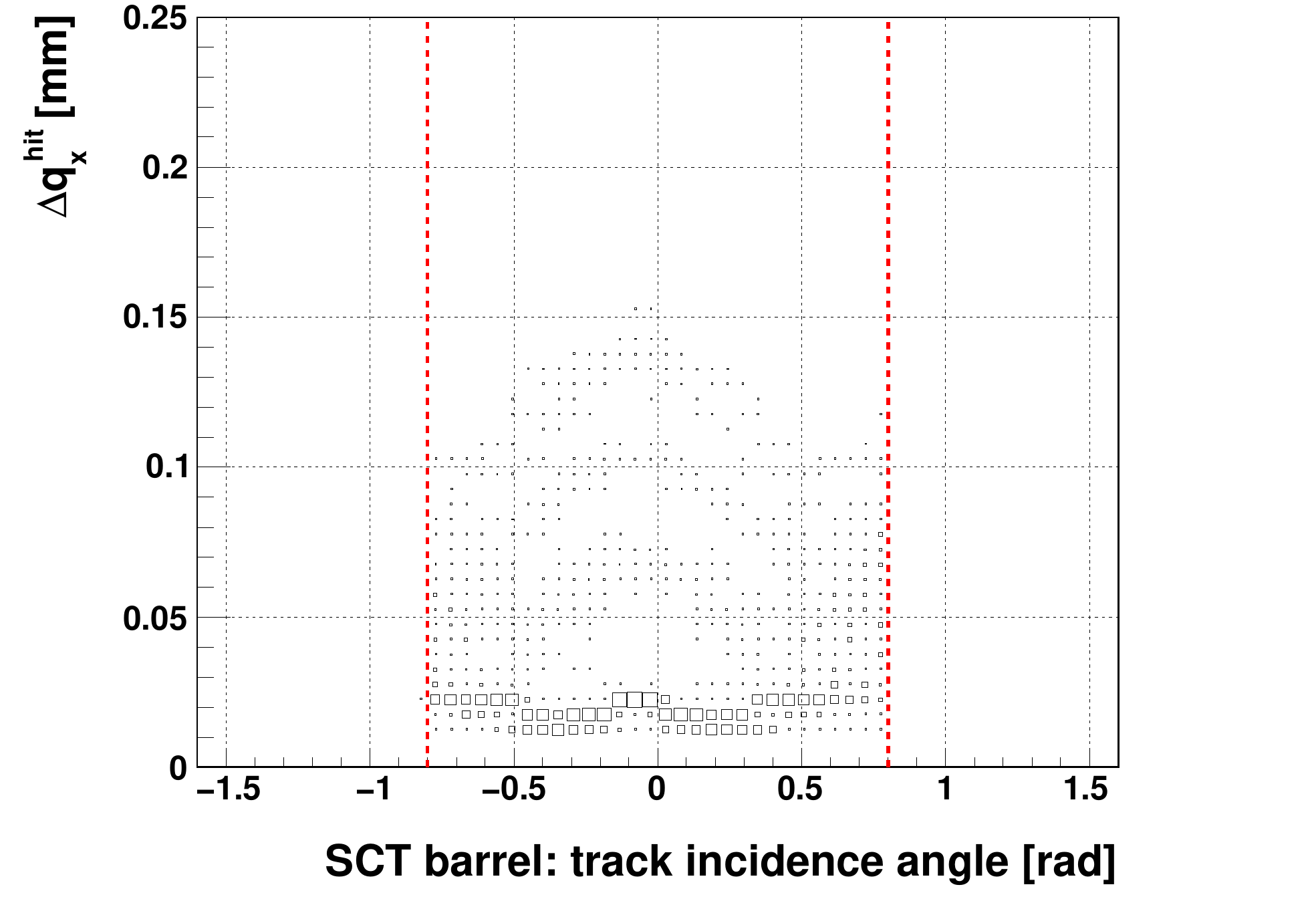}
\vspace{\cDist}
\end{center}
\caption[Hit-on-track uncertainty $\dqhit_x$ versus transverse track incidence angle \incAngleT\ with $B$-field on in M8+]{\label{fig:hitQual_VS_incidAngle}
Hit-on-track uncertainty $\dqhit_x$ versus transverse track incidence angle \incAngleT\  before any hit quality requirements as found in M8+ for the pixel {\bf (left)} and SCT {\bf (right)} barrel detectors with $B$-field on. Analogous distributions with hit quality requirements are shown below for pixel barrel {\bf (bottom left)} and SCT barrel {\bf (bottom right)}. The cut $|\incAngleT|<0.8$\,rad is indicated as red dashed lines. For details see text.
}
\end{figure}

The population plot of $\dqhit_x$ versus $\nclust$ is shown in Figure~\ref{fig:hitQual_VS_clusterSz} for  modules in the barrel of the pixel and the SCT detectors with $B$-field on:
\begin{description}
 \item[Pixel sensors:] the plot before hit quality requirements shows a large accumulation of entries at $\dqhit_x\simeq13\,\mum$ which correspond to hits with $\incAngleT$ near \lorentz. Further, there are some hits with $\dqhit_x\simeq5\,\mum$ for $\nclust=2$ hits where the particle is believed to have passed through the boundary between two pixels. The entries at about 45\,\mum\ are due to hits with negative incidence angles $\incAngleT\lesssim-0.2\,\rad$, for which no dedicated error parameterisation was derived, as they are not expected in $p$-$p$ collisions~\cite{bib:privateAttilio}. Notably, there is an almost uniform low-density population beyond $\dqhit_x\gtrsim100\,\mum$, whcih is due to hits with $|\incAngleT|\gtrsim1\,\rad$ (cf. Figure~\ref{fig:hitQual_VS_incidAngle}). After the application of cuts in Equations~\ref{eqn:incAngleT}, \ref{eqn:nclust} 
% which is shown in the bottom row of the Figure, 
not only the entries beyond the \nclust\ cut are (trivially) cleared away, but also the ones at $\dqhit_x\gtrsim100\,\mum$;%, which supports the hypothesis above;
 \item[SCT sensors:] the situation is qualitatively similar to the pixel case: there is a strong population of entries at about $\dqhit_x\simeq20\,\mum$ with small $\nclust$ values which correspond to well-understood \incAngleT\ values. However, there is also a notable accumulation of values with $\dqhit_x\sim100\,\mum$ and small \nclust, which is due to high transverse incidence angles (cf. Figure~\ref{fig:hitQual_VS_incidAngle}). Moreover, there is a distinct tail with $\dqhit_x\propto\nclust$: for  $\nclust\gtrsim5$ the error parameteresation $\dqhit_x=d_{\rm pitch}/\sqrt{12}\cdot\nclust$ is used due to the binary nature of the SCT readout~\cite{bib:privateRichard}. The application of hit quality cuts results not only in a (trivial) removal of values with $\nclust\geq6$, but also the clearing of values with $\dqhit_x\sim100\,\mum$ which are due to $|\incAngleT|\gtrsim1\,\rad$.
\end{description}

%\nopagebreak[5]

The interplay of $\dqhit_x$ and $\incAngleT$ is shown as a population plot in Figure~\ref{fig:hitQual_VS_incidAngle} for  modules in the barrel of the pixel and the SCT detectors with $B$-field on:
\begin{description}
 \item[Pixel sensors:] the plot before hit quality requirements displays some familiar features in $\dqhit_x$ already discussed above (cf. Figure~\ref{fig:hitQual_VS_clusterSz}). The entries at small values of $\dqhit_x$ tend to stay inside the selected range $\incAngleT\in[-0.8\,\rad,\,0.8\,\rad]$. The effective path length $\effPath$ is different for $\incAngleT~^>\!\!\!_<~\lorentz$ as one moves away from the minimum in $\dqhit_x$ around $\incAngleT\equiv\lorentz$, which results in an assymmetric population of the plot~\cite{bib:privateAttilio}. 
 %As already mentioned, entries at $\dqhit_x\simeq45\,\mum$ and $\incAngleT\lesssim-0.2\,\rad$ indicate a lack of a dedicated error parameteresation for these hits. 
 Most notably, there is a parabola-shaped enhancement of population beyond $\dqhit_x\gtrsim110\,\mum$, $|\incAngleT|\gtrsim0.9\,\rad$ which is cleared away by the hit quality requirements in Equations~\ref{eqn:incAngleT} and \ref{eqn:nclust};
 \item[SCT sensors:] similarly to the pixel case, there is a strong population of entries at about $\dqhit_x\simeq20\,\mum$ with $\incAngleT\in[-0.8\,\rad,\,0.8\,\rad]$. Strikingly, a distribution of $\dqhit_x$ entries is observed above $\dqhit\gtrsim100\,\mum$ which is equidistant in $\dqhit_x$ and almost uniformely distributed in $\incAngleT$ (cf. Figure~\ref{fig:hitQual_VS_clusterSz} and discussion above). These entries are removed by the hit quality requirements specified in Equations~\ref{eqn:incAngleT} and \ref{eqn:nclust}, as are the flanks at $|\incAngleT|>0.8\,\rad$.
\end{description}

\begin{figure}
\begin{center}
\includegraphics[width=7.9cm,clip=true]{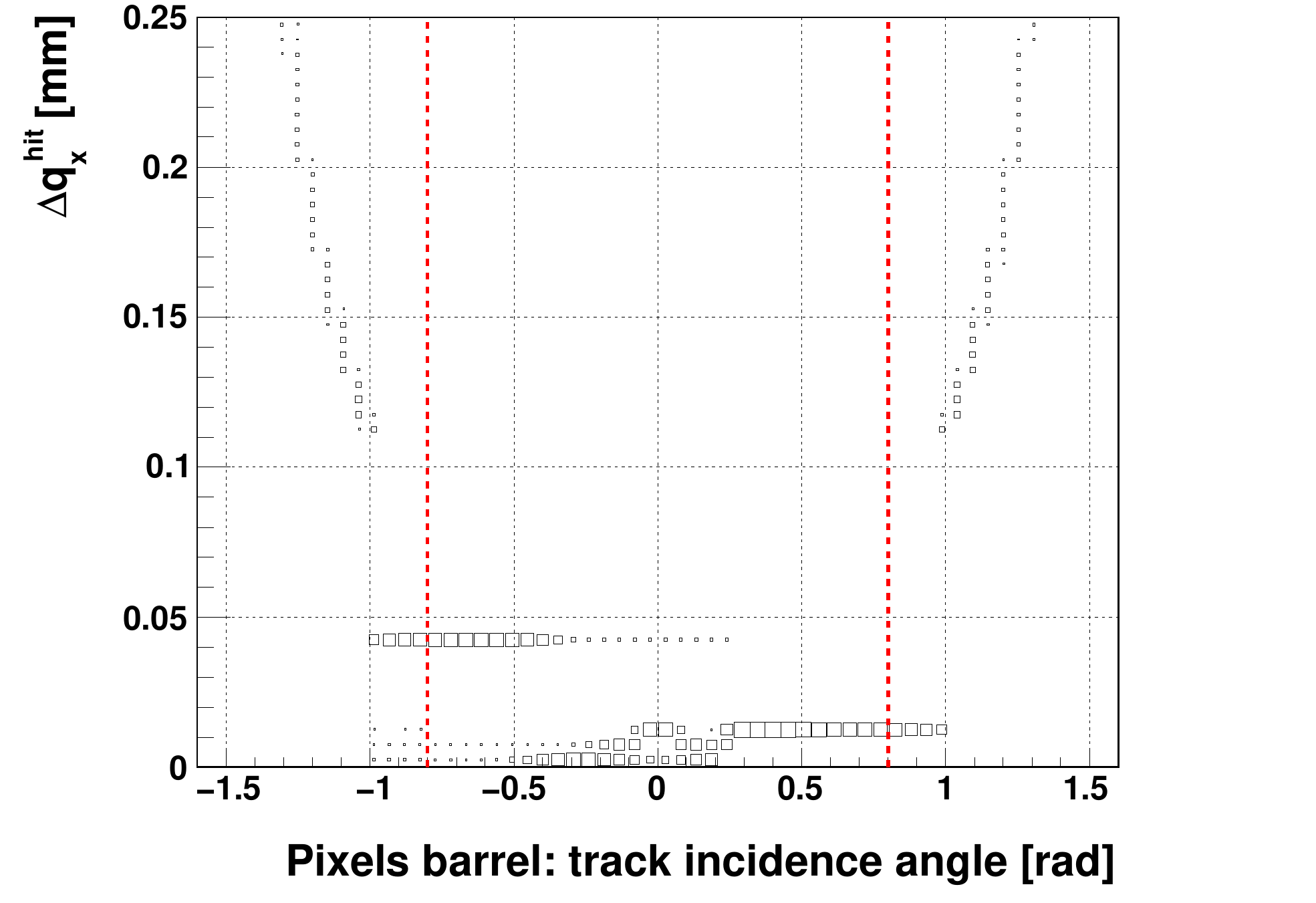}
\includegraphics[width=7.9cm,clip=true]{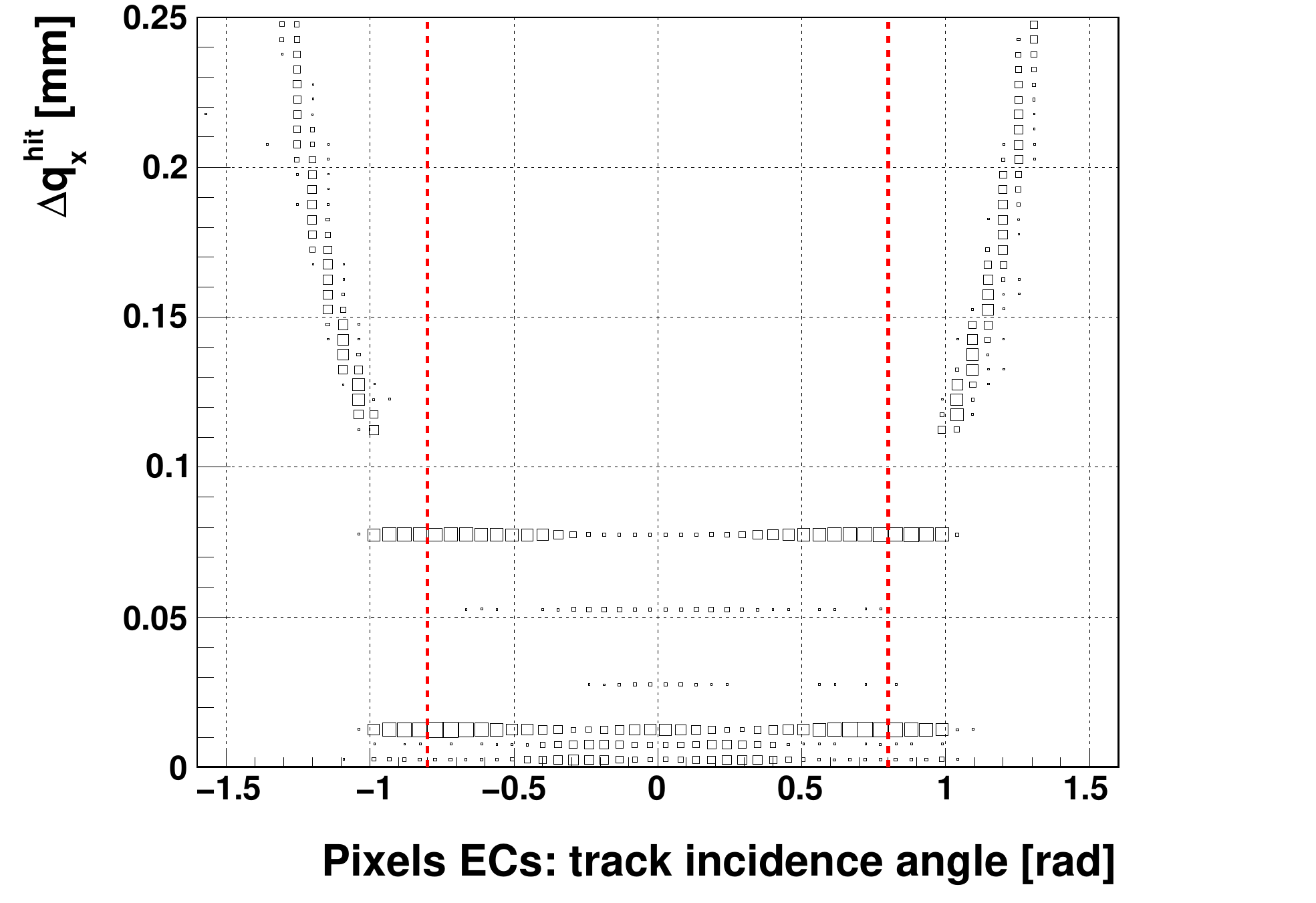}
\vspace{\cDist}
\end{center}
\caption[Hit-on-track uncertainty $\dqhit_x$ versus transverse track incidence angle \incAngleT\ with $B$-field off in M8+]{\label{fig:hitQual_VS_incidAngle_B0}
Hit-on-track uncertainty $\dqhit_x$ versus transverse track incidence angle \incAngleT\  before any hit quality requirements as found in M8+ for the pixel barrel {\bf (left)} and end-cap {\bf (right)} detectors with $B$-field off. The cut $|\incAngleT|<0.8$\,rad is indicated as red dashed lines. For details see text.
}
\end{figure}%\nopagebreak[5]

The case with $B$-field off was also investigated. While the distribution in the SCT sensors matched the expectations from Figures~\ref{fig:hitQual_VS_clusterSz} and \ref{fig:hitQual_VS_incidAngle}, an inconsistency was observed in the $\dqhit_x$ versus $\incAngleT$ population plot for the pixel sensors, which was found to be {\em asymmetric}. The corresponding plot is shown in Figure~\ref{fig:hitQual_VS_incidAngle_B0}~(left). This observation was pointed out to the pixel experts: the reason is the fact that the same error calibration is applied for $B$-field off hits which was derived for the $B$-field on case~\cite{bib:privateAttilio}. As a preliminary solution, it was suggested to use the error calibration derived for the pixel end-caps, where the $B$-field vector is parallel to the ionisation charge travel direction inside the sensors. The corresponding $\dqhit_x$ versus $\incAngleT$ population plot is shown in Figure~\ref{fig:hitQual_VS_incidAngle_B0}~(right).

\subsubsection{The Results of the Hit Quality Requirements via {\tt InDetAlignHitQualSelTool}}

\begin{figure}
\begin{center}
\includegraphics[width=7.9cm,clip=true]{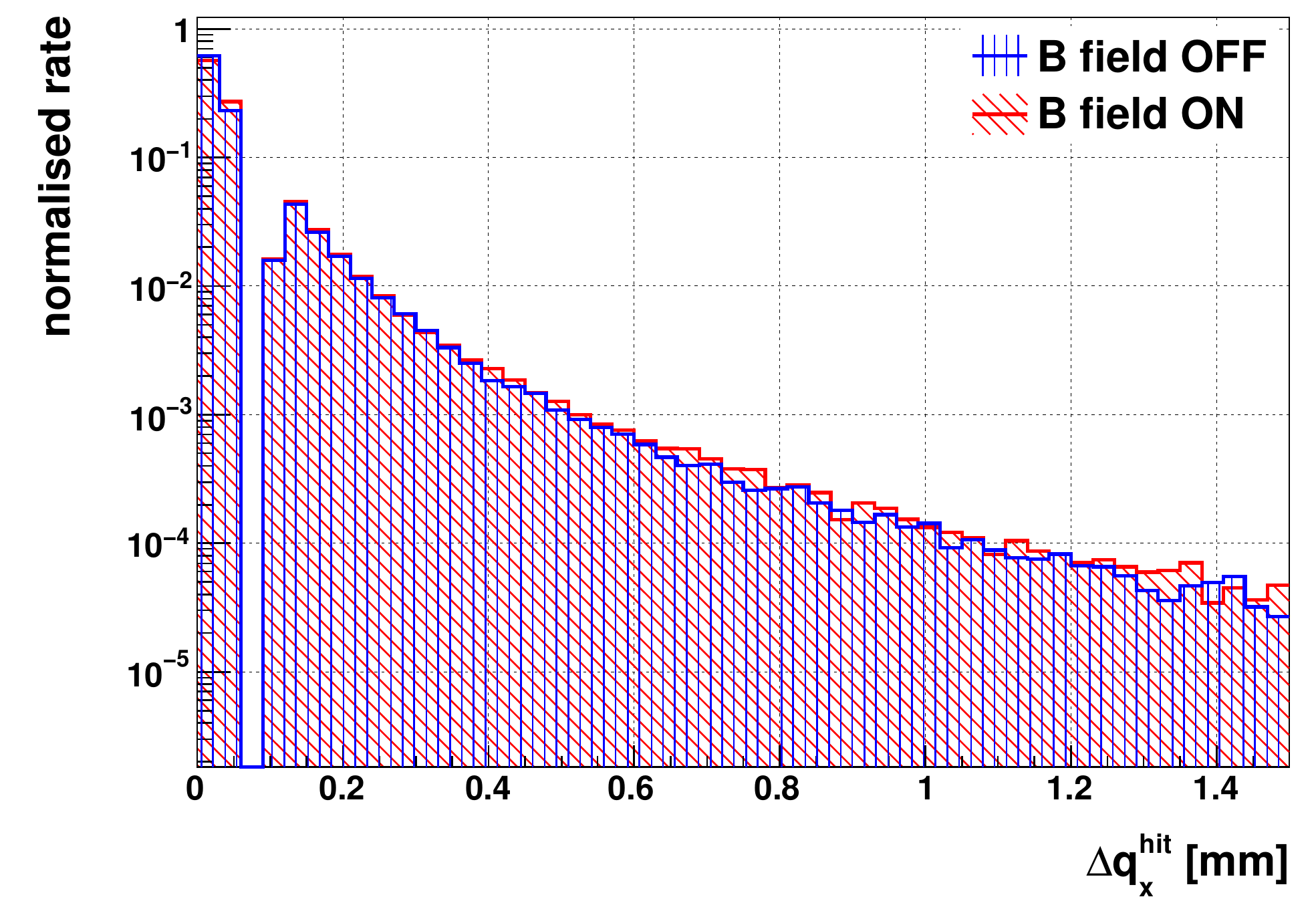}
\includegraphics[width=7.9cm,clip=true]{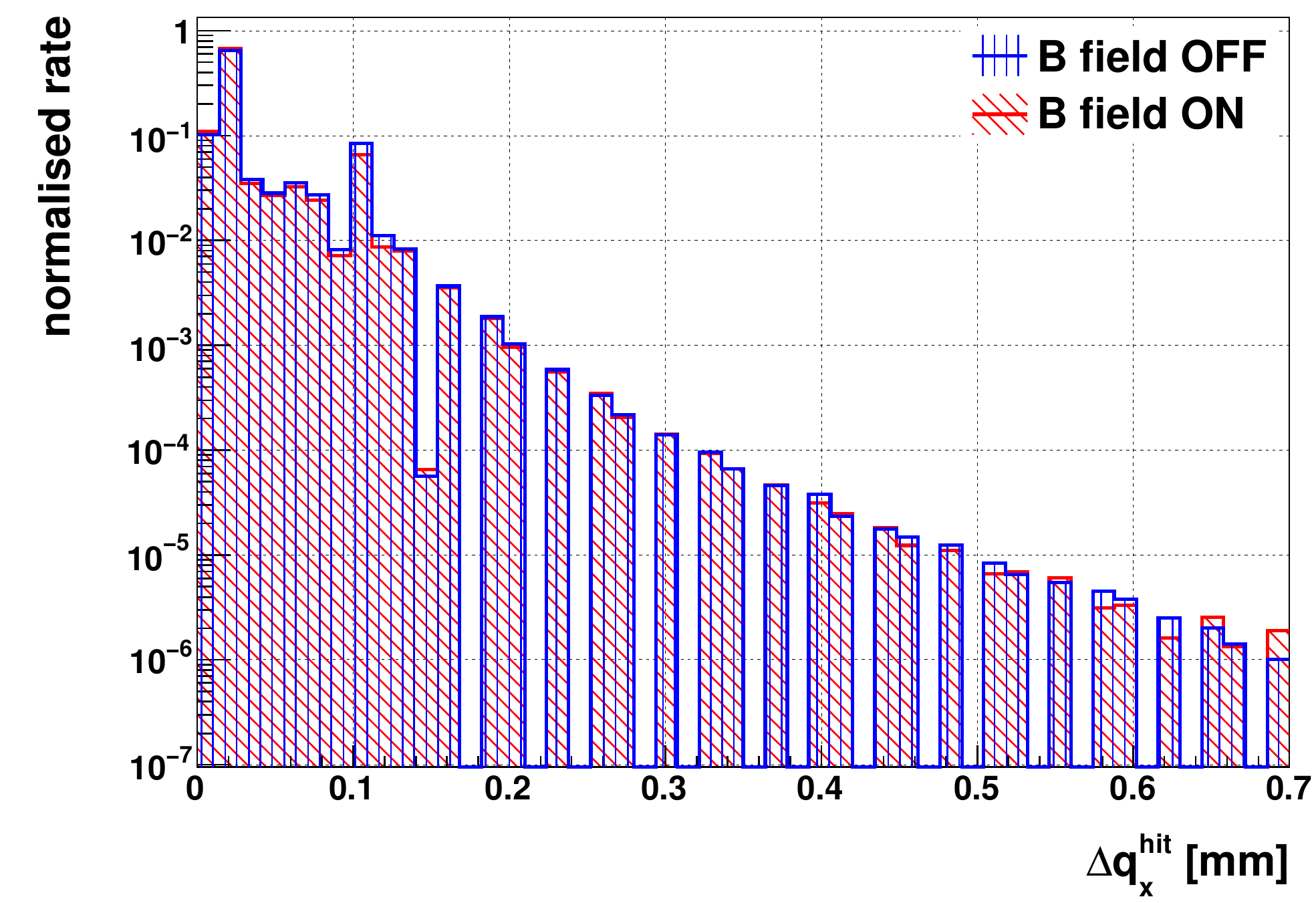}
\includegraphics[width=7.9cm,clip=true]{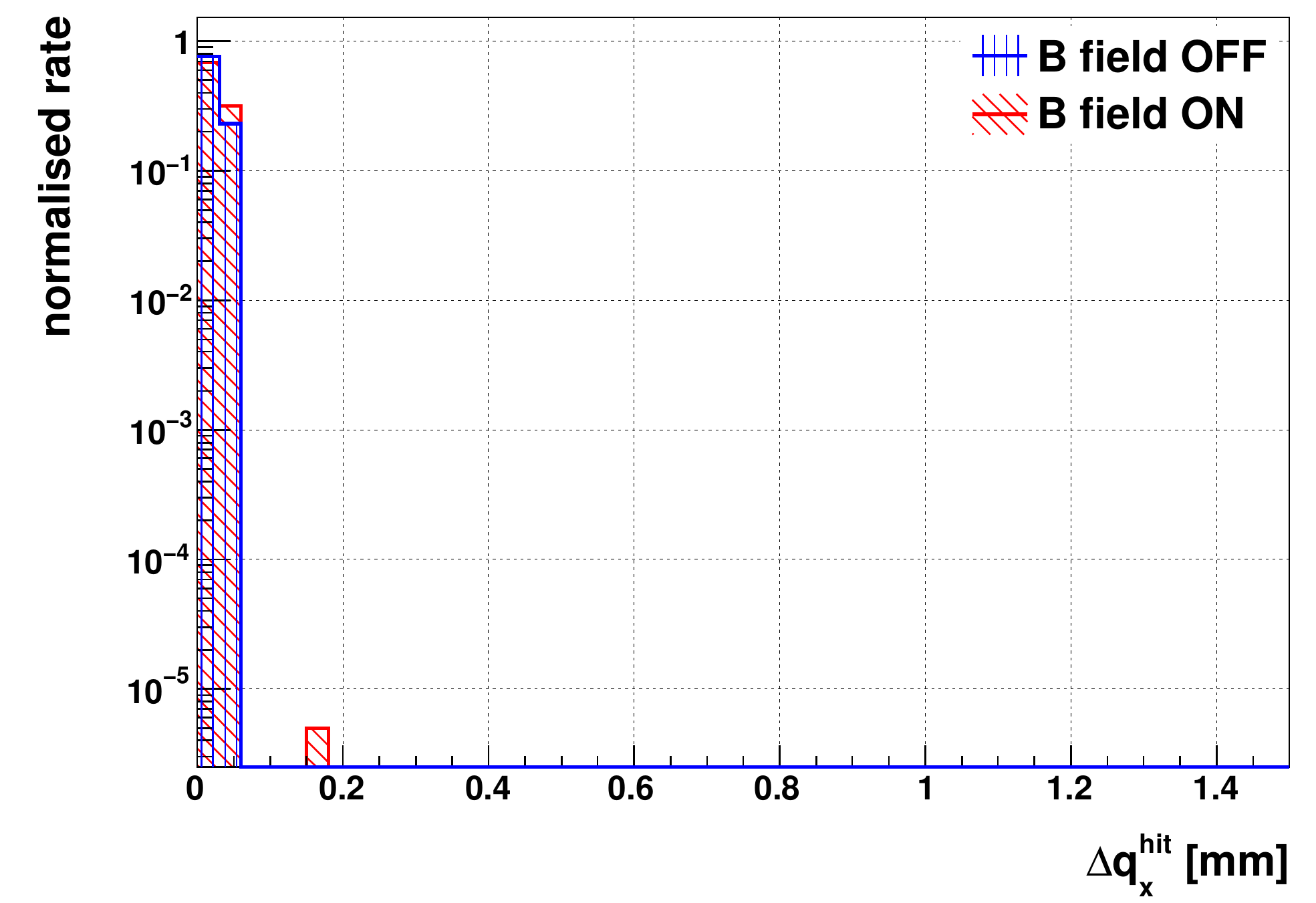}
\includegraphics[width=7.9cm,clip=true]{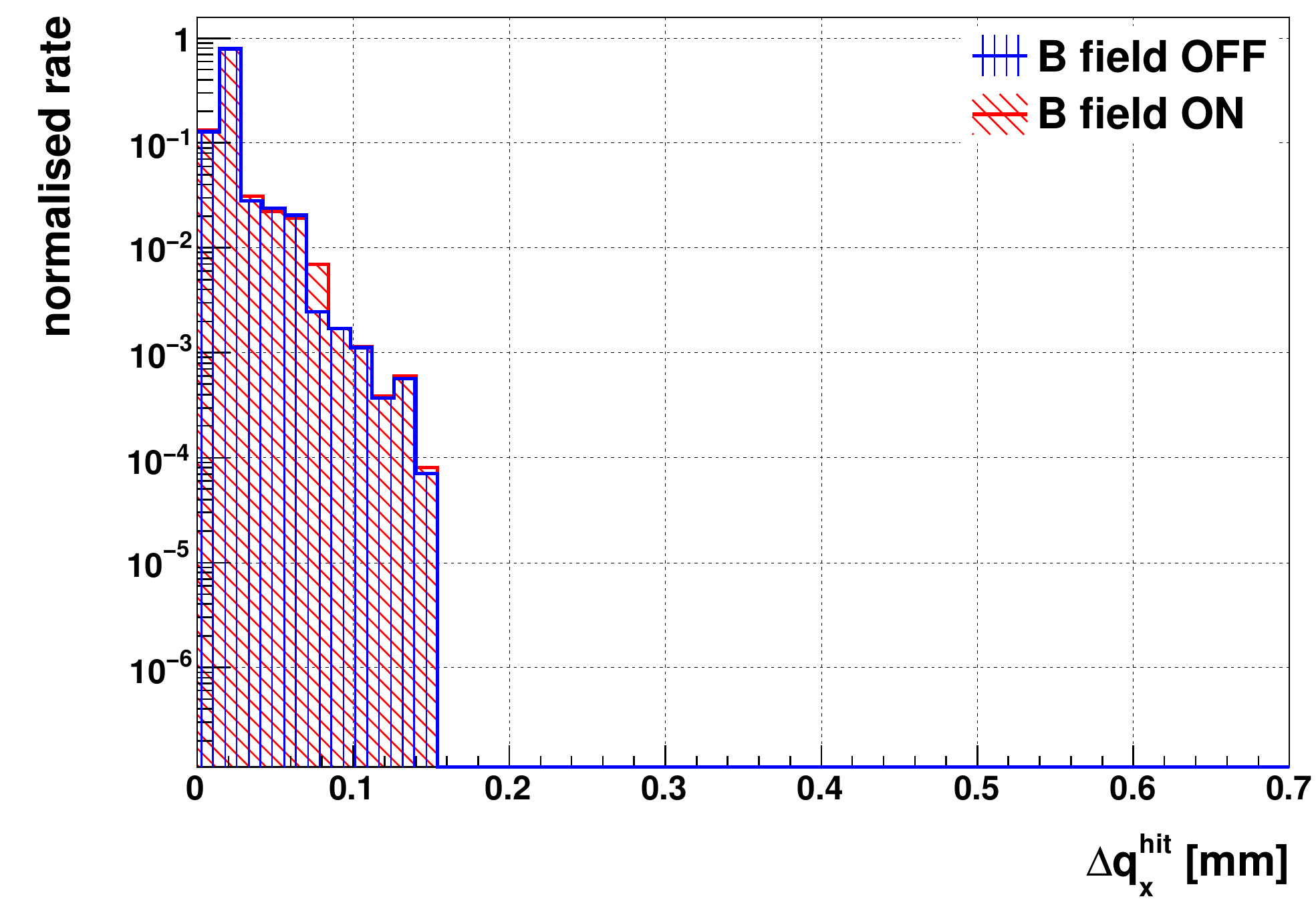}
\vspace{\cDist}
\end{center}
\caption[Hit-on-track uncertainty $\dqhit_x$ before and after hit quality requirements in M8+]{\label{fig:hitErr}
Hit-on-track uncertainty $\dqhit_x$ before any hit quality requirements for the barrel modules of the pixel {\bf (top left)} and SCT {\bf (top right)} detectors  in M8+ after alignment with the \RA\ algorithm. Analogous distributions {\it with} the requirements $\nclust<6$, $|\incAngleT|<0.8$\,rad are shown below for pixel {\bf (bottom left)} and SCT {\bf (bottom right)} modules. For details see text.
}
\end{figure}%\nopagebreak[5]

The results of the application of hit quality requirements on the transverse incidence angle \incAngleT\ and the cluster size \nclust\ specified in Equations~\ref{eqn:incAngleT} and \ref{eqn:nclust} are illustrated in Figure~\ref{fig:hitErr} for $\dqhit_x$ in M8+:
\begin{description}
 \item[Pixel sensors:] Besides the pronounced peak at low $\dqhit_x$ values below 50\,\mum\ a substantial, though exponentially falling tail extending beyond 1\.\mm\ is observed before any cuts. The application of the hit quality requirements results in a clean rejection of hits with  $\dqhit_x\gtrsim50\,\mum$.
 \item[SCT sensors:] The $\dqhit_x$ distribution before any cuts displays two main features: a bulk of values below $\dqhit_x\lesssim150\mum$ with a pronounced peak at about $\dqhit_x\simeq20\,\mum$, and a series of $\delta$-function-like errors which begins at about $\dqhit_x\gtrsim100\,\mum$. While the former is preserved, the latter is cleanly removed by the hit quality requirements.
\end{description}
The extraordinarily clean removal of hits with high errors in case of both the pixel and the SCT sensor is a clear argument for the appropriateness of the hit quality requirements applied with the {\tt InDetAlignHitQualSelTool}.

Certainly, the above ``cleaning'' of the $\dqhit_x$ distribution could have been achieved by a direct cut on this quantity. However, this estimate provided by the reconstruction with \Athena\ {\em relies} on the fact that the underlying hit shows no hidden pathologies and thus cannot be expected be correct in all cases. The basic philosophy of the {\tt InDetAlignHitQualSelTool} is to provide a robust way to reject {\em potentially} pathological hits as they may be harmful for the alignment procedure.

%% file: M8plus/SelRes.tex
Since some minimum quality of the track fit was already established beforehand, and the measured hits were carefully discriminated against any hardware-related pathological cases and any potential problems in reconstruction, the selection of residuals is straightforward. There remains only one important criterion to be verified: the association of residuals with the track. This can be done with a cut on the maximum magnitude of residuals:
\begin{eqnarray} \label{eqn:resCut}
 |r_x| &<& 1.5\,\mm\nonumber\\
 |r_y| &<& 5.0\,\mm\,.
\end{eqnarray}
These cuts were chosen such that they
\begin{itemize}
 \item are not too tight in order not to become ignorant about large L1 misalignments, which in case of the pixel and silicon detectors are of $\order{1\,\mm}$;
 \item do {\it not} discriminate too harshly against residuals from low-$p_T$ tracks with a lot of Coulomb multiple scattering: on average, they bear information relevant for track-based alignment;
 \item discriminate against obviously failed cases of track reconstruction: in the the earliest $14.2.0.X$ versions of \Athena\ releases utilised at the very beginning of M8+, extreme cases of residuals up to 10\,cm (!) were observed. While this is not any more an issue for the recent well-debugged 14.5.2 \Athena\ release used to obtain the results shown in this document. Nevertheless, it was decided to keep that cut since it has an efficiency very close to unity.
\end{itemize}
The appropriateness of the cuts defined in Equation~\ref{eqn:resCut} can be verified by investigating residual distributions before any alignment (in red) in Figures~\ref{fig:r_x_PIXB_L1}, \ref{fig:r_x_SCTB_L1}, \ref{fig:r_x_PIXA_L1}, and \ref{fig:r_x_PIXC_L1} in Subsection~\ref{ssec:l1M8}.

There are no specific selection criteria for overlap residuals implemented, as an overlap residual is formed from a pair of residuals, each of which was selected having regard to the residual quality criteria.

%% file: M8plus/Procedure.tex
As already indicated in the introduction to this Chapter, the \RA\ was used to obtain a set of alignment constants for the ATLAS silicon tracker using M8+ data. While the \RA\ algorithm itself was introduced in Chapter~\ref{chp:ra}, the exact procedure followed to obtain that set of alignment constants is detailed in this Section. Its organisation  reflects the sequence of the alignment steps applied:
\begin{description}
 \item[L1:] 8 iterations with $B$-field off (500k events);
 \item[L2:] 14 iterations with $B$-field off (500k events);
 \item[Pixel stave bow:] 9 iterations with $B$-field off (full statistics);
 \item[L3:] 8 iterations with $B$-field on and off (full statistics for both).
\end{description}
Despite the fact that there still are some unresolved discrepancies in the residual distributions for the $B$-field on and off cases\footnote{A brief summary of the $B$-field on and off discrepancies is given in Subsection~\ref{sec:B0minusB1M8}}, the exclusive usage of $B$-field off data up to L3 should not be a concern. The discrepancies observed in the residuals are small of \order{1\,\mum}, and can easily be corrected for at the final stage of alignment -- L3.

%% file: M8plus/L1.tex
The algorithm to calculate L1 alignment corrections with \RA\ is described Subsection~\ref{ssec:l1}, and its application to M8+ data shall be the subject of this Subsection. As already mentioned above, L1 alignment was performed using $B$-field off data only, which is naturally  more sensitive to global misalignments. A classical example is a shift in the $X$-$Y$ plane between the pixel and SCT subdetectors: in the $B$-field on case it can be compensated for by a change in the curvature of the track, which is fixed to infinity in the $B$-field off case. Only 0.5M of events corresponding to \order{50\rm k} tracks were used, since the number of residuals they yield is enough to provide for a sufficiently small statistical error on the constants. This is because {\em all} residuals collected by the given barrel or end-cap are used {\em coherently} for alignment at L1.% to calculate the alignment constants at L1.

\begin{figure}
\begin{center}
\vspace{\cDistHalf}
\includegraphics[width=15.8cm,clip=true]{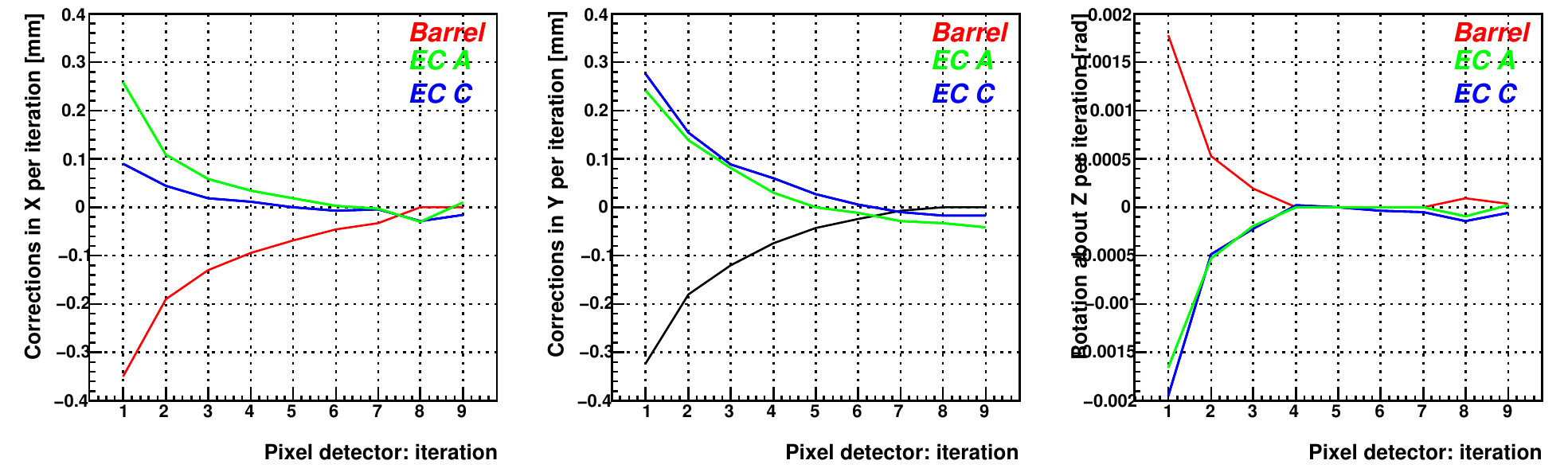}
\vspace{\cDist}
\end{center}
\caption[Convergence of alignment corrections at L1 in M8+]{\label{fig:convergenceL1}
Alignment corrections per iteration with the \RA\ algorithm at {\bf L1} in M8+. Three bodies were aligned for the $X${\bf~(left)}, $Y${\bf~(middle)}, and $\Gamma${\bf~(right)} degrees of freedom: the barrel and the two end-caps of the pixel detector. An exponential-like asymptotic convergence is observed.
\vspace{\cDistHalf}
}
\end{figure}%\nopagebreak[5]

{
\begin{figure}
\vspace{-0.7cm}
\begin{picture}(15.5,24)(0.1,0.1) \large
\put(0.8,16)	{\includegraphics[width=14.5cm,height=8cm,clip=true]{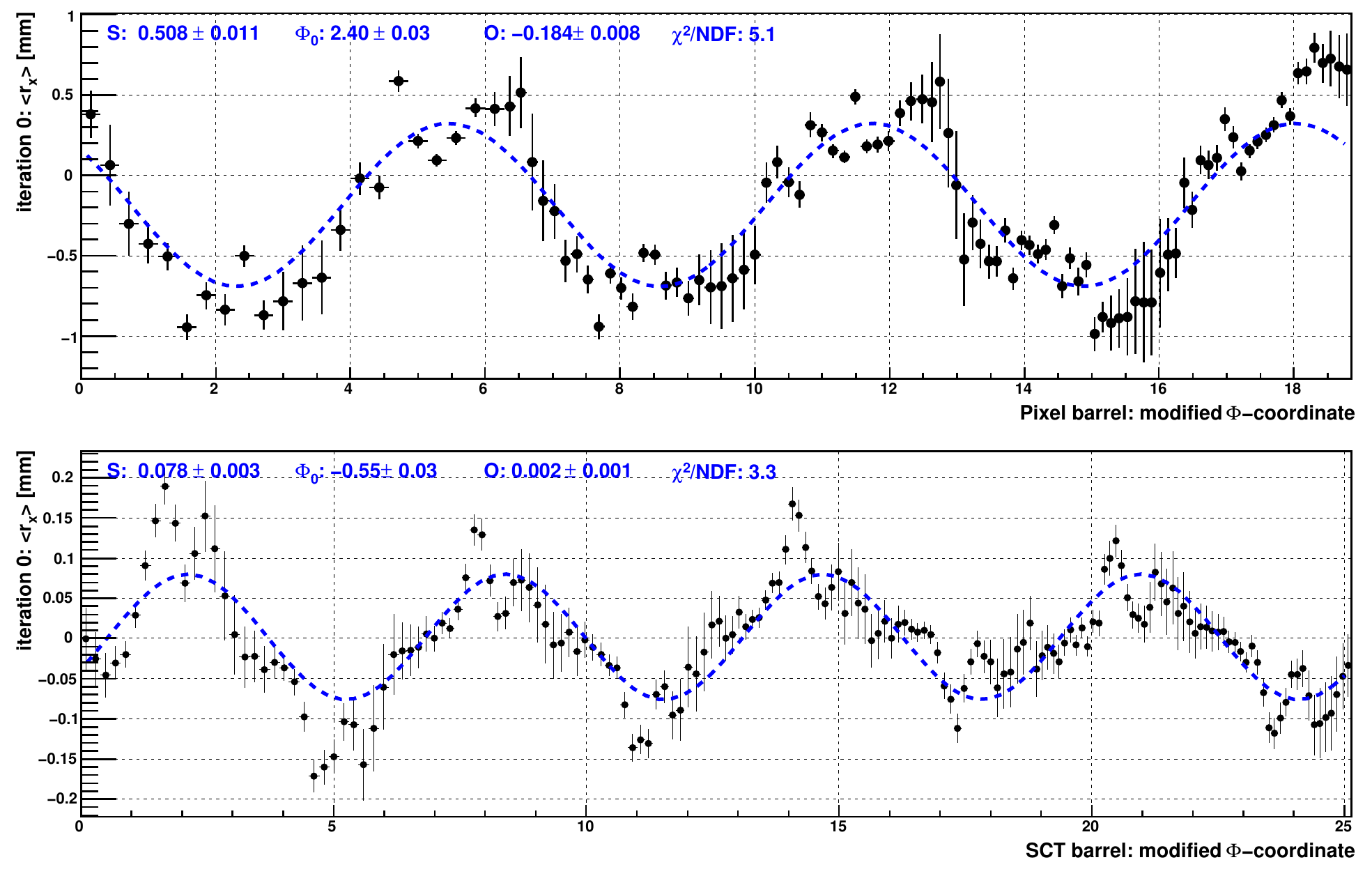}}
\put(0.8,8)	{\includegraphics[width=14.5cm,height=8cm,clip=true]{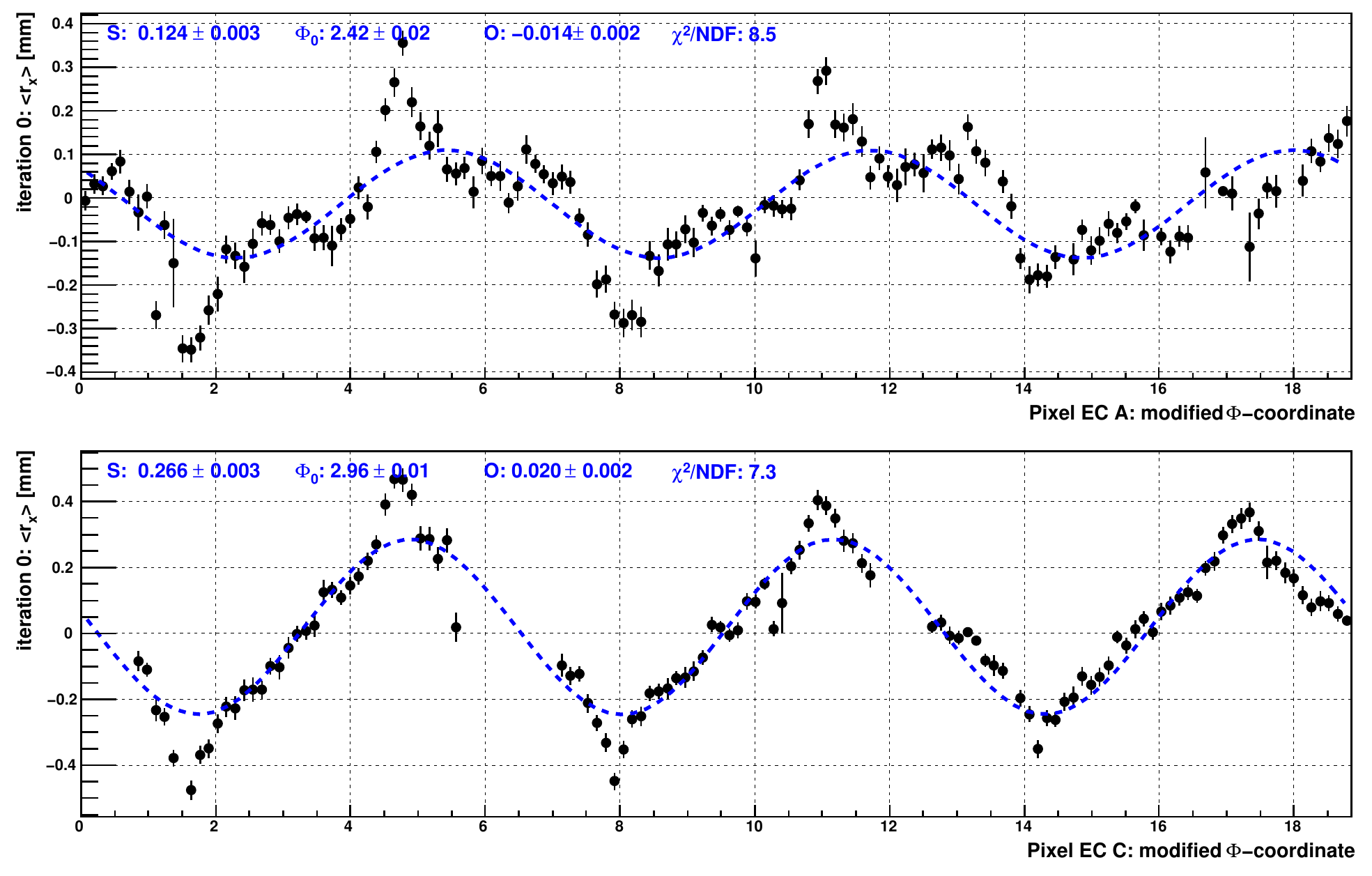}}
\put(0.8,0)	{\includegraphics[width=14.5cm,height=8cm,clip=true]{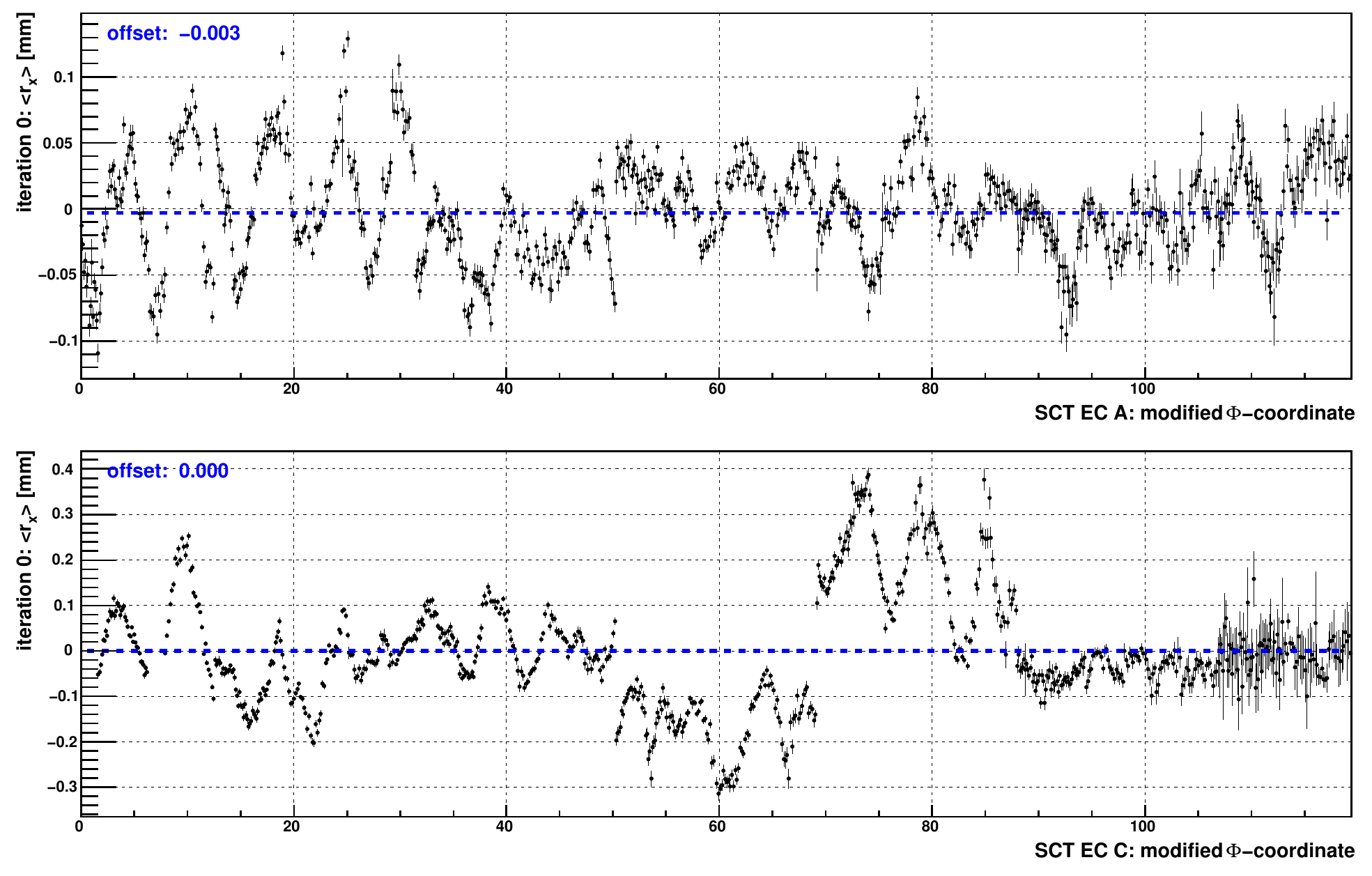}}
\put(0,22.2){\sf (a)}
\put(0,18.2){\sf (b)}
\put(0,14.2){\sf (c)}
\put(0,10.2){\sf (d)}
\put(0, 6.2){\sf (e)}
\put(0, 2.2){\sf (f)}
\normalsize
\end{picture}
\vspace{-0.5cm}
\caption[The distribution $\langle r_x\rangle_{\rm stave}(\Phi)$ for the barrel of pixels and SCT before alignment]{\label{fig:m8_r_x_vs_Phi_L1_brl}
The $\rmean x(\Phi)$ distribution for the barrel part of the pixel detector{\bf~(a)} and of the SCT{\bf~(b)}; for pixel EC~A{\bf~(c)} and EC~C{\bf~(d)}; for SCT EC~A{\bf~(e)} and EC~C{\bf~(f)} using the full $B$-field off M8+ dataset {\bf before} any alignment. The fit results with a sine of the form specified in Equation~\ref{eqn:sineL2} are shown in blue.
}
\end{figure}%\nopagebreak[5]

\begin{figure}
\vspace{-0.7cm}
\begin{picture}(15.5,24)(0.1,0.1) \large
\put(0.8,16)	{\includegraphics[width=14.5cm,height=8cm,clip=true]{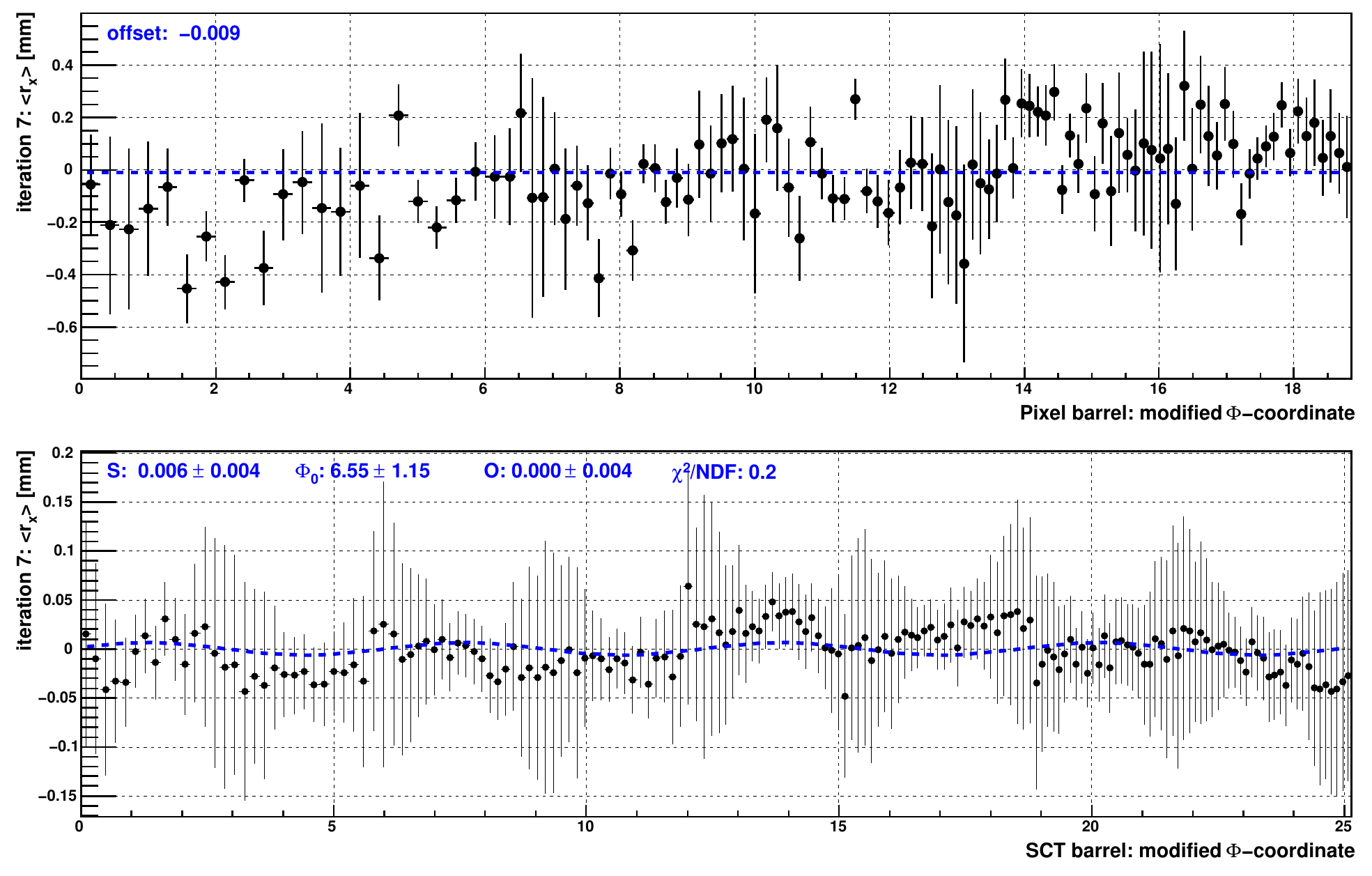}}
\put(0.8,8)	{\includegraphics[width=14.5cm,height=8cm,clip=true]{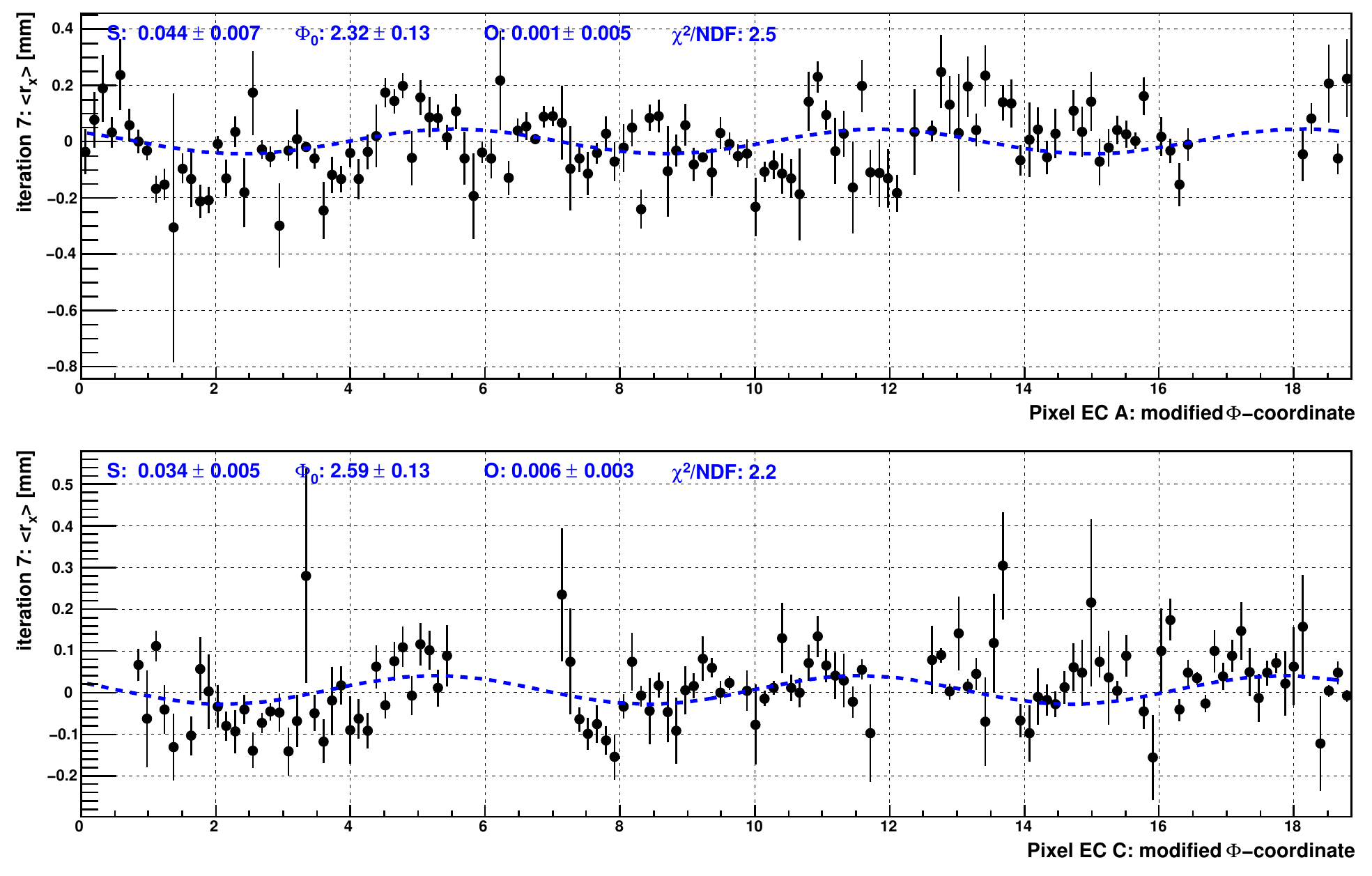}}
\put(0.8,0)	{\includegraphics[width=14.5cm,height=8cm,clip=true]{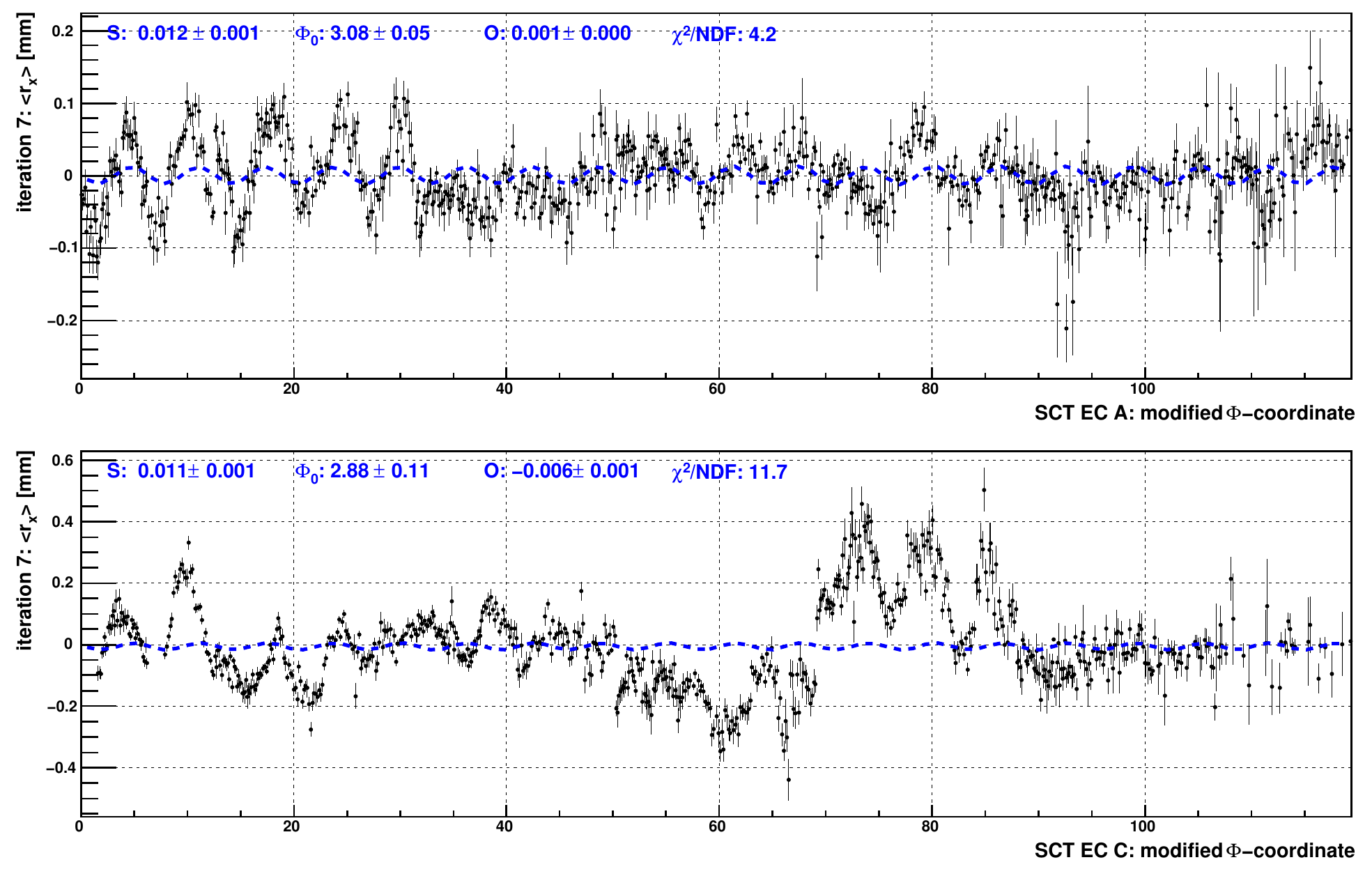}}
\put(0,22.2){\sf (a)}
\put(0,18.2){\sf (b)}
\put(0,14.2){\sf (c)}
\put(0,10.2){\sf (d)}
\put(0, 6.2){\sf (e)}
\put(0, 2.2){\sf (f)}
\normalsize
\end{picture}
\vspace{-0.5cm}
\caption[The distribution $\langle r_x\rangle_{\rm stave}(\Phi)$ for the barrel of pixels and SCT after alignment at L1]{\label{fig:m8_r_x_vs_Phi_L1_brl_after}
The $\rmean x(\Phi)$ distribution for the barrel part of the pixel detector{\bf~(a)} and of the SCT{\bf~(b)}; for pixel EC~A{\bf~(c)} and EC~C{\bf~(d)}; for SCT EC~A{\bf~(e)} and EC~C{\bf~(f)} using the full $B$-field off M8+ dataset {\bf after} the \RA\ procedure at L1. The fit results with a sine of the form specified in Equation~\ref{eqn:sineL2} are shown in blue. For details see text.
}
\end{figure}%\nopagebreak[5]
}

Alignment corrections were calculated for the $X$, $Y$, and $\Gamma$ degrees of freedom\footnote{No alignment constants in global $Z$ were provided by the \RA\ algorithm due to convergence issues: see discussion below Equation~\ref{eqn:c_Z_L2}  in Subsection~\ref{ssec:l2}.} (DoFs) of three bodies: the barrel and the two end-caps of the pixel detector, which were aligned with respect to the SCT. Their convergence over eight iterations is shown in Figure~\ref{fig:convergenceL1}. For a correct interpretation of the figure it is essential to keep in mind that L2 corrections for the pixel ECs are applied ``on top'' of the L1 correction of the barrel, such that for a null transformation of the ECs the negative L1 barrel alignment constants will be written to the L2 section of the database and appear in the Figure. The reason for this is explained in more detail in Subsection~\ref{ssec:l1} on Page~\pageref{par:pixelECL2}.
%
%that in the alignment constant database structure adopted by ATLAS the entire pixel detector is treated as a rigid body at L1. Thus the alignment corrections for the pixel ECs are written to the L2 section of the database, as explained in more detail in Subsection~\ref{ssec:l1} on Page~\pageref{par:pixelECL2}. The consequence of this is, that L2 corrections for the pixel ECs are applied ``on top'' of the L1 correction of the barrel, such that for a null transformation of the ECs the negative L1 barrel alignment constants will be written to the database and appear in the Figure.
%
%, that in the alignment constant database structure adopted by ATLAS the entire pixel detector is treated as a rigid body at L1. Therefore, the alignment corrections for a given pixel EC are written as one and the same L2 correction for each of the three disks it constitutes. Since the L1 correction derived for the pixel barrel is valid for the entire pixel detector and L2 corrections are added on top of L1 per definitionem, the ECs have to ``compensate'' for L1 constants of the barrel. For example, a null translation for an end-cap will be written as exactly the negative L1 corrections of the barrel to database.
All three degrees of freedom are strongly constrained in the underlying experimental setup, which results in a stable, exponential-like convergence for both barrel and ECs, since the alignment corrections per iteration asymptotically approach zero, especially for the $\Gamma$ DoF. This is a clear indication for the robustness of the procedure. The resulting alignment corrections after 8 iterations of the \RA\ algorithm are summarised in Table~\ref{tab:alignL1}.

%Alignment constants:\\
%2 1 0 0 0 0 0 -0.91123 -0.774019 0 0 0 2.5931\quad  Barrel\\
%2 1 -2 0 0 0 0 0.123754 0.584003 0 0 0 -2.8648\quad EC C\\
%2 1 2 0 0 0 0 0.448581 0.417969 0 0 0 -2.47411\quad EC A\\
\begin{table}
\small
\begin{center}
\begin{tabular}{lrrr}
\hline
 & ~~$X$ [$\mu$m]\!\!\!\! & ~~$Y$ [$\mu$m]\!\!\!\! & ~~$\Gamma$ [mrad]
\\\hline\hline
Barrel & $-$911 & $-$774 & 2.59~~~ \\
EC A   & 448 & 417  & $-$2.47~~~ \\ % -463	-357	0.12
EC C   & 123 & 584  & $-$2.86~~~ \\ % -788	-190	-0.27
all    & $-$864 & $-$691 & 2.15~~~ \\ % -863.853211	-691.3486239	2.149908257
\hline
\vspace{\cDistHalf}
\end{tabular}
\caption[The alignment contstants at L1 for the three parts of the pixel detector in M8+]{\label{tab:alignL1}
The alignment contstants derived at {\bf L1} with the \RA\ alignment algorithm for the three parts of the pixel detector and for the entire pixel detector. The latter was found by combining the barrel and EC numbers weighted by their number of modules, id est 1456/1744 and 144/1744. The barrel and entire pixel detector numbers are in the ATLAS global frame.
%
%, ergo with respect to the SCT whose position was not changed by the alignment procedure. 
The alignment corrections for the ECs are understood to be applied on top of the barrel corrections, as explained in the text.
\vspace{\cDist}
}
\end{center}
\end{table}

The $\rmean x(\Phi)$ distributions which were used to calculate the alignment corrections $c_X$, $c_Y$, and $c_\Gamma$ are displayed in Figure~\ref{fig:m8_r_x_vs_Phi_L1_brl} {\em before} any alignment. The analogous distributions {\em after} eight iterations are presented in Figure~\ref{fig:m8_r_x_vs_Phi_L1_brl_after}. The fits which were used to calculate the alignment constants are shown as blue dashed lines. The intepretation of both Figures is given in the following:
\begin{description}
\item[Barrel (Pixel+SCT):] The $\rmeanstave{x}(\Phi)$\ distributions for the barrel regions of the pixel~(a) and SCT~(b) detectors {\em before} any alignment display the expected sine-like dependence defined in Equation~\ref{eqn:sineL2}. 
%with a magnitude of $S\simeq500\,\mum/80\,\mum$, respectively. 
This implies that the rigid-body hypothesis holds for the barrels of both subdetectors on the one hand, and that the magnitude of misalignments at L1 dominates the one at L2 on the other hand. Both are indespensable requirements for credible alignment results, as discussed in Subsection~\ref{ssec:l1}. The magnitude of the sine fit is $S\simeq500\,\mum~(80\,\mum)$ for pixel~(SCT). {\em After} L1 alignment, virtually no sinusoidal modulation remains: while the fit defaults to an offset function for the pixel case, a sine with a tiny magnitude of $S=6\,\mum$ is fitted to the $\rmeanstave{x}(\Phi)$\ distribution in the barrel of the SCT. Keep in mind that no alignment corrections were applied to the SCT barrel, which remained stationary\footnote{This was done because a displacement at L1 will be primarily visible in the residuals of the pixel detector given their statistically smaller weight in the individual track fits.};
%, as can be seen from $S\simeq500\,\mum/80\,\mum$ for the pixel/SCT before any alignment.};
\item[ECs (Pixel):] {\em Before} any alignment, both end-cap~A~(c) and end-cap~C~(d) of the pixel detector show a periodic dependance in $\rmean{x}(\Phi)$. It can be fitted by a sine 
%of the form in Equation~\ref{eqn:sineL2} and 
with a magnitude of $S\simeq124\,\mum~(266\,\mum)$ for EC~A~(C). The $\rmean{x}(\Phi)$\ distribution in EC~A is somewhat more irregular than EC~C. This may be due to the projection of misalignments from other parts of the detector which is hinted at by the fact that the shape of the distribution is about the same for each of the disks. {\em After} alignment, the sinusoidal dependance disappears almost completely, as expected.
\item[ECs (SCT):] The $\rmean{x}(\Phi)$\ distribution for the EC~A~(e) and EC~C~(f) of the SCT are a problematic case as already discussed in detail using the example of EC~A at the end of Subsection~\ref{ssec:l1} on page~\pageref{ssec:l1}~ff. Even though most of the disks of a given EC display a sinusoidal dependance in $\rmean{x}(\Phi)$, their individual offsets $O$, amplitudes $S$, and most importantly, phases $\Phi_0$ appear to be different, which prevents a common sine fit from converging. %This may well be because a treatment of SCT ECs as mechanically rigid bodies is not appropriate, although the hit topology of M8+ cosmic ray tracks may be the reason, too. 
\end{description}

In the following, the $r_x$ residual distribution, the main benchmark of the alignment procedure, are described briefly:
\begin{description}
\item[Pixel Barrel:] The distribution of residuals in the barrel of the pixel detector before and after L1~alignment are shown by layers in Figure~\ref{fig:r_x_PIXB_L1}. The effect of the alignment procedure is striking. Before alignment, all three distributions are with $\rsig x$ of \order{600\,\mum}\ very broad and irregular. After alignment, their shapes begin to resemble a Gaussian, and the widths improve dramatically to \order{250\,\mum}. The means of the individual layers are not centred at zero yet, since layer-to-layer alignment is to be performed in the next step of the procedure -- at L2. However, as can be seen from the summary in Table~\ref{tab:m8plusAlignL1}, $\rmean x$ over all layers is only about 4\,\mum\ away from zero.
\item[SCT Barrel:] The analogous residual distributions in the four layers of the SCT barrel are shown in Figure~\ref{fig:r_x_SCTB_L1}. The picture before alignment is notably different from the pixel detector: a clear, relatively sharp peak is observed in the centre of the distribution, and distinctive shoulders extend from about $|r_x|\simeq200\,\mum$ onwards. The interpretation is straightforward: the peak is comprised of tracks reconstructed only by the SCT barrel, and its realtively small width reflects the superb assembly precision of that subdetector; 
%, which exceeded the expectations by almost one order of magnitude; 
the shoulders are due to tracks going through the pixel detector, whose large misalignment of \order{1\,\mm} with respect to the SCT produces the large magnitude of residuals. After L1 alignment the shoulders disappear, and $\rsig x$  improves from $\sim\!240\,\mum$ to $\sim\!175\,\mum$.
\item[Pixel End-Caps:]
The situation in the end-caps of the pixel detector is not much different from the SCT barrel as shown in Figures~\ref{fig:r_x_PIXA_L1} and \ref{fig:r_x_PIXC_L1} for EC~A and C, respectively. Before alignment, a distribution with wide shoulders is observed, which is due to L1 misalignments between pixel end-caps and SCT barrel bracketing them, cf.\!\! Figure~\ref{fig:inDetTechnical}. EC~C displays somewhat more pronounced shoulders than EC~A, which is due to a larger misalignment with respect to the SCT, as can be verified from Table~\ref{tab:alignL1}. The shoulders disappear after alignment, and $\rsig x$ improves from $\sim\!350\,\mum$ to $\sim\!300\,\mum$ for EC~A and from $\sim\!400\,\mum$ to $\sim\!230\,\mum$ for EC~C. The larger $\rsig x$ in EC~A than in EC~C after alignment is likely due to L3 misalignments;
\item[SCT End-Caps:]
As explained above, SCT end-caps are not aligned at L1, so only marginal changes in the residual distributions result. Therefore, these are not shown here explicitly.
\end{description}

The residual means \rmean{x}, the uncertainties on the residual means $\delta r_x$, and the standard deviation of residuals $\rsig x$ are summarised by layers in Table~\ref{tab:m8plusAlignL1} before and after alignment. All numbers behave as expected, and demostrate the merit of L1 alignment with the \RA\ algorithm.
\newpage
{
\begin{figure}
%\begin{center}
\vspace{\cDistHalf}
\includegraphics[width=15.8cm,clip=true]{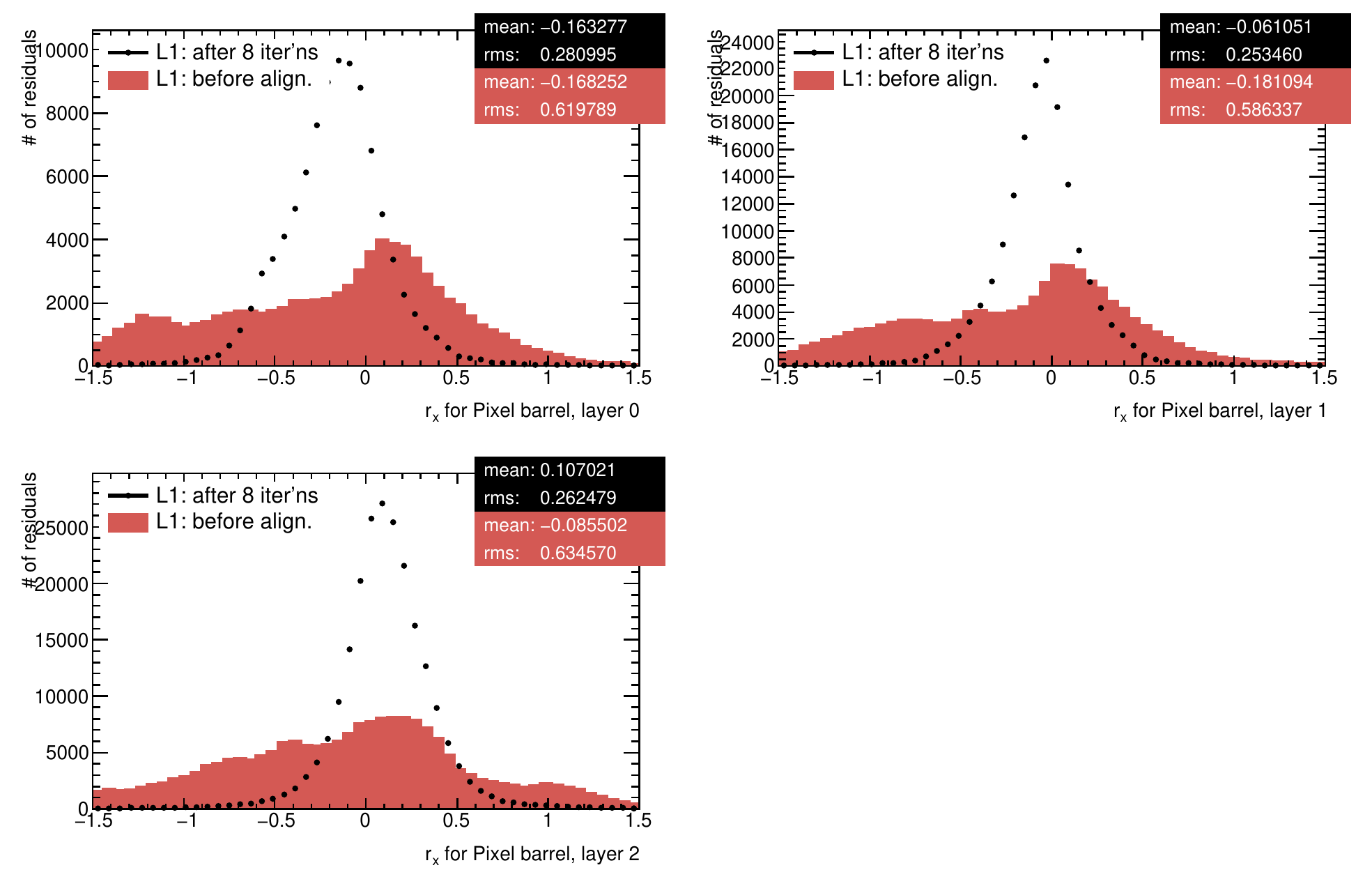}
\vspace{\cDist}
%\end{center}
\caption[$r_x$ residual distribution in the barrel of the pixel detector before and after alignment corrections at L1 in M8+]{\label{fig:r_x_PIXB_L1}
The $r_x$ {\bf residual} distribution in the barrel of the pixel detector by layers before and after alignment corrections at {\bf L1} in M8+. A dramatic improvement in the residual width is observed, which indicates initial misalignments of \order{1\,\rm mm} in the $X$-$Y$ plane. Values are in~mm.
%
%The residual means are not re-centered around 0, as individual layers are not aligned for.
\vspace{\cDistHalf}
}
\end{figure}%\nopagebreak[5]

\begin{figure}
%\begin{center}
\includegraphics[width=15.8cm,clip=true]{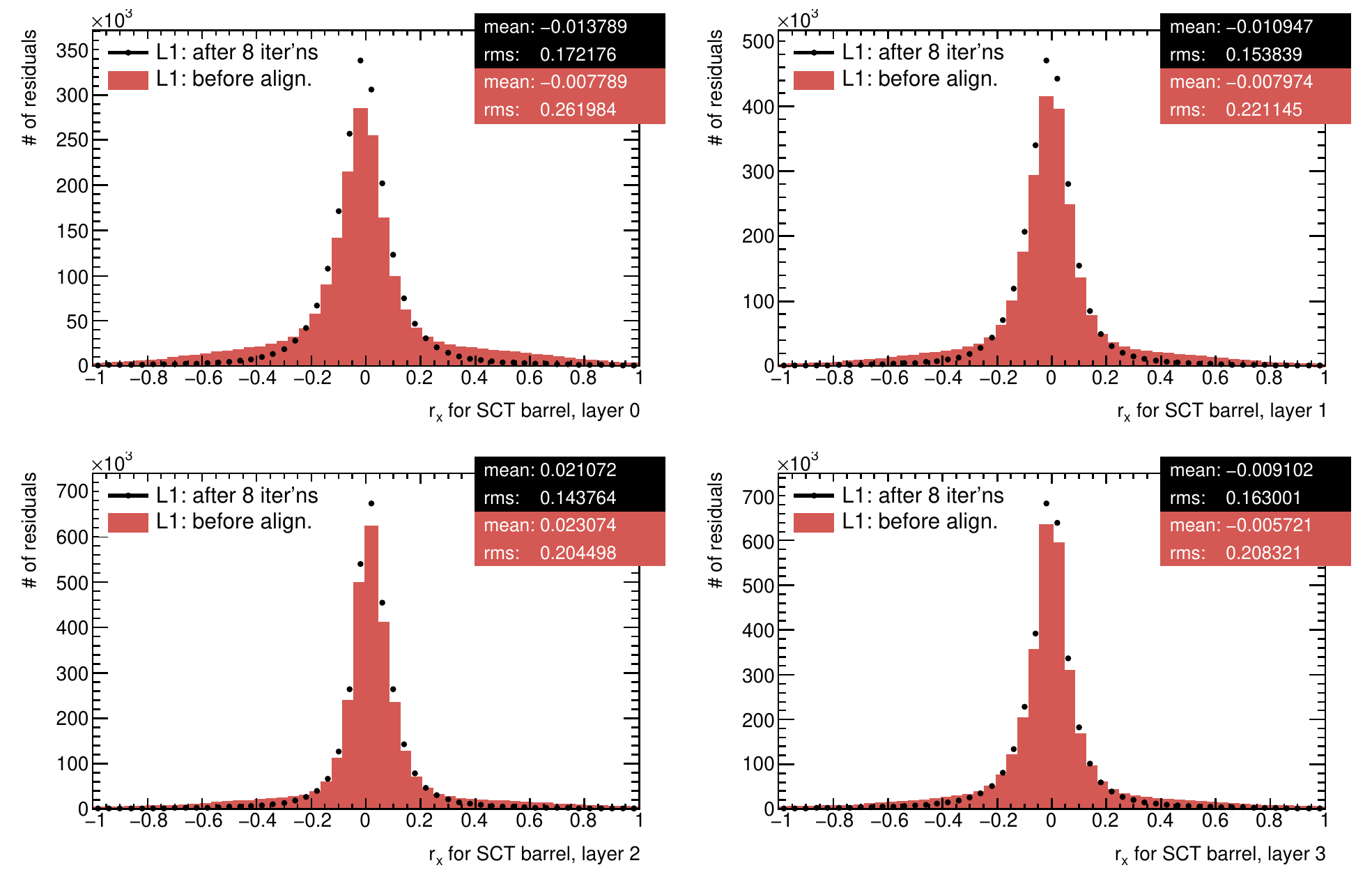}
\vspace{\cDist}
%\end{center}
\caption[$r_x$ residual distribution in the barrel of the SCT detector before and after alignment corrections at L1 in M8+]{\label{fig:r_x_SCTB_L1}
The $r_x$ {\bf residual} distribution in the barrel of the SCT detector by layers before and after alignment corrections at {\bf L1} in M8+. The distribution before alignment has a distinct transition between a narrow peak and wide shoulders, indicating a good internal alignment of the SCT and a large misalignment of the SCT with respect to pixels, respectively. Consequently, the shoulders disappear after the L1 alignment procedure. Values are in mm.
\vspace{\cDistHalf}
}
\end{figure}%\nopagebreak[5]
}

\begin{comment}
\begin{figure}
\begin{center}
\includegraphics[width=15.8cm,clip=true]{M8plus/fig/L1/PIX_BA_Ry}
\vspace{\cDist}
\end{center}
\caption[$r_y$ residual distribution in the barrel of the pixel detector before and after alignment corrections at L1 in M8+]{\label{fig:r_y_PIXB_L1}
The $r_y$ residual distribution in the barrel of the pixel detector by layers before and after alignment corrections at L1 in M8+. Similarly to the $r_x$ case, a dramatic improvement in the residual width and the shoulders of the distributin is observed, indicating initial misalignments of \order{1\,\rm mm} in the $X$-$Y$ plane.  Values are in mm.
%The residual means are not re-centered around 0, as individual layers are not aligned for.
}
\end{figure}%\nopagebreak[5]
\end{comment}

{
\begin{figure}
\begin{center}
\vspace{\cDistHalf}
\includegraphics[width=15.8cm,clip=true]{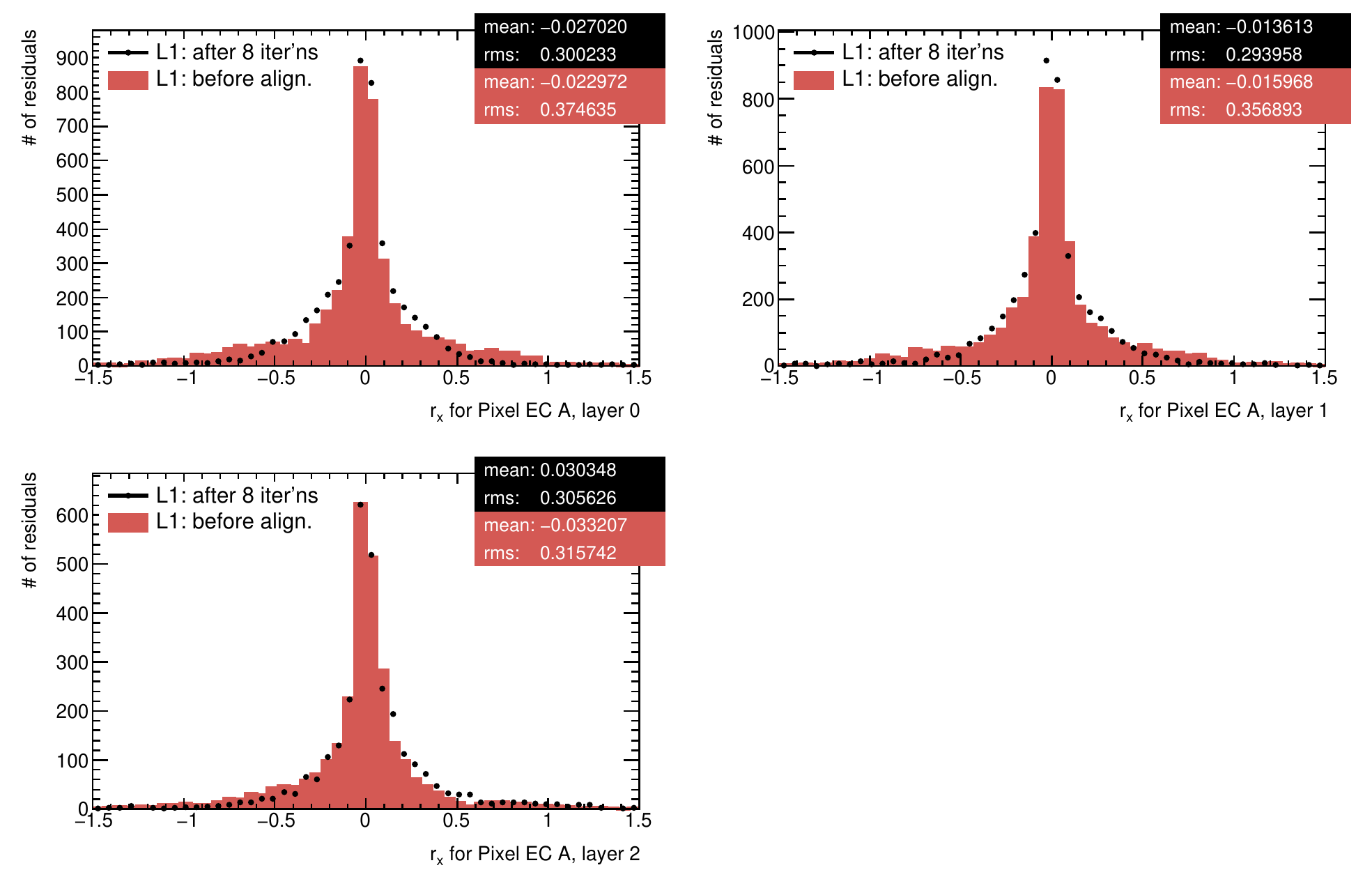}
\vspace{\cDist}
\end{center}
\caption[$r_x$ residual distribution in the EC~A of the pixel detector before and after alignment corrections at L1 in M8+]{\label{fig:r_x_PIXA_L1}
The $r_x$ {\bf residual} distribution in the EC~A of the pixel detector by disk layers before and after alignment corrections at {\bf L1} in M8+. A substantial improvement in the residual width is observed, indicating initial misalignments of \order{\oneOverX 4\,\rm mm} in the $X$-$Y$ plane.  Values are in mm.
%The residual means are not re-centered around 0, as individual layers are not aligned for.
}
\end{figure}%\nopagebreak[5]

\begin{figure}
\begin{center}
\vspace{\cDistHalf}
\includegraphics[width=15.8cm,clip=true]{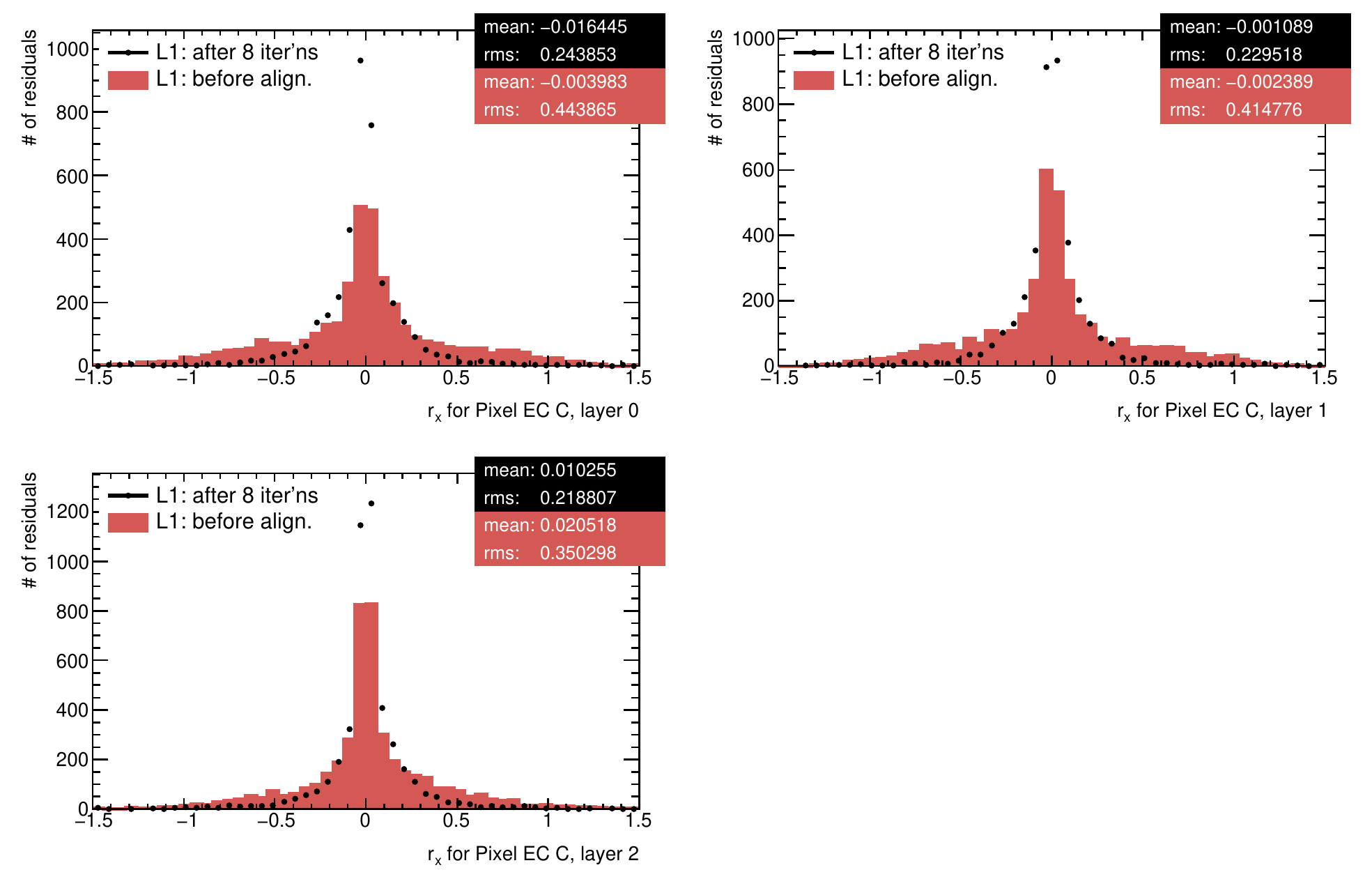}
\vspace{\cDist}
\end{center}
\caption[$r_x$ residual distribution in the EC~C of the pixel detector before and after alignment corrections at L1 in M8+]{\label{fig:r_x_PIXC_L1}
The $r_x$ {\bf residual} distribution in the EC~C of the pixel detector by disk layers before and after alignment corrections at {\bf L1} in M8+. A dramatic improvement in the residual width is observed and the wide shoulders of the distribution disappear, indicating initial misalignments of \order{1\,\rm mm} in the $X$-$Y$ plane. Values are in mm.
%The residual means are not re-centered around 0, as individual layers are not aligned for.
\vspace{\cDist}
}
\end{figure}%\nopagebreak[5]
}

\begin{table}
\small
\begin{center}
\vspace{\cDistHalf}
\begin{tabular}{l|rrr|rrr}
\hline
 & \multicolumn{3}{c|}{{\bf Before} alignment} & \multicolumn{3}{c}{{\bf After} alignment}\\
%\hline
 & $\langle r_x\rangle$ & $\delta r_x$ & $\sigma(r_x)$ & $\langle r_x\rangle$ & $\delta r_x$ & $\sigma(r_x)$  \\
\hline\hline
Pixel barrel layer 0 & $-168.25$ & $2.21$ & $619.8$ & $-163.28$ & $0.92$ & $281.0$\\
Pixel barrel layer 1 & $-181.09$ & $1.56$ & $586.3$ & $-61.05$ & $0.62$ & $253.5$\\
Pixel barrel layer 2 & $-85.50$ & $1.45$ & $634.6$ & $107.02$ & $0.56$ & $262.5$\\
Pixel barrel {\bf (all)} & $-134.18$ & $0.96$ & $617.3$ & $-3.89$ & $0.41$ & $284.7$\\
\hline
SCT barrel layer 0 & $-6.75$ & $0.20$ & $283.3$ & $-14.01$ & $0.14$ & $192.8$\\
SCT barrel layer 1 & $-7.51$ & $0.15$ & $236.7$ & $-11.01$ & $0.11$ & $169.6$\\
SCT barrel layer 2 & $23.88$ & $0.13$ & $218.1$ & $21.09$ & $0.09$ & $157.0$\\
SCT barrel layer 3 & $-5.44$ & $0.13$ & $224.3$ & $-9.38$ & $0.10$ & $180.5$\\
SCT barrel {\bf (all)} & $1.92$ & $0.07$ & $238.0$ & $-2.22$ & $0.05$ & $174.8$\\
\hline
Pixel EC A layer 0 & $-22.97$ & $5.59$ & $374.6$ & $-27.02$ & $4.45$ & $300.2$\\
Pixel EC A layer 1 & $-15.97$ & $5.35$ & $356.9$ & $-13.61$ & $4.37$ & $294.0$\\
Pixel EC A layer 2 & $-33.21$ & $5.94$ & $315.7$ & $30.35$ & $5.70$ & $305.6$\\
Pixel EC A {\bf (all)} & $-22.78$ & $3.27$ & $354.6$ & $-8.13$ & $2.75$ & $300.0$\\
\hline
Pixel EC C layer 0 & $-3.98$ & $7.27$ & $443.9$ & $-16.45$ & $3.93$ & $243.9$\\
Pixel EC C layer 1 & $-2.39$ & $6.78$ & $414.8$ & $-1.09$ & $3.69$ & $229.5$\\
Pixel EC C layer 2 & $20.52$ & $5.30$ & $350.3$ & $10.26$ & $3.27$ & $218.8$\\
Pixel EC C {\bf (all)} & $5.57$ & $3.70$ & $402.2$ & $-1.75$ & $2.09$ & $230.6$\\
\hline
%SCT EC A layer 0 & $-6.25$ & $0.68$ & $213.0$ & $-4.58$ & $0.75$ & $235.2$\\
%SCT EC A layer 1 & $12.38$ & $0.58$ & $192.9$ & $12.11$ & $0.63$ & $210.1$\\
%SCT EC A layer 2 & $-30.05$ & $0.56$ & $167.4$ & $-29.60$ & $0.56$ & $168.7$\\
%SCT EC A layer 3 & $12.68$ & $0.60$ & $159.0$ & $13.25$ & $0.60$ & $159.1$\\
%SCT EC A layer 4 & $1.22$ & $0.64$ & $157.0$ & $2.18$ & $0.63$ & $156.1$\\
%SCT EC A layer 5 & $-19.75$ & $0.96$ & $159.5$ & $-18.88$ & $0.92$ & $155.1$\\
%SCT EC A layer 6 & $14.32$ & $1.47$ & $168.7$ & $14.82$ & $1.44$ & $166.3$\\
%SCT EC A layer 7 & $-5.23$ & $2.27$ & $184.0$ & $-5.79$ & $2.23$ & $181.7$\\
%SCT EC A layer 8 & $-12.65$ & $3.25$ & $193.3$ & $-12.67$ & $3.13$ & $187.2$\\
SCT EC A {\bf (all)} & $-2.86$ & $0.26$ & $182.0$ & $-2.24$ & $0.28$ & $191.4$\\
\hline
%SCT EC C layer 0 & $56.84$ & $0.77$ & $220.8$ & $58.69$ & $0.79$ & $227.6$\\
%SCT EC C layer 1 & $-57.99$ & $0.63$ & $210.5$ & $-59.39$ & $0.64$ & $215.2$\\
%SCT EC C layer 2 & $25.01$ & $0.73$ & $219.5$ & $25.48$ & $0.69$ & $207.8$\\
%SCT EC C layer 3 & $-159.77$ & $1.03$ & $262.4$ & $-159.86$ & $1.01$ & $259.5$\\
%SCT EC C layer 4 & $201.15$ & $1.31$ & $311.3$ & $201.36$ & $1.31$ & $312.4$\\
%SCT EC C layer 5 & $-42.70$ & $1.37$ & $219.9$ & $-41.17$ & $1.36$ & $220.3$\\
%SCT EC C layer 6 & $-6.30$ & $4.61$ & $380.8$ & $-5.71$ & $4.59$ & $382.1$\\
%SCT EC C layer 7 & $6.97$ & $3.43$ & $234.0$ & $5.85$ & $3.42$ & $235.3$\\
%SCT EC C layer 8 & $23.21$ & $5.74$ & $263.7$ & $22.84$ & $5.77$ & $267.1$\\
SCT EC C {\bf (all)} & $0.41$ & $0.39$ & $262.1$ & $0.60$ & $0.39$ & $262.3$\\
\hline
\end{tabular}
\caption[Main residual characteristics for the silicon tracker by layers in M8+ before and after alignment at L1]{\label{tab:m8plusAlignL1}
Main {\bf residual} characteristics for the silicon tracker by layers in M8+ {\em before} and {\em after} alignment at {\bf L1}: the residual mean $\rmean x$, the uncertainty on the residual mean $\delta r_x$, and the standard deviation of the residual $\sigma(r_x)$. The range used for the calculation of the quantities above is $r_x\in[-1.5\,\mm,\,1.5\,\mm]$. All values are given in $\mum$.
\vspace{\cDistHalf}
}
\end{center}
\end{table}

%% file: M8plus/L2.tex
From a conceptual point of view, the \RA\ algorithm at L2 is a special case of alignment at~L1 and is documented in Subsection~\ref{ssec:l2}. In the following, the derivation of L2 alignment constants in M8+ is described. The starting point is the nominal geometry including L1 alignment, whose calculation was documented in the preceding Subsection. 
Like L1, L2 alignment was performed using 0.5M $B$-field off events only. Again, alignment corrections for the three strongly constrained degrees of freedom --- $X$, $Y$, and $\Gamma$ --- were derived. The bodies aligned were the barrel layers of the the pixel and SCT detectors ($3+4$), as well as the disks of their end-caps ($2\times3+2\times9$). 

\begin{figure}
\begin{center}
\includegraphics[width=15.8cm,clip=true]{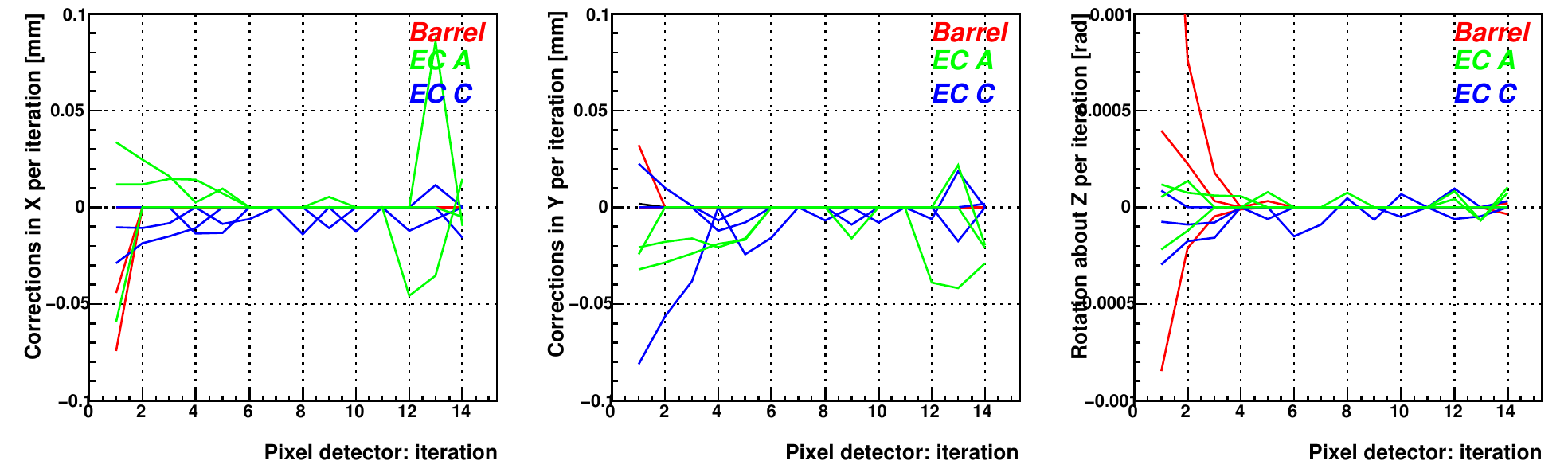}
\end{center}
\vspace{\cDist}
\caption[Convergence of alignment corrections for the pixel detector at L2 in M8+]{\label{fig:convergenceL2_PIX}
Alignment corrections per iteration for the pixel detector with the \RA\ algorithm at {\bf L2} in M8+. $3+3+3$ bodies were aligned for the $X${\bf~(left)}, $Y${\bf~(middle)}, and $\Gamma${\bf~(right)} degrees of freedom: 3 barrel layers, and 3 end-cap disk layers of EC A and C each. An exponential-like asymptotic convergence is observed in the barrel, and the magnitude of corrections is small. The end-caps display somewhat less robust convergence.
}
\end{figure}%\nopagebreak[5]

\begin{figure}
\begin{center}
\includegraphics[width=15.8cm,clip=true]{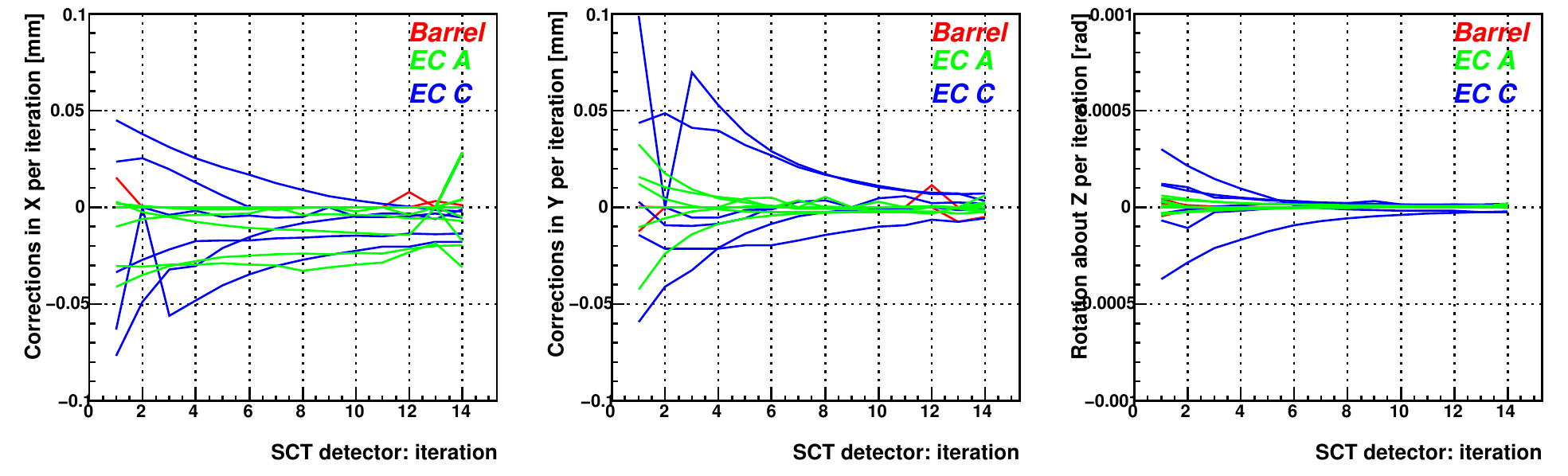}
\end{center}
\vspace{\cDist}
\caption[Convergence of alignment corrections for the SCT detector at L2 in M8+]{\label{fig:convergenceL2_SCT}
Alignment corrections per iteration for the SCT detector with the \RA\ algorithm at {\bf L2} in M8+. $4+9+9$ bodies were aligned for the $X${\bf~(left)}, $Y${\bf~(middle)}, and $\Gamma${\bf~(right)} degrees of freedom: 4 barrel layers, and 9 end-cap disk layers of EC A and C each. An exponential-like asymptotic convergence is observed in the barrel, and the magnitude of corrections is small. The end-caps display somewhat less robust convergence.
}
\end{figure}%\nopagebreak[5]

The convergence of alignment corrections per iteration for the {\bf pixel} detector is shown in Figure~\ref{fig:convergenceL2_PIX}. An almost instantaneous convergence in all three DoFs is observed in the barrel of the pixel detector. The convergence in the ECs is somewhat slower and numerically less stable. However, the sign of the corrections for a given layer is typically the same over the iterations and thus is not of concern to be a indication of oscillatory behaviour. The $c_\Gamma$ corrections in the barrel display a comparably large magnitude, especially for the $b$-layer, which is due to its small lever arm $R$. The increase in magnitude of $c_X$, $c_Y$ corrections in iteration~12 is due to a suddenly converging sine fit in layer~0 of EC~A, which triggers further alignment corrections in iteration~13. The {\bf SCT} displays a similar general picture: while the convergence of the barrel layers is almost instantaneous, it is much less stable in the ECs. 
%The magnitude of the barrel corrections is much smaller than in the pixel due to a better assembly precision. 
The corrections in EC~C are typically larger than in EC~A, which may be due to different macro assembly site procedures~\cite{bib:sctEJINST}.
%which is not surprising given that one of the ECs was assembled at NIKHEF, and the other at Liverpool~\cite{bib:privateHaywood}. 
The convergence behaviour of the $\Gamma$ degree of freedom is stable and robust for both ECs. $c_Y$ corrections are more stable than $c_X$, which is not surprising given the hit topology of cosmic ray tracks.

\begin{figure}
\begin{center}
\vspace{\cDistHalf}
\includegraphics[width=15.8cm,clip=true]{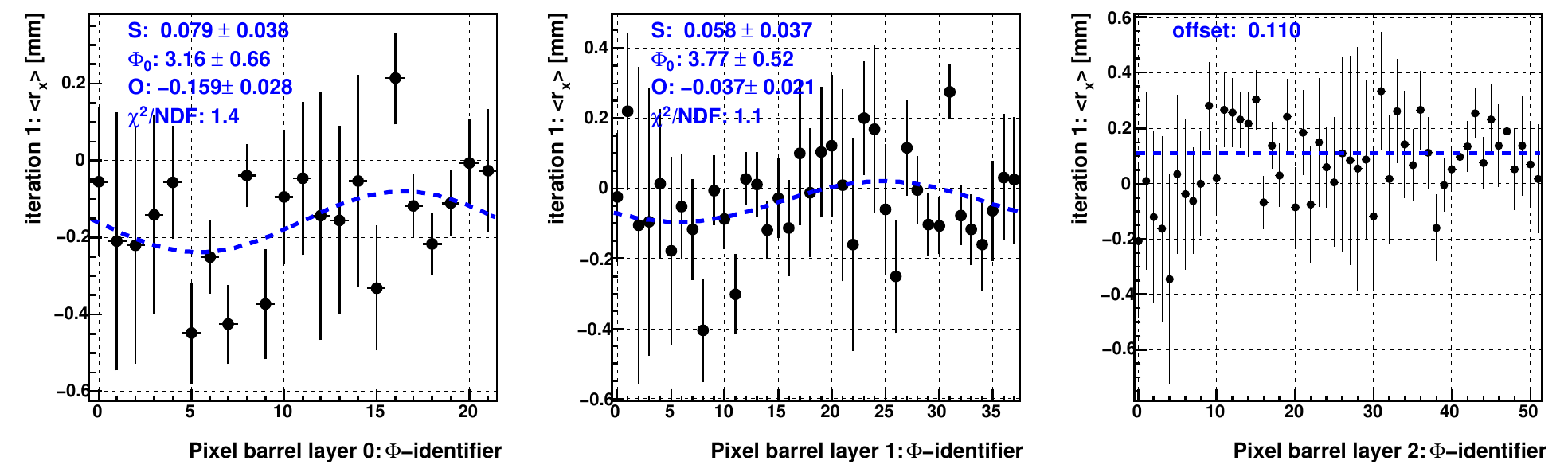}
\includegraphics[width=15.8cm,clip=true]{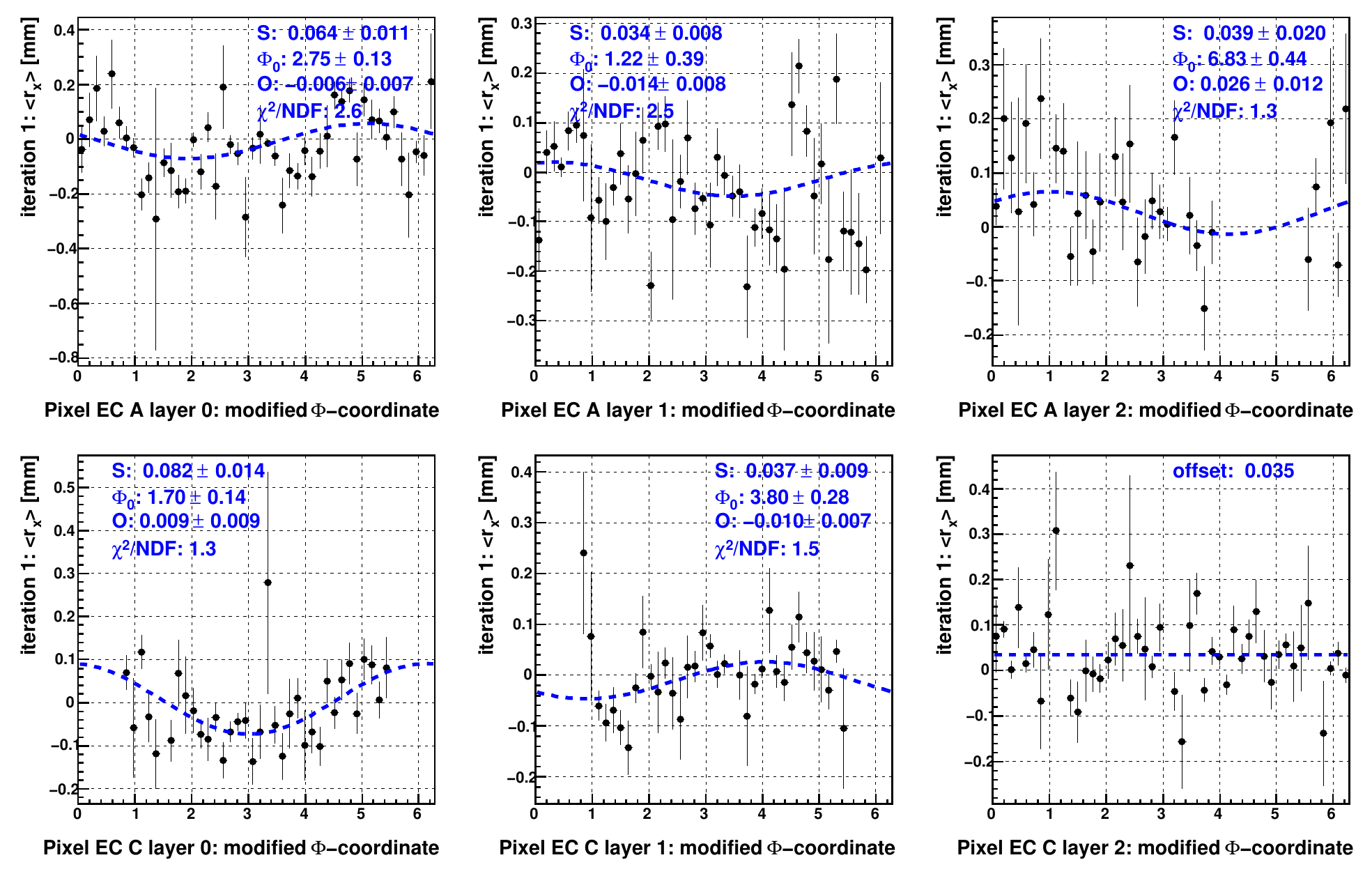}
\end{center}
\vspace{\cDist}
\caption[The distribution $\langle r_x\rangle_{\rm stave}(\Phi)$ for the pixel detector before alignment at~L2]{\label{fig:m8_r_x_vs_Phi_L2_PIX_before}
The $\rmean x(\Phi)$ distribution for the pixel detector using the full $B$-field off M8+ dataset {\bf before} alignment at~L2. The results are shown split by layers ({\bf left-middle-right} column for layer 0, 1, 2, respectively) and by detector parts: barrel{\bf~(top row)}, EC~A{\bf~(middle row)}, and EC~C{\bf~(bottom row)}. The fit results with a sine of the form specified in Equation~\ref{eqn:sineL2} are shown in blue.
\vspace{\cDistHalf}
}
\end{figure}

\begin{figure}
\begin{center}
\vspace{\cDistHalf}
\includegraphics[width=15.8cm,clip=true]{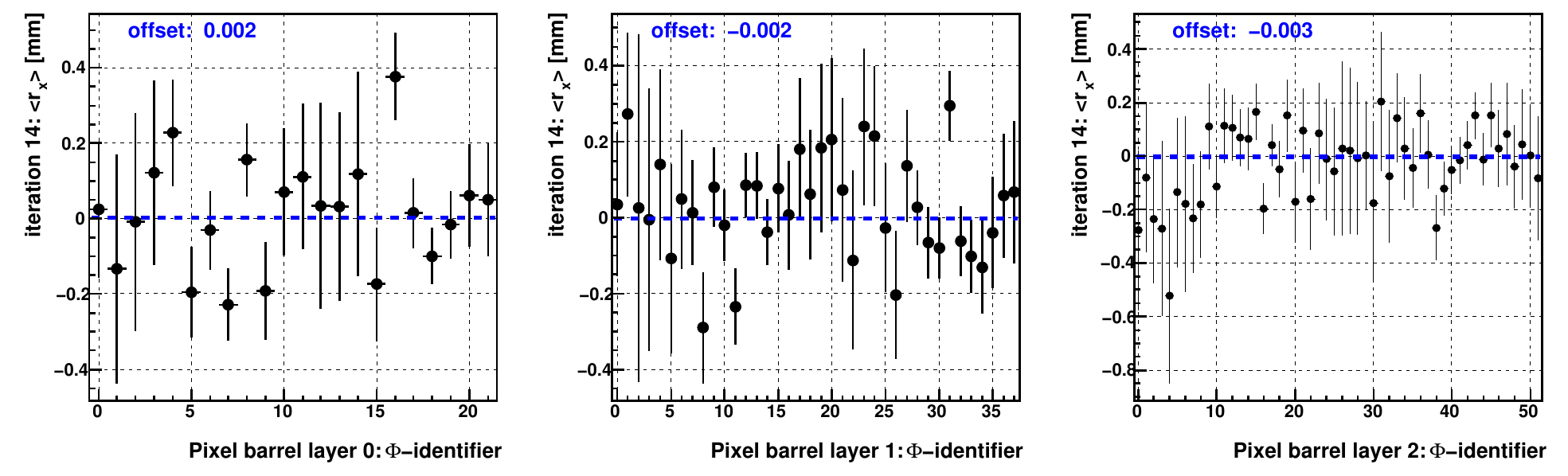}
\includegraphics[width=15.8cm,clip=true]{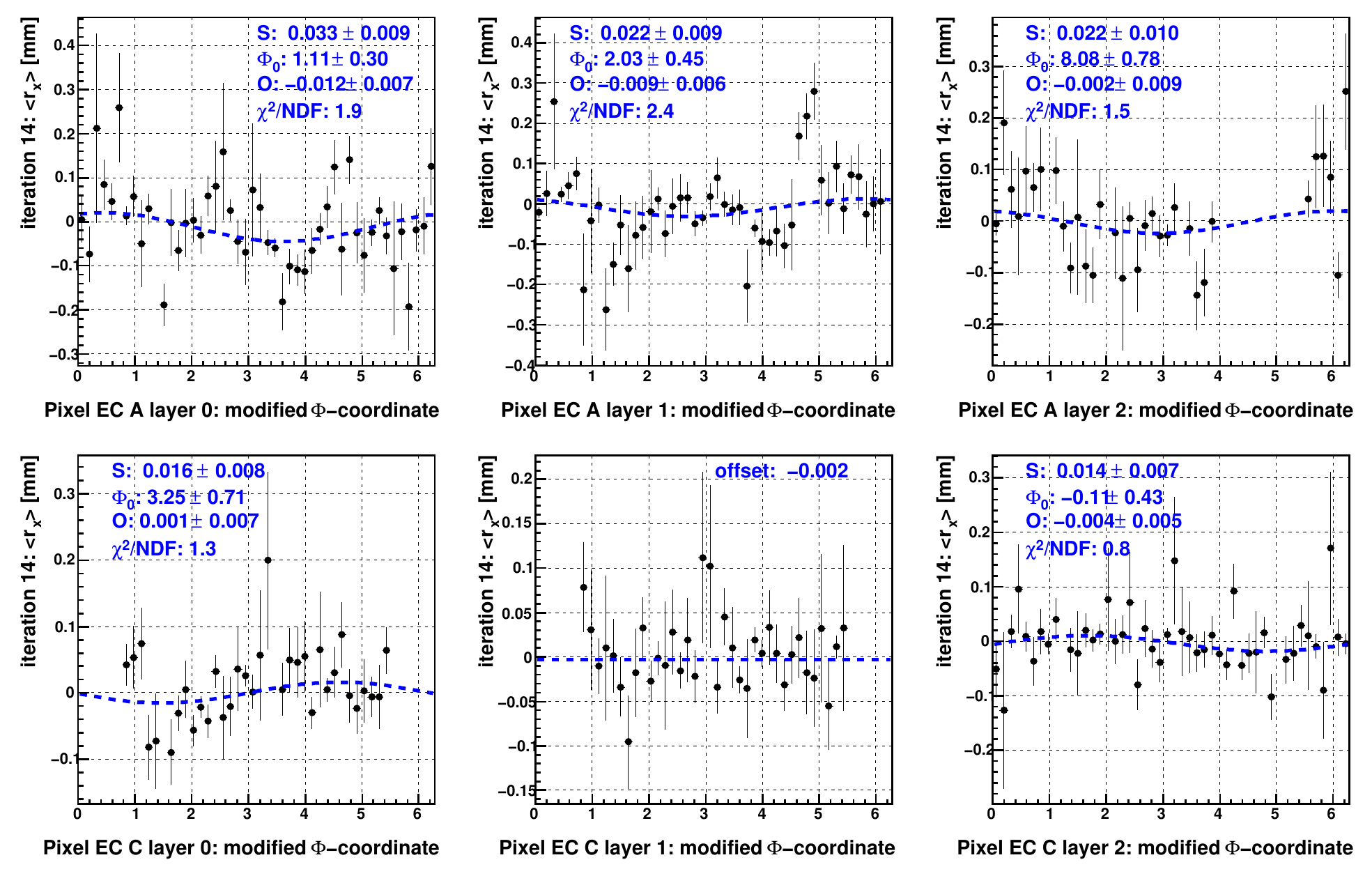}
\end{center}
\vspace{\cDist}
\caption[The distribution $\langle r_x\rangle_{\rm stave}(\Phi)$ for the pixel detector after alignment at~L2]{\label{fig:m8_r_x_vs_Phi_L2_PIX_after}
The $\rmean x(\Phi)$ distribution for the pixel detector using the full $B$-field off M8+ dataset {\bf after} alignment at~L2. The results are shown split by layers ({\bf left-middle-right} column for layer 0, 1, 2, respectively) and by detector parts: barrel{\bf~(top row)}, EC~A{\bf~(middle row)}, and EC~C{\bf~(bottom row)}. The fit results with a sine of the form specified in Equation~\ref{eqn:sineL2} are shown in blue.
\vspace{\cDistHalf}
}
\end{figure}

\begin{figure}
\begin{center}
\includegraphics[width=10.5cm,clip=true]{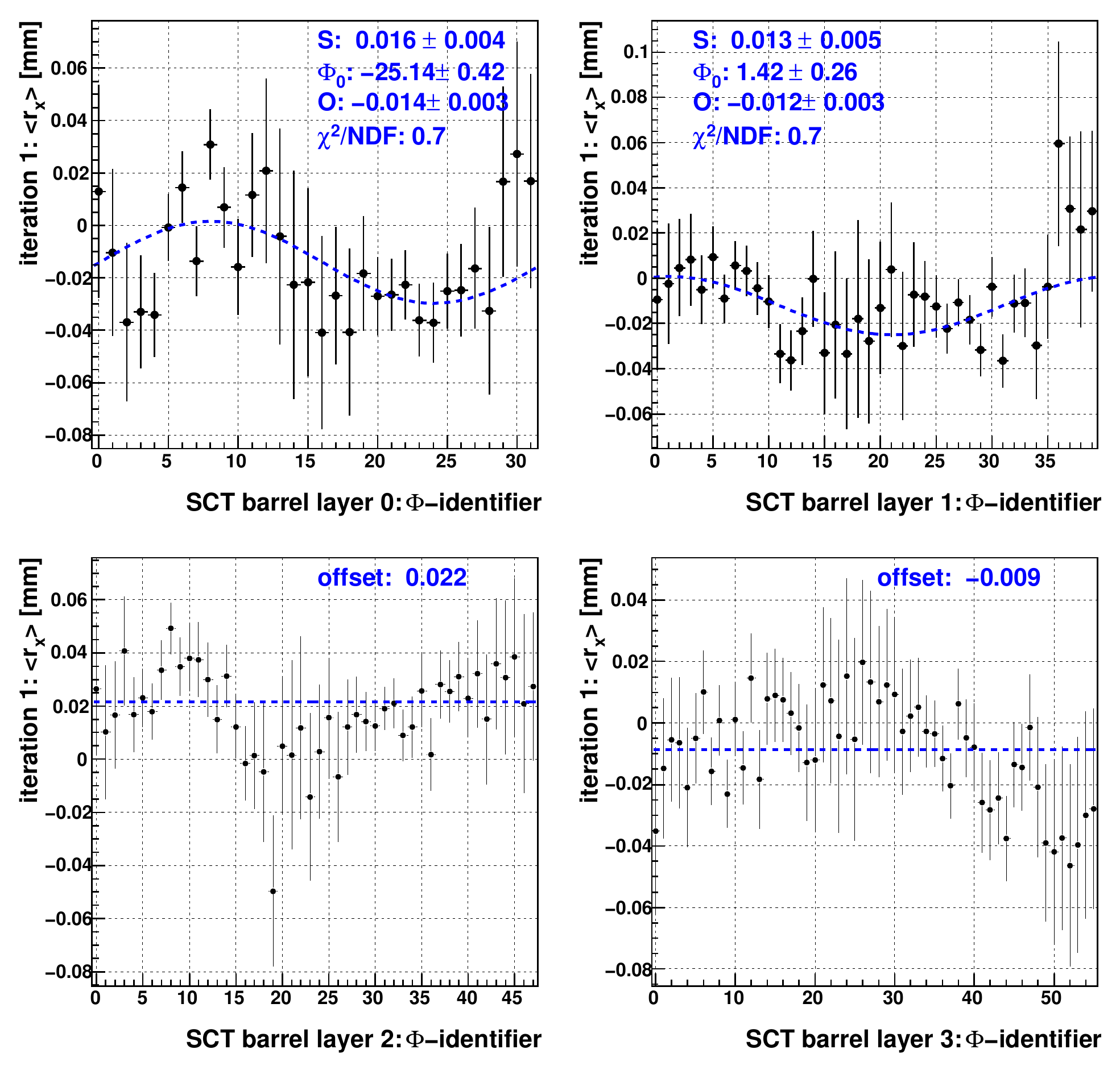}
\end{center}
\vspace{\cDist}
\caption[The distribution $\langle r_x\rangle_{\rm stave}(\Phi)$ for the barrel of the SCT detector before alignment at~L2]{\label{fig:m8_r_x_vs_Phi_L2_SCT_Brl_before}
The $\rmean x(\Phi)$ distribution for the barrel of the SCT detector by layers using the full $B$-field off M8+ dataset {\bf before} alignment at~L2. The fit results with a sine of the form specified in Equation~\ref{eqn:sineL2} are shown in blue.
%\vspace{-0.5cm}
}
\end{figure}

\begin{figure}
\begin{center}
\includegraphics[width=10.5cm,clip=true]{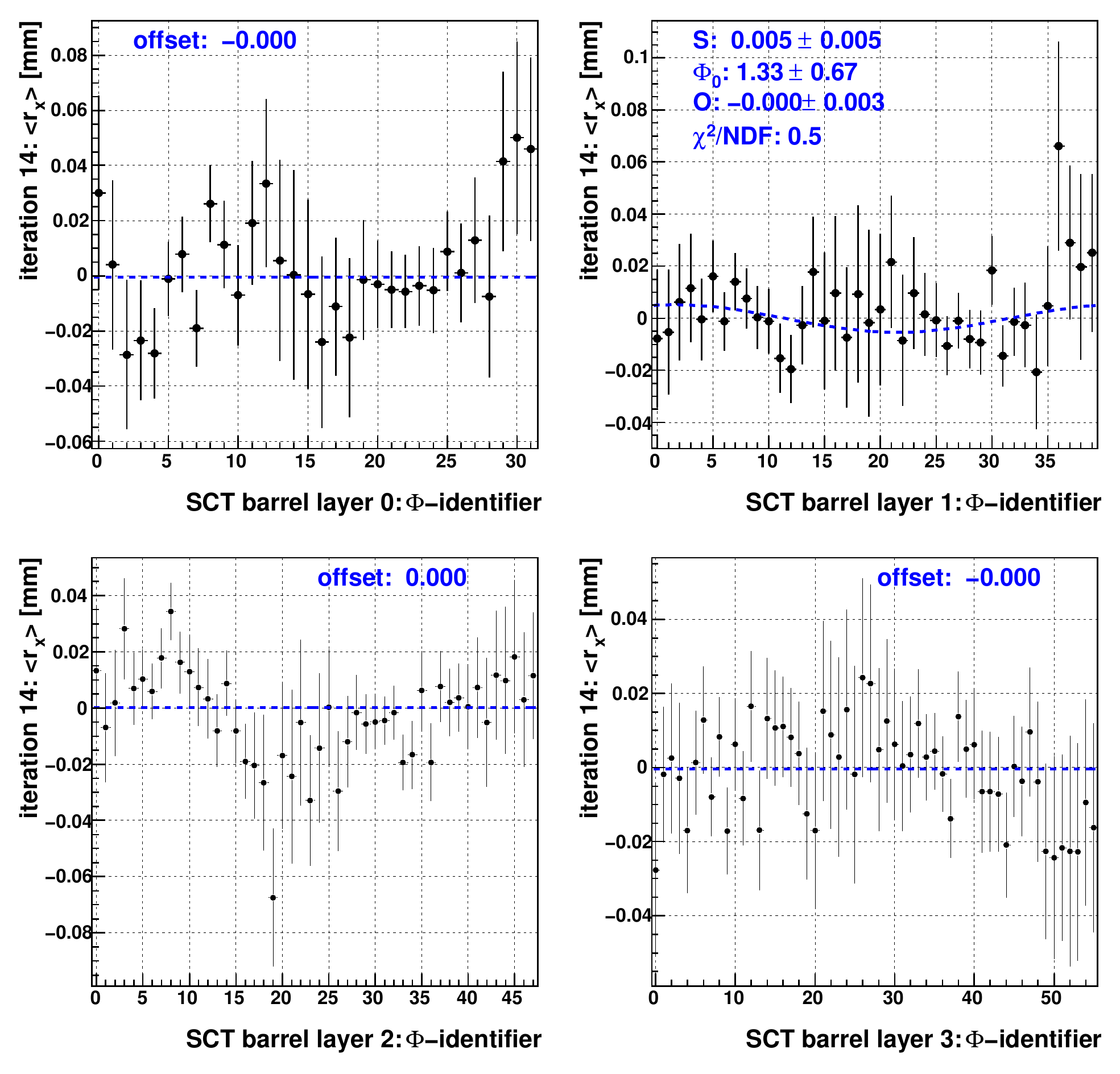}
\end{center}
\vspace{\cDist}
\caption[The distribution $\langle r_x\rangle_{\rm stave}(\Phi)$ for the barrel of the SCT detector after alignment at~L2]{\label{fig:m8_r_x_vs_Phi_L2_SCT_Brl_after}
The $\rmean x(\Phi)$ distribution for the barrel of the SCT detector by layers using the full $B$-field off M8+ dataset {\bf after} alignment at~L2. The fit results with a sine of the form specified in Equation~\ref{eqn:sineL2} are shown in blue.
%\vspace{-0.5cm}
}
\end{figure}

\begin{figure}
\begin{center}
\includegraphics[width=10.5cm,clip=true]{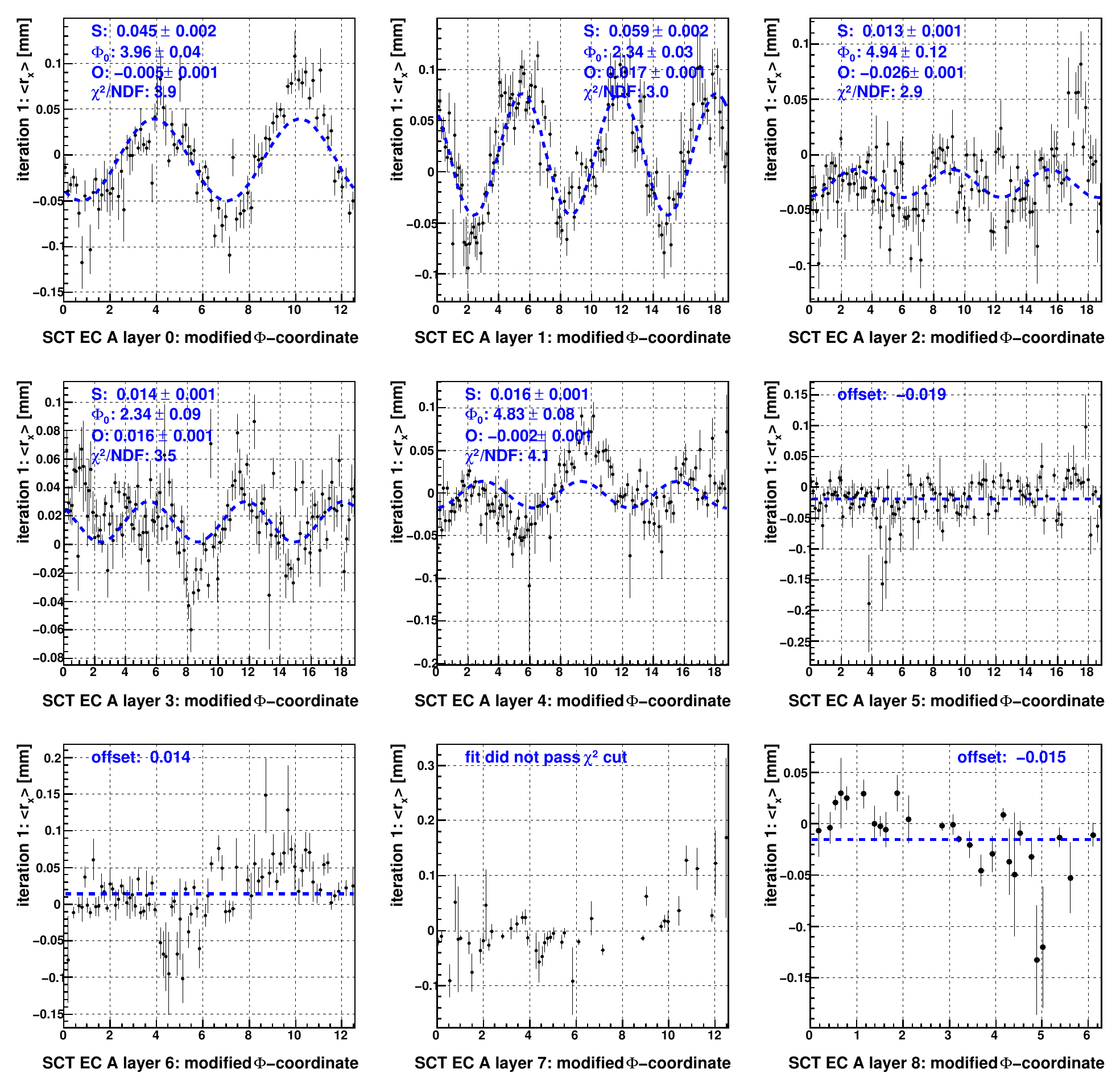}
\end{center}
\vspace{\cDist}
\caption[The distribution $\langle r_x\rangle_{\rm stave}(\Phi)$ for the EC~A of the SCT detector before alignment at~L2]{\label{fig:m8_r_x_vs_Phi_L2_SCT_ECA_before}
The $\rmean x(\Phi)$ distribution for the EC~A of the SCT detector by disk layers using the full $B$-field off M8+ dataset {\bf before} alignment at~L2. The fit results with a sine of the form specified in Equation~\ref{eqn:sineL2} are shown in blue.
%\vspace{-0.5cm}
}
\end{figure}

\begin{figure}
\begin{center}
\includegraphics[width=10.5cm,clip=true]{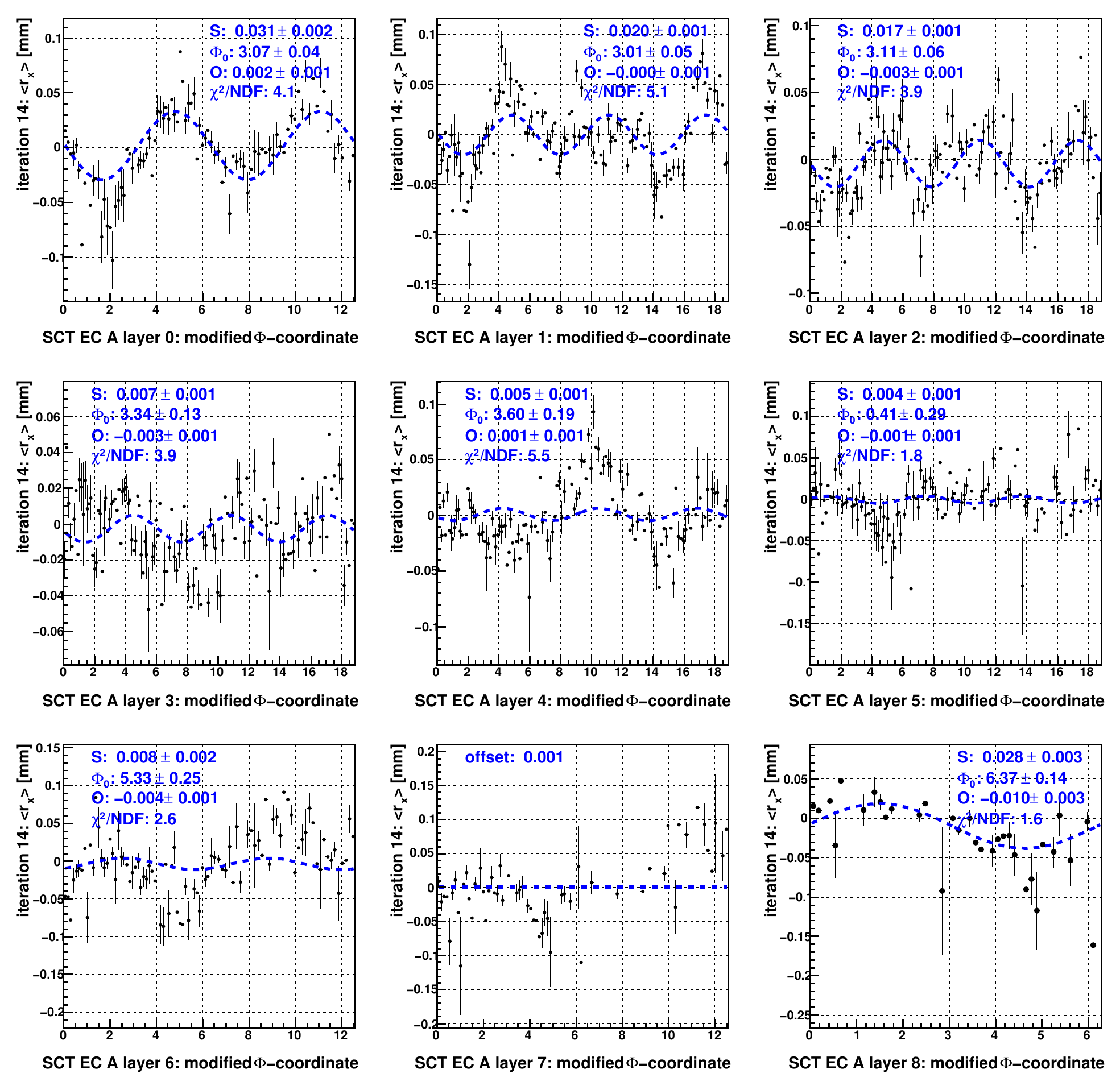}
\end{center}
\vspace{\cDist}
\caption[The distribution $\langle r_x\rangle_{\rm stave}(\Phi)$ for the EC~A of the SCT detector after alignment at~L2]{\label{fig:m8_r_x_vs_Phi_L2_SCT_ECA_after}
The $\rmean x(\Phi)$ distribution for the EC~A of the SCT detector by disk layers using the full $B$-field off M8+ dataset {\bf after} alignment at~L2. The fit results with a sine of the form specified in Equation~\ref{eqn:sineL2} are shown in blue.
%\vspace{-0.5cm}
}
\end{figure}

The 31 
%($3+2\times3+4+2\times9$) 
$\rmean x(\Phi)$ distributions which were the basis for the calculation of the alignment corrections $c_X$, $c_Y$, and $c_\Gamma$ for the individual barrel layers and end-cap disks are displayed in a sequence of Figures on pages~\pageref{fig:m8_r_x_vs_Phi_L2_PIX_before}--\pageref{fig:m8_r_x_vs_Phi_L2_SCT_ECC_after}. The distributions are shown {\em before}\footnote{Here and in the rest of the Subsection, ``before alignment'' refers strictly speaking to the geometry before L2 but after L1~alignment unless stated otherwise.} and {\em after} alignment to monitor the performance of the \RA\ algorithm. The fits which were used to calculate the alignment corrections are shown as blue dashed lines. In the following, an interpretation of the Figures shall be given by subdetectors:
\begin{description}
\item[Pixel barrel:] The $\rmeanstave x(\Phi)$ distributions {\em before} alignment shown in the top row of Figure~\ref{fig:m8_r_x_vs_Phi_L2_PIX_before} display a sinusoidal dependance for layer~0 and 1. The magnitude $S$ is  rather small of $\order{100\,\mum}$, which is owing to a preceding alignment at L1. All three layers show offsets $O$ of $\order{100\,\mum}$, which reflect a rotation with respect to each other about the $Z$-axis. {\em After} alignment the sinusoidal dependance disappears as shown in the top row of Figure~\ref{fig:m8_r_x_vs_Phi_L2_PIX_after}, and only micron-sized offsets remain;
\item[Pixel end-caps:] For EC~A/C, the $\rmean x(\Phi)$ distributions {\em before} alignment are presented in the middle/bottom row of Figure~\ref{fig:m8_r_x_vs_Phi_L2_PIX_before}, respectively. Both ECs show common features: besides disk 2 of EC~C, their $\rmean x(\Phi)$ distributions can be fitted by a sinusoid with a relatively small $S$ of \order{50\,\mum}, which indicates a good assembly precision of the ECs. The phases $\Phi_0$ of the fits vary in $[0,2\pi)$ since misalignments are distributed in the $X$-$Y$ plane. The offsets $O$ do not exceed 26\,\mum, which indicates a good disk-to-disk alignment in $\Gamma$. The analogous distributions {\em after} alignment are shown in the middle/bottom row of Figure~\ref{fig:m8_r_x_vs_Phi_L2_PIX_after}. Only small offsets of \order{5\,\mum}\ remain. The distributions still can be fitted by sine curves, however with a much smaller magnitude of $\order{20\,\mum}$. A close look reveals that those fits are mostly driven by few modules with small uncertainties, and the remaining misalignments should better be dealt with at L3.\\
The missing entries at $\Phi\in[0,\,0.8]\cap[5.5,\,2\pi)\,\rad$ are mostly due to cooling loop failures;
\item[SCT barrel:] The situation {\em before} alignment is relatively well-behaved, as can be seen from the $\rmeanstave x(\Phi)$ distributions in Figure~\ref{fig:m8_r_x_vs_Phi_L2_SCT_Brl_before}, in particular the small range of the $y$-axis. The offsets $O$ are rather small of \order{15\,\mum}, and the magnitudes $S$ of layer~0 and 1, which can be fitted by a sine curve in the 1$^{\rm st}$ iteration, are in the same ball park. All this demonstrates a good assembly precision of the SCT layers with respect to each other, but also at L3. The weakly pronounced sinusoidal dependance in layer~0 is diminished {\em after} alignment, as shown in  Figure~\ref{fig:m8_r_x_vs_Phi_L2_SCT_Brl_after}. For the only layer which can be fitted by a sine the magnitude $S=5\pm5\,\mum$ is consistent with zero. All offsets $O$ are consistent with $0\pm0.5\,\mum$;
\item[SCT end-cap A:] The $\rmean x(\Phi)$ distribution {\em before} alignment is displayed in Figure~\ref{fig:m8_r_x_vs_Phi_L2_SCT_ECA_before}. While the outermost four disks do not display a sinusoidal dependance, it is observed in the innermost five disks with an increasing magnitude from $S\simeq15\,\mum$ in disk~4 to $S\simeq50\,\mum$ in disks~0 and 1. It is worthwhile noting that the phase $\Phi_0$ is similar to within 1\,rad in the innermost five disks, which is a sign of L1 misalignment with respect to the rest of the silicon tracker (which was chosen not to be corrected for as argued in Subsection~\ref{ssec:l1M8}). The offsets are with $\order{15\,\mum}$ small across the disks. {\em After} alignment, the magnitude $S$ reduces by about $\oneOverTwo$, and $O$ is almost consistent with 0, as shown in Figure~\ref{fig:m8_r_x_vs_Phi_L2_SCT_ECA_after}. An interesting observation is the remainder of some residual sinusoidal dependance and the fact that $\Phi_0\simeq3.20\,\rad$ in the innermost five disks despite alignment. This is not fully understood yet. However, the residual distributions undoubtedly improve as can be seen from Figure~\ref{fig:r_x_SCTA_L2} and the summary in Table~\ref{tab:m8plusAlignL2}.\\
Note that the all disks which display a sinusoidal dependance can be described by one single sine fit, which indicates that ring-to-ring misalignments within a disk are indeed smaller than disk-to-disk misalignments;
\end{description}

%\nopagebreak[5]

\begin{figure}
\begin{center}
\includegraphics[width=10.5cm,clip=true]{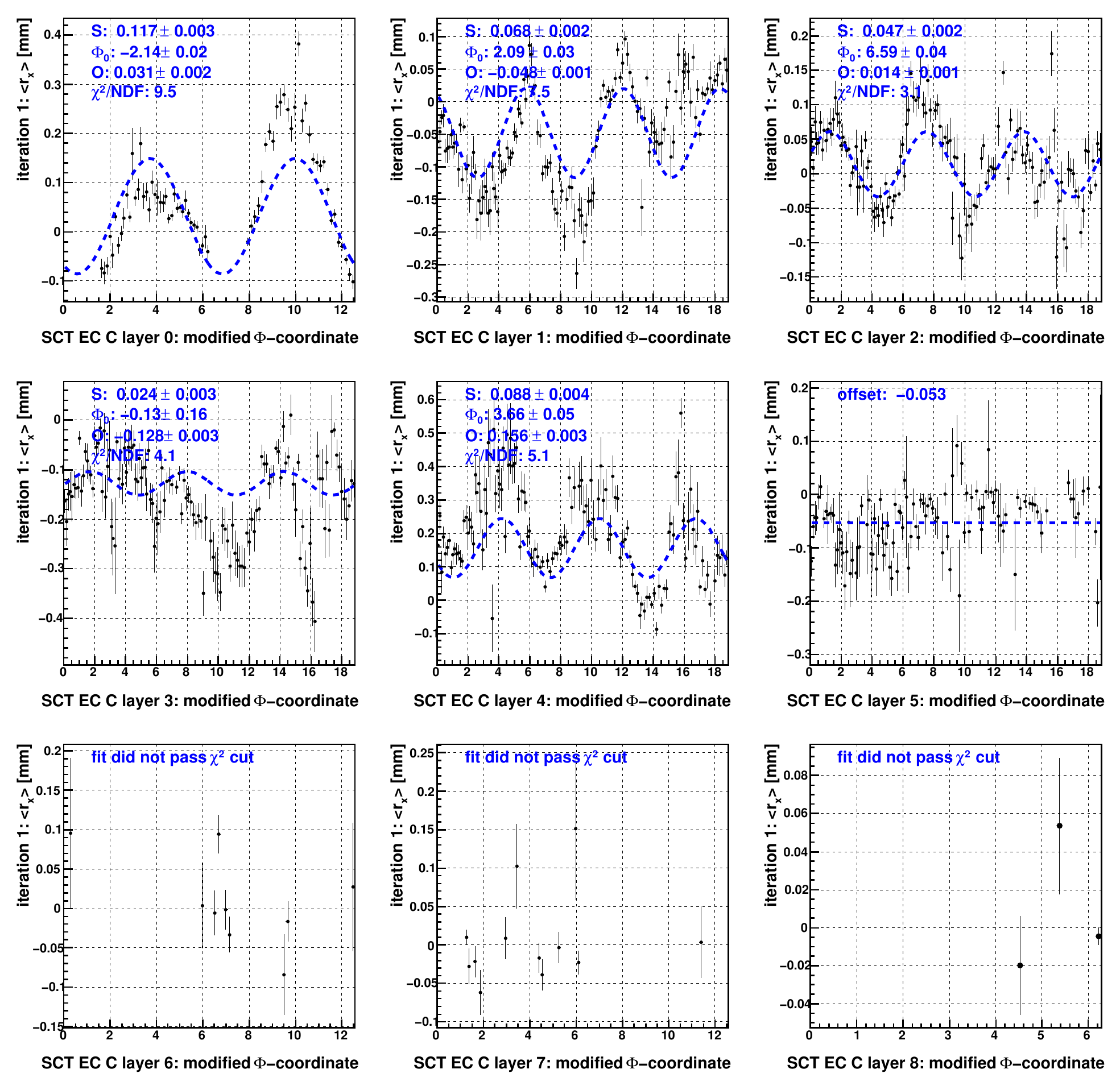}
\end{center}
\vspace{\cDist}
\caption[The distribution $\langle r_x\rangle_{\rm stave}(\Phi)$ for the EC~C of the SCT detector before alignment at~L2]{\label{fig:m8_r_x_vs_Phi_L2_SCT_ECC_before}
The $\rmean x(\Phi)$ distribution for the EC~C of the SCT detector by disk layers using the full $B$-field off M8+ dataset {\bf before} alignment at~L2. The fit results with a sine of the form specified in Equation~\ref{eqn:sineL2} are shown in blue.
}
\end{figure}

\begin{figure}
\begin{center}
\includegraphics[width=10.5cm,clip=true]{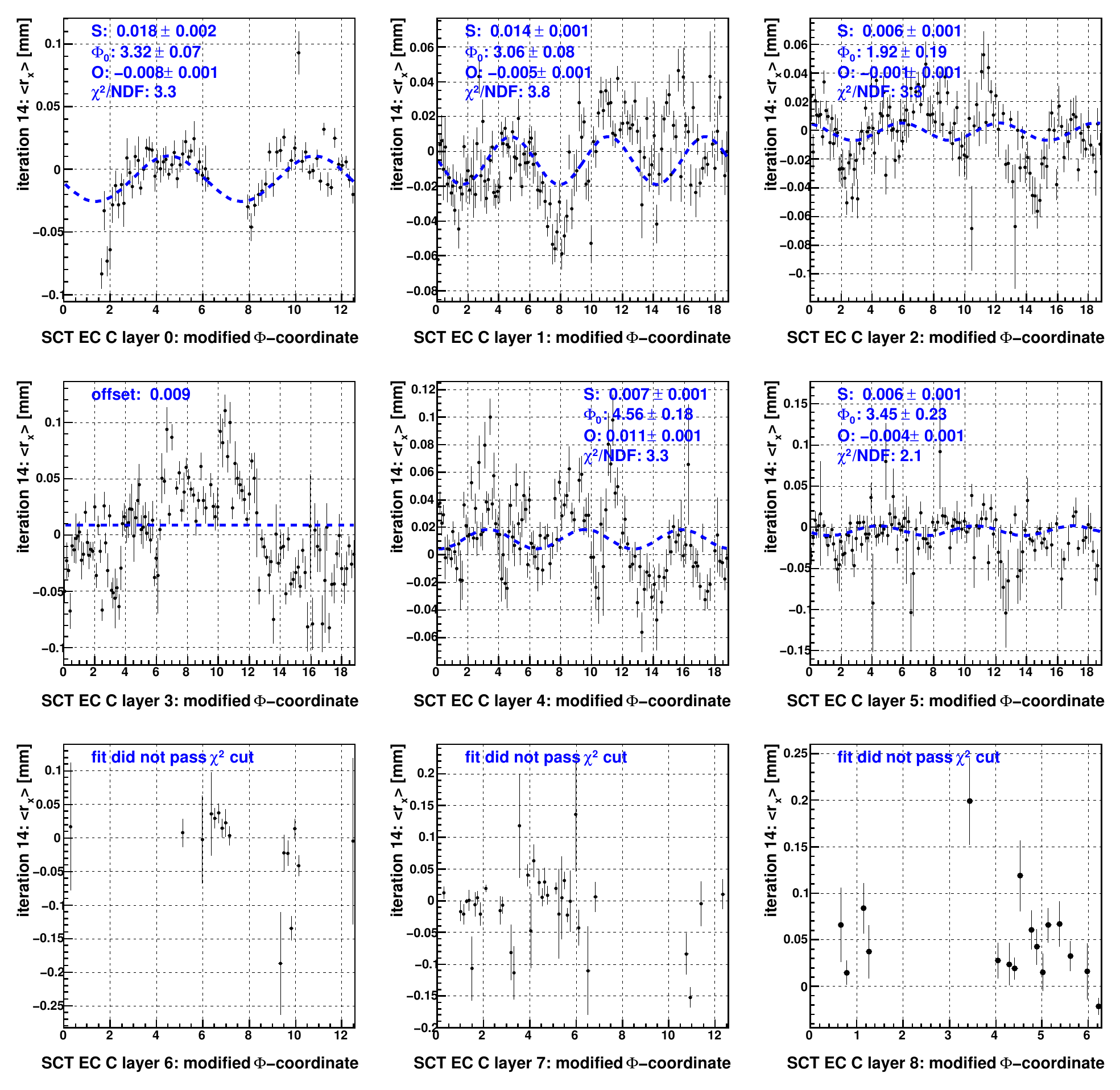}
\end{center}
\vspace{\cDist}
\caption[The distribution $\langle r_x\rangle_{\rm stave}(\Phi)$ for the EC~C of the SCT detector after alignment at~L2]{\label{fig:m8_r_x_vs_Phi_L2_SCT_ECC_after}
The $\rmean x(\Phi)$ distribution for the EC~C of the SCT detector by disk layers using the full $B$-field off M8+ dataset {\bf after} alignment at~L2. The fit results with a sine of the form specified in Equation~\ref{eqn:sineL2} are shown in blue.
}
\end{figure}

\begin{description}
\item[SCT end-cap C:] As can be seen from the $\rmean x(\Phi)$ distribution {\em before} alignment in Figure~\ref{fig:m8_r_x_vs_Phi_L2_SCT_ECC_before}, the L2 alignment of EC~C is the least trivial one. The most striking observation for the innermost five disks is that the $\rmean x(\Phi)$ values in the individual rings constituting a disk are not consistent with a common sine fit hypothesis. This is quite different from EC~A and poses a strong indication for substantial ring-to-ring misalignments in disks. Further, the offset values $O$ range from $-128\,\mum$ to 156\,\mum\ for the 1$^{\rm st}$ iteration. Moreover, $\Phi_0\in[0,\,2\pi)\,\rad$ and $S$ of $\order{100\,\mum}$ are observed. These are symptoms of larger disk-to-disk misalignments, which given the underlying distributions raises doubts about the hierarchy of L1 and L2 alignment in the EC~C in the sense of Equation~\ref{eqn:magMisalLevel}, as discussed in Subsection~\ref{ssec:l1} on page~\pageref{par:pixelECL2} ff.. The ultimate judgement will be rendered by collision data. Nevertheless, the \RA\ procedure converges to a good degree as can be seen from Figure~\ref{fig:m8_r_x_vs_Phi_L2_SCT_ECC_after} {\em after} alignment: $S\simeq10\,\mum$ and $O\simeq5\,\mum$ are obtained. The convergence hypothesis is further supported by Figure~\ref{fig:r_x_SCTC_L2} and the summary in Table~\ref{tab:m8plusAlignL2}. The three outermost disks are not well-illuminated, and therefore were not aligned. Nonetheless, they collect more hits after the alignment procedure due to the alignment of neighbouring disks.
\end{description}

%\nopagebreak[5]

\renewcommand{\floatpagefraction}{0.7}
\renewcommand{\textfraction}{0.1}

%\nopagebreak[5]
%\nopagebreak[5]

%\nopagebreak[5]
%\nopagebreak[5]

%\nopagebreak[5]
%\nopagebreak[5]

\begin{table}
\small
\begin{center}
\vspace{\cDistHalf}
\begin{tabular}{llc|rrr}
\hline
Detector & Subdet. & Layer & ~~~~$X$ [$\mu$m] & ~~~~$Y$ [$\mu$m] & ~~$\Gamma$ [mrad]\\
\hline\hline
		&	& 0 & $-74.5$ & $1.6$   & $3.91$ \\
		&Barrel	& 1 & $-44.3$ & $32.0$  & $0.68$ \\
		&	& 2 & $\nill$&$\nill$&$-1.10$\\
\cmidrule{2-6}
		&	& 0 & $322.6$ & $283.9$ & $-2.06$\\
{\bf Pixel}	&EC~A	& 1 & $584.2$ & $299.7$ & $-2.07$\\
		&	& 2 & $535.6$ & $289.1$ & $-2.81$\\
\cmidrule{2-6}
		&	& 0 & $64.2$  & $369.8$ & $-3.17$\\
		&EC~C	& 1 & $51.0$  & $584.0$ & $-2.94$\\
		&	& 2 & $52.7$  & $561.5$ & $-3.56$\\
\hline\hline
\multirow{22}*{\bf SCT} & \multirow{4}*{Barrel}
			& 0 & $15.3$  &       $0.2$   &       $0.046$ \\
		&	& 1 & $6.5$   &       $-24.9$&        $0.048$ \\
		&	& 2 & $\nill$&     $\nill$&     $-0.071$ \\
		&	& 3 & $7.8$   &       $11.3$  &       $0.031$ \\
\cmidrule{2-6}
		&	& 0 & $-404.0$&       $53.3$  &       $0.025$ \\
		&	& 1 & $-366.4$&       $-115.4$&       $-0.055$ \\
		&	& 2 & $-126.9$&       $-4.9$  &       $0.231$ \\
		&	& 3 & $-66.3$&        $-15.4$&        $-0.091$ \\
		&EC~A	& 4 & $-12.9$&        $52.7$  &       $-0.020$ \\
		&	& 5 & $3.9$   &       $-1.7$  &       $0.228$ \\
		&	& 6 & $1.5$   &       $21.0$  &       $-0.092$ \\
		&	& 7 & $\nill$&        $\nill$&        $\nill$ \\
		&	& 8 & $28.2$  &       $-2.4$  &       $0.050$ \\
\cmidrule{2-6}
		&	& 0 & $-423.8$&       $378.5$&        $-0.141$ \\
		&	& 1 & $-251.6$&       $-188.9$&       $0.512$ \\
		&	& 2 & $205.9$&        $-200.5$&       $0.089$ \\
		&	& 3 & $87.1$&         $-30.6$&        $0.690$ \\
		&EC~C	& 4 & $-268.5$&       $324.1$&        $-1.582$ \\
		&	& 5 & $-48.9$&        $9.9$   &       $0.506$ \\
		&	& 6 & $\nill$&        $\nill$&        $\nill$ \\
		&	& 7 & $\nill$&        $\nill$&        $\nill$ \\
		&	& 8 & $\nill$&        $\nill$&        $\nill$ \\
\hline
\vspace{\cDistHalf}
\end{tabular}
\caption[The alignment contstants at L2 for the pixel and SCT detectors in M8+]{\label{tab:alignL2}
The alignment contstants derived at {\bf L2} with the \RA\ alignment algorithm for the $3+4$ barrel layers and $2\times(3+9)$ end-cap disks of the pixel and SCT detectors. The corrections are to be applied {\em on top} of the L1 correction in Table~\ref{tab:alignL1}. The offset fit in disk~7 in EC~A of the SCT is consistent with 0 and therefore no corrections are applied. Disks 6, 7, 8 in EC~C in the SCT are not well illuminated and thus no alignment constants were calculated. Note the small magnitude of correction in the barrel layers of both pixel and SCT, which are due to elimination of L1 misalignments documented in Subsection~\ref{ssec:l1M8}.
\vspace{\cDist}
}
\end{center}
\end{table}

The alignment corrections derived with the \RA\ algorithm at L2 are summarised in Table~\ref{tab:alignL2}. They basically mirror the discussion above, which shall not be repeated here. Keep in mind that they are supposed to be applied {\em on top} of the L1 corrections derived for the pixel barrel in the first line of Table~\ref{tab:alignL1}. On the contrary, the alignment corrections for the pixel ECs in Table~\ref{tab:alignL2} {\em replace} the ones in line~2 and 3 of Table~\ref{tab:alignL1}.

\begin{figure}
\begin{center}
\vspace{\cDistHalf}
\includegraphics[width=15.8cm,clip=true]{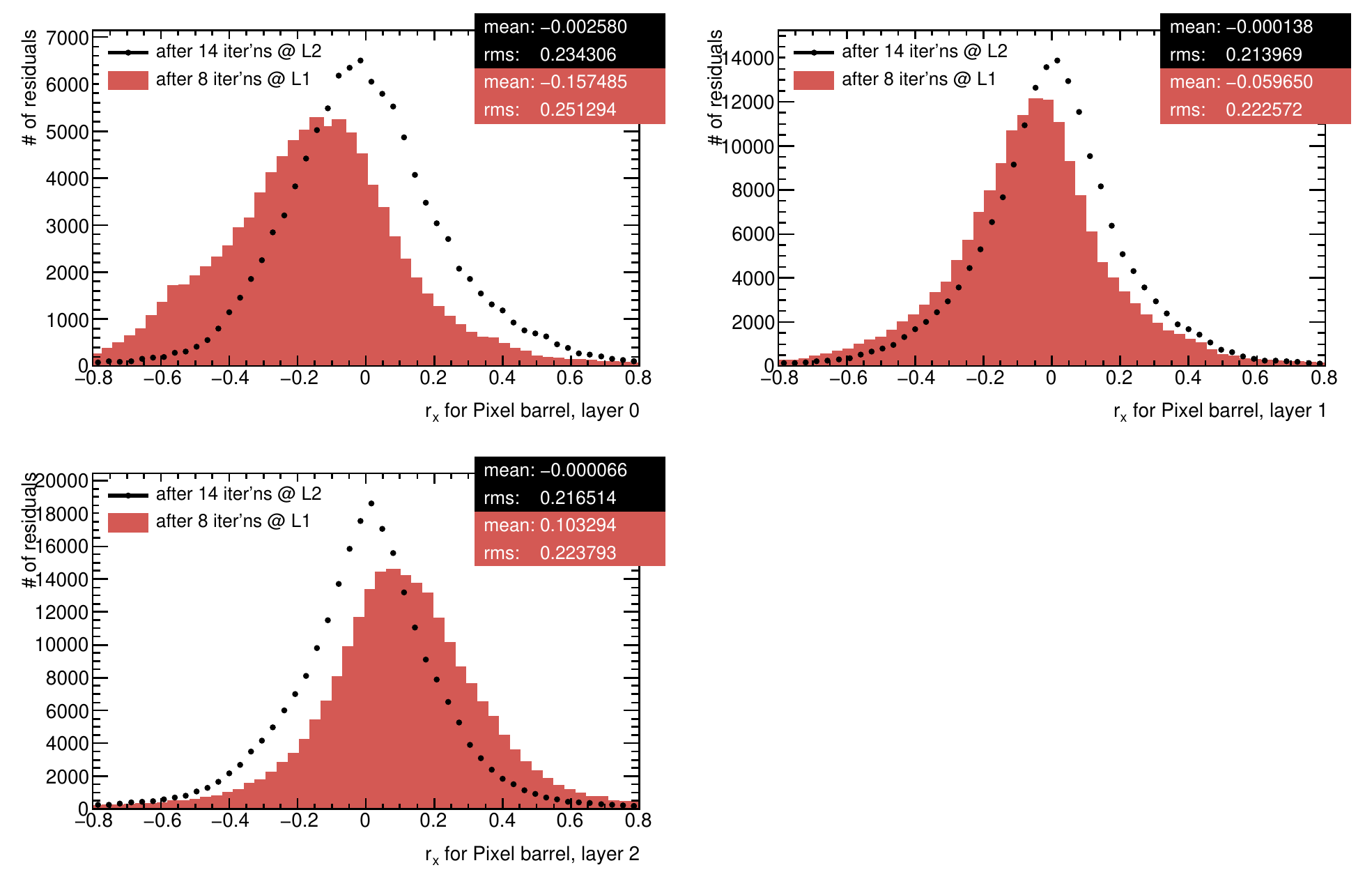}
\vspace{\cDist}
\end{center}
\caption[$r_x$ residual distribution in the barrel of the pixel detector before and after alignment corrections at L2 in M8+]{\label{fig:r_x_PIXB_L2}
The $r_x$ {\bf residual} distribution in the barrel of the pixel detector by layers {\em before} and {\em after} alignment corrections at {\bf L2} in M8+. The residual means dramatically are re-centered around zero to within a micron from $\order{100\,\mum}$, while the residual widths improve by circa~12\%. 
%: initial layer-to-layer misalignments in the $X$-$Y$ plane of \order{50\,\mum} are dominated by misalignments in $\Gamma$ of \order{0.5\,\rm mrad}.
This indicates that initial layer-to-layer misalignments in the $X$-$Y$ plane of \order{50\,\mum} are dominated by the misalignments in $\Gamma$ of \order{0.5\,\rm mrad}. All values are in~mm.
}
\end{figure}

\begin{figure}
\begin{center}
\vspace{\cDistHalf}
\includegraphics[width=15.8cm,clip=true]{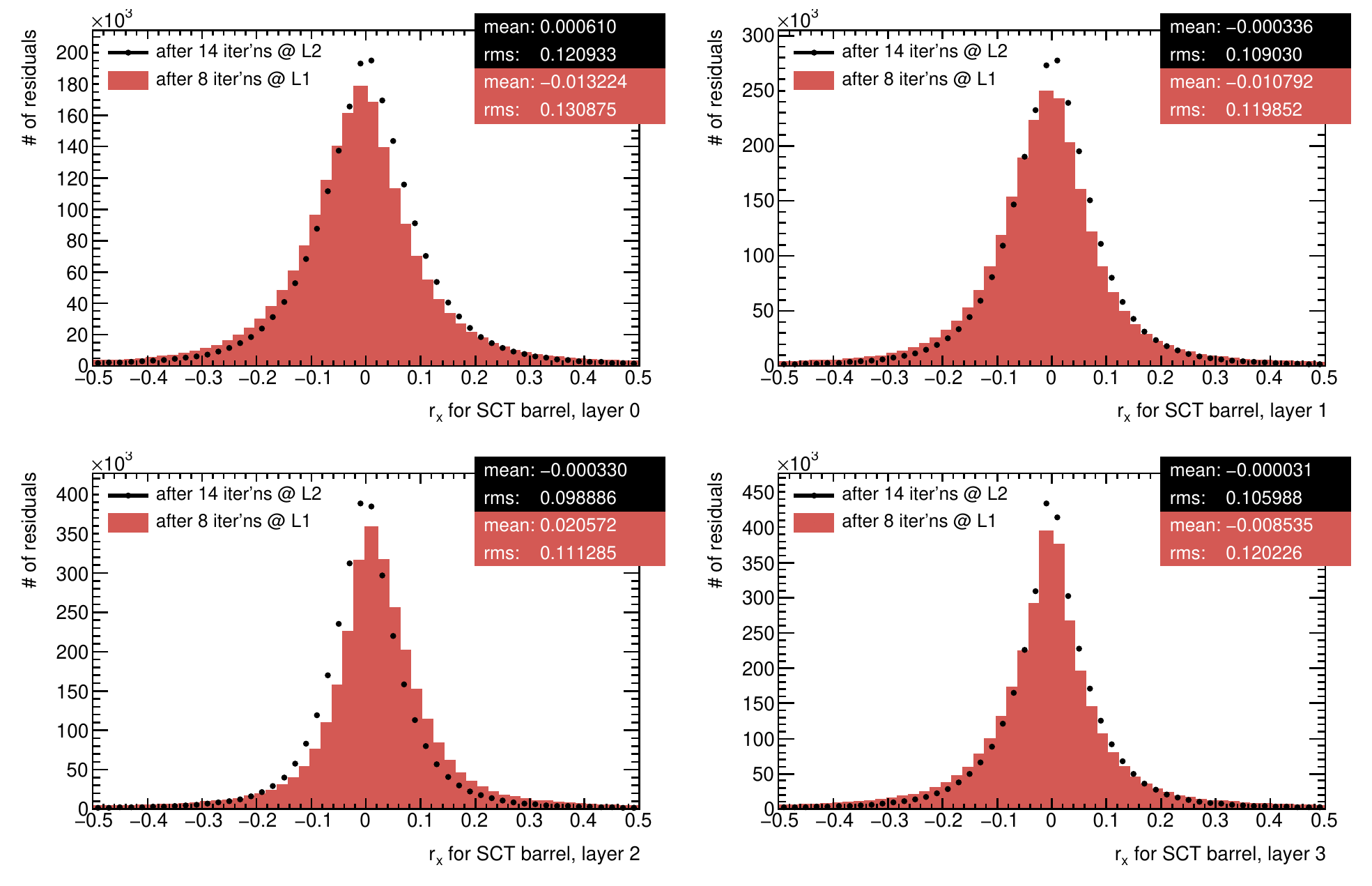}
\vspace{\cDist}
\end{center}
\caption[$r_x$ residual distribution in the barrel of the SCT detector before and after alignment corrections at L2 in M8+]{\label{fig:r_x_SCTB_L2}
The $r_x$ {\bf residual} distribution in the barrel of the SCT detector by layers {\em before} and {\em after} alignment corrections at {\bf L2} in M8+. A qualitatively similar picture to the pixel barrel can be observed. $\rsig x$ refines by about 20\%, while $\rmean x=0\pm0.5\,\mum$. All values are in~mm.
\vspace{\cDist}
}
\end{figure}

\begin{figure}
\begin{center}
\vspace{\cDistHalf}
\includegraphics[width=15.8cm,clip=true]{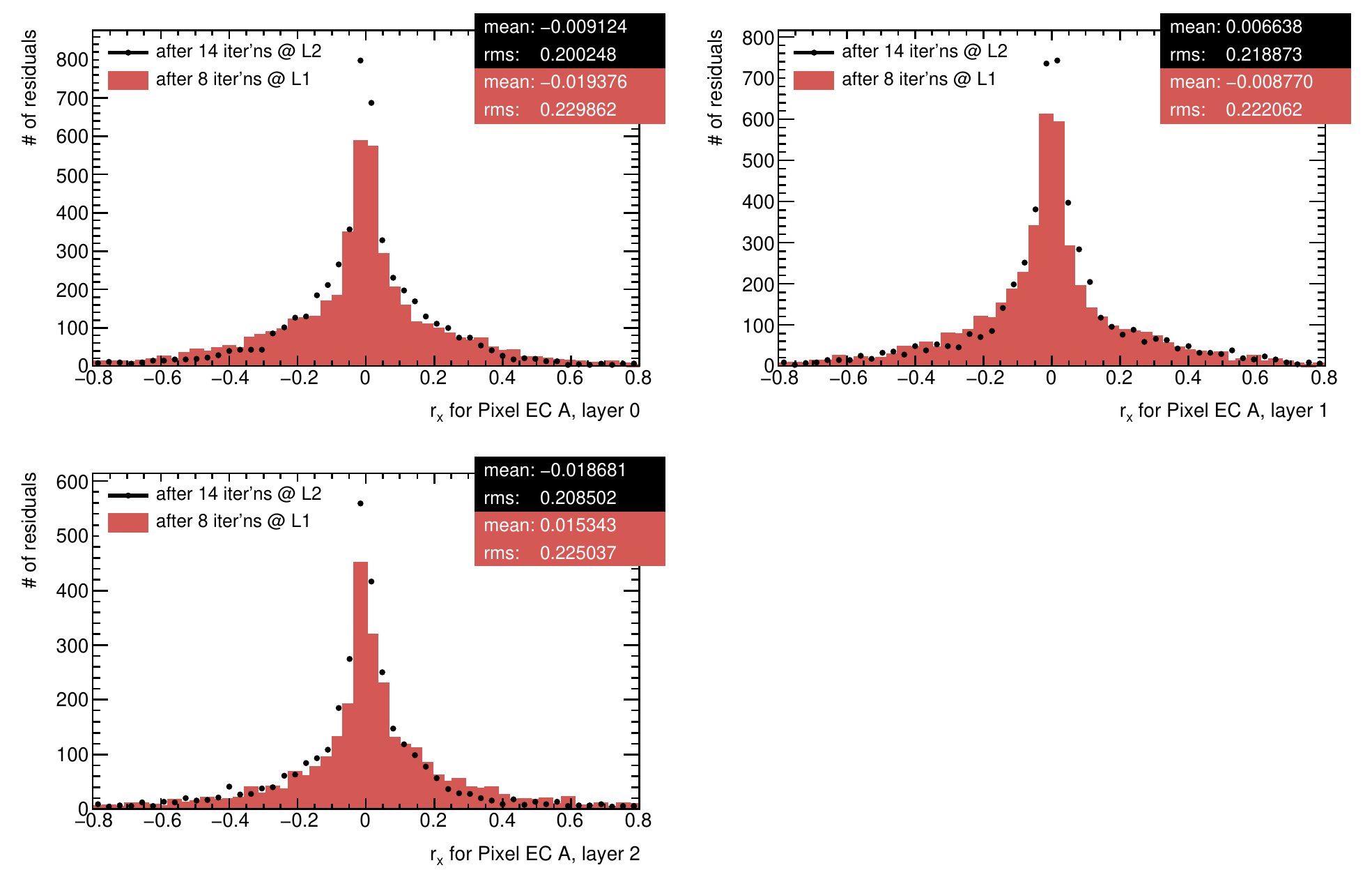}
\vspace{\cDist}
\end{center}
\caption[$r_x$ residual distribution in the EC~A of the pixel detector before and after alignment corrections at L2 in M8+]{\label{fig:r_x_PIXA_L2}
The $r_x$ {\bf residual} distribution in the EC~A of the pixel detector by disk layers {\em before} and {\em after} alignment corrections at {\bf L2} in M8+. The residual means are approaching zero while the widths improve by about 12\% in the process of alignment.  All values are in~mm.
}
\end{figure}

\begin{figure}
\begin{center}
\vspace{\cDistHalf}
\includegraphics[width=15.8cm,clip=true]{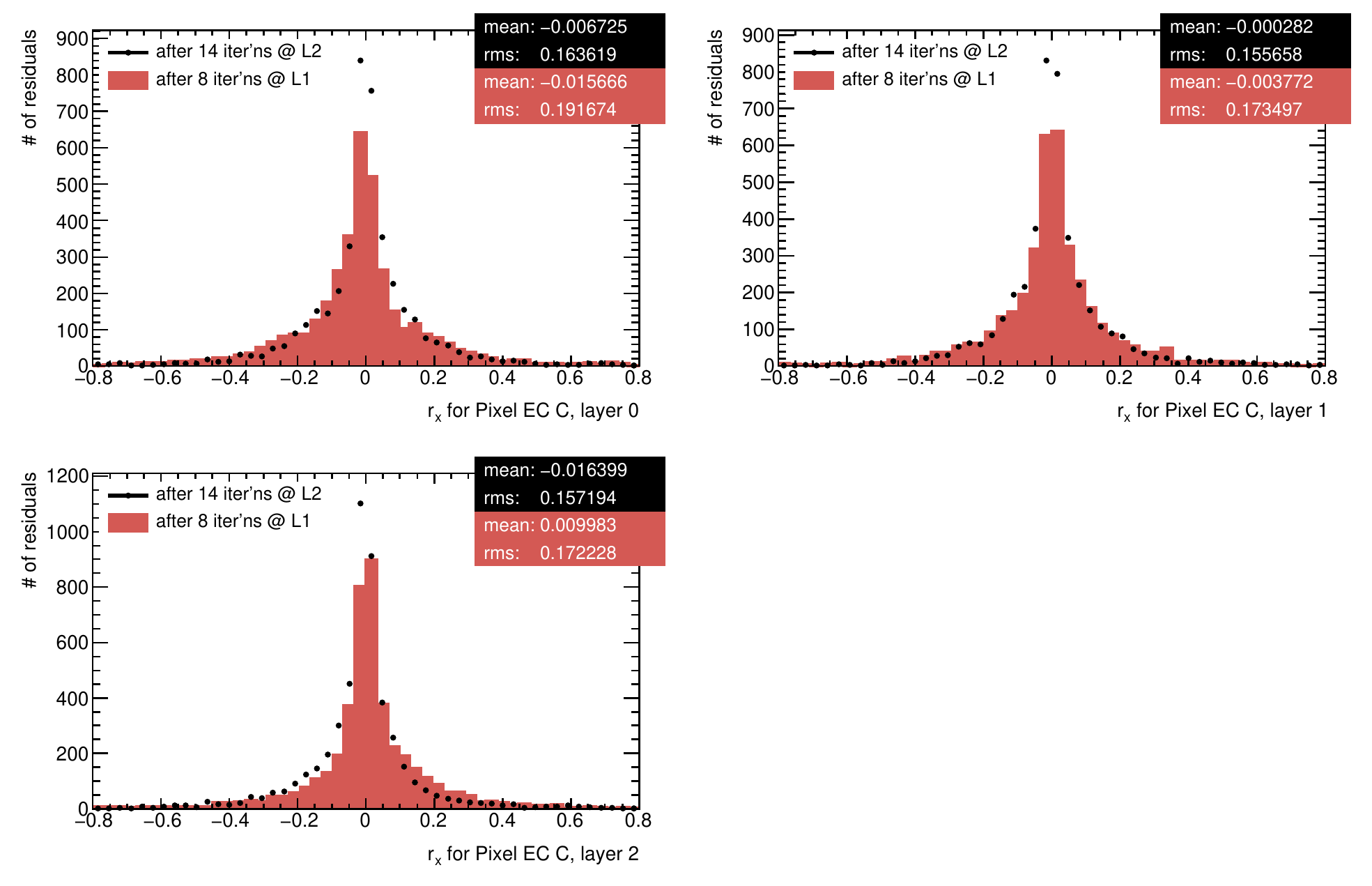}
\vspace{\cDist}
\end{center}
\caption[$r_x$ residual distribution in the EC~C of the pixel detector before and after alignment corrections at L2 in M8+]{\label{fig:r_x_PIXC_L2}
The $r_x$ {\bf residual} distribution in the EC~C of the pixel detector by disk layers {\em before} and {\em after} alignment corrections at {\bf L2} in M8+. A similar picutre to EC~A is observed: the residual means are approaching zero while the widths improve by circa 15\%.  All values are in~mm.
}
\end{figure}

The main figure of merit for track-based alignment -- the residual distributions --- are listed for monitoring purposes {\em before} and {\em after} L2 alignment with the \RA\ algorithm on pages~\pageref{fig:r_x_PIXB_L2}--\pageref{fig:r_x_SCTC_L2}. The main features to be learned from these plots are briefly summarised by subdetectors in the following. For comparison of statistical quantities the figures in the summary Table~\ref{tab:m8plusAlignL2} are used, since they were calculated using the full range $r_x\in[-1.5,\,1.5]\,\mm$:
\begin{description}
\item[Pixel barrel:] The shape of the $r_x$ distributions shown in Figure~\ref{fig:r_x_PIXB_L2} does not change much over the L2 alignment procedure due to preceding L1 alignment. However, the residuals are dramatically re-centred around 0 to within a micron from $\rmean x$ of \order{100\,\mum}. The residual width improves by about 12\% to $\rsig x\simeq250\,\mum$. This indicates that intitial layer-to-layer misalignments in the $X$-$Y$ plane of \order{50\,\mum}\ are dominated by misalignments in $\Gamma$ of \order{0.5\,\mrad};
\item[SCT barrel:] The situation before alignment is well behaved due to a good assembly precision as can be seen from Figure~\ref{fig:r_x_SCTB_L2}. Here, the residual means $\rmean x$ improve from $\order{15\,\mum}$ to 0 within 0.5\,\mum\ and $\rsig x$ refines by about 20\% to circa 140\,\mum;
\item[Pixel end-caps:] For EC~A/C, the $r_x$ distributions are presented in Figures~\ref{fig:r_x_PIXA_L2}/\ref{fig:r_x_PIXC_L2}, respectively. Their peaks get more pronounced, and the residual means approach 0 to within about 10\,\mum. The residual widths improve by about 12.5\% to $\sim\!\!265/195\,\mum$ for EC~A/C, respectively. The $\rsig x$ values after alignment are quite different for EC~A and C, which is due to larger L3 misalignments in case of EC~A;
\item[SCT end-cap A:] The $r_x$ distribution before alignment, as displayed in Figure~\ref{fig:r_x_SCTA_L2}, indicates a good assembly tolerance. The residual means improve from $\order{20\,\mum}$ to 0 to within microns, while the widths refine by a factor of about $\xOverY13$ to circa $\rsig x\simeq130\,\mum$. The residual widths after alignment are smaller than in the SCT barrel, which is due to track reconstruction cuts; 
\item[SCT end-cap C:] The most dramatic improvement in $r_x$ distributions in the process of L2 alignment is undoubtedly found in SCT EC~C, as shown in Figure~\ref{fig:r_x_SCTC_L2}. Distributions in the six innermost disks which are initially highly skewed approach a Gaussian-like shape. The residual means $\rmean x$, which spanned a range of more than 300\,\mum\ before alignment, approach 0 to within about 5\,\mum. It comes as no surspise that the residual widths improve by more than a factor of \xOverY12\ to about $\rsig x\simeq120\,\mum$. Note that disk~6 suffers from a cable swap problem~\cite{bib:privateGiorgio}.%, which in this case affects mostly the monitoring data before L2~alignment.
\end{description}
Overall, a satisfactory improvement is achieved in the $r_x$ distributions, which is further supported by the summary in Table~\ref{tab:m8plusAlignL2}.

\newpage
%\nopagebreak[5]

%\nopagebreak[5]
\clearpage

%\nopagebreak[5]

%\nopagebreak[5]
\clearpage

{
\begin{figure}
\begin{center}
\vspace{\cDist}
\includegraphics[width=15.8cm,clip=true]{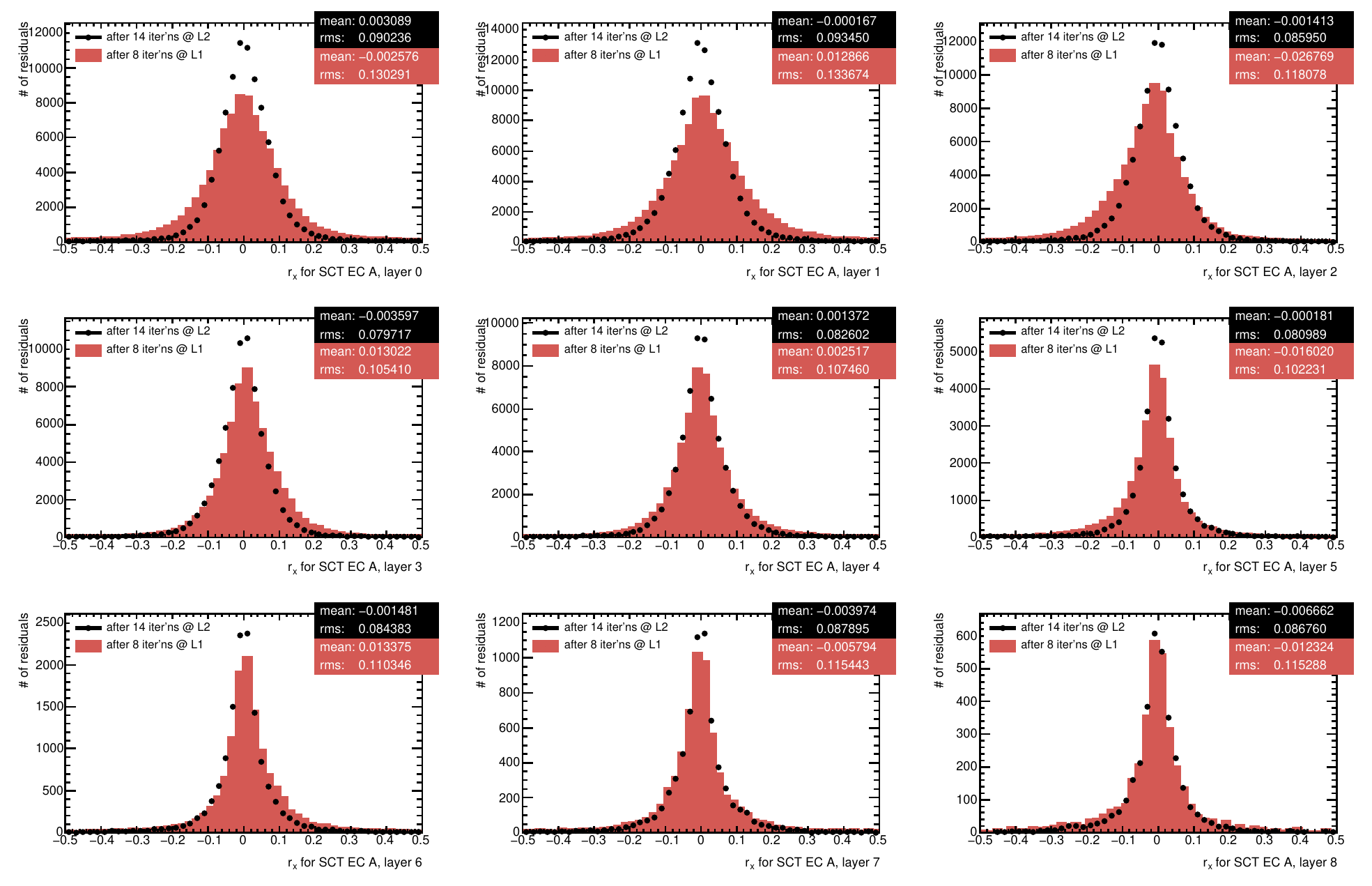}
\vspace{\cDist}
\end{center}
\caption[$r_x$ residual distribution in the EC~A of the SCT detector before and after alignment corrections at L2 in M8+]{\label{fig:r_x_SCTA_L2}
The $r_x$ {\bf residual} distribution in the EC~A of the SCT detector by disk layers {\em before} and {\em after} alignment corrections at {\bf L2} in M8+. The residual means are centered about zero, and the residual widths improve by about \xOverY13\ in the process of alignment. Smaller residual widths after alignment compared to the barrel are due to track reconstruction cuts. All values are in~mm.
}
\end{figure}%\nopagebreak[5]

\begin{figure}
\begin{center}
\vspace{\cDistHalf}
\includegraphics[width=15.8cm,clip=true]{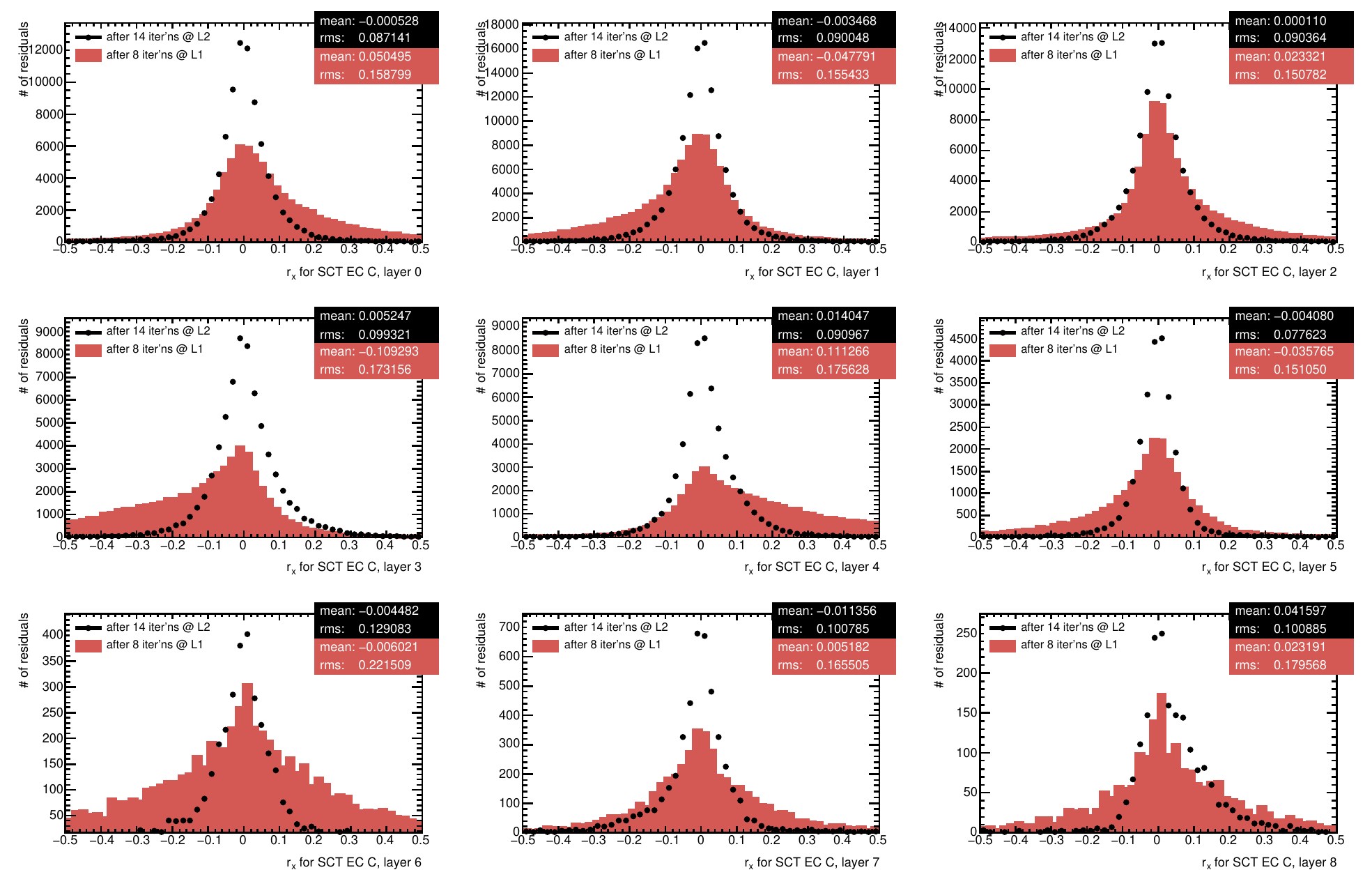}
\vspace{\cDist}
\end{center}
\caption[$r_x$ residual distribution in the EC~C of the SCT detector before and after alignment corrections at L2 in M8+]{\label{fig:r_x_SCTC_L2}
The $r_x$ {\bf residual} distribution in the EC~C of the SCT detector by disk layers {\em before} and {\em after} alignment corrections at {\bf L2} in M8+. The residual means are centered about zero, and the residual widths improve dramatically by more than a factor of two in the process of alignment. The initial distributions indicate a high degree of misalignment. All values are in~mm.
%The wide symmetric flanks of disk 6 are partly due to a noisy module.
%Smaller residual widths after alignment compared to the barrel are due to track reconstruction cuts.
\vspace{\cDist}
}
\end{figure}%\nopagebreak[5]
}
\clearpage

\begin{table}
\small
\begin{center}
\vspace{\cDistHalf}
\begin{tabular}{l|rrr|rrr}
\hline
 & \multicolumn{3}{c|}{{\bf Before} alignment} & \multicolumn{3}{c}{{\bf After} alignment}\\
%\hline
 & $\langle r_x\rangle$ & $\delta r_x$ & $\sigma(r_x)$ & $\langle r_x\rangle$ & $\delta r_x$ & $\sigma(r_x)$  \\
\hline\hline
Pixel barrel layer 0 & $-163.28$ & $0.92$ & $281.0$ & $-0.96$ & $0.81$ & $262.0$\\
Pixel barrel layer 1 & $-61.05$ & $0.62$ & $253.5$ & $-0.08$ & $0.57$ & $243.9$\\
Pixel barrel layer 2 & $107.02$ & $0.56$ & $262.5$ & $0.04$ & $0.52$ & $255.1$\\
Pixel barrel {\bf (all)} & $-3.89$ & $0.41$ & $284.7$ & $-0.20$ & $0.35$ & $252.7$\\
\hline
SCT barrel layer 0 & $-14.01$ & $0.14$ & $192.8$ & $0.44$ & $0.11$ & $161.4$\\
SCT barrel layer 1 & $-11.01$ & $0.11$ & $169.6$ & $-0.21$ & $0.08$ & $136.4$\\
SCT barrel layer 2 & $21.09$ & $0.09$ & $157.0$ & $-0.42$ & $0.07$ & $122.9$\\
SCT barrel layer 3 & $-9.38$ & $0.10$ & $180.5$ & $-0.05$ & $0.08$ & $138.5$\\
SCT barrel {\bf (all)} & $-2.22$ & $0.05$ & $174.8$ & $-0.10$ & $0.04$ & $138.6$\\
\hline
Pixel EC A layer 0 & $-27.02$ & $4.45$ & $300.2$ & $-11.08$ & $3.72$ & $264.8$\\
Pixel EC A layer 1 & $-13.61$ & $4.37$ & $294.0$ & $8.97$ & $3.87$ & $274.3$\\
Pixel EC A layer 2 & $30.35$ & $5.70$ & $305.6$ & $-18.37$ & $4.40$ & $249.3$\\
Pixel EC A {\bf (all)} & $-8.13$ & $2.75$ & $300.0$ & $-5.26$ & $2.30$ & $265.1$\\
\hline
Pixel EC C layer 0 & $-16.45$ & $3.93$ & $243.9$ & $-5.18$ & $3.11$ & $201.6$\\
Pixel EC C layer 1 & $-1.09$ & $3.69$ & $229.5$ & $1.68$ & $2.94$ & $191.3$\\
Pixel EC C layer 2 & $10.26$ & $3.27$ & $218.8$ & $-11.42$ & $2.74$ & $192.7$\\
Pixel EC C {\bf (all)} & $-1.75$ & $2.09$ & $230.6$ & $-5.33$ & $1.69$ & $195.2$\\
\hline
SCT EC A layer 0 & $-4.58$ & $0.75$ & $235.2$ & $1.18$ & $0.53$ & $160.4$\\
SCT EC A layer 1 & $12.11$ & $0.63$ & $210.1$ & $-0.84$ & $0.43$ & $139.7$\\
SCT EC A layer 2 & $-29.60$ & $0.56$ & $168.7$ & $-1.86$ & $0.39$ & $113.9$\\
SCT EC A layer 3 & $13.25$ & $0.60$ & $159.1$ & $-3.77$ & $0.39$ & $105.6$\\
SCT EC A layer 4 & $2.18$ & $0.63$ & $156.1$ & $0.79$ & $0.42$ & $104.0$\\
SCT EC A layer 5 & $-18.88$ & $0.92$ & $155.1$ & $-1.13$ & $0.62$ & $103.9$\\
SCT EC A layer 6 & $14.82$ & $1.44$ & $166.3$ & $-1.61$ & $0.95$ & $109.1$\\
SCT EC A layer 7 & $-5.79$ & $2.23$ & $181.7$ & $-3.08$ & $1.58$ & $127.9$\\
SCT EC A layer 8 & $-12.67$ & $3.13$ & $187.2$ & $-7.94$ & $1.97$ & $114.4$\\
SCT EC A {\bf (all)} & $-2.24$ & $0.28$ & $191.4$ & $-0.99$ & $0.19$ & $127.6$\\
\hline
SCT EC C layer 0 & $58.69$ & $0.79$ & $227.6$ & $-3.72$ & $0.40$ & $116.6$\\
SCT EC C layer 1 & $-59.39$ & $0.64$ & $215.2$ & $-5.09$ & $0.34$ & $113.9$\\
SCT EC C layer 2 & $25.48$ & $0.69$ & $207.8$ & $0.50$ & $0.37$ & $111.5$\\
SCT EC C layer 3 & $-159.86$ & $1.01$ & $259.5$ & $5.48$ & $0.44$ & $117.2$\\
SCT EC C layer 4 & $201.36$ & $1.31$ & $312.4$ & $14.69$ & $0.46$ & $113.0$\\
SCT EC C layer 5 & $-41.17$ & $1.36$ & $220.3$ & $-3.35$ & $0.72$ & $117.4$\\
SCT EC C layer 6 & $-5.71$ & $4.59$ & $382.1$ & $-5.40$ & $4.64$ & $278.2$\\
SCT EC C layer 7 & $5.85$ & $3.42$ & $235.3$ & $-12.38$ & $1.98$ & $134.8$\\
SCT EC C layer 8 & $22.84$ & $5.77$ & $267.1$ & $40.44$ & $3.01$ & $131.5$\\
SCT EC C {\bf (all)} & $0.60$ & $0.39$ & $262.3$ & $0.71$ & $0.17$ & $117.2$\\
\hline
\end{tabular}
\caption[Main residual characteristics for the silicon tracker by layers in M8+ before and after alignment at L2]{\label{tab:m8plusAlignL2}
Main residual characteristics for the silicon tracker by layers in M8+ {\em before} and {\em after} alignment at {\bf L2}: the residual mean $\rmean x$, the uncertainty on the residual mean $\delta r_x$, and the standard deviation of the residual $\sigma(r_x)$. The range used for the calculation of the quantities above is $r_x\in[-1.5\,\mm,\,1.5\,\mm]$. All values are given in $\mum$. The residual widths in SCT ECs are smaller than in the SCT barrel due to track reconstruction cuts.
\vspace{\cDistHalf}
}
\end{center}
\end{table}
\clearpage

\subsubsection{Residual Discrepancies by Module Sides in SCT End-Caps}

\begin{figure}[h]
\begin{center}
\vspace{\cDistHalf}
\includegraphics[width=15.8cm,height=15.0cm,clip=true]{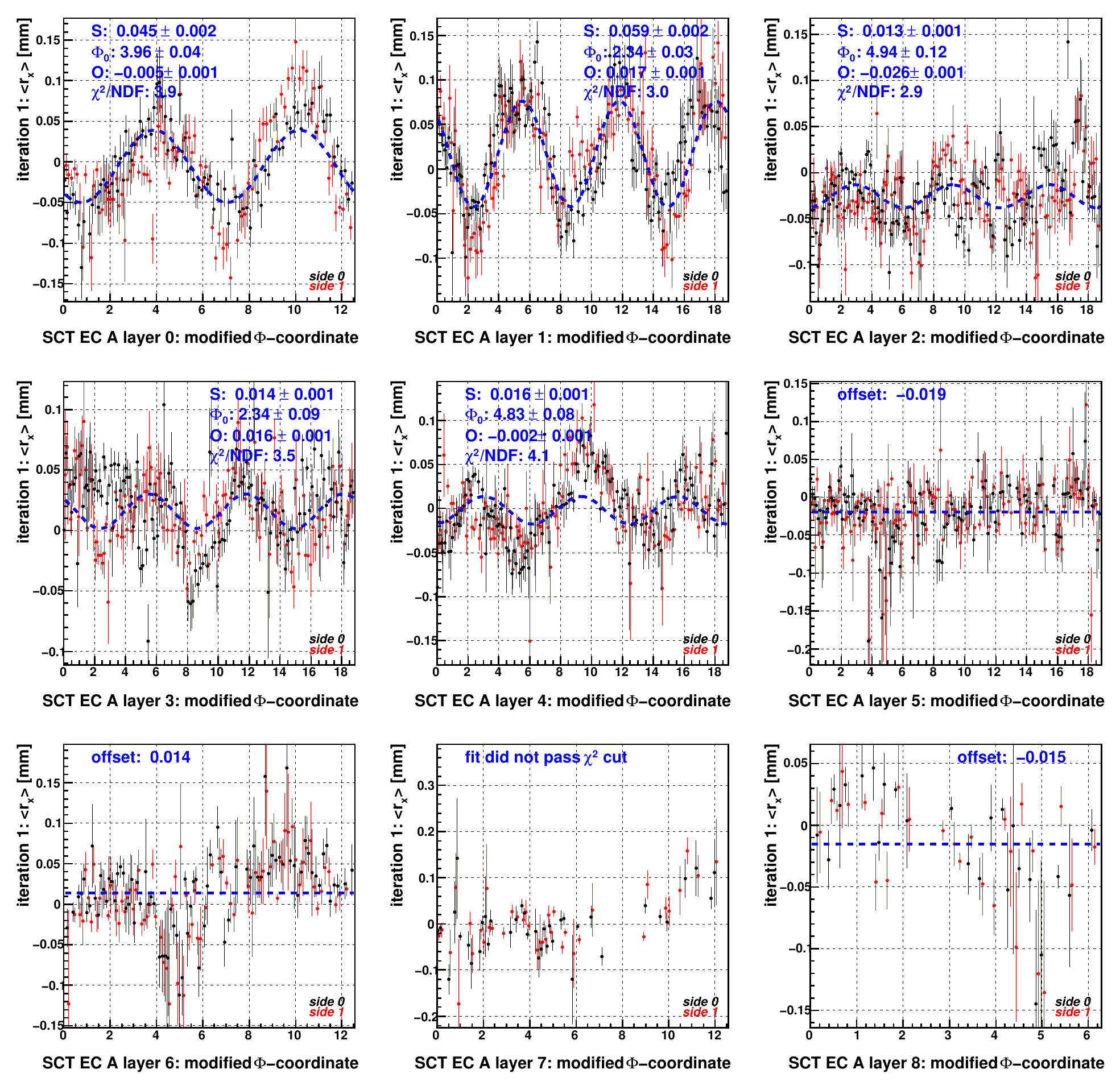}
\end{center}
\vspace{\cDist}
\caption[The distribution $\langle r_x\rangle(\Phi)$ for the EC~A of the SCT detector split by module sides before alignment at~L2]{\label{fig:m8_r_x_vs_Phi_L2_SCT_ECA_sides}
The $\rmean x(\Phi)$ distribution split by module sides for the EC~A of the SCT detector using the full $B$-field off M8+ dataset {\bf before} alignment at~L2. The distributions for side~{\em0} and side~{\em1} display a difference of up to about 50\,\mum. Moreover, in intervals $[n\!\cdot\!2\pi+\oneOverTwo\pi,\,n\!\cdot\!2\pi+\threeOverTwo\pi]$, $n\in\mathbb N$, side~{\em0} tends to have higher $\rmean x$ values than side~{\em1} for rings~0 and 2, whereas in remaining intervals the sides typically change their roles. The whole picture is inverted for ring~1.
\vspace{\cDist}
}
\end{figure}

Working on ATLAS Silicon Tracker alignment, a recurring pattern was found by the author in the $\rmean x(\Phi)$ distributions for the two module sides in the end-caps of the SCT detector. This is explained using EC~A as an example. Its $\rmean x(\Phi)$ distribution by layers is shown in Figure~\ref{fig:m8_r_x_vs_Phi_L2_SCT_ECA_sides}. The crucial difference to Figure~\ref{fig:m8_r_x_vs_Phi_L2_SCT_ECA_before} is that now each side of any given module corresponds to one bin of the histogram. The sides are referred to as {\em 0} and {\em 1}, and are shown in black and red, respectively. Overlaid with random fluctuations from individual module misalignments which do not exceed 100\,\mum~\cite{bib:sctEJINST}, a clear difference between $\langle r_x^{\rm side\,0}\rangle$  and $\langle r_x^{\rm side\,1}\rangle$ of typically several tens of microns is found. More importantly, a uniform pattern is observed in the five innermost disks\footnote{There is not enough statistics to make a similar statement for the outermost three disks.}. It is summarised in Table~\ref{tab:sideDiscrepancy} using disk layer~1 as a typical example. In rings~{\em0} and {\em2}, $\langle r_x^{\rm side\,0}\rangle < \langle r_x^{\rm side\,1}\rangle$ is found in $[0,\,\oneOverTwo\pi]$, $[\threeOverTwo\pi,\,2\pi]$ neglecting multiples of $2\pi$, and $\langle r_x^{\rm side\,0}\rangle > \langle r_x^{\rm side\,1}\rangle$ in $[\oneOverTwo\pi,\,\threeOverTwo\pi]$. The same observation, alas inverted is made in ring {\em1}.

The described pattern can be interpreted as a systematic deviation $d+\delta d$ in the distance between the two sides of any given module from the nominal value of $d=1.25\,\mm$. Assume a scenario where there is a systematic shift $\delta d>0$ across the entire subdetector. %where side {\em1} faces the predominant origin direction of cosmic ray particles coming in at steep incidence angles.
For cosmic ray tracks going through one of the ECs of the SCT, there is an asymmetry in $\theta$ due to the access shafts. Without loss of generality, assume that side {\em1} faces the direction where cosmic ray particles predominantly come from. For modules at $\Phi\simeq0$ where local $x$ is oriented upwards, one will observe an average residual $\langle r_x^{\rm side\,1}\rangle$ which is larger than $\langle r_x^{\rm side\,0}\rangle$. Contrariwise, for $\Phi\simeq\pi$ and $x$ oriented downwards, the opposite trend is expected. The smooth transition between the two secenarios will take place at $\Phi=\xOverY12\pi,\xOverY32\pi$. Again, the entire picture will be inverted if side {\em0} faces the predominant origin direction of cosmic ray particles. Indeed, this matches the pattern summarised in Table~\ref{tab:sideDiscrepancy} for all the three rings\footnote{Keep in mind, that ring 1 is mounted on the side of the disk away from the DIP. Thus, its side {\em 0} will be further away from the DIP than side {\em1}, whereas the opposite is the case for rings 0 and 2.}.

No quantitative analysis was performed due to an insufficient number of residuals in EC modules at this stage and the first priority being to provide a good overall alignment for the silicon tracker. However, a rough estimate was made confirming that the suspected magnitude of effect would be consistent with the assembly tolerances. It is known~\cite{bib:privateSnow} that front and back surfaces have separate tolerances of $\pm$0.115\,mm, so in principle the module can be up to 0.23 mm thicker or thinner than nominal. However, the modules were made in jigs that defined the total thickness rather accurately and the 115 microns tolerance was set to allow for non-flatness. Nevertheless, it cannot be excluded that on average the local $z$ thickness of a module at a typical point differs by $\delta d\simeq100\,\mum$ from nominal~\cite{bib:privateSnow}. Assuming an incidence angle of 45$^\circ$ common in cosmic ray tracks for the sake of the argument, this would give $|\langle r_x^{\rm side\,0}\rangle - \langle r_x^{\rm side\,1}\rangle| \simeq \frac{100\,\mum}{\sqrt2}$, in accordance with the preliminary findings described above.

\begin{table}
\small
\begin{center}
\vspace{\cDistHalf}
\begin{tabular}{llccc}
\hline
$\Phi$-Interval & Observation & Symbol & Side closest to DIP & Ring \\
\hline\hline
$[0,\,\oneOverTwo\pi]$               & $\langle r_x^{\rm side\,0}\rangle < \langle r_x^{\rm side\,1}\rangle$ & $-$ &        &     \\
$[\oneOverTwo\pi,\,\threeOverTwo\pi]$& $\langle r_x^{\rm side\,0}\rangle > \langle r_x^{\rm side\,1}\rangle$ & $+$ &{\bf 0} & {0} \\
$[\threeOverTwo\pi,\,2\pi]$          & $\langle r_x^{\rm side\,0}\rangle < \langle r_x^{\rm side\,1}\rangle$ & $-$ &        &     \\
\hline
$[2\pi,\,2\oneOverTwo\pi]$           & $\langle r_x^{\rm side\,0}\rangle > \langle r_x^{\rm side\,1}\rangle$ & $+$ &        &     \\
$[2\oneOverTwo\pi,\,3\oneOverTwo\pi]$& $\langle r_x^{\rm side\,0}\rangle < \langle r_x^{\rm side\,1}\rangle$ & $-$ &{\bf 1} & {1} \\
$[3\oneOverTwo\pi,\,4\pi]$           & $\langle r_x^{\rm side\,0}\rangle > \langle r_x^{\rm side\,1}\rangle$ & $+$ &        &     \\
\hline
$[4\pi,\,4\oneOverTwo\pi]$           & $\langle r_x^{\rm side\,0}\rangle < \langle r_x^{\rm side\,1}\rangle$ & $-$ &        &     \\
$[4\oneOverTwo\pi,\,5\oneOverTwo\pi]$& $\langle r_x^{\rm side\,0}\rangle > \langle r_x^{\rm side\,1}\rangle$ & $+$ &{\bf 0} & {2} \\
$[5\oneOverTwo\pi,\,6\pi]$           & $\langle r_x^{\rm side\,0}\rangle < \langle r_x^{\rm side\,1}\rangle$ & $-$ &        &     \\
\hline
\end{tabular}
\caption[The typical pattern of $\langle r_x^{\rm side\,0}\rangle,\,\langle r_x^{\rm side\,1}\rangle$ residuals]{\label{tab:sideDiscrepancy}
The typical pattern of $\langle r_x^{\rm side\,0}\rangle,\,\langle r_x^{\rm side\,1}\rangle$ residuals using the concreate example of disk~1 of SCT EC~A displayed in Figure~\ref{fig:m8_r_x_vs_Phi_L2_SCT_ECA_sides}. To spotlight the pattern, the symbol ``$-$'' is used for the $\langle r_x^{\rm side\,0}\rangle < \langle r_x^{\rm side\,1}\rangle$ case, and ``$+$'' for the other case. Further discussion is in the text.
\vspace{\cDistHalf}
}
\end{center}
\end{table}

\clearpage

\begin{comment}
The endcap module thickness is nominally 1.25 mm . Front and
back surfaces have seperate tolerances of +- 0.115 mm so in
principle the module can be up to 0.23 mm thicker or thinner than
nominal, but in practice this does not happen. The modules were
made in jigs that defined the total thickness rather accurately and the 115 microns tolerance was set to allow for non-flatness.

I would guess that the local Z thickness of a module at a typical
point does not differ from 1.25 mm by more than 0.1 mm. There is a
tendency for all modules made at one site to be similar. So it could
be that all modules made at Manchester are on average 0.1 mm thicker than nominal and all made at Geneva could be 0.1 mm thinner than
nominal. This "0.1 mm" is just a guess - the data exists and I
can find the real values if necessary. I have no idea what the angles of the cosmics are that you are using. If they are high
angles , could this 0.1 mm be enough to account for the
50 micron effect you see ?
If it is the right magnitude let me know and I will
find a more accurate value.

There could also be an effect if the detectors were not fully depleted
but I would guess that this was not the case. I'm not an expert on
where the hit should be within the detector thickness
but I believe that for unirradiated modules it is pretty close to the
mid plane.
\end{comment}

%% file: M8plus/L4.tex
The algorithm to calculate alignment corrections for pixel stave bow misalignments with \RA\ is described Subsection~\ref{ssec:pixelStaveBow} on page~\pageref{ssec:pixelStaveBow}~ff., and its application to M8+ data shall be the subject of this Subsection. Due to the small acceptance of the pixel detector, pixel stave bow alignment was performed utilising the full $B$-field off M8+ dataset introduced in Subsection~\ref{sec:datasetM8plus}. The nominal ATLAS ID geometry corrected for L1 and L2 misalignments, as obtained in Subsections~\ref{ssec:l1M8} and \ref{ssec:l2M8}, was used as a starting point. 

\begin{figure}
\begin{center}
\includegraphics[width=15.8cm,clip=true]{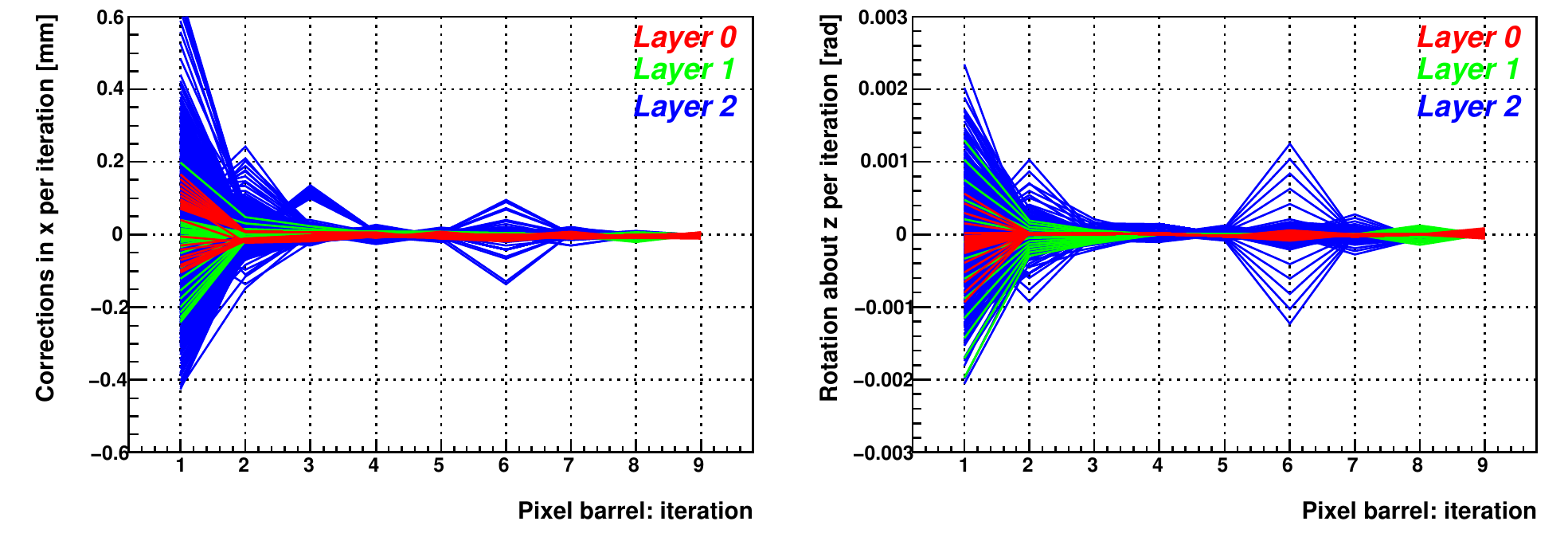}
\end{center}
\caption[Convergence of pixel stave bow alignment corrections in M8+]{\label{fig:convergenceL4}
Alignment corrections per iteration for the {\bf pixel stave bow} alignment with the \RA\ algorithm in M8+. Each of the lines corresponds to one pixel barrel module ({\it not} stave). The local $x${\bf~(left)} and $\gamma${\bf~(right)} degrees of freedom were aligned for. Overall, an exponential-like asymptotic convergence is observed except for stave \#33 in layer~2, which is discussed in the text.
}
\end{figure}%\nopagebreak[5]

Parabolic, linear, and offset fits were performed for each of the 112~staves in the barrel of the pixel detector, as described in Subsection~\ref{ssec:pixelStaveBow}. From these fits, alignment corrections for both degrees of freedom accessible to pixel stave bow alignment, $c_x$ and $c_\gamma$ at the individual module level, were derived. These corrections, being defined in the {\em local} frame of the modules, were written to the L3 section of the alignment database. 

The alignment corrections per module per iteration are shown in Figure~\ref{fig:convergenceL4}. The maximum magnitude of corrections in $c_x$ of beyond 600\,\mum\ in the 1$^{\rm st}$ iteration is remarkable and further supports the need for a coherent pixel stave bow alignment as discussed in general terms in Subsection~\ref{ssec:alignmentStructures}. An almost instantaneous convergence in both DoFs is observed. The sudden jump in the magnitude of corrections in the 6$^{\rm th}$ iteration is due to a parabolic fit for stave with $\Phi$-identifier~33 in layer~2 with a $\chisq$ close to the maximum allowed value of 15: $\chisq<15$ in 6$^{\rm th}$ iteration, and $\chisq>15$ before. This is triggered by about $\xOverY13$ of the \RA\ subjobs crashing in 6$^{\rm th}$ iteration due to LXBATCH~\cite{bib:lxbatch} computing problems. This does not pose a concern and should rather be regarded as a statistical fluctuation: after all, $\xOverY23$ of the M8+ statistics was retained, and only a single stave with a fit close to fulfilling the condition in Equation~\ref{eqn:l4sigmaCut} displayed this behaviour.

\begin{figure}
\begin{center}
\includegraphics[width=5.2cm,clip=true]{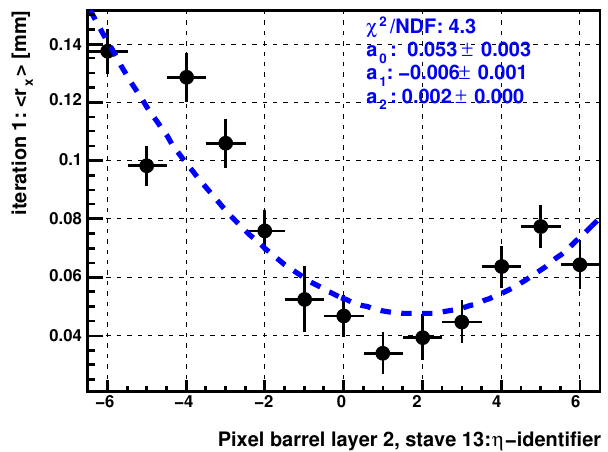}
\includegraphics[width=5.2cm,clip=true]{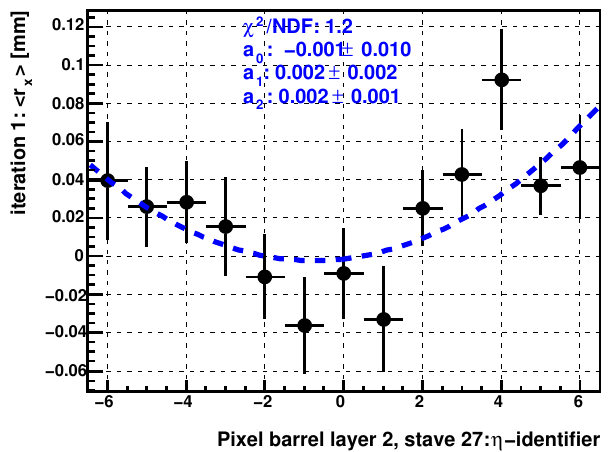}
\includegraphics[width=5.2cm,clip=true]{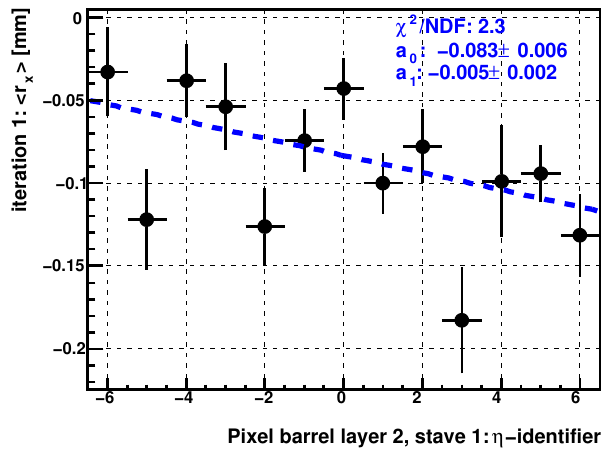}
\includegraphics[width=5.2cm,clip=true]{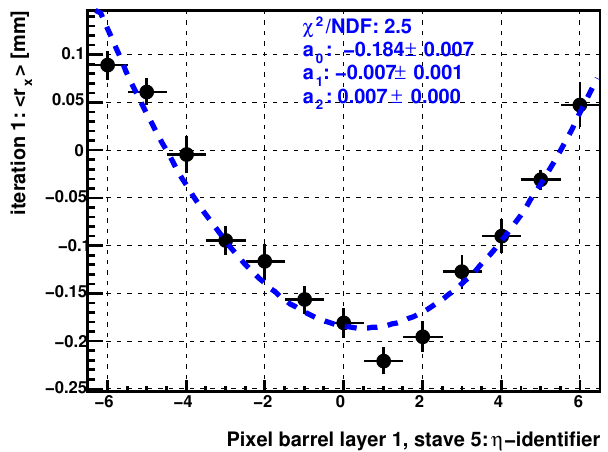}
\includegraphics[width=5.2cm,clip=true]{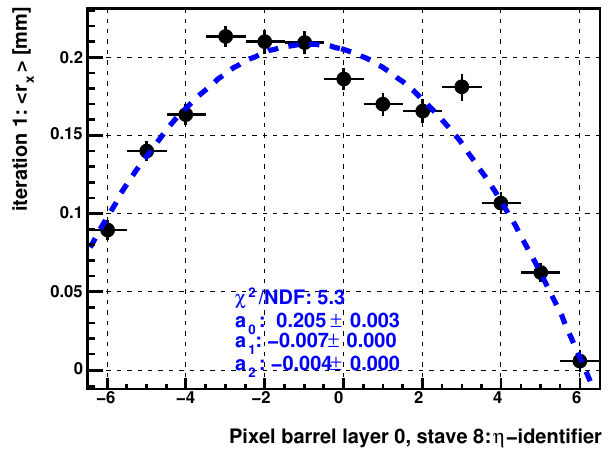}
\includegraphics[width=5.2cm,clip=true]{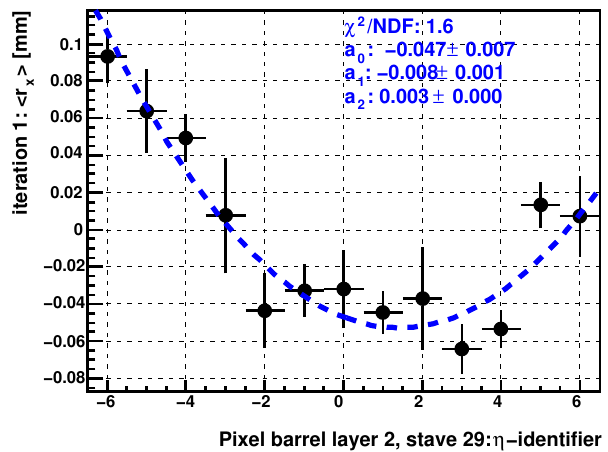}
\vspace{\cDistHalf}
\end{center}
\caption[The $\langle r_x\rangle(\eta)$ distribution before pixel stave bow alignment in M8+]{\label{fig:m8_r_x_vs_eta_L4_before}
The $\rmean x(\eta)$ distribution for six typical staves of the pixel detector (\# 74, 88, 62, 28, 9, 90) using the full $B$-field off M8+ dataset {\bf before} alignment for pixel stave bow. The fit results with a parabola of the form specified in Equation~\ref{eqn:staveBow} are shown in blue. Stave 62 {\bf (top right)} shows a linear dependence, while the rest of the staves are best described by a parabola. The typical magnitude of corrections is of $\order{\oneOverX{4}\,\rm mm}$. The complete set of $\rmean x(\eta)$ distributions before and after alignment is shown in Appendix~\ref{chp:l4}.
%
%random numbers used
%74	88
%62	28
%9	90
}
\end{figure}%\nopagebreak[5]

\begin{figure}
\begin{center}
\includegraphics[width=5.2cm,clip=true]{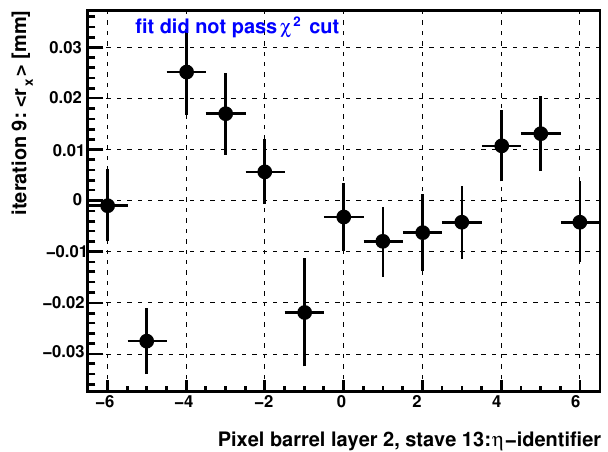}
\includegraphics[width=5.2cm,clip=true]{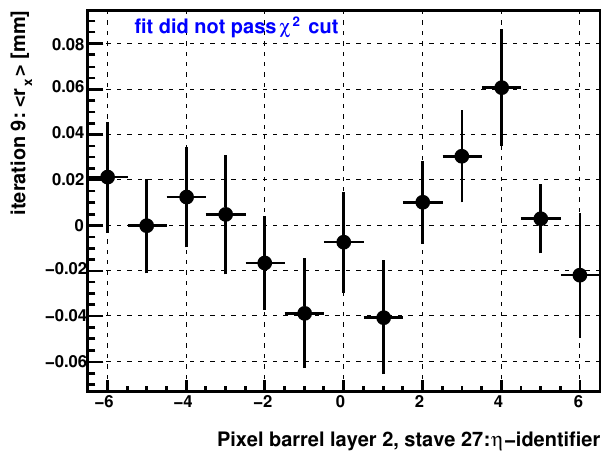}
\includegraphics[width=5.2cm,clip=true]{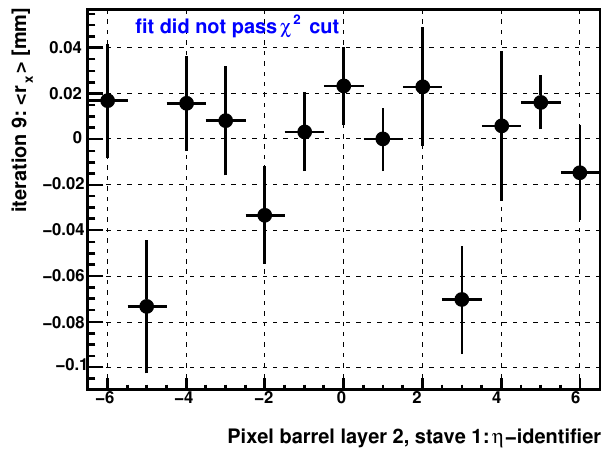}
\includegraphics[width=5.2cm,clip=true]{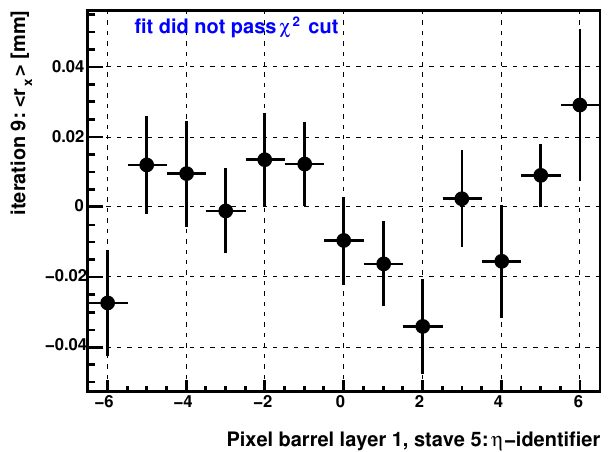}
\includegraphics[width=5.2cm,clip=true]{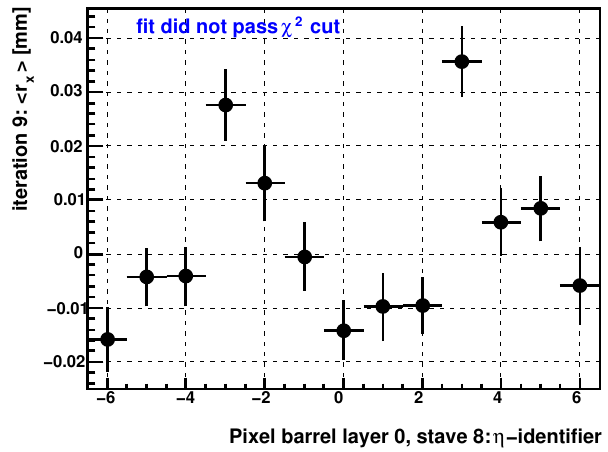}
\includegraphics[width=5.2cm,clip=true]{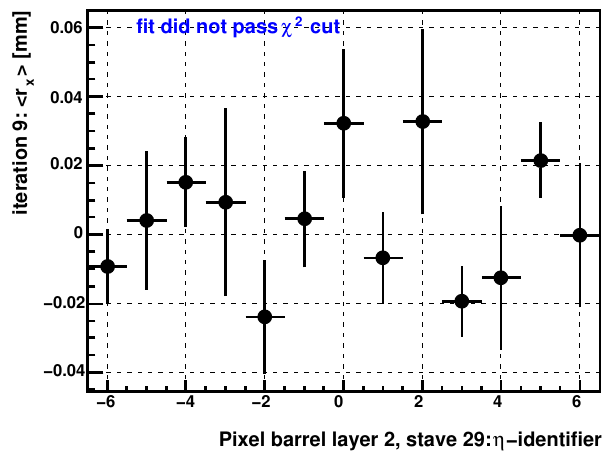}
\vspace{\cDistHalf}
\end{center}
\caption[The $\langle r_x\rangle(\eta)$ distribution before pixel stave bow alignment in M8+]{\label{fig:m8_r_x_vs_eta_L4_after}
The $\rmean x(\eta)$ distribution for six typical staves of the pixel detector (\# 74, 88, 62, 28, 9, 90) using the full $B$-field off M8+ dataset {\bf after} alignment for pixel stave bow. Note the reduction in the range of the $y$-axis compared to Figure~\ref{fig:m8_r_x_vs_eta_L4_before}. None of the fits with a parabola of the form specified in Equation~\ref{eqn:staveBow} passes the $\chisqNDF$ cut, which is an indication for the convergence of the alignment procedure. The remaining random misalignments at L3 are of $\order{25\,\mum}$.  The complete set of  $\rmean x(\eta)$ distributions before and after alignment is shown in Appendix~\ref{chp:l4}.
}
\end{figure}%\nopagebreak[5]

\begin{figure}
\begin{center}
\vspace{-0.3cm}
\vspace{\cDistHalf}
\includegraphics[width=15.8cm,clip=true]{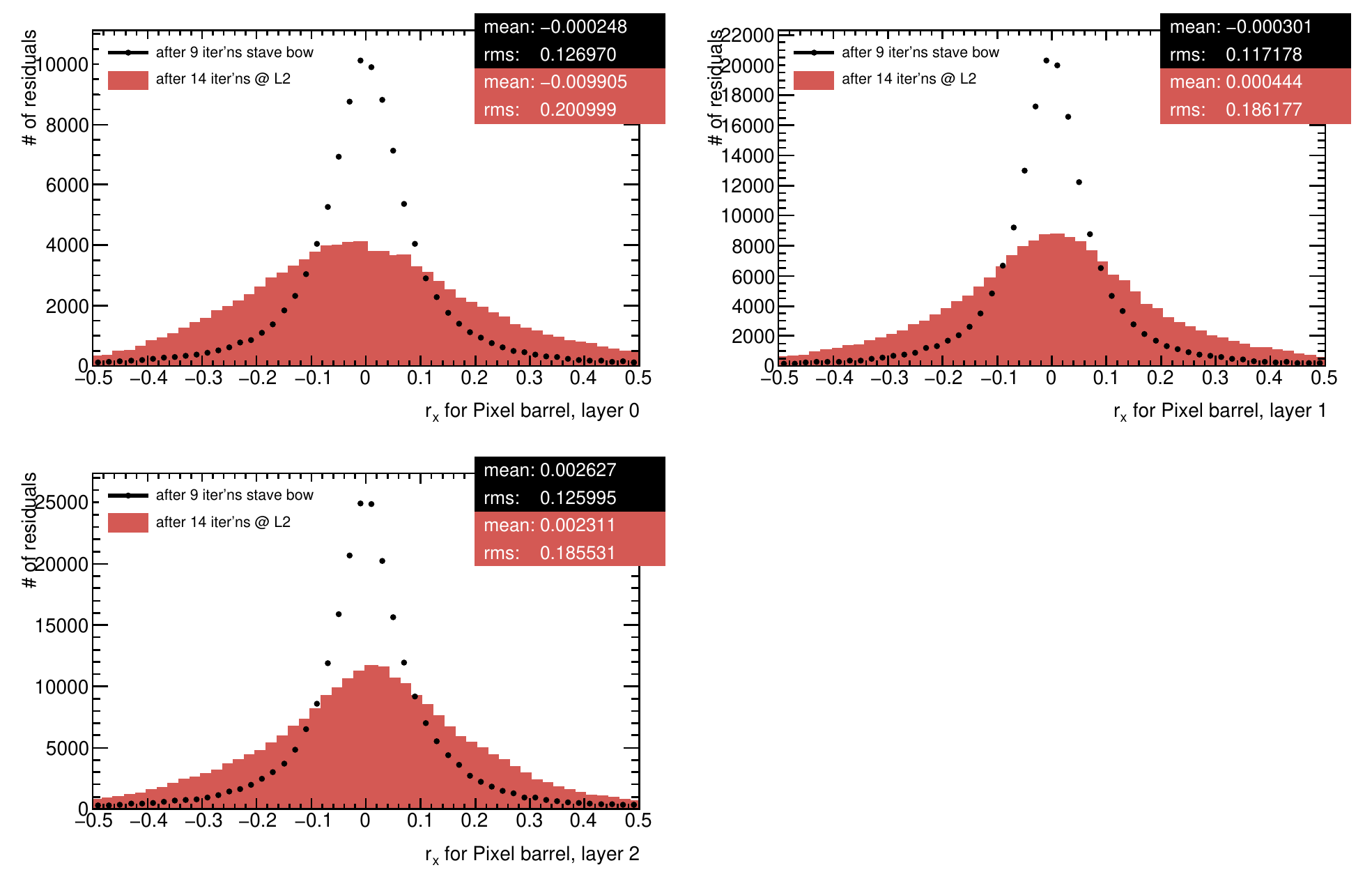}
\vspace{\cDist}
\end{center}
\caption[The $r_x$ residual distribution in the barrel of the pixel detector before and after pixel stave bow alignment corrections in M8+]{\label{fig:m8_r_x_L4}
The $r_x$ {\bf residual} distribution in the barrel of the pixel detector {\em before} and {\em after} {\bf pixel stave bow} alignment corrections in M8+. The residual width improves dramatically -- almost by a factor of two, and resembles more the typical residual curve shape for perfect alignment. The alignment of layer 0 is somewhat more statistically limited than that of the other layers. All values are in~mm.
%\vspace{\cDistHalf}
}
\end{figure}%\nopagebreak[5]

\begin{figure}
\begin{center}
\vspace{-0.3cm}
\vspace{\cDistHalf}
\includegraphics[width=15.8cm,clip=true]{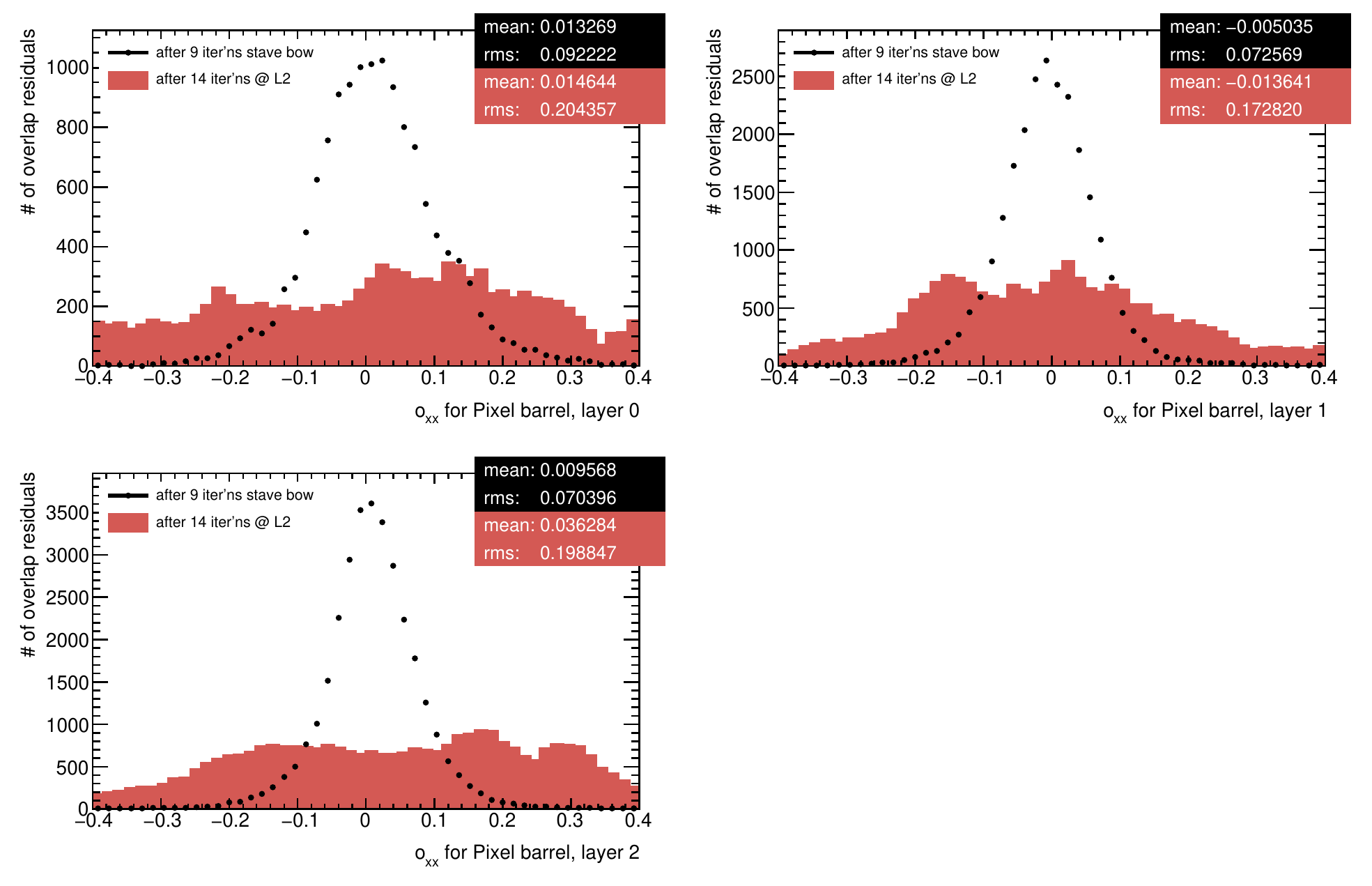}
\vspace{\cDist}
\end{center}
\caption[The $o_{xx}$ overlap residual distribution in the barrel of the pixel detector before and after pixel stave bow alignment corrections in M8+]{\label{fig:m8_r_x_L4_ovres}
The $o_{xx}$ {\bf overlap residual} distribution in the barrel of the pixel detector {\em before} and {\em after} {\bf pixel stave bow} alignment corrections in M8+. The overlap residual width improves dramatically, which indicates that a good alignment precision between neighbouring modules in the same layer is achieved. All values are in~mm.
\vspace{\cDistHalf}
}
\end{figure}%\nopagebreak[5]

%\clearpage

The $\rmean x(\eta)$ residual distributions are displayed for six typical pixel staves before and after alignment in Figures~\ref{fig:m8_r_x_vs_eta_L4_before} and \ref{fig:m8_r_x_vs_eta_L4_after}, respectively. The fit results are indicated as blue dashed lines. All $\rmean x(\eta)$ distributions before alignment shown can be fitted with a parabola except one, for which a linear fit is preferred by the \RA\ algorithm. After alignment, no fit hypothesis passes the \sigmaCut\ cut in Equation~\ref{eqn:l4sigmaCut} for any of the staves, and their alignment can be considered as converged. The complete set of analogous distributions for all 112 pixel barrel staves before and after alignment is documented in Appendix~\ref{chp:l4} for reference.

The $r_x$ residual distributions for the pixel barrel layers are shown in Figure~\ref{fig:m8_r_x_L4} before and after alignment. The shape of the distributions improves, but most importantly, its width refines from $\rsig x\simeq250\,\mum$ to $\rsig x\simeq190\,\mum$, id est by about 25\%. The improvement is most visible in the $b$-layer, since tracks going through it are likely to produce hits in the initially poorly aligned outer pixel layers. This will adversely affect the track fit and on average result in larger residuals for layer~0 before alignment. The residual means in layer~0 and 1 are consistent with 0. In layer~2, $\rmean x=2.87\,\mum$ is observed. This is not surprising given the magnitude of pixel stave bow corrections, which can have an effect on the residual mean. Given the small magnitude of effect it is felt that a dedicated L2 alignment step is not necessary after stave bow alignment, since it can be absorbed in L3 alignment constants for individual modules.

The analogous $o_{xx}$ overlap residual distributions are presented in Figure~\ref{fig:m8_r_x_L4_ovres} before and after alignment. The change in the shape of the distributions is dramatic: while almost flat and uniform distributions are observed before alignment, more Gaussian-like shapes are obtained thereafter. Logically, also the width refines dramatically by more than a factor of $\xOverY12$ to $\rsig x\simeq110\,\mum$. Such a dramatic improvement is rather remarkable: since $o_{xx}$ overlap residuals sharply reflect the alignment quality between modules from the same layer neighbouring each other in local $x$, one can conclude that stave-to-stave misalignments in the same layer are small after the alignment procedure. It should be noted that overlap residuals are not explicitly used for the pixel stave bow alignment procedure.

The improvement in the $r_x$ and $o_{xx}$ distributions is documented in statistical summary Tables~\ref{tab:m8plusAlignL4} and \ref{tab:m8plusAlignL4_ovres}, respectively. It should be mentioned that in those tables the 8$^{\rm th}$ rather than the 9$^{\rm th}$ iteration of pixel stave bow alignment is shown. This is because the alignment constants at L3 were produced with a different cut on the $\chisqNDF$ of reconstructed tracks, which mainly affects residuals with $|r_x|\gtrsim0.75\,\mm$. Thus, a direct benchmark comparison with the monitoring output from the 1$^{\rm st}$ iteration at L3 (which is produced with constants obtained in the last pixel stave bow iteration, id est iteration 9) is not possible. The results presented in the table nevertheless give a fair representation of the pixel stave bow alignment convergence, as can be justified with Figure~\ref{fig:convergenceL4}.

\begin{table}
\small
\begin{center}
\vspace{\cDistHalf}
\begin{tabular}{l|rrr|rrr}
\hline
 & \multicolumn{3}{c|}{{\bf Before} alignment} & \multicolumn{3}{c}{{\bf After} alignment}\\
	& $\langle r_x\rangle$	& $\delta r_x$	& $\sigma(r_x)$	& $\langle r_x\rangle$	& $\delta r_x$	& $\sigma(r_x)$ \\
\hline\hline
Pixel barrel layer 0	& $-0.96$	& $0.81$	& $262.0$	& $0.34$	& $0.57$	& $184.1$\\
Pixel barrel layer 1	& $-0.08$	& $0.57$	& $243.9$	& $0.27$	& $0.42$	& $180.2$\\
Pixel barrel layer 2	& $0.04$	& $0.52$	& $255.1$	& $2.87$	& $0.41$	& $202.1$\\
Pixel barrel {\bf (all)}& $-0.20$	& $0.35$	& $252.7$	& $1.48$	& $0.26$	& $191.3$\\
\hline
%SCT barrel layer 0	& $0.44$	& $0.11$	& $161.4$	& $0.16$	& $0.11$	& $158.9$\\
%SCT barrel layer 1	& $-0.21$	& $0.08$	& $136.4$	& $-0.40$	& $0.08$	& $134.1$\\
%SCT barrel layer 2	& $-0.42$	& $0.07$	& $122.9$	& $-0.53$	& $0.07$	& $120.6$\\
%SCT barrel layer 3	& $-0.05$	& $0.08$	& $138.5$	& $-0.15$	& $0.08$	& $137.0$\\
SCT barrel {\bf (all)}	& $-0.10$	& $0.04$	& $138.6$	& $-0.26$	& $0.04$	& $136.5$\\
\hline
%Pixel EC A layer 0	& $-11.08$	& $3.72$	& $264.8$	& $-12.37$	& $3.65$	& $261.0$\\
%Pixel EC A layer 1	& $8.97$	& $3.87$	& $274.3$	& $8.84$	& $3.85$	& $274.1$\\
%Pixel EC A layer 2	& $-18.37$	& $4.40$	& $249.3$	& $-18.18$	& $4.38$	& $248.9$\\
Pixel EC A {\bf (all)}	& $-5.26$	& $2.30$	& $265.1$	& $-5.77$	& $2.28$	& $263.4$\\
\hline
%Pixel EC C layer 0	& $-5.18$	& $3.11$	& $201.6$	& $-5.28$	& $3.08$	& $200.4$\\
%Pixel EC C layer 1	& $1.68$	& $2.94$	& $191.3$	& $1.92$	& $2.92$	& $190.2$\\
%Pixel EC C layer 2	& $-11.42$	& $2.74$	& $192.7$	& $-11.67$	& $2.73$	& $193.1$\\
Pixel EC C {\bf (all)}	& $-5.33$	& $1.69$	& $195.2$	& $-5.38$	& $1.68$	& $194.6$\\
\hline
%SCT EC A layer 0	& $1.18$	& $0.53$	& $160.4$	& $0.99$	& $0.53$	& $161.3$\\
%SCT EC A layer 1	& $-0.84$	& $0.43$	& $139.7$	& $-0.95$	& $0.43$	& $139.9$\\
%SCT EC A layer 2	& $-1.86$	& $0.39$	& $113.9$	& $-1.96$	& $0.38$	& $113.6$\\
%SCT EC A layer 3	& $-3.77$	& $0.39$	& $105.6$	& $-3.75$	& $0.39$	& $106.2$\\
%SCT EC A layer 4	& $0.79$	& $0.42$	& $104.0$	& $0.71$	& $0.42$	& $104.4$\\
%SCT EC A layer 5	& $-1.13$	& $0.62$	& $103.9$	& $-0.96$	& $0.62$	& $104.3$\\
%SCT EC A layer 6	& $-1.61$	& $0.95$	& $109.1$	& $-1.36$	& $0.95$	& $109.4$\\
%SCT EC A layer 7	& $-3.08$	& $1.58$	& $127.9$	& $-3.17$	& $1.58$	& $127.9$\\
%SCT EC A layer 8	& $-7.94$	& $1.97$	& $114.4$	& $-8.17$	& $2.01$	& $117.3$\\
SCT EC A {\bf (all)}	& $-0.99$	& $0.19$	& $127.6$	& $-1.06$	& $0.19$	& $128.0$\\
\hline
%SCT EC C layer 0	& $-3.72$	& $0.40$	& $116.6$	& $-3.43$	& $0.40$	& $117.4$\\
%SCT EC C layer 1	& $-5.09$	& $0.34$	& $113.9$	& $-5.10$	& $0.34$	& $114.6$\\
%SCT EC C layer 2	& $0.50$	& $0.37$	& $111.5$	& $0.45$	& $0.37$	& $112.2$\\
%SCT EC C layer 3	& $5.48$	& $0.44$	& $117.2$	& $5.52$	& $0.44$	& $117.6$\\
%SCT EC C layer 4	& $14.69$	& $0.46$	& $113.0$	& $14.84$	& $0.46$	& $113.0$\\
%SCT EC C layer 5	& $-3.35$	& $0.72$	& $117.4$	& $-3.06$	& $0.72$	& $117.9$\\
%SCT EC C layer 6	& $-5.40$	& $4.64$	& $278.2$	& $-5.37$	& $4.62$	& $278.3$\\
%SCT EC C layer 7	& $-12.38$	& $1.98$	& $134.8$	& $-12.51$	& $1.97$	& $135.0$\\
%SCT EC C layer 8	& $40.44$	& $3.01$	& $131.5$	& $40.58$	& $3.00$	& $131.4$\\
SCT EC C {\bf (all)}	& $0.71$	& $0.17$	& $117.2$	& $0.79$	& $0.17$	& $117.8$\\
\hline
\end{tabular}
\caption[Main residual characteristics for the silicon tracker by layers in M8+ before and after pixel stave bow alignment]{\label{tab:m8plusAlignL4}
Main {\bf residual} characteristics for the silicon tracker by layers in M8+ {\em before} and {\em after} 8 iterations of {\bf pixel stave bow} alignment: the residual mean $\rmean x$, the uncertainty on the residual mean $\delta r_x$, and the standard deviation of the residual $\sigma(r_x)$. The range used for the calculation of the quantities above is $r_x\in[-1.5\,\mm,\,1.5\,\mm]$. All values are given in $\mum$. See text for discussion.
\vspace{\cDistHalf}
}
\end{center}
\end{table}

\begin{table}
\small
\begin{center}
\vspace{\cDistHalf}
\begin{tabular}{l|rrr|rrr}
\hline
 & \multicolumn{3}{c|}{{\bf Before} alignment} & \multicolumn{3}{c}{{\bf After} alignment}\\
	& $\langle o_{xx}\rangle$	& $\delta o_{xx}$	& $\sigma(o_{xx})$	& $\langle o_{xx}\rangle$	& $\delta o_{xx}$	& $\sigma(o_{xx})$ \\
\hline\hline
Pixel barrel layer 0	& $42.28$	& $2.61$	& $300.7$	& $16.47$	& $1.03$	& $119.6$\\
Pixel barrel layer 1	& $27.28$	& $1.63$	& $256.1$	& $-1.65$	& $0.69$	& $109.2$\\
Pixel barrel layer 2	& $24.92$	& $1.37$	& $244.9$	& $12.75$	& $0.60$	& $107.5$\\
Pixel barrel {\bf (all)}& $29.04$	& $0.98$	& $260.3$	& $8.36$	& $0.42$	& $110.7$\\
\hline
%SCT barrel layer 0	& $3.37$	& $0.28$	& $113.6$	& $3.39$	& $0.28$	& $113.7$\\
%SCT barrel layer 1	& $-1.26$	& $0.23$	& $114.6$	& $-1.39$	& $0.23$	& $114.9$\\
%SCT barrel layer 2	& $-0.81$	& $0.22$	& $127.2$	& $-0.64$	& $0.22$	& $127.4$\\
%SCT barrel layer 3	& $-28.92$	& $0.21$	& $131.7$	& $-28.92$	& $0.21$	& $131.6$\\
SCT barrel {\bf (all)}	& $-9.87$	& $0.12$	& $125.0$	& $-9.85$	& $0.12$	& $125.1$\\
\hline
%Pixel EC A layer 0	& $33.12$	& $7.14$	& $145.2$	& $31.74$	& $7.08$	& $144.2$\\
%Pixel EC A layer 1	& $30.40$	& $5.01$	& $103.6$	& $30.44$	& $5.02$	& $104.1$\\
%Pixel EC A layer 2	& $27.85$	& $5.77$	& $92.8$	& $29.39$	& $5.77$	& $93.2$\\
Pixel EC A {\bf (all)}	& $30.83$	& $3.58$	& $118.9$	& $30.68$	& $3.56$	& $118.6$\\
\hline
%Pixel EC C layer 0	& $28.18$	& $7.49$	& $141.8$	& $27.55$	& $7.48$	& $141.8$\\
%Pixel EC C layer 1	& $20.47$	& $4.97$	& $86.3$	& $19.04$	& $5.00$	& $87.0$\\
%Pixel EC C layer 2	& $33.53$	& $5.99$	& $117.1$	& $33.91$	& $5.99$	& $117.3$\\
Pixel EC C {\bf (all)}	& $27.91$	& $3.68$	& $118.8$	& $27.42$	& $3.68$	& $119.1$\\
\hline
%SCT EC A layer 0	& $10.09$	& $1.98$	& $117.2$	& $9.85$	& $1.96$	& $116.5$\\
%SCT EC A layer 1	& $1.87$	& $1.68$	& $110.2$	& $1.59$	& $1.66$	& $109.5$\\
%SCT EC A layer 2	& $-4.39$	& $1.52$	& $96.8$	& $-3.72$	& $1.51$	& $96.8$\\
%SCT EC A layer 3	& $4.16$	& $1.78$	& $104.9$	& $3.62$	& $1.76$	& $104.8$\\
%SCT EC A layer 4	& $5.12$	& $2.77$	& $154.6$	& $5.48$	& $2.73$	& $153.7$\\
%SCT EC A layer 5	& $-11.08$	& $3.13$	& $127.6$	& $-9.37$	& $3.20$	& $131.7$\\
%SCT EC A layer 6	& $16.89$	& $4.65$	& $131.1$	& $16.27$	& $4.68$	& $131.5$\\
%SCT EC A layer 7	& $-6.46$	& $4.85$	& $98.2$	& $-6.44$	& $4.87$	& $98.4$\\
%SCT EC A layer 8	& $-13.25$	& $7.74$	& $125.9$	& $-13.22$	& $7.51$	& $124.2$\\
SCT EC A {\bf (all)}	& $2.08$	& $0.80$	& $118.1$	& $2.17$	& $0.80$	& $118.0$\\
\hline
%SCT EC C layer 0	& $-5.21$	& $2.51$	& $143.3$	& $-5.89$	& $2.44$	& $140.5$\\
%SCT EC C layer 1	& $-2.10$	& $1.70$	& $116.7$	& $-2.30$	& $1.68$	& $116.0$\\
%SCT EC C layer 2	& $-5.48$	& $2.14$	& $136.8$	& $-5.11$	& $2.14$	& $136.6$\\
%SCT EC C layer 3	& $-18.08$	& $2.12$	& $124.7$	& $-17.99$	& $2.12$	& $124.8$\\
%SCT EC C layer 4	& $-0.29$	& $2.53$	& $142.9$	& $-0.83$	& $2.48$	& $141.7$\\
%SCT EC C layer 5	& $-3.52$	& $3.43$	& $134.0$	& $-3.60$	& $3.40$	& $133.7$\\
%SCT EC C layer 6	& $39.40$	& $19.66$	& $157.3$	& $39.40$	& $19.66$	& $157.3$\\
%SCT EC C layer 7	& $-10.73$	& $7.06$	& $110.8$	& $-10.23$	& $7.02$	& $110.5$\\
%SCT EC C layer 8	& $4.12$	& $16.07$	& $186.0$	& $8.59$	& $15.25$	& $181.8$\\
SCT EC C {\bf (all)}	& $-5.70$	& $0.92$	& $132.7$	& $-5.78$	& $0.91$	& $131.8$\\
\hline
\end{tabular}
\caption[Main overlap residual characteristics for the silicon tracker by layers in M8+ before and after pixel stave bow alignment]{\label{tab:m8plusAlignL4_ovres}
Main {\bf overlap residual} characteristics for the silicon tracker by layers in M8+ {\em before} and {\em after} 8 iterations of  {\bf pixel stave bow} alignment: the overlap residual mean $\ormean{xx}$, the uncertainty on the residual mean $\delta\ormean{xx}$, and the standard deviation of the residual $\orsig{xx}$. The (implicit) range used for the calculation of the quantities above is $r_x\in[-1.5\,\mm,\,1.5\,\mm]$ for each residual constituting an overlap residual. All values are given in $\mum$. See text for discussion.
\vspace{\cDistHalf}
}
\end{center}
\end{table}

%% file: M8plus/L3.tex
Historically, the L3~alignment was the only approach implemented in the \RA\ algorithm, and superstructure alignment was introduced in the context of M8+ by the author. The calculation prescription for L3 alignment corrections is detailed in Subsection~\ref{ssec:l3} on page~\pageref{ssec:l3}~ff., and earlier results with MC simulations are referenced from there. In this Subsection, the derivation of L3 alignment constants with M8+ cosmic ray data is presented. The entire M8+ dataset is used, id est both $B$-field on and off.\\
Individual modules are aligned in local $x$ and $y$ in case of the pixel detector, and in $x$ in case of the SCT\footnote{no alignment for the local $y$ DoF is performed for reasons discussed in Subsection~\ref{ssec:l3}}. To achieve a sufficiently low statistical uncertainty on alignment corrections, at least 50 residuals are to be collected by a module before being aligned. Similarly, 25 or more overlap residuals of the respective type are required in order to contribute to the calculation of the alignment constants via Equation~\ref{eqn:c_xtotal}. To avoid oscillatory or even chaotic behaviour, any given alignment correction with its pull $\frac{c_x}{\delta c_x}<0.7$ is rejected.

\begin{figure}
\begin{center}
\vspace{\cDistHalf}
\includegraphics[width=15.8cm,clip=true]{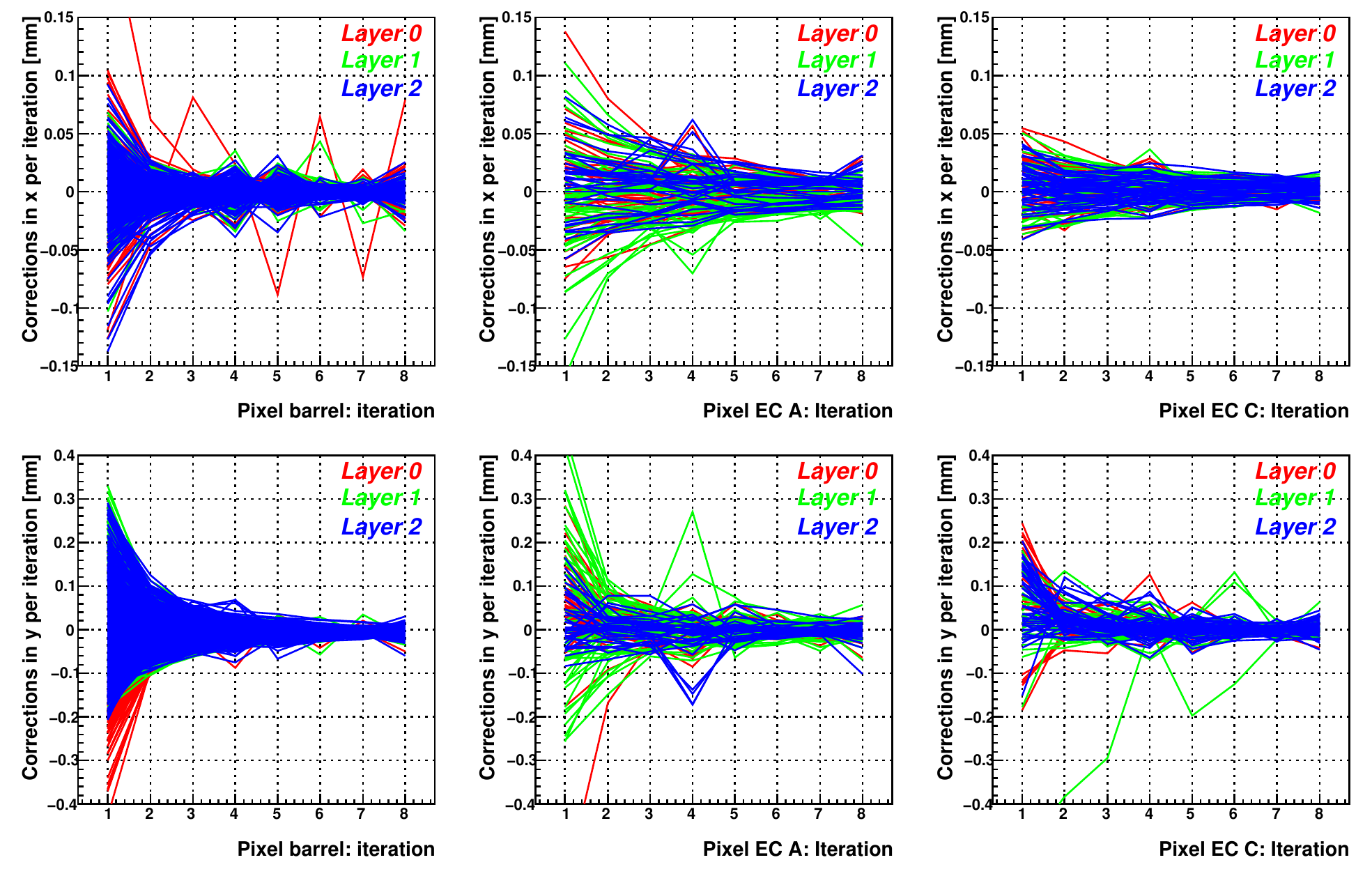}
\vspace{\cDist}
\end{center}
\caption[Convergence of alignment corrections for the pixel detector at L3 in M8+]{\label{fig:convergenceL3_PIX}
Alignment corrections per iteration for the {\bf pixel} detector with the \RA\ algorithm at {\bf L3} in M8+. A dominating fraction of its 1744 modules were aligned for the $x${\bf~(top row)} and $y${\bf~(bottom row)} degrees of freedom. Each line corresponds to the alignment correction of one module. The results for the barrel{\bf~(left column)}, EC~A{\bf~(middle column)}, and EC~C{\bf~(right column)} are shown. An exponential-like asymptotic convergence is observed. The magnitude of corrections in $x$ is smaller due to preceding stave bow alignment.
\vspace{\cDistHalf}
}
\end{figure}%\nopagebreak[5]

\begin{figure}
\begin{center}
\includegraphics[width=15.8cm,clip=true]{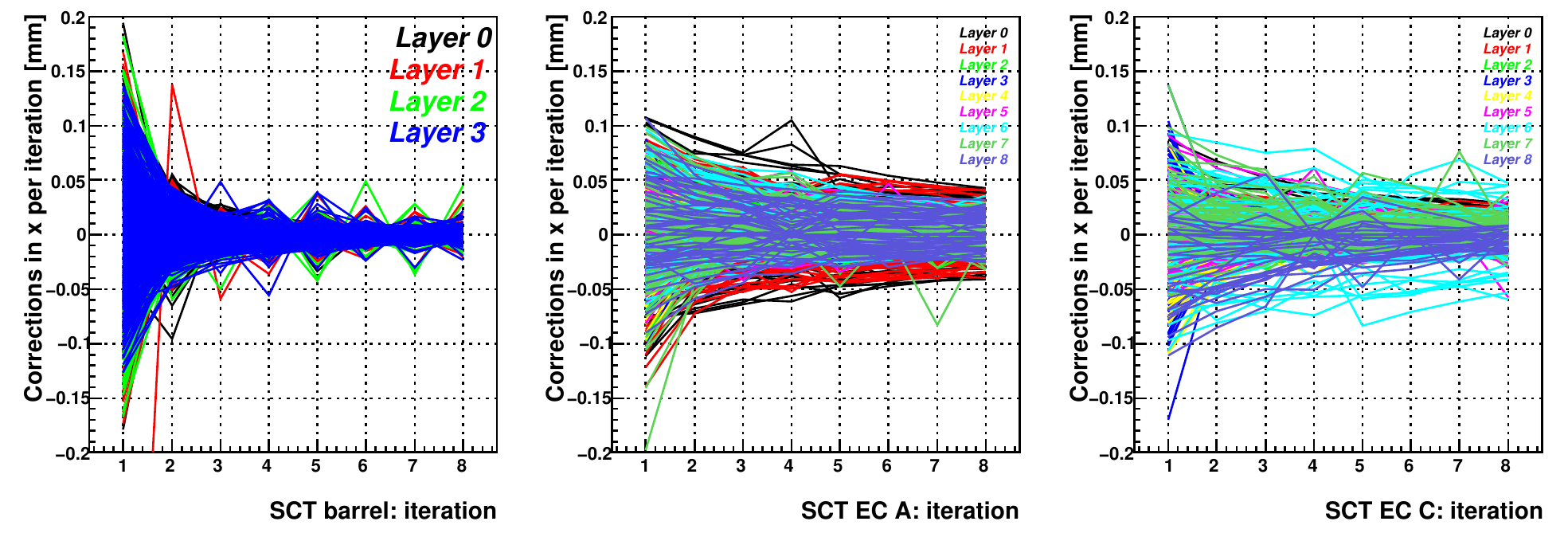}
\vspace{\cDist}
\end{center}
\caption[Convergence of alignment corrections for the SCT detector at L3 in M8+]{\label{fig:convergenceL3_SCT}
Alignment corrections per iteration for the {\bf SCT} detector with the \RA\ algorithm at {\bf L3} in M8+. A dominating fraction of its 4088 modules were aligned for the $x$~degree of freedom. Each line corresponds to the alignment correction of one module. The results for the barrel{\bf~(left)}, EC~A{\bf~(middle)}, and EC~C{\bf~(right)} are shown. An exponential-like asymptotic convergence is observed in the barrel. Due to the limited convergence in the end-caps no alignment corrections at L3 are provided at the end of the alignment procedure.
%
%\vspace{\cDistHalf}
}
\end{figure}%\nopagebreak[5]

The convergence of alignment corrections per iteration for the {\bf pixel} detector is shown in Figure~\ref{fig:convergenceL3_PIX}. A rapid convergence in both DoFs is observed in the barrel of the pixel detector. The magnitude of corrections in $x$ is notably smaller than in $y$ due to the preceding pixel stave bow alignment. Only a single module exhibits oscillatory behaviour but does not diverge.
The convergence in the ECs is somewhat slower and numerically less stable, although it displays a picture similar to the barrel. The magnitude of corrections in EC~A is almost by a factor of 2 larger than in EC~C, which is a sign for a less accurate mounting precision of its modules. This is further confirmed by Tables~\ref{tab:m8plusAlignL2}, \ref{tab:m8plusAlignL4}, and \ref{tab:m8plusAlignL3}. The disparity in the resolution between the two pixel ECs does not disappear since not all the modules fulfil the criteria to be aligned due to a lack of illumination.\\
The analogous distribution for the {\bf SCT} detector in Figure~\ref{fig:convergenceL3_SCT} displays a similar picture in the barrel: an asymptotic, exponential-like convergence is observed. Some corrections of about $c_x\simeq30\,\mum$ suggest oscillatory behaviour. However, these are literally a couple of modules, and  the magnitude is fairly small. The vast bulk of 2112 modules is well-behaved. In the ECs, convergence is clearly an issue. Even though none of the corrections per iteration exceeds 200\,\mum\ and the majority stays within 50\,\mum, the plot is not satisfactory. This is because of a very slow decrease in the magnitude of $c_x$ alignment corrections per iteration for the vast majority of modules. Partly, this may be caused by stricter track reconstruction cuts in the end-caps which prevent the inclusion of large residuals in the track fit. This biases the $\rmean x$ measurement for EC modules, so that the correct alignment constant $c_x$ cannot be determined over one iteration. However, the magnitude of the observed lack of convergence is too big to be exclusively explained by such effects and to be tolerated. Based on the observation in Figure~\ref{fig:convergenceL3_SCT}~(b) and (c), {\em no} alignment corrections at L3 are provided for the SCT ECs by the \RA\ algorithm.

\begin{figure}
\begin{center}
\vspace{\cDistHalf}
\includegraphics[width=15.8cm,clip=true]{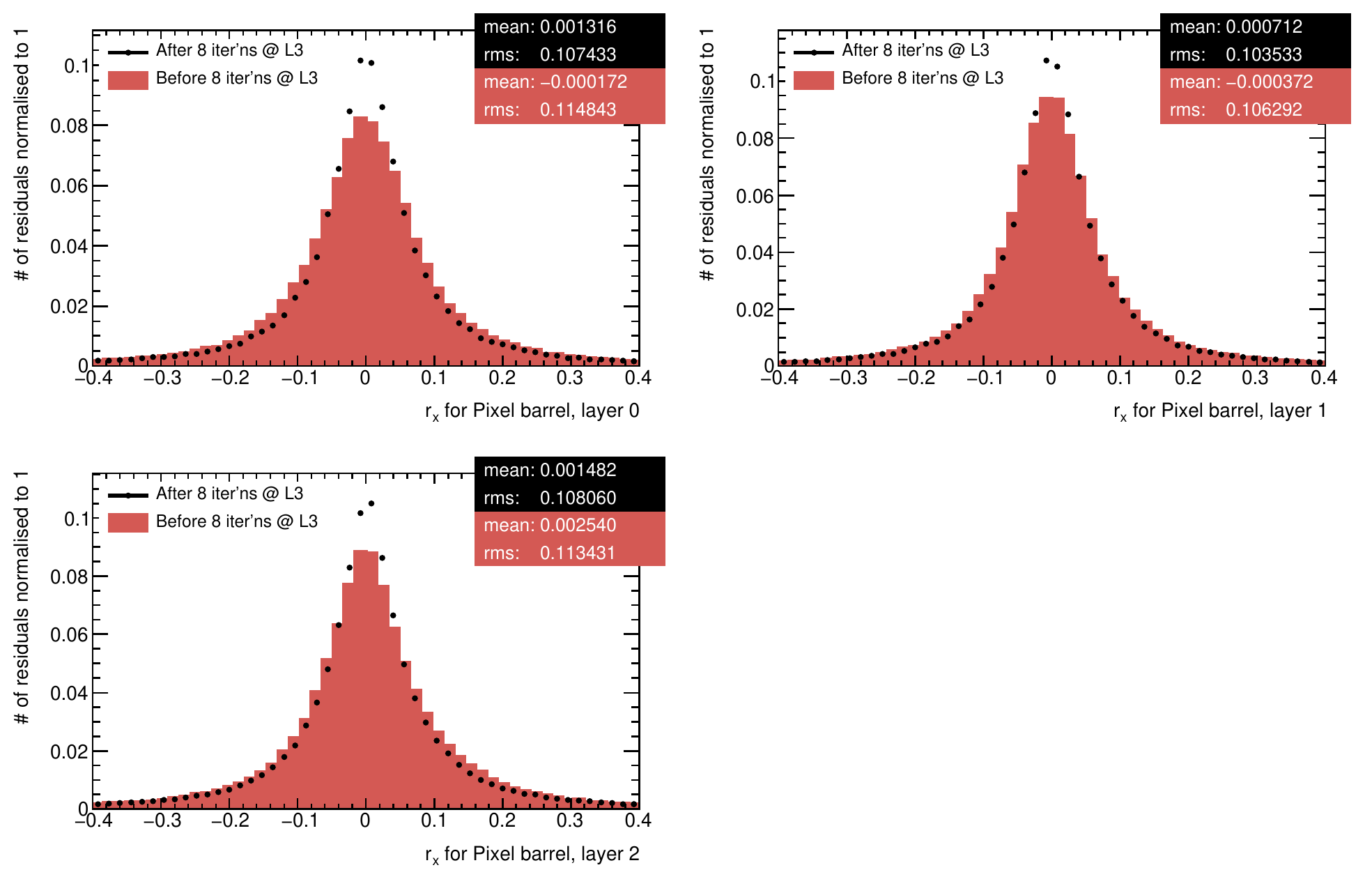}
\vspace{\cDist}
\end{center}
\caption[$r_x$ residual distribution in the barrel of the pixel detector before and after alignment corrections at L3 in M8+]{\label{fig:r_x_PIXB_L3}
The $r_x$ {\bf residual} distribution with {\bf $B$-field off} in the {\bf barrel} of the {\bf pixe}l detector by layers {\em before} and {\em after} alignment corrections at {\bf L3} in M8+. The $\rsig x$ values refine by about 5\% for $r_x\in[-0.4,\,0.4]\,\mm$. The improvement in the peak region is even more pronounced. All values are in~mm.
}
\end{figure}

\begin{figure}
\begin{center}
\vspace{\cDistHalf}
\includegraphics[width=15.8cm,clip=true]{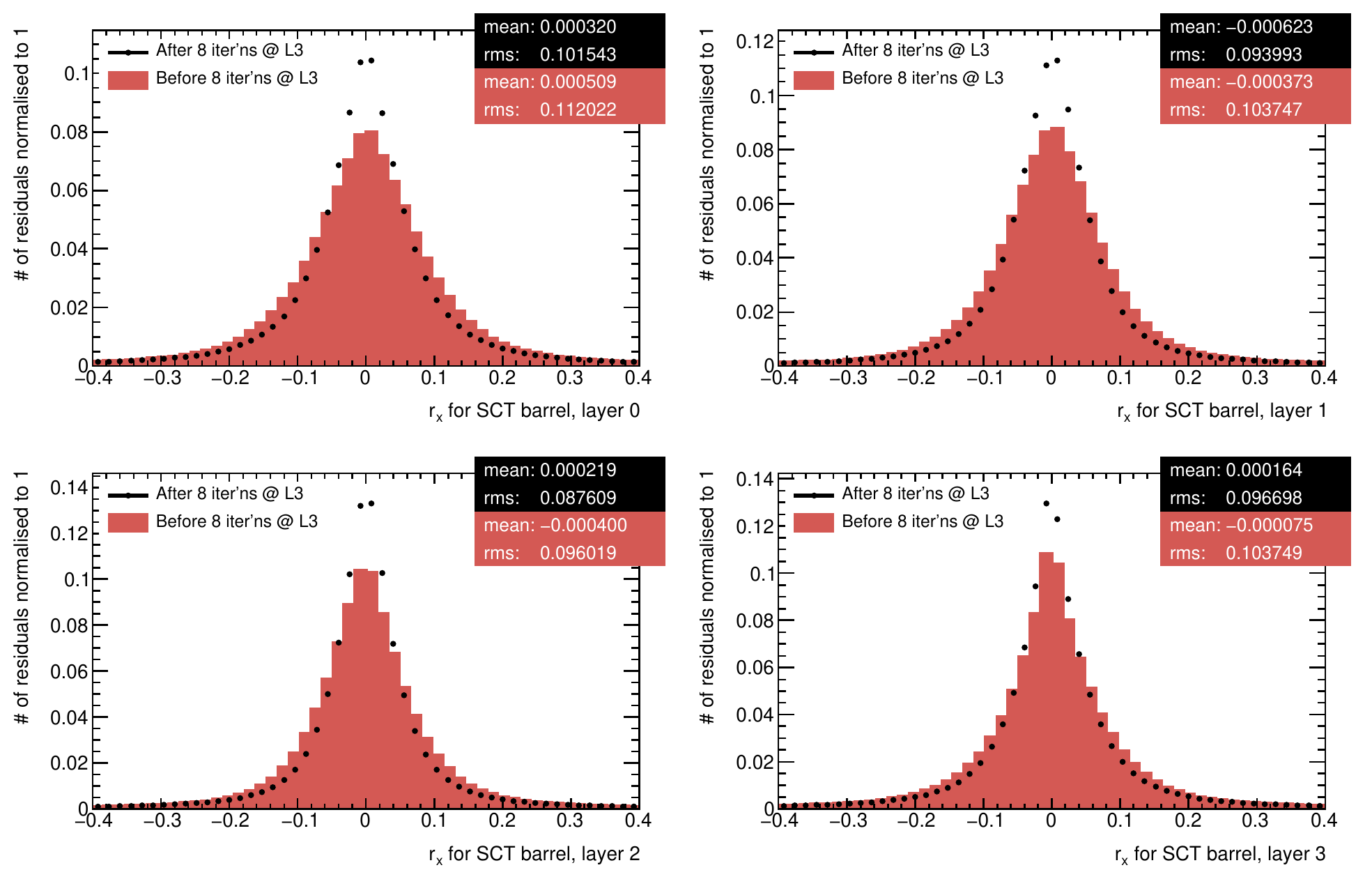}
\vspace{\cDist}
\end{center}
\caption[$r_x$ residual distribution in the barrel of the SCT detector before and after alignment corrections at L3 in M8+]{\label{fig:r_x_SCTB_L3}
The $r_x$ {\bf residual} distribution with {\bf $B$-field off} in the {\bf barrel} of the {\bf SCT} detector by layers {\em before} and {\em after} alignment corrections at {\bf L3} in M8+. A qualitatively similar picture to the pixel barrel can be observed: the $\rsig x$ values refine by about 10\% for $r_x\in[-0.4,\,0.4]\,\mm$. The improvement in the peak region is even more pronounced. All values are in~mm.
\vspace{\cDist}
}
\end{figure}

The {\bf $r_x$ residual} distributions with $B$-field off {\em before} and {\em after} L3 alignment with the \RA\ algorithm are presented for monitoring purposes on page~\pageref{fig:r_x_PIXB_L3} and their statistical benchmarks are summarised in Tables~\ref{tab:m8plusAlignL3}/\ref{tab:m8plusAlignL3_B1} for solenoid off/on:
\begin{description}
\item[Pixel barrel:] The Gaussian-shaped $r_x$ distributions shown in Figure~\ref{fig:r_x_PIXB_L3} are further sharpened by L3 alignment. The residual width improves by about 2\% to $\rsig x=187.4\,\mum$ over the full range $r_x\in[-1.5\,\mm,\,1.5\,\mm]$, and even more so in the peak region. A residual mean of $\rmean x=1.62\pm0.26\,\mum$ is observed, and the means of the three layers are tightly scattered about this value. This is a manifestation of the known~\cite{bib:alignmentWorkshop} discrepancy in residual means between $B$-field off and on data observed in the barrel of the pixel detector ($\rmean x=-2.12\pm0.12\,\mum$ with solenoid on). $B$-field off and on discrepancies are discussed in Section~\ref{sec:B0minusB1M8};
\item[SCT barrel:] The situation is qualitatively similar to the barrel of the pixel detector as can be seen from Figure~\ref{fig:r_x_SCTB_L3}. The sharpening of the peak region is somewhat more pronounced. Nominally, the residual width improves by about 4\% to $\rsig x=123.8\,\mum$ with a more pronounced refinement in the peak region. The residual mean is consistent with 0 to within 0.05\,\mum;
\item[Pixel End-Caps:] The distributions remain fairly similar to the ones presented in Figures~\ref{fig:r_x_PIXA_L2}, \ref{fig:r_x_PIXC_L2} and shall not be shown here separately. Mainly, the shoulders of the distributions are improved. The residual width refines by about 6\% to $\rsig x=241.5/184.5\,\mum$ in EC~A/C. The residual mean in EC~A is with $\rmean x=-4.2\pm2.1\,\mum$ somewhat away from 0;
\item[SCT End-Caps:] Since the SCT ECs were not aligned at L3, their distributions remain almost unchanged from the ones presented in Figures~\ref{fig:r_x_SCTA_L2}, \ref{fig:r_x_SCTC_L2} and are therefore not shown. Due to a better alignment of other parts of the silicon tracker, the residual widths still improve by some percents in the innermost disks.
\end{description}

%\nopagebreak[5]

%\nopagebreak[5]
\newpage

\begin{figure}
\begin{center}
\vspace{\cDistHalf}
\includegraphics[width=15.8cm,clip=true]{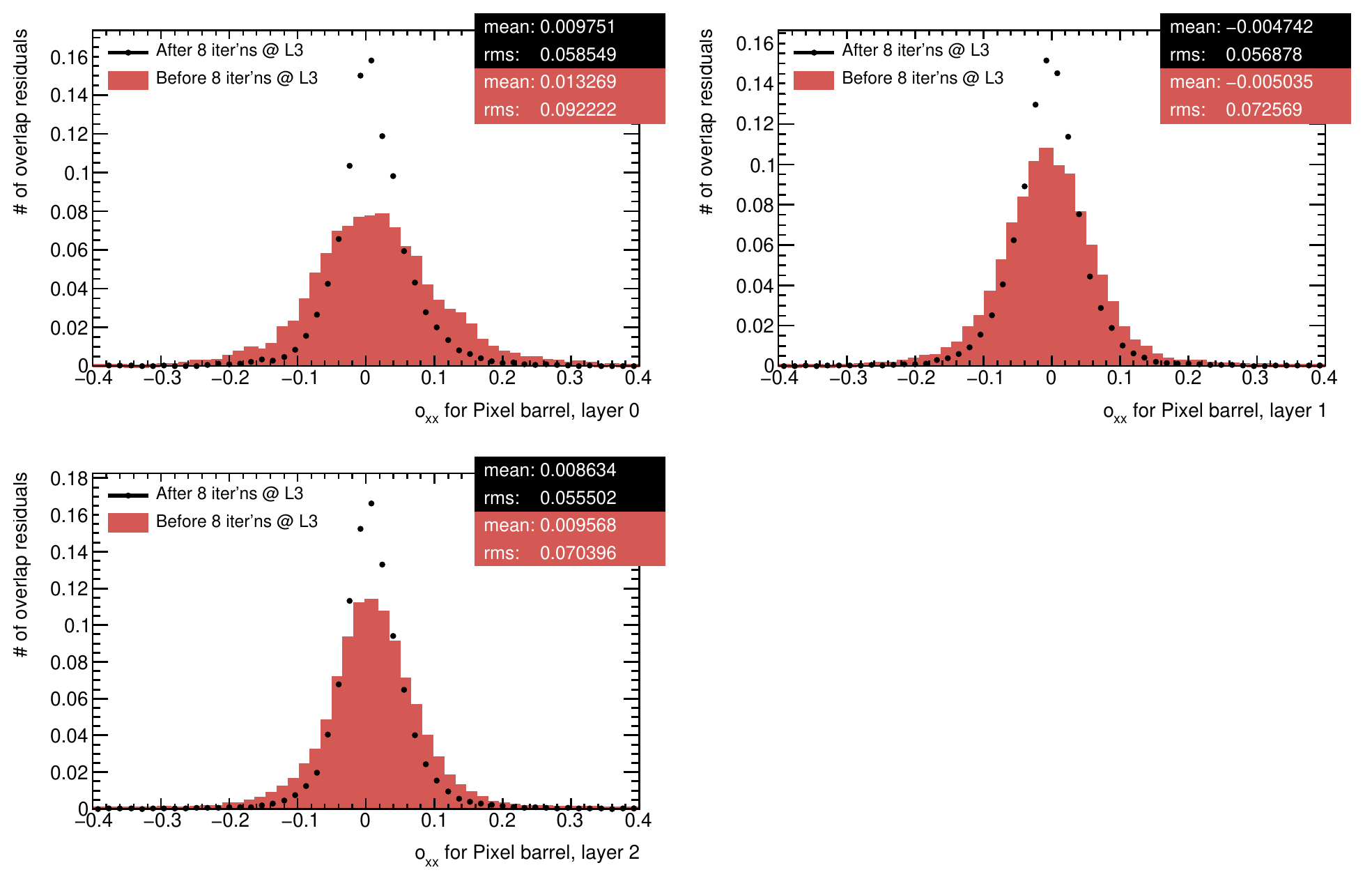}
\vspace{\cDist}
\end{center}
\caption[The $o_{xx}$ overlap residual distribution in the barrel of the pixel detector before and after alignment corrections at L3 in M8+]{\label{fig:o_xx_PIXB_L3}
The $o_{xx}$ {\bf overlap residual} distribution with {\bf $B$-field off} in the {\bf barrel} of the {\bf pixel} detector by layers {\em before} and {\em after} alignment corrections at {\bf L3} in M8+. The overlap residual width in the outer two layers improves by about 20\% for $o_{xx}\in[-0.4,\,0.4]\,\mm$, while the $b$-layer shows a dramatic refinement of about 40\%. Both benchmarks indicate that a good alignment of neighbouring modules in the same layer is achieved. All values are in~mm.
}
\end{figure}

\begin{figure}
\begin{center}
\vspace{\cDistHalf}
\includegraphics[width=15.8cm,clip=true]{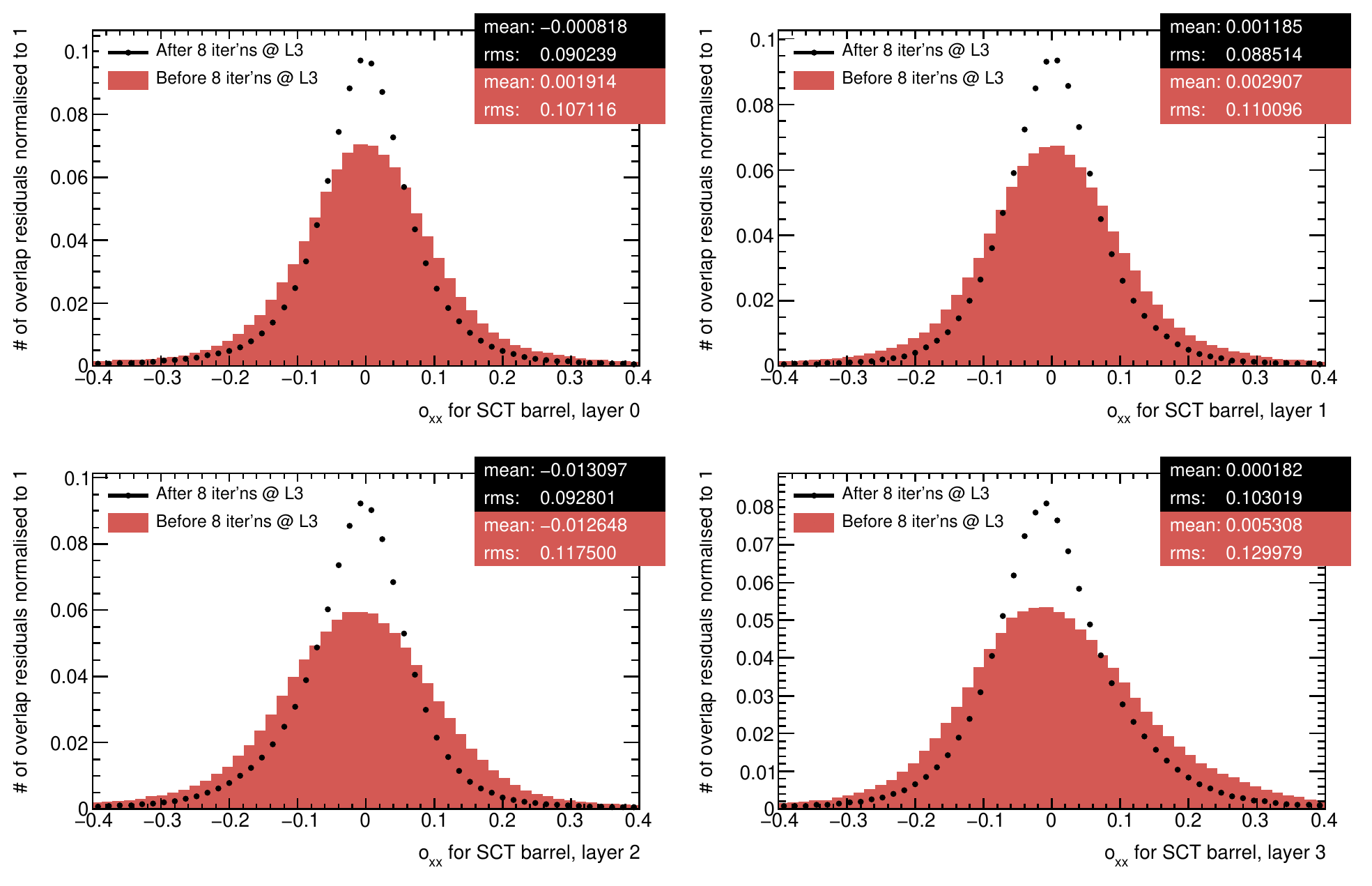}
\vspace{\cDist}
\end{center}
\caption[The $o_{xx}$ overlap residual distribution in the barrel of the SCT detector before and after alignment corrections at L3 in M8+]{\label{fig:o_xx_SCTB_L3}
The $o_{xx}$ {\bf overlap residual} distribution with {\bf $B$-field off} in the {\bf barrel} of the {\bf SCT} detector by layers {\em before} and {\em after} alignment corrections at {\bf L3} in M8+. The overlap residual widths improve by about 20\% for $o_{xx}\in[-0.4,\,0.4]\,\mm$, which indicates that a good alignment of neighbouring modules in the same layer is achieved.  All values are in~mm.
\vspace{\cDist}
}
\end{figure}

The {\bf $o_{xx}$ overlap residual} distributions with $B$-field off {\em before} and {\em after} L3 alignment are shown on page~\pageref{fig:o_xx_PIXB_L3} and their statistical benchmarks are summarised in Tables~\ref{tab:m8plusAlignL3_ovres}/\ref{tab:m8plusAlignL3_B1_ovres} for solenoid off/on:
\begin{description}
\item[Pixel barrel:] The shape of the $o_{xx}$ distributions shown in Figure~\ref{fig:o_xx_PIXB_L3} improves towards a Gaussian, and the peaks are notably sharpened. The overlap residual width refines by about 12\% to $\orsig{xx}=97.7\,\mum$ with an even more pronounced improvement in the $b$-layer~(i.e. layer~0). The $o_{xx}$ overlap residual means in layer~0/1/2 notably deviate from 0 without $B$-field (especially for layers~0 and~2):
\[ \ormean{xx}=13.3\pm0.8{\,/}-\!1.4\pm0.6{\,/\,}11.7\pm0.5\,\mum\]
The same picture is confirmed with $B$-field on:
\[ \ormean{xx}=11.8\pm0.9{\,/}-\!0.2\pm0.7{\,/\,}10.3\pm0.6\,\mum\,.\]
A radial expansion of a given layer, i.e. as an artifact of the assembly procedure, would result in a proportional increase of its circumference and thus \ormean{xx}, which could explain the observation. At this stage, no explicit correction for this effect is applied. No correlation between the overlap residual means $\ormean{xx}$ of various layers is expected as they are a layer-internal construct;
\item[SCT barrel:] The improvement is not as dramatic as in case of the barrel of the pixel detector, but still impressive as can be seen from Figure~\ref{fig:o_xx_SCTB_L3}. Nominally, the overlap residual width improves by about 7\% to $\orsig{xx}=175.1\,\mum$. Again, $o_{xx}$ overlap residual means significantly differing from 0 are observed for layer~0/1/2/3 without $B$-field:
\[ \ormean{xx}=4.0\pm0.3{\,/\,}4.4\pm0.3{\,/\,}4.9\pm0.3{\,/}-\!18.3\pm0.3\,\mum\,.\]
Further, there is a statistically significant difference between the $B$-field off and on cases, which is discussed in Section~\ref{sec:B0minusB1M8};
\item[Pixel End-Caps:] Despite an overlap region of $\order{10\%}$ between pixel EC modules, their illumination is not sufficient to provide for a contribution to the alignment procedure. Nevertheless, the overlap residual widths improve by 5\%\,/\,9\% for end-cap~A/C owing to corrections from regular residuals. Sizable deviations from 0 are found in the overlap residual means of both ECs: $\ormean{xx}=28.0\pm3.4\,/\,22.6\pm3.4\,\mum$ for $B$-field off, which could be due to a radial expansion of the EC disks;
\item[SCT End-Caps:] No effects worth mentioning are observed in the SCT ECs besides statistical deviations of $\ormean{xx}$ from 0 which are summarised in in Tables~\ref{tab:m8plusAlignL3_ovres} and \ref{tab:m8plusAlignL3_B1_ovres}.
\end{description}

Overall, a satisfactory improvement is achieved in the $r_x$ and $o_{xx}$ distributions, which is further supported by summary Tables~\ref{tab:m8plusAlignL3}, \ref{tab:m8plusAlignL3_ovres}, \ref{tab:m8plusAlignL3_B1}, and \ref{tab:m8plusAlignL3_B1_ovres}.

%\nopagebreak[5]

%\nopagebreak[5]
\clearpage

\begin{comment}
\begin{figure}
\begin{center}
\vspace{\cDistHalf}
\includegraphics[width=15.8cm,clip=true]{M8plus/fig/L3/PIX_ECA_Rx}
\vspace{\cDist}
\end{center}
\caption[$r_x$ residual distribution in the EC~A of the pixel detector before and after alignment corrections at L3 in M8+]{\label{fig:r_x_PIXA_L3}
The $r_x$ {\bf residual} distribution with {\bf $B$-field off} in the {\bf end-cap~A} of the {\bf pixel} detector by layers {\em before} and {\em after} alignment corrections at {\bf L3} in M8+. The residual widths improve by $\order{10\%}$ in the process of alignment. {\bf don't show?}
}
\end{figure}%\nopagebreak[5]

\begin{figure}
\begin{center}
\vspace{\cDistHalf}
\includegraphics[width=15.8cm,clip=true]{M8plus/fig/L3/PIX_ECC_Rx}
\vspace{\cDist}
\end{center}
\caption[$r_x$ residual distribution in the EC~C of the pixel detector before and after alignment corrections at L3 in M8+]{\label{fig:r_x_PIXC_L3}
The $r_x$  {\bf residual} distribution with {\bf $B$-field off} in the {\bf end-cap~C} of the {\bf pixel} detector by layers {\em before} and {\em after} alignment corrections at {\bf L3} in M8+. A similar picutre to EC~A is observed: the residual widths improve by $\order{10\%}$. {\bf don't show?}
}
\end{figure}%\nopagebreak[5]
\end{comment}
\clearpage

\begin{comment}
\begin{figure}
\begin{center}
\vspace{\cDistHalf}
\includegraphics[width=15.8cm,clip=true]{M8plus/fig/L3/SCT_ECA_Rx}
\vspace{\cDist}
\end{center}
\caption[$r_x$ residual distribution in the EC~A of the SCT detector before and after alignment corrections at L3 in M8+]{\label{fig:r_x_SCTA_L3}
The $r_x$ residual distribution in the EC~A of the SCT detector by disk layers before and after alignment corrections at L3 in M8+. The residual widths improve by about 5\% in the process of alignment. Due to the unstable convergence displayed in Figure~\ref{fig:convergenceL3_SCT} and the marginal improvement of residuals no L3 alignment corrections for the SCT ECs are provided. All values are in~mm.
}
\end{figure}%\nopagebreak[5]

\begin{figure}
\begin{center}
\vspace{-0.3cm}
%\vspace{\cDistHalf}
\includegraphics[width=15.8cm,clip=true]{M8plus/fig/L3/SCT_ECC_Rx}
\vspace{\cDist}
\end{center}
\caption[$r_x$ residual distribution in the EC~C of the SCT detector before and after alignment corrections at L3 in M8+]{\label{fig:r_x_SCTC_L3}
The $r_x$ residual distribution in the EC~C of the SCT detector by disk layers before and after alignment corrections at L3 in M8+. The residual widths improve by about 5\% in the process of alignment. Due to the unstable convergence displayed in Figure~\ref{fig:convergenceL3_SCT} and the marginal improvement of residuals no L3 alignment corrections for the SCT ECs are provided. All values are in~mm.
\vspace{\cDistHalf}
}
\end{figure}%\nopagebreak[5]
\clearpage
\end{comment}

\begin{table}
\small
\begin{center}
\begin{tabular}{l|rrr|rrr}
\hline
 & \multicolumn{3}{c|}{{\bf Before} alignment} & \multicolumn{3}{c}{{\bf After} alignment}\\
%\hline
 & $\langle r_x\rangle$ & $\delta r_x$ & $\sigma(r_x)$ & $\langle r_x\rangle$ & $\delta r_x$ & $\sigma(r_x)$  \\
\hline\hline
Pixel barrel layer 0	& $0.25$	& $0.57$	& $184.3$	& $1.86$	& $0.55$	& $177.0$\\
Pixel barrel layer 1	& $0.19$	& $0.42$	& $180.5$	& $1.43$	& $0.41$	& $177.2$\\
Pixel barrel layer 2	& $2.85$	& $0.41$	& $202.5$	& $1.66$	& $0.40$	& $198.9$\\
Pixel barrel {\bf (all)}& $1.43$	& $0.26$	& $191.6$	& $1.62$	& $0.26$	& $187.4$\\
\hline
SCT barrel layer 0	& $0.00$	& $0.13$	& $187.5$	& $0.22$	& $0.12$	& $180.1$\\
SCT barrel layer 1	& $-0.12$	& $0.10$	& $165.3$	& $-0.67$	& $0.09$	& $158.0$\\
SCT barrel layer 2	& $-0.48$	& $0.09$	& $152.8$	& $0.12$	& $0.08$	& $146.9$\\
SCT barrel layer 3	& $-0.15$	& $0.09$	& $176.9$	& $0.16$	& $0.09$	& $172.2$\\
SCT barrel {\bf (all)}	& $-0.20$	& $0.05$	& $170.0$	& $-0.04$	& $0.05$	& $163.8$\\
\hline
Pixel EC A layer 0	& $-12.06$	& $3.67$	& $261.7$	& $-11.03$	& $3.42$	& $245.8$\\
Pixel EC A layer 1	& $9.31$	& $3.89$	& $275.0$	& $-3.07$	& $3.44$	& $245.8$\\
Pixel EC A layer 2	& $-18.87$	& $4.43$	& $250.5$	& $4.89$	& $4.24$	& $242.4$\\
Pixel EC A {\bf (all)}	& $-5.65$	& $2.29$	& $264.5$	& $-4.18$	& $2.11$	& $245.1$\\
\hline
Pixel EC C layer 0	& $-3.69$	& $3.03$	& $196.7$	& $-0.49$	& $2.89$	& $188.2$\\
Pixel EC C layer 1	& $1.81$	& $2.94$	& $191.0$	& $-0.59$	& $2.82$	& $184.7$\\
Pixel EC C layer 2	& $-11.50$	& $2.77$	& $194.4$	& $0.59$	& $2.55$	& $181.2$\\
Pixel EC C {\bf (all)}	& $-4.84$	& $1.68$	& $194.1$	& $-0.12$	& $1.58$	& $184.5$\\
\hline
SCT EC A layer 0	& $-3.23$	& $0.63$	& $207.1$	& $-2.70$	& $0.61$	& $201.7$\\
SCT EC A layer 1	& $-2.39$	& $0.53$	& $186.3$	& $-2.37$	& $0.52$	& $184.2$\\
SCT EC A layer 2	& $-3.08$	& $0.50$	& $155.8$	& $-2.67$	& $0.49$	& $153.8$\\
SCT EC A layer 3	& $-3.79$	& $0.54$	& $152.7$	& $-3.51$	& $0.53$	& $150.6$\\
SCT EC A layer 4	& $1.25$	& $0.58$	& $149.3$	& $1.65$	& $0.57$	& $148.3$\\
SCT EC A layer 5	& $-1.96$	& $0.85$	& $151.6$	& $-1.67$	& $0.84$	& $150.3$\\
SCT EC A layer 6	& $-1.15$	& $1.32$	& $160.6$	& $-1.12$	& $1.31$	& $159.7$\\
SCT EC A layer 7	& $-5.14$	& $2.10$	& $180.8$	& $-5.15$	& $2.09$	& $180.1$\\
SCT EC A layer 8	& $-8.69$	& $3.01$	& $188.5$	& $-8.28$	& $2.99$	& $188.2$\\
SCT EC A {\bf (all)}	& $-2.46$	& $0.24$	& $174.0$	& $-2.17$	& $0.23$	& $171.4$\\
\hline
SCT EC C layer 0	& $-4.75$	& $0.53$	& $162.7$	& $-4.84$	& $0.53$	& $162.7$\\
SCT EC C layer 1	& $-5.32$	& $0.45$	& $159.4$	& $-5.97$	& $0.44$	& $159.4$\\
SCT EC C layer 2	& $1.63$	& $0.48$	& $153.4$	& $1.15$	& $0.48$	& $153.1$\\
SCT EC C layer 3	& $5.36$	& $0.57$	& $156.1$	& $5.13$	& $0.57$	& $155.8$\\
SCT EC C layer 4	& $14.51$	& $0.62$	& $159.8$	& $14.39$	& $0.62$	& $159.7$\\
SCT EC C layer 5	& $-1.17$	& $1.05$	& $181.1$	& $-1.35$	& $1.04$	& $181.0$\\
SCT EC C layer 6	& $-3.01$	& $4.47$	& $385.7$	& $-2.90$	& $4.45$	& $385.3$\\
SCT EC C layer 7	& $-10.47$	& $2.78$	& $205.3$	& $-10.27$	& $2.76$	& $204.0$\\
SCT EC C layer 8	& $33.87$	& $4.73$	& $233.9$	& $35.14$	& $4.62$	& $229.8$\\
SCT EC C {\bf (all)}	& $0.74$	& $0.23$	& $166.4$	& $0.41$	& $0.23$	& $166.3$\\
\hline
\end{tabular}
\caption[Main residual characteristics for the silicon tracker by layers with $B$-field off in M8+ before and after alignment at L3 with $B$-field off]{\label{tab:m8plusAlignL3}
Main {\bf residual} characteristics for the silicon tracker by layers with {\bf $B$-field off} in M8+ {\em before} and {\em after} alignment at {\bf L3}: the residual mean $\rmean x$, the uncertainty on the residual mean $\delta r_x$, and the standard deviation of the residual $\sigma(r_x)$. The range used for the calculation of the quantities above is $r_x\in[-1.5\,\mm,\,1.5\,\mm]$. All values are given in $\mum$. See text for discussion.
}
\end{center}
\end{table}

\begin{table}
\small
\begin{center}
\begin{tabular}{l|rrr|rrr}
\hline
 & \multicolumn{3}{c|}{{\bf Before} alignment} & \multicolumn{3}{c}{{\bf After} alignment}\\
	& $\langle o_{xx}\rangle$	& $\delta o_{xx}$	& $\sigma(o_{xx})$	& $\langle o_{xx}\rangle$	& $\delta o_{xx}$	& $\sigma(o_{xx})$ \\
\hline\hline
Pixel barrel layer 0	& $16.41$	& $1.03$	& $118.7$	& $13.29$	& $0.81$	& $94.6$\\
Pixel barrel layer 1	& $-1.53$	& $0.70$	& $109.5$	& $-1.38$	& $0.62$	& $98.8$\\
Pixel barrel layer 2	& $12.85$	& $0.60$	& $107.8$	& $11.66$	& $0.54$	& $97.6$\\
Pixel barrel {\bf (all)}& $8.44$	& $0.42$	& $110.8$	& $7.36$	& $0.37$	& $97.7$\\
\hline
SCT barrel layer 0	& $8.00$	& $0.36$	& $174.1$	& $3.97$	& $0.34$	& $164.8$\\
SCT barrel layer 1	& $3.43$	& $0.31$	& $179.4$	& $4.37$	& $0.29$	& $169.9$\\
SCT barrel layer 2	& $5.01$	& $0.28$	& $189.7$	& $4.93$	& $0.26$	& $173.9$\\
SCT barrel layer 3	& $-18.79$	& $0.27$	& $196.9$	& $-18.29$	& $0.25$	& $182.8$\\
SCT barrel {\bf (all)}	& $-2.92$	& $0.15$	& $188.1$	& $-3.18$	& $0.14$	& $175.1$\\
\hline
Pixel EC A layer 0	& $32.10$	& $7.10$	& $144.5$	& $31.46$	& $6.31$	& $129.8$\\
Pixel EC A layer 1	& $30.63$	& $5.06$	& $104.5$	& $28.13$	& $4.91$	& $102.3$\\
Pixel EC A layer 2	& $29.08$	& $5.79$	& $93.2$	& $22.20$	& $5.96$	& $96.8$\\
Pixel EC A {\bf (all)}	& $30.82$	& $3.59$	& $118.9$	& $27.99$	& $3.36$	& $112.4$\\
\hline
Pixel EC C layer 0	& $27.37$	& $7.53$	& $142.2$	& $20.39$	& $5.76$	& $108.9$\\
Pixel EC C layer 1	& $18.64$	& $4.98$	& $86.8$	& $19.89$	& $5.04$	& $88.5$\\
Pixel EC C layer 2	& $33.69$	& $6.04$	& $117.8$	& $26.74$	& $6.19$	& $122.5$\\
Pixel EC C {\bf (all)}	& $27.13$	& $3.70$	& $119.3$	& $22.60$	& $3.35$	& $108.9$\\
\hline
SCT EC A layer 0	& $20.82$	& $3.15$	& $249.8$	& $20.53$	& $3.12$	& $248.8$\\
SCT EC A layer 1	& $8.95$	& $2.85$	& $238.6$	& $8.29$	& $2.82$	& $237.1$\\
SCT EC A layer 2	& $14.06$	& $2.79$	& $222.1$	& $14.69$	& $2.75$	& $219.8$\\
SCT EC A layer 3	& $9.19$	& $2.74$	& $199.3$	& $9.00$	& $2.72$	& $198.7$\\
SCT EC A layer 4	& $20.83$	& $3.73$	& $256.2$	& $19.64$	& $3.69$	& $254.7$\\
SCT EC A layer 5	& $30.34$	& $6.49$	& $327.6$	& $29.65$	& $6.46$	& $326.8$\\
SCT EC A layer 6	& $32.79$	& $7.98$	& $276.5$	& $33.28$	& $7.96$	& $276.2$\\
SCT EC A layer 7	& $-11.48$	& $10.77$	& $265.2$	& $-13.82$	& $10.78$	& $266.2$\\
SCT EC A layer 8	& $73.58$	& $17.73$	& $364.7$	& $72.52$	& $17.57$	& $363.1$\\
SCT EC A {\bf (all)}	& $16.58$	& $1.33$	& $246.5$	& $16.21$	& $1.32$	& $245.2$\\
\hline
SCT EC C layer 0	& $15.43$	& $3.79$	& $278.6$	& $16.28$	& $3.77$	& $278.7$\\
SCT EC C layer 1	& $5.25$	& $2.54$	& $221.0$	& $5.67$	& $2.54$	& $222.6$\\
SCT EC C layer 2	& $3.10$	& $3.07$	& $245.1$	& $3.51$	& $3.07$	& $245.2$\\
SCT EC C layer 3	& $-3.54$	& $2.95$	& $213.6$	& $-3.59$	& $2.94$	& $213.5$\\
SCT EC C layer 4	& $24.38$	& $3.72$	& $258.4$	& $25.72$	& $3.75$	& $261.2$\\
SCT EC C layer 5	& $15.83$	& $6.24$	& $298.7$	& $17.23$	& $6.22$	& $298.3$\\
SCT EC C layer 6	& $155.72$	& $48.83$	& $483.4$	& $155.70$	& $48.83$	& $483.4$\\
SCT EC C layer 7	& $79.38$	& $23.61$	& $506.9$	& $78.96$	& $23.42$	& $505.0$\\
SCT EC C layer 8	& $41.44$	& $21.99$	& $341.3$	& $40.99$	& $21.91$	& $340.1$\\
SCT EC C {\bf (all)}	& $10.46$	& $1.41$	& $255.1$	& $11.05$	& $1.41$	& $255.7$\\
\hline
\end{tabular}
\caption[Main overlap residual characteristics for the silicon tracker by layers in M8+ before and after alignment at L3 with $B$-field off]{\label{tab:m8plusAlignL3_ovres}
Main {\bf overlap residual} characteristics for the silicon tracker with {\bf $B$-field off} by layers in M8+ {\em before} and {\em after} alignment at {\bf L3}: the overlap residual mean $\ormean{xx}$, its uncertainty~$\delta o_{xx}$, and the standard deviation of the overlap residual $\orsig{xx}$. The (implicit) range used for the calculation of the quantities above is $r_x\in[-1.5\,\mm,\,1.5\,\mm]$ for each residual constituting an overlap residual. All values are given in $\mum$. See text for discussion.
}
\end{center}
\end{table}

\clearpage

\begin{table}
\small
\begin{center}
\begin{tabular}{l|rrr|rrr}
\hline
 & \multicolumn{3}{c|}{{\bf Before} alignment} & \multicolumn{3}{c}{{\bf After} alignment}\\
	& $\langle r_x\rangle$	& $\delta r_x$	& $\sigma(r_x)$	& $\langle r_x\rangle$	& $\delta r_x$	& $\sigma(r_x)$ \\
\hline\hline
Pixel barrel layer 0	& $-5.08$	& $0.28$	& $78.4$	& $-4.26$	& $0.23$	& $66.1$\\
Pixel barrel layer 1	& $-2.63$	& $0.21$	& $77.8$	& $-1.36$	& $0.20$	& $72.7$\\
Pixel barrel layer 2	& $-2.35$	& $0.20$	& $84.7$	& $-1.75$	& $0.18$	& $78.2$\\
Pixel barrel {\bf (all)}& $-2.99$	& $0.13$	& $81.1$	& $-2.12$	& $0.12$	& $74.0$\\
\hline
SCT barrel layer 0	& $0.46$	& $0.06$	& $80.9$	& $0.09$	& $0.06$	& $72.4$\\
SCT barrel layer 1	& $0.76$	& $0.05$	& $76.1$	& $-0.05$	& $0.05$	& $67.1$\\
SCT barrel layer 2	& $-1.40$	& $0.05$	& $72.1$	& $-0.35$	& $0.04$	& $64.0$\\
SCT barrel layer 3	& $0.68$	& $0.04$	& $69.0$	& $0.63$	& $0.04$	& $62.4$\\
SCT barrel {\bf (all)}	& $0.10$	& $0.02$	& $73.9$	& $0.10$	& $0.02$	& $66.0$\\
\hline
%Pixel EC A layer 0	& $2.18$	& $1.88$	& $133.8$	& $2.74$	& $1.75$	& $124.6$\\
%Pixel EC A layer 1	& $7.68$	& $2.13$	& $144.7$	& $0.86$	& $1.85$	& $125.8$\\
%Pixel EC A layer 2	& $-12.15$	& $2.01$	& $114.0$	& $-4.94$	& $1.95$	& $111.0$\\
Pixel EC A {\bf (all)}	& $0.58$	& $1.17$	& $133.5$	& $0.16$	& $1.07$	& $121.8$\\
\hline
%Pixel EC C layer 0	& $-11.10$	& $1.44$	& $92.2$	& $-4.95$	& $1.20$	& $76.7$\\
%Pixel EC C layer 1	& $-2.03$	& $1.61$	& $101.6$	& $2.13$	& $1.40$	& $88.5$\\
%Pixel EC C layer 2	& $-5.14$	& $1.66$	& $112.3$	& $1.99$	& $1.49$	& $100.6$\\
Pixel EC C {\bf (all)}	& $-6.09$	& $0.91$	& $102.8$	& $-0.22$	& $0.80$	& $89.6$\\
\hline
%SCT EC A layer 0	& $-1.29$	& $0.39$	& $116.5$	& $-1.43$	& $0.38$	& $113.1$\\
%SCT EC A layer 1	& $-0.33$	& $0.31$	& $104.0$	& $-0.32$	& $0.31$	& $103.3$\\
%SCT EC A layer 2	& $0.02$	& $0.27$	& $81.1$	& $-0.03$	& $0.27$	& $80.4$\\
%SCT EC A layer 3	& $-0.80$	& $0.32$	& $83.8$	& $-0.75$	& $0.32$	& $82.9$\\
%SCT EC A layer 4	& $1.89$	& $0.36$	& $89.8$	& $1.93$	& $0.36$	& $89.4$\\
%SCT EC A layer 5	& $0.73$	& $0.44$	& $76.7$	& $0.73$	& $0.44$	& $76.4$\\
%SCT EC A layer 6	& $0.66$	& $0.55$	& $69.0$	& $0.57$	& $0.52$	& $65.3$\\
%SCT EC A layer 7	& $0.33$	& $0.87$	& $77.1$	& $0.53$	& $0.86$	& $76.1$\\
%SCT EC A layer 8	& $-0.17$	& $1.18$	& $73.8$	& $-0.08$	& $1.16$	& $72.7$\\
SCT EC A {\bf (all)}	& $-0.12$	& $0.14$	& $94.9$	& $-0.14$	& $0.14$	& $93.6$\\
\hline
%SCT EC C layer 0	& $-1.32$	& $0.31$	& $88.1$	& $-1.72$	& $0.30$	& $86.3$\\
%SCT EC C layer 1	& $-2.80$	& $0.24$	& $80.3$	& $-2.97$	& $0.24$	& $79.7$\\
%SCT EC C layer 2	& $1.65$	& $0.25$	& $75.8$	& $1.70$	& $0.25$	& $75.7$\\
%SCT EC C layer 3	& $2.91$	& $0.32$	& $82.2$	& $2.86$	& $0.32$	& $81.9$\\
%SCT EC C layer 4	& $3.79$	& $0.31$	& $76.3$	& $3.79$	& $0.31$	& $75.7$\\
%SCT EC C layer 5	& $-2.36$	& $0.61$	& $98.8$	& $-2.30$	& $0.61$	& $99.0$\\
%SCT EC C layer 6	& $-0.84$	& $1.69$	& $167.1$	& $-0.75$	& $1.70$	& $167.8$\\
%SCT EC C layer 7	& $-4.94$	& $1.27$	& $94.4$	& $-4.87$	& $1.26$	& $93.9$\\
%SCT EC C layer 8	& $4.58$	& $1.94$	& $95.7$	& $4.11$	& $1.94$	& $95.4$\\
SCT EC C {\bf (all)}	& $0.13$	& $0.13$	& $84.9$	& $0.02$	& $0.12$	& $84.3$\\
\hline
\end{tabular}
\caption[Main residual characteristics for the silicon tracker by layers with $B$-field off in M8+ before and after alignment at L3 with $B$-field on]{\label{tab:m8plusAlignL3_B1}
Main {\bf residual} characteristics for the silicon tracker by layers with {\bf $B$-field on} in M8+ {\em before} and {\em after} alignment at {\bf L3}: the residual mean $\rmean x$, the uncertainty on the residual mean $\delta r_x$, and the standard deviation of the residual $\sigma(r_x)$. The range used for the calculation of the quantities above is $r_x\in[-1.5\,\mm,\,1.5\,\mm]$. All values are given in $\mum$. See text for discussion.
}
\end{center}
\end{table}

\begin{table}
\small
\begin{center}
\begin{tabular}{l|rrr|rrr}
\hline
 & \multicolumn{3}{c|}{{\bf Before} alignment} & \multicolumn{3}{c}{{\bf After} alignment}\\
	& $\langle o_{xx}\rangle$	& $\delta o_{xx}$	& $\sigma(o_{xx})$	& $\langle o_{xx}\rangle$	& $\delta o_{xx}$	& $\sigma(o_{xx})$ \\
\hline\hline
Pixel barrel layer 0	& $14.42$	& $1.13$	& $112.2$	& $11.85$	& $0.88$	& $87.6$\\
Pixel barrel layer 1	& $-0.03$	& $0.74$	& $99.3$	& $-0.12$	& $0.67$	& $89.9$\\
Pixel barrel layer 2	& $11.63$	& $0.64$	& $94.8$	& $10.34$	& $0.57$	& $84.8$\\
Pixel barrel {\bf (all)}& $7.97$	& $0.45$	& $100.2$	& $6.86$	& $0.39$	& $87.4$\\
\hline
SCT barrel layer 0	& $7.59$	& $0.32$	& $131.5$	& $5.09$	& $0.30$	& $120.2$\\
SCT barrel layer 1	& $4.71$	& $0.25$	& $128.7$	& $5.65$	& $0.23$	& $116.1$\\
SCT barrel layer 2	& $6.47$	& $0.25$	& $138.6$	& $5.84$	& $0.21$	& $117.3$\\
SCT barrel layer 3	& $-13.82$	& $0.25$	& $145.0$	& $-13.85$	& $0.22$	& $127.2$\\
SCT barrel {\bf (all)}	& $-0.37$	& $0.13$	& $137.7$	& $-0.72$	& $0.12$	& $121.1$\\
\hline
%Pixel EC A layer 0	& $36.96$	& $5.42$	& $109.9$	& $28.08$	& $4.21$	& $85.3$\\
%Pixel EC A layer 1	& $44.83$	& $7.47$	& $140.7$	& $41.56$	& $7.58$	& $142.2$\\
%Pixel EC A layer 2	& $24.38$	& $3.43$	& $57.2$	& $17.05$	& $4.21$	& $70.3$\\
Pixel EC A {\bf (all)}	& $36.30$	& $3.45$	& $111.5$	& $29.68$	& $3.27$	& $105.4$\\
\hline
%Pixel EC C layer 0	& $31.27$	& $2.40$	& $43.5$	& $30.29$	& $2.86$	& $51.8$\\
%Pixel EC C layer 1	& $30.29$	& $7.56$	& $136.2$	& $32.82$	& $7.03$	& $126.5$\\
%Pixel EC C layer 2	& $25.72$	& $4.52$	& $85.2$	& $26.09$	& $4.69$	& $88.4$\\
Pixel EC C {\bf (all)}	& $29.00$	& $3.01$	& $95.7$	& $29.62$	& $2.95$	& $93.7$\\
\hline
%SCT EC A layer 0	& $19.78$	& $2.61$	& $174.3$	& $20.56$	& $2.63$	& $175.5$\\
%SCT EC A layer 1	& $13.20$	& $2.51$	& $194.6$	& $12.44$	& $2.47$	& $192.0$\\
%SCT EC A layer 2	& $6.86$	& $2.16$	& $155.3$	& $7.26$	& $2.18$	& $156.6$\\
%SCT EC A layer 3	& $17.70$	& $3.04$	& $198.1$	& $17.11$	& $3.02$	& $196.7$\\
%SCT EC A layer 4	& $24.60$	& $3.37$	& $205.3$	& $25.62$	& $3.38$	& $206.0$\\
%SCT EC A layer 5	& $24.71$	& $4.37$	& $188.4$	& $24.69$	& $4.38$	& $188.2$\\
%SCT EC A layer 6	& $-5.03$	& $4.43$	& $130.2$	& $-5.03$	& $4.42$	& $129.9$\\
%SCT EC A layer 7	& $20.70$	& $8.61$	& $191.4$	& $20.05$	& $8.62$	& $191.1$\\
%SCT EC A layer 8	& $-21.46$	& $8.29$	& $134.7$	& $-22.92$	& $8.05$	& $132.8$\\
SCT EC A {\bf (all)}	& $15.36$	& $1.12$	& $183.9$	& $15.39$	& $1.12$	& $183.5$\\
\hline
%SCT EC C layer 0	& $7.87$	& $3.12$	& $206.1$	& $7.00$	& $3.08$	& $203.4$\\
%SCT EC C layer 1	& $12.55$	& $2.16$	& $165.4$	& $12.83$	& $2.19$	& $167.8$\\
%SCT EC C layer 2	& $1.04$	& $2.08$	& $154.5$	& $2.60$	& $2.15$	& $159.6$\\
%SCT EC C layer 3	& $5.94$	& $3.27$	& $200.9$	& $6.13$	& $3.26$	& $200.5$\\
%SCT EC C layer 4	& $9.29$	& $2.57$	& $154.1$	& $9.93$	& $2.57$	& $154.0$\\
%SCT EC C layer 5	& $18.19$	& $5.58$	& $217.9$	& $18.04$	& $5.58$	& $217.5$\\
%SCT EC C layer 6	& $110.55$	& $30.82$	& $373.6$	& $110.53$	& $30.81$	& $373.5$\\
%SCT EC C layer 7	& $-3.78$	& $10.42$	& $189.7$	& $-4.08$	& $10.46$	& $189.9$\\
%SCT EC C layer 8	& $139.05$	& $35.99$	& $440.8$	& $140.53$	& $35.97$	& $440.5$\\
SCT EC C {\bf (all)}	& $9.23$	& $1.16$	& $183.8$	& $9.59$	& $1.16$	& $184.7$\\
\hline
\end{tabular}
\caption[Main overlap residual characteristics for the silicon tracker by layers in M8+ before and after alignment at L3 with $B$-field on]{\label{tab:m8plusAlignL3_B1_ovres}
Main {\bf overlap residual} characteristics for the silicon tracker with {\bf $B$-field on} by layers in M8+ {\em before} and {\em after} alignment at {\bf L3}: the overlap residual mean $\ormean{xx}$, its uncertainty~$\delta o_{xx}$, and the standard deviation of the overlap residual $\orsig{xx}$. The (implicit) range used for the calculation of the quantities above is $r_x\in[-1.5\,\mm,\,1.5\,\mm]$ for each residual constituting an overlap residual. All values are given in $\mum$. See text for discussion.
}
\end{center}
\end{table}

\clearpage

%% file: M8plus/Results.tex
%Check mean residual for SCT rings (should alternate positive-negative!) [done, ranges in +/- 1 um]
%
%Check whether $c_{yx}$ is implemented with the correct sign by looking at the corrections and overlap residuals!
% [in principle possible since overlap percentage not negligible]
%

As described in the preceding Section~\ref{sec:alignM8}, 
%``\nameref{sec:alignM8}'',
the alignment of the ATLAS silicon tracker was performed with the \RA\ algorithm using the full datased of cosmic ray data collected in the M8+ run. These final results and the performance of the \RA\ shall be reviewed briefly in this Section. The official validation results of the alignment constants provided by the \RA\ algorithm are presented in Section~\ref{sec:validationM8}. 
%``\nameref{sec:validationM8}''. 
The final set of alignment constants can be found in~\cite{bib:raConst, bib:raConstDB}. 

The improvement of the main benchmark of track-based alignment -- the cumulative $r_x$~residual and $o_{xx}$~overlap residual distributions (typically for the individual layers of a given subdetector) -- after the application of the \RA\ algorithm are briefly described here:
\begin{description}
\item[Pixel Barrel:]
%The distribution of $r_x$ residuals in the barrel of the pixel detector is shown by layers in Figure~\ref{fig:r_x_PIXB}. 
The effect of the alignment procedure is striking. Before alignment, all three distributions are with $\rsig x$ of \order{600\,\mum}\ very broad and shapeless. After alignment, their shapes approach a Gaussian, and the widths refine dramatically by~70\% to about~190\,\mum. The residual mean improves from $\order{100\,\mum}$ to $\rmean x\simeq1.6\pm0.3\,\mum$. As already mentioned in Subsection~\ref{ssec:l3M8}, this is a manifestation of the residual mean discrepancy between solenoid on and off data, and will be discussed in more detail in Section~\ref{sec:B0minusB1M8}.\\
The dramatic improvement of the $r_x$ residuals is equally matched by the $o_{xx}$ overlap residual distribution: %shown in Figure~\ref{fig:o_xx_PIXB}: 
a clear peak emerges in the process of \RA\ from a broad mass of histogram entries. $\orsig{xx}$ improves by~70\% to about 100\,\mum, which demonstrates that a high degree of module-to-module alignment precision is achieved. $\ormean{xx}$ of the individual layers are distributed within $\sim$10\,\mum\ of~0. As discussed in Subsection~\ref{ssec:l3M8}, this may be an indication of a radial expansion/shrinking of the pixel barrel layers;
\item[SCT Barrel:]
The improvement in the $r_x$ residual distributions in the barrel of the SCT detector is less dramatic but still impressive. %, as shown in Figure~\ref{fig:r_x_SCTB}. 
A clear, relatively sharp peak is observed in the centre of the distribution before alignment, and distinctive shoulders extend from about $|r_x|\simeq200\,\mum$ outwards. The interpretation is straightforward: the peak is comprised of tracks reconstructed only by the SCT barrel, and its realtively small width reflects the superb assembly precision of that subdetector; the shoulders are due to tracks going through the pixel detector, whose macroscopic misalignment of \order{1\,\mm} with respect to the SCT produces the large magnitude of residuals. The shoulders disappear, and $\rsig x$  refines by 40\% to $\sim\!164\,\mum$ in the process of alignment. The residual means of the individual layers, being scattered around 0 by $\order{10\,\mum}$, approach 0 to within $\sim$0.5\,\mum.\\
The improvement in the $o_{xx}$ overlap residual distribution %shown in Figure~\ref{fig:o_xx_SCTB} 
is less pronounced since it is mostly sensitive to the relative alignment of neighbouring modules, which is beyond expectations in the SCT barrel. The overlap residual width improves by about 8\% to $\orsig{xx}\simeq175\,\mum$. This values is mostly driven by the flanks of the distribution, and for $o_{xx}\in[-0.4,\,0.4]\,\mm$ enclosing the peak of the distribution a width of about 110\,\mum\ is achieved. Similarly to the pixel barrel, overlap residual means are significantly away from 0 in all layers;
\item[Pixel End-Caps:]
%The $r_x$ distribution in the end-caps of the pixel detector is shown in Figures~\ref{fig:r_x_PIXA}/\ref{fig:r_x_PIXC} for end-cap~A/C. 
Wide shoulders are observed before alignment in both end-caps, which is due to L1 misalignments with respect to the SCT barrel bracketing them, cf.\!\! Figure~\ref{fig:inDetTechnical} on page~\pageref{fig:inDetTechnical}. End-cap~C displays somewhat more pronounced shoulders than end-cap~A, which is due to a larger misalignment with respect to the SCT, as can be verified from Table~\ref{tab:alignL1}. The shoulders disappear after alignment, and $\rsig x$ improves by 40\% to $\sim\!245\,\mum$ for EC~A and 55\% to $\sim\!185\,\mum$ for EC~C. As detailed in the discussion of L3 alignment in Subsection~\ref{ssec:l3M8}, the smaller $\rsig x$ in EC~C than in EC~A after alignment is likely due to a higher assembly precision and lacking illumination. Residual means of $\rmean x=-4.2\pm2.1$\,/\,$-0.1\pm1.6\,\mum$ are achieved.\\
The overlap residual widths notably improve with alignment: by 25\% to $\orsig{xx}\simeq112\,\mum$ in case of EC~A and by 39\% to $\orsig{xx}\simeq109\,\mum$ in case of EC~C. Coherent deviations of the overlap residual mean from 0 are found in both end-cap~A/C: $\ormean{xx}=28.0\pm3.4\,\mum$\,/\,$22.6\pm3.4\,\mum$. This could be due to a systematic radial displacement of pixel modules towards higher $R$;
%\item[SCT End-Caps:]
\item[SCT end-cap A:] The $r_x$ distribution before alignment % displayed in Figure~\ref{fig:r_x_SCTA}
indicates a reasonable assembly tolerance. The residual means approach 0 to within microns, while the widths refine by a factor of 5\% to circa $\rsig x\simeq171\,\mum$ over the full $r_x\in[-1.5,\,1.5]\,\mm$ range. This refinement is even more pronounced in the peak region. Despite lacking a L3 alignment, the residual widths after alignment are similar to the SCT barrel, which is mostly due to tighter track reconstruction cuts in the end-caps. The overlap residual widths do not change much since no module-to-module alignment was performed for the SCT ECs. Even though the $o_{xx}$ overlap residual means are not consistent with 0 within their errors for most of the disks, no conclusions can be made at this stage, since each of the disks comprises up to three rings;
\item[SCT end-cap C:] The most dramatic improvement in $r_x$ distributions in the process of L2 (only) alignment is undoubtedly found in SCT EC~C. %, as shown in Figure~\ref{fig:r_x_SCTC}. 
Distributions in the six innermost disks, which are initially highly skewed, approach a Gaussian-like shape. The residual means $\rmean x$, which spanned a range of more than 300\,\mum\ before alignment, approach 0 to within about 5\,\mum\ for those disks. It comes as no surspise that the residual widths improve by 35\% to about $\rsig x\simeq166\,\mum$. Note that disk~6 suffers from a cable swap problem~\cite{bib:privateGiorgio}.
\end{description}
The basic statistical quantities for $r_x$ residuals and $o_{xx}$ overlap residuals, cumulatively per-layer of a given subdetector and per-subdetector, are summarised in Tables~\ref{tab:m8plusAlign} and \ref{tab:m8plusAlign_ovres} for the entire M8+ dataset without solenoidal magnetic field.

\begin{table}
\small
\begin{center}
\begin{tabular}{l|rrr|rrr}
\hline
 & \multicolumn{3}{c|}{{\bf Before} alignment} & \multicolumn{3}{c}{{\bf After} alignment}\\
%\hline
 & $\langle r_x\rangle$ & $\delta r_x$ & $\sigma(r_x)$ & $\langle r_x\rangle$ & $\delta r_x$ & $\sigma(r_x)$  \\
\hline\hline
Pixel barrel layer 0	& $-168.25$	& $2.21$	& $619.8$	& $1.86$	& $0.55$	& $177.0$\\
Pixel barrel layer 1	& $-181.09$	& $1.56$	& $586.3$	& $1.43$	& $0.41$	& $177.2$\\
Pixel barrel layer 2	& $-85.50$	& $1.45$	& $634.6$	& $1.66$	& $0.40$	& $198.9$\\
Pixel barrel {\bf (all)}& $-134.18$	& $0.96$	& $617.3$	& $1.62$	& $0.26$	& $187.4$\\
\hline
SCT barrel layer 0	& $-6.75$	& $0.20$	& $283.3$	& $0.22$	& $0.12$	& $180.1$\\
SCT barrel layer 1	& $-7.51$	& $0.15$	& $236.7$	& $-0.67$	& $0.09$	& $158.0$\\
SCT barrel layer 2	& $23.88$	& $0.13$	& $218.1$	& $0.12$	& $0.08$	& $146.9$\\
SCT barrel layer 3	& $-5.44$	& $0.13$	& $224.3$	& $0.16$	& $0.09$	& $172.2$\\
SCT barrel {\bf (all)}	& $1.92$	& $0.07$	& $238.0$	& $-0.04$	& $0.05$	& $163.8$\\
\hline
Pixel EC A layer 0	& $-22.97$	& $5.59$	& $374.6$	& $-11.03$	& $3.42$	& $245.8$\\
Pixel EC A layer 1	& $-15.97$	& $5.35$	& $356.9$	& $-3.07$	& $3.44$	& $245.8$\\
Pixel EC A layer 2	& $-33.21$	& $5.94$	& $315.7$	& $4.89$	& $4.24$	& $242.4$\\
Pixel EC A {\bf (all)}	& $-22.78$	& $3.27$	& $354.6$	& $-4.18$	& $2.11$	& $245.1$\\
\hline
Pixel EC C layer 0	& $-3.98$	& $7.27$	& $443.9$	& $-0.49$	& $2.89$	& $188.2$\\
Pixel EC C layer 1	& $-2.39$	& $6.78$	& $414.8$	& $-0.59$	& $2.82$	& $184.7$\\
Pixel EC C layer 2	& $20.52$	& $5.30$	& $350.3$	& $0.59$	& $2.55$	& $181.2$\\
Pixel EC C {\bf (all)}	& $5.57$	& $3.70$	& $402.2$	& $-0.12$	& $1.58$	& $184.5$\\
\hline
SCT EC A layer 0	& $-6.25$	& $0.68$	& $213.0$	& $-2.70$	& $0.61$	& $201.7$\\
SCT EC A layer 1	& $12.38$	& $0.58$	& $192.9$	& $-2.37$	& $0.52$	& $184.2$\\
SCT EC A layer 2	& $-30.05$	& $0.56$	& $167.4$	& $-2.67$	& $0.49$	& $153.8$\\
SCT EC A layer 3	& $12.68$	& $0.60$	& $159.0$	& $-3.51$	& $0.53$	& $150.6$\\
SCT EC A layer 4	& $1.22$	& $0.64$	& $157.0$	& $1.65$	& $0.57$	& $148.3$\\
SCT EC A layer 5	& $-19.75$	& $0.96$	& $159.5$	& $-1.67$	& $0.84$	& $150.3$\\
SCT EC A layer 6	& $14.32$	& $1.47$	& $168.7$	& $-1.12$	& $1.31$	& $159.7$\\
SCT EC A layer 7	& $-5.23$	& $2.27$	& $184.0$	& $-5.15$	& $2.09$	& $180.1$\\
SCT EC A layer 8	& $-12.65$	& $3.25$	& $193.3$	& $-8.28$	& $2.99$	& $188.2$\\
SCT EC A {\bf (all)}	& $-2.86$	& $0.26$	& $182.0$	& $-2.17$	& $0.23$	& $171.4$\\
\hline
SCT EC C layer 0	& $56.84$	& $0.77$	& $220.8$	& $-4.84$	& $0.53$	& $162.7$\\
SCT EC C layer 1	& $-57.99$	& $0.63$	& $210.5$	& $-5.97$	& $0.44$	& $159.4$\\
SCT EC C layer 2	& $25.01$	& $0.73$	& $219.5$	& $1.15$	& $0.48$	& $153.1$\\
SCT EC C layer 3	& $-159.77$	& $1.03$	& $262.4$	& $5.13$	& $0.57$	& $155.8$\\
SCT EC C layer 4	& $201.15$	& $1.31$	& $311.3$	& $14.39$	& $0.62$	& $159.7$\\
SCT EC C layer 5	& $-42.70$	& $1.37$	& $219.9$	& $-1.35$	& $1.04$	& $181.0$\\
SCT EC C layer 6	& $-6.30$	& $4.61$	& $380.8$	& $-2.90$	& $4.45$	& $385.3$\\
SCT EC C layer 7	& $6.97$	& $3.43$	& $234.0$	& $-10.27$	& $2.76$	& $204.0$\\
SCT EC C layer 8	& $23.21$	& $5.74$	& $263.7$	& $35.14$	& $4.62$	& $229.8$\\
SCT EC C {\bf (all)}	& $0.41$	& $0.39$	& $262.1$	& $0.41$	& $0.23$	& $166.3$\\
\hline
\end{tabular}
\caption[Main residual characteristics for the silicon tracker by layers in M8+ before and after alignment]{\label{tab:m8plusAlign}
Main {\bf residual} characteristics for the silicon tracker by layers in M8+ with \mbox{$B$-{\bf field off}} {\em before} and {\em after} the full alignment procedure with the \RA\ algorithm: the residual mean $\rmean x$, the uncertainty on the residual mean $\delta r_x$, and the standard deviation of the residual $\sigma(r_x)$. The values are given per-layer of a given subdetector, or per-subdetector in lines marked with ``(all)''. The range used for the calculation of the quantities above is $r_x\in[-1.5\,\mm,\,1.5\,\mm]$. All values are given in $\mum$.
}
\end{center}
\end{table}

\begin{table}
\small
\begin{center}
\begin{tabular}{l|rrr|rrr}
\hline
 & \multicolumn{3}{c|}{{\bf Before} alignment} & \multicolumn{3}{c}{{\bf After} alignment}\\
%\hline
\hline
	& $\langle o_{xx}\rangle$	& $\delta o_{xx}$	& $\sigma(o_{xx})$	& $\langle o_{xx}\rangle$	& $\delta o_{xx}$	& $\sigma(o_{xx})$ \\
\hline\hline
Pixel barrel layer 0	& $56.11$	& $3.11$	& $296.6$	& $13.29$	& $0.81$	& $94.6$\\
Pixel barrel layer 1	& $40.51$	& $2.03$	& $269.2$	& $-1.38$	& $0.62$	& $98.8$\\
Pixel barrel layer 2	& $42.41$	& $1.65$	& $250.0$	& $11.66$	& $0.54$	& $97.6$\\
Pixel barrel {\bf (all)}& $44.25$	& $1.19$	& $266.0$	& $7.36$	& $0.37$	& $97.7$\\
\hline
SCT barrel layer 0	& $7.46$	& $0.38$	& $175.8$	& $3.97$	& $0.34$	& $164.8$\\
SCT barrel layer 1	& $2.86$	& $0.33$	& $180.9$	& $4.37$	& $0.29$	& $169.9$\\
SCT barrel layer 2	& $4.43$	& $0.30$	& $191.5$	& $4.93$	& $0.26$	& $173.9$\\
SCT barrel layer 3	& $-19.22$	& $0.29$	& $199.7$	& $-18.29$	& $0.25$	& $182.8$\\
SCT barrel {\bf (all)}	& $-3.44$	& $0.16$	& $190.1$	& $-3.18$	& $0.14$	& $175.1$\\
\hline
Pixel EC A layer 0	& $26.81$	& $8.99$	& $172.5$	& $31.46$	& $6.31$	& $129.8$\\
Pixel EC A layer 1	& $29.79$	& $7.09$	& $138.5$	& $28.13$	& $4.91$	& $102.3$\\
Pixel EC A layer 2	& $14.67$	& $8.71$	& $130.6$	& $22.20$	& $5.96$	& $96.8$\\
Pixel EC A {\bf (all)}	& $25.17$	& $4.83$	& $150.7$	& $27.99$	& $3.36$	& $112.4$\\
\hline
Pixel EC C layer 0	& $56.76$	& $9.56$	& $168.9$	& $20.39$	& $5.76$	& $108.9$\\
Pixel EC C layer 1	& $67.81$	& $11.58$	& $188.8$	& $19.89$	& $5.04$	& $88.5$\\
Pixel EC C layer 2	& $47.03$	& $9.84$	& $180.9$	& $26.74$	& $6.19$	& $122.5$\\
Pixel EC C {\bf (all)}	& $56.38$	& $5.93$	& $179.5$	& $22.60$	& $3.35$	& $108.9$\\
\hline
SCT EC A layer 0	& $22.75$	& $3.59$	& $267.2$	& $20.53$	& $3.12$	& $248.8$\\
SCT EC A layer 1	& $8.41$	& $3.12$	& $246.6$	& $8.29$	& $2.82$	& $237.1$\\
SCT EC A layer 2	& $14.28$	& $3.03$	& $229.1$	& $14.69$	& $2.75$	& $219.8$\\
SCT EC A layer 3	& $10.04$	& $2.90$	& $200.2$	& $9.00$	& $2.72$	& $198.7$\\
SCT EC A layer 4	& $16.25$	& $3.87$	& $249.7$	& $19.64$	& $3.69$	& $254.7$\\
SCT EC A layer 5	& $26.28$	& $6.82$	& $318.0$	& $29.65$	& $6.46$	& $326.8$\\
SCT EC A layer 6	& $36.20$	& $8.37$	& $275.0$	& $33.28$	& $7.96$	& $276.2$\\
SCT EC A layer 7	& $-2.51$	& $12.48$	& $289.4$	& $-13.82$	& $10.78$	& $266.2$\\
SCT EC A layer 8	& $52.29$	& $16.34$	& $326.3$	& $72.52$	& $17.57$	& $363.1$\\
SCT EC A {\bf (all)}	& $16.04$	& $1.43$	& $250.2$	& $16.21$	& $1.32$	& $245.2$\\
\hline
SCT EC C layer 0	& $20.53$	& $4.40$	& $303.6$	& $16.28$	& $3.77$	& $278.7$\\
SCT EC C layer 1	& $7.37$	& $2.96$	& $243.2$	& $5.67$	& $2.54$	& $222.6$\\
SCT EC C layer 2	& $1.02$	& $3.47$	& $257.2$	& $3.51$	& $3.07$	& $245.2$\\
SCT EC C layer 3	& $0.16$	& $4.03$	& $271.4$	& $-3.59$	& $2.94$	& $213.5$\\
SCT EC C layer 4	& $20.80$	& $4.37$	& $280.5$	& $25.72$	& $3.75$	& $261.2$\\
SCT EC C layer 5	& $13.30$	& $6.85$	& $308.7$	& $17.23$	& $6.22$	& $298.3$\\
SCT EC C layer 6	& $176.38$	& $57.03$	& $522.7$	& $155.70$	& $48.83$	& $483.4$\\
SCT EC C layer 7	& $78.10$	& $24.56$	& $499.6$	& $78.96$	& $23.42$	& $505.0$\\
SCT EC C layer 8	& $31.66$	& $26.99$	& $368.1$	& $40.99$	& $21.91$	& $340.1$\\
SCT EC C {\bf (all)}	& $11.27$	& $1.66$	& $279.2$	& $11.05$	& $1.41$	& $255.7$\\
\hline
\end{tabular}
\caption[Main overalap residual characteristics for the silicon tracker by layers in M8+ before and after alignment]{\label{tab:m8plusAlign_ovres}
Main {\bf overlap residual} characteristics for the silicon tracker by layers in M8+ with $B$-{\bf field off} {\em before} and {\em after} the full alignment procedure with the \RA\ algorithm: the residual mean $\ormean{xx}$, the uncertainty on the residual mean $\delta o_{xx}$, and the standard deviation of the residual $\orsig{xx}$. The values are given per-layer of a given subdetector, or per-subdetector in lines marked with ``(all)''. The range (implicitly) used for the calculation of the quantities above is $r_x\in[-1.5\,\mm,\,1.5\,\mm]$. All values are given in $\mum$.
}
\end{center}
\end{table}
\clearpage

\begin{figure}
\begin{center}
\vspace{\cDistHalf}
\includegraphics[height=15.6cm,width=9.5cm,angle=-90,clip=true]{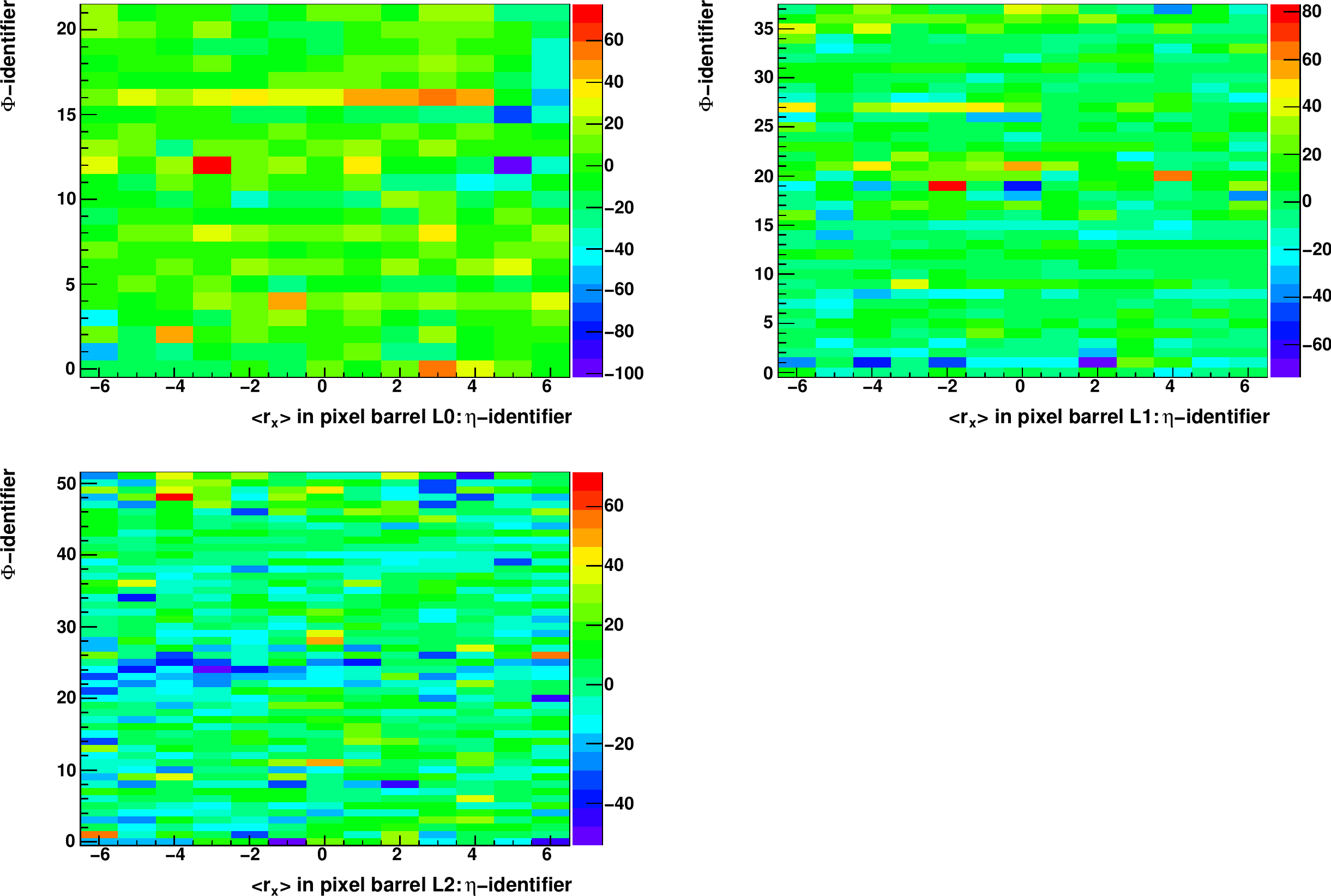}
\vspace{\cDistHalf}
\end{center}
\caption[The per-layer $\rmean x(\eta,\,\Phi)$ distribution for the barrel of the pixel detector in M8+ $B$-field off cosmic ray data after alignment]{\label{fig:mean_rx_PIXB}
The per-layer $\rmean x(\eta,\,\Phi)$ distribution for the {\bf barrel} of the {\bf pixel} detector in M8+ $B$-{\bf field off} cosmic ray data {\em after} alignment. Modules which did not collect any hits are shown with $\rmean x\equiv0$. The distribution is very uniform in well-illuminated regions around $\Phi=\oneOverTwo\pi,\,\xOverY32\pi$. Note the small range of the $z$-axis, which is mostly determined by few `extreme' modules with large $\rsig x$ values, as can be seen from Figure~\ref{fig:pull_rx_PIXB}. Stave~16 in layer~0 displays a weak parabolic dependence. However, its modules are well below the $\sigmaCut=0.7$ value used.
}
\end{figure}

\begin{figure}
\begin{center}
\vspace{\cDistHalf}
\includegraphics[height=15.6cm,width=9.5cm,angle=-90,clip=true]{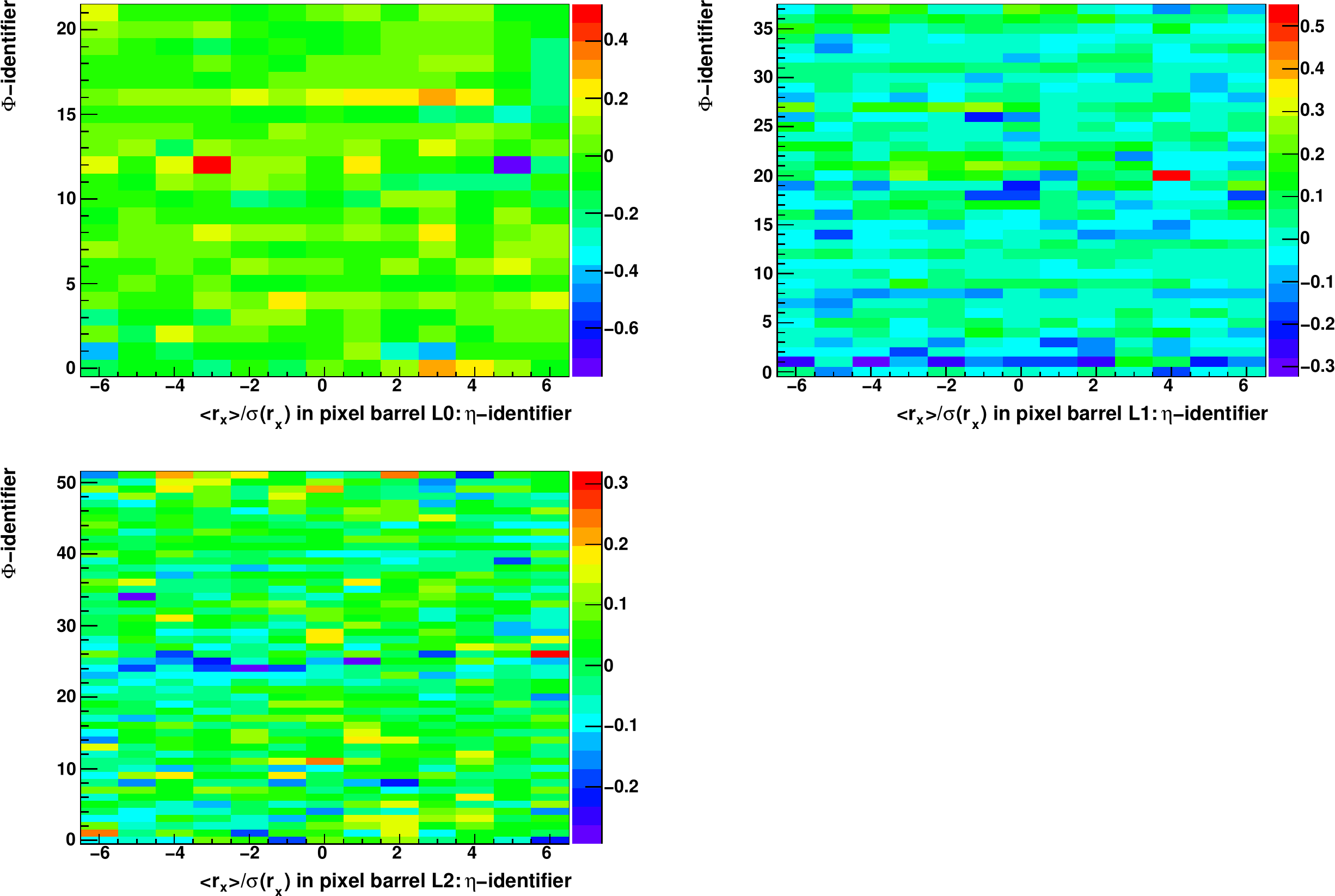}
\vspace{\cDistHalf}
\end{center}
\caption[The per-layer $\frac{\rmean x}{\rsig x}(\eta,\,\Phi)$ pull distribution for the barrel of the pixel detector in M8+ $B$-field off cosmic ray data after alignment]{\label{fig:pull_rx_PIXB}
The per-layer $\frac{\rmean x}{\rsig x}(\eta,\,\Phi)$ pull distribution for the {\bf barrel} of the {\bf pixel} detector in M8+ $B$-{\bf field off} cosmic ray data {\em after} alignment. Modules which did not collect any hits are shown with $\frac{\rmean x}{\rsig x}\equiv0$. The distribution is very uniform, with some `extreme' modules typically in not well-illuminated regions. Note the small range of the $z$-axis.
\vspace{\cDist}
}
\end{figure}

\begin{figure}
\begin{center}
\vspace{\cDistHalf}
\includegraphics[height=15.6cm,width=9.5cm,angle=-90,clip=true]{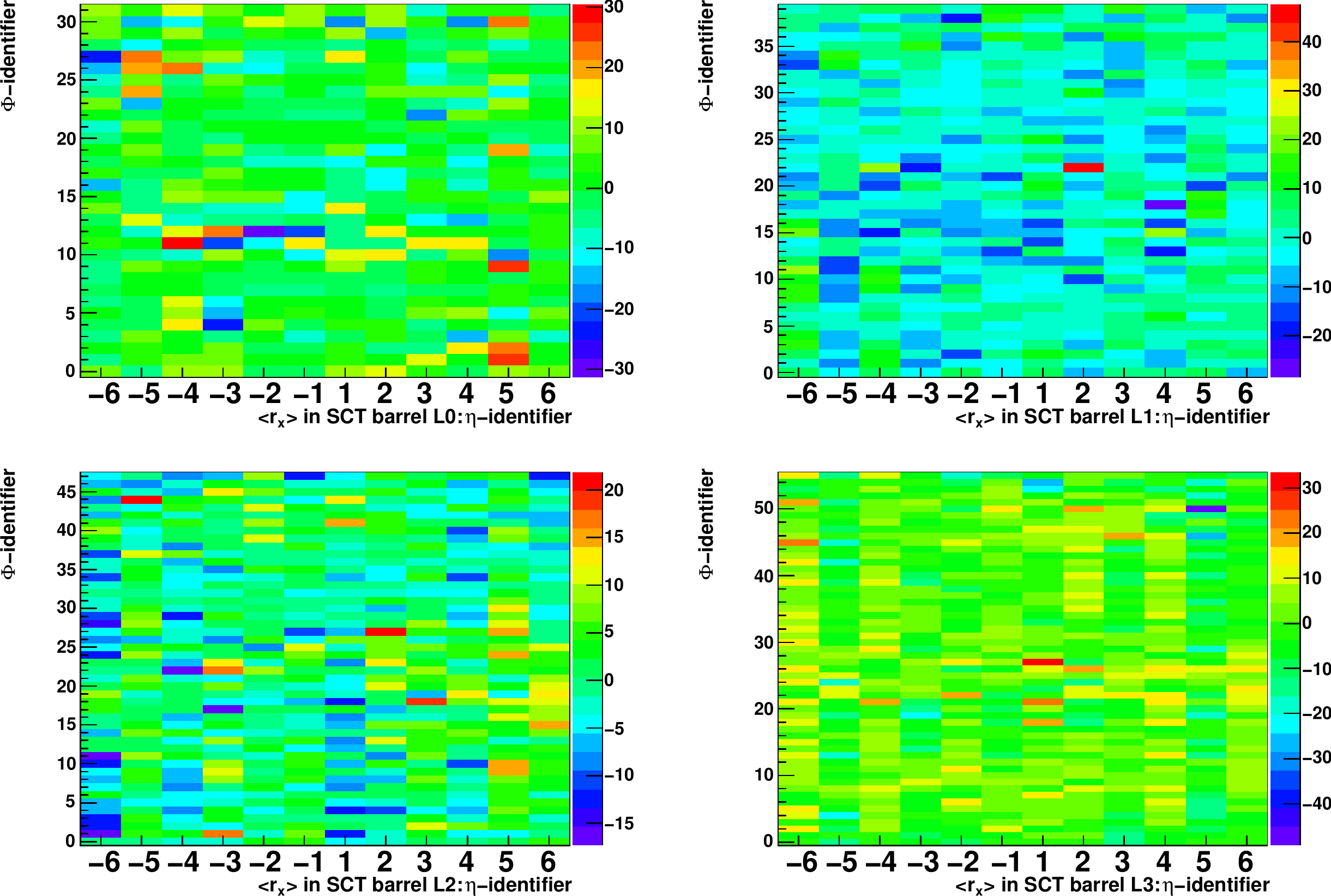}
\vspace{\cDistHalf}
\end{center}
\caption[The per-layer $\rmean x(\eta,\,\Phi)$ distribution for the barrel of the SCT detector in M8+ $B$-field off cosmic ray data after alignment]{\label{fig:mean_rx_SCTB}
The per-layer $\rmean x(\eta,\,\Phi)$ distribution for the {\bf barrel} of the {\bf SCT} detector in M8+ $B$-{\bf field off} cosmic ray data {\em after} alignment. Modules which did not collect any hits are shown with $\rmean x\equiv0$. The distribution is very uniform in well-illuminated regions around $\Phi=\oneOverTwo\pi,\,\xOverY32\pi$. Note the small range of the $z$-axis. %, which is mostly determined by few `extreme' modules with large $\rsig x$ values, as can be seen from Figure~\ref{fig:pull_rx_SCTB}. 
A barely pronounced alternation between rings of the same layer with neighbouring $\Phi$-identifiers is visible. It is well below the alignment threshold for single modules. This might justify a coherent ring alignment in the SCT barrel.
}
\end{figure}

\begin{figure}
\begin{center}
\vspace{\cDistHalf}
\includegraphics[height=15.6cm,width=9.5cm,angle=-90,clip=true]{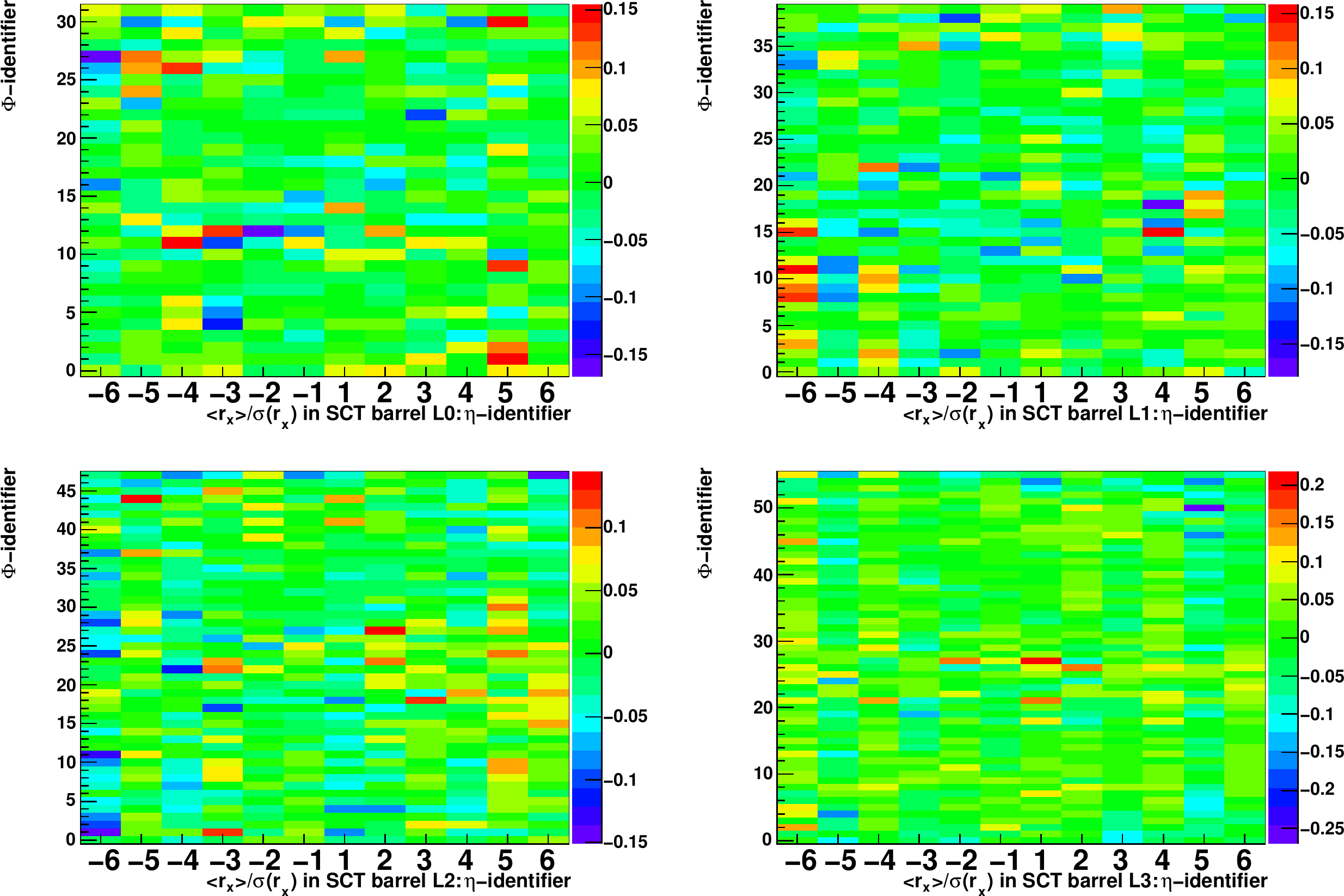}
\vspace{\cDistHalf}
\end{center}
\caption[The per-layer $\frac{\rmean x}{\rsig x}(\eta,\,\Phi)$ pull distribution for the barrel of the SCT detector in M8+ $B$-field off cosmic ray data after alignment]{\label{fig:pull_rx_SCTB}
The per-layer $\frac{\rmean x}{\rsig x}(\eta,\,\Phi)$ pull distribution for the {\bf barrel} of the {\bf SCT} detector in M8+ $B$-{\bf field off} cosmic ray data {\em after} alignment. Modules which did not collect any hits are shown with $\frac{\rmean x}{\rsig x}\equiv0$. The distribution is very uniform, with some `extreme' modules typically in not well-illuminated regions. Note the small range of the $z$-axis.
\vspace{\cDist}
}
\end{figure}

\begin{figure}
\begin{center}
\vspace{\cDistHalf}
\includegraphics[height=15.6cm,width=9.5cm,angle=-90,clip=true]{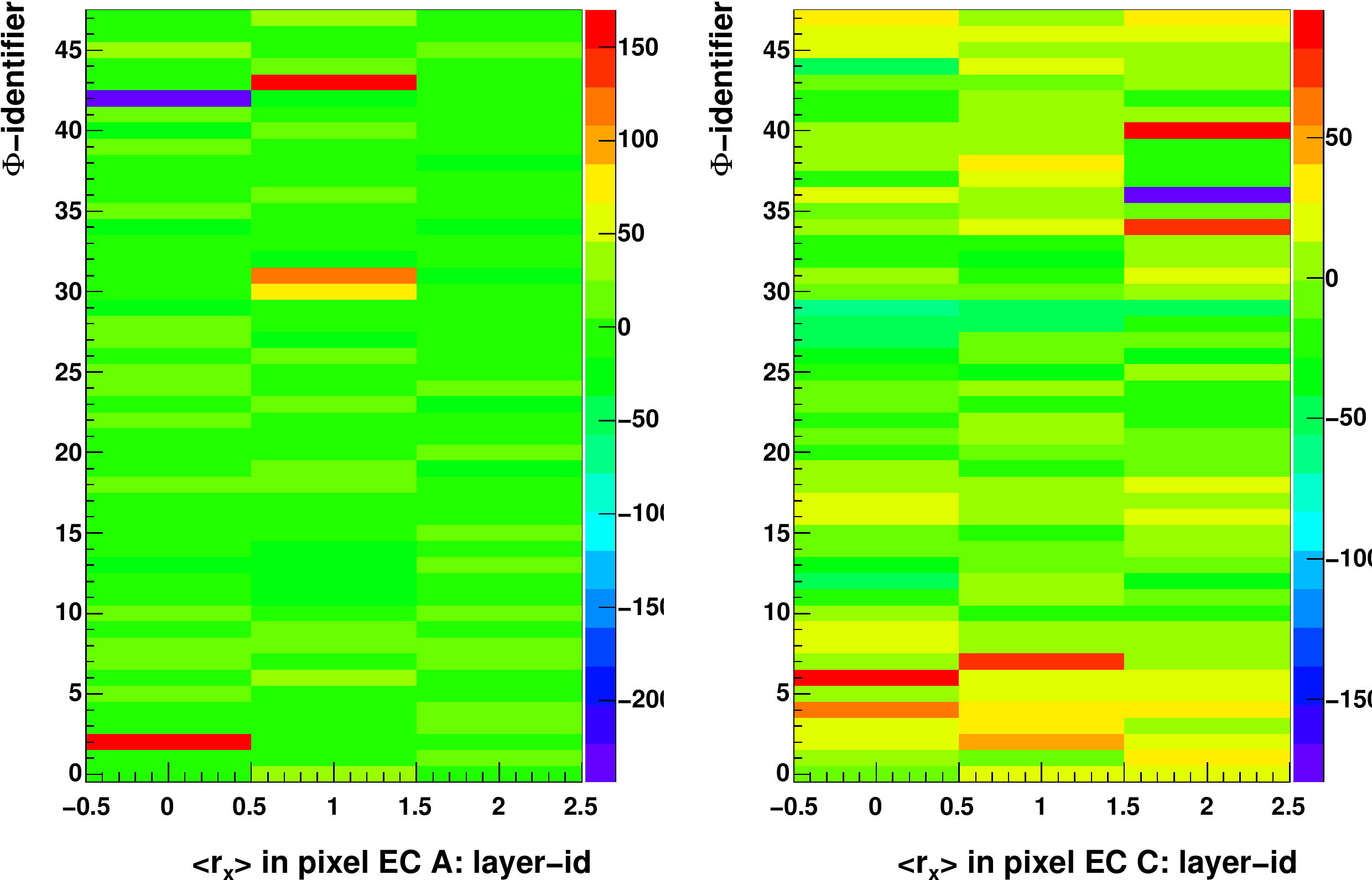}
\vspace{\cDistHalf}
\end{center}
\caption[The per-layer $\rmean x(\eta,\,\Phi)$ distribution for the end-caps of the pixel detector in M8+ $B$-field off cosmic ray data after alignment]{\label{fig:mean_rx_PIXE}
The per-layer $\rmean x(\eta,\,\Phi)$ distribution for the {\bf end-caps} of the {\bf pixel} detector in M8+ $B$-{\bf field off} cosmic ray data {\em after} alignment. Modules which did not collect any hits are shown with $\rmean x\equiv0$. The distribution is very uniform with few `extreme' modules. These fall below the alignment threshold $\sigmaCut=0.7$ because of their large $\rsig x$ values, as can be seen from Figure~\ref{fig:pull_rx_PIXE}.
}
\end{figure}

\begin{figure}
\begin{center}
\vspace{\cDistHalf}
\includegraphics[height=15.6cm,width=9.5cm,angle=-90,clip=true]{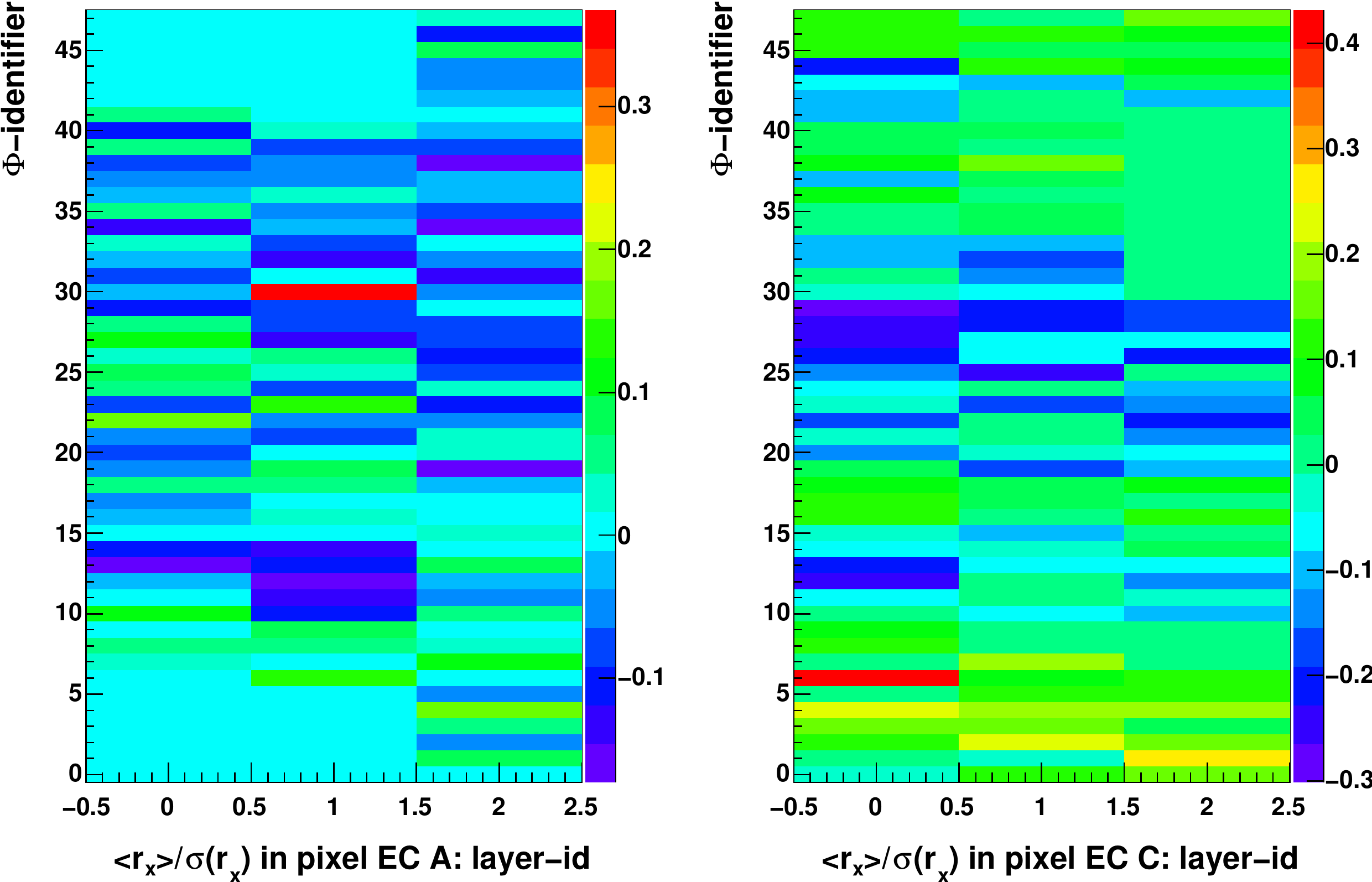}
\vspace{\cDistHalf}
\end{center}
\caption[The per-layer $\frac{\rmean x}{\rsig x}(\eta,\,\Phi)$ pull distribution for the end-caps of the pixel detector in M8+ $B$-field off cosmic ray data after alignment]{\label{fig:pull_rx_PIXE}
The per-layer $\frac{\rmean x}{\rsig x}(\eta,\,\Phi)$ pull distribution for the {\bf end-caps} of the {\bf pixel} detector in M8+ $B$-{\bf field off} cosmic ray data {\em after} alignment. Modules which did not collect any hits are shown with $\frac{\rmean x}{\rsig x}\equiv0$. Note the small range of the $z$-axis.
\vspace{\cDist}
}
\end{figure}

The maps of residual means $\rmean x$ and pulls $\rpull x$ versus $\eta$ and $\Phi$-identifiers of modules are shown by layers on pages~\pageref{fig:mean_rx_PIXB}--\pageref{fig:pull_rx_PIXE} {\em after} alignment. They are briefly discussed below:
\begin{description}
\item[Pixel Barrel:] The $\rmean x(\eta,\Phi)$ maps displayed in Figure~\ref{fig:mean_rx_PIXB} are fairly uniform, especially in well-illuminated regions around $\Phi=\xOverY12\pi,\,\xOverY32\pi$. The $z$-range is relatively small in all layers and is driven by few modules with $|\rmean x|$ of $\order{50\,\mum}$. A comparison with the corresponding $\rpull x(\eta,\Phi)$ maps in Figure~\ref{fig:pull_rx_PIXB} reveals that these modules typically have high residual pulls. However, none of them fulfills the condition\footnote{The satisfaction of this condition {\em and} $|\rmean x|\simeq50\,\mum$ trivially implies a large (statistical) error.} $|\rpull x|>\sigmaCut\simeq0.7$ and is therefore not further aligned at L3 to avoid oscillatory or even chaotic behaviour. Stave 16 in layer~0 displays what seems to be a weak parabolic dependence. This is not an artifact of a failed pixel stave bow alignment, as confirmed by the corresponding $\rmean{x}(\Phi)$ distribution in Figure~\ref{fig:appL4p2after};
\item[SCT Barrel:] The $z$-ranges of $\rmean x(\eta,\Phi)$ maps presented in Figure~\ref{fig:mean_rx_SCTB} are  notably smaller of $\order{30\,\mum}$ than in case of the pixel barrel. Again, modules with high $|\rmean x|$ values tend to have large errors, as can be verified with the corresponding $\rpull x(\eta,\Phi)$ distribution displayed in Figure~\ref{fig:pull_rx_SCTB}. In both maps a slightly pronounced stripe structure is observed for odd/even $\eta$-identifiers. Averaged over one ring, the magnitude of the alternation typically stays within some microns. This is a remainder of initial odd/even ring-to-ring misalignments of 17\,\mum\ on average due to modules not fulfilling the condition to be aligned. The stripe structure is expected to disappear with more data.  Note the small $z$-range of the $\rpull x(\eta,\Phi)$ map.
\item[Pixel End-Caps:] Figure~\ref{fig:mean_rx_PIXE} demonstrates that the fluctuations of residual means in pixel end-cap modules are much larger than for the barrel region. Simultaneously, the $z$-range of the $\rmean x(\eta,\Phi)$ map is also notably larger. Both are due to statistical limitations, as can be inferred from the corresponding $\rpull x(\eta,\Phi)$ map  in Figure~\ref{fig:pull_rx_PIXE}. None of the modules satisfies the $|\rpull x|>\sigmaCut\equiv0.7$ cut to be further aligned.
\item[SCT End-Caps:]
Since no L3 alignment was performed for the SCT end-caps, their distributions do not change much in the process of alignment\footnote{Of course, with the exception of the absolute per-disk scale.} and are therefore not shown here.
\end{description}

Summa summarum, a very high degree of alignment precision of the silicon tracker has been achieved with the \RA\ algorithm using M8+ cosmic ray data. A credible set of alignment constants has been provided~\cite{bib:raConst, bib:raConstDB}. The validation of the alignment constant set with official ATLAS monitoring tools can be found in the following Section~\ref{sec:validationM8}.

\pagebreak
%\nopagebreak[5]

%\nopagebreak[5]
\clearpage

%\nopagebreak[5]

%\nopagebreak[5]
\clearpage

%\nopagebreak[5]

%\nopagebreak[5]
\clearpage

%% file: M8plus/Validation.tex
The alignment results for the silicon tracker obtained with the \RA\ algorithm were validated and compared against the alignment constant sets derived with the \GX\ and \LX\ algorithms. These checks were performed using the official ATLAS monitoring tools, and the results were presented to the alignment working group~\cite{bib:validation}. A selection of the plots is shown below. Their discussion should be prefaced by the disclaimer that the \LX\ results were obtained with an \Athena\ release which is known to contain a software bug affecting the cluster formation in the pixel detector.

Clearly, any monitoring results related to track reconstruction and thus residuals will strongly depend on the dataset and the reconstruction cuts used. The monitoring distributions are shown for five M8+ runs with $B$-field on: 91885, 91888, 91890, 91891, and 91900. These representative runs make up for about 25\% of the dataset with the solenoid on. {\tt NewTracking} and release 14.5.2.1 of \Athena\ are used. To reduce the amount of Coulomb multiple scattering, tracks with 
\[\pt>1\,\GeV\]
are analysed. Further, cuts on the transverse and longitudinal impact parameters are applied:
\begin{eqnarray*}
 |d_0| &<& 50\,\mm\\
 |z_0| &<& 400\,\mm\,,
\end{eqnarray*}
in other words tracks are required to go through the $b$-layer of the pixel detector. 
%Finally, 8 or more hits in the silicon tracker are required. 
All fits are performed with one single Gaussian. Its mean $\mu$ and sigma $\sigma$ parameters are reported on the histograms.

\begin{figure}
\begin{center}
\vspace{\cDistHalf}
\includegraphics[width=7.9cm,clip=true]{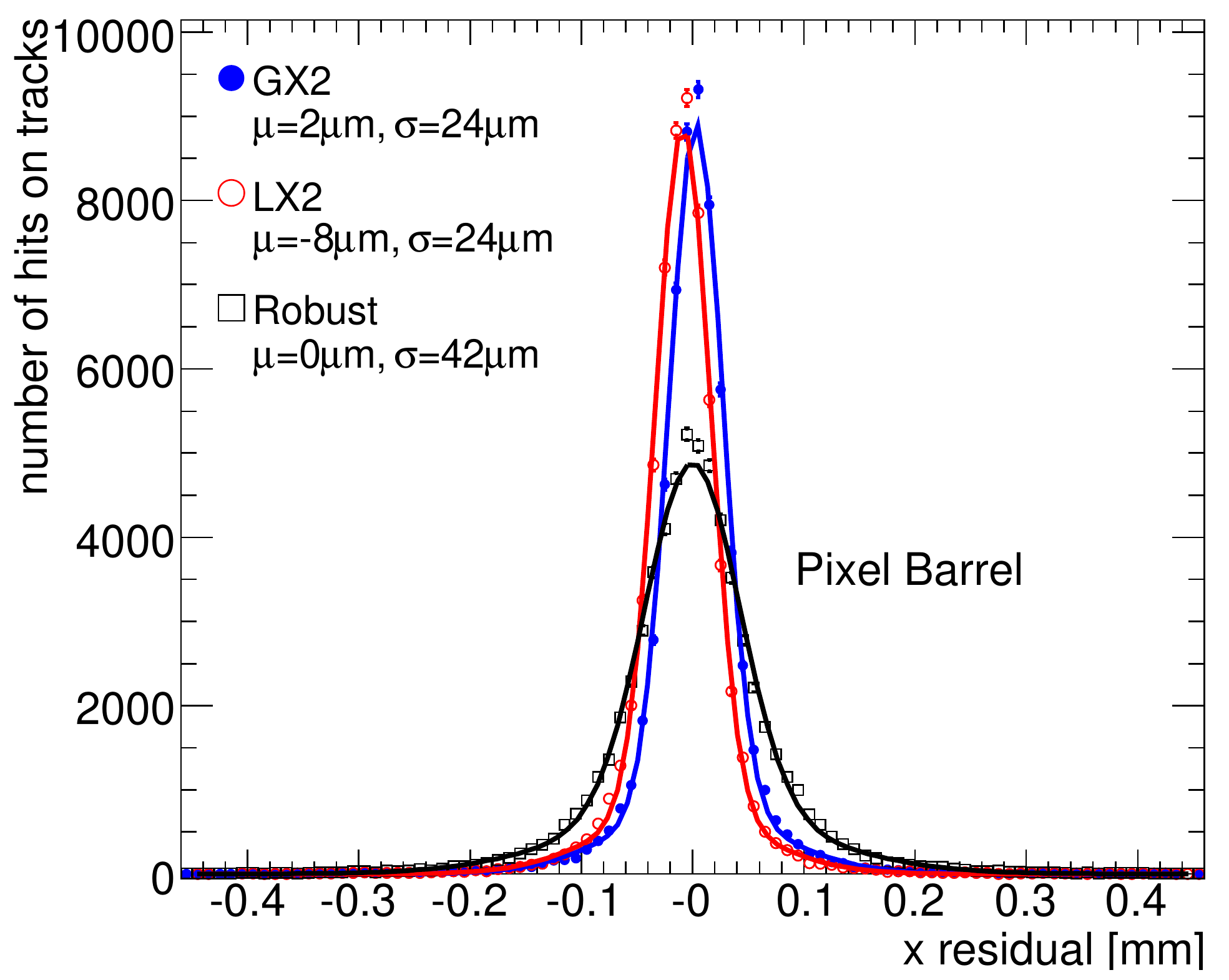}
\includegraphics[width=7.9cm,clip=true]{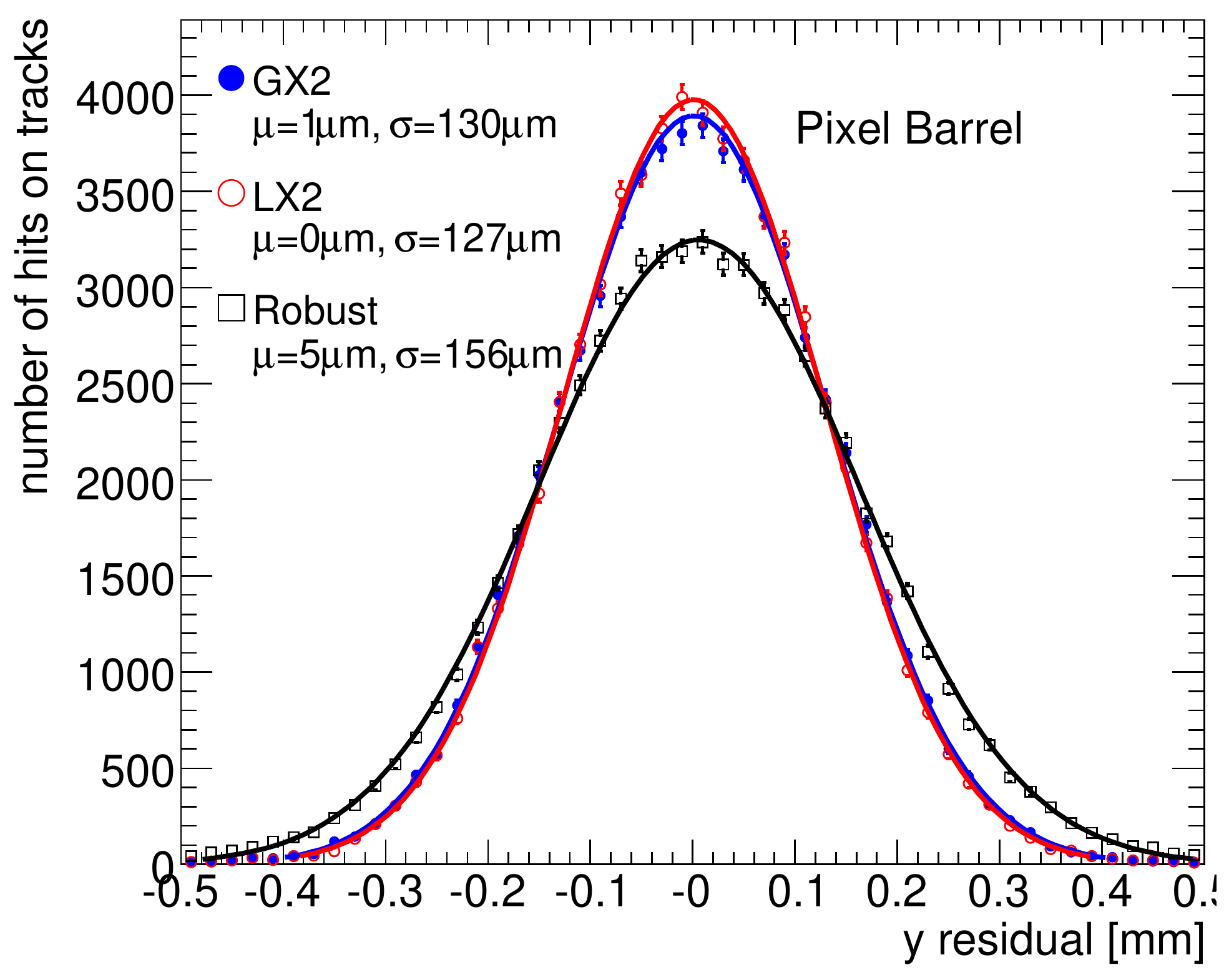}
\vspace{\cDist}
\vspace{\cDist}
%\vspace{-0.2cm}
\end{center}
\caption[The $r_x$, $r_y$ residual distributions in the barrel of the pixel detector after {\it all} alignment corrections with the three alignment algorithms in M8+ using official monitoring.]{\label{fig:r_PIX_mon}
The $r_x$, $r_y$ residual distributions in the barrel of the pixel detector after {\it all} alignment corrections with the \GX\ (GX2), \LX\ (LX2), and \RA~(Robust)  algorithms in M8+ with official monitoring. Plots courtesy T.~Golling.
}
\end{figure}%\nopagebreak[5]

\begin{figure}
\begin{center}
\includegraphics[width=7.9cm,clip=true]{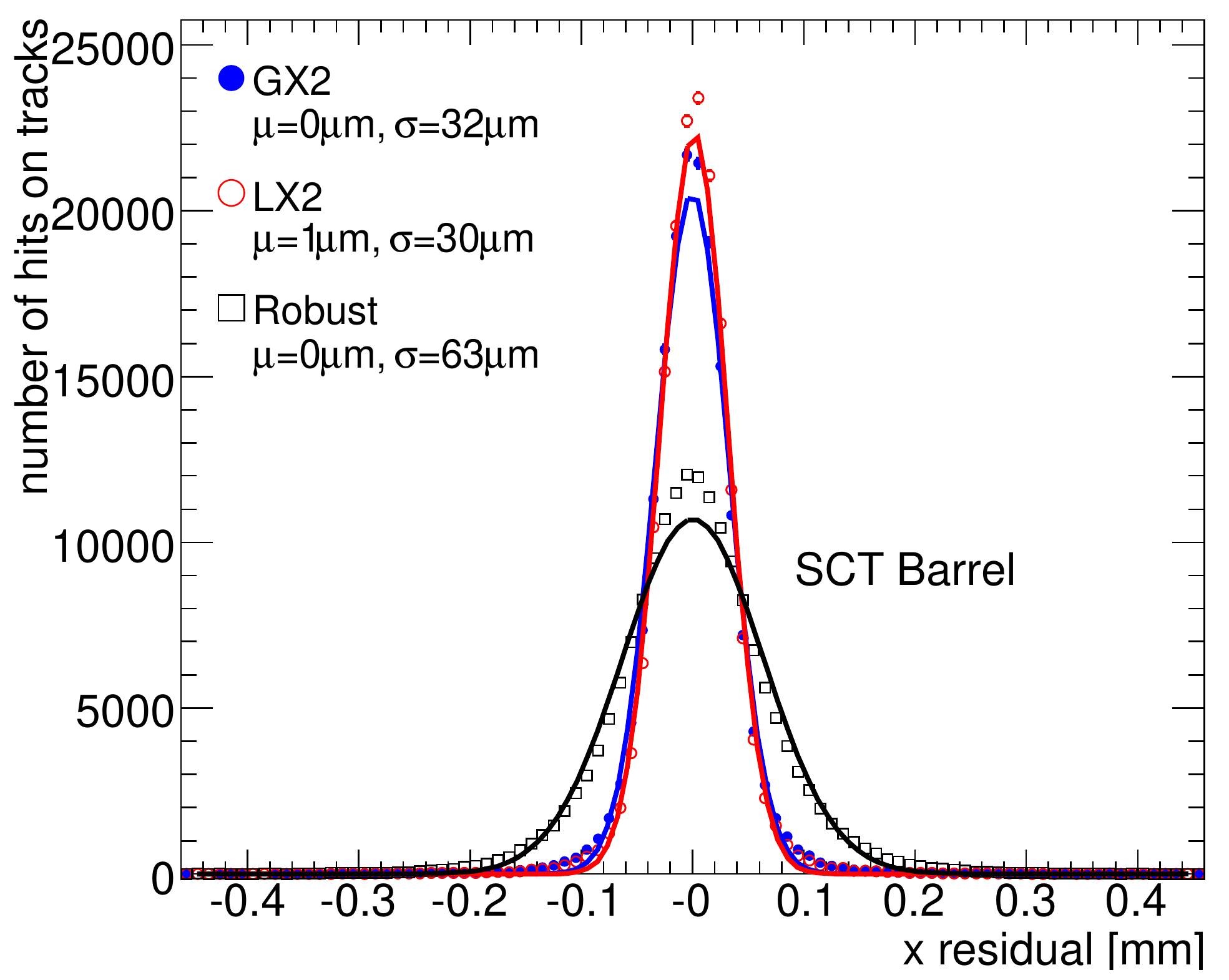}
\vspace{\cDist}
%\vspace{-0.2cm}
\end{center}
\caption[The $r_x$ residual distributions in the barrel of the SCT detector after {\it all} alignment corrections with the three alignment algorithms in M8+ using official monitoring.]{\label{fig:r_SCT_mon}
The $r_x$ residual distribution in the barrel of the SCT detector after {\it all} alignment corrections with the \GX\ (GX2), \LX\ (LX2), and \RA\ (Robust)  algorithms in M8+ with official monitoring. Plot courtesy T.~Golling.
}
\end{figure}%\nopagebreak[5]

The $r_x$ and $r_y$ residual distribution in the barrel of the pixel detector are shown in Figure~\ref{fig:r_PIX_mon}. Figure~\ref{fig:r_SCT_mon} shows the $r_x$ distribution for the barrel of the SCT. The shape of all distributions for all three algorithms is well-approximated by a Gaussian. For all three plots the $\sigma$-parameter of the Gaussian found for the \RA\ constants is somewhat larger than for the $\chisq{\mbox{-}\rm based}$ algorithms. This difference is mostly due to the lack of alignment for the local~$\gamma$ DoF in \RA: it is a ``strong'' alignment parameter in the sense that it describes an {\em in-plane} rotation of the sensor, and thus has a strong effect on the residuals. Thus, its alignment can dramatically refine the width of the residual distributions which can be detrimentally affected by a limited module mounting precision. Meanwhile, the local~$\gamma$ DoF has been incorporated in the \RA\ algorithm, and results compatible with the \GX\ and \LX\ alignment algorithms have been obtained~\cite{bib:localGammaRA}. It is remarkable that \RA\ is the only alignment approach with the mean of the $r_x$ distribution consistent with 0 for both subdetectors. This is an indication that the algorithm is working correctly and delivering credible and reliable results. Moreover, this finding somewhat supports the claim that the lack of alignment for the $\gamma$ rotation of individual modules is the major, if not the only limiting factor precluding \RA\ to enter a direct competition with the \chisq-based algorithms. Indeed, randomly distributed local~$\gamma$ misalignments in an otherwise perfectly aligned detector will not affect the means of the distributions, but merely their widths. Studies with simulated Monte Carlo events reconstructed with perfectly known detector geometry result in $\mu_{r_x}=0\,\mum$, $\sigma_{r_x}=16\,\mum$ and $\mu_{r_y}=2\,\mum$, $\sigma_{r_y}=127\,\mum$ for the barrel of the pixel detector, while $\mu_{r_x}=0\,\mum$, $\sigma_{r_x}=24\,\mum$ are obtained for the  barrel of the SCT.

\begin{figure}
\begin{center}
\vspace{\cDistHalf}
\includegraphics[width=7.9cm,clip=true]{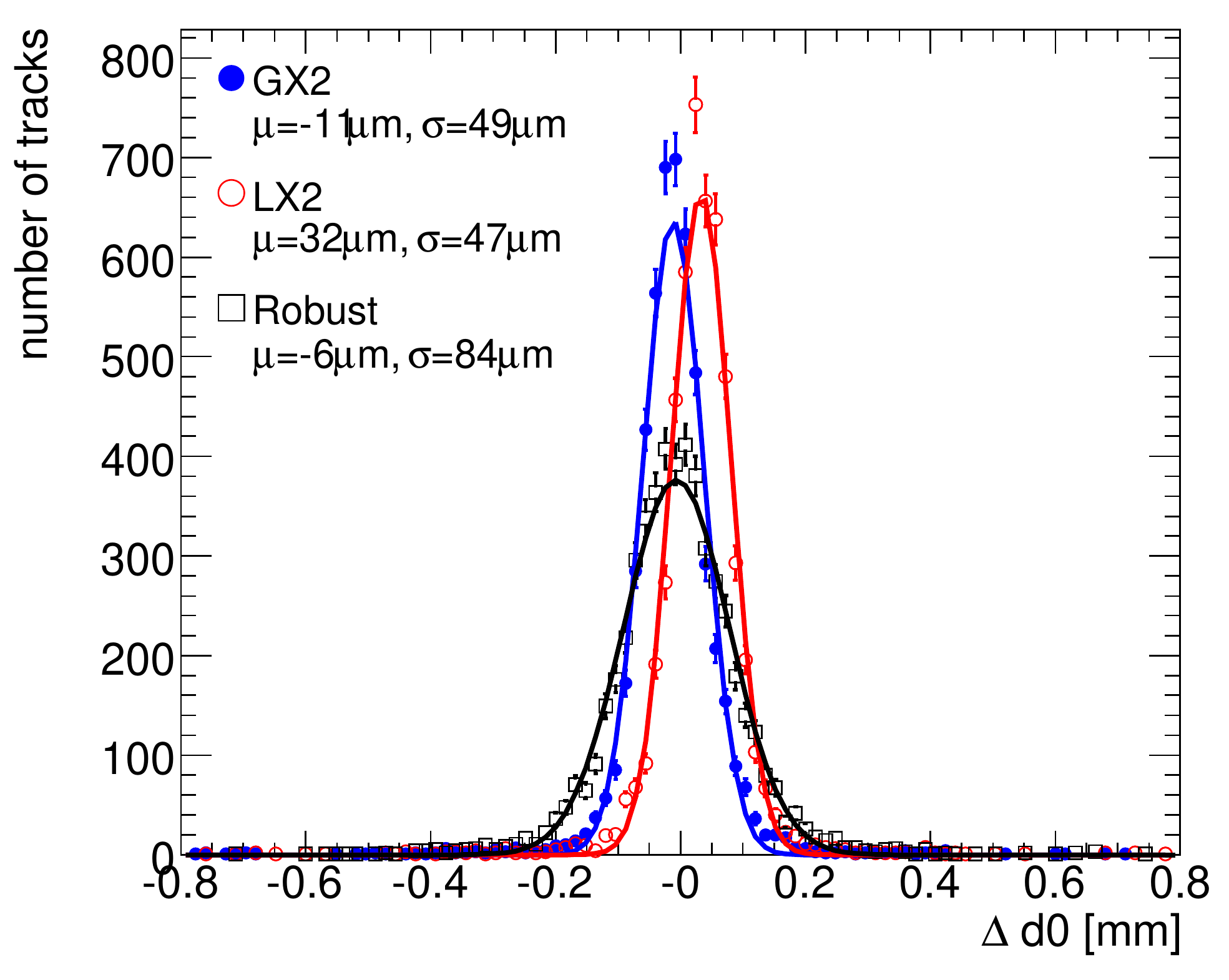}
\includegraphics[width=7.9cm,clip=true]{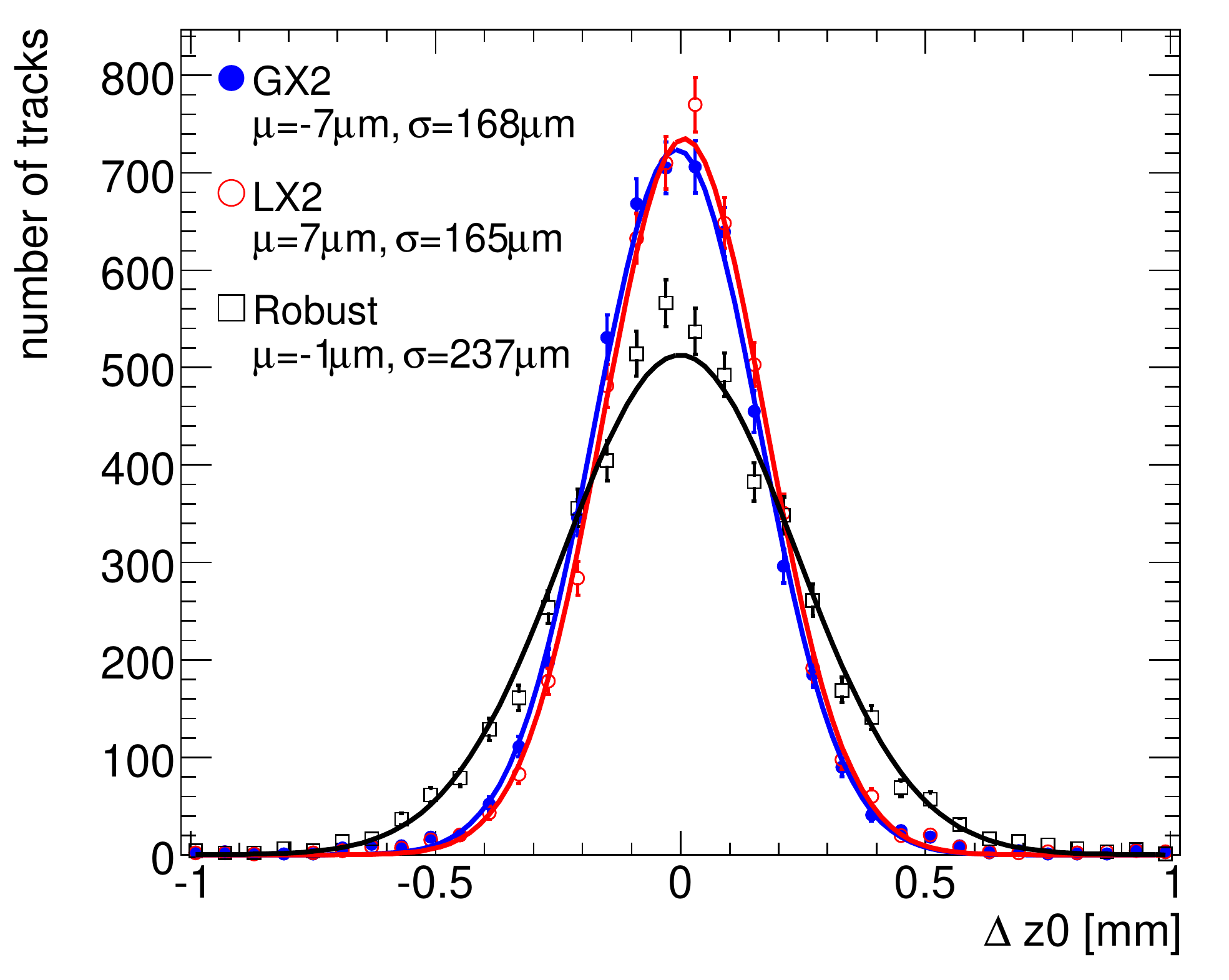}
\includegraphics[width=7.9cm,clip=true]{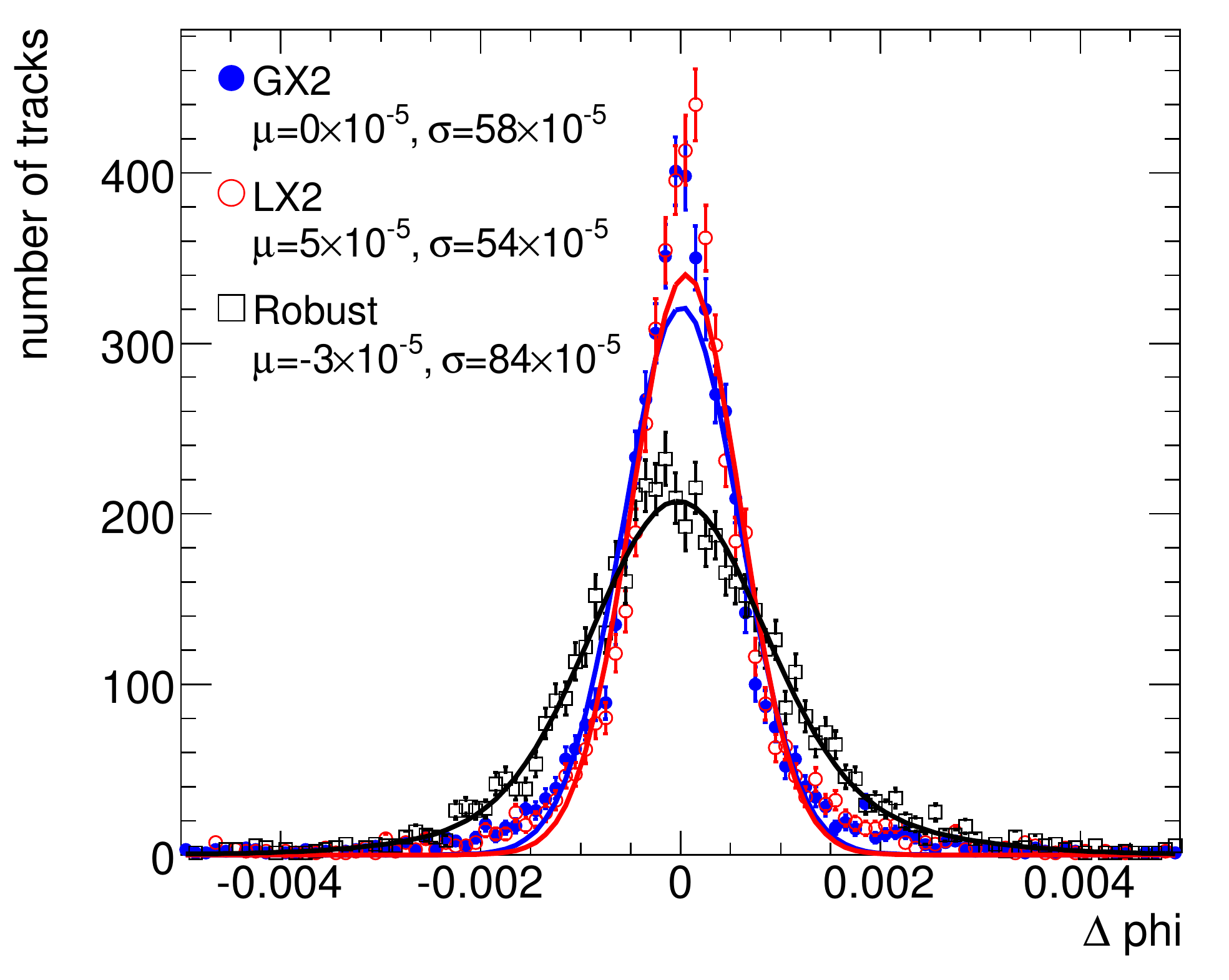}
\includegraphics[width=7.9cm,clip=true]{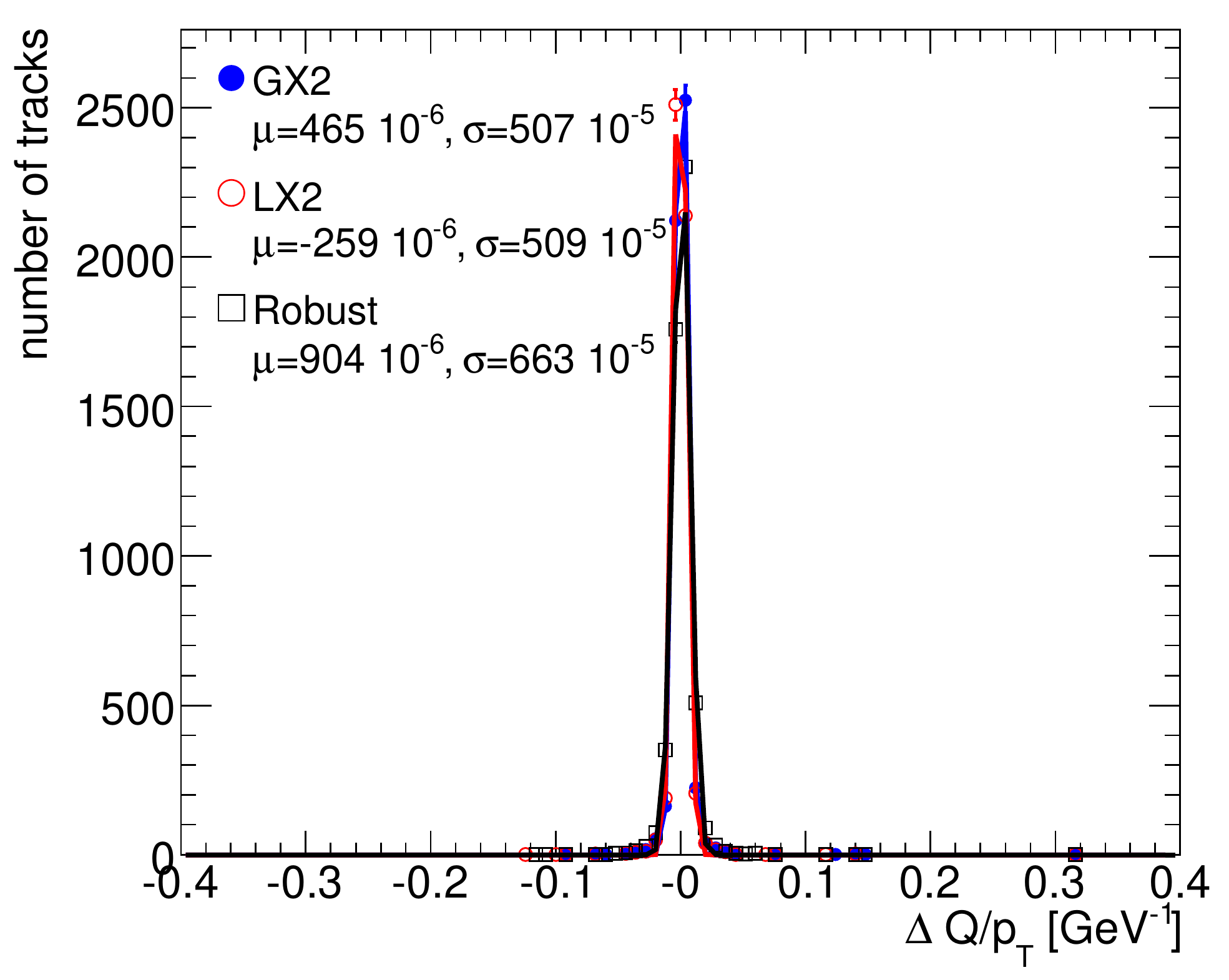}
\vspace{\cDistHalf}
\vspace{\cDist}
%\vspace{-0.2cm}
\end{center}
\caption[The resolution of the $d_0$, $z_0$, $\phi_0$, and $\frac q{p_T}$ track parameters in the barrel of the silicon tracker after {\it all} alignment corrections with the three alignment algorithms in M8+ using official monitoring.]{\label{fig:trkpar_mon}
The difference $\Delta\pi$ defined in Equation~\ref{eqn:trkParUncert} for the $d_0${\bf~(top~left)}, $z_0${\bf~(top~right)}, $\phi_0${\bf~(bottom~left)}, and $\frac q{\pt}${\bf~(bottom~right)} track parameters in the barrel of the silicon tracker after {\it all} alignment corrections with the \GX\ (GX2), \LX\ (LX2), and \RA\ (Robust) algorithms in M8+ with official monitoring. See text for discussion. Plots courtesy T.~Golling.
\vspace{\cDistHalf}
}
\end{figure}%\nopagebreak[5]

Certainly, the real acid test and raison d'\^etre of inner detector alignment is a precise and bias-free track parameter reconstruction. The resolution~$\sigma(\pi)$ for a given track parameter~$\pi$ is estimated in M8+ cosmic ray data in the following way: tracks which traverse the entire inner detector are split near the DIP in two halves -- the upper and the lower hemisphere part. Each of the parts is refitted independently. The difference~$\Delta$ between the track parameters is then calculated as:
\begin{equation} \label{eqn:trkParUncert}
  \Delta\pi\equiv\pi_{\rm upper}-\pi_{\rm lower}\,,
\end{equation} 
where $\pi_{\rm upper}$, $\pi_{\rm lower}$ stand for the track parameter $\pi$ of the upper and lower hemisphere part of the track, respectively. A single Gaussian fit is performed to the resulting $\Delta\pi$~distribution, and its $\sigma$-parameter is quoted for the resolution~$\sigma(\pi)$. In addition to the selection applied in case of residual distributions, at least one hit in the $b$-layer and at least three hits in total are required in the pixel detector, and seven hits or more in the entire silicon tracker.

The resolutions for the $d_0$, $z_0$, $\phi_0$ and $\frac q\pt$ track parameters for the three alignment algorithms are presented in Figure~\ref{fig:trkpar_mon}. Overall, a similar picture as discussed in the context of residuals is observed: while the $\mu$-parameters of the distributions are very close to 0 in case of \RA\ (in fact they are closest to 0 for $d_0$ and $z_0$), their $\sigma$-parameters tend to be somewhat larger than for the $\chisq$-based algorithms. The same argumentation as above can be employed. It is worthwhile to point out that the resolution found by \RA\ for the $\frac q\pt$ track parameter is closest to the \GX\ and \LX\ results, while the $d_0$ parameter is furthest. Consider that all modules traversed by a track contribute about equally to the measurement of the former, while the latter is mostly determined by the inner layers of the pixel detector being closest to the perigee. Thefore, one expects the $\frac q\pt$ resolution to be notably less affected by the missing of $c_\gamma$ corrections than $d_0$, which is in accordance with the experimental observation. It is planned to implement $c_\gamma$ alignment corrections in the \RA\ algorithm, as will be outlined in Chapter~\ref{chp:conclusionAlignment} ``\nameref{chp:conclusionAlignment}''. In studies with simulated Monte Carlo events reconstructed using the full knowledge of the detector geometry the following values have been found: $\mu_{d_0}=-1\,\mum$ and $\sigma_{d_0}=32\,\mum$, $\mu_{z_0}=4\,\mum$ and $\sigma_{z_0}=151\,\mum$, $\mu_{\phi}=0\times10^{-4}$ and $\sigma_{\phi}=3\times10^{-4}$, $\mu_{\frac q\pt}=0\,{\rm TeV}^{-1}$ and $\sigma_{\frac q\pt}=4\,{\rm TeV}^{-1}$.

It is truly remarkable how well the \RA\ is able to compete with the \chisq-based algorithms given the main cornerstones of its philosophy: transparency and robustness. One should keep in mind that for {\em superstructure} aligment, all it utilises are topological residual distributions and intuitive Ansatzes which are easy to understand and monitor. For {\em local} alignment at individual module level it employs the most basic and transparent approach possible: re-centring of residual distributions to compensate for in-plane translative misalignments only. Despite its robustness and transparency, the \RA\ algorithm delivered reliable and credible results, and is able to provide for an excellent track parameter reconstruction demonstrated in Figure~\ref{fig:trkpar_mon}.

\begin{comment}
\begin{figure}
\begin{center}
\vspace{\cDistHalf}
\includegraphics[width=7.9cm,clip=true]{M8plus/fig/val/Approved_DeltaD0}
\includegraphics[width=7.9cm,clip=true]{M8plus/fig/val/Approved_DeltaZ0}
\vspace{\cDist}
%\vspace{-0.2cm}
\end{center}
\caption[The $d_0$, $z_0$ track parameter distributions in the barrel of the silicon tracker after {\it all} alignment corrections with the three alignment algorithms in M8+ using official monitoring.]{\label{fig:d0z0_mon}
The $d_0$, $z_0$ track parameter distributions in the barrel of the silicon tracker after {\it all} alignment corrections with the \GX, \LX, and \RA\ algorithms in M8+ with official monitoring. Plots courtesy T.~Golling.
}
\end{figure}%\nopagebreak[5]
\begin{figure}
\begin{center}
\vspace{\cDistHalf}
\includegraphics[width=7.9cm,clip=true]{M8plus/fig/val/Approved_DeltaPhi}
\includegraphics[width=7.9cm,clip=true]{M8plus/fig/val/Approved_DeltaQoPT}
\vspace{\cDist}
%\vspace{-0.2cm}
\end{center}
\caption[The $\phi$, $\frac q{p_T}$ track parameter distributions in the barrel of the silicon tracker after {\it all} alignment corrections with the three alignment algorithms in M8+ using official monitoring.]{\label{fig:d0z0_mon}
The $\phi$, $\frac q{p_T}$ track parameter distributions in the barrel of the silicon tracker after {\it all} alignment corrections with the \GX, \LX, and \RA\ algorithms in M8+ with official monitoring. Plots courtesy T.~Golling.
}
\end{figure}%\nopagebreak[5]
\end{comment}
\clearpage

%% file: M8plus/B0vsB1.tex
As already touched upon in Subsection~\ref{ssec:l3M8} and Section~\ref{sec:resultsM8} ``\nameref{sec:resultsM8}'', significant discrepancies between $B$-field on and off data are found by the \RA\ algorithm in residual means in the barrel of the pixel detector. This is a long standing and yet unresolved problem, and analogous observations have been reported by the \GX\ and \LX\ alignment algorithms~\cite{bib:alignmentWorkshop}. In this Section, a brief review of the findings in $r_x$ residual and $o_{xx}$ overlap residual distributions shall be given.

Table~\ref{tab:B0vsB1} summarises the differences in the mean of the $r_x$ residuals and $o_{xx}$ overlap residuals defined as:
\begin{eqnarray} \label{eqn:diffRes}
 \Delta r_x &\equiv& \langle r_x^{B~\rm off}\rangle-\langle r_x^{B~\rm on}\rangle\nonumber \\
 \Delta o_{xx} &\equiv& \langle o_{xx}^{B~\rm off}\rangle-\langle o_{xx}^{B~\rm on}\rangle
\end{eqnarray}
between the $B$-field off and on cases after the \RA\ procedure. The uncertainties on these values are conventionally defined as the uncorrelated Gaussian sum of their individual statistical errors:
\begin{eqnarray} \label{eqn:diffResErr}
 \delta(\Delta r_x) &\equiv& \left\{\left(\delta r_x^{B~\rm off}\right)^2 + \left(\delta r_x^{B~\rm on}\right)^2\right\}^{-\frac12}\nonumber \\
 \delta(\Delta o_{xx}) &\equiv& \left\{\left(\delta o_{xx}^{B~\rm off}\right)^2 + \left(\delta o_{xx}^{B~\rm on}\right)^2\right\}^{-\frac12}\,,
\end{eqnarray}
where the individual statistical errors are determined according to Equation~\ref{eqn:residualErrAll}. For the calculation of the table the full M8+ dataset and $r_x\in[-1.5\,\mm,\,1.5\,\mm]$ were used. No explicit cut on overlap residuals was applied.

As already mentioned above, $\Delta r_x$ value in the pixel barrel displays a difference of $3.74\pm0.28\,\mum$, which constitutes a deviation from 0 with $\sim$13$\sigma$ significance. As expected, a shift in the opposite direction by $-0.13\pm0.05\,\mum$ ($\sim$3$\sigma$) is observed in the SCT barrel. Although the first conclusion from these findings may be that the origin of the discrepancy is to be found in the pixel detector since it displays a larger discrepancy, this reasoning is not applicable: the statistical weight of residuals found in the SCT is much larger due to more hits on average, but most importantly, because of many tracks not traversing the pixel detector at all.\\
No $\delta r_x$ discrepancies are observed in the ECs other than in SCT~EC~A, where $\delta r_x=-2.03\pm0.27\,\mum$~($7\sigma$) is found. According to our best knowledge, there is no reason for such a difference between the two ECs of the SCT.

The overlap residuals in all subdetectors are consistent for $B$-field off and on with the exception of the SCT barrel. There, $\Delta o_{xx}=-2.46\pm0.18\,\mum$ is observed with 13$\sigma$ significance, which was first reported by the author~\cite{bib:overlapDiscepancy}. As of now, there is no explanation for this behaviour.

\begin{table}
\small
\begin{center}
\begin{tabular}{l|rr|rr}
\hline
 & $\Delta r_x$ & $\delta(\Delta r_x)$ & $\Delta o_{xx}$ & $\delta(\Delta o_{xx})$ \\
\hline\hline
Pixel barrel layer 0	& $6.13$	& $0.59$& $1.44$	& $1.20$\\
Pixel barrel layer 1	& $2.79$	& $0.46$& $-1.26$	& $0.91$\\
Pixel barrel layer 2	& $3.41$	& $0.44$& $1.32$	& $0.79$\\
Pixel barrel {\bf (all)}& $3.74$	& $0.28$& $0.50$	& $0.54$\\
\hline
SCT barrel layer 0	& $0.13$	& $0.13$& $-1.12$	& $0.45$\\
SCT barrel layer 1	& $-0.62$	& $0.11$& $-1.29$	& $0.37$\\
SCT barrel layer 2	& $0.47$	& $0.09$& $-0.91$	& $0.33$\\
SCT barrel layer 3	& $-0.46$	& $0.10$& $-4.44$	& $0.33$\\
SCT barrel {\bf (all)}	& $-0.13$	& $0.05$& $-2.46$	& $0.18$\\
\hline
Pixel EC A layer 0	& $-13.77$	& $3.84$& $3.38$	& $7.59$\\
Pixel EC A layer 1	& $-3.93$	& $3.90$& $-13.43$	& $9.03$\\
Pixel EC A layer 2	& $9.83$	& $4.67$& $5.15$	& $7.30$\\
Pixel EC A {\bf (all)}	& $-4.34$	& $2.36$& $-1.69$	& $4.68$\\
\hline
Pixel EC C layer 0	& $4.46$	& $3.12$& $-9.90$	& $6.43$\\
Pixel EC C layer 1	& $-2.72$	& $3.15$& $-12.93$	& $8.65$\\
Pixel EC C layer 2	& $-1.39$	& $2.96$& $0.65$	& $7.77$\\
Pixel EC C {\bf (all)}	& $0.10$	& $1.77$& $-7.03$	& $4.46$\\
\hline
SCT EC A layer 0	& $-1.27$	& $0.72$& $-0.04$	& $4.08$\\
SCT EC A layer 1	& $-2.06$	& $0.60$& $-4.15$	& $3.75$\\
SCT EC A layer 2	& $-2.64$	& $0.56$& $7.43$	& $3.51$\\
SCT EC A layer 3	& $-2.76$	& $0.62$& $-8.12$	& $4.06$\\
SCT EC A layer 4	& $-0.28$	& $0.68$& $-5.98$	& $5.00$\\
SCT EC A layer 5	& $-2.40$	& $0.95$& $4.96$	& $7.80$\\
SCT EC A layer 6	& $-1.68$	& $1.41$& $38.30$	& $9.10$\\
SCT EC A layer 7	& $-5.68$	& $2.26$& $-33.87$	& $13.80$\\
SCT EC A layer 8	& $-8.19$	& $3.21$& $95.43$	& $19.33$\\
SCT EC A {\bf (all)}	& $-2.03$	& $0.27$& $0.83$	& $1.72$\\
\hline
SCT EC C layer 0	& $-3.12$	& $0.61$& $9.29$	& $4.87$\\
SCT EC C layer 1	& $-3.00$	& $0.50$& $-7.16$	& $3.36$\\
SCT EC C layer 2	& $-0.56$	& $0.54$& $0.91$	& $3.74$\\
SCT EC C layer 3	& $2.27$	& $0.65$& $-9.72$	& $4.39$\\
SCT EC C layer 4	& $10.60$	& $0.69$& $15.80$	& $4.55$\\
SCT EC C layer 5	& $0.95$	& $1.21$& $-0.80$	& $8.36$\\
SCT EC C layer 6	& $-2.15$	& $4.77$& $45.17$	& $57.74$\\
SCT EC C layer 7	& $-5.40$	& $3.03$& $83.04$	& $25.65$\\
SCT EC C layer 8	& $31.03$	& $5.01$& $-99.54$	& $42.12$\\
SCT EC C {\bf (all)}	& $0.38$	& $0.26$& $1.46$	& $1.83$\\
\hline
\end{tabular}
\caption[$\langle r_x^{B~\rm off}\rangle-\langle r_x^{B~\rm on}\rangle$ and $\langle o_{xx}^{B~\rm off}\rangle-\langle o_{xx}^{B~\rm on}\rangle$ of the silicon tracker by layers in M8+]{\label{tab:B0vsB1}
$\Delta r_x\equiv\langle r_x^{B~\rm off}\rangle-\langle r_x^{B~\rm on}\rangle$ and $\Delta o_{xx}\equiv \langle o_{xx}^{B~\rm off}\rangle-\langle o_{xx}^{B~\rm on}\rangle$ of the silicon tracker by layers in M8+. The notation used in this table is defined in Equations~\ref{eqn:diffRes}, \ref{eqn:diffResErr}. All values are in \mum, and were calculated with $r_x\in[-1.5\,\mm,\,1.5\,\mm]$.\\
The $\Delta r_x$ value in the pixel barrel displays a difference of 3.74\,\mum\ with $\sim$13$\sigma$ significance, and a shift in the opposite direction by $-0.13\,\mum$ ($\sim$3$\sigma$) is found in the SCT barrel. No effect in the ECs is observed other than in SCT~EC~A, where $\Delta r_x=-2.03\,\mum$~($7\sigma$).\\
The overlap residuals in all subdetectors are consistent for $B$-field off and on with the exception of the SCT barrel. There, $\Delta o_{xx}=-2.46\,\mum$ is observed with 13$\sigma$ significance.
}
\end{center}
\end{table}

%% file: ConclusionAlign/ConclusionAlign.tex
The precise knowledge of silicon tracker modules positions and orientation in space is vital for a major fraction of the ambitious ATLAS physics program. The \RA\ algorithm was successfully employed by the author to determine the alignment constants for the silicon tracker using about 2M cosmic ray particle tracks collected by ATLAS in autumn 2008. In the {\em barrel} region of the silicon tracker, module-level alignment was provided for the critical local $x$ DoF. Additionally, local $y$ and $\gamma$ were aligned for in the barrel part of the pixel detector\footnote{Strictly speaking, local $\gamma$ was not aligned for at individual module level, but determined from pixel stave bow alignment.}. Due to a lacking {\em end-cap} module illumination in both pixel and SCT detectors with cosmic rays, only a disk-level alignment could be derived.

Overall, an improvement of about 70\%~(40\%) was achieved in terms of residual width in the barrel of the pixel (SCT) detector, and of about 50\%~(20\%) in the end-caps of the pixel~(SCT). In the barrel of the silicon tracker as well as in the end-caps of the pixel detector, the residual means approach zero to within microns. Similar improvements are observed in figures of merit based on overlap residuals.

The performance of the derived alignment constants in the barrel of the silicon tracker was critically investigated with official ATLAS monitoring tools~\cite{bib:monitoring}. Within the limitations of the experimental setup, a reasonably bias-free track parameter reconstruction was found. The track parameter resolution is somewhat behind the \chisq-based alignment algorithms in use at ATLAS.

To achieve the demonstrated level of alignment precision with cosmic ray data, the \RA\ was substantially extended by the author to perform coherent superstructure alignment using topological $\rmean x(\eta,\Phi)$ distributions. This intuitive and transparent alignment procedure serves as a valuable cross-check for the alignment constants provided by \chisq-based alignment algorithms. 

In the course of this research, various new observations were made:
\begin{itemize}
\vspace{-3mm}
\item
the $\Phi$-dependent modulations in residual widths for odd and even module sides in the end-caps of the SCT{\bf *};
\item
the incoherent $\rmean x(\Phi)$ distributions in end-cap~C of the SCT which could be partly explained by large misalignments of the SCT end-cap disks in global $Z$;
\item
the discrepancy in residual means between solenoid on and off data in the barrel of the pixel detector;
\item
the unexpectedly large bowing of pixel barrel staves in the transverse plane approximately along local $x$ of the modules;
\item
the increased average overlap residual means in all of pixel end-cap disks in accordance with the optical survey of pixel end-cap~A\!{\bf *}
\vspace{-3mm}
\end{itemize}
to name a few. Items marked with a ``{\bf *}'' were first observed by the author using track-based techniques.

The width of the pixel end-cap~A disks was measured for the first time using a track-based approach, and found to deviate by $\delta(\Delta z)=57.9\pm10.7\,\mu{\rm m}$ from the nominal value, in accordance with the optical survey. \vspace{\cDistHalf}

\subsubsection{Outlook}
\vspace{\cDistHalf}
In view of the recently recommenced cosmic ray data taking and the first $p$-$p$ collisions expected in autumn 2009, the immediate goal of the \RA\ algorithm is clear: it needs to be checked whether the detector geometry has changed in course of maintenance works during the winter shut-down, and if need be a new set of alignment constants derived based on the 2008 experience. The alignment precision -- which is still statistically limited in all parts of the detector -- can be increased by deriving alignment constants with larger data samples. With the much increased rate-to-tape of cosmic ray particles traversing the silicon tracker, time-dependent alignment studies can be attacked and correlated with the results of the FSI, which will be continuously operated during data taking for the first time. Tracks from beam-gas interactions, beam halo, and eventually collision data are expected to constrain the alignment of the end-caps.

Clearly, all the yet-unresolved alignment anomalies described above need to be addressed with new 2009 data.

Once the statistical alignment precision with collision data will have reached the level of residual systematic effects (which is expected to happen about \xOverY12\ year into data taking), additional constraints to alignment similar to the ones described in~\cite{bib:weakModes} can be applied in order to guarantee a bias-free track parameter reconstruction.

On a more technical note, the logical next step in the development of the \RA\ algorithm appears to be the implementation of the local $\gamma$ DoF at individual module level, which -- alongside with local $x$ -- is a critical degree of freedom with a particular impact on $d_0$ and $\frac q\pt$ track parameter resolution. It is expected that once this has happened, the \RA\ algorithm will become fully competitive with \chisq-based algorithms. In the spirit of  \RA, local $\gamma$ alignment could be implemented by analysing $r_x$ residuals of a given module versus their position along local $y$: $\gamma\simeq{\rm slope}\left\{r_x(y)\right\}$ using a tangent expansion to first order.

%% file: Appendix/Spectra/Spectra.tex
In the following, sparticle mass spectra will be presented for the ATLAS SU$x$ mSUGRA benchmark points used in the trilepton search analysis in Part~\ref{prt:susy}, as defined in Table~\ref{tab:susy_points}. The $y$-axis gives the mass scale of sparticles in GeV. 
%The $x$-axis gives the production cross section of a given sparticle in arbitrary units. 
The transition lines between sparticles indicate possible decays: solid black lines stand for a branching ratio of more than $10^{-1}$, dashed blue are in the range between $[10^{-1}\,,10^{-2}]$, and dotted red in $[10^{-2}\,,10^{-3}]$.

For the sake of convenience, Table~\ref{tab:susy_points} on page~\pageref{tab:susy_points} is reproduced below (Table~\ref{tab:susy_pointsA}).

\begin{table}[h]
\begin{center}
\begin{small}
\begin{tabular}{lccccccr}
\hline
Process & $M_0$ [GeV]	& \Mhalf [GeV]	& $A_0$ [GeV]	& $\tan\beta$	& $\arg\mu$	& $\sigma$ [pb] & Region \\
\hline\hline	
SU1 	& 70		& 350		& 0 		& 10 	 	& +		& 7.43	& Coannihilation\\
SU2 	& 3550  	& 300		& 0 		& 10 	 	& +		& 4.86	& Focus\\
SU3 	& 100  		& 300		& $-300$	& 6 	 	& +		& 18.59 & Bulk\\
SU4 	& 200  		& 160		& $-400$	& 10 	 	& +		& 262 	& Low Mass\\
SU8 	& 210  		& 360		& 0		& 40 	 	& +		& 6.44 	& Coannihilation\\
\hline
\end{tabular}
\end{small}
\end{center}
\vspace{\cDistHalf}
\caption[The SU$x$ ATLAS benchmark points, together with mSUGRA parameters and total production cross-sections at LO.]{\small\label{tab:susy_pointsA}
The SU$x$ ATLAS benchmark points, together with mSUGRA parameters and total production cross-sections~\cite{bib:msugraPoints} at LO.
\vspace{\cDistHalf}
}
\end{table}

\begin{figure}[h]
\begin{center}
\includegraphics[height=7.9cm,angle=90]{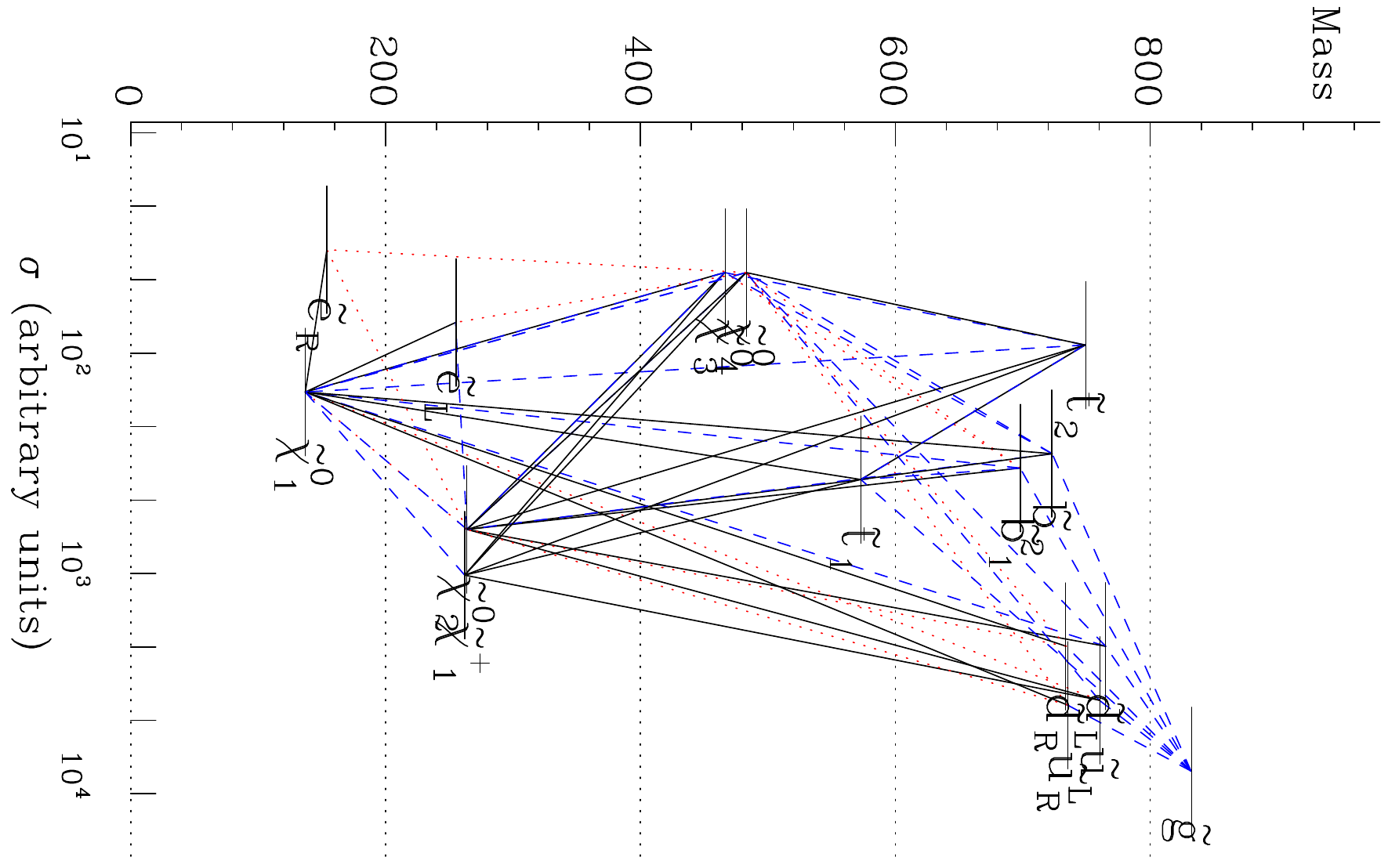}
\includegraphics[height=7.9cm,angle=90]{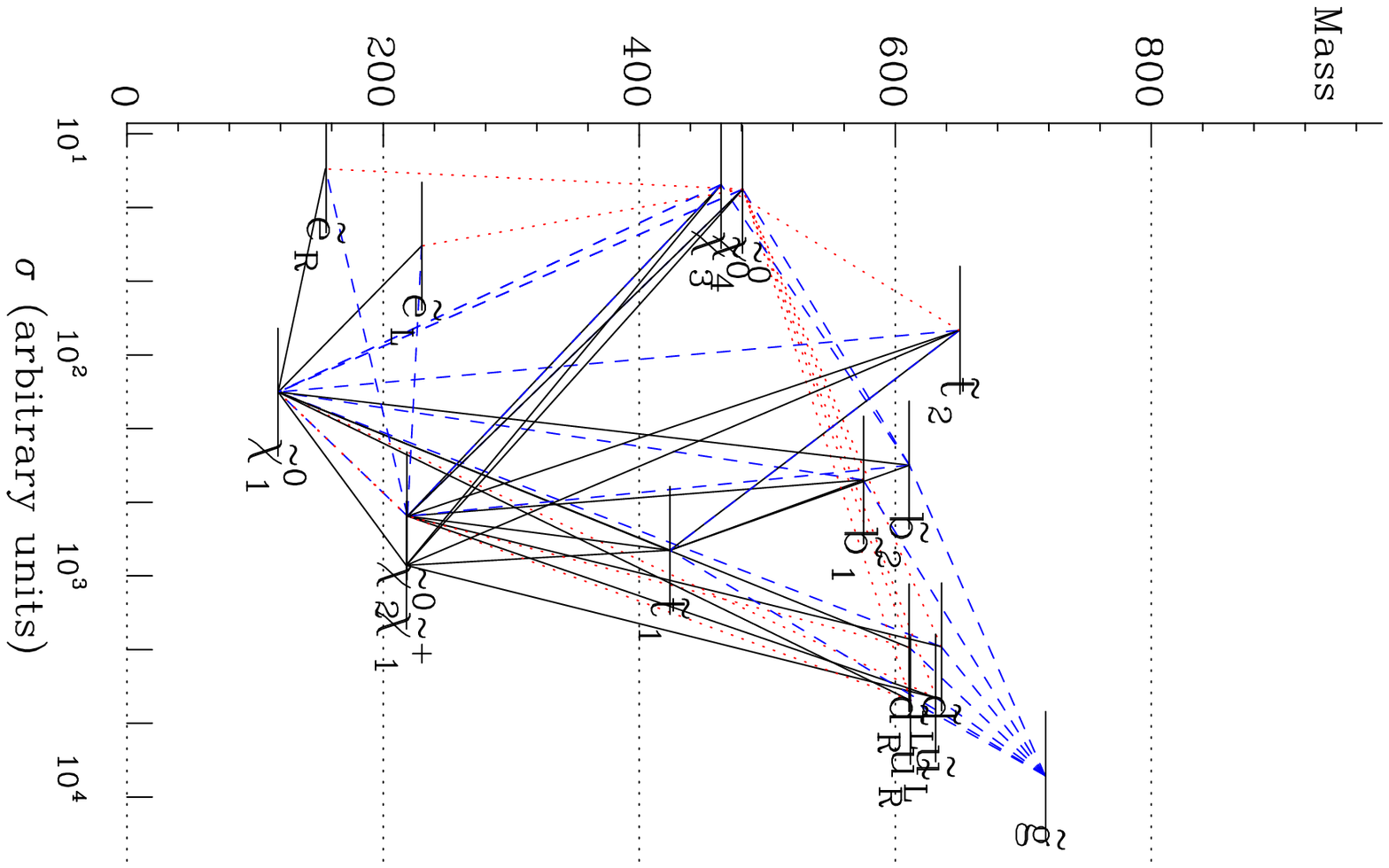}
\end{center}\vspace{-4mm}
\caption[Sparticle mass spectrum for the ATLAS mSUGRA benchmark points SU1 and SU3]{\label{fig:SU13}
Sparticle mass spectrum for the ATLAS mSUGRA benchmark point SU1{\bf~(left)} and SU3{\bf~(right)}. The mass scale is given in GeV.
}\vspace{-2mm}
\end{figure}

\begin{figure}
\begin{center}
\includegraphics[height=7.9cm,angle=90]{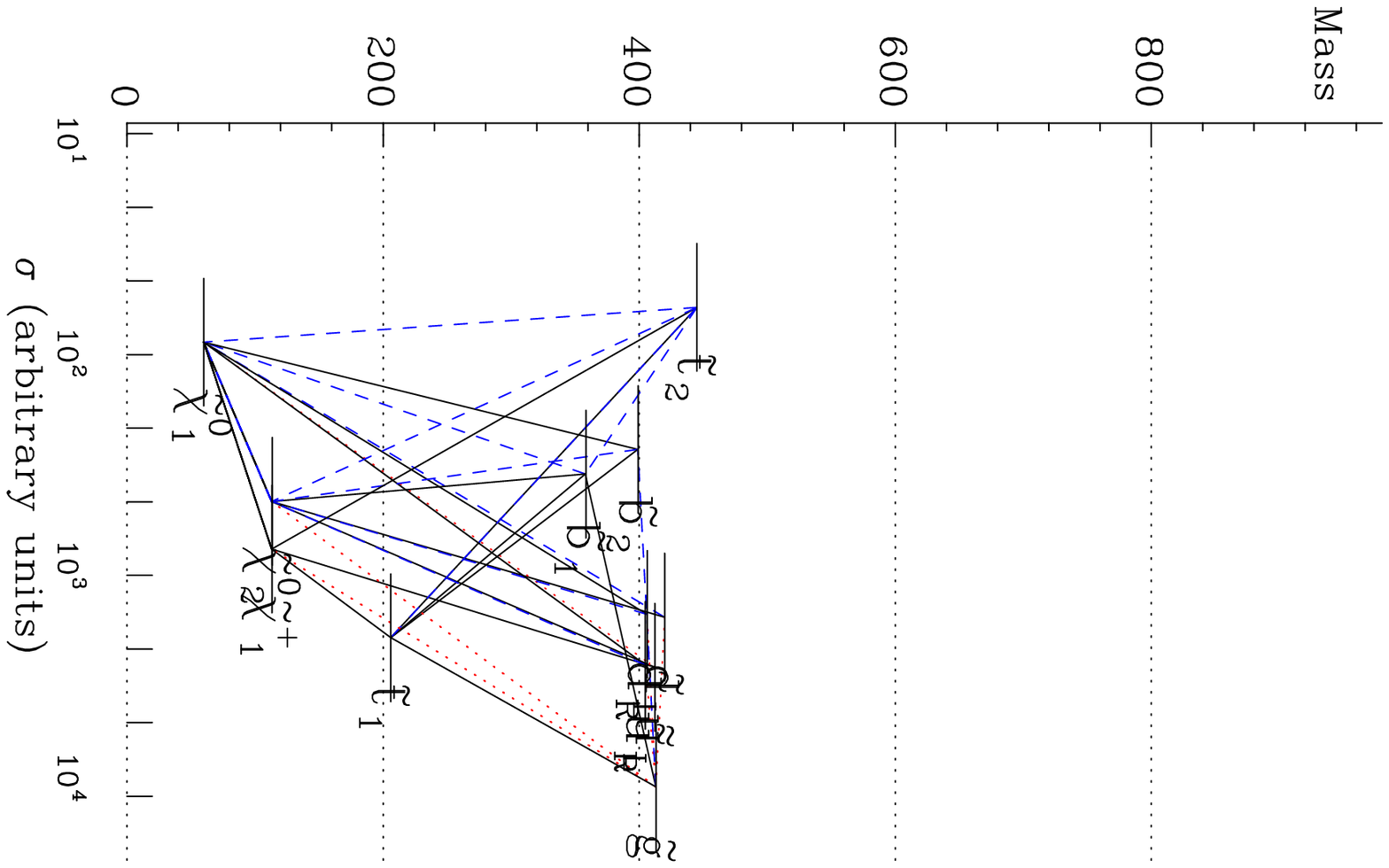}
\includegraphics[height=7.9cm,angle=90]{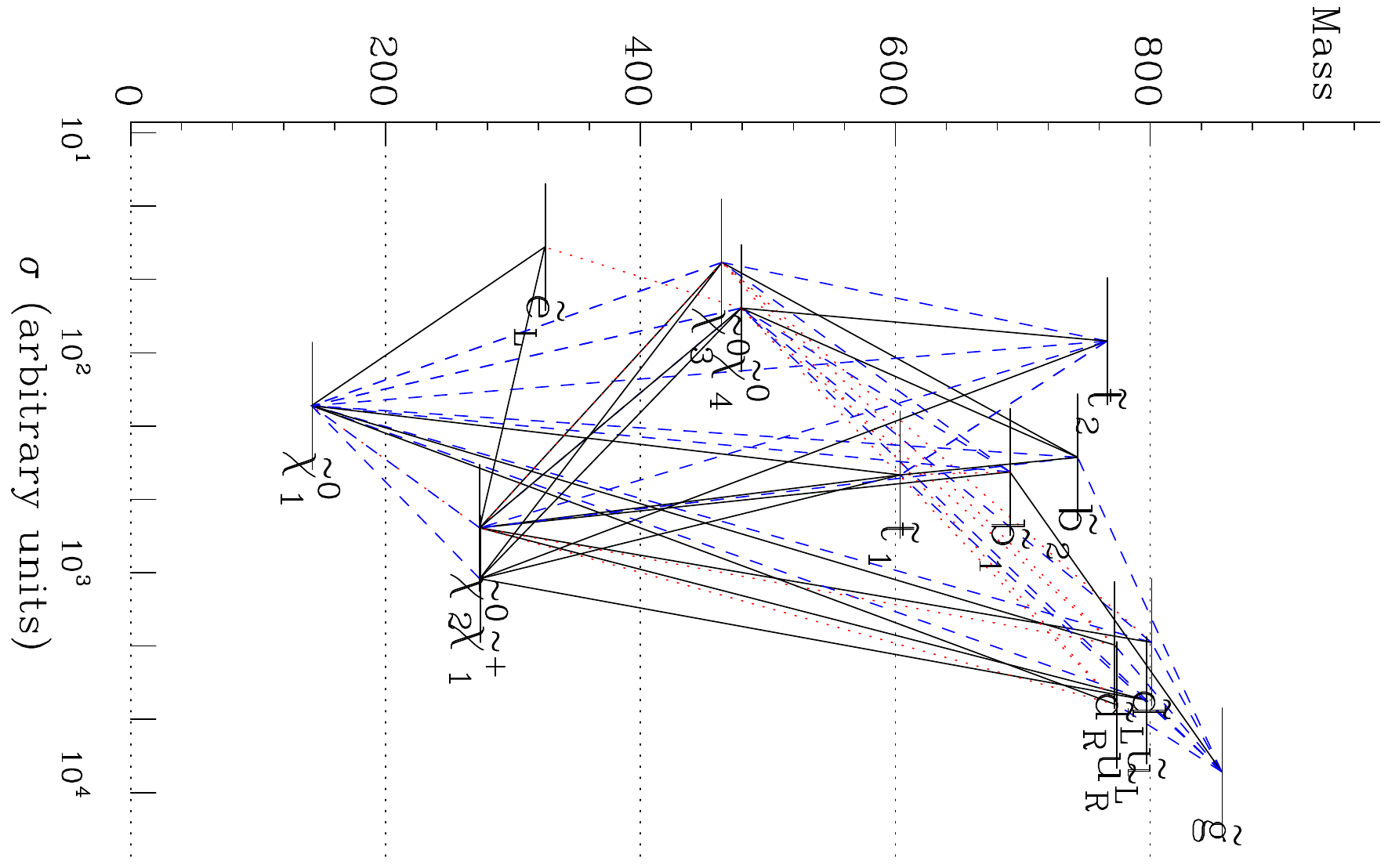}
\end{center}\vspace{-4mm}
\caption[Sparticle mass spectrum for the ATLAS mSUGRA benchmark points SU4 and SU8]{\label{fig:SU48}
Sparticle mass spectrum for the ATLAS mSUGRA benchmark points SU4{\bf~(left)} and SU8{\bf~(right)}. The mass scale is given in GeV.
}\vspace{-2mm}
\end{figure}

\begin{figure}
\begin{center}
\includegraphics[height=12cm,angle=90]{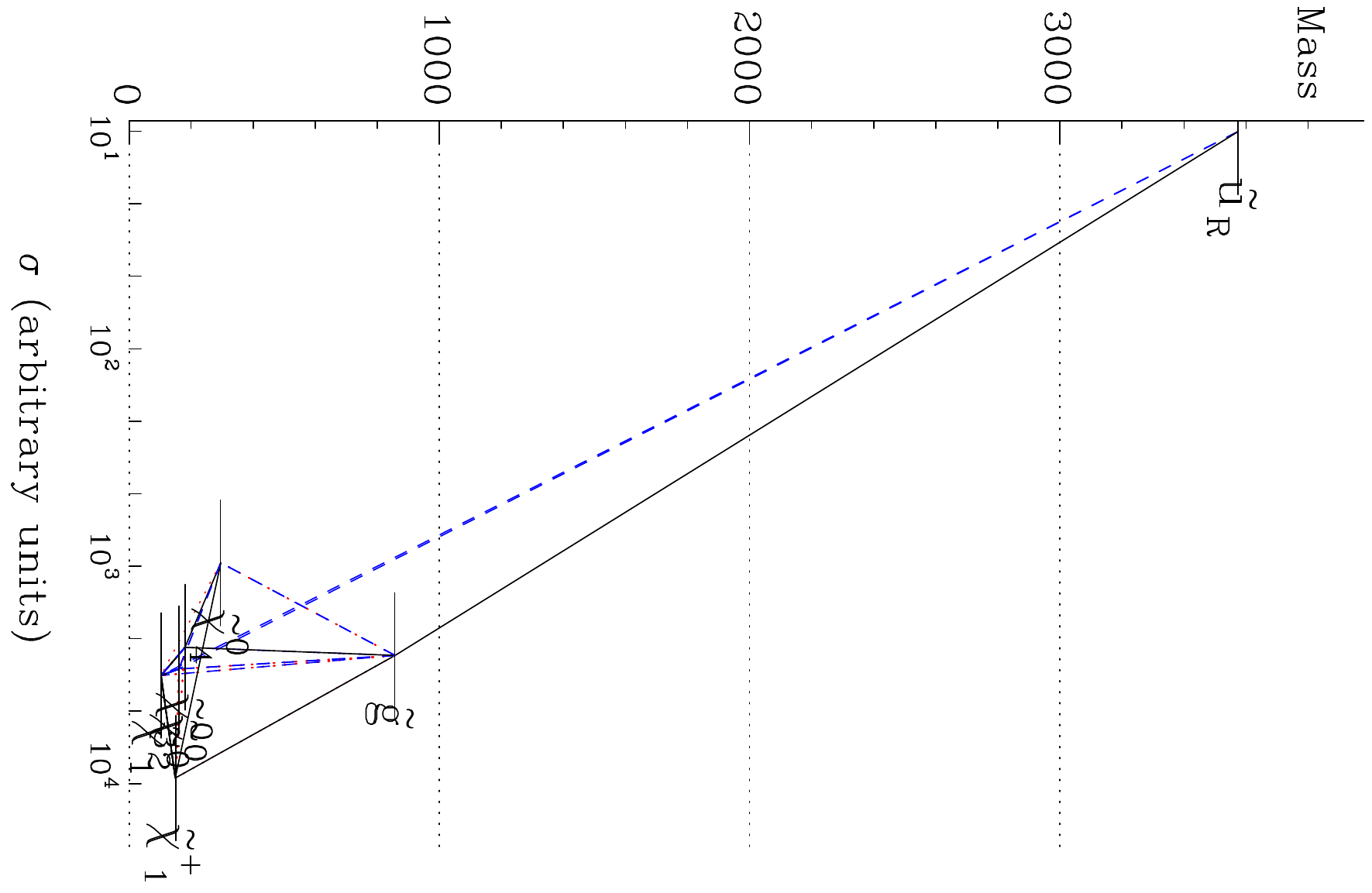}
\end{center}\vspace{-4mm}
\caption[Sparticle mass spectrum for the ATLAS mSUGRA benchmark point SU2]{\label{fig:SU2}
Sparticle mass spectrum for the ATLAS mSUGRA benchmark point SU2. The mass scale is given in GeV.
}\vspace{-2mm}
\end{figure}

%% file: Appendix/L4/L4.tex
In the following, the $\rmean x(\eta)$ distributions are shown for all 112 pixel barrel staves, as found in M8+. They serve as a reference for Subsection~\ref{ssec:pixelStaveBowM8}, and complete the Figures~\ref{fig:m8_r_x_vs_eta_L4_before} and \ref{fig:m8_r_x_vs_eta_L4_after}. This part of the appendix is organised as follows: the $\rmean x(\eta)$ distributions are shown in groups of 15 per figure. Each of these groups is presented in two separate figures: {\em before} and {\em after} alignment. Note the difference in the $y$-axis scale before and after alignment. The entire discussion is in Subsection~\ref{ssec:pixelStaveBowM8}.

\begin{figure}
\begin{center}
\includegraphics[width=15.8cm,clip=true]{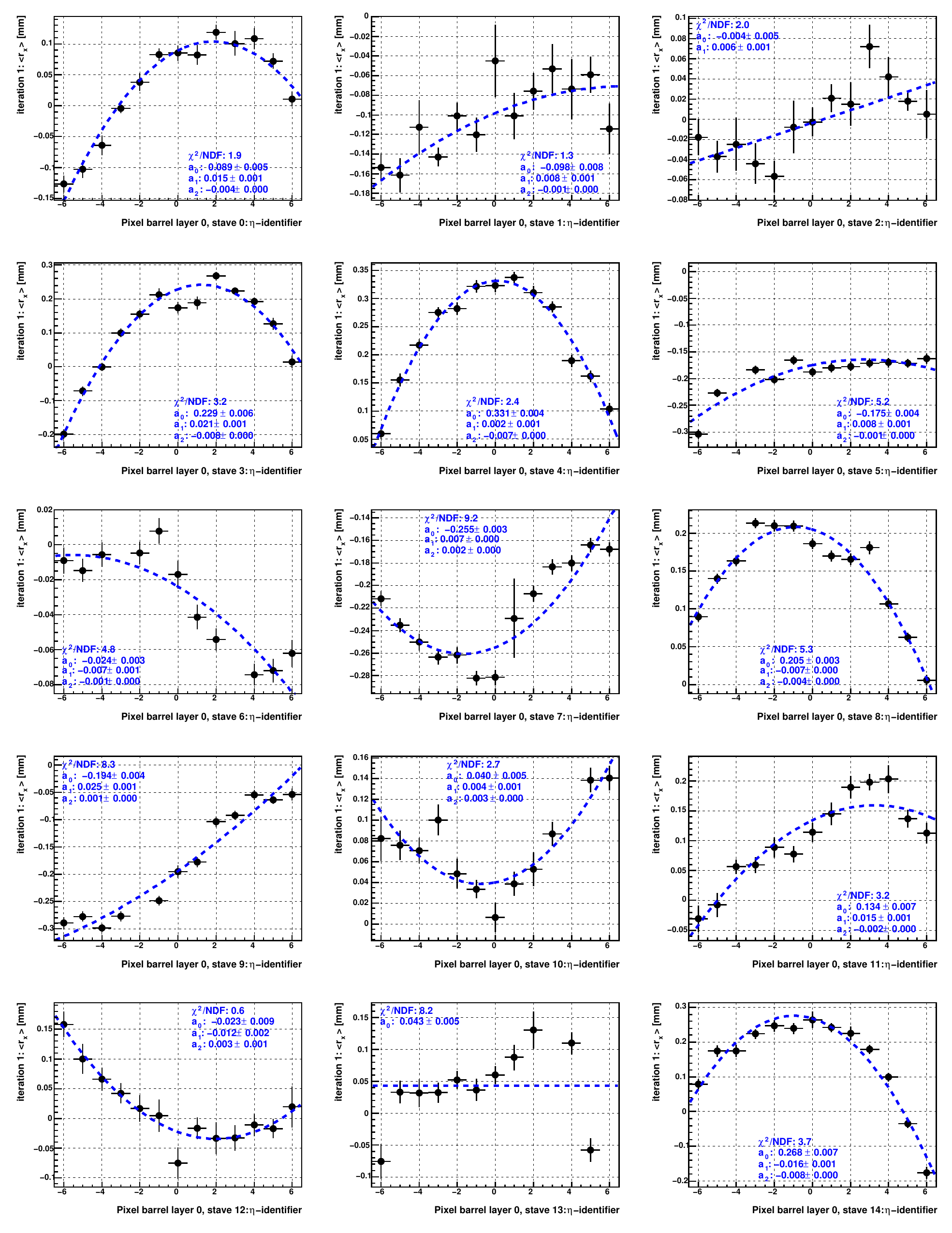}
\end{center}
\vspace{\cDist}
\caption[The $\langle r_x\rangle(\eta)$ distribution before pixel stave bow alignment in M8+ (page~1)]{
The $\rmean x(\eta)$ distribution for staves 1 to 15 of the pixel detector using the full $B$-field off M8+ dataset {\bf before} alignment for pixel stave bow. 
%The results on this plot . 
The fit results with a parabola of the form specified in Equation~\ref{eqn:staveBow} are shown in blue.
}
\end{figure}%\nopagebreak[5]

\begin{figure}
\begin{center}
\includegraphics[width=15.8cm,clip=true]{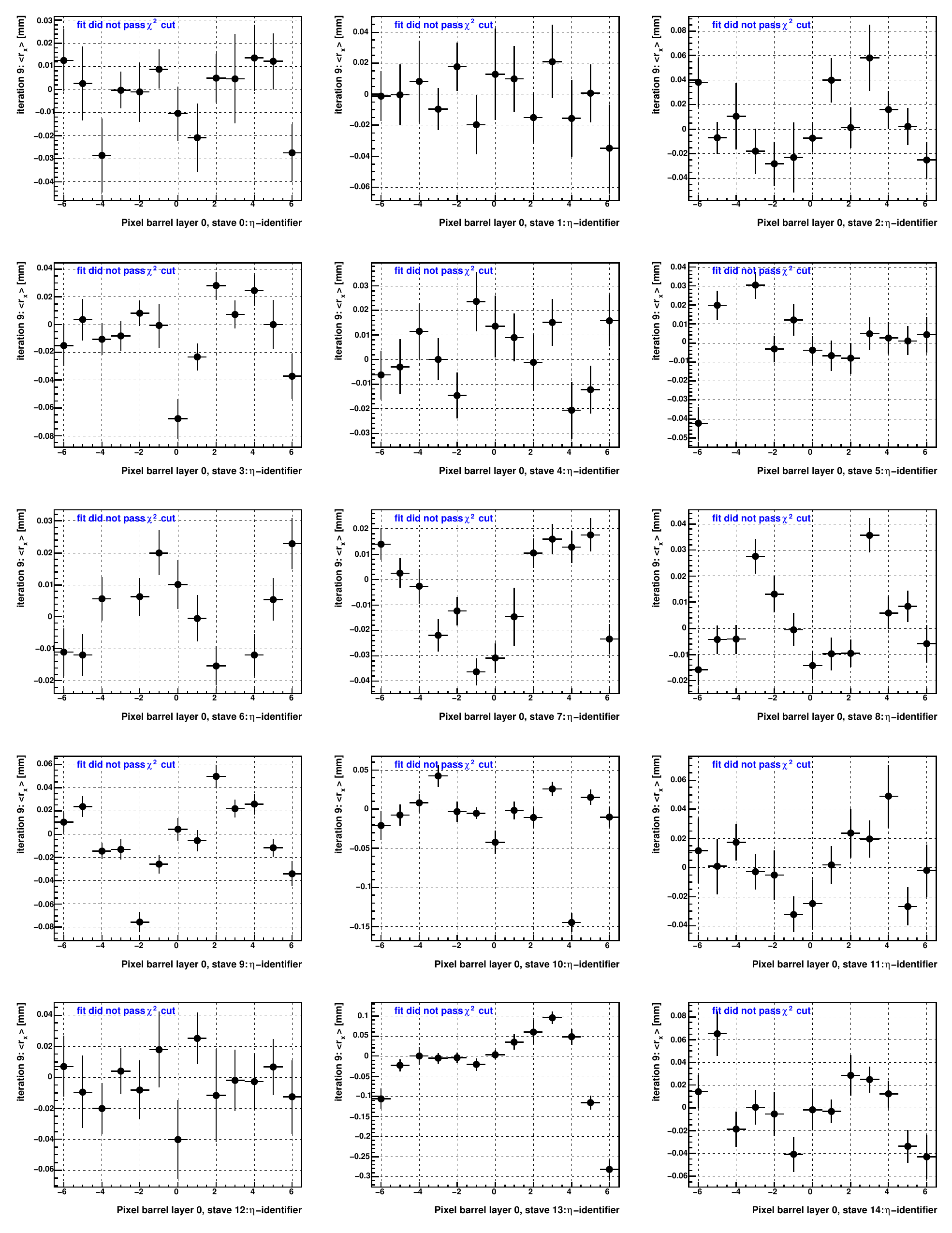}
\end{center}
\vspace{\cDist}
\caption[The $\langle r_x\rangle(\eta)$ distribution after pixel stave bow alignment in M8+ (page~1)]{
The $\rmean x(\eta)$ distribution for staves 1 to 15 of the pixel detector using the full $B$-field off M8+ dataset {\bf after} alignment for pixel stave bow. Note the difference in the $y$-axis scale before and after alignment.
%The results on this plot . 
The fit results with a parabola of the form specified in Equation~\ref{eqn:staveBow} are shown in blue.
}
\end{figure}%\nopagebreak[5]

\begin{figure}
\begin{center}
\includegraphics[width=15.8cm,clip=true]{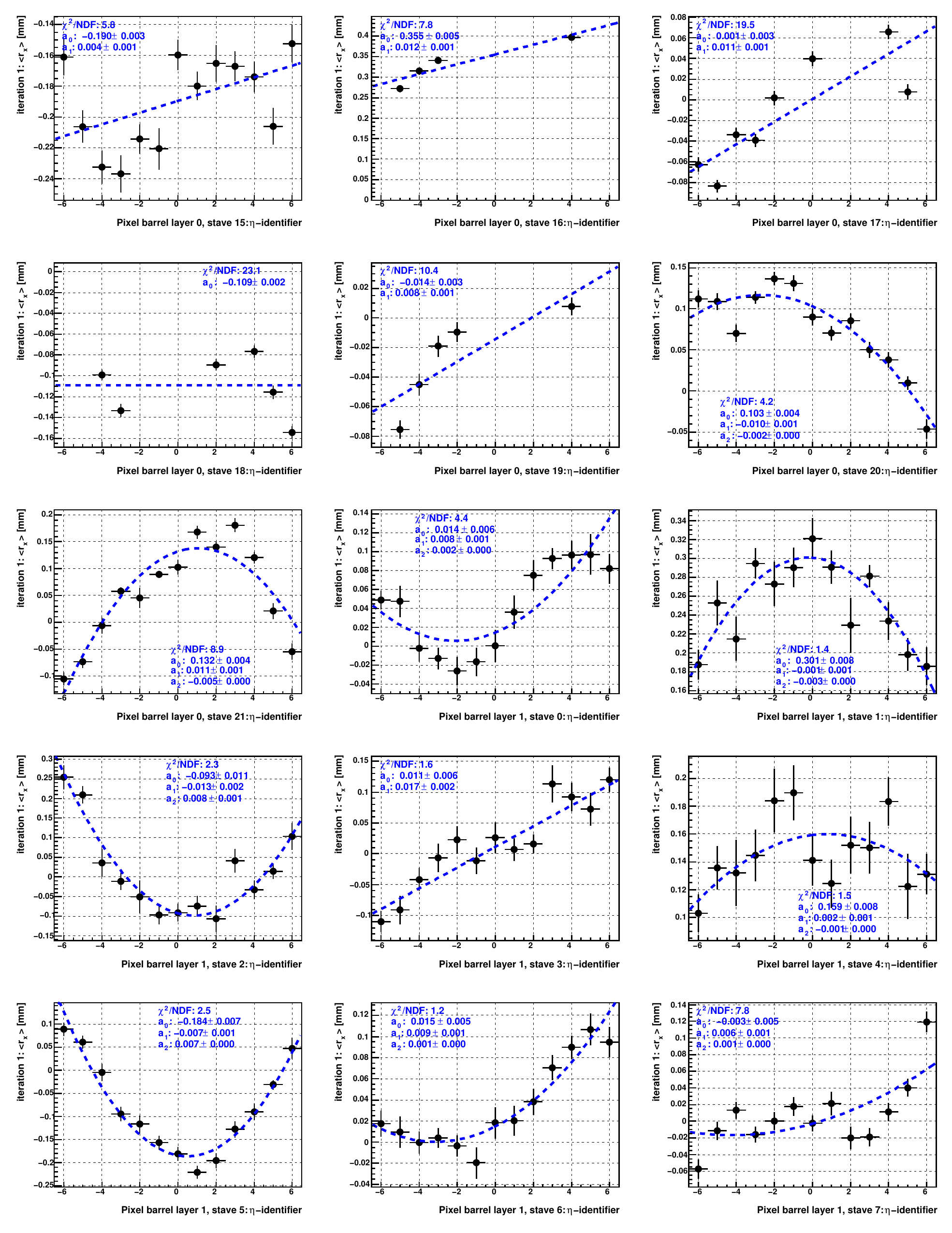}
\end{center}
\vspace{\cDist}
\caption[The $\langle r_x\rangle(\eta)$ distribution before pixel stave bow alignment in M8+ (page~2)]{
The $\rmean x(\eta)$ distribution for staves 16 to 30 of the pixel detector using the full $B$-field off M8+ dataset {\bf before} alignment for pixel stave bow. 
%The results on this plot . 
The fit results with a parabola of the form specified in Equation~\ref{eqn:staveBow} are shown in blue.
}
\end{figure}%\nopagebreak[5]

\begin{figure}
\begin{center}
\includegraphics[width=15.8cm,clip=true]{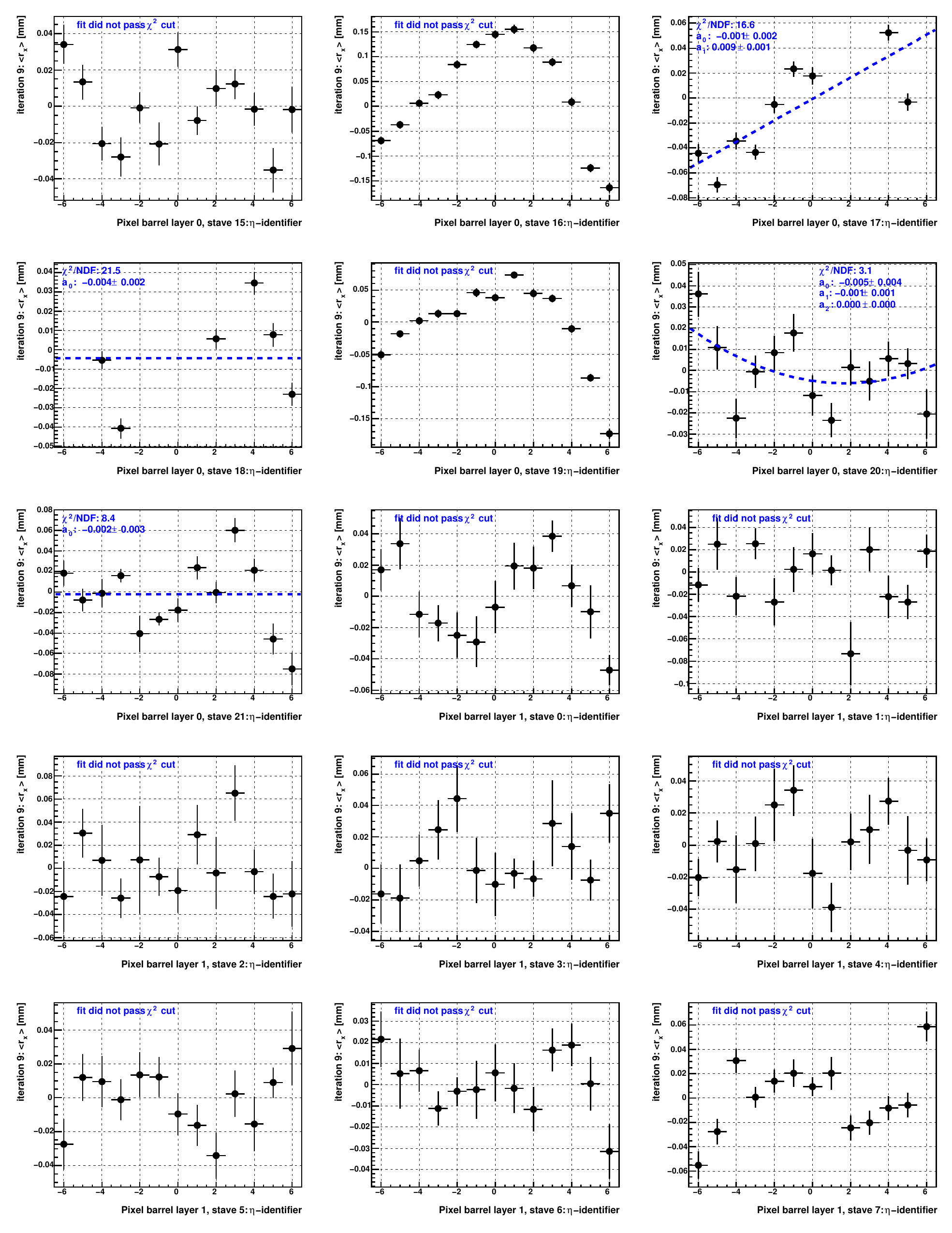}
\end{center}
\vspace{\cDist}
\caption[The $\langle r_x\rangle(\eta)$ distribution after pixel stave bow alignment in M8+ (page~2)]{\label{fig:appL4p2after}
The $\rmean x(\eta)$ distribution for staves 16 to 30 of the pixel detector using the full $B$-field off M8+ dataset {\bf after} alignment for pixel stave bow. Note the difference in the $y$-axis scale before and after alignment.
%The results on this plot . 
The fit results with a parabola of the form specified in Equation~\ref{eqn:staveBow} are shown in blue.
}
\end{figure}%\nopagebreak[5]

\begin{figure}
\begin{center}
\includegraphics[width=15.8cm,clip=true]{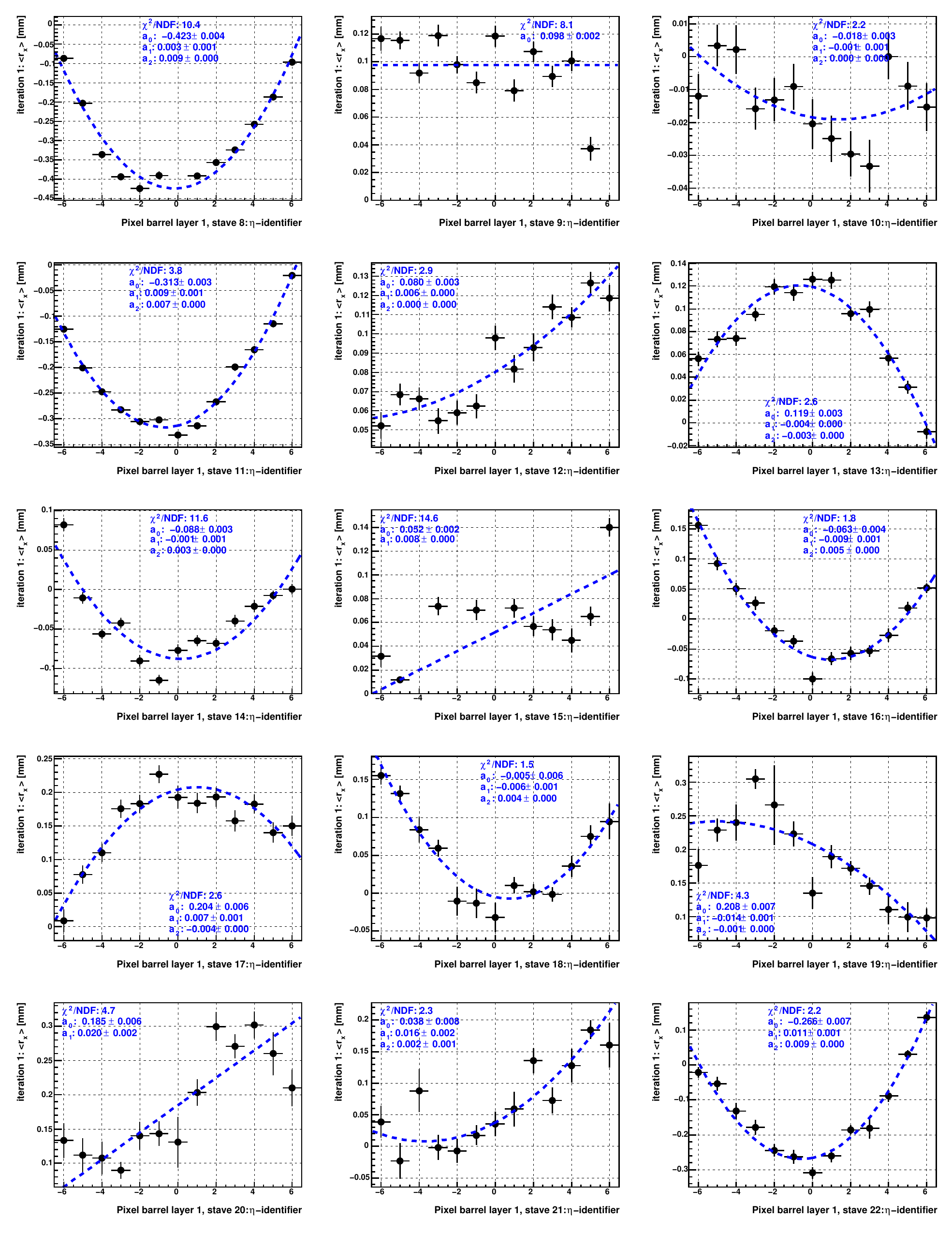}
\end{center}
\vspace{\cDist}
\caption[The $\langle r_x\rangle(\eta)$ distribution before pixel stave bow alignment in M8+ (page~3)]{
The $\rmean x(\eta)$ distribution for staves 31 to 45 of the pixel detector using the full $B$-field off M8+ dataset {\bf before} alignment for pixel stave bow. 
%The results on this plot . 
The fit results with a parabola of the form specified in Equation~\ref{eqn:staveBow} are shown in blue.
}
\end{figure}%\nopagebreak[5]

\begin{figure}
\begin{center}
\includegraphics[width=15.8cm,clip=true]{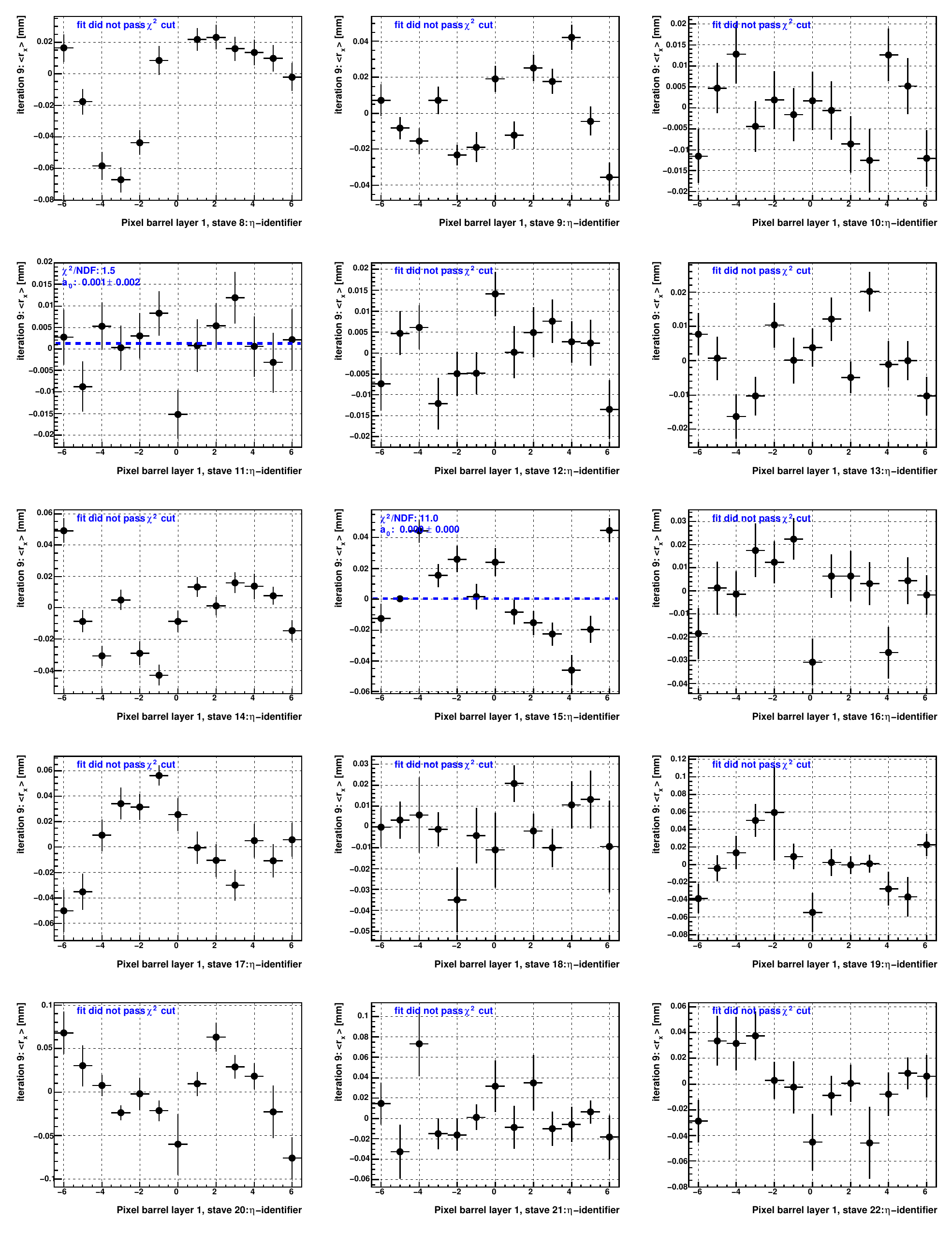}
\end{center}
\vspace{\cDist}
\caption[The $\langle r_x\rangle(\eta)$ distribution after pixel stave bow alignment in M8+ (page~3)]{
The $\rmean x(\eta)$ distribution for staves 31 to 45 of the pixel detector using the full $B$-field off M8+ dataset {\bf after} alignment for pixel stave bow. Note the difference in the $y$-axis scale before and after alignment.
%The results on this plot . 
The fit results with a parabola of the form specified in Equation~\ref{eqn:staveBow} are shown in blue.
}
\end{figure}%\nopagebreak[5]

\begin{figure}
\begin{center}
\includegraphics[width=15.8cm,clip=true]{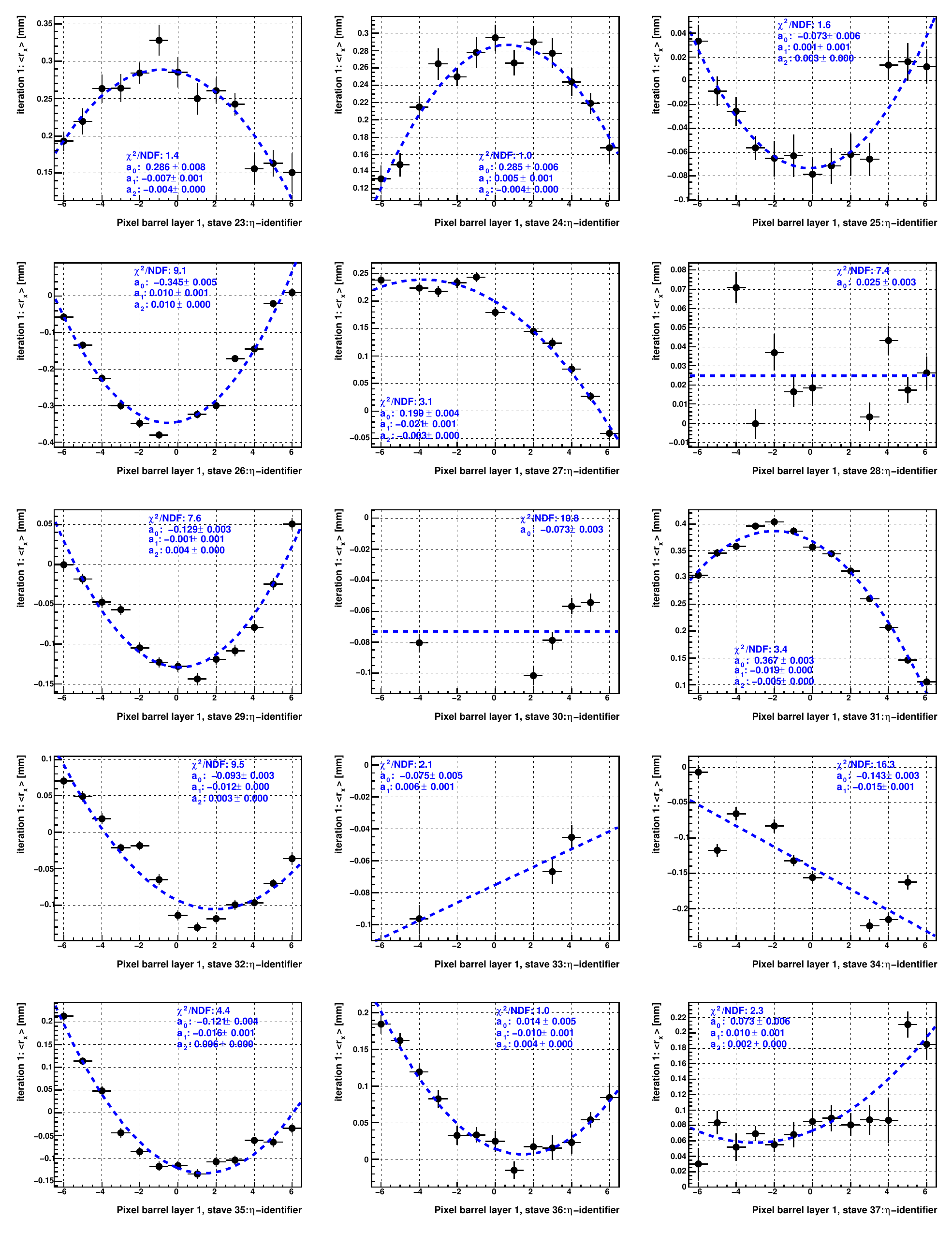}
\end{center}
\vspace{\cDist}
\caption[The $\langle r_x\rangle(\eta)$ distribution before pixel stave bow alignment in M8+ (page~4)]{
The $\rmean x(\eta)$ distribution for staves 46 to 60 of the pixel detector using the full $B$-field off M8+ dataset {\bf before} alignment for pixel stave bow. 
%The results on this plot . 
The fit results with a parabola of the form specified in Equation~\ref{eqn:staveBow} are shown in blue.
}
\end{figure}%\nopagebreak[5]

\begin{figure}
\begin{center}
\includegraphics[width=15.8cm,clip=true]{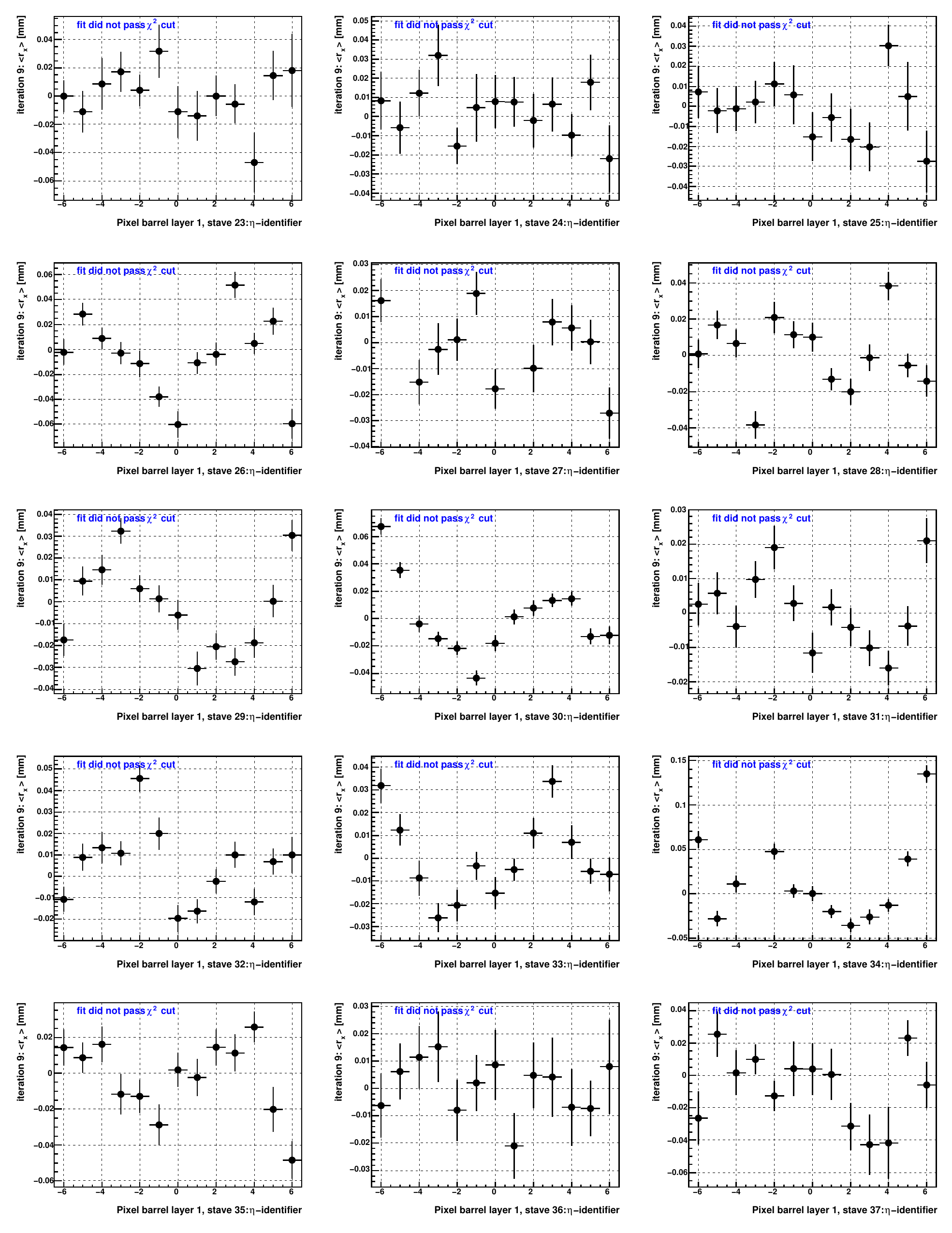}
\end{center}
\vspace{\cDist}
\caption[The $\langle r_x\rangle(\eta)$ distribution after pixel stave bow alignment in M8+ (page~4)]{
The $\rmean x(\eta)$ distribution for staves 46 to 60 of the pixel detector using the full $B$-field off M8+ dataset {\bf after} alignment for pixel stave bow. Note the difference in the $y$-axis scale before and after alignment.
%The results on this plot . 
The fit results with a parabola of the form specified in Equation~\ref{eqn:staveBow} are shown in blue.
}
\end{figure}%\nopagebreak[5]

\begin{figure}
\begin{center}
\includegraphics[width=15.8cm,clip=true]{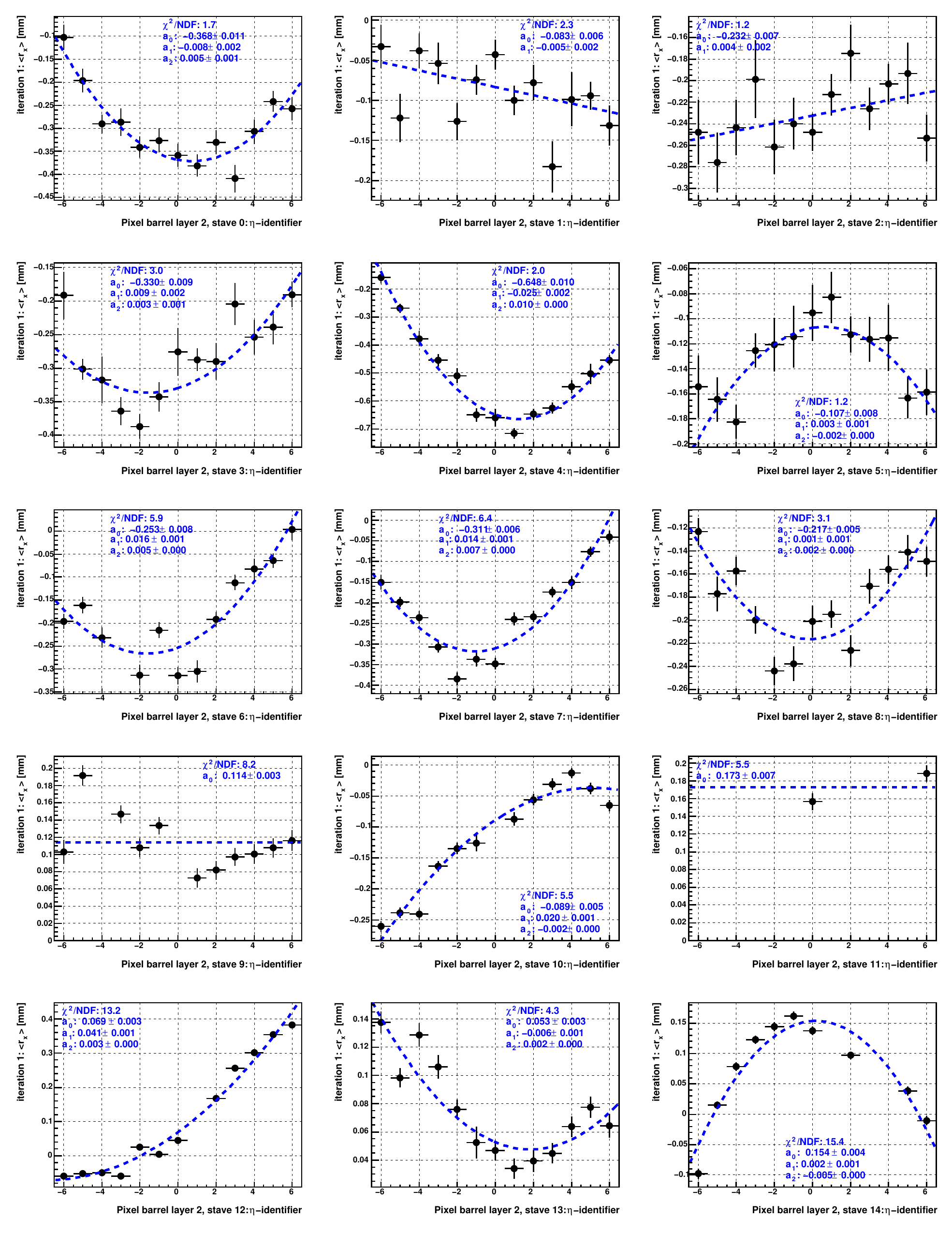}
\end{center}
\vspace{\cDist}
\caption[The $\langle r_x\rangle(\eta)$ distribution before pixel stave bow alignment in M8+ (page~5)]{
The $\rmean x(\eta)$ distribution for staves 61 to 75 of the pixel detector using the full $B$-field off M8+ dataset {\bf before} alignment for pixel stave bow. 
%The results on this plot . 
The fit results with a parabola of the form specified in Equation~\ref{eqn:staveBow} are shown in blue.
}
\end{figure}%\nopagebreak[5]

\begin{figure}
\begin{center}
\includegraphics[width=15.8cm,clip=true]{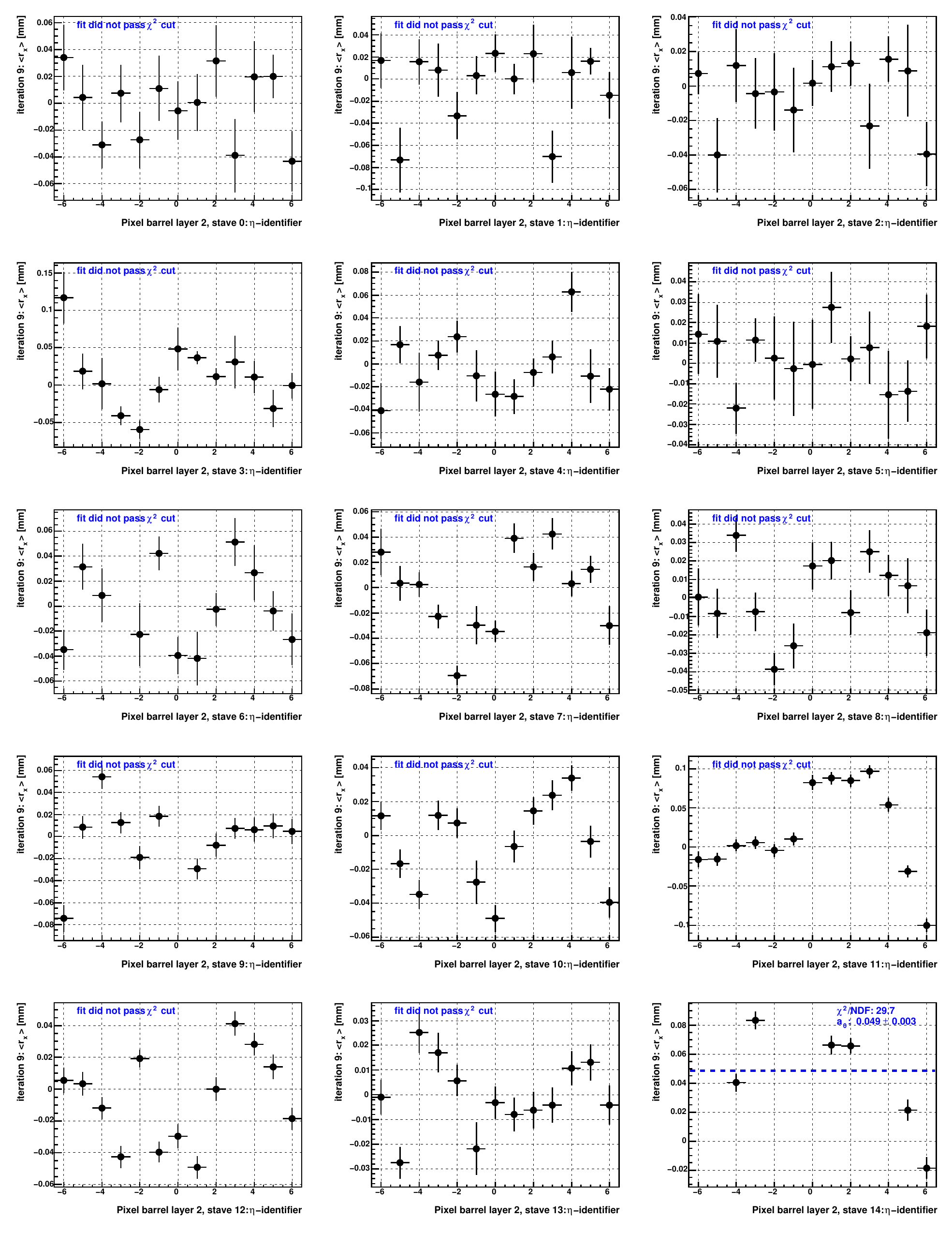}
\end{center}
\vspace{\cDist}
\caption[The $\langle r_x\rangle(\eta)$ distribution after pixel stave bow alignment in M8+ (page~5)]{
The $\rmean x(\eta)$ distribution for staves 61 to 75 of the pixel detector using the full $B$-field off M8+ dataset {\bf after} alignment for pixel stave bow. Note the difference in the $y$-axis scale before and after alignment.
%The results on this plot . 
The fit results with a parabola of the form specified in Equation~\ref{eqn:staveBow} are shown in blue.
}
\end{figure}%\nopagebreak[5]

\begin{figure}
\begin{center}
\includegraphics[width=15.8cm,clip=true]{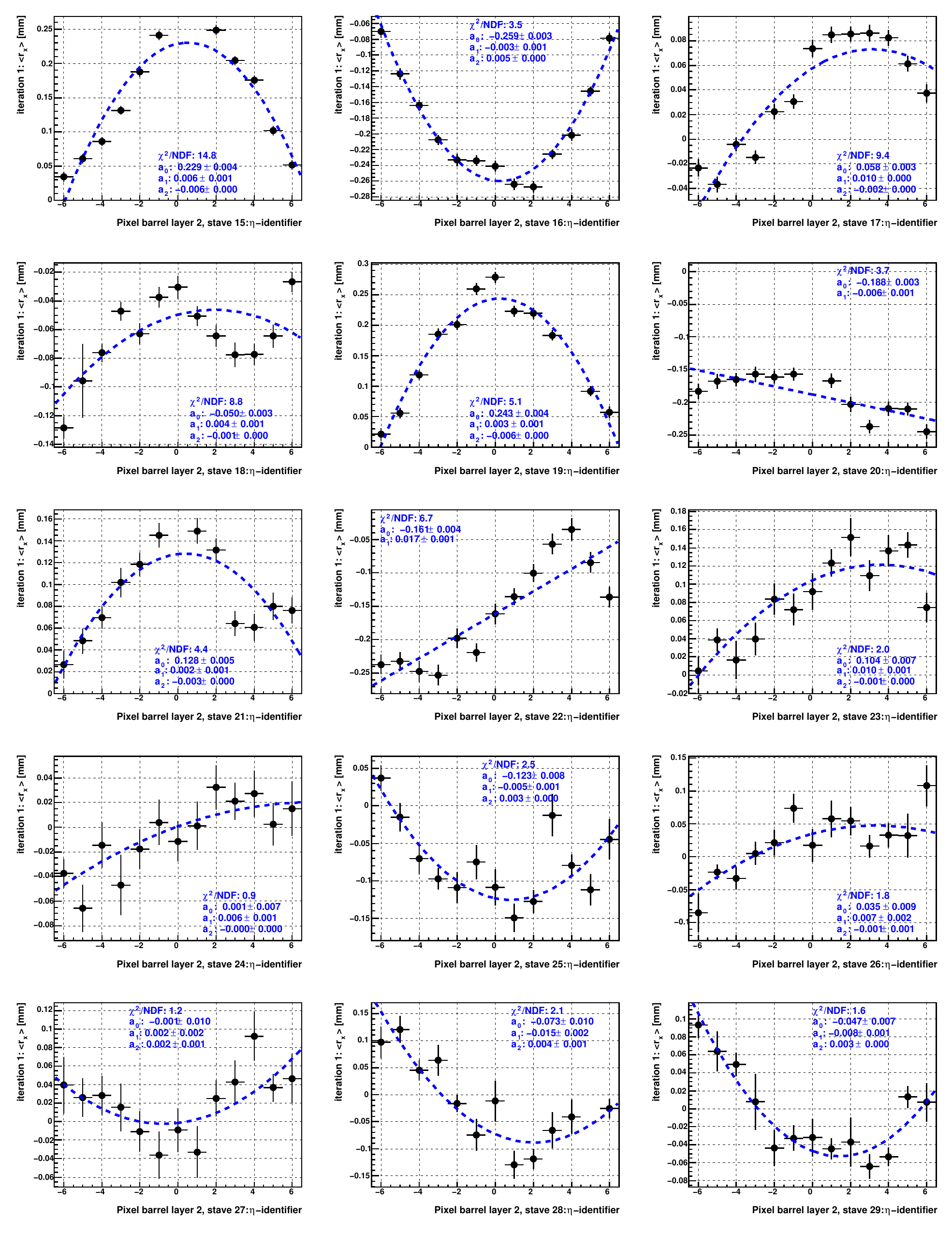}
\end{center}
\vspace{\cDist}
\caption[The $\langle r_x\rangle(\eta)$ distribution before pixel stave bow alignment in M8+ (page~6)]{
The $\rmean x(\eta)$ distribution for staves 76 to 90 of the pixel detector using the full $B$-field off M8+ dataset {\bf before} alignment for pixel stave bow. 
%The results on this plot . 
The fit results with a parabola of the form specified in Equation~\ref{eqn:staveBow} are shown in blue.
}
\end{figure}%\nopagebreak[5]

\begin{figure}
\begin{center}
\includegraphics[width=15.8cm,clip=true]{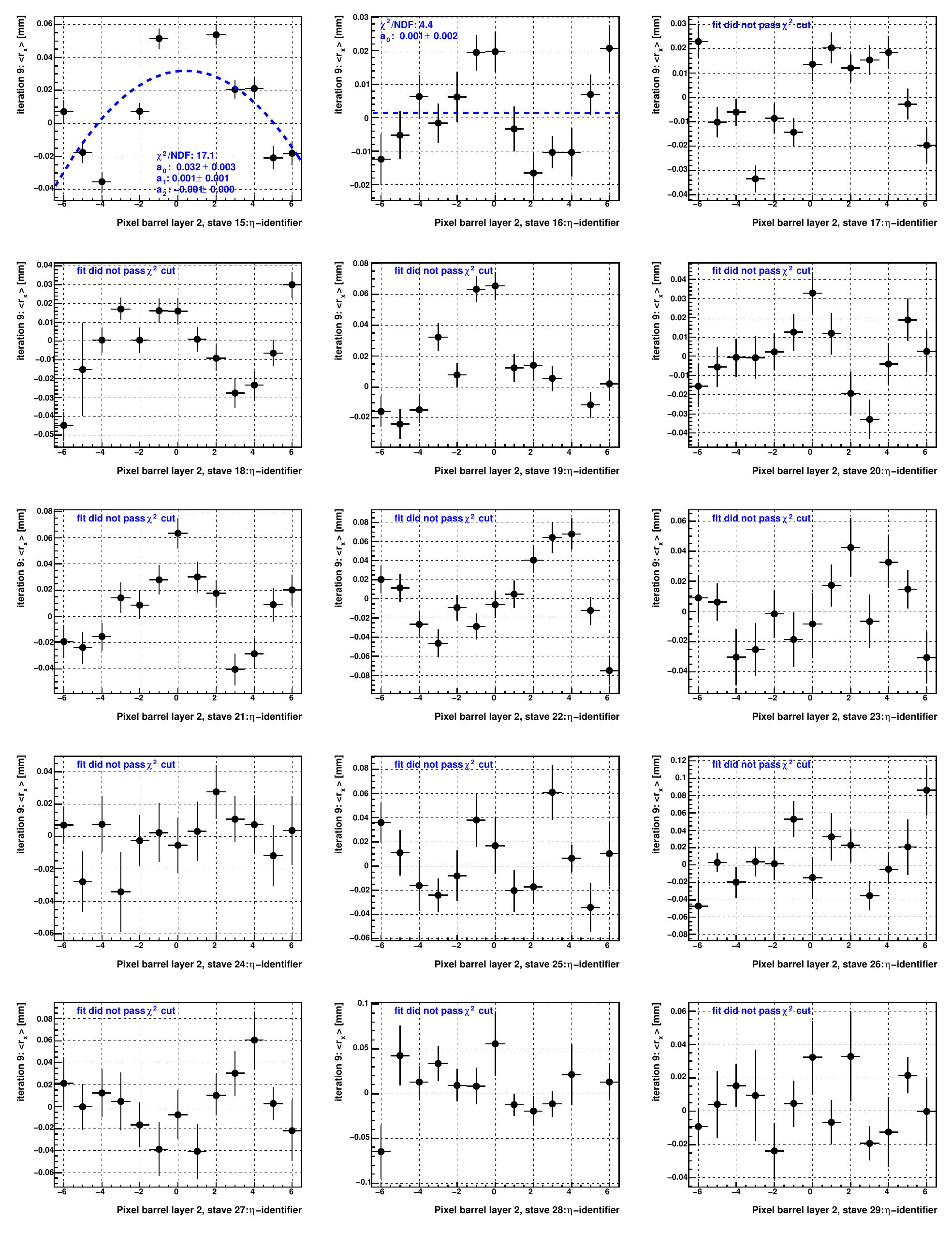}
\end{center}
\vspace{\cDist}
\caption[The $\langle r_x\rangle(\eta)$ distribution after pixel stave bow alignment in M8+ (page~6)]{
The $\rmean x(\eta)$ distribution for staves 76 to 90 of the pixel detector using the full $B$-field off M8+ dataset {\bf after} alignment for pixel stave bow. Note the difference in the $y$-axis scale before and after alignment.
%The results on this plot . 
The fit results with a parabola of the form specified in Equation~\ref{eqn:staveBow} are shown in blue.
}
\end{figure}%\nopagebreak[5]

\begin{figure}
\begin{center}
\includegraphics[width=15.8cm,clip=true]{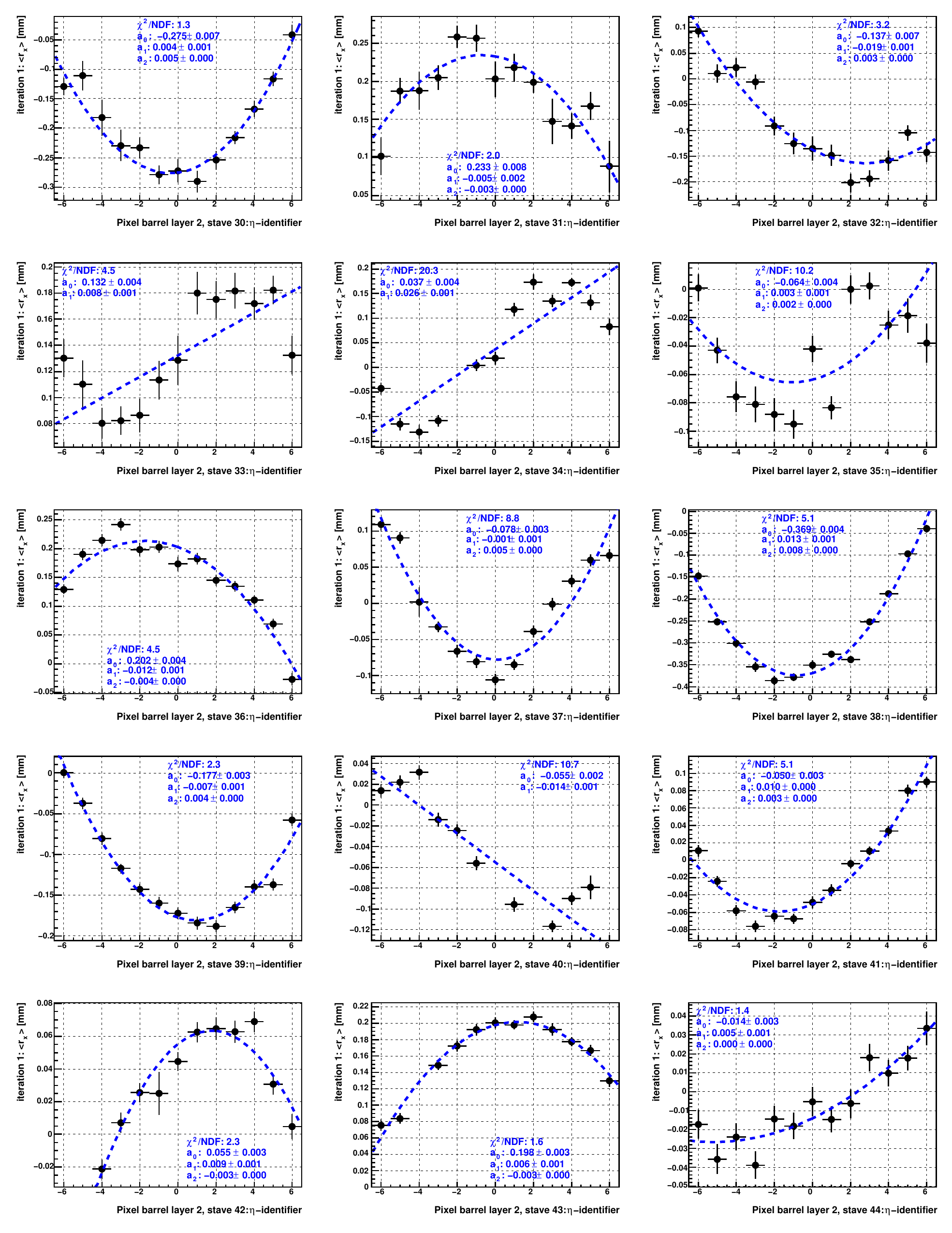}
\end{center}
\vspace{\cDist}
\caption[The $\langle r_x\rangle(\eta)$ distribution before pixel stave bow alignment in M8+ (page~7)]{
The $\rmean x(\eta)$ distribution for staves 91 to 105 of the pixel detector using the full $B$-field off M8+ dataset {\bf before} alignment for pixel stave bow. 
%The results on this plot . 
The fit results with a parabola of the form specified in Equation~\ref{eqn:staveBow} are shown in blue.
}
\end{figure}%\nopagebreak[5]

\begin{figure}
\begin{center}
\includegraphics[width=15.8cm,clip=true]{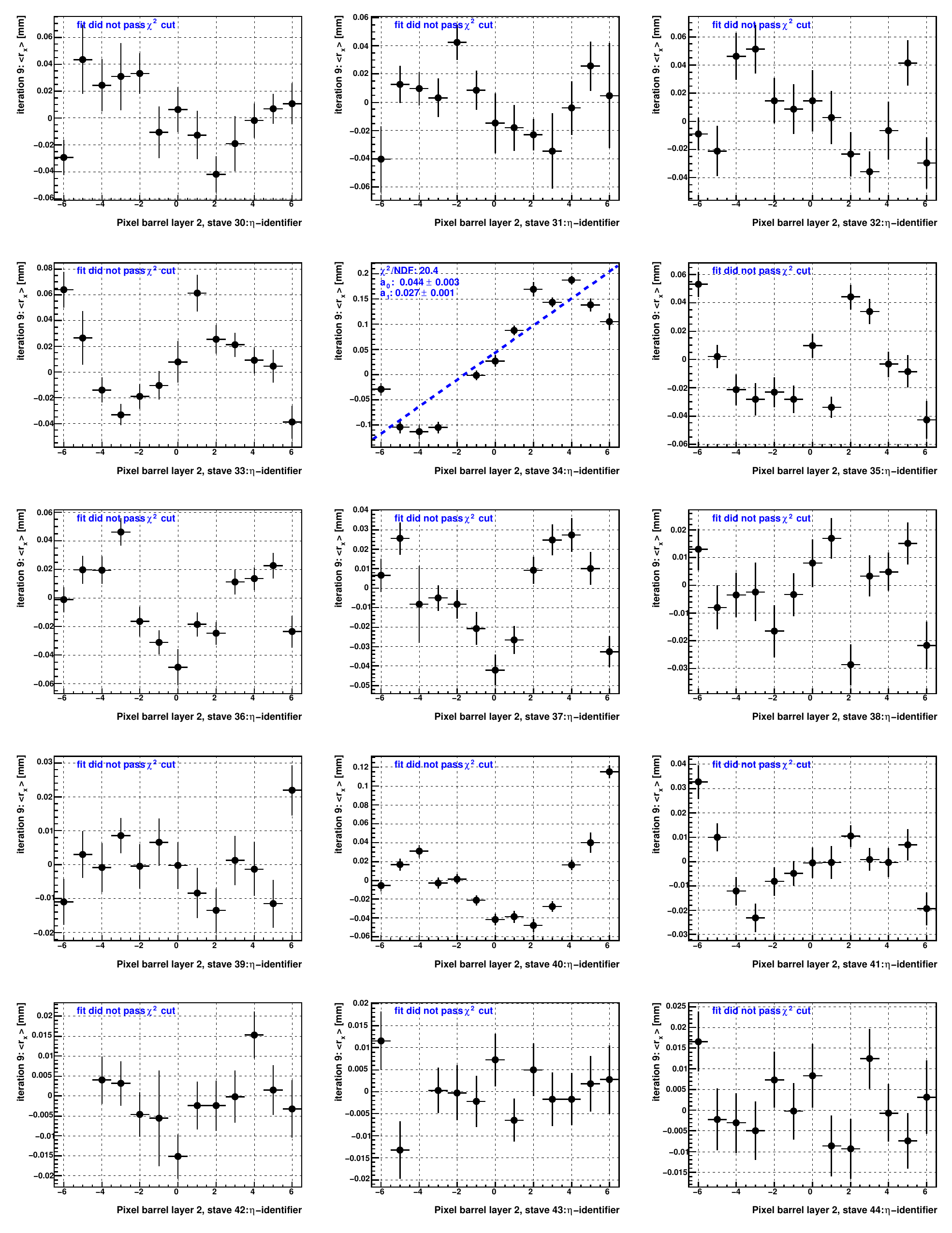}
\end{center}
\vspace{\cDist}
\caption[The $\langle r_x\rangle(\eta)$ distribution after pixel stave bow alignment in M8+ (page~7)]{
The $\rmean x(\eta)$ distribution for staves 91 to 105 of the pixel detector using the full $B$-field off M8+ dataset {\bf after} alignment for pixel stave bow. Note the difference in the $y$-axis scale before and after alignment.
%The results on this plot . 
The fit results with a parabola of the form specified in Equation~\ref{eqn:staveBow} are shown in blue.
}
\end{figure}%\nopagebreak[5]

\begin{figure}
\begin{center}
\includegraphics[width=15.8cm,clip=true]{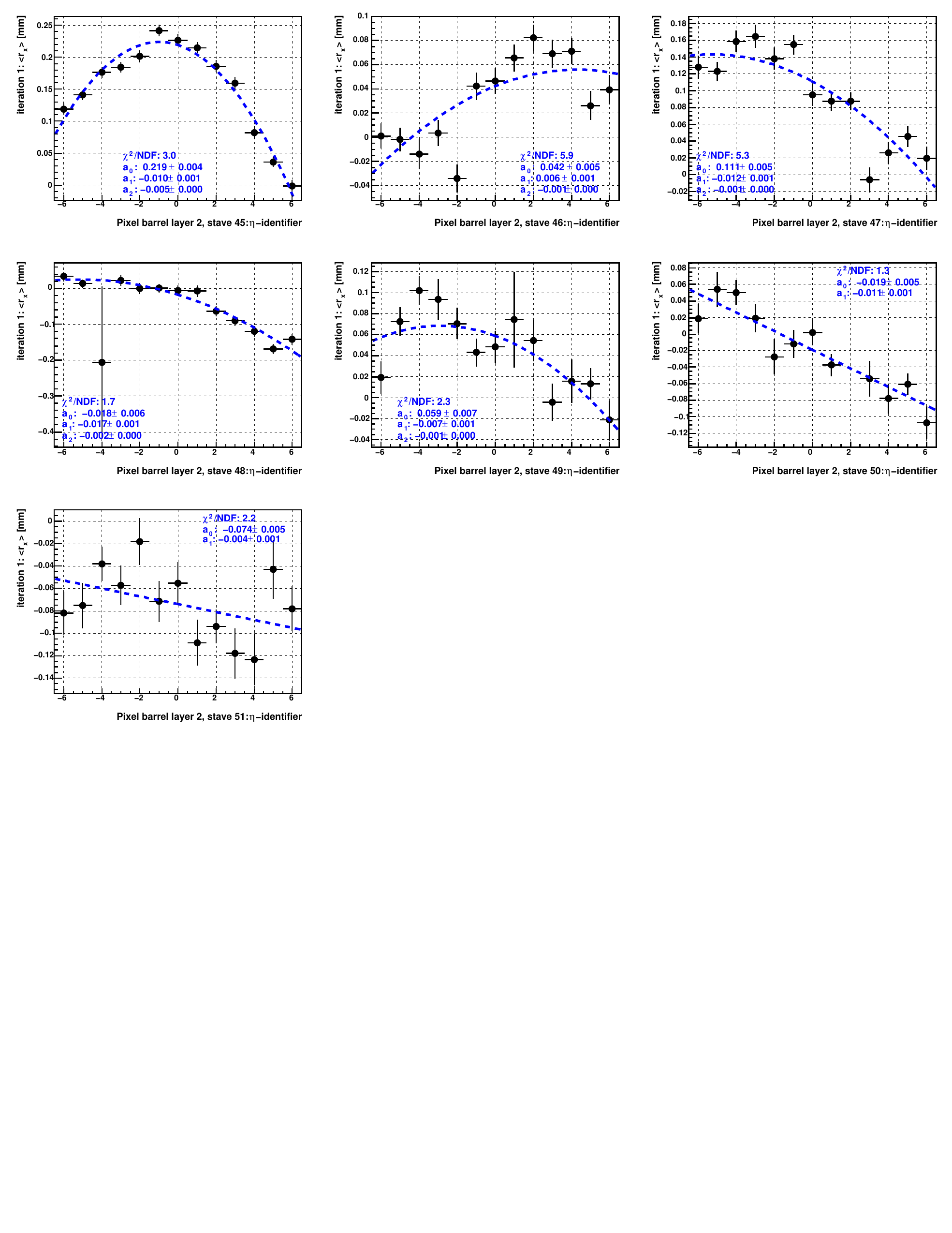}
\end{center}
\vspace{\cDist}
\caption[The $\langle r_x\rangle(\eta)$ distribution before pixel stave bow alignment in M8+ (page~8)]{
The $\rmean x(\eta)$ distribution for staves 106 to 112 of the pixel detector using the full $B$-field off M8+ dataset {\bf before} alignment for pixel stave bow. 
%The results on this plot . 
The fit results with a parabola of the form specified in Equation~\ref{eqn:staveBow} are shown in blue.
}
\end{figure}%\nopagebreak[5]

\begin{figure}
\begin{center}
\includegraphics[width=15.8cm,clip=true]{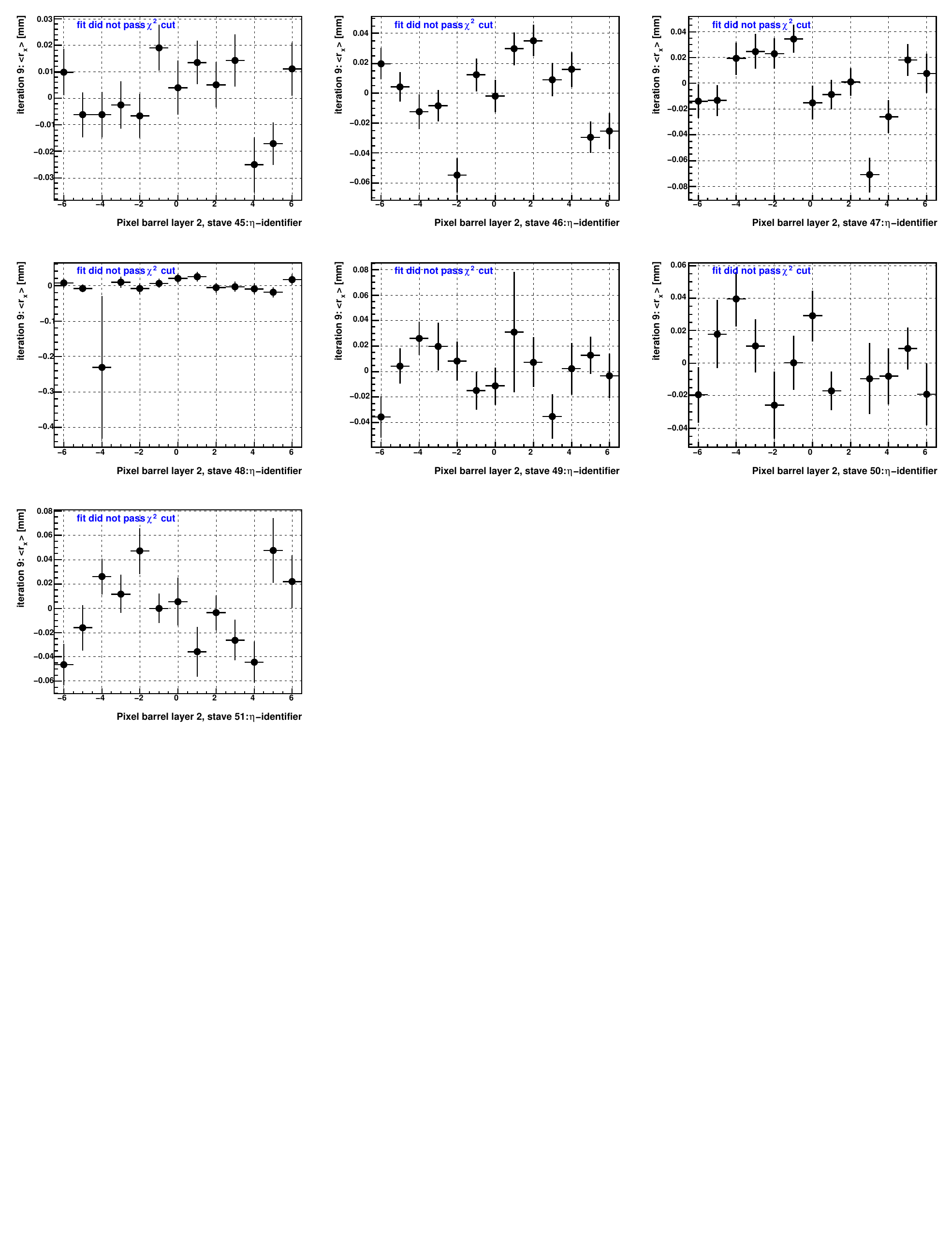}
\end{center}
\vspace{\cDist}
\caption[The $\langle r_x\rangle(\eta)$ distribution after pixel stave bow alignment in M8+ (page~8)]{
The $\rmean x(\eta)$ distribution for staves 106 to 112 of the pixel detector using the full $B$-field off M8+ dataset {\bf after} alignment for pixel stave bow. Note the difference in the $y$-axis scale before and after alignment.
%The results on this plot . 
The fit results with a parabola of the form specified in Equation~\ref{eqn:staveBow} are shown in blue.
}
\end{figure}%\nopagebreak[5]